RÉPUBLIQUE FRANÇAISE
*Liberté*
*Égalité*
*Fraternité*

Université de Toulouse

# Doctorat de l'Université de Toulouse

## Nonlocal Games Through Communication Complexity and Quantum Cryptography

Thèse présentée et soutenue, le 9 juillet 2025 par

**Pierre BOTTERON**

**École doctorale**
EDMITT - Ecole Doctorale Mathématiques, Informatique et Télécommunications de Toulouse

**Spécialité**
Mathématiques et Applications

**Unité de recherche**
IMT : Institut de Mathématiques de Toulouse

**Thèse dirigée par**
Clément PELLEGRINI et Ion NECHITA

**Composition du jury**
M. Salman BEIGI, Rapporteur, Institute for Research in Fundamental Sciences
M. Simon PERDRIX, Rapporteur, LORIA, INRIA, Université de Lorraine
M. Marco TOMAMICHEL, Rapporteur, CQT, National University of Singapore
Mme Mirjam WEILENMANN, Examinatrice, INRIA, Télécom Paris
M. Pierre PUJOL, Examinateur, LPT, Université de Toulouse
M. Clément PELLEGRINI, Directeur de thèse, IMT, Université de Toulouse
M. Ion NECHITA, Co-directeur de thèse, LPT, CNRS, Université de Toulouse

**Membres invités**
Mme Anne BROADBENT, Co-encadrante de thèse, Université d'Ottawa

# Summary (English)

This thesis explores foundational aspects of quantum information theory and quantum cryptography.

First, we investigate quantum correlations in interactive settings, including the CHSH and graph isomorphism games. We aim to distinguish quantum correlations from non-signaling correlations by leveraging the principle of communication complexity. To this end, we employ techniques such as distributed computation, majority-function-based distillation protocols, the algebraic and geometric properties of nonlocal box wirings, and variations of some graph properties such as isomorphism, transitivity, and equitable partitions. This inquiry advances our understanding of non-physical correlations.

Second, we address a key open problem in cryptography: the feasibility of unclonable encryption. We aim to construct an encryption scheme that prevents two distant parties from simultaneously obtaining information about a shared encrypted message. We introduce a candidate for unclonable encryption in the plain model, *i.e.* without assumptions, in working towards an unconditional proof. Our protocol is based on Clifford algebra, utilizing complex Hermitian unitary matrices that anti-commute. For small key sizes, we rigorously prove security using sum-of-squares methods, while for larger key sizes, we provide strong numerical evidence via the NPA hierarchy.

# Résumé (French)

Cette thèse explore des aspects fondamentaux de la théorie de l'information quantique et de la cryptographie quantique.

D'une part, nous étudions les corrélations quantiques dans des contextes interactifs, notamment les jeux de CHSH et d'isomorphisme de graphes. Notre objectif est de distinguer les corrélations quantiques des corrélations non-signalantes en nous appuyant sur le principe de complexité de la communication. Pour cela, nous utilisons des techniques telles que le calcul distribué, l'amplification de biais grâce à la fonction majorité, les propriétés algébriques et géométriques des câblages de boîtes non-locales, ainsi que des variantes de certaines propriétés de graphes comme l'isomorphisme, la transitivité et les partitions équitables. Cette étude fait progresser notre compréhension des corrélations non-physiques.

D'autre part, nous abordons un problème ouvert majeur en cryptographie : la faisabilité du chiffrement non-clonable. Notre objectif est de construire un schéma de chiffrement qui empêche deux réceptionneurs distants l'un de l'autre d'obtenir simultanément de l'information sur un message chiffré partagé. Nous introduisons un candidat au chiffrement non-clonable dans le modèle standard, c'est-à-dire sans hypothèse, en vue d'obtenir une preuve inconditionnelle de la sécurité. Notre protocole repose sur l'algèbre de Clifford et utilise des matrices unitaires hermitiennes à coefficients complexes qui anti-commutent. Pour des tailles de clés réduites, nous prouvons rigoureusement la sécurité à l'aide de méthodes de sommes de carrés, tandis que pour des tailles de clés plus grandes, nous fournissons des validations numériques solides via la hiérarchie NPA.

# List of Manuscripts

Here are the author's manuscripts included in this thesis:

**Manuscript 1 —** Pierre Botteron, Anne Broadbent, and Marc-Olivier Proulx. "Extending the Known Region of Nonlocal Boxes that Collapse Communication Complexity". In: *Physical Review Letters* 132 (Feb. 2024), p. 070201. DOI: 10.1103/PhysRevLett.132.070201.

**Manuscript 2 —** Pierre Botteron, Anne Broadbent, Reda Chhaibi, Ion Nechita, and Clément Pellegrini. "Algebra of Nonlocal Boxes and the Collapse of Communication Complexity". In: *Quantum* 8 (July 2024), p. 1402. ISSN: 2521-327X. DOI: 10.22331/q-2024-07-10-1402.

**Manuscript 3 —** Pierre Botteron and Moritz Weber. *Communication Complexity of Graph Isomorphism, Coloring, and Distance Games*. 2024. arXiv: 2406.02199 [quant-ph].

**Manuscript 4 —** Pierre Botteron, Anne Broadbent, Eric Culf, Ion Nechita, Clément Pellegrini, and Denis Rochette. *Towards Unconditional Uncloneable Encryption*. 2024. arXiv: 2410.23064 [quant-ph].

# Dedication

## *To my supervisors*

First of all, I am profoundly grateful to my advisory team—Dr. Anne Broadbent, Dr. Ion Nechita, and Dr. Clément Pellegrini—without whose guidance this work would not have been possible. Our international collaboration enriched me in countless ways, offering a diversity of expertise, perspectives, methodologies, and networks.

Together, you demonstrated unwavering patience, clear mentorship, and invaluable advice—qualities essential to the completion of this work.

Individually, each of you has influenced me beyond the scientific domain. Anne, thank you for exemplifying the art of collaboration, for teaching me the social aspects of research, and for your generous spirit in every interaction. Clément, thank you for managing the organizational challenges, for being involved at every stage of my degree, and for unifying our discussions with your insightful approach. Ion, thank you for inspiring me with your efficiency, for your kind and encouraging words, and for your committed engagement in all our projects.

I am truly grateful for the growth and learning I have experienced during this journey.

## *To the researchers who supported me*

First, I wish to express my gratitude to Dr. Moritz Weber. Since our meeting during the Focus Semester in Saarbrucken, I have greatly enjoyed our collaborative project. Your support felt like having an extra supervisor by my side. Moreover, I fondly recall the unique collection of teas in your office and the many delightful songs you produced—one even in French!

I also extend my thanks to Dr. Reda Chhaibi. Our coding sessions in your office were inspiring—I appreciated your guidance on creating a clean Python package, and I fondly remember your humorous coffee room jokes.

My sincere thanks also go to Dr. Guillaume Aubrun and Dr. Francesco Costantino for participating in the “comité de suivi de thèse” and providing valuable advice that helped my thesis flourish.

Finally, I am grateful to the many researchers who spent time discussing various aspects of research life with me: Dr. Jean-Daniel Bancal, Dr. Andreas Bluhm, Dr. Tristan Benoist, Dr. Michael Brannan, Dr. Benoît Collins, Dr. Jason Crann, Dr. Mariami Gachechiladze, Dr. Maria Jivulescu, Dr. Cécilia Lancien, Dr. Leevi Leppajarvi, Dr. Faedi Loulidi, Dr. Arthur Mehta, Dr. Roberto H. Palomares, Dr. Connor Paddock, Dr. Sang-Jun Park, Dr. Denis Rochette, and Dr. Mirjam Weilenmann.

## *To my colleagues*

I extend my sincere thanks to my colleagues, whose camaraderie and lively discussions made this journey all the more enjoyable. I especially liked our conversations about quirky arXiv papers and math-inspired jokes.

From the TIQ-TOQS group in Toulouse, I thank Aabhas, Andreina, Anna, Arnaud, Denis, Faedi, Gian Luca, Jan Luka, Kieran, Laxmi-Prasad, Linda, Miao, Reshmi, Sang-Jun, Tristan B., and Tristan K.

From the QUASAR group in Ottawa, I thank Allan, Arthur, Bennett, Connor, Daniel, Denis, Eric, Jason, Joshua, Jyoti, Laura, Martti, Monica, Nagisa, Omar, Oren, Peter, Sébastien, Sherry, Sohrab, Upendra, and Yasin.

From the Focus Semester group in Saarbrucken, I thank Adina, Akihiro, Alexander, Atsuya, Ayesha, Håkon, Jennifer, Junichiro, Katsunori, Manuel, Nagisa, Nina, Pádraig, Roberto, Sherry, and the local participants.

Finally, I am also grateful to Ayoub, Butian, Cong, Élisa, Flore, Florian, Jiaqi, Jonathan, Mathis, Max, Médard, Moussadek, Nathanaël, Nicolas, Paul, Prabhav, Ravi, and Robin, with whom I hang out at conferences or in the lab.

## *To my friends*

I am deeply grateful to all my friends who supported me throughout this thesis. While many of you reacted with amusement when I attempted to explain my research topic, you nonetheless understood the challenges of this journey and provided countless moments of joy and entertainment outside the office.

From my music band, AwaCœurs, I thank Péky and Samuel. I truly cherished our weekly gatherings and the meaningful time we spent together on various projects.

From Toulouse Ouest, I thank Abraham, Adriel, Albertine, Alicia, Alizée, Andriana, Angèle, Anne, Aristide, Armandine, Arnaud, Bernadette, Bunnary, Carmen, Chan, Chantal, Charlène, Charly, Céline, Clifford, Colombe, Constant, Corine, Corinne, Cynthia, Darline, David, Delphine, Denis, Diego, Dominique, Edilenespo, Élisabeth, Éliam, Éliora, Élodie, Emmanuel, Emmanuella, Éric, Esméralda, Esteban, Esther, Éthan, Eugénie, Exaucée, Eyram, Filipe, Florian, Françoise, Gaby, Gaëlle, Geneviève, Gertrude, Gilbert, Havila, Héléna, Henri-Pierre, Hermine, Issifou, Jacob, Jamila, Jean-Barthélémy, Jean-Calvin, Jean-Lou, Jean-Paul, Jean-Philippe, Jérémy, Jessica, Joël B., Joël D., John, Jonathan, Josaï, Joseph, Joseph-Yvan, Julia, Justine, Kevin, Laetitia, Lahatra, Laurent, Léa, Lémis, Lesly, Leyna, Louise, Luc, Lucía, Magali, Marlène, Marie, Marie-Jeanne, Marion, Max, Mégane, Merry, Merveille, Mikaël, Michelène, Myriam, Narcisse, Natàlia, Nicole, Niouma, Noah, Olivia, Olivier, Patrick, Péky, Phanuelle, Philippe, Rachel, Raphaël, Rémy, Rodrigue, Rose-Marie, Sabine, Sacha, Samara, Sarah, Sidonie, Stéphane, Stéphanie, Suzanne, Sylvie, Tsiory, Valérie, Vanny, Victor, Victoria, Viarine, Violaine, Wilky, Yohan, Yon, and Yuli. A special mention to the young adults' group, the "groupe de maison," and the music band, with whom I shared many wonderful moments.

From Bethel Ottawa, I thank Ben, Dotun, Enoch, Ezra, Fabrice, Fabiola, Jaydon, Kamoi, Koffi, Raphael, Sheril, Tasha, and Véronique. A special thanks to the "Pieds poudrés" group.

From Saarbrücken, I thank Aurélie, Andreas, Benjamin, Concetta, David, Elya, Evens, Josia, Manuella, Mélissa, Mika, Silas, Yannick, Yoshua, and Yona.

From Chicago, I thank Alvaro, Amaranta, Andy, Angel, Araceli, Blanca, Chuy, Elian, Emmily, Ezechiel, Iliana, Isabel, Israel, Jorge, Jose, Katherine, Martha, Mauro, Miguel, Miriam, Moises, Naty, Norma, Pamela, Rachel, Ramon, Samantha, Santiago, and Vidalia.

From the extended group of former "Licence Parcours Spécial" students, I thank Adrian, Élohan, Elsa, Grégoire, Julien, Malou, Nicolas, Olivier, and Paul.

From the "Vauquelinois" group, I thank Arnaud, Carolina, Gaspard,

Guillaume, Mathilda, Ritchie, and Sara.

And of course, I do not forget to thank Allan, Aurore, Barbara, Chahinez, Chloé, Christivie, Eunice, Ézéchiel, Florian, Julie, Michaël, Miché, Steyvan, and Thomas.

## *To my family*

Last but not least, I want to express my deepest gratitude to my closest relatives: my father Jean-Marc, my mother Laurence, and my siblings, Sarah and Samuel. Thank you for your unwavering support, for believing in me, and for being an integral part of this journey.

I am also grateful to Christine, Claude, Fanny-Anne, François, Françoise, Hugo, Jordan, Marie, Maryse, Matthias, Michaël, Michel, Mireille, Olivier, Sylvain, and Yvan.

*Thank you all for who you are to me!*

# Graphical Contents

## Manuscript Map

We illustrate in Figure 1 the dependencies between chapters.

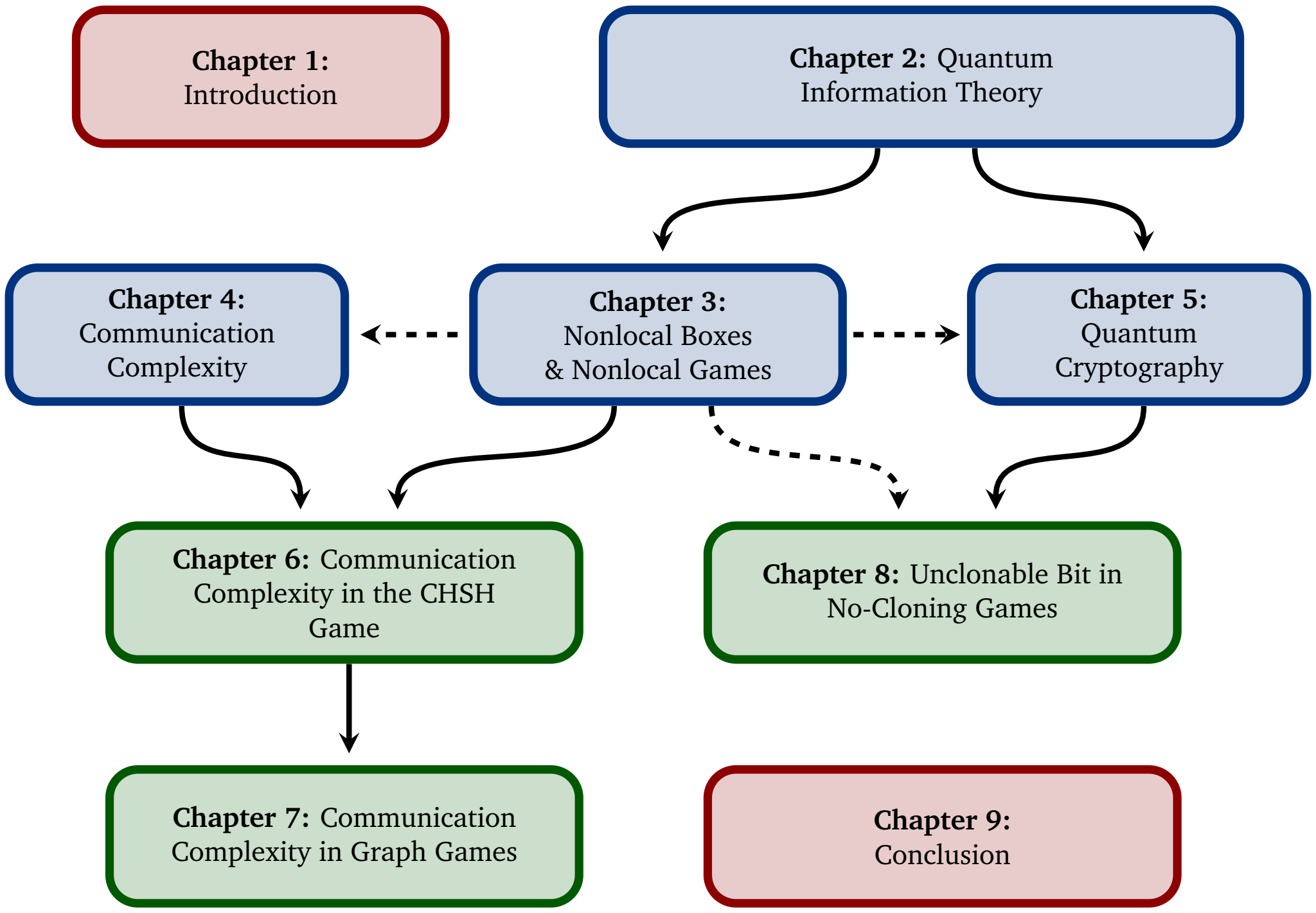

**Figure 1 —** *Solid arrows indicate strong connections, while dashed arrows represent more subtle links. Following the introductory chapter (Chapter 1, shown in red), this manuscript consists of two main parts. The first part comprises four background chapters (Chapters 2 to 5, shown in blue), which provide the necessary foundations and related topics but do not present our results. The second part contains three contribution chapters (Chapters 6 to 8, shown in green), where we present our four manuscripts. Finally, we conclude with perspectives and discussions in Chapter 9 (shown in red).*

## How To Read this Thesis

We suggest two reading approaches depending on the reader's familiarity with the concepts presented.

For readers new to quantum information theory, we recommend a linear reading of this manuscript, as each chapter builds on previous definitions, following the dependencies outlined in the diagram above.

For readers already familiar with the background material, it is possible to skip the background chapters (blue) after reading the introduction. They can proceed directly to the contribution chapters (green), where references to relevant background concepts are provided as needed.

# Contents

# Lists of Symbols and Acronyms

## General Mathematics

| Symbol | Description | Page |
|---|---|---|
| $\mathbf{0}$ | Zero matrix or vector | 27 |
| $\mathbb{1}_C$ | Indicator function, taking value $1$ *if, and only if,* condition $C$ is satisfied, and $0$ otherwise | 59 |
| $\oplus$ | Sum modulo $2$ | 67 |
| $\otimes$ | Tensor product | 29 |
| $\|\alpha\|$ | Absolute value, complex modulus | 26 |
| $A^\top$ | Transposition of a matrix $A$ | 50 |
| $A^*$ | Adjoint or trans-conjugate of a complex matrix $A$ | 27 |
| $\mathrm{Tr}$ | Trace, sum of the eigenvalues | 27 |
| $\mathbb{N}$ | Set of natural numbers $\{0, 1, 2, \dots\}$ | 31 |
| $\mathbb{Z}$ | Ring of integer numbers | 93 |
| $\mathbb{R}$ | Field of real numbers | 28 |
| $\mathbb{C}$ | Field of complex numbers | 26 |
| $\mathcal{M}_d(\mathbb{C})$ | Space of $d \times d$ complex matrices | 27 |
| $\mathbb{I}$ | Identity matrix or identity operator | 28 |
| $\lambda_i$ | Eigenvalue | 28 |
| $\mathcal{H}, \mathcal{K}$ | Generic Hilbert spaces | 26 |
| $\mathcal{B}(\mathcal{H})$ | Space of bounded operators on $\mathcal{H}$ | 27 |
| $\mathbb{P}$ | Probability measure | 42 |
| $\mathbb{E}$ | Expectation | 80 |
| $\mathbb{V}$ | Variance | 85 |
| $\mathfrak{S}_k$ | Set of permutations of $\{1, .., k\}$ | 63 |
| $\sigma_x, \sigma_y, \sigma_z$ | Pauli matrices | 28 |
| $\mathrm{Conv}$ | Convex hull | 67 |
| PSD | Positive semi-definite | 27 |
| SDP | Semi-definite programming | 72 |
| SoS | Sum-of-square decomposition | 72 |

# Quantum Information Theory (Chapter 2)

| Symbol | Description | Page |
|---|---|---|
| A, B, C, E, P, R | Parties' generic name: Alice, Bob, Charlie, Eve, Pirate, Referee | 30 |
| $d$ | Dimension of the Hilbert space | 26 |
| $n$ | Number of parties, number of tensor components in $\mathcal{H}$ | 29 |
| $\lvert\psi\rangle, \lvert\varphi\rangle$ | Generic pure states, "kets $\psi$ and $\varphi$" | 26 |
| $\langle\psi\rvert$ | Adjoint of $\lvert\psi\rangle$, "bra $\psi$" | 27 |
| $\lvert 0\rangle, \lvert 00\rangle$ | Ket-0 and ket-00 | 26 |
| $\lvert\Omega\rangle$ | Maximally entangled state, "ket $\Omega$" (pure version) | 32 |
| $\lvert\mathrm{GHZ}\rangle$ | Greenberger–Horne–Zeilinger state [GHZ89] | 32 |
| $\lvert\mathrm{W}\rangle$ | W state, $\big(\lvert 001\rangle + \lvert 010\rangle + \lvert 100\rangle\big)/\sqrt{3}$ | 32 |
| $\mathcal{D}(\mathcal{H})$ | Set of density matrices over $\mathcal{H}$, mixed states | 27 |
| $\rho, \sigma$ | Generic mixed states, density matrices | 27 |
| $\rho^{\Gamma}$ | Partial transposition of $\rho$ | 38 |
| $\omega$ | Maximally entangled state (mixed version) | 33 |
| $\Phi$ | Generic quantum channel | 50 |
| CPTP | Completely-positive trace-preserving (linear) map | 50 |
| LOCC | Local operations and classical communication | 35 |
| MoE | Monogamy-of-entanglement | 39 |
| PVM | Projection-valued measure | 44 |
| POVM | Positive operator-valued measure | 45 |
| QIT | Quantum Information Theory | 25 |

## Nonlocal Boxes (Section 3.1)

| Symbol | Description | Page |
|---|---|---|
| $n$ | Number of parties | 58 |
| $N$ | Number of inputs | 58 |
| $M$ | Number of outputs | 58 |
| $\mathbf{P}, \mathbf{Q}, \mathbf{R}$ | Generic nonlocal boxes | 59 |
| $\mathbf{P_{00}}$ | Deterministic box, always outputting $(0, 0)$ | 59 |
| $\mathbf{P_{11}}$ | Deterministic box, always outputting $(1, 1)$ | 59 |
| **SR** | Shared-randomness box, outputting $(a, b)$ s.t. $a = b$ | 60 |
| $\mathbf{I}$ | Fully mixed box, outputting uniformly random tuples $(a, b)$ | 60 |
| $\mathbf{P}_{\mathbf{L}}^{\alpha,\beta,\gamma,\delta}$ | Extreme local box | 67 |
| $\mathbf{P}_{\mathbf{NL}}^{\alpha,\beta,\gamma}$ | Extreme nonlocal box | 67 |
| **PR** | Popescu–Rohrlich box [PR94] | 63 |
| $M_{\mathbf{P}}$ | Correlation table of **P** | 66 |
| W | Generic wiring between nonlocal boxes | 77 |
| $\mathbf{P} \boxtimes_{\mathsf{W}} \mathbf{Q}$ | Nonlocal box obtained from wiring **P** and **Q** by W | 77 |
| $\mathcal{W}$ | Set of wirings | 80 |

## Correlation Sets (Section 3.1.1)

| Symbol | Description | Page |
|---|---|---|
| $\mathcal{L}_{\text{det}}$ | Set of deterministic correlations | 59 |
| $\mathcal{L}$ | Set of local/classical correlations | 60 |
| $\mathcal{Q}_{\text{finite}}$ | Set of finite quantum correlations | 60 |
| $\mathcal{Q}_{\text{infinite}}$ | Set of infinite quantum correlations | 62 |
| $\mathcal{Q}$ | Set of quantum (tensor) correlations | 60 |
| $\mathcal{Q}_c$ | Set of quantum commuting correlations | 62 |
| $\widetilde{\mathcal{Q}}$ | Set of almost quantum correlations | 63 |
| $\mathcal{NL}$ | Set of non-signaling correlations “above” the CHSH hyperplane | 71 |
| $\mathcal{NS}_{\text{corr}}$ | Set of non-signaling correlators | 70 |
| $\mathcal{NS}$ | Set of non-signaling correlations | 63 |
| NPA | Navascués–Pironio–Acín hierarchy [NPA07; NPA08] | 72 |

## Nonlocal Games (Section 3.2)

| Symbol | Description | Page |
|---|---|---|
| $\mathcal{A}$ | Set of answers $a$ on Alice's side | 87 |
| $\mathcal{B}$ | Set of answers $b$ on Bob's side | 87 |
| $\mathcal{X}$ | Set of questions $x$ on Alice's side | 87 |
| $\mathcal{Y}$ | Set of questions $y$ on Bob's side | 87 |
| $\pi(x, y)$ | Probability distribution of the questions $(x, y)$ | 87 |
| $\mathscr{V}(a, b, x, y)$ | Rule or predicate of the game | 87 |
| $\mathsf{G}$ | Generic nonlocal game | 87 |
| $\mathfrak{w}(\mathsf{G})$ | Value of the game $\mathsf{G}$, best winning probability | 89 |
| $\mathcal{S}$ | Generic strategy | 87 |
| CHSH | Clauser–Horne–Shimony–Holt game [CHSH69] | 90 |

## Graph Theory (Section 3.2.3)

| Symbol | Description | Page |
|---|---|---|
| $\sim$ | Adjacency relation of vertices | 95 |
| $\not\simeq$ | Non-equality-and-non-adjacency relation | 95 |
| $\cong$ | Graph isomorphism | 96 |
| $\mathcal{G}, \mathcal{H}$ | Generic graphs | 94 |
| $A_{\mathcal{G}}$ | Adjacency matrix of $\mathcal{G}$ | 96 |
| $E(\mathcal{G})$ | Edge set of $\mathcal{G}$ | 229 |
| $V(\mathcal{G})$ | Vertices set of $\mathcal{G}$ | 94 |
| $\mathcal{G}^c$ | Complement graph of $\mathcal{G}$ | 228 |
| $\mathcal{K}_M$ | Complete graph with $M$ vertices | 98 |
| $\mathcal{C}_M$ | Cycle graph with $M$ vertices | 219 |
| $\mathcal{P}_M$ | Path graph with $M$ vertices | 219 |
| $\mathcal{G} \to \mathcal{H}$ | Graph homomorphism from $\mathcal{G}$ to $\mathcal{H}$ | 98 |
| $\operatorname{diam}(\mathcal{G})$ | Diameter of the graph $\mathcal{G}$, *i.e.* larger *finite* distance between two vertices | 219 |

## Communication Complexity (Chapter 4)

| Symbol | Description | Page |
|---|---|---|
| CC | Communication complexity | 116 |
| $X$ | String $X = (x_1, .., x_n) \in \{0,1\}^n$ | 116 |
| $Y$ | String $Y = (y_1, .., y_m) \in \{0,1\}^m$ | 116 |
| $f$ | Boolean function $f : \{0,1\}^n \times \{0,1\}^m \to \{0,1\}$ | 116 |
| $\mathfrak{p}$ | Universal probability value for the collapse of CC | 122 |
| $\mathrm{IP}_n$ | Inner product function | 122 |
| $\mathrm{EQ}_n$ | Equality function | 121 |

## Cryptography (Chapter 5)

| Symbol | Description | Page |
|---|---|---|
| Gen | Key-generating algorithm | 144 |
| Enc | Encoding algorithm | 144 |
| Dec | Decoding algorithm | 144 |
| $\mathcal{K}$ | Key set | 144 |
| $\mathcal{M}$ | Message set | 144 |
| $\mathcal{C}$ | Ciphertext set | 144 |
| $\ell$ | Message length | 146 |
| $\perp$ | Error message | 147 |
| $\lambda$ | Security parameter | 146 |
| $\mathrm{negl}(\lambda)$ | Negligible function in $\lambda$ | 147 |
| PPT | Probabilistic polynomial-time algorithm | 147 |
| QKD | Quantum key distribution | 151 |
| QECM | Quantum encryption of a classical message | 155 |

# Chapter 1

# Introduction

In this introductory chapter, we provide a broad overview of the thesis, covering all its key aspects, including a summary of our contributions and an outline of the document.

## Chapter Contents

## 1.1 Context

In this section, we briefly introduce the necessary background to set the stage for our contributions. We begin with an overview of quantum information theory (Section 1.1.1), followed by discussions on nonlocal games (Section 1.1.2) and nonlocal boxes (Section 1.1.3). We then present communication complexity (Section 1.1.4) and conclude with the unclonable bit problem (Section 1.1.5).

> « *Each answer raises new questions, completely different in nature from the ones one started with; this, more than anything else, indicates that finally we might be on the right track.* » — Popescu [Pop14]

### *1.1.1 Quantum Information Theory*

Quantum information theory is the study of information processes through the postulates of quantum mechanics. Examples of famous foundational papers in this field include the Einstein–Podolsky–Rosen paradox [EPR35], Bell's inequalities [Bel64], and Tsirelson's bound [Tsi80]. Below, we briefly present the notions of quantum states, quantum measurements, and quantum channels. Find details in Chapter 2.

**Quantum State.** The basic objects are *quantum states*, defined as density matrices $\rho$ in a Hilbert space $\mathcal{H}$, forming a convex set as follows:

$$\mathcal{D}(\mathcal{H}) := \left\{ \rho \in \mathcal{B}(\mathcal{H}) \,:\, \rho \succcurlyeq \mathbf{0},\ \mathrm{Tr}[\rho] = 1 \right\},$$

where $\mathcal{B}(\mathcal{H})$ is the set of bounded operators over $\mathcal{H}$. Interestingly, quantum states can be *entangled*, meaning that they have very correlated behaviors. In a Hilbert space $\mathcal{H}_\mathsf{A} \otimes \mathcal{H}_\mathsf{B}$, they are defined as the density matrices $\rho$ that cannot be decomposed as follows:

$$\rho = \sum_i \alpha_i \left( \sigma_\mathsf{A}^{(i)} \otimes \sigma_\mathsf{B}^{(i)} \right),$$

where the index $i$ is finite, where the coefficients $\alpha_i \in \mathbb{R}_{\geqslant 0}$ are non-negative and sum to $\sum_i \alpha_i = 1$, and where $\sigma_\mathsf{A}^{(i)} \in \mathcal{D}(\mathcal{H}_\mathsf{A})$ and $\sigma_\mathsf{B}^{(i)} \in \mathcal{D}(\mathcal{H}_\mathsf{B})$.

**Quantum Measurement.** To extract classical information from a quantum state, one needs to perform a *quantum measurement*. These are called *positive operator-valued measure* (POVM) and are defined as finite sets $\{E_i\}_i$ of bounded operators $E_i \in \mathcal{B}(\mathcal{H})$ that are positive semi-definite and that sum to the identity:

$$E_i \succcurlyeq \mathbf{0} \qquad \text{and} \qquad \sum_i E_i = \mathbb{I}_d \,.$$

The act of measuring is intrisically probabilistic and gives a random output according to the following probability law:

$$\mathbb{P}\big(\text{obtaining "}i\text{"}\big) \;=\; \mathrm{Tr}(E_i\,\rho)\,.$$

**Quantum Channel.** Finally, we mention that the most general way to transform a quantum state into another one is via *quantum channels*. These are linear maps $\Phi : \mathcal{B}(\mathcal{H}) \to \mathcal{B}(\mathcal{K})$ that are completely-positive (CP):

$$\forall \mathcal{H}' \text{ Hilbert space}\,,\ \forall X \succcurlyeq \mathbf{0} \text{ in } \mathcal{B}(\mathcal{H}\otimes\mathcal{H}')\,, \quad \big[\Phi\otimes\mathbb{I}_{\mathcal{H}'}\big](X) \succcurlyeq \mathbf{0}\,,$$

and trace-preserving (TP):

$$\forall X \in \mathcal{B}(\mathcal{H})\,, \qquad \mathrm{Tr}\big[\Phi(X)\big] \;=\; \mathrm{Tr}[X]\,.$$

### *1.1.2 Nonlocal Games*

As detailed in Chapter 3, a two-player *nonlocal game* is a cooperative game played by two characters, commonly named Alice and Bob, who agree on a common strategy $\mathcal{S}$ beforehand, but who are space-like separated during the game, meaning that communication is forbidden. Each of the players is provided with a "question" $x$ (resp. $y$) by a Referee, and the players prepare their "answer" $a$ (resp. $b$) based on the chosen strategy $\mathcal{S}$, possibly using a "shared resource" (like a pair of entangled quantum states). Finally, the Referee verifies whether the players win the game: he computes a deterministic Boolean function depending on the questions and answers, called the "rule" of the game. The goal of Alice and Bob is to maximize their winning probability. If they win almost surely, *i.e.* with probability $1$, then we say that they have a *perfect strategy*. Find a representation in Figure 1.1.

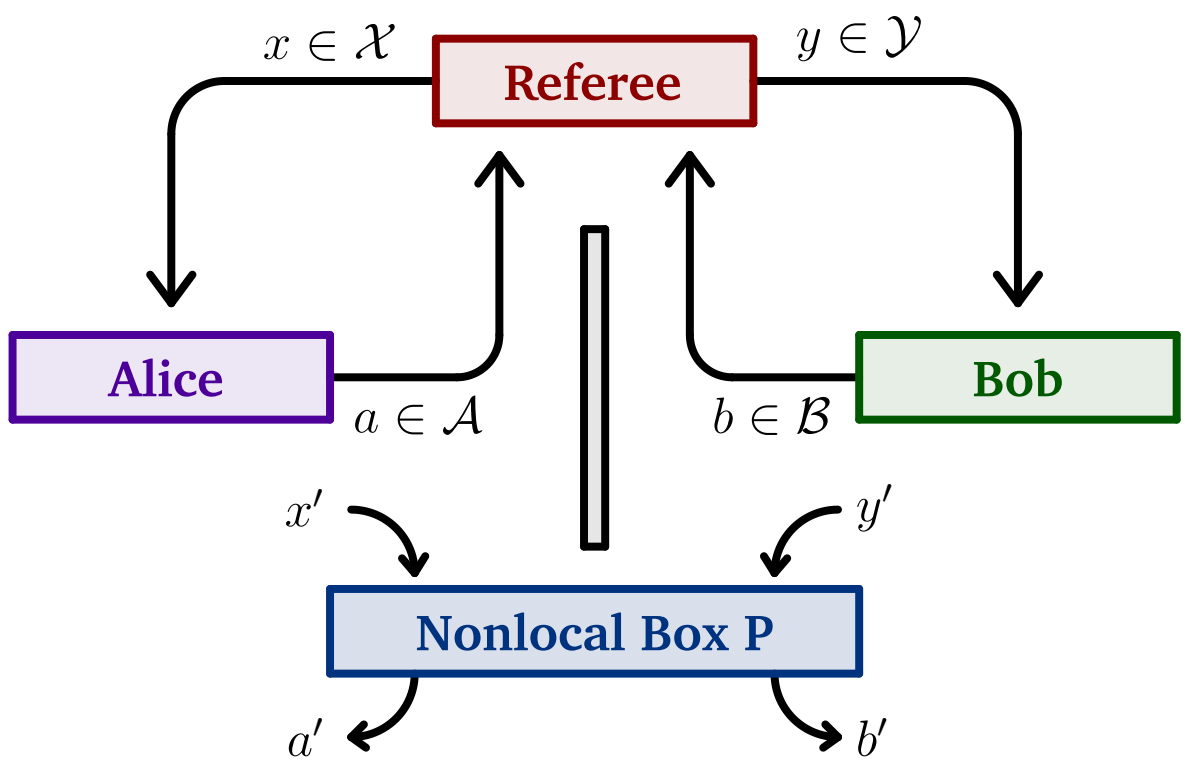

**Figure 1.1** — *Representation of a generic nonlocal game. A similar diagram also appears in [BBP24; Bot22].*

**Example 1.1** (CHSH Game) — The best-known example of a nonlocal game is the *CHSH game*, named after Clauser, Horne, Shimony, and Holt [CHSH69]. In this game, the questions $x, y$ and the answers $a, b$ are classical bits in $\{0, 1\}$. Alice and Bob win the CHSH game if and only if

$$a \oplus b = x \cdot y \,,$$

where the symbol "$\oplus$" is the sum modulo $2$ and "$\cdot$" is the product. We comment on the perfect strategies for this game below.

### *1.1.3 Nonlocal Boxes*

As mentioned above, Alice and Bob are allowed to use shared resources in their strategy $\mathcal{S}$. Depending on the type of allowed resources, the strategy will belong to a certain set of correlations. For instance, quantum mechanics allows two particles to be entangled, meaning that they have very correlated behaviors even if they are separated and used far away from each other. We present this topic more in detail in Chapter 3.

**Nonlocal Box.** In a device-independent approach, we do not focus our attention on the resource itself, but rather on the statistics that one can obtain from a resource. The device is viewed as a black box, called *nonlocal box*, taking some classical strings as inputs and outputs. Formally, a nonlocal box is defined by its associated conditional probability distribution:

$$\mathbf{P}(a, b \,|\, x, y) \,,$$

which tells the probability of obtaining the tuple $(a,b)$ when $(x,y)$ is input.

**Sets of Boxes.** In this thesis, we mainly study three types of boxes:

(1) *Classical boxes* or *local boxes*, arising from classical mechanics, forming a set $\mathcal{L}$, of the following form:

$$\mathbf{P}\big(a,b\,\big|\,x,y\big) \;=\; \int_{\lambda\in\Lambda} \mathbb{P}_{\mathsf{A}}\big(a\,\big|\,x,\lambda\big)\,\mathbb{P}_{\mathsf{B}}\big(b\,\big|\,y,\lambda\big)\,\mu(\lambda)\,,$$

for some probability measures $\mathbb{P}_{\mathsf{A}}$, $\mathbb{P}_{\mathsf{B}}$, and $\mu$;

(2) *Quantum boxes*, coming from quantum mechanics, forming a set $\mathcal{Q}$, of the following form:

$$\mathbf{P}\big(a,b\,\big|\,x,y\big) \;=\; \mathrm{Tr}\Big[\big(E_{a|x}\otimes F_{b|y}\big)\,\rho\Big]\,,$$

for some quantum state $\rho$ and some quantum measurements $\{E_{a|x}\}_a$ and $\{F_{b|y}\}_b$;

(3) *Non-signaling boxes*, obeying the no-faster-than-light-communication principle. They form a set $\mathcal{NS}$ that satisfies the following linear relations:

$$\begin{aligned}
\forall a,b,x,y\,, &\qquad \mathbf{P}\big(a,b\,\big|\,x,y\big) \;\geqslant\; 0\\
\forall x,y\,, &\qquad \sum_{a,b}\mathbf{P}\big(a,b\,\big|\,x,y\big) \;=\; 1\,,\\
\forall b,x,x',y\,, &\qquad \sum_{a}\mathbf{P}\big(a,b\,\big|\,x,y\big) \;=\; \sum_{a}\mathbf{P}\big(a,b\,\big|\,x',y\big) \;=:\; \mathbf{P}(b\,|\,y)\,,\\
\forall a,x,y,y'\,, &\qquad \sum_{b}\mathbf{P}\big(a,b\,\big|\,x,y\big) \;=\; \sum_{b}\mathbf{P}\big(a,b\,\big|\,x,y'\big) \;=:\; \mathbf{P}(a\,|\,x)\,.
\end{aligned}$$

These three sets are compact and convex, and relate as follows:

$$\mathcal{L} \subsetneq \mathcal{Q} \subsetneq \mathcal{NS}\,.$$

**Nonlocal Boxes in Nonlocal Games.** In a nonlocal game, Alice and Bob's strategy $\mathcal{S}$ often simply consists in inputting the questions $x$ and $y$ in the nonlocal box and using its outputs $a$ and $b$ as answers to the Referee.

**Example 1.2** (Strategy in the CHSH Game) — For the CHSH game mentioned above, an optimal classical strategy from $\mathcal{L}$ yields a winning probability of $75\%$, achieved for instance with the nonlocal box $\mathbf{P_{00}}$ that always outputs $(0,0)$. As for quantum boxes in $\mathcal{Q}$, the best quantum strategy yields a winning probability of $\cos^2(\pi/8) \approx 85\%$. It is achieved by the so-called *EPR-pair* of entangled particles, named after Einstein, Podolsky, and Rosen [EPR35]. Lastly, in the non-signaling set $\mathcal{NS}$, the best winning strategy is the **PR** box, named after Popescu and Rohrlich [PR94], winning with probability $100\%$, *i.e.* it is a perfect strategy.

### *1.1.4 Communication Complexity*

*Communication complexity* (CC) is a notion that was introduced by Yao in [Yao79] and later reviewed in [KN96; RY20]. It quantifies the difficulty of performing a distributed computation. We detail the principle of communication complexity in Chapter 4.

**Communication Complexity.** Assume that we have access to two distant computers and that we want to compute the value of a Boolean function $f : \{0,1\}^n \times \{0,1\}^m \to \{0,1\}$ evaluated at some strings $(X,Y)$, where $X \in \{0,1\}^n$ is given to the first computer and $Y \in \{0,1\}^m$ to the other, with the constraint of minimizing the number of communication bits between the computers. The CC of the function $f$ is then defined as the minimal number of bits that the computers need to communicate so that the first computer is able to compute the value $f(X,Y) \in \{0,1\}$.

**Example 1.3** — For instance, when $n = m = 2$, $X = (x_1, x_2)$, $Y = (y_1, y_2)$, the communication complexity of

$$f_1(x_1, x_2, y_1, y_2) := x_1 \cdot (y_1 \oplus y_2)$$

equals $1$, sending the communication bit $y_1 \oplus y_2$ from the second computer to the first one. However, it can be shown that the communication complexity of

$$f_2(x_1, x_2, y_1, y_2) := (x_1 \cdot y_1) \oplus (x_2 \cdot y_2)$$

equals $2$, using as communication bits the two input bits $y_1$ and $y_2$ of the second computer. Thus, it means that $f_2$ is more complex than $f_1$ in the sense of CC.

**Collapse of CC.** The notion of CC is connected with nonlocal boxes. We say that a nonlocal box **P** *collapses communication complexity* if there exists a universal constant $\mathfrak{p} > 1/2$ such that for all strings $X, Y$ and all Boolean functions $f$, the first computer outputs the correct value of $f(X, Y)$ with probability at least $\mathfrak{p}$ with only one bit of communication. Note that, in this definition, an arbitrary number of copies of the box **P** can be used to achieve the task. Such a collapse is strongly believed to be unachievable in nature since it would imply the absurdity that a single bit of communication is sufficient to distantly estimate any value of any Boolean function $f$ with arbitrary large input size [BG15; Bra+06; BS09; EWC23a; vD99]. This gives rise to the following open question:

**Open Question 1.4 —** *What are all non-signaling boxes that collapse CC?*

The **PR** box is known to collapse CC [vD99], so this correlation is physically unfeasible according to the principle of communication complexity. More generally, it is known that some noisy versions of the **PR** box also collapse CC for different types of noise [BBP24; Bot+24a; Bra+06; Bri+19; BS09; BW24; EWC23a]. On the other hand, it is known that quantum correlations do *not* collapse communication complexity [Cle+99], and neither does a slightly wider set named "almost quantum correlations" [Nav+15]. To this day, the question is still open whether the remaining non-signaling boxes are collapsing, meaning that there is still a gap to be filled.

### *1.1.5 Unclonable Bit*

In quantum cryptography, the *unclonable bit* is a protocol that would allow a sender, Alice, to encrypt a single bit $m$ into a quantum state $\rho$ in such a way that it is not possible to clone it and retrieve $m$ in multiple locations once the key is revealed. Its existence in the plain model, *i.e.* without assumptions on the adversaries, is an open question in the strong security regime. This follows the pioneering work of Broadbent and Lord [BL20] on unclonable cryptography and is detailed in Chapter 5.

**No-Cloning Games.** This notion can be modeled by a family of (extended) nonlocal games, called the *no-cloning games*. In this game, Alice (A) plays against adversaries called Pirate (P), Bob (B), and Charlie (C). Alice encrypts a uniformly random message $m \in \{0, 1\}$ using a classical key $k \in$ $\{0, .., K\}$ into a quantum state $\rho_{m,k}$. She transmits it to the Pirate who tries to clone the quantum state in two copies with a quantum channel. We know from quantum mechanics postulates that it is not possible to perfectly clone an unknown state, but he can accept some noise and prepare "bad" copies of $\rho_{m,k}$. After this, the Pirate sends one copy for each of Bob and Charlie. Finally, Bob and Charlie receive the encryption key $k$ and they try to retrieve the initial message $m$ from their imperfect copies: they produce guesses $m_\mathsf{B}$ and $m_\mathsf{C}$ in $\{0, 1\}$ respectively, and *win* the game if and only if $m = m_\mathsf{B} = m_\mathsf{C}$. The game is depicted in Figure 1.2.

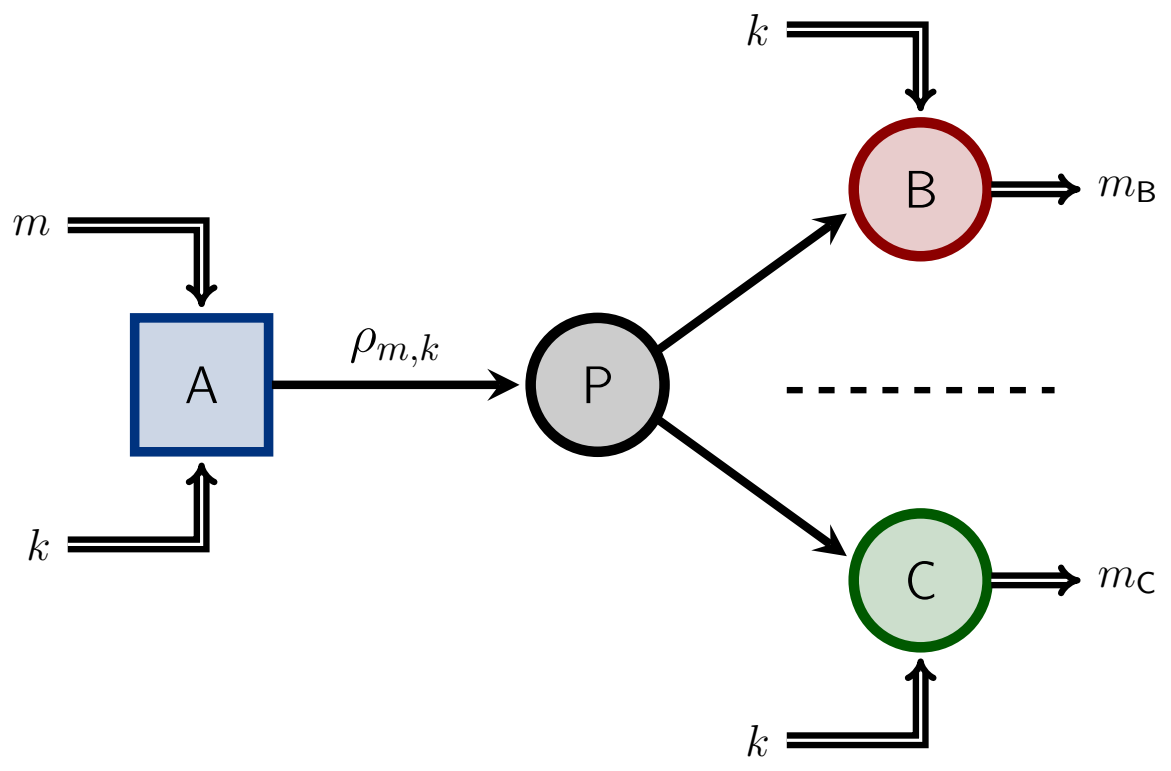

**Figure 1.2 —** *No-cloning game. A similar diagram appears in [Bot+24b].*

**Security.** We say that Alice's encryption scheme satisfies the *unclonable-indistinguishable security* if the winning probability of any adversary $(\mathsf{P}, \mathsf{B}, \mathsf{C})$ is upper-bounded by:

$$\mathbb{P}\Big((\mathsf{P}, \mathsf{B}, \mathsf{C}) \text{ win}\Big) \leqslant 1/2 + \mathrm{negl}(\lambda)\,,$$

for any security parameter $\lambda \in \mathbb{N}$, where $\mathrm{negl}(\lambda)$ is a negligible function (vanishing to $0$ at infinity faster than the inverse of any positive polynomial). Note that $1/2$ corresponds to the trivial winning probability, where the pirate sends the whole state $\rho_{m,k}$ to Bob, thus able to obtain correct output $m_\mathsf{B} = m$ using the key $k$, and where Charlie produces a uniformly random guess $m_\mathsf{C} \in \{0, 1\}$. An *unclonable bit protocol* is an encryption scheme that achieves unclonable-indistinguishable security. It gives rise to the following open question:

**Open Question 1.5 —** *Does the unclonable bit exist?*

As detailed in Chapter 5, a sequence of papers studied this question or related ones, but to this day none of them comes up with a proof of strong security in the plain model, *i.e.* without assumptions.

## 1.2 Summary of Contributions

In this section, we summarize the main contributions of this thesis.

### *1.2.1 First Contribution: CC in the CHSH Game*

In this section, we present the results from the following manuscript:

**[BBP24]** Pierre Botteron, Anne Broadbent, and Marc-Olivier Proulx. "Extending the Known Region of Nonlocal Boxes that Collapse Communication Complexity". In: *Physical Review Letters* 132 (Feb. 2024), p. 070201. DOI: 10.1103/PhysRevLett.132.070201

This work enters into the CHSH scenario: two players, bit inputs $x, y \in \{0,1\}$, and bit outputs $a, b \in \{0,1\}$. Recall that the CHSH game was described in Example 1.1. As indicated below, this work builds on preliminary results reported in the M.Sc. thesis of our co-author Marc-Olivier Proulx [Pro18], which are not exposed as new results here. Find more details in Chapter 6.

**Bias of a Nonlocal Box.** Given a nonlocal box $\mathbf{P} \in \mathcal{NS}$ and some inputs $x, y \in \{0,1\}$, the *bias* of $\mathbf{P}$ in the CHSH game is defined as the only real number $\eta_{x,y}(\mathbf{P})$ satisfying the following relation:

$$\sum_{a,b} \mathbf{P}\big(a, b \,|\, x, y\big)\, \mathbb{1}_{a\oplus b = xy} = \frac{1+\eta_{x,y}(\mathbf{P})}{2}\,.$$

Using this notion of bias, we find a sufficient explicit condition for a nonlocal box $\mathbf{P} \in \mathcal{NS}$ to collapse communication complexity:

**Result 1** (Theorem 6.1) — *Any nonlocal box* $\mathbf{P} \in \mathcal{NS}$ *satisfying the following condition collapses communication complexity:*

$$\Big(\sum_{x,y} \eta_{x,y}(\mathbf{P})\Big)^2 + 2\,\eta_{0,0}(\mathbf{P})^2 + 4\,\eta_{0,1}(\mathbf{P})\,\eta_{1,0}(\mathbf{P}) + 2\,\eta_{1,1}(\mathbf{P})^2 \;>\; 16\,.$$

The proof is a generalization of the protocol from Brassard, Buhrman, Linden, Méthot, Tapp, and Unger [Bra+06], based on bias amplification using majority functions.

**Expressions in Some Slices of $\mathcal{NS}$.** In the CHSH scenario, the convex set $\mathcal{NS}$ is 8-dimensional. As such, it is hard to represent it in our minds. So, one can rather study some of its 2-dimensional slices, like we would do with a cake to observe its inner layers. A 2-dimensional slice is given by three non-aligned nonlocal boxes (forming an affine basis of the underlying hyperplane). Consider the following nonlocal boxes:

$$\mathbf{PR}(a,b\,|\,x,y) := \tfrac{1}{2}\,\mathbb{1}_{a\oplus b = xy}\,, \qquad \mathbf{I}(a,b\,|\,x,y) := \tfrac{1}{4}\,,$$
$$\mathbf{PR}'(a,b\,|\,x,y) := \tfrac{1}{2}\,\mathbb{1}_{a\oplus b = (x\oplus 1)(y\oplus 1)}\,, \qquad \mathbf{SR}(a,b\,|\,x,y) := \tfrac{1}{2}\,\mathbb{1}_{a=b}\,.$$

As already reported in [Pro18], in the slice of $\mathcal{NS}$ passing through $\mathbf{PR}$, $\mathbf{PR}'$, and $\mathbf{I}$, any box $\mathbf{P}$ satisfying one of the following two conditions collapses communication complexity:

$$\Big(\eta_{0,0}(\mathbf{P}) + \eta_{0,1}(\mathbf{P})\Big)^2 + \frac{1}{3}\Big(-\eta_{0,0}(\mathbf{P}) + \eta_{0,1}(\mathbf{P})\Big)^2 \;>\; \frac{8}{3}\,,$$
$$\frac{1}{3}\Big(\eta_{0,0}(\mathbf{P}) + \eta_{0,1}(\mathbf{P})\Big)^2 + \Big(-\eta_{0,0}(\mathbf{P}) + \eta_{0,1}(\mathbf{P})\Big)^2 \;>\; \frac{8}{3}\,.$$

Similarly in the slice of $\mathcal{NS}$ passing through $\mathbf{PR}$, $\mathbf{SR}$, and $\mathbf{I}$ with the following condition:

$$\Big(3\,\eta_{0,0}(\mathbf{P}) + \eta_{1,1}(\mathbf{P})\Big)^2 + \frac{1}{3}\Big(\eta_{0,0}(\mathbf{P}) - \eta_{1,1}(\mathbf{P})\Big)^2 \;>\; \frac{32}{3}\,.$$

These two results are represented in Figure 1.3 and correspond to particular cases of Result 1.

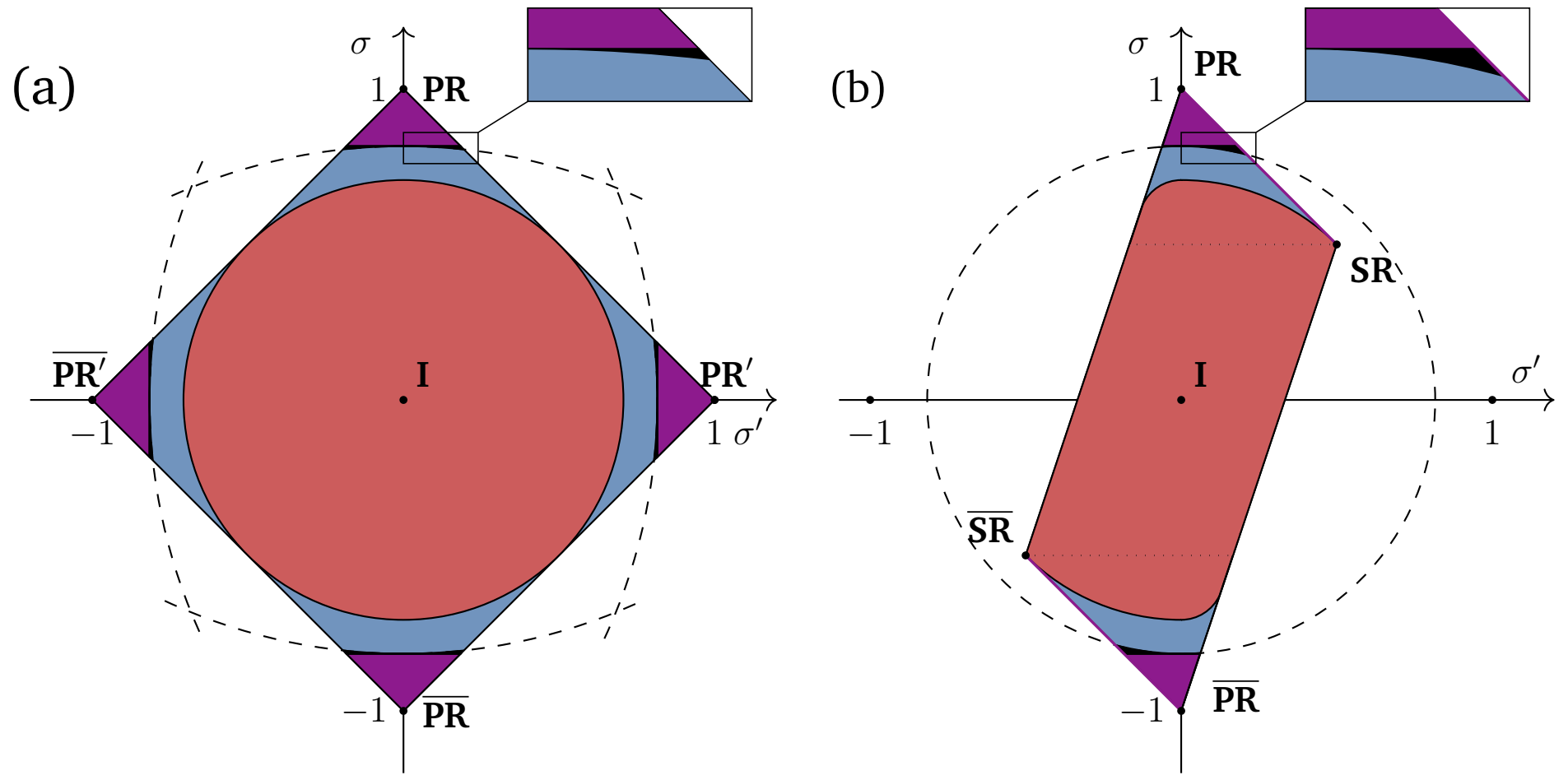

**Figure 1.3 —** *In* **purple** *is drawn the prior (analytically) known collapsing region. We extend it as follows: the* **black** *area is the new analytic collapsing region. The* **red** *area corresponds to the area of non-collapsing boxes. The* **blue** *area is the gap to be filled in red or purple (open problem). Diagrams (a) and (b) represent the slices of* $\mathcal{NS}$ *passing through respectively* $\{\mathbf{PR}, \mathbf{PR}', \mathbf{I}\}$ *with* $\sigma = \eta_{0,0} + \eta_{0,1}$ *and* $\sigma' = -\eta_{0,0} + \eta_{0,1}$ *(improving [Bra11]) and* $\{\mathbf{PR}, \mathbf{SR}, \mathbf{I}\}$ *with* $\sigma = 3\,\eta_{0,0} + \eta_{1,1}$ *and* $\sigma' = \eta_{0,0} - \eta_{1,1}$ *(improving [BS09]). We use the convention* $\overline{\mathbf{P}} := 1 - \mathbf{P}$.

### *1.2.2 Second Contribution: Algebra of Boxes*

In this section, we present the results from the following manuscript:

**[Bot+24a]** Pierre Botteron, Anne Broadbent, Reda Chhaibi, Ion Nechita, and Clément Pellegrini. "Algebra of Nonlocal Boxes and the Collapse of Communication Complexity". In: *Quantum* 8 (July 2024), p. 1402. ISSN: 2521-327X. DOI: 10.22331/q-2024-07-10-1402

We provide a new mathematical framework and algorithms in working towards finding nonlocal boxes that collapse communication complexity. As indicated below, some of the ideas are based on the M.Sc. thesis of the author [Bot22], thus are not exposed as new results here. Find more details in Chapter 6.

**Algebra of Boxes.** We introduce a new framework that we call the *algebra of boxes*. Given two boxes **P** and **Q**, it is possible to combine them via a *wiring* W to obtain a new box denoted $\mathbf{P} \boxtimes_{\mathsf{W}} \mathbf{Q}$. This gives rise to the notion of *product of boxes* $\boxtimes_{\mathsf{W}}$ and of an algebra over nonlocal boxes, the algebra of boxes $\mathcal{B}_{\mathsf{W}}$, for which we have the following result:

**Result 2** (Proposition 6.3) — *We characterize the associativity and commutativity of the algebra* $\mathcal{B}_{\mathsf{W}}$ *depending on the wiring* W.

This gives an algebraic perspective on protocols for correlation distillation—for instance, the non-associativity of the algebra of boxes tells us that the order in which the boxes are wired matters.

**Orbit of a Box.** This framework gives rise to the notion of *orbit of a box*. The orbit of $\mathbf{P} \in \mathcal{NS}$ is roughly defined as the set of all possible boxes that can be produced by wiring arbitrarily many copies of $\mathbf{P}$. As already observed in [Bot22], this allows for interesting visualizations of the hidden structure of boxes, for example, that these orbits satisfy strong alignment and parallelism properties. Moreover, it is possible to derive the expression of the highest CHSH-valued box of the orbit, which explains the numerical observation reported in [EWC23a, Supplementary Material, II], and from which one can derive an insightful linear-time algorithm that is exponentially more efficient compared to the naive exponential-time computation of the entire orbit. In addition, this technique allows one to recover a similar result as in [EWC23a] stating that those methods lead to finding collapsing boxes via the recursive application of the multiplication $\cdot \boxtimes \mathbf{P}$ on the right.

**Numerical Results.** We provide algorithms in our GitHub page [BC23a] for the following task: given a box $\mathbf{P}$ that we want to show is collapsing, find an appropriate wiring W such that the orbit contains a collapsing box. The idea is to repeat several times in parallel a variant of the Gradient Descent Algorithm in order to find the most appropriate wiring W. These algorithms allow us to find new collapsing areas (concurrent and independent of [EWC23a, Figure 3]):

**Result 3** (Figure 6.10) — *We numerically find new collapsing boxes.*

**Analytical Results.** Finally, we show that our framework also allows us to recover an analytical result from [Bri+19][1] with different methods:

> **Result 4** (Theorem 6.18 & Corollary 6.20) — *With a new proof based on the algebra of boxes, we find that some triangular areas of boxes in the boundary of $\mathcal{NS}$ are collapsing.*

Moreover, we also retrieve a result from [Rai+19], again with a different method:

> **Result 5** (Corollary 6.22) — *With a new proof based on communication complexity, we find that some triangular areas of boxes in the boundary of $\mathcal{NS}$ are* quantum voids, i.e. *faces of $\mathcal{NS}$ for which all quantum boxes are local.*

Hence, a strength of this framework is that it allows us to unify the perspectives of [Bri+19] and [Rai+19], as well as of the concurrent and independent work [EWC23a].

### 1.2.3 Third Contribution: CC in Graph Games

In this section, we present the results from the following manuscript:

**[BW24]** Pierre Botteron and Moritz Weber. *Communication Complexity of Graph Isomorphism, Coloring, and Distance Games*. 2024. arXiv: 2406.02199 [quant-ph]

In this work, we study three nonlocal games related to graphs: the graph isomorphism game, the graph coloring game, and the vertex distance game, a new game depending on a parameter $D \in \mathbb{N}$ that we define in this manuscript. All graphs are always assumed to be non-empty, finite, undirected, and loopless. In the sequel, let $\mathcal{G}$ be a given graph, with vertex set $V(\mathcal{G})$, and write $g \sim g'$ if two vertices $g, g' \in V(\mathcal{G})$ are linked by an edge. Let $\mathcal{H}$ be a graph with disjoint vertex set $V(\mathcal{G}) \cap V(\mathcal{H}) = \emptyset$. We refer to [GR01] for more background on graph theory. Find more details in Chapter 7.

[1] This result was brought to our attention in the finalization of the paper.

**Graph Isomorphism Game [Ats+19].** For the well-known *graph isomorphism game* $(\mathcal{G},\mathcal{H})$, Alice and Bob receive vertices $x_\mathsf{A}, x_\mathsf{B} \in V = V(\mathcal{G}) \cup V(\mathcal{H})$ and they respond with vertices $y_\mathsf{A}, y_\mathsf{B} \in V$. The first winning condition is that the set $\{x_\mathsf{A}, y_\mathsf{A}\}$ consists in exactly one vertex from $V(\mathcal{G})$, that we call $g_\mathsf{A} \in V(\mathcal{G})$, and the other from $V(\mathcal{H})$, called $h_\mathsf{A} \in V(\mathcal{H})$; and similarly for $\{x_\mathsf{B}, y_\mathsf{B}\}$ giving rise to $g_\mathsf{B} \in V(\mathcal{G})$ and $h_\mathsf{B} \in V(\mathcal{H})$. The second winning condition condition is that $g_\mathsf{A}$ and $g_\mathsf{B}$ are related in the same way as $h_\mathsf{A}$ and $h_\mathsf{B}$ are related, in the sense that:

(i) if $g_\mathsf{A} = g_\mathsf{B}$, then $h_\mathsf{A} = h_\mathsf{B}$;
(ii) if $g_\mathsf{A} \sim g_\mathsf{B}$, then $h_\mathsf{A} \sim h_\mathsf{B}$;
(iii) if $g_\mathsf{A} \not\simeq g_\mathsf{B}$, then $h_\mathsf{A} \not\simeq h_\mathsf{B}$;

where the symbol "$\simeq$" means equal or linked by an edge. Note that the three implications in items (i), (ii), and (iii) are actually equivalences. For the graph isomorphism game, we prove the following result:

**Result 6** (Theorem 7.2, Corollary 7.4, Theorem 7.9, Theorem 7.16, & Corollary 7.17) — *Given $\mathcal{G}$ and $\mathcal{H}$ for the graph isomorphism game, we have:*

*(1) If the diameter of $\mathcal{G}$ satisfies $\mathrm{diam}(\mathcal{G}) \geqslant 2$ and if $\mathcal{H}$ has exactly two connected components which are both complete, then any perfect non-signaling strategy collapses CC. This may be weakened to strategies winning with probability $p > \frac{3+\sqrt{6}}{6} \approx 0.91$.*

*(2) If $\mathrm{diam}(\mathcal{G}) \geqslant 2$, if $\mathcal{H}$ is not connected, and if there is a common equitable partition with an additional technical assumption (H), then there is a non-signaling strategy which collapses CC.*

*(3) With the same conditions as in the previous item, and if $\mathcal{H}$ is additionally strongly transitive (a generalization of the notion of* transitivity *from graph automorphism theory) and $d$-regular, and if Alice and Bob share randomness, then any perfect non-signaling strategy collapses CC. As a consequence, these strategies cannot be quantum.*

**Graph Coloring Game [Cam+07a].** The graph isomorphism game can be relaxed to a *graph homomorphism game* $\mathcal{G} \to \mathcal{H}$ omitting item (iii) in the above game and with questions $x_\mathsf{A}, x_\mathsf{B}$ always lying in $V(\mathcal{G})$. If $\mathcal{H} = \mathcal{K}_N$, the complete graph on $N$ points, then the graph homomorphism game $\mathcal{G} \to \mathcal{K}_N$ is called the *graph coloring game*. For this game, we show the following result:

**Result 7** (Theorem 7.20 & Theorem 7.23) — *Given $\mathcal{G}$ and $\mathcal{H}$ for the graph homomorphism game (resp. the graph coloring game), we have:*

*(1) For any non-signaling strategy winning the homomorphism game $\mathcal{K}_3 \to \mathcal{G}$ with probability $p$, together with another non-signaling strategy winning the graph coloring game $\mathcal{G} \to \mathcal{K}_2$ with probability $q$ such that $pq > \frac{3+\sqrt{6}}{6} \approx 0.91$, there is a collapse of CC.*

*(2) If $\mathrm{diam}(\mathcal{G}) \geqslant 2$ and if $\mathcal{H}$ has exactly $N$ connected components all of which are complete, then for any non-signaling strategy winning the graph isomorphism game $\mathcal{G} \to \mathcal{H}$ with probability $p$, combined with any non-signaling strategy winning the coloring game $\mathcal{K}_N \to \mathcal{K}_2$ with probability $q$, such that $pq > \frac{3+\sqrt{6}}{6} \approx 0.91$, there is a collapse of CC.*

**Vertex Distance Game.** We introduce a new game that we call *vertex $D$-distance game*, with a parameter $D \in \mathbb{N}$. This is a generalization of the graph isomorphism game, changing the winning conditions (i), (ii), and (iii) into:

(i) if $d(g_\mathsf{A}, g_\mathsf{B}) = t \leqslant D$, then $d(h_\mathsf{A}, h_\mathsf{B}) = t$;

(ii) if $d(g_\mathsf{A}, g_\mathsf{B}) > D$, then $d(h_\mathsf{A}, h_\mathsf{B}) > D$.

For $D = 0$, if we consider the graphs $\mathcal{G} = \mathcal{K}_M$ and $\mathcal{H} = \mathcal{K}_N$, this exactly corresponds to the $N$-coloring game of $\mathcal{K}_M$. For $D = 1$, this is the graph isomorphism game. We show that neither classical nor quantum strategies may distinguish the vertex $D$-distance games for different parameters $D$:

**Result 8** (Proposition 7.27 & Theorem 7.46) — *For any $D \geqslant 1$, perfect classical and quantum strategies are precisely the same for the $D$-distance game as for the isomorphism game. However, in the non-signaling setting, the perfect strategies for the two games differ.*

More precisely, for any $D \in \mathbb{N}$, there is a pair of graphs which admits a perfect non-signaling strategy for the vertex $D$-distance game but not for the vertex $(D+1)$-distance game, see Proposition 7.43. So, non-signaling strategies provide a finer tool for distinguishing nonlocal games.

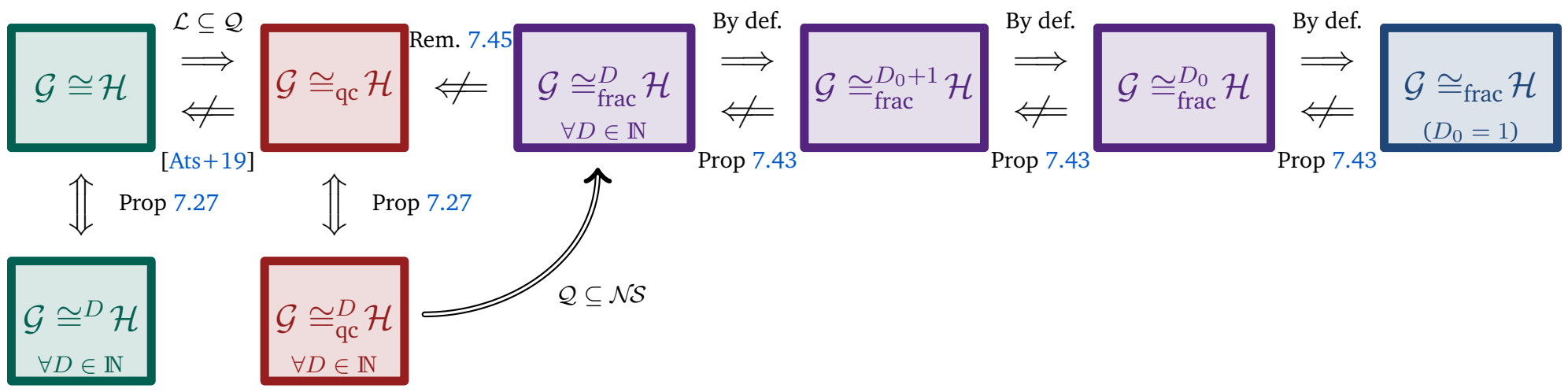

**Figure 1.4** — *Chain of strict implications, with $D_0 \geqslant 2$ fixed.*

Moreover, our definition of a vertex $D$-distance game produces a notion of *$D$-fractional isomorphism*, see details in Section 7.3.3, and we obtain the chain of strict implications drawn in Figure 1.4.

We also characterize perfect non-signaling strategies for the vertex $D$-distance game by adapting results from [Ats+19; RSU94]:

**Result 9** (Theorem 7.30) — *For any $D \geqslant 0$, the followings are equivalent:*

*(i) $\mathcal{G}$ and $\mathcal{H}$ are $\mathcal{NS}$-isomorphic in the sense of the $D$-distance game;*

*(ii) $\mathcal{G}$ and $\mathcal{H}$ are $D$-fractionally isomorphic;*

*(iii) There exists a $D$-common equitable partition of $\mathcal{G}$ and $\mathcal{H}$.*

This characterization allows us to finally study the collapse of CC for this game:

**Result 10** (Theorem 7.49, Proposition 7.50, Proposition 7.52, & Theorem 7.54) — *Given $\mathcal{G}$ and $\mathcal{H}$ for the vertex $D$-distance game:*

*(1) If $1 \leqslant D < \mathrm{diam}(\mathcal{G})$, if $\mathcal{H}$ is not connected, and if the graphs $(\mathcal{G}, \mathcal{H})$ admit a $D$-common equitable partition with technical assumption (H′), then there exists a perfect strategy collapsing CC.*

*(2) If $1 \leqslant D \leqslant \mathrm{diam}(\mathcal{H}) < \mathrm{diam}(\mathcal{G})$ and if $\mathcal{H}$ admits exactly two connected components, then any perfect non-signaling strategy collapses CC. This may be weakened to strategies winning with probability $p > \frac{3+\sqrt{6}}{6} \approx 0.91$.*

(3) If $1 \leqslant D \leqslant \text{diam}(\mathcal{H}) < \text{diam}(\mathcal{G})$ and if $\mathcal{H}$ admits exactly $N$ connected components, then for any perfect non-signaling strategy for the $D$-distance game, combined with a perfect non-signaling strategy for the coloring game $\mathcal{K}_N \to \mathcal{K}_2$, there is a collapse of CC. This may be weakened to strategies winning with respective probabilities $p$ and $q$ such that $pq > \frac{3+\sqrt{6}}{6} \approx 0.91$.

(4) If $2 \leqslant \text{diam}(\mathcal{G})$, if $\mathcal{H}$ is not connected, if the graphs $(\mathcal{G}, \mathcal{H})$ admit a $D$-common equitable partition with technical assumption (H′), and if $\mathcal{H}$ is strongly transitive and regular, then any perfect strategy collapses CC.

**Related Work.** Recent works by Assadi, Chakrabarti, Ghosh, and Stoeckl [Ass+23] and then by Flin and Mittal [FM25] also study the link between the graph coloring game and CC: the authors upper-bound the minimal number of communication bits required to compute a coloration of a graph with two distant parties. There is a similar approach for the graph isomorphism game by Loukas [Lou20] and by Chakraborty, Ghosh, Mishra, and Sen [Cha+21].

### 1.2.4 *Fourth Contribution: Towards the Unclonable Bit*

In this section, we present the results from the following manuscript:

**[Bot+24b]** Pierre Botteron, Anne Broadbent, Eric Culf, Ion Nechita, Clément Pellegrini, and Denis Rochette. *Towards Unconditional Uncloneable Encryption*. 2024. arXiv: 2410.23064 `[quant-ph]`

Our work focuses on the achievability of an encryption scheme that realizes an *unclonable bit* in the statistical model, thus without any computational or setup assumptions. However, given the apparent difficulty of achieving this task, we relax the security requirement, and we ask that the success probability of the adversaries be no more than $\frac{1}{2} + f(\lambda)$, for some function $f : \mathbb{R} \to \mathbb{R}$ such that $\lim_\lambda f(\lambda) = 0$. This is a relaxation of the usual requirement that $f$ be a negligible function. Find more details in Chapter 8.

Our contribution is a new candidate scheme for this relaxation of the unclonable bit question. We prove security for some small security parame-ters and provide strong numerical evidence that a security conjecture holds exactly.

**Candidate Scheme for an Unclonable bit.** Our candidate scheme is based on a family of pairwise anti-commuting $n$-qubit Pauli strings, of the following form for $n$ even:

$$\sigma_x^{\otimes(i-1)} \otimes \sigma_y \otimes \mathbb{I}^{\otimes(n-i)} \quad \text{and} \quad \sigma_x^{\otimes(i-1)} \otimes \sigma_z \otimes \mathbb{I}^{\otimes(n-i)}, \quad i \in \{1, \ldots, n\}\,.$$

Note that there are $2n$ such strings. When $n$ is odd, we add to the above set $\sigma_x^{\otimes n}$ and obtain $2n+1$ strings. We index these strings as $\Gamma_k$, and based on this, define the following candidate (see details in Section 8.2).

**Definition** (Candidate Scheme for an Unclonable Bit, Definition 8.1) — *Consider $\Gamma_1, ..., \Gamma_K$ Hermitian unitaries that anti-commute (*e.g.* pairwise anti-commuting Pauli strings), of dimension $d = 2^\lambda$, with $K = 2\lambda$ for even $\lambda$. Sample uniformly at random a key $k \in \{1, \ldots, K\}$. From this key, encrypt a classical message $m \in \{0, 1\}$ into the following quantum state:*

$$\rho_{m,k} = \frac{2}{d}\,\frac{\mathbb{I}_d + (\text{-}1)^m \Gamma_k}{2}\,,$$

*which is the normalized projector onto the eigenvalue $(\text{-}1)^m$ of $\Gamma_k$. For the decryption scheme of a quantum state $\rho$, measure it in the eigenbasis of $\Gamma_k$. The outcome is the decrypted message.*

One can easily check the correctness of this protocol, *i.e.* decrypting an encrypted message recovers the initial message with probability 1.

Our scheme can be seen as a generalization of the basic unclonable encryption scheme [BL20] that encodes a bit $m$ into $H^k|m\rangle$ using a key $k \in \{0, 1\}$. This encoding corresponds to the case $K = 2$ of our candidate. It is already known that, in the case $K = 2$, using the proof techniques from monogamy-of-entanglement games [Tom+13], the best success probability of the adversary at the no-cloning game is $\frac{1}{2} + \frac{1}{2\sqrt{2}}$, which is consistent with the following conjecture:

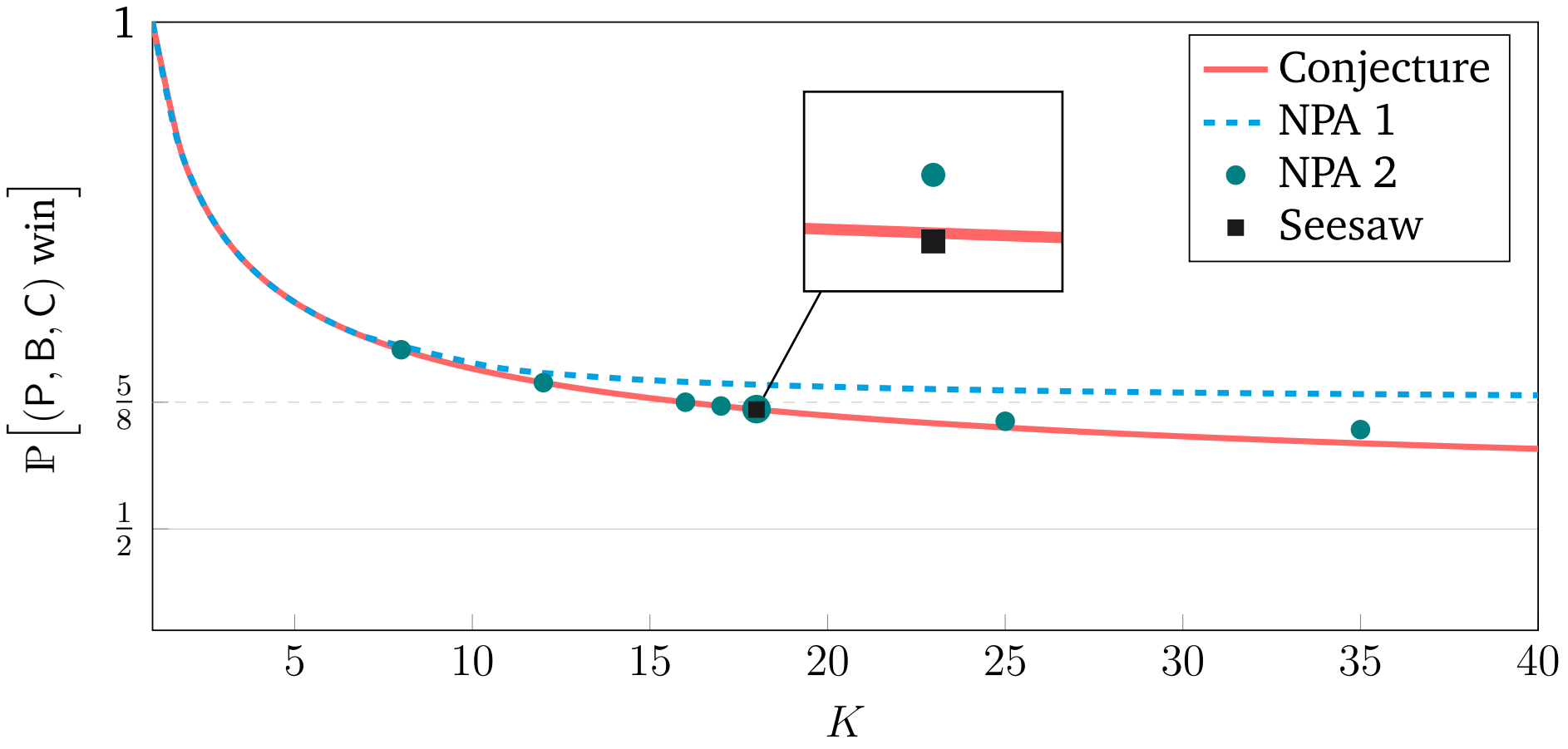

**Figure 1.5** — *Upper bounds on the winning probability in the no-cloning game involving three adversaries* $(\mathsf{P}, \mathsf{B}, \mathsf{C})$ *for our candidate scheme for Uncloneable Encryption with $K$ keys. The solid line (**red**) is the conjectured security bound, the dashed line (**cyan**) corresponds to the upper bound derived from NPA level 1, the circles (**teal**) are the numerical upper bounds obtained from NPA level 2, and the square (**black**) is the numerical result obtained using the seesaw optimization method on* $K = 18$.

**Conjecture** (Conjecture 8.5) — *Our candidate encryption scheme is unclonable-indistinguishable secure with the following upper bound:*

$$\mathbb{P}\left[(\mathsf{P}, \mathsf{B}, \mathsf{C}) \textit{ win the no-cloning game}\right] \leqslant \frac{1}{2} + \frac{1}{2\sqrt{K}} \,.$$

The contributions of this work focus on this conjecture and are divided into several analytical and numerical results summarized in Figure 1.5. The source code for our numerical methods is available on GitHub[2].

**Result 11** (Section 8.3.2, Cyan in Figure 1.5) — *The conjecture is confirmed for $K \leqslant 7$ using both an explicit Sum-of-Squares decomposition, and an analytical level-1 NPA proof.*

[2] https://github.com/denis-rochette/Towards-Unconditional-Unclonable-Encryption

**Result 12** (Section 8.3.3, Teal in Figure 1.5) — *The conjecture is numerically confirmed for $K \leqslant 17$ using a level-2 NPA optimization.*

For larger $K$, we would need the next levels of the NPA hierarchy—we were stopped by lack of computational power. Then, using the analytic solution of the first level of the NPA hierarchy and taking its limit in $K$, we also derive the following bound:

**Result 13** (Theorem 8.17, Cyan in Figure 1.5) — *Asymptotically, the winning probability of the no-cloning game for our candidate scheme is upper bounded by $5/8$.*

Additionally, using an alternating-optimization problem algorithm known as the *seesaw method*, we explore the case $K = 18$ and find no instance that violate the conjecture:

**Result 14** (Section 8.3.4, Black in Figure 1.5) — *For $K = 18$, no violation of the conjecture are found using the seesaw method.*

Finally, although we are unable to prove full unclonable-indistinguishability, we prove that conventional indistinguishability holds, in a relaxed sense, for our candidate scheme (see Section 8.2.5).

## 1.3 Outline

We now present an outline of this thesis. The dependencies between chapters are illustrated in the Graphical Contents (see page xiii).

**First Part: Background.** The first part introduces the necessary background and related results. In Chapter 2, we present fundamental tools from quantum information theory, including quantum states, entanglement, measurement, and channels—concepts used throughout this thesis. In Chapter 3, we discuss nonlocal boxes, nonlocal games, and their applications across various domains. In Chapter 4, we explore communication complexity, its known results and limitations, and its relevance compared to other principles. Finally, in Chapter 5, we introduce the basics of cryptography, discuss some quantum cryptographic schemes, and present the unclonable bit problem.

**Second Part: Contributions.** The second part presents our contributions, with links to background concepts where relevant. In Chapter 6, we discuss our first two contributions, both set in the CHSH scenario (two players, bit inputs, bit outputs), where we develop new techniques to collapse communication complexity. In Chapter 7, we extend these results to graph games, corresponding to our third contribution. In Chapter 8, we present our progress on the unclonable bit problem, covered in our fourth contribution.

**Conclusion.** Finally, in Chapter 9, we conclude the thesis with discussions and a list of open questions.

## 1.4 Acknowledgements

We are grateful to Dr. Anne Broadbent, Dr. Ion Nechita, and Dr. Clément Pellegrini for their insightful comments on preliminary versions of this thesis. We also thank Dr. Denis Rochette and Dr. Faedi Loulidi for many instructive comments.

This work was partially supported by the Institute for Quantum Technologies in Occitanie. We acknowledge funding from the Natural Sciences and Engineering Research Council of Canada (NSERC) [Reference Nos. ALLRP/580876-2022 and DGDND-2022-05167], as well as support from the MITACS grant FR113029 and the ANR project ESQuisses (Grant No. ANR-20-CE47-0014-01).

# Part I

# Background

# Chapter 2

# Quantum Information Theory

This chapter is devoted to a mathematical introduction to quantum information theory (QIT). It follows standard references such as [AS17; NC00; Wat18].

## Chapter Contents

## 2.1 Quantum States

Quantum states are fundamental objects in quantum information theory, manipulated to perform various tasks.

In this section, we begin by introducing the most basic example of a non-trivial quantum state, the *qubit* (Section 2.1.1). We then extend this concept in two steps: first to *pure states* (Section 2.1.2), and later to *mixed states* (Section 2.1.3).

### 2.1.1 Qubits

*Qubits* are the quantum analogous of classical bits $0$ and $1$. They are defined as unit vectors in the Hilbert space $\mathcal{H} = \mathbb{C}^2$ equipped with the usual inner product. A common way to write them is through *Dirac notation*. The standard basis vectors of $\mathcal{H}$ are written $|0\rangle := \left[\begin{smallmatrix}1\\0\end{smallmatrix}\right]$ and $|1\rangle := \left[\begin{smallmatrix}0\\1\end{smallmatrix}\right]$ and are respectively called *ket*-$0$ and *ket*-$1$. Therefore, by linearity, a general qubit $|\psi\rangle$ is of the following form:

$$|\psi\rangle \;=\; \alpha|0\rangle + \beta|1\rangle\,,$$

where the complex coefficients $\alpha, \beta \in \mathbb{C}$ are such that $|\alpha|^2 + |\beta|^2 = 1$. We will see later that this notation is particularly handy for computations.

### 2.1.2 Pure States

A natural generalization of qubits leads to the notion of *pure quantum states*. These are again defined in terms of unit vectors, but here we do not specify what is the Hilbert space $\mathcal{H}$ other than being a finite-dimensional complex space. As for qubits, we can express an orthonormal basis of $\mathcal{H}$ in Dirac notation as $|0\rangle, \ldots, |d-1\rangle$ where $d = \dim(\mathcal{H})$, called *computational basis*. Then a pure state $|\psi\rangle$ is of the following form:

$$|\psi\rangle \;=\; \alpha_0|0\rangle + \cdots + \alpha_{d-1}|d-1\rangle\,,$$

where $\alpha_i \in \mathbb{C}$ and $\sum_i |\alpha_i|^2 = 1$. With this notation, we say that the state $|\psi\rangle$ is a *quantum superposition* of the states $|0\rangle, \ldots, |d-1\rangle$.

### 2.1.3 Mixed States

*Mixed states* represent the most general definition of a quantum state. They are defined as *density operators* over a Hilbert space $\mathcal{H}$, *i.e.* positive semi-definite (PSD) operators $\rho$ with unit trace, forming the following set:

$$\mathcal{D}(\mathcal{H}) := \left\{\rho \in \mathcal{B}(\mathcal{H}) \,:\, \rho \succcurlyeq \mathbf{0},\, \mathrm{Tr}[\rho] = 1\right\},$$

where $\mathcal{B}(\mathcal{H})$ is the set of bounded operators over $\mathcal{H}$. Recall that a matrix $M \in \mathcal{M}_d(\mathbb{C})$ is said to be positive semi-definite, denoted $M \succcurlyeq \mathbf{0}$, if it is Hermitian (or self-adjoint, *i.e.* $M^* = M$) and $z^* M z \geqslant 0$ for all non-zero vectors $z \in \mathbb{C}^d$. In particular, all of its eigenvalues are non-negative and can be expressed as $M = B^* B$ for some matrix $B \in \mathcal{M}_d(\mathbb{C})$.

**Pure States as Density Matrices.** Mixed states generalize the former two sections in the sense that a pure state $|\psi\rangle \in \mathcal{H}$ may be viewed as the projector $|\psi\rangle\langle\psi| \in \mathcal{D}(\mathcal{H})$, where $\langle\psi|$ is the Dirac notation for the adjoint of $|\psi\rangle$ and is called *bra-*$\psi$. For instance, in $\mathcal{H} = \mathbb{C}^2$, we have:

$$\langle 0| := |0\rangle^* = \begin{bmatrix} 1 & 0 \end{bmatrix} \quad \text{therefore} \quad |0\rangle\langle 0| := \begin{bmatrix} 1 \\ 0 \end{bmatrix} \begin{bmatrix} 1 & 0 \end{bmatrix} = \begin{bmatrix} 1 & 0 \\ 0 & 0 \end{bmatrix}.$$

Note that $|\psi\rangle\langle\varphi| = |\psi\rangle \cdot \langle\varphi|$ is the matrix product of the column vector $|\psi\rangle$ and the row vector $\langle\varphi|$. See also that, in the other way, we retrieve the usual scalar product $\langle\varphi| \cdot |\psi\rangle = \langle\varphi, \psi\rangle$, which is often denoted $\langle\varphi|\psi\rangle$. This is the reason why $\langle\varphi|$ and $|\psi\rangle$ are conveniently called *bra* and *ket*: put together they form a *braket* $\langle\varphi|\psi\rangle$.

**Extreme Mixed States.** The set $\mathcal{D}(\mathcal{H})$ of mixed states is convex, and its extreme points are exactly the pure states $|\psi\rangle\langle\psi|$. Actually, it is not hard to see that any mixed state is a convex mixture of pure states (hence their name!), using the Spectral Theorem:

**Theorem 2.1** (Spectral Theorem) — *A Hermitian matrix $M \in \mathcal{M}_d(\mathbb{C})$ only admits real eigenvalues $\lambda_1, \ldots, \lambda_d \in \mathbb{R}$ (possibly equal each other), and it can be decomposed in an orthonormal eigenbasis:*

$$M = \sum_{i=1}^{d} \lambda_i \, |\psi_i\rangle\langle\psi_i| \,,$$

*where the $|\psi_i\rangle \in \mathbb{C}^d$ are orthogonal eigenvectors of $M$ with unit norm.*

**Example 2.2** (Maximally Mixed State) — A common example is the *maximally mixed state*, defined as the normalized identity:

$$\frac{1}{d}\,\mathbb{I}_d = \frac{1}{d}\Big(|0\rangle\langle 0| + |1\rangle\langle 1| + \cdots + |d-1\rangle\langle d-1|\Big)\,.$$

Note that this state is not pure, but like any mixed state, it can be seen as the partial trace of a pure state. See the definition of the partial trace in Section 2.2.1.

**Bloch Sphere.** In the simplest non-trivial setting, when $\mathcal{H} = \mathbb{C}^2$, the mixed states can be viewed as points of the three-dimensional unit ball of $\mathbb{R}^3$. This representation is called *Bloch sphere*. It uses the identity matrix and Pauli matrices:

$$\mathcal{D}(\mathbb{C}^2) = \left\{ \frac{\mathbb{I}_2 + r_x\,\sigma_x + r_y\,\sigma_y + r_z\,\sigma_z}{2} \;:\; r = (r_x, r_y, r_z) \in \mathbb{R}^3,\; \|r\| \leqslant 1 \right\}.$$

Recall that the Pauli matrices express as follows:

$$\sigma_x := \begin{bmatrix} 0 & 1 \\ 1 & 0 \end{bmatrix}, \qquad \sigma_y := \begin{bmatrix} 0 & -i \\ i & 0 \end{bmatrix}, \qquad \text{and} \quad \sigma_z := \begin{bmatrix} 1 & 0 \\ 0 & -1 \end{bmatrix}. \tag{2.1}$$

The set of pure states $|\psi\rangle\langle\psi|$ exactly corresponds to the points of the sphere, *i.e.* when $\|r\| = 1$.

## 2.2 Quantum Entanglement

One of the many astonishing properties of quantum mechanics is the existence of *quantum entanglement*. This is mathematically described using the tensor product of Hilbert spaces and physically interpreted as the presence of two particles that are highly correlated, with numerous experimental verifications [ADR82; Hen+15; Wei+98].

In this section, we first present the general framework of multipartite systems (Section 2.2.1). We then define entanglement for pure and mixed states (Sections 2.2.2 and 2.2.3), describe various characterizations and criteria for entanglement (Section 2.2.4), and conclude with the monogamy-of-entanglement principle (Section 2.2.5).

> « *I cannot seriously believe in [quantum entanglement] because the theory cannot be reconciled with the idea that physics should represent a reality in time and space, free from spooky action at a distance.* » — Einstein (1947) [Hei73]

## 2.2.1 *Multipartite System*

A system $\mathcal{H}$ is said to be *multipartite* (or *compound*) if it can be written as the tensor product of $n \geqslant 2$ Hilbert spaces:

$$\mathcal{H} = \mathcal{H}_1 \otimes \cdots \otimes \mathcal{H}_n := \operatorname{span}\Big\{ v_1 \otimes \cdots \otimes v_n \ :\ v_k \in \mathcal{H}_k \text{ for all } k \Big\}.$$

Below, after presenting the formula of the tensor product for matrices, we list some useful facts for computation.

**Matrix Setting.** As we work in finite dimension, given a fixed basis, all the elements of $\mathcal{H}$ may be viewed as column vectors, and operators on $\mathcal{H}$ as matrices. Therefore it is convenient to have an explicit formula to compute the tensor product of matrices, also called *Kronecker product*. In the bipartite matrix setting $\mathcal{H}_1 \otimes \mathcal{H}_2$, *i.e.* when $n = 2$ and $\mathcal{H}_1, \mathcal{H}_2$ are matrix spaces, the tensor product $A \otimes B$ of two complex matrices $A$ and $B$ of respective sizes $p \times q$ and $r \times s$ is the complex matrix of size $pr \times qs$ resulting from the following block-decomposition:

$$A \otimes B := \begin{bmatrix} a_{11}\, B & \dots & a_{1q}\, B \\ \vdots & & \vdots \\ a_{p1}\, B & \dots & a_{pq}\, B \end{bmatrix} \in \mathcal{M}_{pr,qs}(\mathbb{C}). \tag{2.2}$$

In particular, the same computation also holds for vectors viewed as column matrices. This formula generalizes to the multipartite setting by associativity: $A \otimes B \otimes C = (A \otimes B) \otimes C = A \otimes (B \otimes C)$.

**Basic Computational Rules.** The tensor product "commutes" with the usual matrix product in the following sense:

$$\big(A \otimes B\big)\big(C \otimes D\big) \;=\; \big(AC\big) \otimes \big(BD\big)\,,$$

where $A, B, C, D$ are matrices with suitable sizes for matrix products. Note also that the trace and the adjoint behave well with the tensor product: $\mathrm{Tr}(A \otimes B) = \mathrm{Tr}(A)\,\mathrm{Tr}(B)$ and $(A \otimes B)^* = A^* \otimes B^*$. The space $\mathcal{H}_1 \otimes \mathcal{H}_2$ is again a Hilbert space, whose inner product is induced by the ones of $\mathcal{H}_1$ and $\mathcal{H}_2$ respectively:

$$\langle v_1 \otimes v_2 \,|\, w_1 \otimes w_2 \rangle_{\mathcal{H}_1 \otimes \mathcal{H}_2} \;=\; \langle v_1 \,|\, w_1 \rangle_{\mathcal{H}_1}\, \langle v_2 \,|\, w_2 \rangle_{\mathcal{H}_2}\,,$$

for vectors $v_1, w_1 \in \mathcal{H}_1$ and $v_2, w_2 \in \mathcal{H}_2$. As a consequence, norms follow the same rule:

$$\|v_1 \otimes v_2\|_{\mathcal{H}_1 \otimes \mathcal{H}_2} \;=\; \|v_1\|_{\mathcal{H}_1}\, \|v_2\|_{\mathcal{H}_2}\,.$$

It is also worth noting that the tensor product is bilinear:

$$\Big(\textstyle\sum_i \alpha_i v_i\Big) \otimes \Big(\textstyle\sum_j \beta_j w_j\Big) \;=\; \displaystyle\sum_{i,j} \alpha_i \beta_j \left(v_i \otimes w_j\right),$$

for finite indices $i, j$, coefficients $\alpha_i, \beta_j \in \mathbb{C}$, and vectors $v_i \in \mathcal{H}_1$, $w_j \in \mathcal{H}_2$.

**Partial Trace.** In a multipartite system $\mathcal{H} = \mathcal{H}_1 \otimes \cdots \otimes \mathcal{H}_n$, the partial trace on the first component of a mixed state $\rho \in \mathcal{D}(\mathcal{H})$ is defined as:

$$\mathrm{Tr}_1\big(\rho\big) \;:=\; \big(\mathrm{Tr} \otimes \mathbb{I}_{d_2} \otimes \cdots \otimes \mathbb{I}_{d_n}\big)(\rho)\,.$$

In contrast with the usual trace, note that the partial trace of a state is not a scalar, it is a quantum state in $\mathcal{D}(\mathcal{H}_2 \otimes \cdots \otimes \mathcal{H}_n)$. Similarly, the $i$-th partial trace $\mathrm{Tr}_i$ is when we trace out on the $i$-th Hilbert space $\mathcal{H}_i$. More generally, one can have partial trace over any subset $S$ of $\{1, .., n\}$, and we obtain $\mathrm{Tr}_S$. Notice that if we compose all the partial traces, we retrieve the usual trace: $\mathrm{Tr}_1 \circ \mathrm{Tr}_2 \circ \cdots \circ \mathrm{Tr}_n = \mathrm{Tr}$. If the parties are denoted with letters, *e.g.* $\mathcal{H} = \mathcal{H}_{\mathsf{A}} \otimes \mathcal{H}_{\mathsf{B}} \otimes \mathcal{H}_{\mathsf{C}}$ named after Alice, Bob, and Charlie, then we denote the partial traces as $\mathrm{Tr}_{\mathsf{A}}$, $\mathrm{Tr}_{\mathsf{B}}$, and $\mathrm{Tr}_{\mathsf{C}}$. Using the linearity and cyclicity of the trace, one has $\mathrm{Tr}(\rho) = \mathrm{Tr}(\mathbb{I}_d\, \rho) = \mathrm{Tr}\big(\sum_{i=1}^d |i\rangle\langle i|\rho\big) = \sum_{i=1}^d \mathrm{Tr}(\langle i|\rho|i\rangle) = \sum_{i=1}^d \langle i|\rho|i\rangle$ (it also simply follows from the fact that the trace is the sum

of the diagonal elements). Therefore, one has the following expressing for the partial trace:

$$\mathrm{Tr}_1(\rho) = \sum_{i=1}^{d} \Big( \langle i| \otimes \mathbb{I}_{d_2} \otimes \cdots \otimes \mathbb{I}_{d_n} \Big) \, \rho \, \Big( |i\rangle \otimes \mathbb{I}_{d_2} \otimes \cdots \otimes \mathbb{I}_{d_n} \Big) .$$

This operator $\mathrm{Tr}_1(\rho)$ can also be characterized as the only operator in $\mathcal{B}(\mathcal{H}_2 \otimes \cdots \otimes \mathcal{H}_n)$ such that:

$$\forall A \in \mathcal{B}(\mathcal{H}_2 \otimes \cdots \otimes \mathcal{H}_n) , \qquad \langle \mathrm{Tr}_1(\rho), A \rangle_{\mathcal{H}_2 \otimes \cdots \otimes \mathcal{H}_n} = \langle \rho, \mathbb{I}_{d_1} \otimes A \rangle_{\mathcal{H}} ,$$

with the Frobenius inner product $\langle A, B\rangle_{\mathcal{H}} := \mathrm{Tr}(AB^*)$ on $\mathcal{B}(\mathcal{H})$. Examples of computations of partial traces are given in Example 2.10, and it will be particularly useful for defining local measurements in Section 2.3.3.

**Computational Basis.** Given computational bases $\{|i_1\rangle\}_{i_1}, .., \{|i_n\rangle\}_{i_n}$ of $\mathcal{H}_1, .., \mathcal{H}_n$ respectively, the computational basis of $\mathcal{H} = \mathcal{H}_1 \otimes \cdots \otimes \mathcal{H}_n$ is defined as:

$$\Big\{ |i_1\rangle \otimes \cdots \otimes |i_n\rangle \Big\}_{i_1,..,i_n} .$$

This is again an orthonormal basis of $\mathcal{H}$ and it is composed of pure states. In Dirac notation, a common shorthand for the state $|0\rangle \otimes |0\rangle$ is $|00\rangle$, called *ket*-$00$, and similarly for the other states of the computational basis.

## 2.2.2 *Entanglement of Pure States*

The notion of entanglement is closely linked with the one of tensor rank.

**Tensor Rank.** By definition of computational basis, any state $|\psi\rangle \in \mathcal{H}_1 \otimes \cdots \otimes \mathcal{H}_n$ can be expressed as a linear combination of simple tensors:

$$|\psi\rangle = \sum_{i=1}^{r} \lambda_i \left( |\varphi_i^{(1)}\rangle \otimes \cdots \otimes |\varphi_i^{(n)}\rangle \right) ,$$

for some integer $r \in \mathbb{N}$, coefficients $\lambda_i \in \mathbb{C}$, and states $|\varphi_i^{(k)}\rangle \in \mathcal{H}_k$. The *rank* of $|\psi\rangle$ is then defined as the minimal such $r \in \mathbb{N}$. This definition naturally extends the one of matrix rank, which is, for a matrix $A$, the minimal integer $r \in \mathbb{N}$ such that $A = \sum_{i=1}^{r} x_i y_i^*$ for some vectors $x_i, y_i$. Nevertheless, as opposed to the matrix rank that can be efficiently computed via

Gaussian elimination, the tensor rank becomes NP-hard to compute when $n \geqslant 3$ [Hås90]. It is easy to see that the rank is sub-multiplicative:

$$\operatorname{rank}\Big(|\psi\rangle \otimes |\varphi\rangle\Big) \leqslant \operatorname{rank}\Big(|\psi\rangle\Big) \operatorname{rank}\Big(|\varphi\rangle\Big).$$

Moreover, if $d_k$ denotes the dimension of $\mathcal{H}_k$ for all $k$, then we can always find $r \leqslant d_1 \cdots d_n / \max\{d_1, .., d_n\}$ [BFŻ23, Thm. 5.1]. Interestingly, as opposed to the matrix rank, the sublevel sets $\big\{|\psi\rangle : \operatorname{rank}(\psi) \leqslant k\big\}$ of the tensor rank for fixed $k \in \mathbb{N}$ are not topologically closed in general: for instance, one may find a sequence of rank-$2$ tensors converging to a rank-$3$ tensor [BFŻ23, eq. (20)].

**Entanglement.** A pure state $|\psi\rangle$ is said to be *entangled* if its rank satisfies $\operatorname{rank}(|\psi\rangle) \geqslant 2$. Otherwise, it is of the form $|\psi\rangle = |\varphi^{(1)}\rangle \otimes \cdots \otimes |\varphi^{(n)}\rangle$ and it is called *separable*.

**Important Fact 2.3 —** *Pure entangled states exist.*

**Example 2.4** (Maximally Entangled State) **—** In $\mathcal{H} = \mathbb{C}^2 \otimes \mathbb{C}^2$, a common example of entangled state is the *maximally entangled state*[1]:

$$|\Omega\rangle := \frac{|00\rangle + |11\rangle}{\sqrt{2}},$$

whose rank is precisely $2$. In contrast, the states $|00\rangle$ and $|11\rangle$ are examples of separable states. Note that, sometimes, one needs to factorize to see that the state is separable:

$$\frac{|00\rangle + |01\rangle}{\sqrt{2}} = \frac{|0\rangle \otimes \big(|0\rangle + |1\rangle\big)}{\sqrt{2}},$$

which has rank $1$.

**Example 2.5** (GHZ and W States) **—** In $\mathcal{H} = \mathbb{C}^2 \otimes \mathbb{C}^2 \otimes \mathbb{C}^2$, two standard examples of entangled states are the following ones:

$$|\mathrm{GHZ}\rangle := \frac{|000\rangle + |111\rangle}{\sqrt{2}} \quad \text{and} \quad |\mathrm{W}\rangle := \frac{|001\rangle + |010\rangle + |100\rangle}{\sqrt{3}}.$$

[1] See a discussion in the paragraph about LOCC at page 35 to justify why we use the article “the” before “maximally entangled state”.

The GHZ state is named after Greenberger, Horne, and Zeilinger [GHZ89]. These states have respective ranks of $2$ and $3$ and thus are entangled.

**Example 2.6** (Maximally Entangled State) — In $\mathcal{H} = \mathbb{C}^d \otimes \mathbb{C}^d$ with $d \geqslant 2$, the maximally entangled state is generalized to:

$$|\Omega\rangle := \frac{|00\rangle + |11\rangle + \cdots + |d-1, d-1\rangle}{\sqrt{d}},$$

whose rank is precisely $d$.

### 2.2.3 Entanglement of Mixed States

For mixed states, we differentiate three types of states: product states, separable states, and entangled states.

**Definition 2.7** (Product, Separable, and Entangled States) — *Consider $\mathcal{H} = \mathcal{H}_1 \otimes \cdots \otimes \mathcal{H}_n$. A mixed state $\rho \in \mathcal{D}(\mathcal{H})$ is said to be* product *if it is of the form*

$$\rho = \sigma_1 \otimes \cdots \otimes \sigma_n .$$

*where the $\sigma_k \in \mathcal{D}(\mathcal{H}_k)$ are quantum states. It is said to be* separable *if it is a convex combination of product states,* i.e. *of the following form:*

$$\rho = \sum_i \alpha_i \left(\sigma_1^{(i)} \otimes \cdots \otimes \sigma_n^{(i)}\right),$$

*where the index $i$ is finite, and the coefficients $\alpha_i \in \mathbb{R}_{\geqslant 0}$ are non-negative and verify $\sum_i \alpha_i = 1$. Finally, the state is said to be* entangled *if it is not separable.*

**Important Fact 2.8** — *Mixed entangled states exist.*

**Example 2.9** (Maximally Entangled State) — When $\mathcal{H} = \mathbb{C}^2 \otimes \mathbb{C}^2$, the maximally entangled state $|\Omega\rangle$ from the previous subsection translates as follows in terms of density operators:

$$\omega := |\Omega\rangle\langle\Omega| = \frac{|00\rangle\langle 00| + |00\rangle\langle 11| + |11\rangle\langle 00| + |11\rangle\langle 11|}{2} = \frac{1}{2}\begin{bmatrix} 1 & 0 & 0 & 1 \\ 0 & 0 & 0 & 0 \\ 0 & 0 & 0 & 0 \\ 1 & 0 & 0 & 1 \end{bmatrix}.$$

This state $\omega$ is entangled, whereas the *maximally mixed state* $\frac{\mathbb{I}_4}{4}$ is separable. Convex combinations of the two form the *isotropic states*, defined as:

$$\alpha\,\omega + (1-\alpha)\,\frac{\mathbb{I}_4}{4} \in \mathcal{D}(\mathcal{H})\,, \tag{2.3}$$

for a real coefficient $\alpha \in [0,1]$.[2] It is known that isotropic states are entangled exactly for coefficient $\alpha$ ranging in $(\frac{1}{3},1]$. More generally, when $\mathcal{H} = \mathbb{C}^d \otimes \mathbb{C}^d$ with $d \geqslant 2$, the maximally entangled state is $\omega = \frac{1}{d}\sum_{i,j=0}^{d-1} |ii\rangle\langle jj|$ and the maximally mixed state $\mathbb{I}_{d^2}/d^2$. In this case, the isotropic states are entangled exactly for the following range of $\alpha$ [HH99]:

$$\alpha > \frac{1}{d+1}\,.$$

**Example 2.10** (Computations of the Partial Trace) — Let us compute the first partial trace of the maximally entangled state $\omega \in \mathcal{D}(\mathbb{C}^2 \otimes \mathbb{C}^2)$ defined in Example 2.9. We have:

$$\mathrm{Tr}_{\mathsf{A}}\big(\omega\big) \;=\; \big(\mathrm{Tr}\otimes\mathbb{I}_2\big)(\omega) \;=\; \sum_{i=0}^{1}\Big(\langle i|\otimes\mathbb{I}_2\Big)\,\omega\,\Big(|i\rangle\otimes\mathbb{I}_2\Big) \;=\; \sum_{i=0}^{1}\frac{|0\rangle\langle 0| + |1\rangle\langle 1|}{2} \;=\; \frac{\mathbb{I}_2}{2}\,,$$

where we used the fact that the computational basis is orthonormal: $\langle 0|0\rangle = \langle 1|1\rangle = 1$ and $\langle 0|1\rangle = \langle 1|0\rangle = 0$. We obtain the same result for the second partial trace $\mathrm{Tr}_2$. Hence, interestingly, the partial trace of the maximally entangled state is the maximally mixed state. Here are some computations of the partial trace of the GHZ and W states lying in $\mathcal{D}(\mathbb{C}^2 \otimes \mathbb{C}^2 \otimes \mathbb{C}^2)$:

$$\begin{aligned}
\mathrm{Tr}_{\mathsf{A}}\Big(|\mathsf{GHZ}\rangle\langle\mathsf{GHZ}|\Big) &= \frac{|00\rangle\langle 00| + |11\rangle\langle 11|}{2} = \frac{1}{2}\begin{bmatrix} 1 & 0 & 0 & 0 \\ 0 & 0 & 0 & 0 \\ 0 & 0 & 0 & 0 \\ 0 & 0 & 0 & 1 \end{bmatrix}, \\
\mathrm{Tr}_{\mathsf{A}}\Big(|\mathsf{W}\rangle\langle\mathsf{W}|\Big) &= \frac{|00\rangle\langle 00| + |01\rangle\langle 01| + |01\rangle\langle 10| + |10\rangle\langle 01| + |10\rangle\langle 10|}{3} \\
&= \frac{1}{3}\begin{bmatrix} 1 & 0 & 0 & 0 \\ 0 & 1 & 1 & 0 \\ 0 & 1 & 1 & 0 \\ 0 & 0 & 0 & 0 \end{bmatrix}, \\
\mathrm{Tr}_{\{\mathsf{A},\mathsf{B}\}}\Big(|\mathsf{W}\rangle\langle\mathsf{W}|\Big) &= \frac{2\,|0\rangle\langle 0| + |1\rangle\langle 1|}{3} = \frac{1}{3}\begin{bmatrix} 2 & 0 \\ 0 & 1 \end{bmatrix}.
\end{aligned}$$

[2] One can even extend the coefficients $\alpha$ to a larger set $\left[-\frac{1}{d^2-1},1\right]$ so that the isotropic state in eq. (2.3) remains in the density set $\mathcal{D}(\mathcal{H})$.

Finally, for product states, note that $\mathrm{Tr}_{\mathsf{A}}(\rho_{\mathsf{A}} \otimes \rho_{\mathsf{B}}) = \rho_{\mathsf{B}}$.

**Comparing Entanglements via LOCC.** Given two states $\rho, \sigma \in \mathcal{D}(\mathcal{H})$, one may wonder which of the two is more entangled. A standard approach consists in verifying if one state, say $\rho$, can be transformed into the other, $\sigma$, by *Local Operations and Classical Communication* (LOCC), in which case we write:

$$\rho \xrightarrow[\text{LOCC}]{} \sigma\,.$$

As the name suggests, in such a protocol, one can perform local operations on $\rho$ on each subsystem $\mathcal{H}_i$ (these local operations are described and defined in Section 2.3.3), as well as communicating an unlimited amount of classical bits between the $n$ subsystems. If indeed $\rho$ can be transformed into $\sigma$ with such a protocol, then we can say that $\rho$ is more entangled than $\sigma$, because such operations cannot increase the entanglement of a state. In other words, anything we can do with $\sigma$ and LOCC operations, we can also achieve with $\rho$ and LOCC operations. In the review [PV14], the authors even state that "entanglement may be defined as the sort of correlations that may not be created by LOCC alone." Now that we can compare the amount of entanglement between two states, one may wonder if there is a state more entangled than any other. In the bipartite setting $\mathcal{H} = \mathbb{C}^d \otimes \mathbb{C}^d$, this is indeed true: the maximally entangled state $\omega = \frac{1}{d}\sum_{i,j=0}^{d-1} |ii\rangle\langle jj|$ is the unique state, up to unitary transformation, that is more entangled than any other pure or mixed state (which is the reason why we can use the article "the" before maximally entangled state). In contrast, the maximally mixed state $\mathbb{I}_{d^2}/d^2$ is less entangled than any other state since it is not entangled, hence we have:

$$\forall \rho \in \mathcal{D}(\mathbb{C}^d \otimes \mathbb{C}^d)\,, \qquad \omega \xrightarrow[\text{LOCC}]{} \rho \xrightarrow[\text{LOCC}]{} \frac{\mathbb{I}_{d^2}}{d^2}\,.$$

Nevertheless, in the more general multipartite setting $\mathcal{H} = \mathcal{H}_1 \otimes \cdots \otimes \mathcal{H}_n$, we do not have such a simple statement. Many non-equivalent definitions of maximal entanglement exist depending on the choice of entanglement measure, which is one of the reasons why multipartite entanglement is significantly more challenging to study. Another limitation is that the LOCC order is partial, even in the bipartite scenario. More precisely, there exist some states, for instance $|\psi\rangle = \frac{1}{\sqrt{2}}|00\rangle + \frac{2}{\sqrt{10}}|11\rangle + \frac{1}{\sqrt{10}}|22\rangle$ and $|\phi\rangle = \frac{3}{\sqrt{15}}|00\rangle + \frac{1}{\sqrt{5}}|11\rangle + \frac{1}{\sqrt{5}}|22\rangle$ [Nie99], that cannot be compared:

$$|\psi\rangle \underset{\text{LOCC}}{\not\longrightarrow} |\phi\rangle \qquad \text{and} \qquad |\phi\rangle \underset{\text{LOCC}}{\not\longrightarrow} |\psi\rangle\,,$$

suggesting that they lie in different entanglement classes. These entanglement classes have been studied and characterized using majorization techniques of the Schmidt coefficients of the states involved [Har99; JP99; Nie99; Vid99]. Note that LOCC transformation was also studied in the noisy case [VJN00] and in the asymptotic regime $\rho^{\otimes k} \rightarrow_{\text{LOCC}} \sigma^{\otimes m}$ [Ben+96a].

### 2.2.4 *Characterizing Entanglement*

The problem of determining whether a given state $\rho \in \mathcal{D}(\mathcal{H})$ is separable or entangled is fundamental in QIT because entangled states can bring advantages over their classical counterpart in terms of information processing tasks. This problem is called *Quantum Separability Problem* (QUSEP), and it is known to be hard to solve:

> **Fact 2.11** ([Gha10; Gur03]) — *Determining whether a state $\rho$ is separable or entangled is NP-hard in general.*

Below, we present some characterizations of entanglement in terms of tensor norms, and then provide examples of sufficient conditions, called *entanglement criteria*, that are easier to compute. Find a review on this topic in [GT09].

**Characterization of Pure Entangled States.** Consider a vector $x \in \mathcal{H} = \mathcal{H}_1 \otimes \cdots \otimes \mathcal{H}_n$, and define two norms, namely the *injective norm* $\|\cdot\|_\varepsilon$ and the *projective norm* $\|\cdot\|_\pi$, which are dual from each other:

$$\|x\|_\varepsilon := \sup\Big\{\langle \alpha_1 \otimes \cdots \otimes \alpha_n \,|\, x\rangle \;:\; \alpha_k \in \mathcal{H}_k^*,\, \|\alpha_k\| \leqslant 1\Big\},$$
$$\|x\|_\pi := \inf\Big\{\sum_{i=1}^r \|a_1^{(i)}\| \cdots \|a_n^{(i)}\| \;:\; r \in \mathbb{N},\, a_k^{(i)} \in \mathcal{H}_k,\, x = \sum_{i=1}^r a_1^{(i)} \otimes \ldots \otimes a_n^{(i)}\Big\}.$$

They are examples of tensor norms because they satisfy the relation $\|x_1 \otimes \cdots \otimes x_n\|_\varepsilon = \|x_1 \otimes \cdots \otimes x_n\|_\pi = \|x_1\| \cdots \|x_n\|$ for rank-1 tensors. Moreover, they are extremal in the sense that, for any other tensor norm $\|\cdot\|$ on $\mathcal{H}$, we have [Rya02, page 127]:

$$\forall x \in \mathcal{H}, \qquad \|x\|_\varepsilon \leqslant \|x\| \leqslant \|x\|_\pi .$$

Using these definitions, there is the following characterization of entanglement for a pure state $|\psi\rangle \in \mathcal{H}$:

$$|\psi\rangle \text{ is separable} \qquad \Longleftrightarrow \qquad \Big\||\psi\rangle\Big\|_\varepsilon = \Big\||\psi\rangle\Big\|_\pi = 1 .$$

In addition to characterizing entanglement, these norms are used to quantify it, notably with the *geometric measure of entanglement*, which is defined as $G\big(|\psi\rangle\big) := -2\log\big\||\psi\rangle\big\|_\varepsilon$ [SHI95; WG03; ZCH10]. In the bipartite scenario when $n = 2$, these two norms are efficiently computable by performing a singular value decomposition (in polynomial time), because they express respectively as the largest singular value and the sum of all the singular values of the corresponding matrix [GVL96]. Nevertheless, they are NP-hard to compute in the general multipartite setting [FL17; HL13].

**Characterization of Mixed Entangled States.** We denote by $\mathcal{H}_1 \otimes_\pi \cdots \otimes_\pi \mathcal{H}_n$ the Banach space $\mathcal{H}_1 \otimes \cdots \otimes \mathcal{H}_n$ endowed with the norm $\|\cdot\|_\pi$. As we work in finite dimension, we have $\mathcal{B}(\mathcal{H}_k) \cong \mathcal{M}_{d_k}(\mathbb{C})$ and we set the structure induced by the *Schatten* 1*-norm, a.k.a. nuclear norm*:

$$\|A\|_1 := \operatorname{Tr}\sqrt{A^*A}\,.$$

This Banach space is denoted $S_1^d$. One can also restrict it to the set of self-adjoint matrices, with the same norm, to give rise to $S_{1,sa}^d$. We obtain the following characterization of entanglement for a mixed state $\rho \in$

$\mathcal{D}(\mathcal{H})$ [Rud00]:

$$\rho \text{ is separable} \iff \|\rho\|_{S_{1,sa}^{d_1}\otimes_\pi\cdots\otimes_\pi S_{1,sa}^{d_n}} = 1 \iff \|\rho\|_{S_1^{d_1}\otimes_\pi\cdots\otimes_\pi S_1^{d_n}} = 1\,.$$

However, this norm is NP-hard to compute in general [FL17; HL13]. A common technique to tackle this issue is to consider other tensor norms, easier to compute, leading to sufficient or necessary conditions for entanglement [JLN22].

**Entanglement Criterion #1: Bell Inequalities.** One of the first historical entanglement criteria was formulated by Bell [Bel64] and later by Clauser, Horne, Shimony, and Holt [CHSH69]. Given a multipartite quantum state $\rho \in \mathcal{D}(\mathcal{H}_1 \otimes \cdots \otimes \mathcal{H}_n)$, it is possible to measure it according to some mathematical rules, see Section 2.3. This gives rise to the statistics of obtaining certain outputs $a_1,..,a_n$, one for each party, given some inputs $x_1,..,x_n$. These statistics are studied in the context of *nonlocal boxes*, see Section 3.1 for more details. Now, the criterion of *Bell inequalities* consists in linear inequalities in terms of the $a_i$'s and $x_j$'s that only entangled states can violate. Therefore, it is an efficient and sufficient way of checking entanglement. Nevertheless, it is not a necessary condition since there exist entangled states that do not violate any of the Bell inequalities [Aug+15; Bar02; Wer89b].

**Entanglement Criterion #2: PPT.** In the bipartite scenario, *i.e.* when $n = 2$, Peres introduced a simple tool to detect entanglement [Per96]. It relies on the notion *partial transposition* of a state $\rho \in \mathcal{D}(\mathcal{H}_1 \otimes \mathcal{H}_2)$, defined as the operation turning $\rho$ into $\rho^\Gamma := (\mathbb{I}_{\mathcal{H}_1} \otimes \mathrm{transpose}_{\mathcal{H}_2})(\rho)$, *i.e.* leaving the first component the same and applying transposition on the second component. The operator $\rho^\Gamma$ has again unit trace, and Peres showed that for separable states it is again positive semi-definite. This leads to the *PPT criterion* (Positive Partial Transpose):

$$\rho \text{ is separable} \implies \rho^\Gamma \text{ is a quantum state}\,.$$

In other words, we have the following sufficient condition: if $\rho^\Gamma$ admits a negative eigenvalue, then $\rho$ must be entangled. For instance, consider $\rho = \omega$ the maximally entangled state in $\mathcal{D}(\mathbb{C}^d \otimes \mathbb{C}^d)$. By definition, we have:

$$\omega = \frac{1}{d}\sum_{i,j=0}^{d-1} |ii\rangle\langle jj|\,, \qquad \text{so} \qquad \omega^\Gamma = \frac{1}{d}\sum_{i,j=0}^{d-1} |ij\rangle\langle ji|\,.$$

This operator $\omega^\Gamma$ is called *flip operator* since on simple tensors $x \otimes y$, it "flips" the components: $\omega^\Gamma(x \otimes y) = \frac{1}{d}(y \otimes x)$. Therefore, the difference $x \otimes y - y \otimes x$ is an eigenvector associated with the eigenvalue $-\frac{1}{d}$, and from the PPT criterion we retrieve the fact that $\omega$ is entangled. Later, Horodecki, Horodecki, and Horodecki proved a refinement of the criterion: in addition to being sufficient in the general case, this condition is also necessary when $\dim(\mathcal{H}_1 \otimes \mathcal{H}_2) \leqslant 6$, but not in higher dimensions [HHH96].

**Other Characterizations and Criteria.** There exist many other characterizations and criteria for entanglement. Here is a non-exhaustive list: the computable cross norm or realignment criterion [CW03; Rud04], the $k$-extendibility criterion [DPS04; Lan16], the range criterion [Hor97], majorization criterion [NK01], the symmetric extensions methods [DPS02; Wer89a], based on covariance matrices [Güh+07; HT03], using the expectation value matrix [SV05], via entanglement testers [JLN22], and based on SIC-POVMs [Sha+18]. Note that most of them are designed for the bipartite case, only a few of them are well-suited for genuinely multipartite entanglement.

### *2.2.5 Monogamy of Entanglement*

The principle of *monogamy of entanglement* (MoE) was introduced by Coffman, Kundu, and Wootters in [CKW00], based on the following idea:

> **Intuition 2.12** (Monogamy of Entanglement) — *Assume that three parties Alice, Bob, and Charlie share a state $\rho \in \mathcal{D}(\mathcal{H}_\mathsf{A} \otimes \mathcal{H}_\mathsf{B} \otimes \mathcal{H}_\mathsf{C})$. If this state is very entangled between two parties, say, Alice and Bob, then it cannot be significantly entangled between two other parties, like Alice and Charlie.*

This phenomenon is intrinsic to quantum information theory since it does not occur in classical theory, where a system can be strongly correlated with several others. It has applications in nonlocal games and quantum cryptography, see examples in Chapters 3 and 5. Now, when it comes to formalizing this principle, there is no one standard formulation. Indeed, as the definition of maximal entanglement is not unique in the general multipartite setting [PV14], it gives rise to different versions of the MoE principle, all with the same underlying intuition but with different viewpoints. Below, we present three non-equivalent formalizations of this principle.

**Formalization #1: Tangle Inequality [CKW00].** This is the original version of the MoE principle. Given a tri-qubit state $\rho_{\mathsf{ABC}} \in \mathcal{D}(\mathbb{C}^2 \otimes \mathbb{C}^2 \otimes \mathbb{C}^2)$ shared by Alice, Bob, and Charlie, the principle is formulated in terms of the *tangle*, denoted $\tau_{\mathsf{AB}}$, which is a bipartite measure of entanglement (also known as *concurrence* in earlier work when taking its square root [Woo98]). This measure $\tau_{\mathsf{AB}}$ is defined as follows for pairs of qubits $|\psi\rangle \in \mathbb{C}^2 \otimes \mathbb{C}^2$:

$$\tau_{\mathsf{AB}}\big(|\psi\rangle\big) \;:=\; \big|\langle\psi\,|\,\tilde{\psi}\rangle\big|^2 \,,$$

where $|\tilde{\psi}\rangle := (\sigma_y \otimes \sigma_y)\,|\psi^*\rangle$ and $|\psi^*\rangle$ is the complex conjugate of $|\psi\rangle$. The expression of $\tau_{\mathsf{AB}}$ can be extended to mixed states of $\mathcal{D}(\mathbb{C}^2 \otimes \mathbb{C}^2)$ and to pure tripartite qubits of $\mathbb{C}^2 \otimes \big(\mathbb{C}^2 \otimes \mathbb{C}^2\big)$. It has minimal value $0$, achieved precisely by separable states, and maximal value $1$, attained exactly by maximally entangled states. Interestingly, Coffman, Kundu, and Wootters prove the following relation between the measures, giving a first formalization of the MoE principle [CKW00]:

$$\tau_{\mathsf{AB}}(\rho) + \tau_{\mathsf{AC}}(\rho) \;\leqslant\; \inf_{\rho=\sum_i p_i\,|\psi_i\rangle\langle\psi_i|} \sum_i p_i\,\tau_{\mathsf{A(BC)}}(\psi_i)\,.$$

This formula can be understood in that if the entanglement between Alice and Bob is high, then it can only be small between Alice and Charlie. Note that in this formalization of MoE, the definition of maximal entanglement is taken from the *entanglement-of-formation* measure, which is intended to quantify the amount of quantum communication required to create a given state [Ben+96b].

**Formalization #2: State Extendibility [DPS04; Ter04].** In the bipartite setting, a property of the maximally entangled state $\omega = |\Omega\rangle\langle\Omega| \in \mathcal{D}(\mathcal{H}_{\mathsf{A}} \otimes \mathcal{H}_{\mathsf{B}})$, with $|\Omega\rangle = \frac{1}{\sqrt{2}}\big(|00\rangle + |11\rangle\big)$, is that it can be viewed the partial trace of a state $\omega' \in \mathcal{D}(\mathcal{H}_{\mathsf{A}} \otimes \mathcal{H}_{\mathsf{B}} \otimes \mathcal{H}_{\mathsf{C}})$ in the tripartite setting:

$$\omega = \mathrm{Tr}_{\mathsf{C}}\big(\omega'\big)\,.$$

For instance, consider $\omega' = \omega \otimes \sigma$ for any state $\sigma \in \mathcal{D}(\mathcal{H}_{\mathsf{C}})$. Actually, one can show that any state $\omega'$ satisfying the above equation is necessary of the form $\omega' = \omega \otimes \sigma$, and it yields that $\mathrm{Tr}_{\mathsf{B}}\big(\omega'\big) = \frac{\mathbb{I}}{2} \otimes \sigma \neq \omega$. So the state $\omega$ is said to be not $2$-extendible. More generally, a bipartite state $\rho \in \mathcal{D}(\mathcal{H}_{\mathsf{A}} \otimes \mathcal{H}_{\mathsf{B}})$ is said to be $k$-*extendible*, for an integer $k \geqslant 2$, if there exists a state $\rho' \in \mathcal{D}(\mathcal{H}_\mathsf{A} \otimes \mathcal{H}_{\mathsf{B}_1} \otimes \cdots \otimes \mathcal{H}_{\mathsf{B}_k})$, called *extension* of $\rho$, such that we always retrieve $\rho$ after computing the partial trace on all parties but $\mathsf{B}_i$ and A:

$$\forall i \in \{1, .., k\}\,, \qquad \mathrm{Tr}_{\{\mathsf{B}_1,..,\mathsf{B}_k\}\setminus\{\mathsf{B}_i\}}(\rho') \;=\; \rho\,.$$

This gives rise to the following formalization of MoE: A bipartite state $\rho$ cannot be both entangled and $k$-extendible for all $k \geqslant 2$. More precisely, the state $\rho$ is entangled *if, and only if,* it is not $k$-extendible for some $k \geqslant 2$ [DPS04]. Find a generalization to graph-extendibility in [All+24].

**Formalization #3: Reversable Unitary Evolution [Cul22].** This formalization builds on the following characterization of maximal entanglement: maximally entangled states $|\psi\rangle \in \mathcal{H}_\mathsf{A} \otimes \mathcal{H}_\mathsf{B}$ are those for which unitary evolution on one system can be reversed by an operation on the other system, *i.e.* for all unitary $U_\mathsf{A} \in \mathcal{U}(\mathcal{H}_\mathsf{A})$, there exists another unitary $U_\mathsf{B} \in \mathcal{U}(\mathcal{H}_\mathsf{B})$ such that $(U_\mathsf{A} \otimes U_\mathsf{B})\,|\psi\rangle = |\psi\rangle$. Then, the arising MoE principle is as follows: When $\mathcal{H}_\mathsf{A} = \mathcal{H}_\mathsf{B} = \mathcal{H}_\mathsf{C} = \mathbb{C}^d$, there does not exist any state $|\psi\rangle \in \mathcal{H}_\mathsf{A}$ such that, for all $U \in \mathcal{U}_\mathsf{A}(\mathcal{H}_\mathsf{A})$, there exist $U_\mathsf{B} = U_\mathsf{C} \in \mathcal{U}(\mathbb{C}^d)$ such that:

$$(U_\mathsf{A} \otimes \mathbb{I}_\mathsf{B} \otimes \mathbb{I}_\mathsf{C})\,|\psi\rangle \;=\; (\mathbb{I}_\mathsf{A} \otimes U_\mathsf{B} \otimes U_\mathsf{C})\,|\psi\rangle\,.$$

## 2.3 Quantum Measurements

A *measurement* is the process by which classical information is extracted from a quantum state. This is a specific instance of a *quantum channel*, *i.e.* a quantum state transformation, which we present in Section 2.4.

In this section, we first introduce the measurement of quantum observables (Section 2.3.1), then generalize to PVMs, POVMs, and other types of measurements (Section 2.3.2), and finally, we explore local measurements in multipartite scenarios (Section 2.3.3).

### *2.3.1 Measuring a Quantum Observable*

Given a quantum state $\rho \in \mathcal{D}(\mathbb{C}^d)$, the physical quantities that we can measure are called *observables*. They are mathematically modeled as follows:

**Postulate 2.13** (Quantum Observable) — *A* quantum observable *on* $\mathbb{C}^d$ *is a Hermitian matrix* $A = A^* \in \mathcal{M}$.

Using the Spectral Theorem (Theorem 2.1), an observable can be decomposed as $A = \sum_{i=1}^{d} \lambda_i \, |\psi_i\rangle\langle\psi_i|$. To avoid repeating eigenvalues we may rewrite it as follows:

$$A = \sum_{i=1}^{k} \lambda_i \, P_i \,, \tag{2.4}$$

where $k \leqslant d$, where the $\lambda_i$ are the pair-wise distinct real eigenvalues of $A$, and the $P_i \in \mathcal{M}$ are the orthogonal projection on eigen-subspaces $E_{\lambda_i}$ of $A$. These matrices $P_i$ are called *spectral projections*. Notice that $P_i^2 = P_i$ and $P_i P_j = \mathbf{0}$ for all $i \neq j$, and that $\sum_i P_i = \mathbb{I}_d$, which characterize what we call below projection-valued measures (PVMs). The set of possible values that the observable $A$ can take is finite, it is defined in terms of its spectrum:

**Postulate 2.14** (Values of an Observable) — *The* values *of an observable* $A$ *are all its distinct eigenvalues* $\lambda_1, .., \lambda_k$.

**Example 2.15** (Spin) — The Pauli-$\sigma_z$ operator, introduced at page 28, is the observable corresponding to the measurement of the spin along the $z$-axis for a spin-$1/2$ particle:

$$\sigma_z = \begin{bmatrix} 1 & 0 \\ 0 & -1 \end{bmatrix} .$$

The values of this observable are $+1$ and $-1$, associated to the projectors $|0\rangle\langle 0|$ and $|1\rangle\langle 1|$ respectively. The value $+1$ corresponds to the particle being in spin "up", and $-1$ to spin "down."

Quantum phenomena are intrinsically probabilistic. As a consequence, in contrast with classical results which are deterministic, quantum measurements lead to stochastic results:

**Postulate 2.16** (Stochastic Outcomes) — *The values* $\lambda_i$ *of an observable* $A$ *are distributed according to the following law:*

$$\mathbb{P}\Big(\textit{obtaining "}\lambda_i\textit{" when measuring } A \textit{ on } \rho\Big) = \mathrm{Tr}\big(P_i \, \rho\big) .$$

It means that the value cannot be physically predicted in advance: we can only know its probability distribution.

**Collapse of the Wave Packet.** A fascinating phenomenon occurs automatically right after performing a measurement on a state: the state changes. This is again typical of quantum mechanics, it never happens in the classical setting. This phenomenon is called the *collapse of the wave packet* and was first observed by experimentalists in [Bru+96; Ton+89]. It is formalized as follows:

**Postulate 2.17** (Collapse of the Wave Packet) — *If the value "$\lambda_i$" is observed when measuring $A = \sum_i \lambda_i P_i$ on a quantum state $\rho$, the state is immediately transformed into the following new state:*

$$\tilde{\rho} := \frac{P_i \, \rho \, P_i}{\mathrm{Tr}\big(P_i \, \rho\big)}\,.$$

**Remark 2.18** (Measurement Incompatibility) — If two observables $A$ and $B$ do not commute, *i.e.* if $[A, B] := AB - BA \neq \mathbf{0}$, then the matrices are not diagonalizable in the same basis and we cannot measure them simultaneously. We say that these observables induce *incompatible measurements*. More precisely, one has the following inequality called *Heisenberg's uncertainty principle* [Hei27; Rob29]:

$$\Delta_\psi A \cdot \Delta_\psi B \geqslant \frac{1}{2} \left| \big\langle [A, B] \big\rangle_\psi \right|,$$

where $\Delta_\psi A := \sqrt{\langle A^2 \rangle_\psi - \langle A \rangle_\psi^2}$ is the standard deviation of $A$ and $\langle A \rangle_\psi := \langle \psi | \, A \, | \psi \rangle$ is the expectation value of $A$ over the quantum state $|\psi\rangle$. A canonical example is given by the position $\hat{x}$ and momentum $\hat{p}$ operators, which satisfy $[\hat{x}, \hat{p}] = i\,\hbar$, and therefore:

$$\Delta \hat{x} \cdot \Delta \hat{p} \geqslant \frac{\hbar}{2}\,.$$

This explains the famous *wave-particle duality*: if we know the position of a particle with precision, then we have large uncertainty on the momentum, and vice-versa, independently of the accuracy of the measurement tool. Find in [LN21] a study of the largest Hilbert space dimension for which measurements are compatible.

### 2.3.2 General Measurements

In quantum information theory, we often ignore the observable $A$ and replace the value "$\lambda_i$" by its index "$i$", thus obtaining the following more abstract definition of measurement:

> **Definition 2.19** (PVM) — *A* projection-valued measure *(*PVM*) is a finite set $\{P_i\}_i$ of bounded operators $P_i \in \mathcal{B}(\mathcal{H})$ that are orthogonal projections and that sum to the identity:*
>
> $$P_i P_j = \delta_{ij} P_i \,, \qquad \textit{and} \qquad \sum_i P_i = \mathbb{I}_d \,,$$
>
> *where $\delta_{ij}$ is the* Dirac delta*, taking value $1$ if $i = j$, and $0$ otherwise.*

Notice that, from the first condition, one can deduce that the eigenvalues of each $P_i$ are either $0$ or $1$, so we always have $P_i \succcurlyeq \mathbf{0}$ and $P_i$ Hermitian.

**Example 2.20** (Spectral Projectors) — As highlighted below eq. (2.4), the spectral projectors $P_i$ of the former subsection satisfy all these conditions, so they form a PVM. Conversely, any PVM can be viewed as the set of the spectral projectors of a certain observable $A$. Therefore, we obtain the same formulae as in Postulates 2.16 and 2.17:

$$\mathbb{P}\big(\text{obtaining "}i\text{"}\big) = \mathrm{Tr}(P_i\,\rho) \qquad \text{and} \qquad \tilde{\rho} := \frac{P_i\,\rho\,P_i}{\mathrm{Tr}\big(P_i\,\rho\big)} \,.$$

**Example 2.21** (Basis Measurement) — Fix an orthonormal basis $\big\{|e_i\rangle\big\}_i$ of $\mathcal{H}$ and consider the PVM $P_i := |e_i\rangle\!\langle e_i|$, called *basis measurement*. Using the trace cyclicity, we obtain $\mathbb{P}(\text{"}i\text{"}) = \langle e_i \,|\, \rho \,|\, e_i\rangle$. For instance, consider $\rho = |0\rangle\!\langle 0| \in \mathcal{D}(\mathbb{C}^2)$. We can measure it in the $\sigma_x$-basis $\{|+\rangle, |-\rangle\}$, which is actually a set of eigenvectors of the Pauli-$\sigma_x$ operator introduced at page 28. We obtain outcomes uniformly at random:

$$\mathbb{P}(\text{obtaining "}+\text{"}) = \mathbb{P}(\text{obtaining "}-\text{"}) = 1/2 \,.$$

Similarly, we can measure $\rho$ in the $\sigma_z$-basis $\{|0\rangle, |1\rangle\}$ composed of eigenvectors of the Pauli-$\sigma_z$ operator. It yields deterministic results:

$$\mathbb{P}(\text{obtaining "0"}) = 1 \qquad \text{and} \qquad \mathbb{P}(\text{obtaining "1"}) = 0 \,.$$

Now, for some reasons detailed in Section 2.3.3 about local measurements, it is convenient to have a generalization of these projective measurements to the following class of measurements:

**Definition 2.22** (POVM) — *A* positive operator-valued measure *(*POVM*) is a finite set $\{E_i\}_i$ of bounded operators $E_i \in \mathcal{B}(\mathcal{H})$ that are semi-definite positive and that sum to the identity:*

$$E_i \succcurlyeq \mathbf{0} \qquad \textit{and} \qquad \sum_i E_i = \mathbb{I}_d \,.$$

When measuring a state $\rho$ with the POVM $\{E_i\}_i$, we obtain the value "$i$" with probability:

$$\mathbb{P}\big(\text{obtaining "}i\text{"}\big) \;=\; \mathrm{Tr}(E_i\, \rho)\,.$$

**Example 2.23** (Trivial Measurement) — Fix a basis $\big\{|e_i\rangle\big\}_i$ of $\mathcal{H}$ and consider the POVM $E_i := p_i\, \mathbb{I}_d$, called *trivial measurement*, where $p_i \geqslant 0$ for all $i$ and $\sum_i p_i = 1$. From the above formula and using the normalization of $\rho$, we have $\mathbb{P}(\text{"}i\text{"}) = \mathrm{Tr}(p_i\, \mathbb{I}_d\, \rho) = p_i$. It is *trivial* because it does not depend on the quantum state $\rho$.

Nevertheless, this level of abstraction does not allow us to describe the new state $\tilde{\rho}$ after the collapse of the wave packet. This is why we need to generalize this notion once again as follows:

**Definition 2.24** (General Measurement) — *A* general measurement *is a finite set $\{M_i\}_i$ of bounded operators $M_i \in \mathcal{B}(\mathcal{H})$ such that:*

$$\sum_i M_i^* M_i \;=\; \mathbb{I}_d \,.$$

Each general measurement $\{E_i\}_i$ gives rise to a POVM using the following relation:

$$E_i = M_i^* M_i \,.$$

When measuring a state $\rho$ with the general measurement $\{M_i\}_i$, we obtain the value "$i$" with probability:

$$\mathbb{P}\big(\text{obtaining "}i\text{"}\big) \;=\; \mathrm{Tr}(M_i\, \rho\, M_i^*)\,,$$

and the state collapses to the following one:

$$\tilde{\rho} := \frac{M_i\, \rho\, M_i^*}{\mathrm{Tr}(M_i\, \rho\, M_i^*)} .$$

Although these three notions of measurement seem disparate, there is a strong connection between them all, and one can often reduce to the easiest case of projective measurements:

**Theorem 2.25** (Equivalence Between Measurements) — *Up to unitary transformation and ancillary systems, the three notions of measurements are equivalent:*

- ***Naimark's Dilation [Nai40]:*** *Let* $\{E_i\}_i$ *be a* POVM *on a Hilbert space* $\mathcal{H}$*. Then, there exist an auxiliary Hilbert space* $\mathcal{K}$*, an isometry* $V : \mathcal{H} \to \mathcal{H} \otimes \mathcal{K}$*, and a* PVM $\{P_i\}_i$ *on the extended space* $\mathcal{H} \otimes \mathcal{K}$ *such that the original* POVM *is recovered by "compressing" these projectors via the isometry:*
$$\forall i\,, \qquad E_i \,=\, V^*\, P_i\, V\,.$$

- ***Polar Decomposition [RS80]:*** *Let* $\{M_i\}_i$ *be a general measurement over* $\mathcal{H}$*, and consider* $E_i = M_i^* M_i$ *forming a* POVM*. Then, there exist unitaries* $U_i \in \mathcal{U}(\mathcal{H})$ *for all* $i$ *such that:*
$$\forall i\,, \qquad M_i \,=\, U_i \sqrt{E_i}\,.$$

**Remark 2.26** (Quantum Instrument) — Another generalization of measurement can be phrased in terms of *quantum instruments*. A quantum instrument is a collection $\{\mathcal{I}_i\}_i$ of completely-positive trace-non-increasing maps $\mathcal{I}_i : \mathcal{B}(\mathcal{H}) \to \mathcal{B}(\mathcal{H})$ that sum to a quantum channel $\Phi = \sum_i \mathcal{I}_i$ (see definition in Section 2.4). They encode both the outcome probabilities and the post-measurement state as follows:

$$\mathbb{P}\big(\text{obtaining “}i\text{”}\big) \,=\, \mathrm{Tr}\big(\mathcal{I}_i(\rho)\big) \qquad \text{and} \qquad \tilde{\rho} := \frac{\mathcal{I}_i(\rho)}{\mathrm{Tr}\big(\mathcal{I}_i(\rho)\big)} .$$

From a quantum instrument, one can define a POVM by:

$$E_i \,=\, \mathcal{I}_i^*(\mathbb{I}_d)\,,$$

where $\mathcal{I}_i^*$ is the dual map of $\mathcal{I}_i$ given by the Frobenius inner product $\langle A, B\rangle := \mathrm{Tr}(AB^*)$ on $\mathcal{B}(\mathcal{H})$. For more on quantum instruments, we refer to [DL70; Hol11; Oza84].

### 2.3.3 *Local Measurements*

For the sake of simplicity, consider the bipartite setting $\mathcal{H} = \mathcal{H}_\mathsf{A} \otimes \mathcal{H}_\mathsf{B}$, keeping in mind that the same reasoning holds for any number of parties. We can see this space as an *open quantum system*, where we take the viewpoint of Alice's measurement and where Bob's subsystem is viewed as an unknown *environment*, in contrast with the previous section where the quantum system was *closed* because we took the viewpoint of measuring the whole space at once.

**Importance of** POVMs**.** Given a product state $\rho = \rho_\mathsf{A} \otimes \rho_\mathsf{B}$, we want to measure the values "$\lambda_i$" of an observable $A = \sum_i \lambda_i P_i$. As before, the probability of obtaining "$\lambda_i$" is described by:

$$\begin{aligned} \mathbb{P}\big(\text{obtaining "}\lambda_i\text{"}\big) &= \mathrm{Tr}\big[P_i\,(\rho_\mathsf{A} \otimes \rho_\mathsf{B})\big] \\ &= \mathrm{Tr}\big[P_i\,(\mathbb{I}_\mathsf{A} \otimes \rho_\mathsf{B})(\rho_\mathsf{A} \otimes \mathbb{I}_\mathsf{B})\big] \\ &= \mathrm{Tr}\Big[\underbrace{\mathrm{Tr}_\mathsf{B}\big[P_i\,(\mathbb{I}_\mathsf{A} \otimes \rho_\mathsf{B})\big]}_{=:\,E_i\,\in \mathcal{B}(\mathcal{H}_\mathsf{A})}\ \rho_\mathsf{A}\Big]\,, \end{aligned}$$

where $\mathrm{Tr}_\mathsf{B}$ is the partial trace defined at page 30. From this computation, two observations are natural: first, Alice's perspective of the global measurement can be understood in terms of the partial trace over Bob's subsystem; second, the operators $E_i$ are positive semi-definite and sum to the identity, so they form a POVM. Although the global measurement was a PVM, the local measurement $\{E_i\}_i$ is not necessarily projective, which is why the notion of POVM is so important in the multipartite setting. Moreover, the formalism of POVMs is better suited to work with because it does not require us to actually know or even think of the other parties. Note that it is not surprising that the POVM $\{E_i\}_i \subseteq \mathcal{B}(\mathcal{H}_\mathsf{A})$ comes from a PVM in a larger set $\{P_i\}_i \subseteq \mathcal{B}(\mathcal{H}_\mathsf{A} \otimes \mathcal{H}_\mathsf{B})$ thanks to Naimark's Dilation Theorem (Theorem 2.25).

> **Postulate 2.27** (Local Measurement) — *Given a quantum state* $\rho \in \mathcal{D}(\mathcal{H}_{\mathsf{A}} \otimes \mathcal{H}_{\mathsf{B}})$*, Alice's local measurement is described by a* POVM $\{E_i\}_i \subseteq \mathcal{B}(\mathcal{H}_{\mathsf{A}})$ *applied on the state* $\mathrm{Tr}_{\mathsf{B}}(\rho) \in \mathcal{D}(\mathcal{H}_{\mathsf{A}})$*, and the probability of outcome "*$i$*" is described by:*
>
> $$\mathbb{P}\big(\textit{Alice obtains ``}i\textit{''}\big) = \mathrm{Tr}\Big[E_i \ \mathrm{Tr}_{\mathsf{B}}[\rho]\Big] = \mathrm{Tr}\Big[(E_i \otimes \mathbb{I}_{\mathsf{B}})\,\rho\Big] .$$

**When Bob Measures After Alice.** Note that it is consistent with the formula of the collapse of the wave packet. Indeed, if Alice's POVM is turned into a general measurement $M_i := \sqrt{E_i}$, and if Bob uses a POVM $\{F_j\}_j$ on his subsystem, then the probability that Bob gets the value "$j$" knowing that Alice obtained the value "$i$" is:

$$\begin{aligned}
\mathbb{P}\big(``j" \,\big|\, ``i"\big) &= \mathrm{Tr}\left[(\mathbb{I}_{\mathsf{A}} \otimes F_j) \cdot \frac{(M_i \otimes \mathbb{I}_{\mathsf{B}})\,\rho\,(M_i^* \otimes \mathbb{I}_{\mathsf{B}})}{\mathbb{P}(\text{Alice obtains } ``i")}\right] \\
&= \frac{1}{\mathbb{P}(``i")}\,\mathrm{Tr}\Big[(M_i^* \otimes \mathbb{I}_{\mathsf{B}}) \cdot (\mathbb{I}_{\mathsf{A}} \otimes F_j) \cdot (M_i \otimes \mathbb{I}_{\mathsf{B}})\,\rho\Big] \\
&= \frac{1}{\mathbb{P}(``i")}\,\mathrm{Tr}\Big[(E_i \otimes F_j)\,\rho\Big] ,
\end{aligned}$$

where we used the linearity and cyclicity of the trace in the second last equality, and the fact that the operators commute in the last equality. This is the reason why, if Alice and Bob measure their respective subsystem of $\rho$ with POVMs $\{E_i\}_i$ and $\{F_j\}_j$, we have the following formula:

$$\mathbb{P}\big(\text{Alice obtains } ``i" \text{ \& Bob obtains } ``j"\big) = \mathrm{Tr}\Big[(E_i \otimes F_j)\,\rho\Big] ,$$

which does not depend on who measured first. This will be particularly useful to describe *quantum correlations* in Section 3.1.

**Local Measurements on Entangled States.** When $\mathcal{H} = \mathbb{C}^2 \otimes \mathbb{C}^2$, consider the maximally entangled state:

$$\omega := |\Omega\rangle\langle\Omega| = \frac{|00\rangle\langle 00| + |00\rangle\langle 11| + |11\rangle\langle 00| + |11\rangle\langle 11|}{2} .$$

If Alice measures her qubit $\mathrm{Tr}_{\mathsf{B}}[\omega]$ in the $\sigma_z$-basis, *i.e.* if $\{E_i\}_i = \{|0\rangle\langle 0|, |1\rangle\langle 1|\}$ (note that this is a PVM in this case), then she obtains each of the outcomes

"0" and "1" with probability $1/2$, uniformly at random. Now, we know that her measurement causes a collapse of the wave packet into either:

$$\tilde{\omega} = |00\rangle\langle 00| \qquad \text{or} \qquad \tilde{\omega} = |11\rangle\langle 11|\,,$$

depending on the outcome "0" or "1" she obtains. After that, if Bob choses to measure his qubit $\mathrm{Tr}_{\mathsf{A}}[\tilde{\omega}]$ in the same basis, *i.e.* if $\{F_j\}_j = \{|0\rangle\langle 0|, |1\rangle\langle 1|\}$, then he gets a deterministic outcome, either "0" or "1" depending on what Alice obtained. Interestingly, the two parties Alice and Bob always obtain the same outcome, but we do not know which one in advance: it is uniformly random. This shows that quantum entanglement allows us to have strongly correlated outcomes *without communicating*. Moreover, note that only the first outcome is random since the second one becomes deterministic, but from both Alice's and Bob's perspectives the result is uniformly random. Nevertheless, until now, this behavior can be simulated by classical correlations—the full strength of quantum correlations is revealed for instance in quantum teleportation, see next paragraph, or in nonlocal games, see Section 3.2.

**Application: Quantum Teleportation.** A famous example of application is *quantum teleportation*, originally presented in [Ben+93] and experimentally proved in [Bou+97]. This protocol allows Alice to "transfer" an unknown qubit to Bob utilizing local measurement and classical communication (LOCC, see page 35). Beforehand, Alice and Bob share the maximally entangled state $|\Omega\rangle$, one qubit each, and Alice has an additional unknown qubit $|\psi\rangle = \alpha|0\rangle + \beta|1\rangle$. Alice performs a local measurement on her pair of qubits: she applies the PVM induced by the orthonormal basis $\big\{\big(|00\rangle \pm |11\rangle\big)/\sqrt{2}, \big(|01\rangle \pm |10\rangle\big)/\sqrt{2}\big\}$ of $\mathbb{C}^2 \otimes \mathbb{C}^2$, called *Bell-state measurement*. With the collapse of the wave packet, Bob's qubit is instantly changed into $|\psi\rangle$ up to a unitary transformation. Finally, Alice sends two classical bits to Bob, encoding the result she obtained from the Bell-state measurement, so that Bob can reverse the unitary with a good choice of Pauli matrices. At the end of the protocol, Bob's qubit is precisely $|\psi\rangle$. Note that the state was not physically transported, but rather reconstructed on Bob's subsystem, while "destroyed" on Alice's location, which ensures no-cloning (see Theorem 2.37). Note that this teleportation protocol does not violate relativity, since it requires classical communication in order to reconstruct the state.

## 2.4 Quantum Channels

Quantum channels provide the most general framework for describing valid transformations of quantum states.

In this section, we begin by defining quantum channels and providing several examples (Section 2.4.1). We then present three characterizations of quantum channels (Section 2.4.2) and conclude the chapter with an application to the quantum no-cloning theorem (Section 2.4.3).

### *2.4.1 Definition and Examples*

Quantum channels are linear maps, often denoted $\Phi$, such that the image of any mixed state is again a mixed state:

**Definition 2.28** (Quantum Channel) — *A* quantum channel *is a linear map* $\Phi : \mathcal{B}(\mathcal{H}) \to \mathcal{B}(\mathcal{K})$ *that is completely-positive* (CP)*:*

$$\forall \mathcal{H}' \textit{ Hilbert space}\,,\ \forall X \succcurlyeq \mathbf{0} \textit{ in } \mathcal{B}(\mathcal{H}\otimes\mathcal{H}')\,, \quad \big[\Phi\otimes\mathbb{I}_{\mathcal{H}'}\big](X) \succcurlyeq \mathbf{0}\,,$$

*and trace-preserving* (TP)*:*

$$\forall X \in \mathcal{B}(\mathcal{H})\,, \qquad \mathrm{Tr}\big[\Phi(X)\big] \;=\; \mathrm{Tr}[X]\,.$$

*Quantum channels are often abbreviated in* CPTP maps.

**Remark 2.29** (Why Complete Positivity?) — All completely-positive maps are positive, meaning that they map PSD operators $X \succcurlyeq \mathbf{0}$ to PSD operators $f(X) \succcurlyeq \mathbf{0}$, but the converse is false. For instance, consider the transposition map $X \mapsto X^\top$, which is positive but not completely-positive. As such, this map preserves quantum states if we apply it globally, *i.e.* $\mathrm{transpose}(\rho)$ is a quantum state, but it is no longer true if we apply this map only on a subsystem. For instance, with the maximally entangled state $\omega \in \mathcal{D}(\mathbb{C}^2 \otimes \mathbb{C}^2)$, we have:

$$\big[\mathbb{I}_{\mathbb{C}^2}\otimes \mathrm{transpose}_{\mathbb{C}^2}\big](\omega) \;=\; \omega^\Gamma \;\not\succcurlyeq\; \mathbf{0}\,,$$

which is the partial transposition used in the PPT criterion, see page 38. Hence, to be able to apply quantum channels on subsystems only, we need to have complete-positivity.

**Example 2.30** (Classical Channels) — Classical channels are embedded in the set of quantum channels. A *classical channel* $p$ from an alphabet $\mathcal{X}$ to an alphabet $\mathcal{A}$ is modeled by a conditional probability distribution of receiving an output $a \in \mathcal{A}$ when sending an input $x \in \mathcal{X}$:

$$\forall a, x\,, \qquad p(a\,|\,x) \geqslant 0 \qquad \text{and} \qquad \sum_a p(a\,|\,x) = 1\,.$$

The quantum version of a classical channel is as follows:

$$\Phi : \rho \mapsto \sum_{a,x} p(a\,|\,x)\, \langle x|\rho|x\rangle\, |a\rangle\!\langle a|\,,$$

where each state $|x\rangle\!\langle x|$ is turned into the state $|a\rangle\!\langle a|$ with probability $p(a\,|\,x)$.

**Example 2.31** (Depolarizing Channel) — The *depolarizing channel* is a channel modeling the presence of noise in the transmission:

$$\Delta_\lambda(X) \;:=\; (1-\lambda)\, X + \lambda \frac{\mathbb{I}_d}{d}\, \mathrm{Tr}[X]\,,$$

for any coefficient $\lambda \in [0,1]$ (or even until $1 + \frac{1}{d^2-1}$). From a quantum state $\rho$, it results in a convex combination of this state with the maximally mixed state.

**Example 2.32** (Unitary Channel) — The *unitary channel* is the conjugation by a unitary operator $U \in \mathcal{U}(\mathcal{H})$:

$$\Phi_U : X \;\mapsto\; U\, X\, U^*\,.$$

According to Schrödinger picture [Sch26], it represents the dynamics of a quantum state in a closed quantum system. Indeed, based on the postulate that quantum evolution is governed by the Schrödinger equation, if we start with a state $\rho_0$ at time $t = 0$, it should evolve unitarily:

$$\rho_t = U_t\, \rho_0\, U_t^*\,,$$

with $U_t = e^{-itH/\hbar}$ and Hamiltonian $H$. In contrast, Heisenberg picture [Hei25] assumes that unitary time evolution occurs rather for observables:

$$A_t \;=\; U_t^*\, A_0\, U_t\,,$$

with the same expression for $U_t$.

**Example 2.33** (Measurement Channel) — Measurements introduced in Section 2.3 are particular cases of quantum channels. For instance, given a POVM $\{E_i\}_i$, one can define its associated *measurement channel*:

$$\Phi_{\{E_i\}}(X) = \sum_i \mathrm{Tr}\big(E_i\, \rho\big)\, |i\rangle\!\langle i|\,.$$

**Example 2.34** (Quantum Circuits) — In quantum computing, a *quantum circuit* is a sequence of quantum gates (unitary operators) and measurements, applied to a set of qubits. So the combination of the former two examples gives rise to quantum circuits as quantum channels.

## 2.4.2 Characterizations

Several equivalent definitions of quantum channels exist, each of them being useful in different contexts. But first, we need to introduce the notion of the Choi matrix:

> **Definition 2.35** (Choi Matrix) — *Given a linear map $\Phi : \mathcal{B}(\mathcal{H}) \to \mathcal{B}(\mathcal{K})$, the* Choi matrix $C_\Phi \in \mathcal{B}(\mathcal{H} \otimes \mathcal{K})$ *is defined as follows:*
>
> $$\begin{aligned} C_\Phi &:= \sum_{i,j} |i\rangle\!\langle j| \otimes \Phi\big(|i\rangle\!\langle j|\big) \\ &= \Big[\mathbb{I}_d \otimes \Phi\Big]\big(\dim(\mathcal{H}) \cdot \omega\big)\,. \end{aligned}$$

Studying this matrix allows us to deduce some properties of the quantum channel. For instance, the quantum channel $\Phi$ is positive *if, and only if,* its Choi matrix $C_\Phi$ is block-positive:

$$\Big(\forall X \succcurlyeq \mathbf{0},\ \ \Phi(X) \succcurlyeq \mathbf{0}\Big) \quad \Leftrightarrow \quad \Big(\forall (x,y) \in \mathcal{H} \times \mathcal{K},\ \ \langle x \otimes y|\, C_\Phi\, |x \otimes y\rangle \geqslant 0\Big)\,.$$

Moreover, the quantum channel $\Phi$ is completely positive *if, and only if,* its Choi matrix $C_\Phi$ is positive semi-definite:

$$\Big(\forall \mathcal{H}', \forall X \succcurlyeq \mathbf{0},\ \ [\Phi \otimes \mathbb{I}_{\mathcal{H}'}](X) \succcurlyeq \mathbf{0}\Big) \quad \Leftrightarrow \quad \Big(\forall z \in \mathcal{H} \otimes \mathcal{K},\ \ \langle z|\, C_\Phi\, |z\rangle \geqslant 0\Big)\,.$$

Note also that, alternatively, one can express a linear map $\Phi$ in terms of its Choi matrix $C_\Phi$ [Wat18]:

$$\Phi(X) = \mathrm{Tr}_{\mathcal{H}}\Big[C_\Phi\big(X^\top \otimes \mathbb{I}_{\mathcal{K}}\big)\Big]\,.$$

Here are standard characterizations of quantum channels:

**Theorem 2.36** (Characterizations of Quantum Channels) — *Let $\Phi : \mathcal{B}(\mathcal{H}) \to \mathcal{B}(\mathcal{K})$ be a linear map. The following are equivalent:*

*(1) $\Phi$ is a quantum channel.*

*(2)* ***Choi Theorem [Cho75]:*** *The Choi matrix $C_\Phi \in \mathcal{B}(\mathcal{H} \otimes \mathcal{K})$ is* PSD *and normalized:*

$$C_\Phi \succcurlyeq \mathbf{0} \qquad \textit{and} \qquad \mathrm{Tr}_{\mathcal{K}}\big[C_\Phi\big] = \mathbb{I}_{\mathcal{H}}\,.$$

*(3)* ***Kraus Decomposition [Kra71]:*** *There exist operators $K_1, .., K_r : \mathcal{H} \to \mathcal{K}$, called* Kraus operators*, such that:*

$$\sum_{i=1}^{r} K_i^* K_i = \mathbb{I}_{\mathcal{H}} \qquad \textit{and} \qquad \forall X\,, \quad \Phi(X) = \sum_{i=1}^{r} K_i\, X\, K_i^*\,.$$

*(4)* ***Stinespring Dilation [Sti55]:*** *There exist a Hilbert space $\mathcal{K}'$ and an isometry $V : \mathcal{H} \to \mathcal{K} \otimes \mathcal{K}'$ such that:*

$$\forall X\,, \qquad \Phi(X) = \mathrm{Tr}_{\mathcal{K}'}\big[V\,X\,V^*\big]\,.$$

*Proof.* We prove the four implications:

(1)⇒(2) If $\Phi$ is a quantum channel, then it is completely positive so $\big[\mathbb{I}_d \otimes \Phi\big](d \cdot \omega) \succcurlyeq \mathbf{0}$, and it is trace-preserving so $\mathrm{Tr}_{\mathcal{K}}\big[C_\Phi\big] = \mathrm{Tr}_{\mathcal{K}}\big[d \cdot \omega\big] = \mathbb{I}_{\mathcal{H}}$.

(2)⇒(3) Assume that the Choi matrix $C_\Phi$ is PSD and normalized. Then, it is diagonalizable $C_\Phi = \sum_{i=1}^r \lambda_i\, |z_i\rangle\langle z_i|$ with $\lambda_i \geqslant 0$ and $r = \mathrm{rank}(C_\Phi)$. For each $i$, consider the unique operator $Z_i : \mathcal{K} \to \mathcal{H}$ such that $\langle x|\, Z_i\, |y\rangle = \langle x \otimes y \,|\, z_i\rangle$ for all $x \in \mathcal{H}$ and $y \in \mathcal{K}$, and define $K_i := \sqrt{\lambda_i} Z_i^*$. Now, computations lead to $\mathrm{Tr}\big(\sum_i K_i^* K_i X\big) = \mathrm{Tr}\big(\mathrm{Tr}_{\mathcal{K}}(C_\Phi) X\big) = \mathrm{Tr}\big(X\big)$ and $\mathrm{Tr}\big(\Phi(X)\, Y^*\big) = \mathrm{Tr}\big(\sum_i K_i\, X\, K_i^*\, Y^*\big)$ for all $X \in \mathcal{B}(\mathcal{H})$ and $Y \in \mathcal{B}(\mathcal{K})$, hence the wanted equalities.

(3)⇒(4) Assume the Kraus decomposition of $\Phi$. Consider the Hilbert space $\mathcal{K}' := \mathbb{C}^r$ and define $V : x \mapsto \sum_{i=1}^r K_i\, x \otimes |i\rangle$. On the one hand, it is an isometry: $V^*V = \sum_{i,j} K_i^* K_j \langle i|j\rangle = \sum_i K_i^* K_i = \mathbb{I}_{\mathcal{H}}$. On the other hand, for all $X \in \mathcal{B}(\mathcal{H})$, we have the wanted equality:

$$\mathrm{Tr}_{\mathcal{K}'}\big(V\,X\,V^*\big) \;=\; \sum_{i,j} \mathrm{Tr}_{\mathcal{K}'}\Big(K_i\,X\,K_j^* \otimes |i\rangle\!\langle j|\Big) \;=\; \sum_i K_i\,X\,K_i^* \;=\; \Phi(X)\,.$$

(4)⇒(1) Assume $\Phi(X) = \mathrm{Tr}_{\mathcal{K}'}\big[V\,X\,V^*\big]$. Then, for any extra Hilbert space $\mathcal{H}'$ and $Z \succcurlyeq \mathbf{0}$ in $\mathcal{B}(\mathcal{H}\otimes\mathcal{H}')$, we have $\big[\Phi\otimes\mathbb{I}_{\mathcal{H}'}\big](Z) = \mathrm{Tr}_{\mathcal{K}'}\big[(V\otimes\mathbb{I}_{\mathcal{H}'})\,Z\,(V\otimes\mathbb{I}_{\mathcal{H}'})^*\big] \succcurlyeq \mathbf{0}$, hence $\Phi$ is completely positive. Moreover, it is trace-preserving by cyclicity of the trace. Thus $\Phi$ is a quantum channel. ∎

### 2.4.3 Non-Existence of a Cloning Channel

**Quantum No-Cloning.** As opposed to the classical setting, another crucial feature of quantum mechanics is the impossibility of duplicating arbitrary unknown qubits. This observation has significant cryptographic applications, see Chapter 5, as well as consequences in quantum computing and quantum error correction. It can be formalized as follows:

**Theorem 2.37** (Quantum No-Cloning [Die82; WZ82]) — *There is no quantum channel $\Phi : \mathcal{B}(\mathcal{H}) \to \mathcal{B}(\mathcal{H}\otimes\mathcal{H})$ such that:*

$$\forall \rho \in \mathcal{D}(\mathcal{H})\,, \qquad \Phi(\rho) \;=\; \rho\otimes\rho\,.$$

*Proof.* It is sufficient to prove the result for pure states. Denote by $|0\rangle$ and $|1\rangle$ the first two vectors of an orthonormal basis of $\mathcal{H}$. Assume by contradiction the existence of a linear map $\Phi$ such that $|\psi\rangle \mapsto |\psi\rangle\otimes|\psi\rangle$. In particular, we have $|0\rangle \mapsto |00\rangle$ and $|1\rangle \mapsto |11\rangle$. Then, by linearity, we obtain:

$$\begin{aligned}\Phi\big(\,|+\rangle\,\big) \;&=\; \frac{\Phi\big(|0\rangle\big)+\Phi\big(|1\rangle\big)}{\sqrt{2}} \;=\; \frac{|00\rangle+|11\rangle}{\sqrt{2}} \\ &\neq\; \frac{|00\rangle+|01\rangle+|10\rangle+|11\rangle}{2} \;=\; |{+}{+}\rangle\,.\end{aligned}$$

Hence the contradiction $\Phi : |+\rangle \not\mapsto |{+}{+}\rangle$, and there is no perfect cloning quantum channel. ∎

**Remark 2.38** (Imperfect Cloning) — Although quantum mechanics does not allow the perfect cloning of an unknown state, one can try to find copies that are not exact but as “close” as possible to the original state. This is called *imperfect cloning* and was long studied [BH96; Sca+05]. Note that it is also possible to clone a state asymmetrically, with different marginals for the different parties [Cer00; NPR21; NPR23].

**Quantum No-Broadcasting.** The *no-broadcasting theorem* is a refinement of the No-Cloning Theorem, in the sense that cloning always implies broadcasting but not the converse. Here, instead of obtaining several copies $\rho \otimes \cdots \otimes \rho$ of an unknown state $\rho$, the task of broadcasting consists in finding a channel $\Phi$ such that each partial trace of the state $\sigma = \Phi(\rho)$ is exactly the unknown state $\rho$:

$$\mathrm{Tr}_{\mathsf{B}}\big(\Phi(\rho)\big) = \rho \qquad \text{and} \qquad \mathrm{Tr}_{\mathsf{C}}\big(\Phi(\rho)\big) = \rho \,,$$

where $\Phi : \mathcal{B}(\mathcal{H}_{\mathsf{A}}) \to \mathcal{B}(\mathcal{H}_{\mathsf{B}} \otimes \mathcal{H}_{\mathsf{C}})$ and $\mathcal{H}_{\mathsf{A}} = \mathcal{H}_{\mathsf{B}} = \mathcal{H}_{\mathsf{C}}$. Note that this is very related to the notion of $k$-extendibility presented in Section 2.2.5 about monogamy-of-entanglement. From the No-Cloning Theorem, we know that it is not possible to broadcast an unknown pure state $|\psi\rangle$, because it amounts to having it in the product form $|\psi\rangle \otimes |\psi\rangle$, *i.e.* to cloning it. Nevertheless, the mixed-state version of the No-Cloning Theorem is not sufficient to demonstrate no-broadcasting in general, for there may be many ways to broadcast $\rho$ without cloning it. For instance, the state $\rho = \frac{|0\rangle\langle 0| + |1\rangle\langle 1|}{2}$ can be broadcasted using the state $\sigma = \frac{|00\rangle\langle 00| + |11\rangle\langle 11|}{2}$, which is not of the form $\rho \otimes \rho$ but whose partial traces are exactly $\rho$. But in most cases, it is not possible to broadcast an unknown state given only a single copy:

> **Theorem 2.39** (Quantum No-Broadcasting [Bar+96]) — *Fix two mixed states $\rho_0, \rho_1 \in \mathcal{D}(\mathcal{H})$. It is not possible to broadcast an unknown state $\rho_s$ $(s \in \{0, 1\})$ unless the two states $\rho_0$ and $\rho_1$ commute.*

**Remark 2.40** (Superbroadcasting) — However, if we are given several copies of the state $\rho$, say $N \geqslant 4$ copies, then broadcasting $\rho$ to $M$ parties with $M > N$ becomes possible. It requires that $\rho$ is sufficiently mixed, and at the end of the process, the marginals are aligned in the Bloch sphere (but not necessarily equal). This is called *superbroadcasting* [DMP05].

**Link with Monogamy-of-Entanglement.** Consider $\mathcal{H}_{\mathsf{A}} = \mathcal{H}_{\mathsf{B}} = \mathcal{H}_{\mathsf{C}}$ of dimension $d$. Assume by contradiction that a perfect cloning channel $\Phi : \mathcal{B}(\mathcal{H}_{\mathsf{A}}) \to \mathcal{B}(\mathcal{H}_{\mathsf{B}} \otimes \mathcal{H}_{\mathsf{C}})$ exists. Then, its marginals would be the identity map $\Phi_{\mathsf{B}} = \Phi_{\mathsf{C}} = \mathrm{id}$ and their Choi matrix would be $C_{\Phi_{\mathsf{B}}} = C_{\Phi_{\mathsf{C}}} = d \cdot \omega$ by definition. But now, if we renormalize the Choi matrix of $\Phi$ in order to have a quantum state $\frac{1}{d} C_\Phi \in \mathcal{D}(\mathcal{H}_{\mathsf{A}} \otimes \mathcal{H}_{\mathsf{B}} \otimes \mathcal{H}_{\mathsf{C}})$, we have that the partial traces satisfy:

$$\mathrm{Tr}_{\mathsf{B}}\left(\frac{C_\Phi}{d}\right) = \frac{C_{\Phi_{\mathsf{C}}}}{d} = \omega \qquad \text{and} \qquad \mathrm{Tr}_{\mathsf{C}}\left(\frac{C_\Phi}{d}\right) = \frac{C_{\Phi_{\mathsf{B}}}}{d} = \omega \,,$$

which contradicts the principle of Monogamy-of-Entanglement (MoE) introduced in Section 2.2.5 (see the “state extendibility” formalization). This shows that, somehow, the MoE principle implies the no-cloning theorem.

# Chapter 3

# Nonlocal Boxes & Nonlocal Games

In this chapter, we continue introducing the background material. After presenting nonlocal boxes and nonlocal games, we provide a few applications.

## Chapter Contents

## 3.1 Nonlocal Boxes

Nonlocal boxes are theoretical tools that describe the statistics of a behavior in a device-independent manner. They have numerous applications, such as in self-testing or device-independent quantum key distribution protocols, as discussed in Section 3.3.

In this section, we first define several types of correlations (Section 3.1.1), then explore some geometric aspects of nonlocal boxes (Section 3.1.2). We next present a method to approximate the boundary of the quantum set (Section 3.1.3), and finally introduce the concepts of wirings and measures of nonlocal boxes (Sections 3.1.4 and 3.1.5). For further details on this topic, we refer to [Bru+14; Pop14; Sca12; Sca19].

### *3.1.1 Correlation Sets*

As detailed in Section 2.3.3, if two non-communicating parties Alice and Bob share an entangled state $\rho$, then their classical outputs $a$ and $b$ are correlated. Now, imagine a Referee provides them with classical instructions $x$ and $y$ determining which local measurements $\{E_{a|x}\}_a$ and $\{F_{b|y}\}_b$ they should apply on their share of the state. Then, we may study the statistics of obtaining $(a, b)$ knowing that they received $(x, y)$ and used their resource:

$$\mathbf{P}(a, b \,|\, x, y) \;:=\; \mathbb{P}\Big(a, b \,\Big|\, x, y, \rho, \{E_{a|x}\}_a, \{F_{b|y}\}_b\Big)\,,$$

also sometimes shortened in $\mathbf{P}(ab|xy)$. These statistics are called *correlation*, more precisely *quantum correlation* in this case because the shared resource is a quantum state. Below, after formalizing the notion of a scenario, we present several types of correlations from the weakest to the strongest, depending on the resources jointly available to the parties.

**Scenario.** An $(n, N, M)$-*scenario* describes the situation where we have:

- $n$ *parties* $\mathsf{A}_1, .., \mathsf{A}_n$ that are space-like separated, meaning that communication between them is impossible;
- $N$ is the number of possible classical *inputs* for each party, *i.e.* each party $i$ is given a possibly different integer $x_i \in \{1, .., N\}$; and

- $M$ is the number of possible classical *outputs* for each party, *i.e.* each party $i$ is given a possibly different integer $a_i \in \{1, .., M\}$.

A standard example is the $(2, 2, 2)$-scenario, where the two parties A and B receive input bits $x, y \in \{0, 1\}$ and answer output bits $a, b \in \{0, 1\}$ (notice that, in this case, the values of $x, y, a, b$ begin at $0$ and not at $1$ to agree with the definition of a bit). This scenario is known as CHSH-*scenario*, referring to the renowned CHSH inequality, see eq. (3.11), and the CHSH-game, see Section 3.2. Note that one can consider more general types of scenarios, with different numbers of inputs for each party, and different numbers of outputs depending on the party and the input [Pir05], but we do not need this generality level in this thesis.

**Definition 3.1** (Correlation) — *In an $(n, N, M)$-scenario where the parties share a resource $R$, a* correlation $\mathbf{P}$ *is the probability of obtaining the outputs $(a_1, .., a_n) \in \{1, .., M\}^n$ given the inputs $(x_1, .., x_n) \in \{1, .., N\}^n$:*

$$\mathbf{P}\big(a_1, .., a_n \,\big|\, x_1, .., x_n\big) \;:=\; \mathbb{P}\big(a_1, .., a_n \,\big|\, x_1, .., x_n, R\big)\,,$$

*and often shortened in $\mathbf{P}(a_1..a_n|x_1..x_n)$.*

**Deterministic Correlations ($\mathcal{L}_{\text{det}}$).** One of the most basic types of correlations is the *deterministic* class. Here, the outputs $a_i$ are computed as the image of some function $f_i : \{1, .., N\} \to \{1, .., M\}$ applied to the input $x_i$:

$$a_i \;=\; f_i(x_i)\,.$$

Hence each party is independent and we have the following expression for a general deterministic correlation:

$$\mathbf{P}\big(a_1, .., a_n \,\big|\, x_1, .., x_n\big) \;=\; \mathbb{1}_{a_1=f_1(x_1)} \times \cdots \times \mathbb{1}_{a_n=f_n(x_n)}\,,$$

where $\mathbb{1}_C$ is the indicator function taking value $1$ if condition $C$ is satisfied and value $0$ otherwise. In the CHSH-scenario when $n = N = M = 2$, examples of deterministic correlations are $\mathbf{P_{00}}(ab|xy) := \mathbb{1}_{a=b=0}$ and $\mathbf{P_{11}}(ab|xy) := \mathbb{1}_{a=b=1}$, always outputting tuples $(0, 0)$ and $(1, 1)$ respectively, independently of $x$ and $y$.

**Local Correlations ($\mathcal{L}$).** The next level of correlation is the set $\mathcal{L}$ of *local correlations*, also known as *classical correlations*. Here, the parties can have access to *shared randomness*. This resource can be seen as a random number generator that produces the same value for all parties at the same time. For instance, it can be more or less modelized by a big die that every party can see at the same time, by the $10^{th}$ letter of today's journal, or by the weather conditions under some assumptions. This common data is often called *local hidden variable*, denoted $\lambda$ in some probability space $(\Lambda, \mu)$, and can be viewed as an additional input for all the parties, so that each individual party's behavior can be described by some probability of the form:

$$\mathbb{P}_{\mathsf{A}_i}\big(a_i \,\big|\, x_i, \lambda\big)\,.$$

Hence, any local correlation is expressed as follows [Fin82]:

$$\mathbf{P}\big(a_1, .., a_n \,\big|\, x_1, .., x_n\big) \;=\; \int_{\lambda\in\Lambda} \mathbb{P}_{\mathsf{A}_1}\big(a_1 \,\big|\, x_1, \lambda\big) \cdots \mathbb{P}_{\mathsf{A}_n}\big(a_n \,\big|\, x_n, \lambda\big)\, \mu(\lambda)\,.$$

In the CHSH-scenario when $n = N = M = 2$, here are two examples of local correlation: $\mathbf{SR}(ab|xy) := \frac{1}{2}\,\mathbb{1}_{a=b}$, called *shared randomness box*, that outputs either $(0,0)$ or $(1,1)$ uniformly at random; and $\mathbf{I}(ab|xy) := \frac{1}{4}$, called *fully mixed box*, that outputs any tuple $(a, b)$ uniformly at random. Note that these two correlations are not deterministic, so we have the strict inclusion $\mathcal{L}_{\mathsf{det}} \subsetneq \mathcal{L}$. Nevertheless, one can show that local correlations are actually convex combinations of deterministic ones, so that $\mathcal{L}$ is the convexified version of $\mathcal{L}_{\mathsf{det}}$, in the sense that $\mathcal{L} = \mathrm{Conv}\big(\mathcal{L}_{\mathsf{det}}\big)$, see Section 3.1.2.

**Remark 3.2** (Why "Local Hidden Variable"?) **—** This parameter $\lambda$ is "locally hidden" because it can represent the lack of knowledge of the parties—in the related *causality* theory, when the relation between a cause and an effect is not clear, it is because there is an unknown global parameter that also impacts this effect. For instance, if we knew with high precision all the parameters of the Universe, we would be able to predict the result of a die in advance. Moreover, the name *local correlation* comes from the *local realism* theory, in which objects can only be influenced by their immediate surroundings, with influences not traveling faster than light, as opposed to what happens with entanglement in quantum mechanics.

**Quantum Correlations ($\mathcal{Q}$).** Now, if the parties can share a quantum state $\rho$, correlations are called *quantum*, forming a set denoted $\mathcal{Q}$. In order to define this set properly, we first need to introduce an auxiliary set, denoted $\mathcal{Q}_{\text{finite}}$ and called the set of *finite quantum correlations*. Assume we have a composite quantum system $\mathcal{H} = \mathcal{H}_{\mathsf{A}_1} \otimes \cdots \mathcal{H}_{\mathsf{A}_n}$. Up to adding auxiliary dimensions to some subsystems $\mathcal{H}_{\mathsf{A}_i}$, one may assume without loss of generality that all the subsystems have the same finite dimension $d \in \mathbb{N}$. Each party is endowed with a collection of local POVMs $\big\{\{E_{a_i|x_i}\}_{a_i}\big\}_{x_i}$ parametrized by the input $x_i$. Upon receiving the instructions $x_i$, each party measures his share of the state $\rho$ according to the rules of quantum mechanics described in Section 2.3.3, and obtains:

$$\mathbb{P}_{\mathsf{A}_i}\Big(a_i \,\Big|\, x_i, \rho, \{E_{a_i|x_i}\}_{a_i}\Big) = \mathrm{Tr}\Big[\big(\mathbb{I}_{\mathsf{A}_1} \otimes \cdots \otimes \mathbb{I}_{\mathsf{A}_{i-1}} \otimes E_{a_i|x_i} \otimes \mathbb{I}_{\mathsf{A}_{i+1}} \otimes \cdots \otimes \mathbb{I}_{\mathsf{A}_n}\big)\, \rho\Big] \,.$$

Then, the general expression for a $d^n$-dimensional finite quantum correlation is as follows:

$$\mathbf{P}\big(a_1, .., a_n \,\big|\, x_1, .., x_n\big) \;=\; \mathrm{Tr}\Big[\big(E_{a_1|x_1} \otimes \cdots \otimes E_{a_n|x_n}\big)\, \rho\Big] \,,$$

forming a set denoted $\mathcal{Q}_d$. Note that as mixed states are always convex mixtures of pure states (see Theorem 2.1) and by the linearity of the trace, it is sufficient to study correlations coming from pure states $|\psi\rangle$:

$$\mathbf{P}\big(a_1, .., a_n \,\big|\, x_1, .., x_n\big) \;=\; \langle\psi|\big(E_{a_1|x_1} \otimes \cdots \otimes E_{a_n|x_n}\big)\, |\psi\rangle \,.$$

Now, when taking the union over all possible finite dimensions $d \in \mathbb{N}$, we obtain the set $\mathcal{Q}_{\text{finite}}$ of *finite quantum correlations*:

$$\mathcal{Q}_{\text{finite}} \;:=\; \bigcup_{d \in \mathbb{N}} \mathcal{Q}_d \,.$$

However, as established by Slofstra in [Slo19], although each individual $\mathcal{Q}_d$ is topologically closed, the union $\mathcal{Q}_{\text{finite}}$ is not closed[1]. Therefore, we define $\mathcal{Q}$ as the topological closure of this set, *i.e.* we add to $\mathcal{Q}_{\text{finite}}$ all the limit points, so that:

$$\mathcal{Q} \;:=\; \overline{\mathcal{Q}_{\text{finite}}} \,. \tag{3.1}$$

The set $\mathcal{Q}$ of quantum correlations contains classical correlations, since the latter one can be seen as quantum correlations with separable states $\rho = \sum_i \alpha_i \, \rho^{(i)}_{\mathsf{A}_1} \otimes \cdots \otimes \rho^{(i)}_{\mathsf{A}_n}$. Actually, as first shown by Bell and later by

[1] Also proved later by Dykema, Paulsen, and Prakash [DPP19] via graph theory.

Clauser, Horne, Shimony, and Holt, the inclusion is strict $\mathcal{L} \subsetneq \mathcal{Q}$ [Bel64; CHSH69]. For instance, there is a quantum correlation $\mathbf{P}$ employing the maximally entangled state $\omega$ and some well-suited collections of POVMs such that the outputs of $\mathbf{P}(ab|xy)$ satisfies the relation $a \oplus b = xy$ with probability $\cos^2(\pi/8) \approx 85\%$, which is impossible using local correlations only as detailed in Section 3.2. This shows that quantum mechanics allows more correlation between parties than its classical counterpart.

**Remark 3.3** (Infinite Quantum Correlations $\mathcal{Q}_{\text{infinite}}$) — To be more precise, there is also a variant $\mathcal{Q}_{\text{infinite}}$ consisting of all quantum correlations with possibly infinite-dimensional Hilbert spaces. It is clear that $\mathcal{Q}_{\text{finite}} \subseteq \mathcal{Q}_{\text{infinite}}$, and it is established that $\mathcal{Q}_{\text{infinite}} \subseteq \mathcal{Q}$ by Scholz and Werner [SW08]. Determining if equality holds between these sets is known as *Tsirelson's problem* [DP15; Tsi06] and was answered to the negative by Coladangelo and Stark in former inclusion [CS18] and by Slofstra in the latter [Slo19]. In particular, neither $\mathcal{Q}_{\text{finite}}$ nor $\mathcal{Q}_{\text{infinite}}$ are closed, and both have the same closure $\mathcal{Q}$.

**Remark 3.4** (Quantum Commuting Correlations $\mathcal{Q}_c$) — A variant of this definition is the set of *quantum commuting correlations*, denoted $\mathcal{Q}_c$, in opposition to the former one that is also called the set of *quantum tensor correlations*. Here, the measurements $\big\{\{F_{a_i|x_i}\}_{a_i}\big\}_{x_i}$ are different: instead of being POVMs defined on each subsystem, they are PVMs on the global system with the additional condition that they commute pairwise:

$$\forall i, k\,, \qquad F_{a_i|x_i} \circ F_{a_k|x_k} \;=\; F_{a_k|x_k} \circ F_{a_i|x_i}\,. \tag{3.2}$$

This way, these measurements are compatible and can be performed simultaneously, see Remark 2.18. A quantum commuting correlation is then expressed as:

$$\mathbf{P}\big(a_1, .., a_n \,\big|\, x_1, .., x_n\big) \;=\; \mathrm{Tr}\Big[\big(F_{a_1|x_1} \circ \cdots \circ F_{a_n|x_n}\big)\,\rho\Big]\,. \tag{3.3}$$

This generalizes the former definition since the induced global measurements of the local POVMs are indeed pairwise commuting. In finite dimension (our case), it can be shown that the two definitions are equivalent, but they differ in infinite dimension: there is a strict inclusion $\mathcal{Q} \subsetneq \mathcal{Q}_c$ [Ji+21], combined with [FNT14]. The former, called "MIP*=RE" in quantum complexity theory, also has deep consequences to $C^*$-algebras theory thanks to [Jun+11; Oza13] and was proved through nonlocal games—see more in Section 3.3. Note that this set $\mathcal{Q}_c$ is topologically closed as demonstrated by Fritz in [Fri12].

**Remark 3.5** (Almost Quantum Correlations $\widetilde{\mathcal{Q}}$) — Another variant of the definition gives rise to the set of *almost quantum correlation*, denoted $\widetilde{\mathcal{Q}}$ [Nav+15]. As for quantum commuting correlations, the measurements are global PVM$s$ but with a weaker commuting condition than eq. (3.2):

$$\left(F_{a_1|x_1} \circ \cdots \circ F_{a_n|x_n}\right) \rho \;=\; \left(F_{a_{\pi(1)}|x_{\pi(1)}} \circ \cdots \circ F_{a_{\pi(n)}|x_{\pi(n)}}\right) \rho\,,$$

for any arbitrary permutation $\pi \in \mathfrak{S}_n$. In other words, we only require that the operator commute when applied to the state. The almost correlations are then expressed as in eq. (3.3). In the bipartite setting $n = 2$, this set can be equivalently defined as the level "$1 + AB$" of the NPA hierarchy presented in Section 3.1.2, which is the feasible set of a semi-definite program, making this set $\widetilde{\mathcal{Q}}$ to be topologically closed. Note that almost quantum correlations generalize quantum commuting correlations $\mathcal{Q}_c \subsetneq \widetilde{\mathcal{Q}}$ [Nav+15], even in finite dimension.

**Non-Signaling Correlations ($\mathcal{NS}$).** The most general type of correlations considered in this thesis is the *non-signaling correlations*, forming a set denoted $\mathcal{NS}$. It generalizes all the former sets by relaxing the constraint as follows: these correlations are conditional probability distributions that do not signal between parties. In other words, here, we keep all correlations that do not violate the *no-faster-than-light communication* principle. Mathematically, these correlations are defined as functions $\mathbf{P} : \{1, .., M\}^n \times \{1, .., N\}^n \to \mathbb{R}$ which are valid conditional probability distributions:

$$\forall a_i, x_i\,, \qquad \mathbf{P}\big(a_1..a_n \,\big|\, x_1..x_n\big) \;\geqslant\; 0 \tag{3.4}$$

$$\forall x_i\,, \qquad \sum_{a_1,..,a_n} \mathbf{P}\big(a_1..a_n \,\big|\, x_1..x_n\big) \;=\; 1\,, \tag{3.5}$$

and such that the marginals over any subset $S \subseteq \{1, .., n\}$ are independent of the inputs from the parties in $S$:

$$\forall a_i, x_i\,, \qquad \sum_{a_i\,:\,i\in S} \mathbf{P}\big(\bar{a} \,\big|\, \bar{x}\big) \;=:\; \mathbf{P}\big(\bar{a}^S \,\big|\, \bar{x}^S\big)\,, \tag{3.6}$$

where $\bar{a} := (a_1, .., a_n)$, where $\bar{a}^S$ is the same as $\bar{a}$ but without the elements $a_i$ whose index $i$ lies in $S$, and similarly for $\bar{x}$ and $\bar{x}^S$. In the simpler bipartite case $n = 2$, the latter formula translates as:

$$\forall b, x, x', y\,, \qquad \sum_a \mathbf{P}\big(a, b \,\big|\, x, y\big) \;=\; \sum_a \mathbf{P}\big(a, b \,\big|\, x', y\big) \;=: \; \mathbf{P}(b \,|\, y)\,, \tag{3.7}$$

$$\forall a, x, y, y'\,, \qquad \sum_b \mathbf{P}\big(a, b \,\big|\, x, y\big) \;=\; \sum_b \mathbf{P}\big(a, b \,\big|\, x, y'\big) \;=: \; \mathbf{P}(a \,|\, x)\,. \tag{3.8}$$

The non-signaling set strictly contains quantum correlations $\mathcal{Q} \subsetneq \mathcal{NS}$ (and actually $\widetilde{\mathcal{Q}} \subsetneq \mathcal{NS}$), as shown by Popescu and Rohrlich with the famous **PR** box bearing their name [PR94]. This correlation is defined as $\mathbf{PR}(ab|xy) := \frac{1}{2}\,\mathbb{1}_{a\oplus b=xy}$, satisfying the relation $a \oplus b = xy$ with full probability.

**Summary of all the Inclusions.** To have a large picture of all these strict inclusions, we summarize them below:

$$\mathcal{L}_{\text{det}} \subsetneq \mathcal{L} \overset{\substack{\text{[Bel64]} \\ \text{[CHSH69]}}}{\subsetneq} \mathcal{Q}_{\text{finite}} \overset{\text{[CS18]}}{\subsetneq} \mathcal{Q}_{\text{infinite}} \overset{\substack{\text{[SW08]} \\ \text{[Slo19]}}}{\subsetneq} \mathcal{Q} \overset{\substack{\text{[FNT14]} \\ \text{[Ji+21]}}}{\subsetneq} \mathcal{Q}_c \overset{\text{[Nav+15]}}{\subsetneq} \widetilde{\mathcal{Q}} \overset{\text{[PR94]}}{\subsetneq} \mathcal{NS}\,.$$

In this thesis, we focus our attention on the sets $\mathcal{L}$, $\mathcal{Q}$, and $\mathcal{NS}$. We may have the following sketch in mind to represent them[2]:

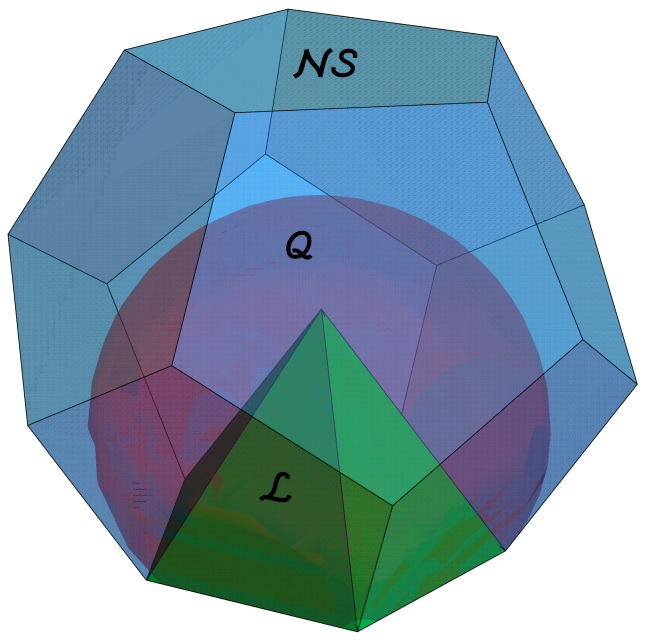

We present some geometric properties of these sets in the next section.

**Nonlocal Boxes.** To harmonize all these notations, we take the point of view of *nonlocal boxes*. These are black boxes with input parameters $x_1, .., x_n$, and output parameters $a_1, .., a_n$, for which we do not know what

[2] This image was also displayed in the M.Sc. thesis of the author [Bot22].

happens inside but whose statistics correspond to a correlation $\mathbf{P} \in \mathcal{NS}$. This is called a *device-independent* approach because instead of relying on a physical theory, it just depends on the statistics. In the bipartite case $n = 2$, one can represent a nonlocal box as follows[3]:

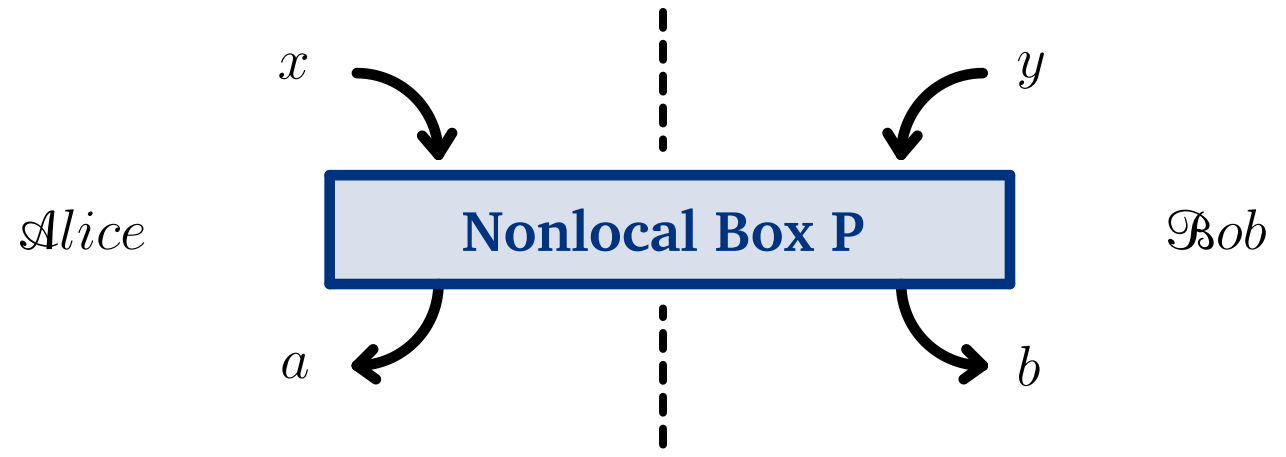

**Remark 3.6** (GPTs) — The framework of *Generalized Probabilistic Theories* (GPTs) aims to generalize classical, quantum, and non-signaling theories, as well as any arbitrary physical theory (real or hypothetical). A GPT comes with four types of mathematical structures, namely a family of state spaces (representing the physical systems); a composition rule (specifying how to gather state spaces); a set of measurements (mapping states to probabilities); and a set of possible physical operations (defining all possible transformations from a state space to another). This idea was already present in the mid-twentieth century, *e.g.* [BN36; Seg47], but GPTs under their current form were introduced by Hardy to offer a new axiomatization of quantum theory with "five very reasonable axioms" [Har01], and later developed by Barrett in the context of nonlocal boxes [Bar07]. This framework reveals that some features of quantum mechanics, like entanglement or teleportation, also hold in much more general theories and therefore are not intrinsic from the quantum theory [Aub+21; Bar+12a]. Find a good review on this subject in [Plá23].

### *3.1.2 Geometry of Nonlocal Boxes*

In this section, we describe some geometric properties of the three main correlation sets: $\mathcal{L}$, $\mathcal{Q}$, and $\mathcal{NS}$. We present several parametrizations of nonlocal boxes, discuss the properties of compactness and convexity, detail the extreme points and CHSH inequalities, and finally present some remarks on the dimensions, faces, and slices of these sets. For more details, we refer to [Bru+14; Goh+18].

[3] A similar diagram also appears in [Bot+24a; BW24].

**Parametrizations of Nonlocal Boxes.** In page 63, we presented non-signaling correlations as functions $\mathbf{P} : \{1,..,M\}^n \times \{1,..,N\}^n \to \mathbb{R}$. One can equivalently view them as real vectors:

$$\Big(\mathbf{P}(1..1|1..1),\, \mathbf{P}(1..1|1..2),\, \dots,\, \mathbf{P}(M..M|N..N)\Big) \in \mathbb{R}^{(MN)^n},$$

or as tensors in:

$$\underbrace{\mathbb{R}^M \otimes \cdots \otimes \mathbb{R}^M}_{n \text{ times}} \otimes \underbrace{\mathbb{R}^N \otimes \cdots \otimes \mathbb{R}^N}_{n \text{ times}},$$

or as real matrices:

$$M_{\mathbf{P}} := \begin{bmatrix} \mathbf{P}(0..00|0..00) & \mathbf{P}(0..01|0..00) & \cdots & \mathbf{P}(M..MM|0..00) \\ \mathbf{P}(0..00|0..01) & \mathbf{P}(0..01|0..01) & \cdots & \mathbf{P}(M..MM|0..01) \\ \vdots & \vdots & \ddots & \vdots \\ \mathbf{P}(0..00|N..NN) & \mathbf{P}(0..01|N..NN) & \cdots & \mathbf{P}(M..MM|N..NN) \end{bmatrix} \in \mathcal{M}_{N^n \times M^n}(\mathbb{R}),$$

called *correlation table* of $\mathbf{P}$, where every row sums to $1$ (used for instance in the proof of Lemma 7.10).

**Compactness.** Let us argue that the three correlation sets $\mathcal{L}$, $\mathcal{Q}$, and $\mathcal{NS}$ are compact subsets of $\mathbb{R}^m$, with $m = (MN)^n$. First, see that $\mathcal{NS}$ is bounded using the non-negativity condition (3.4) and the fact that all coefficients of a vector $\mathbf{P} \in \mathcal{NS}$ sum to $N^n$ by summing eq. (3.5) over all $x_i \in \{1,..,N\}$. Moreover, the set $\mathcal{NS}$ is topologically closed as the pre-image of the closed set defined by eqs. (3.4) to (3.6) under a continuous function. Hence, as we are in finite dimension, the set $\mathcal{NS}$ is compact. Now, using the inclusions $\mathcal{L} \subseteq \mathcal{Q} \subseteq \mathcal{NS}$, the sets $\mathcal{L}$ and $\mathcal{Q}$ are also bounded, and it remains to see that they are closed. This is true for $\mathcal{Q}$ by construction, see eq. (3.1), and although it is not obvious for $\mathcal{L}$ from its definition, it is also closed since it is a polytope, see eq. (3.9).

**Convexity.** All these sets have the very good property of being convex. This means that, given two nonlocal boxes $\mathbf{P}$ and $\mathbf{Q}$, any box of the form:

$$\alpha\,\mathbf{P} + (1-\alpha)\,\mathbf{Q},$$

with $0 \leqslant \alpha \leqslant 1$, is again in the same correlation set, where this real coefficient $\alpha$ can be viewed as additional shared randomness.

Indeed, in the non-signaling case $\mathcal{NS}$, convexity follows from the fact that each constraint in eqs. (3.4) to (3.6) preserves convexity.

In the local setting $\mathcal{L}$, it suffices to define a probability distribution $\mu$ taking value $\mu_{\mathbf{P}}(\lambda_{\mathbf{P}})$ with probability $\alpha$ and value $\mu_{\mathbf{Q}}(\lambda_{\mathbf{Q}})$ with probability $(1-\alpha)$, on a probability space $\Lambda$ consisting of the direct sum of $\Lambda_{\mathbf{P}}$ and $\Lambda_{\mathbf{Q}}$.

As for quantum correlations $\mathcal{Q}$, convexity holds because we can take any dimension for the POVMs and the quantum state. More precisely, the new state $\rho$, aiming to describe the convex combination of **P** and **Q**, is taken to be the direct sum of the unnormalized states $\alpha\,\rho_{\mathbf{P}}$ and $(1-\alpha)\,\rho_{\mathbf{Q}}$, and similarly for the local POVMs, so that the convex combination $\alpha\,\mathbf{P}+(1-\alpha)\,\mathbf{Q}$ is indeed a quantum correlation—find details in [WW01a, Section 5.C]. This means that both $\mathcal{Q}_{\text{finite}}$ and $\mathcal{Q}$ are convex, but that a restriction to some $\mathcal{Q}_d$ with fixed local dimensions $d$ may be non-convex. For instance, surprisingly, it is shown by Goh, Kaniewski, Wolfe, Vértesi, Wu, Cai, Liang, and Scarani that in the $(n,2,2)$-scenario, *i.e.* with $n$ parties and binary input-outputs, we always have $\mathcal{Q}=\mathrm{Conv}(\mathcal{Q}_2)$, meaning that one can restrain the study of quantum correlations to subsystems with dimension $d=2$ without loss of generality [Goh+18].

**Extreme Points.** As any compact convex subset of $\mathbb{R}^m$, these correlation sets can be characterized in terms of their extreme points. This comes from *Krein-Milman Theorem*, stating that any compact convex set of $\mathbb{R}^m$ is equal to the convex hull of its extreme points [KM40]:

$$\mathcal{L} \;=\; \mathrm{Conv}\Big(\mathrm{ext}(\mathcal{L})\Big)\,,$$

and similarly for $\mathcal{Q}$ and $\mathcal{NS}$. Recall that an *extreme point* in a convex set $C$ is a point that cannot be written as the trivial convex combination $\alpha\,x+(1-\alpha)\,y$, with $0<\alpha<1$, of two distinct points $x\neq y$ in $C$. Extreme points are always in the boundary of $C$, but not every boundary point is extreme—for instance, in a square, the boundary points are the four segments, whereas the extreme points are precisely the four vertices.

Let us give a description of the extreme points of $\mathcal{L}$, $\mathcal{Q}$, and $\mathcal{NS}$ in the CHSH-scenario where $n=N=M=2$, coming from [Bar+05]. The extreme points of the local set $\mathcal{L}$ are exactly the 16 deterministic boxes of $\mathcal{L}_{\text{det}}$, parametrized as follows for $\alpha,\beta,\gamma,\delta\in\{0,1\}$:

$$\mathbf{P}_{\mathbf{L}}^{\alpha,\beta,\gamma,\delta}(a,b\,|\,x,y) \;:=\; \begin{cases} 1 & \text{if } a=\alpha\,x\oplus\beta \text{ and } b=\gamma\,y\oplus\delta\,,\\ 0 & \text{otherwise}\,,\end{cases}$$

where the symbol "$\oplus$" denotes the sum modulo 2. As a consequence, it turns out that we have:

$$\mathcal{L} \;=\; \mathrm{Conv}\Big(\mathcal{L}_{\text{det}}\Big) \;=\; \mathrm{Conv}\Big(\big\{\mathbf{P}_{\mathbf{L}}^{\alpha,\beta,\gamma,\delta}\big\}\Big)\,. \tag{3.9}$$

As for the non-signaling set, it admits 24 extreme points: the 16 deterministic boxes and 8 variants of the **PR** box:

$$\mathbf{P}_{\mathbf{NL}}^{\alpha,\beta,\gamma}(a,b\,|\,x,y) \;:=\; \begin{cases} 1/2 & \text{if } a\oplus b = x\,y\oplus \alpha\,x\oplus\beta\,y\oplus\gamma\,, \\ 0 & \text{otherwise}\,, \end{cases}$$

for some parameters $\alpha,\beta,\gamma\in\{0,1\}$. It leads to the following expression of the non-signaling set:

$$\mathcal{NS} \;=\; \mathrm{Conv}\Big(\big\{\mathbf{P}_{\mathbf{L}}^{\alpha,\beta,\gamma,\delta}\big\}\cup\big\{\mathbf{P}_{\mathbf{NL}}^{\alpha,\beta,\gamma}\big\}\Big)\,. \tag{3.10}$$

Hence, as $\mathcal{L}$ and $\mathcal{NS}$ are the convex hull of finitely many points, they are *polytopes*. In contrast, the quantum set admits an infinite amount of extremal points [Bar+05], so although being convex it is not a polytope. Importantly, note also that in the CHSH-scenario, all extreme points of $\mathcal{Q}$ can be achieved via projective measurements on two-qubit pure states [Mas05; Mas06]. Moreover, note that in [Goh+18], the authors show that if a quantum box satisfies some self-testing properties, then it must be an extreme point of $\mathcal{Q}_{\text{finite}}$, and they mention that for similar results in $\mathcal{Q}$, we rather need robust self-testing (find more details about self-testing in Section 3.3.1). Interestingly, in the CHSH scenario, the extreme points of $\mathcal{Q}_{\text{infinite}}$ have recently been characterized by Barizien and Bancal in [BB25]. We present some elements of computation for the boundary of $\mathcal{Q}$ in any bipartite scenario in Section 3.1.3.

For more generality, we refer to [Bar+05] studying extreme points in the $(2,2,M)$- and $(3,2,2)$- scenarios, and to [Bie16] in the $(2,N,2)$-scenario, but from the best of our knowledge a full characterization in the general setting is an open problem to this day.

**Bell Inequalities.** As any closed convex subset of $\mathbb{R}^m$, these correlation sets are characterized in terms of half-spaces. More precisely, they are the intersection of all closed half-spaces containing them:

$$\mathcal{L} \;=\; \bigcap_{H\supseteq\mathcal{L}} H\,,$$

and similarly for $\mathcal{Q}$ and $\mathcal{NS}$, where the sets $H$ are closed half-spaces, *i.e.* sets of the form $H_{a,b} = \{x \in \mathbb{R}^m : \langle x, a \rangle \leqslant b\}$ for some fixed vector $a \in \mathbb{R}^m$ and scalar $b \in \mathbb{R}$. This is a corollary of *Hahn-Banach Theorem* in functional analysis [Ban32; Hah27], and for the polytopes $\mathcal{L}$ and $\mathcal{NS}$ this is a dual approach to the description of the extremal points due to *Minkowski-Weyl Theorem* [Zie95].

For the set of local correlations $\mathcal{L}$, this characterization gives rise to the famous *Bell inequalities* [Bel64]. They are defined as any inequality of the form $\langle x, a \rangle \leqslant b$ satisfied by all local correlations. This is a simple way to detect nonlocality: if a correlation violates this inequality, then it does not belong to $\mathcal{L}$. A particular example in the CHSH-scenario is the *CHSH inequality* [CHSH69], named after its authors Clauser, Horne, Shimony, and Holt, defined as follows:

$$S(\mathbf{P}) := E(A_0, B_0) + E(A_0, B_1) + E(A_1, B_0) - E(A_1, B_1) \leqslant 2\,, \tag{3.11}$$

where $E(A_x, B_y) := \sum_{a,b} (-1)^{a+b}\, \mathbf{P}(ab|xy)$ is the *expected value* of $\mathbf{P}$ at $x, y \in \{0, 1\}$. This equation is satisfied by all local correlations, but violated by some quantum correlations, for instance derived from the maximally entangled state $\omega$ achieving the value $S = 2\sqrt{2} > 2$. Using the symmetries of the local set $\mathcal{L}$, this inequality comes with seven other similar inequalities (also called CHSH inequalities) obtained by considering $S(\mathbf{P}) \geqslant -2$ and/or permutations of the coefficients $E(A_x, B_y)$. All together and with the normalization and non-negativity constraints in eqs. (3.4) and (3.5), they fully characterize the local set in the CHSH-scenario. Note also that there is a one-to-one correspondence between a CHSH inequality and an extreme nonlocal box $\mathbf{P}_{\mathbf{NL}}^{\alpha,\beta,\gamma}$ of $\mathcal{NS}$ [Bar+05]. A connection between Bell inequality violation and measurement incompatibility is studied in [LN22]. See a representation of CHSH inequality in the paragraph about the slices of $\mathcal{NS}$, at page 72, and find reviews on Bell inequalities in [GT09; WW01b].

In the quantum case $\mathcal{Q}$, the analog inequalities are often called *quantum Bell inequalities*. A famous one is *Tsirelson's bound* [Tsi80]:

$$S(\mathbf{P}) \leqslant 2\sqrt{2}\,, \tag{3.12}$$

satisfied by all quantum correlations, but violated by some non-signaling correlations like the **PR** box achieving the value $S = 4$. Note that this inequality is called *self-testing* because it has the remarkable property of being achievable only by the maximally entangled state $\omega \in \mathcal{D}(\mathbb{C}^2 \otimes \mathbb{C}^2)$ up to local isometries [BMR92; PR92; SW87; Tsi80], and it is even *robust* because being close to the value $2\sqrt{2}$ implies being close to $\omega$ [Mag+06]. This notion of self-testing is discussed in greater detail in Section 3.3.1. Find more about quantum Bell inequalities in [Tsi93]. Moreover, elements of computation for the boundary of $\mathcal{Q}$ are given in Section 3.1.3.

As for the set of non-signaling correlations $\mathcal{NS}$, the defining inequalities are explicitly given in its definition eqs. (3.4) to (3.6). An equivalent of the CHSH-inequality in this case is [PR94]:

$$S(\mathbf{P}) \leqslant 4\,, \tag{3.13}$$

satisfied by all non-signaling correlations and maximally achieved by the **PR** box. This is also a self-testing inequality because the only non-signaling box achieving this value is the **PR** box.

**Remark 3.7** (Correlators) — In eq. (3.11), the expected value $E(A_x, B_y) := \sum_{a,b}(-1)^{a+b}\mathbf{P}(ab|xy)$ of $\mathbf{P}$ is a particular example of *correlator* of $\mathbf{P}$. There are also two other correlators:

$$E(A_x) := \sum_{a\in\{0,1\}} (-1)^a\,\mathbf{P}(a\,|\,x) \quad \text{and} \quad E(B_y) := \sum_{b\in\{0,1\}} (-1)^b\,\mathbf{P}(b\,|\,y)\,.$$

In the bipartite setting with binary outputs $n = M = 2$, these $2N + N^2$ correlators completely parametrize correlations, with the following relation:

$$\mathbf{P}\big(a, b\,|\,x, y\big) = \frac{1 + (-1)^a\,E(A_x) + (-1)^b\,E(B_y) + (-1)^{a+b}\,E(A_x, B_y)}{4}\,.$$

The *correlator set* $\mathcal{NS}_{\text{corr}}$ is often considered to be the subset of $\mathcal{NS}$ for which $E(A_x)$ and $E(B_y)$ vanish (unbiased marginals):

$$\mathcal{NS}_{\text{corr}} := \big\{\mathbf{P} \in \mathcal{NS} \,:\, \forall x, y,\; E(A_x) = E(B_y) = 0\big\} \subsetneq \mathcal{NS}\,.$$

The elements of this set are constrained by $-1 \leqslant E(A_x, B_y) \leqslant 1$ only. Note that a generalization with more outputs $M > 2$ was introduced in [BGP10].

**Dimension.** As for any convex set, the notion of dimension is well-defined: it is defined as the dimension of the affine space obtained by spanning the convex set. For correlation sets, it is known that the three dimensions always coincide [Avi+04; Pir05]:

$$\dim \mathcal{L} = \dim \mathcal{Q} = \dim \mathcal{NS} = \Big(N(M-1)+1\Big)^n - 1\,. \tag{3.14}$$

The proof relies on the fact that $\mathcal{L}$ and $\mathcal{NS}$ span exactly the same affine space, defined by the normalization equation (3.5) and the non-signaling conditions (3.6), for which the dimension can be computed by counting the number of independent equations. In the special case with two parties and binary input-outputs, *i.e.* in the CHSH scenario $n = N = M = 2$, notice that $\dim \mathcal{NS} = 8$.

**Faces.** In the CHSH-scenario with two parties and binary input-output, consider the *nonlocal set*, denoted by $\mathcal{NL}$, defined as the convex body that is "above the CHSH hyperplane":

$$\mathcal{NL} := \left\{ \mathbf{P} \in \mathcal{NS} \,:\, S(\mathbf{P}) \geqslant 2 \right\} \subseteq \mathcal{NS}\,.$$

It is shown by Rai, Duarte, Brito, and Chaves that all its faces of dimension $7$ or less are simplexes [Rai+19], meaning that they are polytopes for which the dimension is automatically reduced when an extreme point is removed. In other words, every point of a face has a unique convex decomposition in terms of the vertices. This property allows the authors to obtain a full characterization of faces of $\mathcal{NL}$ [Rai+19]. Other applications of the simplex properties of $\mathcal{NL}$ can be found in [Bie16]. Moreover, some faces of $\mathcal{NS}$ have the special property of containing only local and post-quantum correlations, *i.e.* the only quantum correlations of these faces are actually local. These faces are called *quantum voids*, and they are fully characterized in the CHSH-scenario [Rai+19]. For instance, faces of dimension $\leqslant 4$ are always quantum voids, whereas no face of dimension $7$ is a quantum void (recall that $\dim \mathcal{NS} = 8$ in the CHSH-scenario). See also [Che+23].

Concerning local correlations $\mathcal{L}$ in the CHSH-scenario, its faces of dimension precisely $7$ are characterized by taking the equality case in the eight CHSH inequalities (3.11) together with the non-negativity condition (3.4) and the normalization condition (3.5). Some of its faces of $\mathcal{L}$ are also faces of $\mathcal{NS}$. For instance, as described in eqs. (3.9) and (3.10), its faces of dimension $0$, *i.e.* its extreme points, are the deterministic correlations and are extreme points of $\mathcal{NS}$ as well.

As for the quantum set $\mathcal{Q}$, although it is not a polytope, the notion of a face can still be defined as a set of points belonging to a given exposed hyperplane. An extensive study of faces of $\mathcal{Q}$ can be found in [Goh+18]. In particular, there, the authors identify several flat regions on the boundary of $\mathcal{Q}$ and find extreme points that are not exposed. Moreover, find examples of faces of $\mathcal{Q}$ in common with faces of $\mathcal{L}$ in [Lin+07], or even examples of faces in common of maximal dimension (*i.e.* facets) [Alm+10]. We present elements of computation for the boundary of $\mathcal{Q}$ in Section 3.1.3.

**Slices of $\mathcal{NS}$.** Consider the CHSH-scenario with two parties and binary input-output. A convenient way to visualize the "inside" of the $8$-dimensional polytope $\mathcal{NS}$ is to study some of its $2$-dimensional slices, like we would do for a cake to see its inner layers. Here are two examples[4] of slices containing the **PR** box:

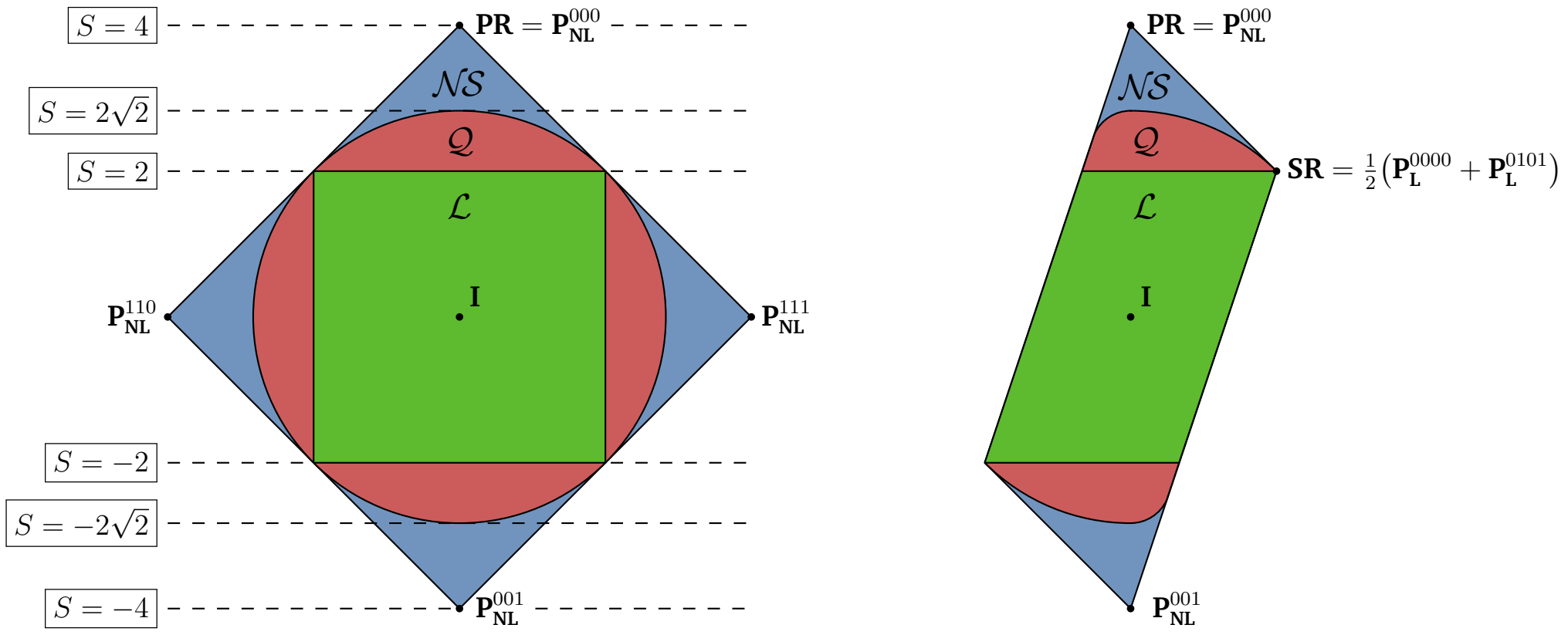

In green is represented the slice of the local set ($\mathcal{L}$), in red the quantum set ($\mathcal{Q}$), and in blue the non-signaling set ($\mathcal{NS}$). See that the inclusion and convexity properties are preserved in each slice. Find other examples of slices in [Goh+18] and in Chapter 6.

### 3.1.3 *Approximating the Boundary of $\mathcal{Q}$*

Understanding the boundary of the quantum set $\mathcal{Q}$ is crucial for both foundational questions, including its description in terms of a physical principle like communication complexity (see Chapter 4), and practical tasks in device-independent quantum information processing, see Section 3.3. Nevertheless, as mentioned in Section 3.1.2, although being convex, this set is inconveniently not a polytope, making the description of its boundary $\partial\mathcal{Q}$ much harder in general:[5]

[4] Similar diagrams appear in our manuscript [BBP24].

[5] Interestingly, in the CHSH scenario, the extreme points of $\mathcal{Q}_{\text{infinite}}$ have recently been characterized by Barizien and Bancal in [BB25].

**Fact 3.8** ([Kem+11]) — *Computing $\partial\mathcal{Q}$ is NP-hard in the three-partite setting.*

In this section, we present a standard way to approximate the boundary of the quantum commuting correlations $\mathcal{Q}_c$ (containing $\mathcal{Q}$ strictly, introduced in Remark 3.4) via a sequence of semidefinite programs (SDPs) called Navascués-Pironio-Acín (NPA) hierarchy [NPA07; NPA08], and its equivalent in terms of Sum-of-Squares (SoS) decompositions [Doh+08]. We use both of them in Chapter 8 [Bot+24b]. Note that these results are restricted to the bipartite setting $\mathcal{H} = \mathcal{H}_\mathsf{A} \otimes \mathcal{H}_\mathsf{B}$. Note that the idea of using SDPs for such problems was already present in [Weh06]. Find a review on this topic in [Bru+14].

**Remark 3.9** (Much Simpler for Correlators) — Before delving into the details of the approximation of the boundary of $\mathcal{Q}_c$, we mention that the study of the quantum boundary is much simpler in the space of correlators $\mathcal{NS}_{\text{corr}}$ introduced in Remark 3.7. In the bipartite case with two outputs, *i.e.* when $n = M = 2$, the boundary of $\mathcal{Q}_c \cap \mathcal{NS}_{\text{corr}}$ is completely characterized in terms of the following equation called Tsirelson–Landau–Masanes bound [Lan88; Mas03; Tsi80; Tsi93]:

$$\Big| \arcsin E(A_0, B_0) + \arcsin E(A_0, B_1) + \arcsin E(A_1, B_0) - \arcsin E(A_1, B_1) \Big| \leqslant \pi \,,$$

combined the symmetric inequalities obtained by permuting the coefficients $E(A_x, B_y)$. Those equations are non-linear and they correspond to taking the $\arcsin$ of each coefficient in Bell inequality and changing the upper bound $2$ into $\pi$. Nevertheless, although it is a precise characterization in $\mathcal{Q}_c \cap \mathcal{NS}_{\text{corr}}$, for quantum commuting correlations $\mathbf{P} \in \mathcal{Q}_c$ these inequalities are only necessary conditions in general, which is why we need more precise conditions as the ones presented below.

**NPA Hierarchy [NPA07; NPA08].** This method was introduced by Navascués, Pironio, and Acín in [NPA07] as a bound of the set $\mathcal{Q}_c$, and therefore of $\mathcal{Q}$ as well. The idea is derived from Lasserre [Las01], stating that any polynomial optimization problem in commutative variables can in principle be solved using a hierarchy of SDPs. Consider a general quantum commuting correlation:

$$\mathbf{P}\big(a, b \,|\, x, y\big) \;=\; \langle\psi|E_{a|x}F_{b|y}|\psi\rangle \;\in\; \mathcal{Q}_c\,,$$

where $\{E_{a|x}\}_a$ and $\{F_{b|y}\}_b$ are PVM$s$ that commute, and where $|\psi\rangle$ is a pure state in $\mathbb{C}^{d_\mathsf{A}} \otimes \mathbb{C}^{d_\mathsf{B}}$. Consider the sequence of sets $\mathcal{O}^{(\ell)}$, $\ell \in \mathbb{N}$, constructed as follows from the operators $E_{a|x}$ and $F_{b|y}$ by induction:

$$\begin{aligned}
\mathcal{O}^{(0)} &:= \{\mathbb{I}\}\,,\\
\mathcal{O}^{(1)} &:= \mathcal{O}^{(0)} \cup \big\{E_{a|x}\big\}_{a,x} \cup \big\{F_{b|y}\big\}_{b,y}\,,\\
\mathcal{O}^{(2)} &:= \mathcal{O}^{(1)} \cup \big\{E_{a|x} \cdot E_{a'|x'}\big\}_{a,a',x,x'} \cup \big\{E_{a|x} \cdot F_{b|y}\big\}_{a,b,x,y} \cup \big\{F_{b|y} \cdot F_{b'|y'}\big\}_{b,b',y,y'}\,,\\
&\dots\\
\mathcal{O}^{(\ell)} &:= \Big\{\text{products of } E_{a|x} \text{ and } F_{b|y} \text{ of length} \leqslant \ell\Big\}\,,\\
&\dots
\end{aligned}$$

Denote by $k_\ell$ the number of elements in $\mathcal{O}^{(\ell)}$, and by $O_i^{(\ell)}$ the elements ($i = 1, .., k_\ell$). Then, for each level $\ell \in \mathbb{N}$, one can define the *moment matrix* $\Gamma^{(\ell)}$ of size $k_\ell \times k_\ell$ whose coefficients are:

$$\Gamma_{ij}^{(\ell)} \;:=\; \Big\langle \psi \,\Big|(O_i^{(\ell)})^* O_i^{(\ell)}\Big|\psi \Big\rangle\,.$$

This matrix satisfies three good properties for any $\ell \geqslant 1$:

- $\Gamma^{(\ell)} \succcurlyeq \mathbf{0}$ is positive semi-definite;
- the entries of $\Gamma^{(\ell)}$ satisfy a series of linear inequalities;
- the values of $\mathbf{P}(ab|xy)$ correspond to a subset of the entries of $\Gamma^{(\ell)}$.

As we can build such a moment matrix $\Gamma^{(\ell)}$ from any quantum commuting correlation $\mathcal{Q}_c$, these three properties become necessary conditions for $\mathcal{Q}_c$. More precisely, if we denote by $\mathcal{Q}^{(\ell)}$ the set of all nonlocal boxes $\mathbf{P} \in \mathcal{NS}$ such that there exists a moment matrix $\Gamma^{(\ell)}$ with the three above requirements, then we have a decreasing sequence of sets:

$$\mathcal{Q}^{(1)} \;\supseteq\; \mathcal{Q}^{(2)} \;\supseteq\; \mathcal{Q}^{(3)} \;\supseteq\; \cdots \;\supseteq\; \mathcal{Q}_c\,.$$

It means that each $\mathcal{Q}^{(\ell)}$ is a relaxation of the quantum commuting set $\mathcal{Q}_c$. Furthermore, what is even more remarkable is that their intersection tends precisely to $\mathcal{Q}_c$ as $\ell \to \infty$ [NPA08]:

$$\bigcap_{\ell=1}^{\infty} \mathcal{Q}^{(\ell)} = \mathcal{Q}_c \,.$$

In other words, quantum commuting correlations are characterized as follows:

$$\mathbf{P} \in \mathcal{Q}_c \qquad \Longleftrightarrow \qquad \forall \ell \geqslant 1, \;\; \mathbf{P} \in \mathcal{Q}^{(\ell)} \,.$$

This is particularly interesting since the three above conditions can be efficiently determined by SDPs. Note that the first level $\mathcal{Q}^{(1)}$ is also *analytically* characterized in [NPA07], and that a more general approach not limited to quantum commuting correlations is developed in [PNA10].

**Remark 3.10** (Almost Quantum Correlations) **—** The almost quantum correlations $\widetilde{\mathcal{Q}}$, introduced in Remark 3.5, correspond to an intermediate level between $\ell = 1$ and $\ell = 2$ in the hierarchy:

$$\mathcal{Q}^{(1)} \supseteq \widetilde{\mathcal{Q}} \supseteq \mathcal{Q}^{(2)} \,.$$

More precisely, the corresponds the the level called "$1 + AB$" defined from the following set:

$$\mathcal{O}^{(1+AB)} := \mathcal{O}^{(1)} \cup \left\{ E_{a|x} \cdot F_{b|y} \right\}_{a,b,x,y} \,.$$

Thus, its boundary can also be efficiently computed with SDP$s$.

**Dual Approach: Sum-of-Squares [Doh+08].** Subsequently, another approach was proposed by Doherty, Liang, Toner, and Wehner following dual methods from Parrilo [Par03] in terms of *Sum-of-Squares* (SoS) decompositions. If we view nonlocal boxes as elements of $\mathbb{R}^m$, we can try to determine the boundary of $\mathcal{Q}_c$ in a certain direction $\vec{s} \in \mathbb{R}^m$ by computing the following minimization:

$$\begin{array}{rl} \text{Minimize} & \beta \,, \\ \text{subject to} & \cdot \; \langle \vec{s}, \mathbf{P} \rangle \leqslant \beta \,, \\ & \cdot \; \mathbf{P} \in \mathcal{Q}_c \,. \end{array}$$

Denote $\beta_*$ the optimal value. We can rephrase the problem as follows:

$$\beta_* \;=\; \max_{\mathbf{P}\in\mathcal{Q}_c} \langle \vec{s}, \mathbf{P}\rangle \;=\; \max_{\{E_{a|x}\}_a,\{F_{b|y}\}_b} \left\| \sum_{abxy} s_{xy}^{ab}\, E_{a|x} F_{b|y} \right\|_{\text{op}} ,$$

where the coefficients $s_{xy}^{ab}$ of fixed by $\vec{s}$, where the optimization is over all commuting PVMs $\{E_{a|x}\}_a$ and $\{F_{b|y}\}_b$, and where $\|\cdot\|_{\text{op}}$ is the operator norm (*i.e.* the larger eigenvalue of the operator).

Now, seeing that the condition $\beta \geqslant \|X\|_{\text{op}}$ for an operator $X$ is implied by $\beta\,\mathbb{I} - X \succcurlyeq \mathbf{0}$, the idea is to proceed as follows:

$$\begin{array}{ll} \text{Minimize} & \beta\,, \\ \text{subject to} & \cdot\ \beta\,\mathbb{I} - \sum_{abxy} s_{xy}^{ab}\, E_{a|x}F_{b|y} \succcurlyeq \mathbf{0}\,, \\ & \cdot\ \ \{E_{a|x}\}_a \text{ and } \{F_{b|y}\}_b \text{ are commuting PVM}s\,. \end{array}$$

But, an operator $Y \succcurlyeq \mathbf{0}$ is positive semi-definite *if, and only if,* it can be written as a sum-of-squares [Par03], meaning that the first constraint may be rewritten as follows:

$$\beta\,\mathbb{I} - \sum_{abxy} s_{xy}^{ab}\, E_{a|x}F_{b|y} \;=\; \sum_{k=1}^{K} \alpha_k\, B_k^* B_k\,,$$

for some operators $B_i$ and some positive coefficients $\alpha_k > 0$. This can be cast to an SDP problem by fixing a maximal degree $K$ in the sum, and we again obtain a sequence of SDPs (here in the parameter $K$) whose solution $\beta^{(K)}$ converges to the quantum commuting value $\beta_*$ in the asymptotic regime [Doh+08].

**Remark 3.11** (Convergence at Finite Level) — In both methods (NPA and SoS), it is possible to prove that a box $\mathbf{P}$ lies in $\mathcal{Q}_c$ (or to find $\beta_*$) at a finite level of the hierarchy. Find examples in [Doh+08; NPA08]. In such a case, it is even sometimes possible to explicit what state $|\psi\rangle$ and PVMs $\{E_{a|x}\}_a$ and $\{F_{b|y}\}_b$ can be used to achieve the decomposition of $\mathbf{P}$.

**Remark 3.12** (Comparison with Lower Bounds) — Lower bounds on $\beta_*$ can be found by exhibiting examples of $\mathbf{P} \in \mathcal{Q}_c$ such that $\langle \vec{s}, \mathbf{P}\rangle$ is high enough. If this lower bound matches a level $\beta^{(K)}$, it means that convergence is already achieved at step at most $K$. Lower bounds can be found by optimizing over quantum states of fixed finite dimension. Using this lower bound method, Pál and Vértesi found the optimal value $\beta_*$ for $221$ examples of $\vec{s}$ matching the result with the level $K = 3$ of the hierarchy [PV09].

**Remark 3.13** (Precision of First Levels) — The first levels of both hierarchies are already very precise. For instance, Kempe, Regev, and Toner proved that, in certain scenarios, the first level $\ell = 1$ (or $K = 1$) already gives the quantum bound. This is in particular the case for the quantum CHSH inequality $S \leqslant 2\sqrt{2}$ from eq. (3.12): all correlations of $\mathcal{Q}^{(1)}$ already satisfy this equation.

### *3.1.4 Wirings of Nonlocal Boxes*

In this section, in the CHSH-scenario, we present a natural way of building a new box $\mathbf{R} := \mathbf{P} \boxtimes_{\mathsf{W}} \mathbf{Q}$ out of two nonlocal boxes $\mathbf{P}$ and $\mathbf{Q}$ and what is called a *wiring* W. This framework leads to the notion of *algebra of boxes* introduced and developed in Chapter 6 [Bot+24a]. Here, we begin by giving an intuition behind wirings, then we describe deterministic and mixed wirings, and finally, we provide examples of wirings used in the literature. A more general framework of wirings can be found in [BG15].

**Intuition Behind Wirings.** Given two non-signaling boxes $\mathbf{P}, \mathbf{Q} \in \mathcal{NS}$, it is possible to build a new box by *wiring* them together. This notion of wiring has found great interest in the last two decades, especially with the following two goals: (i) attempting nonlocality distillation, *i.e.* we want to build a box that is "strongly nonlocal" starting from several copies of a box that is "weakly nonlocal" [BG15; Bri+19; BS09; DW08; EW14; EWC23a; EWC23b; FWW09; HR10; Nai+23]; (ii) finding sets that are closed under wirings, because it is argued that a consistent physical theory should, in principle, be closed under natural simple operations as wirings [All+09a; BG15; LVN14; Nav+15; NW09].

As one might guess, a wiring simply connects the outputs of a box to the inputs of another box under some rules, with the possibility of applying some pre- and post-processing operations to the carried bits. An example of wiring is presented in Figure 3.1 (a). This wiring indeed connects some outputs to some inputs, but might seem counter-intuitive at first since Alice and Bob do not use their share of the boxes in the same order: while Alice uses $\mathbf{P}$ then $\mathbf{Q}$, Bob uses $\mathbf{Q}$ then $\mathbf{P}$. This independence on the choice of box order for each party generalizes quantum mechanics because likewise if Alice and Bob share two entangled pairs of quantum states instead of two nonlocal boxes, Alice would be able to measure her first particle and then the second one, while Bob would be able to do the converse, and they would still receive the outputs "instantaneously" without having to wait that the other party performs a measurement. Hence, as in the quantum case, Alice receives an answer from the box **P** instantaneously even if Bob has not yet inputted a bit in his side of **P**, and she can use the output $a_1$ as a parametrization for the input $x_2$ of the box **Q**; similarly for Bob. This "instantaneous-answer" property of a box is typical of non-signaling correlations, as modeled by eqs. (3.7) and (3.8) saying that Alice's marginal is independent of Bob's input, and vice-versa. Note that a wiring cannot link Alice's side to Bob's side, nor the opposite, since otherwise it could create a signaling box: there would be communication between parties.

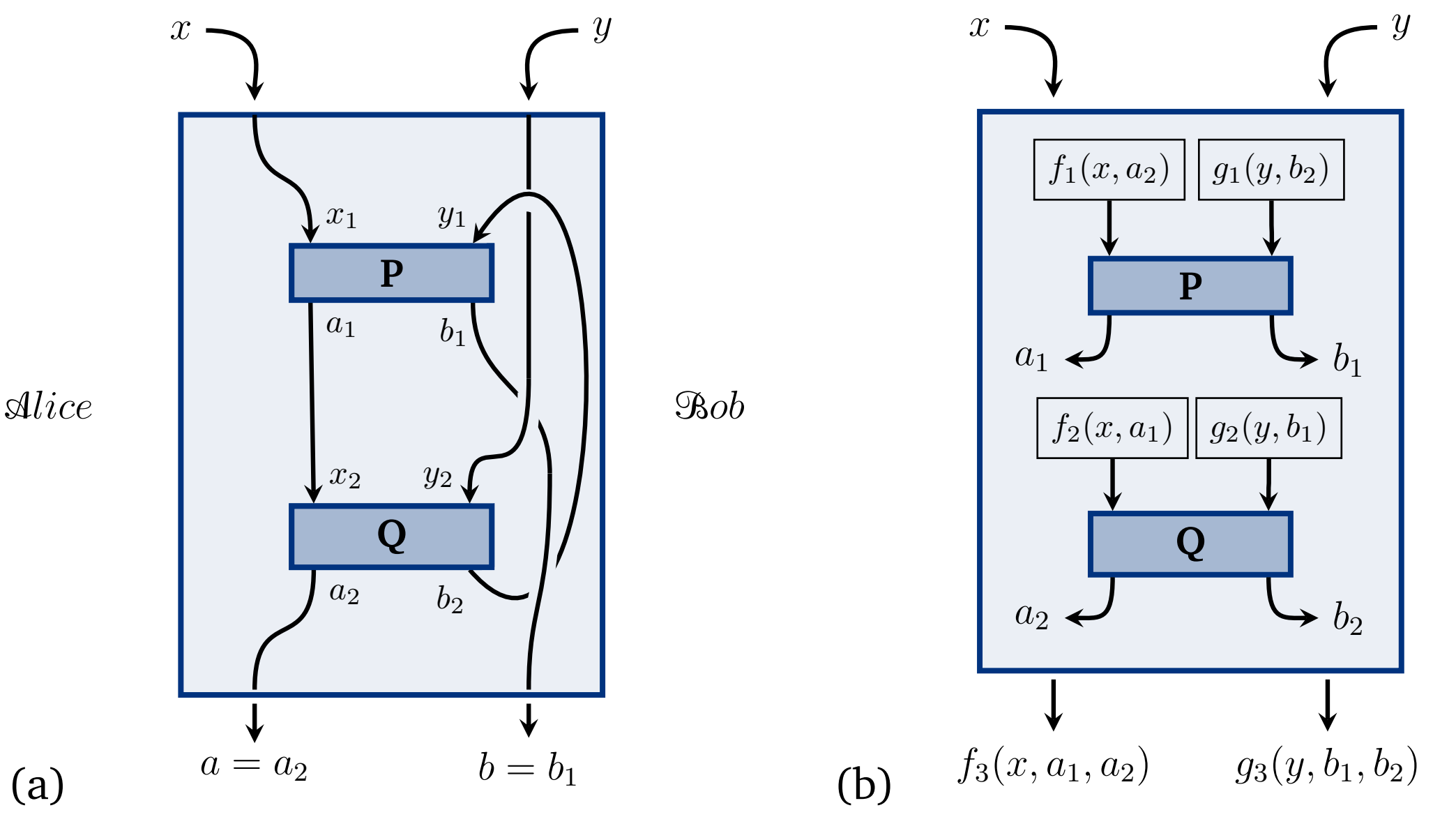

**Figure 3.1** — *(a) Example of a wiring between two boxes* **P** *and* **Q**. *(b) General wiring between two boxes* **P** *and* **Q**. *These diagrams also appear in [Bot+24a].*

**Deterministic Wirings.** More generally, two boxes **P** and **Q** can be wired as in Figure 3.1 (b), using functions $f_i$ and $g_j$ depending on the global entries $x$ and $y$ and on the outputs $a_k$ and $b_\ell$ of the boxes. Nevertheless, to be a valid wiring, the inputs on Alice's side must be in a valid order: the input $x_2$ of **Q** can depend on the output $a_1$ of **P** only if the input $x_1$ of **P** does not depend on the output $a_2$ of **Q**; the same should also hold on Bob's side. In other words, the functions $f_1(x, a_2)$ and $f_2(x, a_1)$ cannot both depend on $a_2$ and $a_1$ respectively for the same value of $x$, and similarly for $g_1(y, b_2)$ and $g_2(y, b_1)$. These conditions are formalized in eqs. (3.15) and (3.16) of the following definition:

**Definition 3.14** (Deterministic wiring) — *A deterministic wiring* W *between two boxes* $\mathbf{P}, \mathbf{Q} \in \mathcal{NS}$ *consists in six Boolean functions* $f_1, f_2, g_1, g_2 : \{0,1\}^2 \to \{0,1\}$ *and* $f_3, g_3 : \{0,1\}^3 \to \{0,1\}$ *satisfying the following* non-cyclicity conditions*:*

$$\forall x, \qquad \Big(f_1(x,0) - f_1(x,1)\Big)\Big(f_2(x,0) - f_2(x,1)\Big) = 0\,, \tag{3.15}$$

$$\forall y, \qquad \Big(g_1(y,0) - g_1(y,1)\Big)\Big(g_2(y,0) - g_2(y,1)\Big) = 0\,. \tag{3.16}$$

Given a wiring W and two boxes $\mathbf{P}, \mathbf{Q} \in \mathcal{NS}$, we obtain a new box that we denote $\mathbf{P} \boxtimes_{\mathsf{W}} \mathbf{Q}$. Formally, this new box is defined as the following conditional probability distribution:

$$\mathbf{P} \underset{\mathsf{W}}{\boxtimes} \mathbf{Q}(a,b\,|\,x,y) := \sum_{a_1,a_2,b_1,b_2} \mathbf{P}\Big(a_1,\, b_1\,|\, f_1(x,a_2),\, g_1(y,b_2)\Big)$$
$$\times\, \mathbf{Q}\Big(a_2,\, b_2\,|\, f_2(x,a_1),\, g_2(y,b_1)\Big) \times \mathbb{1}_{a=f_3(x,a_1,a_2)} \times \mathbb{1}_{b=g_3(y,b_1,b_2)}\,. \tag{3.17}$$

Note that it is important to specify the condition $\mathbf{P}, \mathbf{Q} \in \mathcal{NS}$ since it implies that $\mathbf{P} \boxtimes_{\mathsf{W}} \mathbf{Q}$ is indeed a conditional probability distribution. Otherwise, for instance consider the probability distributions $\mathbf{P} = \mathbf{Q} = \mathbb{1}_{a=y}\mathbb{1}_{b=x}$ not in $\mathcal{NS}$ and the deterministic wiring $\mathsf{W} = (f_1 = x, f_2 = a_1, g_1 = b_2, g_2 = y, f_3 = 0, g_3 = 0)$, then the resulting box $\mathbf{P} \boxtimes_{\mathsf{W}} \mathbf{Q}$ is not a well-defined probability distribution.

**Mixed Wirings.** Using local randomness, one can generalize from deterministic to *mixed wirings*. The difference resides in that the functions $f_i$ and $g_j$ now take values in $[0,1]$ instead of $\{0,1\}$. This can be interpreted as follows: using the notation of Figure 3.1, if $f_1(x, a_1) = p \in [0,1]$ for some fixed bits $x$ and $a_1$, then it means that Alice uses a Bernoulli distribution $\mathcal{B}(p)$ to assign a value to the input bit $x_1 \in \{0,1\}$—the value $x_1 = 1$ with probability $p$ and $x_1 = 0$ with probability $1 - p$. Again, the mixed wiring is denoted $\mathsf{W} = \big(f_1(0,0), f_1(0,1), \ldots, g_3(1,1,1)\big) \in \mathbb{R}^{32}$ and can be expressed as the expected value of the 32 Bernoulli variables, since their realizations $B_i \in \{0,1\}$ form a deterministic wiring $\mathsf{W}_{\text{det}}^{\{B_i\}} = (B_1, \dots, B_{32})$:

$$\mathbf{P} \boxtimes_{\mathsf{W}} \mathbf{Q} = \mathop{\mathbb{E}}_{B_1} \cdots \mathop{\mathbb{E}}_{B_{32}} \left[ \mathbf{P} \boxtimes_{\mathsf{W}_{\text{det}}^{\{B_i\}}} \mathbf{Q} \right] . \tag{3.18}$$

Now, as we use real numbers instead of bits for the inputs of the nonlocal box, we need to adopt the following convention:

$$\begin{aligned} \mathbf{P}(a,b \,|\, \alpha,\beta) \;:=\; & (1-\alpha)(1-\beta)\,\mathbf{P}(ab\,|\,00) + (1-\alpha)\beta\,\mathbf{P}(ab\,|\,01) \\ & + \alpha(1-\beta)\,\mathbf{P}(ab\,|\,10) + \alpha\beta\,\mathbf{P}(ab\,|\,11)\,, \end{aligned} \tag{3.19}$$

for any coefficients $\alpha, \beta \in [0,1]$. Moreover, in order to ensure a well-defined local order for both Alice and Bob, we use again the non-cyclicity conditions introduced in eqs. (3.15) and (3.16), so that there is a dependence relation between the variables $B_i$. It yields the following definition:

**Definition 3.15** (Mixed wiring) — *A* mixed wiring $\mathsf{W}$ *between two boxes* $\mathbf{P}, \mathbf{Q} \in \mathcal{NS}$ *consists of six functions* $f_1, f_2, g_1, g_2 : \{0,1\}^2 \to [0,1]$ *and* $f_3, g_3 : \{0,1\}^3 \to [0,1]$ *satisfying the* non-cyclicity conditions *eqs. (3.15) and (3.16). Mixed wirings form the following set:*

$$\mathcal{W} := \Big\{ \textit{mixed wirings } \mathsf{W} \Big\} .$$

The set of mixed wirings $\mathcal{W}$ is not convex because of the non-affinity of the non-cyclicity conditions eqs. (3.15) and (3.16). For instance, consider the wirings $\mathsf{W}$, $\mathsf{W}'$ with all coefficients $0$ except the one corresponding to $f_1(0,0) = 1$, $f_2'(0,0) = 1$ respectively. Then, each of these wirings satisfies the non-cyclicity conditions, but their average $\mathsf{W}'' = (\mathsf{W} + \mathsf{W}')/2$ does not:

$$\Big( f_1''(0,0) - f_1''(0,1) \Big) \Big( f_2''(0,0) - f_2''(0,1) \Big) \;=\; \big(1/2 - 0\big)\big(1/2 - 0\big) \;\neq\; 0\,.$$

Hence $\mathcal{W}$ is non-convex. Moreover, for mixed wirings, the expression of $\mathbf{P} \boxtimes_{\mathsf{W}} \mathbf{Q}$ can be taken the same as before if we use the above convention

eq. (3.19), and we have:

$$
\begin{aligned}
&\mathbf{P} \boxtimes_{\mathsf{W}} \mathbf{Q}(a,b\,|\,x,y) \\
&= \sum_{a_1,a_2,b_1,b_2\in\{0,1\}} \Big[\mathbf{P}\big(a_1,\,b_1\,|\,0,\,0\big)(1-f_1(x,a_2))\,(1-g_1(y,b_2))+\mathbf{P}\big(a_1,\,b_1\,|\,0,\,1\big)(1-f_1(x,a_2))\,g_1(y,b_2) \\
&\quad + \mathbf{P}\big(a_1,\,b_1\,|\,1,\,0\big)\,f_1(x,a_2)\,(1-g_1(y,b_2)) + \mathbf{P}\big(a_1,\,b_1\,|\,1,\,1\big)\,f_1(x,a_2)\,g_1(y,b_2)\Big] \\
&\times \Big[\mathbf{Q}\big(a_2,\,b_2\,|\,0,\,0\big)(1-f_2(x,a_1))\,(1-g_2(y,b_1))+\mathbf{Q}\big(a_2,\,b_2\,|\,0,\,1\big)(1-f_2(x,a_1))\,g_2(y,b_1) \\
&\quad + \mathbf{Q}\big(a_2,\,b_2\,|\,1,\,0\big)\,f_2(x,a_1)\,(1-g_2(y,b_1)) + \mathbf{Q}\big(a_2,\,b_2\,|\,1,\,1\big)\,f_2(x,a_1)\,g_2(y,b_1)\Big] \\
&\times \Big[(1-f_3(x,a_1,a_2))\,\mathbb{1}_{a=0}+f_3(x,a_1,a_2)\,\mathbb{1}_{a=1}\Big] \times \Big[(1-g_3(y,b_1,b_2))\,\mathbb{1}_{b=0}+g_3(y,b_1,b_2)\,\mathbb{1}_{b=1}\Big]\,.
\end{aligned}
\tag{3.20}
$$

**Closure Under Wirings.** The notion of being *closed under wirings* is introduced in [All+09a] and is presented as a necessary condition for a correlation set to describe a valid physical theory. This notion means that wiring boxes from a certain set does not permit to exit this set, and can be formalized as follows[6]:

> **Definition 3.16** (Closed under Wirings) — *A subset $X \subseteq \mathcal{NS}$ is said to be* closed under wirings *if for all boxes* $\mathbf{P}, \mathbf{Q}$ *in $X$ and all mixed wirings* W, *the new box* $\mathbf{P} \boxtimes_{\mathsf{W}} \mathbf{Q}$ *is in $X$ again.*

**Example 3.17 —** The three correlations sets $\mathcal{L}, \mathcal{Q}, \mathcal{NS}$ are closed under wirings [All+09a], as well as the set of almost quantum correlations $\widetilde{\mathcal{Q}}$ [Nav+15]. Notice that, as mixed wirings are convex combinations of deterministic wirings (see eq. (3.18)) and as these sets are convex, it suffices to show the closure only for deterministic wirings. Find other examples of closed sets under wirings in [BG15; LVN14; NW09].

**Typical Examples of Wirings.** We now review some typical wirings that are studied in the literature. See Figure 3.2 for an illustration of these wirings. Note that all of these wirings are *deterministic*.

[6]There exist more general definitions of being closed under wirings, involving $k$ boxes and $n$ parties. Nevertheless, in this thesis, we restrict the study to the simpler case of $k = n = 2$, which is the reason why we give a weaker definition here. For the general framework, see [All+09a; BG15; LVN14; Nav+15; NW09].

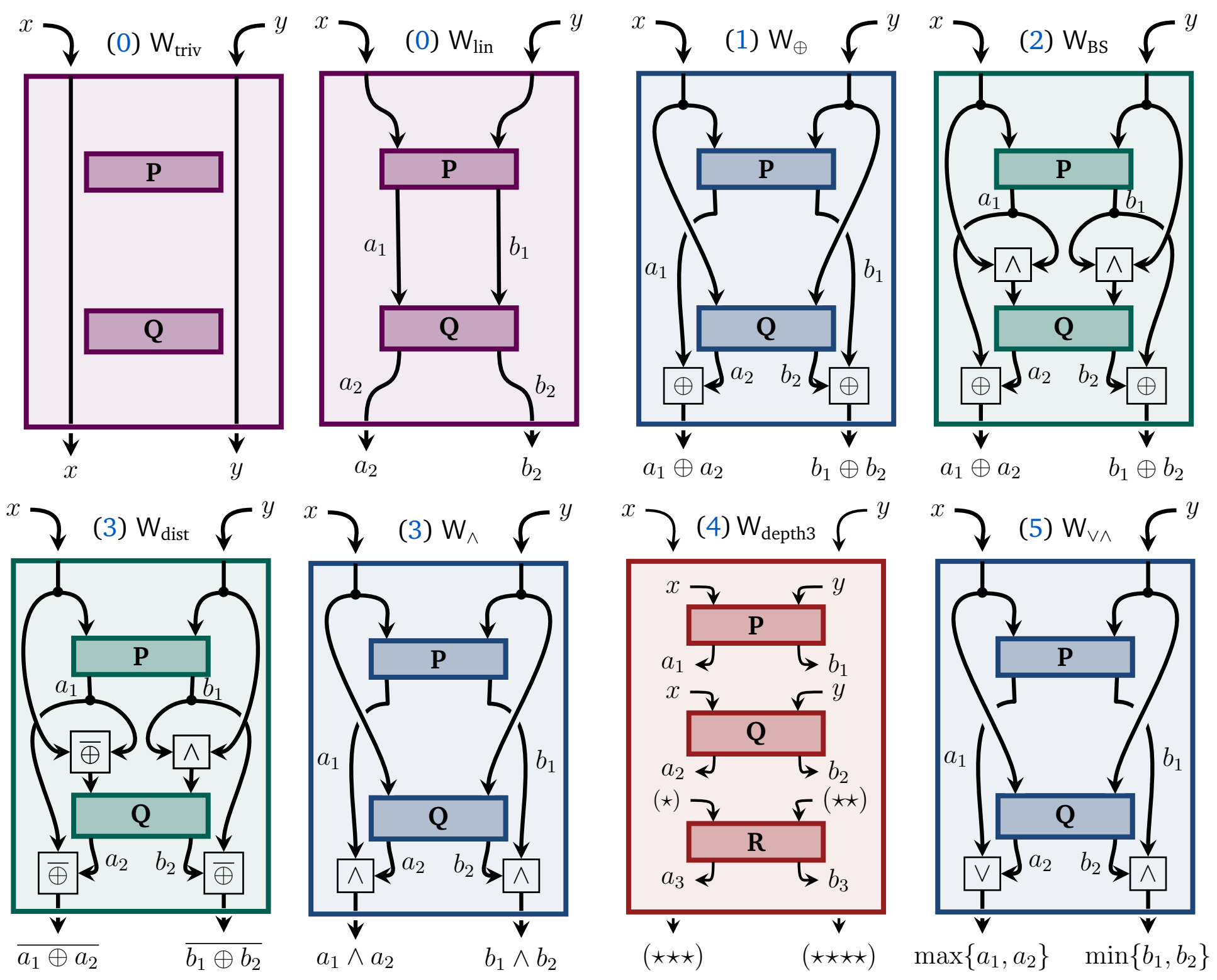

**Figure 3.2 —** *Typical examples of wirings. The same color indicates a similar internal structure. The overline bar denote the NOT gate:* $\overline{x} = x \oplus 1$. *The symbol* $(\star)$ *stands for* $xa_2 \vee x\overline{a_1} \vee \overline{xa_2}a_1$, $(\star\star)$ *for* $yb_2 \vee y\overline{b_1}$, $(\star\star\star)$ *for* $a_3a_2 \vee a_3\overline{a_1} \vee \overline{a_3a_2}a_1$, *and* $(\star\star\star\star)$ *for* $b_3b_2 \vee b_3\overline{b_1} \vee \overline{b_3b_2}b_1$. *Similar diagrams also appear in [Bot+24a].*

(0) The *trivial wiring* $\mathsf{W}_{\text{triv}}$ is defined as the wiring that does "nothing", in the sense that it outputs exactly the global inputs: $(a, b) = (x, y)$. Another one is the *linear wiring* $\mathsf{W}_{\text{lin}}$ that simply connects the output of a box to the input of the box immediately below.

(1) Forster, Winkler, and Wolf introduced in [FWW09] the *XOR wiring* $\mathsf{W}_{\oplus}$ in order to distill nonlocality, also studied in [AIR25]. It consists in setting boxes in parallel and taking the sum mod $2$ of the outputs on each party.

(2) Then, Brunner and Skrzypczyk enhanced the wiring from item (1) in [BS09] in order to distillate nonlocality. Their wiring $\mathsf{W}_{\text{BS}}$ is *adap-* *tive*, in the sense that boxes are no longer in parallel: the second box's inputs $x_2, y_2$ are not simply equal to the previous box's outputs $a_1, b_1$, but they are combined with the general inputs $x, y$. Their new protocol is so powerful that it allows to arbitrarily reduce the noise of any *correlated box* (defined as convex combinations of **PR** and **SR**) so that the **PR** box is almost perfectly simulated. This has significant consequences for the collapse of communication complexity, see Chapter 4.

(3) Then, Allcock, Brunner, Linden, Popescu, Skrzypczyk, and Vértesi studied two variants of the previous wirings in [All+09a]. First, their *distillation wiring* $\mathsf{W}_{\text{dist}}$ is similar to the one in item (2) because it is also adaptive and distills correlated boxes. This wiring, together with $\mathsf{W}_{\text{dist}}$, allows to fully characterize the distillable region of the slice **PR**–**SR**–**I** [Bru+11]. Second, their *AND wiring* $\mathsf{W}_{\wedge}$ resembles the one in item (1) since boxes are set in parallel, but we take the product of the outputs instead of the sum.

(4) After, Høyer and Rashid studied some depth-$k$ generalizations of the wirings from items (1) and (2) in [HR10], *i.e.* they wired $k$ boxes instead of only two. In particular, they found an example of a depth-$3$ protocol that extends the known region of distillable boxes. This idea is improved upon in [EWC23a] by constructing genuine *depth*-$3$ *wirings* like $\mathsf{W}_{\text{depth3}}$ drawn in Figure 3.2. This new wiring is shown to be strictly better than any depth-$2$ wiring in terms of the collapse of communication complexity, because there exist collapsing nonlocal boxes with this wiring that cannot be distilled using depth-$2$ wirings only. Note that in this thesis, the study is limited to depth-$2$ wirings.

(5) More recently, Naik, Sidhardh, Sen, Roy, Rai, and Banik defined the *OR-AND wiring* $\mathsf{W}_{\vee\wedge}$ in order to distill quantum nonlocality [Nai+23]. This wiring is a mixing of the ones in items (1) and (3): it consists in setting boxes in parallel and in taking the maximum (the "OR") of Alice's outputs and the minimum (the "AND") of Bob's outputs.

### *3.1.5 Measures of Nonlocal Boxes*

A measure of nonlocal boxes is a function taking a nonlocal box **P** and assigning to it a real value in $[0, 1]$. If this function has some good properties, then it allows us to infer interesting features of nonlocal boxes, for instance, the impossibility of distilling nonlocality above a certain threshold. Below, after presenting a measure of probability distributions and quantum states, we describe a measure of nonlocal boxes called *maximal correlation* that has the good property of being monotone under wirings. This is the framework for our ongoing work presented in Chapter 9. This subsection is very much related to work from Beigi and Gohari [BG15].

**Measure of Probability Distributions.** Hirschfeld and Gebelein introduced in the mid-twentieth century a measure of correlation called *maximal correlation* of a probability distribution [Geb41; Hir35], which was then developed by Rényi [Rén59a; Rén59b] and many others [Ana+14; KA16; KU11; Pol12; Wit75]. It is defined as follows. If $\mathcal{A}$ and $\mathcal{B}$ are measurable spaces and if $p$ is a probability measure on $\mathcal{A}\times\mathcal{B}$, then the maximal correlation of $p$ is the maximum of Pearson's correlation coefficient:

$$\mu_{\text{prob}}(p) := \max_{A,B} \ \mathbb{E}_p\left[A\,B\right], \qquad \text{s.t.} \ \begin{cases} \mathbb{E}_{p_{\mathsf{A}}}[A] = \mathbb{E}_{p_{\mathsf{B}}}[B] = 0\,, \\ \mathbb{E}_{p_{\mathsf{A}}}[A^2] = \mathbb{E}_{p_{\mathsf{B}}}[B^2] = 1\,, \end{cases}$$

where $A : \mathcal{A} \to \mathbb{R}$ and $B : \mathcal{B} \to \mathbb{R}$ are random variables, and where the last expected values are relative to the marginals $p_{\mathsf{A}}(a) := \sum_b p(a,b)$ and $p_{\mathsf{B}}(b) := \sum_a p(a,b)$ of $p$. Note that $\mu_{\text{prob}}$ is not defined when the support of $p_{\mathsf{A}}$ (or $p_{\mathsf{B}}$) is a singleton, but in this case, we take the convention $\mu_{\text{prob}}(p) = 0$, which makes sense because $A$ and $B$ are completely uncorrelated. This measure has the following good properties:

**Fact 3.18** (Classical Maximal Correlation) — *The maximal correlation $\mu_{\text{prob}}$ satisfies all of the following:*

- $\mu_{\text{prob}}(p) = 0 \quad \Leftrightarrow \quad p(a,b) = p_{\mathsf{A}}(a)\, p_{\mathsf{B}}(b)$.
- [Wit75] $\mu_{\text{prob}}(p) = 1 \quad \Leftrightarrow \quad p_{\mathsf{A}}$, $p_{\mathsf{B}}$ *have "common data".*
- [KU11] *Efficiently computable: $\mu_{\text{prob}}$ is the second singular value of a matrix.*
- [Wit75] *Tensorization:* $\mu_{\text{prob}}(p \times q) = \max\{\mu_{\text{prob}}(p), \mu_{\text{prob}}(q)\}$.
- *Monotony: if $q$ is a local stochastic transformation of $p$, then $\mu_{\text{prob}}(q) \leqslant \mu_{\text{prob}}(p)$.*

Note that the tensorization property contrasts with the usual subadditivity of Shannon entropy $H(X,Y) \leqslant H(X) + H(Y)$.

**Remark 3.19** (Why this Formula?) — We believe that the formula in the definition of $\mu_{\text{prob}}$ is chosen as such for two reasons. First, it is because the measure $\mu_{\text{prob}}$ can be rephrased in terms of the covariance and the variance:

$$\mu_{\text{prob}}(p) = \max_{A,B} \frac{\operatorname{cov}(A,B)}{\sqrt{\mathbb{V}(A)\mathbb{V}(B)}} ,$$

where the covariance is a known measure of correlation between two random variables. Second, it is because such a definition automatically leads to a decrease under wirings. Indeed, a wiring of probability $p(a,b)$ can be seen as the pre-processing of the inputs, giving rise to a new probability distribution $\tilde{p}(a,b)$. This measure is the maximum over all possible pre-processings of $p$, so we necessarily have $\mu_{\text{prob}}(p) \geqslant \mu_{\text{prob}}(\tilde{p})$, *i.e.* a decrease under wirings.

**Measures of Quantum States.** A first extension of $\mu_{\text{prob}}$ was introduced by Beigi to quantum states [Bei13]. The *maximal correlation* of a quantum state $\rho \in \mathcal{D}(\mathcal{H}_{\mathsf{A}} \otimes \mathcal{H}_{\mathsf{B}})$ is:

$$\begin{aligned} \mu_{\text{quant}}(\rho) := \quad & \max_{X,Y} \quad \left| \operatorname{Tr}\big(\rho\,(X \otimes Y^*)\big) \right| , \\ & \text{s.t.} \quad \begin{cases} \operatorname{Tr}\big[\rho\,(X \otimes \mathbb{I}_{\mathsf{B}})\big] = \operatorname{Tr}\big[\rho\,(\mathbb{I}_{\mathsf{A}} \otimes Y)\big] = 0 , \\ \operatorname{Tr}\big[\rho\,(XX^* \otimes \mathbb{I}_{\mathsf{B}})\big] = \operatorname{Tr}\big[\rho\,(\mathbb{I}_{\mathsf{A}} \otimes YY^*)\big] = 1 , \end{cases} \end{aligned}$$

where $X$ lies in the set $\mathcal{B}(\mathcal{H}_{\mathsf{A}})$ of bounded operators acting on the Hilbert space $\mathcal{H}_{\mathsf{A}}$, and likewise $Y \in \mathcal{B}(\mathcal{H}_{\mathsf{B}})$. It has the following good properties:[7]

**Fact 3.20** (Quantum Maximal Correlation [Bei13]) — *The maximal correlation $\mu_{\text{quant}}$ satisfies all of the following:*

- $\mu_{\text{quant}}(\rho) = 0 \quad \Leftrightarrow \quad \rho = \rho_{\mathsf{A}} \otimes \rho_{\mathsf{B}}$ *is a product state.*
- $\mu_{\text{quant}}(\rho) = 1 \quad \Leftrightarrow$ *there exist local operators $X \in \mathcal{B}(\mathcal{H}_{\mathsf{A}})$ and $Y \in \mathcal{B}(\mathcal{H}_{\mathsf{B}})$ different from* $\mathbf{0}$ *and* $\mathbb{I}$ *such that* $\rho\big(X \otimes \mathbb{I}_{\mathsf{B}}\big) = \rho\big(\mathbb{I}_{\mathsf{A}} \otimes Y\big)$.

[7] The characterization of $\mu_{\text{quant}} = 1$ is available on a more recent arXiv version of the paper: https://arxiv.org/pdf/1210.1689v5

- *Efficiently computable:* $\mu_{\text{quant}}$ *is the second Schmidt coefficient of a matrix.*
- *Tensorization:* $\mu_{\text{quant}}(\rho\otimes\sigma) = \max\{\mu_{\text{quant}}(\rho), \mu_{\text{quant}}(\sigma)\}$.
- *Monotony: if* $\sigma$ *is a local transformation of* $\rho^{\otimes k}$ *for some* $k$*, then* $\mu_{\text{quant}}(\sigma) \leqslant \mu_{\text{quant}}(\rho)$.

The value of this measure on isotropic states $\rho^{(\alpha)} := (1-\alpha)\,\mathbb{I}_d/d^2 + \alpha\,\omega$, where $\omega$ is the maximally entangled state and $0 \leqslant \alpha < 1$, is $\mu_{\text{quant}}(\rho^{(\alpha)}) = \alpha$, so using the monotony property the author deduces that entanglement cannot be distilled from these states and that we cannot even extract common randomness from them.

Based on this measure, Beigi also introduced a variant called *maximal entanglement* [Bei14], defined as the quasi-convexification of $\mu_{\text{quant}}$:

$$\nu_{\text{quant}}(\rho) \; := \inf_{\rho=\sum_i \alpha_i \tau_i} \; \max_i \; \mu_{\text{quant}}(\tau_i)\,,$$

where $\alpha_i \geqslant 0$ (they can equal $0$) and the $\tau_i$'s are quantum states, and where the sum has a finite index. By construction, we always have $\nu_{\text{quant}}(\rho) \leqslant \mu_{\text{quant}}(\rho)$, and it can be shown that equality holds precisely when the state $\rho$ is pure. As for $\mu_{\text{quant}}$, there is a characterization of $\nu_{\text{quant}}(\rho) = 0$, it is exactly when $\rho$ is a separable state $\rho = \sum_i \alpha_i \rho_{\mathsf{A}}^{(i)} \otimes \rho_{\mathsf{B}}^{(i)}$. Moreover, it also satisfies the tensorization and monotony properties, and it is quasi-convex:

$$\nu_{\text{quant}}\Big(\textstyle\sum_i \beta_i \rho^{(i)}\Big) \;\leqslant\; \max_i \; \nu_{\text{quant}}\big(\rho^{(i)}\big)\,.$$

Nevertheless, to the best of our knowledge, no characterization of the maximal value $\nu_{\text{quant}}(\rho) = 1$ is known, neither of an efficient way of computing this measure.

**Measure of Non-signaling Boxes.** Another generalization of $\mu_{\text{prob}}$ was proposed by Beigi and Gohari to nonlocal boxes [BG15][8]. The *maximal correlation* of a non-signaling box $\mathbf{P}$ is defined as the maximum over the inputs of the maximal correlation of the outputs:

$$\mu_{\text{box}}(\mathbf{P}) := \max_{x,y} \; \mu_{\text{prob}}\big(\mathbf{P}(\cdot,\cdot\,|\,x,y)\big)\,.$$

This measure has the following good properties:

[8] It was also recently generelized to the "Gaussian maximal correlation" [BRK23].

**Fact 3.21** (Non-Signaling Maximal Correlation [BG15]) — *The maximal correlation $\mu_{\text{box}}$ satisfies all of the following:*

- $\mu_{\text{box}}(\mathbf{P}) = 0 \quad \Leftrightarrow \quad \mathbf{P}(ab|xy) = \mathbf{P}_{\mathsf{A}}(a|x)\,\mathbf{P}_{\mathsf{B}}(b|y)$ *is a product box.*
- *Efficiently computable: $\mu_{\text{box}}$ is the maximum over four values that are the second singular of some matrices.*
- *Monotony: $\forall$ wiring* W, $\mu_{\text{box}}\big(\mathbf{P} \boxtimes_{\mathsf{W}} \mathbf{Q}\big) \leqslant \max\big\{\mu_{\text{box}}(\mathbf{P}), \mu_{\text{box}}(\mathbf{Q})\big\}$.

The last property about monotony is significant. It allows the authors to find new correlation sets that are closed under wiring. It also gives bound on nonlocal distillation via wirings: as the measure cannot increase via wiring, it suffices to compute the sublevel sets $\{\mathbf{P} \in \mathcal{NS} \,:\, \mu_{\text{box}}(\mathbf{P}) \leqslant x\}$ to see how high one can hope to distillate a nonlocal box.

## 3.2 Nonlocal Games

Nonlocal games play a crucial role in demonstrating quantum advantage in various situations, thereby showcasing the power of quantum correlations. They have applications in fields such as complexity theory and quantum cryptography, as detailed in Section 3.3.

In this section, we begin by presenting some generalities about nonlocal games (Section 3.2.1). We then describe several types of games developed in the literature, including but not limited to the CHSH game (Section 3.2.2), graph games (Section 3.2.3), and no-cloning games (Section 3.2.4), which will be used in our contributions (Chapters 6 to 8). For more details on this topic, we refer to [Cle+04; PV16].

### *3.2.1 Generalities*

Let us introduce all the general materials for nonlocal games. For the sake of simplicity, we consider the bipartite scenario where $n = 2$, but everything here can be generalized to the multipartite setting.

**Vocabulary.** In the context of nonlocal games, the two parties, often named Alice (A) and Bob (B), are called the *players* of the game. A third non-playing party is also involved in the process, often called the *Referee* (R), with the ability to communicate with Alice and Bob. The game starts when the Referee samples some inputs $x$ and $y$, called *questions*, from a joint probability distribution $\pi$ on a finite set $\mathcal{X} \times \mathcal{Y}$, and sends them to the players, $x$ for Alice and $y$ for Bob. Upon receiving $x$ and $y$, the players proceed with their strategy and send outputs $a$ and $b$ to the Referee, called *answers*, from finite sets $\mathcal{A}$ and $\mathcal{B}$ respectively. Finally, the Referee uses a Boolean function $\mathcal{V} : \mathcal{A} \times \mathcal{B} \times \mathcal{X} \times \mathcal{Y} \to \{0,1\}$, called *rule of the game* or *predicate of the game*, to determine whether the players win ($\mathcal{V}(a,b,x,y) = 1$) or lose ($\mathcal{V}(a,b,x,y) = 0$).[9] Note that the game is inherently collaborative: either both Alice and Bob win, or both of them lose. Nevertheless, during the game phase time, communication between the players is not allowed—they can only agree on a common strategy beforehand. More precisely, we say that they are *space-like separated*, meaning that no communication traveling at most at the speed of light can physically reach the other player within the expected game time frame. This is one reason why such games are labeled *nonlocal*. In short, a nonlocal game can thus be defined as follows:

> **Definition 3.22** (Nonlocal Game) — *A* nonlocal game $\mathsf{G}$ *is the data of* $(\mathcal{A}, \mathcal{B}, \mathcal{X}, \mathcal{Y}, \pi, \mathcal{V})$*, where:*
>
> - $\mathcal{A}, \mathcal{B}, \mathcal{X}, \mathcal{Y}$ *are finite sets;*
> - $\pi : \mathcal{X} \times \mathcal{Y} \to [0,1]$ *is a probability distribution;*
> - $\mathcal{V} : \mathcal{A} \times \mathcal{B} \times \mathcal{X} \times \mathcal{Y} \to \{0,1\}$ *is a Boolean function.*

**Strategy.** Although Alice and Bob cannot communicate during the game, they can share a resource like entanglement or more generally a nonlocal box, so that their answers can become very correlated. Find a representation in Figure 3.3. For their questions $x$ and $y$, the players can produce some input $x'$ and $y'$ for the nonlocal box, they receive some outputs $a'$ and $b'$, and finally they can produce their answers $a$ and $b$. The transformations $x \mapsto x'$ and $y \mapsto y'$ are called *pre-processings*, while $a' \mapsto a$ and $b' \mapsto b$ *post-processings*. Note that, more generally, their strategy may involve several nonlocal boxes (or several copies of a box), which is why we introduced

[9] A variant of this function $\mathcal{V}$ taking values in $[0,1]$ instead of $\{0,1\}$ is used in [PV16] to allow randomized predicates.

the notion of *wirings* in Section 3.1.4.

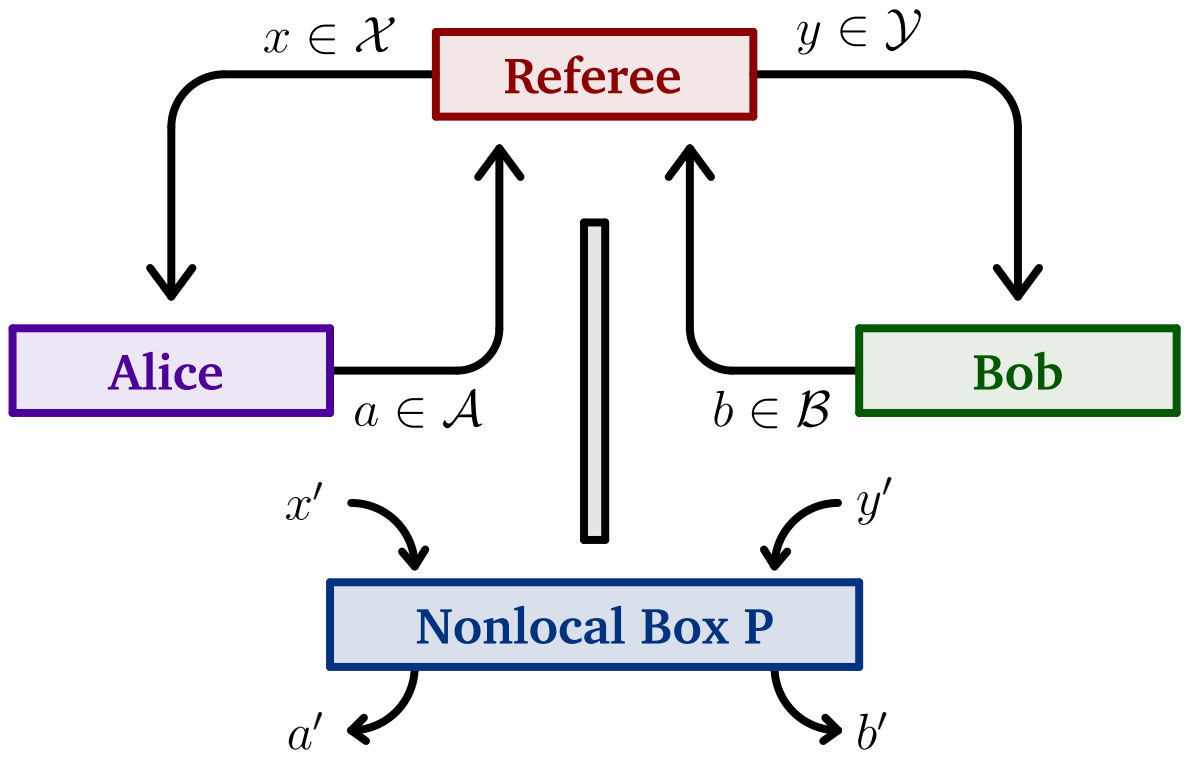

**Figure 3.3** — *Representation of a generic nonlocal game. A similar diagram also appears in [BBP24; Bot22].*

**Winning Probability.** Suppose Alice and Bob are provided with a nonlocal box $\mathbf{P} \in \mathcal{NS}$, and that they do not apply any pre- or post-processing: they use their questions $x,y$ as inputs $x',y'$ for $\mathbf{P}$, and their outputs $a',b'$ from $\mathbf{P}$ as answers to the Referee[10]. This leads to the following expression for the winning probability at the game G:

$$\mathbb{P}\big(\mathbf{P} \text{ wins at } \mathsf{G}\big) = \sum_{a,b,x,y} \pi(x,y)\, \mathbf{P}(a,b\,|\,x,y)\, \mathscr{V}(a,b,x,y)\,. \tag{3.21}$$

The best winning probability for the game G given local, quantum, or non-signaling boxes is called *value* of the game and denoted as follows:

$$\mathfrak{w}_{\mathcal{L}}(\mathsf{G}) := \max_{\mathbf{P}\in\mathcal{L}} \mathbb{P}\big(\mathbf{P} \text{ wins at } \mathsf{G}\big)\,, \tag{3.22}$$

and similarly for $\mathfrak{w}_{\mathcal{Q}}$ and $\mathfrak{w}_{\mathcal{NS}}$. Of course, for any game G, we always have the following relation:

$$\mathfrak{w}_{\mathcal{L}}(\mathsf{G}) \leqslant \mathfrak{w}_{\mathcal{Q}}(\mathsf{G}) \leqslant \mathfrak{w}_{\mathcal{NS}}(\mathsf{G})\,,$$

and we say that there is an *advantage* when one of the inequalities is strict (quantum or non-signaling advantage). Not all games display an advantage, but we will see in Section 3.2.2 an example of a game for which the three values are all distinct, namely the CHSH game. If we have $\mathfrak{w}_{\mathcal{L}} < \mathfrak{w}_{\mathcal{Q}} = 1$, then we say that there is *quantum pseudo-telepathy* [BBT03; BBT05; BCT99]. For instance, the *magic square game* showcases quantum pseudo-telepathy, see Section 3.2.4.

[10] Note that any box with pre- and post-processing gives rise to another box without pre- or post-processing.

**Remark 3.23** (Computing the Value $\mathfrak{w}(\mathrm{G})$) — Note that the maximization problem in eq. (3.22) is a linear optimization over a convex set. As a consequence, the optimal value is achieved at an extreme point of $\mathcal{L}$, *i.e.* at a deterministic box in $\mathcal{L}_{\text{det}}$. Likewise, the quantum and non-signaling values $\mathfrak{w}_{\mathcal{Q}}$ and $\mathfrak{w}_{\mathcal{NS}}$ are achieved at their extreme points. This is one reason why knowing all the points is crucial—find a description of them on page 67. Notice that, however, the optimal value can also be achieved at a non-extreme point, which is the case for instance at the CHSH game for the classical value, see Section 3.2.2.

### 3.2.2 The CHSH Game

The CHSH game is indubitably the best-known nonlocal game. Derived from the CHSH inequality (eq. (3.11)) and named after Clauser, Horne, Shimony, and Holt [CHSH69], this game consists in obtaining the formula $a \oplus b = xy$, where the sign "$\oplus$" denotes the sum modulo 2, given bit questions $x, y \in \{0,1\}$ sampled uniformly at random, and with bit answers $a, b \in \{0,1\}$. More formally, we have:

**Definition 3.24** (CHSH Game) — *The* CHSH game *is defined by the following data:*

$$\mathcal{A} = \mathcal{B} = \mathcal{X} = \mathcal{Y} = \{0,1\}\,, \quad \pi(x,y) = \frac{1}{4}\,, \quad \textit{and} \quad \mathcal{V}(a,b,x,y) = \mathbb{1}_{a\oplus b = xy}\,.$$

As the name suggests, note that this game is played in the CHSH-scenario ($n = N = M = 2$). This game was the first one to show a quantum advantage ($\mathfrak{w}_{\mathcal{L}} < \mathfrak{w}_{\mathcal{Q}}$) [CHSH69], as well as a non-signaling advantage ($\mathfrak{w}_{\mathcal{Q}} < \mathfrak{w}_{\mathcal{NS}}$) [PR94]. Let us study the different values of the game.

**Classical Value.** As mentioned in Remark 3.23, computing the classical value $\mathfrak{w}_{\mathcal{L}}(\mathrm{CHSH})$ amounts to optimizing the winning probability (eq. (3.21)) over the deterministic set $\mathcal{L}_{\text{det}}$. Now, as this set is finite, one can quickly check that the best winning probability is [CHSH69]:

$$\mathfrak{w}_{\mathcal{L}}(\text{CHSH}) = \frac{3}{4}.$$

For instance, it is achieved by the box $\mathbf{P_{00}}$ that always outputs $(a, b) = (0, 0)$ independently of $x$ and $y$. Indeed, such a box satisfies the relation $a \oplus b = xy$ three times out of four, precisely when $x$ or $y$ is zero. Note that optimal value $\frac{3}{4}$ is also achieved by other boxes, like $\mathbf{P_{11}}$ that always outputs $(a, b) = (1, 1)$, or any convex combination of the two (which are no longer deterministic but still optimal). Notice that, similarly, we can show that the lowest classical winning probability at CHSH is $\frac{1}{4}$, so no matter what is their strategy, the players cannot always lose.

**Quantum Value.** Now, using a quantum resource, one can show that the value of the game becomes [Tsi80]:

$$\mathfrak{w}_{\mathcal{Q}}(\text{CHSH}) = \cos^2\!\left(\tfrac{\pi}{8}\right) = \frac{1}{2} + \frac{1}{2\sqrt{2}} \approx 0.8536...$$

Importantly, since $\mathfrak{w}_{\mathcal{L}} < \mathfrak{w}_{\mathcal{Q}}$, we infer that the CHSH game shows a *quantum advantage*. Furthermore, here, the optimal value is uniquely achieved up to local isometries, which is used for self-testing applications, see Section 3.3. This value is achieved by the maximally entangled state over two qubits:

$$\omega = |\Omega\rangle\langle\Omega| \in \mathcal{D}\big(\mathbb{C}^2 \otimes \mathbb{C}^2\big),$$

where $|\Omega\rangle := \frac{1}{\sqrt{2}}\big(|00\rangle + |11\rangle\big)$. The idea of the proof is as follows. Consider the local PVM induced by the following orthonormal basis of $\mathbb{C}^2$:

$$B_\theta := \Big\{\cos(\theta)\,|0\rangle + \sin(\theta)\,|1\rangle,\; -\sin(\theta)\,|0\rangle + \cos(\theta)\,|1\rangle\Big\},$$

for some angular parameter $\theta \in [-\pi, \pi]$. (Recall the definition of a basis measurement in Example 2.21 and of a local measurement in Section 2.3.3.) Upon receiving $x$, Alice performs a measurement of her qubit in the basis $B_\theta$ with $\theta = 0$ if $x = 0$, and $\theta = \frac{\pi}{4}$ otherwise. Likewise, upon receiving $y$, Bob performs a measurement with $\theta = \frac{\pi}{8}$ if $y = 0$, and $\theta = -\frac{\pi}{8}$ otherwise. Then we can compute all the values of the induced quantum box $\mathbf{P}_{\text{quant}}$. For instance:

$$\begin{aligned}\mathbf{P}_{\text{quant}}(0,0\,|\,0,0) &= \operatorname{Tr}\Big[\Big(|0\rangle\langle 0|\otimes|\psi\rangle\langle\psi|\Big)\,\omega\Big] = \frac{\cos^2\left(\frac{\pi}{8}\right)}{2}\,,\\ \mathbf{P}_{\text{quant}}(1,0\,|\,0,0) &= \operatorname{Tr}\Big[\Big(|1\rangle\langle 1|\otimes|\psi\rangle\langle\psi|\Big)\,\omega\Big] = \frac{\sin^2\left(\frac{\pi}{8}\right)}{2}\,,\end{aligned}$$

where $|\psi_\theta\rangle = \cos\left(\frac{\pi}{8}\right)|0\rangle + \sin\left(\frac{\pi}{8}\right)|1\rangle$ is the first vector of the basis $B_\theta$ with $\theta = \frac{\pi}{8}$. Going through the computations for all possibilities of $\mathbf{P}_{\text{quant}}(ab|xy)$, we see that its value is always $\frac{1}{2}\cos^2\left(\frac{\pi}{8}\right)$ when $a \oplus b = xy$. Using eq. (3.21), this leads to the claimed value:

$$\begin{aligned}\mathbb{P}\big(\mathbf{P}_{\text{quant}} \text{ wins at CHSH}\big) &= \sum_{a,b,x,y} \frac{1}{4} \times \mathbf{P}_{\text{quant}}(a,b\,|\,x,y) \times \mathbb{1}_{a\oplus b=xy}\\ &= \frac{1}{4} \times \frac{\cos^2\left(\frac{\pi}{8}\right)}{2} \times \sum_{a,b,x,y} \mathbb{1}_{a\oplus b=xy}\\ &= \cos^2\left(\frac{\pi}{8}\right).\end{aligned}$$

As $\mathbf{P}_{\text{quant}} \in \mathcal{Q}$ by construction, this proves that $\mathfrak{w}_{\mathcal{Q}} \geqslant \cos^2(\frac{\pi}{8})$. Conversely, Tsirelson's bound [Tsi80] gives the other inequality, see Remark 3.25 below. Hence $\mathfrak{w}_{\mathcal{Q}} = \cos^2(\frac{\pi}{8})$ as claimed. Note that this value is also experimentally confirmed [ADR82; Hen+15; Wei+98].

**Non-Signaling Value.** Notice that the **PR** box in $\mathcal{NS}$ is designed to perfectly win the CHSH game since it always outputs $(a,b)$ such that $a \oplus b = xy$. Hence $\mathbb{P}\big(\mathbf{PR} \text{ wins at CHSH}\big) = 1$, which yields [PR94]:

$$\mathfrak{w}_{\mathcal{NS}} = 1\,.$$

As a consequence, in addition to showing a quantum advantage, the CHSH game demonstrates a *non-signaling advantage*. Note that the **PR** box is the only non-signaling box achieving this winning probability, thus it can be self-tested as well.

**Remark 3.25** (Link With the CHSH Inequality) — There is a close relation between the CHSH inequality and the CHSH game. The function $S(\mathbf{P})$ defined in eq. (3.11) characterizes the winning probability of $\mathbf{P}$ at CHSH and vis-versa as follows:

$$\mathbb{P}\big(\mathbf{P} \text{ wins at CHSH}\big) \;=\; \frac{1}{2} + \frac{S(\mathbf{P})}{8}\,.$$

This allows us to retrieve the classical value from the CHSH inequality [CHSH69]:

$$S\big(\mathbf{P}\big) \;\leqslant\; 2 \qquad \iff \qquad \mathfrak{w}_{\mathcal{L}} \;\leqslant\; \frac{3}{4}\,,$$

as well as the quantum value from Tsirelson's bound [Tsi80]:

$$S\big(\mathbf{P}\big) \;\leqslant\; 2\sqrt{2} \qquad \iff \qquad \mathfrak{w}_{\mathcal{Q}} \;\leqslant\; \frac{1}{2} + \frac{1}{2\sqrt{2}}\,,$$

and as well as the non-signaling value from the non-signaling CHSH inequality [PR94]:

$$S\big(\mathbf{P}\big) \;\leqslant\; 4 \qquad \iff \qquad \mathfrak{w}_{\mathcal{NS}} \;\leqslant\; 1\,.$$

**Remark 3.26** (Generalizations of the CHSH Game) — Based on the Chained Bell Inequalities of Braunstein and Caves [BC90b], the CHSH game can be generalized to the *odd cycle game* as follows [Cle+04; Vai01]. Let $m \geqslant 3$ be an odd integer. Questions $x, y$ belong to the ring of integers modulo $m$, *i.e.* $\mathcal{X} = \mathcal{Y} = \mathbb{Z}_m$, and answers $a, b$ are bits, *i.e.* $\mathcal{A} = \mathcal{B} = \{0, 1\}$. The probability distribution $\pi$ is the uniform distribution on:

$$\Big\{(x, y) \in \mathbb{Z}_m \times \mathbb{Z}_m \;:\; x = y \text{ or } x + 1 \equiv y \pmod{m}\Big\}\,,$$

and the predicate is:

$$\mathcal{V}(a, b, x, y) \;:=\; \begin{cases} 1 & a \oplus b = (x + 1 \equiv y \pmod{m})\,, \\ 0 & \text{otherwise}\,, \end{cases}$$

where "true" is associated to "1" and "fasle" to "0." This game can be viewed as the 2-coloring game of the cycle $\mathcal{C}_m$ introduced in Section 3.2.3. The values for this game are [BC90b]:

$$\mathfrak{w}_{\mathcal{L}} \;=\; 1 - \frac{1}{2m}\,, \qquad \mathfrak{w}_{\mathcal{Q}} \;=\; \cos^2\!\Big(\frac{\pi}{4m}\Big) \;\geqslant\; 1 - \Big(\frac{\pi}{4m}\Big)^2\,, \qquad \mathfrak{w}_{\mathcal{NS}} \;=\; 1\,.$$

Note that the quantum value is quadratically close to 1. Find another generalization of the CHSH game with the class of XOR games at page 102, or with roots of unity in [Cui+20].

### 3.2.3 Graph Games

Graph games are nonlocal games where the questions and answers are vertices of some graphs, and where the players want to mimic to the Referee that a certain property holds in the graph by giving the "good answers" to their questions (the notion of good being defined by the rule $\mathscr{V}$ of the game). In this section, we mainly present three graph games, namely the *graph isomorphism game*, the *graph homomorphism game*, and the *graph coloring game*, that we will need for Chapter 7 [BW24]. Then, we list some examples of other graph games studied in the literature. Note that another graph game is introduced in Chapter 7, called the *vertex distance game*, generalizing the isomorphism and coloring games to some extent. Here, all graphs are always assumed to be non-empty, finite, undirected, and loopless.

**Graph Isomorphism Game.** The graph isomorphism game was introduced by Atserias, Mančinska, Roberson, Šámal, Severini, and Varvitsiotis in [Ats+19]. This is a nonlocal game based on two graphs, $\mathcal{G}$ and $\mathcal{H}$, for which the players Alice and Bob try to pretend to the Referee that they are isomorphic in the classical sense. Recall that $\mathcal{G}$ is said to be *isomorphic* to $\mathcal{H}$, denoted $\mathcal{G} \cong \mathcal{H}$, if there exists a bijection map $\varphi$ from the vertex set $V(\mathcal{G})$ to the vertex set $V(\mathcal{H})$ such that adjacency is preserved in both ways:

$$\forall g, g' \in \mathcal{G}, \qquad g \sim g' \quad \Longleftrightarrow \quad \varphi(g) \sim \varphi(g')\,, \tag{3.23}$$

where the symbol "$\sim$" denotes the adjacency relation.

The game consists in the following. The Referee provides the players Alice and Bob with respective questions $x_{\mathsf{A}}, x_{\mathsf{B}} \in V := V(\mathcal{G}) \sqcup V(\mathcal{H})$, where $V(\mathcal{G})$ and $V(\mathcal{H})$ are assumed to be disjoint. In return, Alice and Bob use a predetermined strategy (a nonlocal box) in order to produce some vertices $y_{\mathsf{A}}, y_{\mathsf{B}} \in V$, and they send $y_{\mathsf{A}}, y_{\mathsf{B}}$ to the Referee, who finally verifies if the players won the game. The first condition they need to satisfy is that $x_{\mathsf{A}}$ and $y_{\mathsf{A}}$ have to be in different vertex sets, and similarly for $x_{\mathsf{B}}$ and $y_{\mathsf{B}}$, meaning that

$$x_{\mathsf{A}} \in V(\mathcal{G}) \Leftrightarrow y_{\mathsf{A}} \in V(\mathcal{H}) \quad \text{and} \quad x_{\mathsf{B}} \in V(\mathcal{G}) \Leftrightarrow y_{\mathsf{B}} \in V(\mathcal{H})\,, \tag{3.24}$$

otherwise, they lose the game. Now, assuming that this condition holds, only one vertex among $x_{\mathsf{A}}$ and $y_{\mathsf{A}}$ is in $V(\mathcal{G})$, let us call it $g_{\mathsf{A}} \in V(\mathcal{G})$, and the other $h_A \in V(\mathcal{H})$, and similarly for $g_B \in V(\mathcal{G})$ and $h_B \in V(\mathcal{H})$. The second condition for Alice and Bob to win the game is that $g_A$ has the same relation to $g_B$ as $h_A$ has to $h_B$, *i.e.* the three following equivalences are satisfied:

$$g_A = g_B \Leftrightarrow h_A = h_B\,, \quad g_A \sim g_B \Leftrightarrow h_A \sim h_B\,, \quad g_A \not\simeq g_B \Leftrightarrow h_A \not\simeq h_B\,, \tag{3.25}$$

where the symbol "$\not\simeq$" means neither equal nor adjacent. Find an example in Figure 3.4.

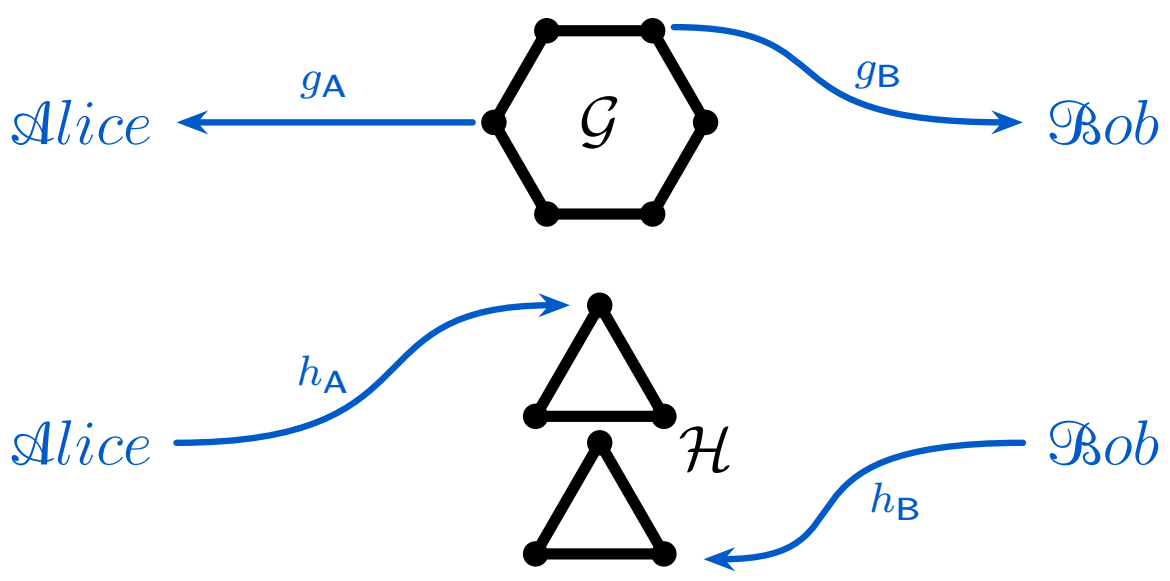

**Figure 3.4** — *Example of what can happen in the graph isomorphism game associated with the* 6*-cycle* $\mathcal{G} = C_6$ *and the disjoint union of two* 3*-cycles* $\mathcal{H} = C_3 \sqcup C_3$*. There, both inputs* $g_A$ *and* $g_B$ *are in* $V(\mathcal{G})$*, and they are non-equal and non-adjacent. The players correctly answer since* $h_A$ *and* $h_B$ *are both in* $V(\mathcal{H})$ *and are non-equal and non-adjacent again. In this example of graphs* $(\mathcal{G}, \mathcal{H})$*, one can show that it is possible to win perfectly with non-signaling resource (*$\mathcal{G} \cong_{ns} \mathcal{H}$*, that is* $\mathfrak{w}_{\mathcal{NS}} = 1$*) but impossible with local or quantum resource only (*$\mathcal{G} \ncong \mathcal{H}$ *and* $\mathcal{G} \ncong_{qc} \mathcal{H}$*, that is* $\mathfrak{w}_{\mathcal{L}}, \mathfrak{w}_{\mathcal{Q}} < 1$*).*

**Definition 3.27** (Graph Isomorphism Game) — *Let* $\mathcal{G}$ *and* $\mathcal{H}$ *be two graphs. The* graph isomorphism game *of* $(\mathcal{G}, \mathcal{H})$ *is defined by the following data:*

- $\mathcal{A}, \mathcal{B}, \mathcal{X}, \mathcal{Y} = V(\mathcal{G}) \sqcup V(\mathcal{H})$*;*
- $\pi(x_A, x_B) = \frac{1}{|V(\mathcal{G})|\,|V(\mathcal{H})|}$ *is the uniform distribution over* $V(\mathcal{G}) \sqcup V(\mathcal{H})$*;*
- $\mathcal{V}(y_A, y_B, x_A, x_B) = 1 \quad \Leftrightarrow \quad$ *both eqs. (3.24) and (3.25) are satisfied.*

**Remark 3.28** — The game is defined by the choice of a pair of graphs $(\mathcal{G}, \mathcal{H})$. If we change a graph, we have a different graph isomorphism game.

**Remark 3.29** (Variant of the Game) — In this work, we use the original definition from [Ats+19]. There exists a simpler variant of the graph isomorphism game, in which $x_\mathsf{A}$ and $x_\mathsf{B}$ are always given in $V(\mathcal{G})$, giving rise to the following different game:

- $\mathcal{A}, \mathcal{B} = V(\mathcal{H})$ and $\mathcal{X}, \mathcal{Y} = V(\mathcal{G})$;
- $\pi(x_\mathsf{A}, x_\mathsf{B}) = \frac{1}{|V(\mathcal{G})|}$ is the uniform distribution over $V(\mathcal{G})$;
- $\mathcal{V}(y_\mathsf{A}, y_\mathsf{B}, x_\mathsf{A}, x_\mathsf{B}) = 1 \quad \Leftrightarrow \quad$ eq. (3.25) is satisfied.

As explained in [RS21, Rem. 2.3], in the classical and quantum settings, if the graphs $\mathcal{G}$ and $\mathcal{H}$ are assumed to have the same number of vertices, then this simpler version is equivalent to the original version. Nevertheless, they differ in the non-signaling setting. Notably, we will need in Chapter 7 a characterization of perfect non-signaling strategies in terms of common equitable partition and fractional isomorphism that holds in the original setting only.

If the graphs $\mathcal{G}$ and $\mathcal{H}$ are actually isomorphic, with bijection $\varphi$, then Alice and Bob can perfectly win the game by simply answering $h_\mathsf{A} = \varphi(g_\mathsf{A})$ and $h_\mathsf{B} = \varphi(g_\mathsf{B})$. Conversely, assume that this game admits a perfect deterministic strategy in $\mathcal{L}_{\text{det}}$. Then the deterministic behavior of Alice to produce $h_\mathsf{A}$ out of $g_\mathsf{A}$ defines an isomorphism $\varphi$ between the graphs $\mathcal{G}$ and $\mathcal{H}$. As a result, we have $\mathcal{G} \cong \mathcal{H}$ *if, and only if,* Alice and Bob can perfectly win the deterministic isomorphism game. More generally, we can extend this result by convexity to the set of classical strategies $\mathcal{L}$, and we have again that $\mathcal{G} \cong \mathcal{H}$ *if, and only if,* Alice and Bob can perfectly win the classical isomorphism game.

Now, even if $\mathcal{G}$ and $\mathcal{H}$ are not truly isomorphic, Alice and Bob can try to mimic it to the Referee using their nonlocal box to correlate the answer. We say that $\mathcal{G}$ and $\mathcal{H}$ are *quantum (commuting) isomorphic*, denoted $\mathcal{G} \cong_{\text{qc}} \mathcal{H}$, if Alice and Bob can perfectly win the game using quantum (commuting) strategies; and similarly $\mathcal{G}$ and $\mathcal{H}$ are *non-signaling isomorphic*, denoted $\mathcal{G} \cong_{\text{ns}} \mathcal{H}$, using non-signaling strategies—see also [Ats+19] for details. These are equivalence relations and they relax the usual isomorphism of graphs $\cong$:

$$\mathcal{G} \cong \mathcal{H} \quad \Longrightarrow \quad \mathcal{G} \cong_{\text{qc}} \mathcal{H} \quad \Longrightarrow \quad \mathcal{G} \cong_{\text{ns}} \mathcal{H}\,. \tag{3.26}$$

Note that, if $\mathcal{G} \cong_s \mathcal{H}$ for some $s \in \{\emptyset, \text{q}, \text{ns}\}$, then the graphs $\mathcal{G}$ and $\mathcal{H}$ must have the same number of vertices [Ats+19], which is why we do not have to require it. Surprisingly, Mančinska and Roberson proved that quantum isomorphism is characterized in terms of counting homomorphisms from planar graphs [MR20], and with a larger team they showed that nonsignaling is equivalent to fractional isomorphism [Ats+19]. These two results, in addition to many others, are summarized in Figure 3.5. For the sake of completeness, we recall that the adjacency matrix $A_{\mathcal{G}}$ of a graph $\mathcal{G}$ with $m$ vertices $g_1, .., g_m$ is an $m \times m$ matrix defined using the set of edges of $\mathcal{G}$, where the coefficient $a_{ij}$ of the matrix is set to $1$ if $g_i \sim g_j$, and to $0$ otherwise. The notion of graph homomorphism is recalled in the next paragraph. Remarkably, note also that [Ats+19] gives examples of two graphs $\mathcal{G}, \mathcal{H}$ such that $\mathcal{G} \cong_{\text{qc}} \mathcal{H}$ but $\mathcal{G} \ncong \mathcal{H}$, and others such that $\mathcal{G} \cong_{\text{ns}} \mathcal{H}$ but $\mathcal{G} \ncong_{\text{qc}} \mathcal{H}$, and they prove that the problem of determining whether $\mathcal{G} \cong_{\text{qc}} \mathcal{H}$ is undecidable. Other related results may be found in [CY24; Fur+25].

| Isom. | Adjacency Matrices | Homomorphism Counts |
|---|---|---|
| $\mathcal{G} \cong \mathcal{H}$ | $\exists$ permutation matrix $u$<br>s.t. $A_{\mathcal{G}} u = u A_{\mathcal{H}}$ [Ats+19, Lem 3.1]<br>(equiv.: $\exists$ quantum permutation matrix $u$<br>with commuting entries<br>s.t. $A_{\mathcal{G}} u = u A_{\mathcal{H}}$ [MR20, Thm II.1] ) | • $\forall$ graph $\mathcal{F}$,<br>$\#\operatorname{Hom}(\mathcal{F},\mathcal{G}) = \#\operatorname{Hom}(\mathcal{F},\mathcal{H})$<br>[Lov67, Eq (5)]<br>• $\forall$ graph $\mathcal{F}$, [CV93]<br>$\#\operatorname{Hom}(\mathcal{G},\mathcal{F}) = \#\operatorname{Hom}(\mathcal{H},\mathcal{F})$ |
| $\mathcal{G} \cong_{\text{qc}} \mathcal{H}$ | $\exists$ quantum permutation matrix $u$<br>s.t. $A_{\mathcal{G}} u = u A_{\mathcal{H}}$ [LMR20, Thm 4.4] | $\forall$ planar graph $\mathcal{P}$,<br>$\#\operatorname{Hom}(\mathcal{P},\mathcal{G}) = \#\operatorname{Hom}(\mathcal{P},\mathcal{H})$<br>[MR20, Main Thm] |
| $\mathcal{G} \cong_{\text{ns}}^{D} \mathcal{H}$<br>[BW24] | $\exists$ bistochastic matrix $u$<br>s.t. $A_{\mathcal{G}}^{(t)} u = u A_{\mathcal{H}}^{(t)}$ $\forall t \leqslant D$ [Chapter 7]<br>(i.e. $D$-fractionally isomorphic) | ? |
| $\mathcal{G} \cong_{\text{ns}} \mathcal{H}$ | $\exists$ bistochastic matrix $u$<br>s.t. $A_{\mathcal{G}} u = u A_{\mathcal{H}}$ [Ats+19, Thm 4.5]<br>(i.e. fractionally isomorphic) | $\forall$ tree $\mathcal{T}$,<br>$\#\operatorname{Hom}(\mathcal{T},\mathcal{G}) = \#\operatorname{Hom}(\mathcal{T},\mathcal{H})$<br>[DGR18, Thm 1] |

**Figure 3.5** — *Characterization of different types of isomorphism. The D-NS-isomorphism $\cong_{\text{ns}}^{D}$ is defined and characterized in Chapter 7. The question mark "?" indicates an open question (to the best of our knowledge). A similar table also appears in [BW24].*

**Graph Homomorphism Game.** Another graph game is the *graph homomorphism game*, introduced by Mančinska and Roberson [MR16]. Fix two graphs $\mathcal{G}$ and $\mathcal{H}$. In the same vein as in the graph isomorphism game, the players try to convince the Referee that they know a homomorphism $\varphi$ from $\mathcal{G}$ to $\mathcal{H}$. Recall that a map $\varphi : V(\mathcal{G}) \to V(\mathcal{H})$ is a *graph homomorphism* if it preserves the adjacency, meaning that an edge $g \sim g'$ in $\mathcal{G}$ is sent to an edge $\varphi(g) \sim \varphi(g')$ in $\mathcal{H}$. Note that it is a relaxation of a graph isomorphism, for which we would additionally require that $\varphi$ is bijective and that $\varphi^{-1}$ preserves the adjacency, as stated in eq. (3.23). Note also that the composition of two graph homomorphisms is again a graph homomorphism itself, thus defining a category.

In this game, Alice and Bob are given some vertices $g_\mathsf{A}, g_\mathsf{B} \in V(\mathcal{G})$ respectively and answer some vertices $h_\mathsf{A}, h_\mathsf{B} \in V(\mathcal{H})$. Then, the Referee declares that they win the game *if, and only if,* they satisfy the following conditions:

$$g_\mathsf{A} = g_\mathsf{B} \Longrightarrow h_\mathsf{A} = h_\mathsf{B}\,, \qquad g_\mathsf{A} \sim g_\mathsf{B} \Longrightarrow h_\mathsf{A} \sim h_\mathsf{B}\,. \tag{3.27}$$

It yields the following formal definition of the game:

**Definition 3.30** (Graph Homomorphism Game) — *Let $\mathcal{G}$ and $\mathcal{H}$ be two graphs. The* graph homomorphism game *of $(\mathcal{G}, \mathcal{H})$ is defined by the following data:*

- *$\mathcal{A}, \mathcal{B} = V(\mathcal{H})$ and $\mathcal{X}, \mathcal{Y} = V(\mathcal{G})$;*
- *$\pi(g_\mathsf{A}, g_\mathsf{B}) = \frac{1}{|V(\mathcal{G})|^2}$ is the uniform distribution over $V(\mathcal{G}) \times V(\mathcal{G})$;*
- *$\mathcal{V}(h_\mathsf{A}, h_\mathsf{B}, g_\mathsf{A}, g_\mathsf{B}) = 1 \quad \Leftrightarrow \quad$ eq. (3.27) is satisfied.*

In contrast with the graph isomorphism game, in this game, it may happen that $g_\mathsf{A} \not\sim g_\mathsf{B}$ but still $h_\mathsf{A} = h_\mathsf{B}$ or $h_\mathsf{A} \sim h_\mathsf{B}$. We denote $\mathcal{G} \to \mathcal{H}$ if Alice and Bob can perfectly win the game using classical resources, which is equivalent to saying that there actually exists a graph homomorphism from $\mathcal{G}$ to $\mathcal{H}$. We similarly denote $\mathcal{G} \to_{\text{qc}} \mathcal{H}$ and $\mathcal{G} \to_{\text{ns}} \mathcal{H}$ when the game can be perfectly won using quantum (commuting) and non-signaling resources respectively.

**Graph Coloring Game.** The *graph coloring game*, introduced by Cameron, Montanaro, Newman, Severini, and Winter [Cam+07b], is a particular case of the graph homomorphism game. Indeed, when $\mathcal{H} = \mathcal{K}_N$ is complete, the game corresponds to proving to the Referee that the graph $\mathcal{G}$ is $N$-colorable,

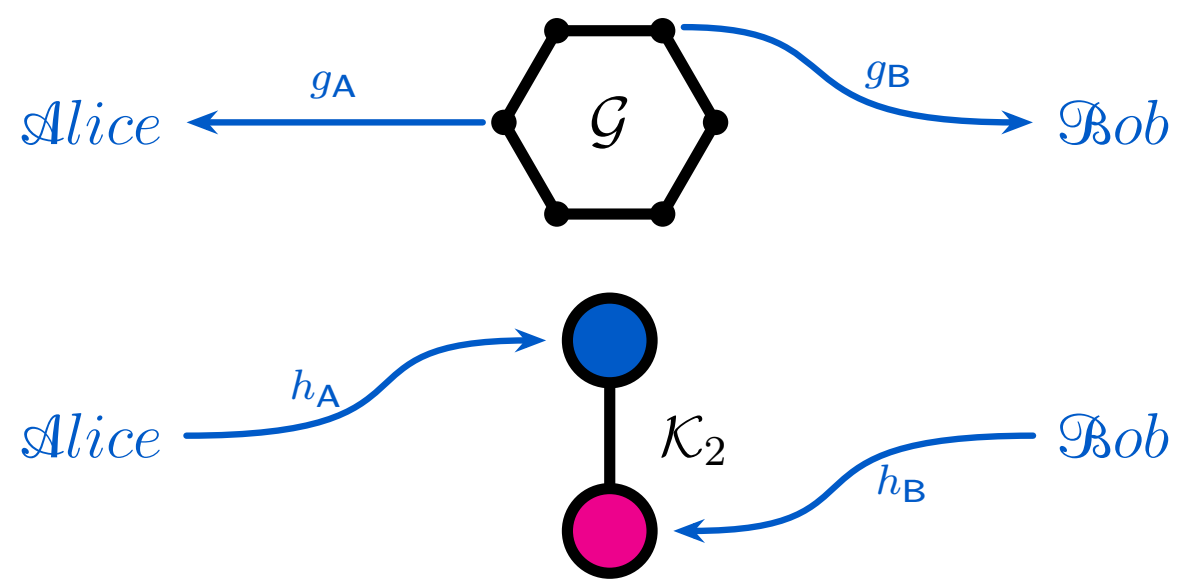

**Figure 3.6** — *Example of what can happen in the graph 2-coloring game associated with the 6-cycle $\mathcal{G} = \mathcal{C}_6$. There, the questions $g_A$ and $g_B$ are non-equal and non-adjacent. So the players can answer any pairs of colors of their choice to win. In this example, the graph $\mathcal{G}$ is actually 2-colorable, so the game can always be won without using any shared resource (that is $\mathfrak{w}_{\mathcal{L}} = \mathfrak{w}_{\mathcal{Q}} = \mathfrak{w}_{\mathcal{NS}} = 1$).*

which explains the name of this particular case. Recall that a graph $\mathcal{G}$ is said to be $N$-colorable if, from a set of $N$ different colors, we can assign a color to each vertex of $\mathcal{G}$ so that no two adjacent vertices have the same color. (Equivalently, in the graph homomorphism game of $(\mathcal{G}, \mathcal{K}_N)$, two adjacent vertices in $\mathcal{G}$ are sent to adjacent vertices in $\mathcal{K}_N$, representing two different "colors"). Find an example in Figure 3.6. Note that this game generalizes the *odd cycle game* introduced in Remark 3.26. Here is the formal definition of the game:

**Definition 3.31** (Graph Coloring Game) — *Let $\mathcal{G}$ be a graph and $N \in \mathbb{N}$ an integer. The* graph $N$-coloring game *of $\mathcal{G}$ is defined by the following data:*

- *$\mathcal{A}, \mathcal{B} = V(\mathcal{K}_N)$ and $\mathcal{X}, \mathcal{Y} = V(\mathcal{G})$;*
- *$\pi(g_A, g_B) = \frac{1}{|V(\mathcal{G})|^2}$ is the uniform distribution over $V(\mathcal{G}) \times V(\mathcal{G})$;*
- *$\mathcal{V}(h_A, h_B, g_A, g_B) = 1 \quad \Leftrightarrow \quad$ eq. (3.27) is satisfied.*

**Example 3.32** (Complete Graphs are Always $\mathcal{NS}$-Colorable) — In the classical setting, we know that the complete graph $\mathcal{K}_M$ is $N$-colorable *if, and only if,* $M \leqslant N$. However, with non-signaling strategies, Alice and Bob are interestingly able to pretend to the Referee that they know an $N$-coloring for $\mathcal{K}_M$ even when $M > N$. Indeed, let us prove that $\mathcal{K}_M \to_{\mathrm{ns}} \mathcal{K}_N$ for any

$M, N \geqslant 2$. Consider the following function:

$$\mathbf{P}\big(h_{\mathsf{A}},\, h_{\mathsf{B}} \,\big|\, g_{\mathsf{A}},\, g_{\mathsf{B}}\big) \;:=\; \begin{cases} 1/N & \text{if } h_{\mathsf{A}} = h_{\mathsf{B}} \text{ and } g_{\mathsf{A}} = g_{\mathsf{B}}, \\ 1/N(N-1) & \text{if } h_{\mathsf{A}} \sim h_{\mathsf{B}} \text{ and } g_{\mathsf{A}} \sim g_{\mathsf{B}}, \\ 0 & \text{otherwise.} \end{cases}$$

Let us prove that it is a well-defined probability distribution. First, it is non-negative by construction. Second, for all fixed $g_{\mathsf{A}}, g_{\mathsf{B}} \in V(\mathcal{K}_M)$, it sums to $1$ over $h_{\mathsf{A}}, h_{\mathsf{B}} \in V(\mathcal{K}_N)$ because (i) if $g_{\mathsf{A}} = g_{\mathsf{B}}$, then necessarily $h_{\mathsf{A}} = h_{\mathsf{B}}$, which happens $N$ times (once for each of the $N$ vertices of $\mathcal{K}_N$), (ii) if $g_{\mathsf{A}} \sim g_{\mathsf{B}}$, then we have $h_{\mathsf{A}} \neq h_{\mathsf{B}}$, and we know that there are exactly $N^2 - N = N(N-1)$ pairs of distinct elements $(h_{\mathsf{A}}, h_{\mathsf{B}}) \in V(\mathcal{K}_N)^2$, and (iii) the case $g_{\mathsf{A}} \not\simeq g_{\mathsf{B}}$ never happens in $\mathcal{K}_M$. Hence $\mathbf{P}$ is indeed a probability distribution. Let us prove that it is non-signaling. We have that Bob's marginal is independent of Alice's input $g_{\mathsf{A}}$:

$$\begin{aligned} \sum_{h_{\mathsf{A}}} \mathbf{P}(h_{\mathsf{A}},\, h_{\mathsf{B}} \,|\, g_{\mathsf{A}},\, g_{\mathsf{B}}) \;&=\; \frac{1}{N} \sum_{h_{\mathsf{A}}} \delta_{h_{\mathsf{A}} = h_{\mathsf{B}}} \delta_{g_{\mathsf{A}} = g_{\mathsf{B}}} + \frac{1}{N(N-1)} \sum_{h_{\mathsf{A}}} \delta_{h_{\mathsf{A}} \sim h_{\mathsf{B}}} \delta_{g_{\mathsf{A}} \sim g_{\mathsf{B}}} \\ &=\; \frac{1}{N}\, \delta_{g_{\mathsf{A}} = g_{\mathsf{B}}} + \frac{1}{N}\, \delta_{g_{\mathsf{A}} \sim g_{\mathsf{B}}} \;=\; \frac{1}{N}\,, \end{aligned}$$

where the last line holds because $\mathcal{K}_N$ is complete so exactly one of the Kronecker deltas is $1$ and the other is zero. Likewise, Alice's marginal is independent of Bob's input. Hence we have $\mathbf{P} \in \mathcal{NS}$. Finally, it satisfies the rules of the homomorphism game (eq. (3.27)) by construction. Hence, this non-signaling box $\mathbf{P}$ perfectly wins at this graph coloring game, and we have $\mathcal{K}_M \to_{\text{ns}} \mathcal{K}_N$ for any $M, N \geqslant 2$ as wanted.

**Other Graph Games.** Here is a non-exhaustive list of other nonlocal games involving graphs that are studied in the literature. There is the *quantum graph homomorphism game*, a generalization of graph homomorphism with quantum inputs [BGH22; Bra+23; TT20], and similarly the *quantum graph coloring game* [TT20]. There is also the *graph bisynchronous game*, in which the players want to have the same answers *if, and only if,* they had the same questions [PR21], a game where the connectivity of a graph needs to be preserved [AG04; Cib+13], and the *vertex distance game*, in which the distance of answer vertices have to be the same as the distance of the question vertices [BW24]. Find also a generalization to hypergraph nonlocal games [Hoe25; HT23; HT25].

### 3.2.4 *Other Examples of Games*

In this section, we delve into a few other nonlocal games studied in the literature, namely the magic square game, the XOR games, the constraint satisfaction problem games, and some generalizations of nonlocal games to obtain the monogamy-of-entanglement games and the no-cloning games.

**Mermin-Peres Magic Square Game.** Based on the notion of *magic square* introduced by Mermin [Mer90a; Mer90b] and Peres [Per90], Aravind defined the *magic square game* [Ara04]. The idea behind this game is based on the following observation. On the one hand, it is not possible to fill in a $3 \times 3$ table with coefficients $\pm 1$ such that each row multiplies to $1$ and each column to $-1$. Here is an example:

$$\begin{array}{|c|c|c|l} \hline +1 & +1 & +1 & \rightarrow +1 \\ \hline +1 & -1 & -1 & \rightarrow +1 \\ \hline -1 & +1 & ?? & \rightarrow +1 \\ \hline \downarrow & \downarrow & \downarrow & \\ -1 & -1 & -1 & \end{array}$$

A simple proof goes by contradiction: If such a table existed, then the product of all its coefficients would be equal to the multiplication of each row product, *i.e.* $1 \times 1 \times 1 = 1$, but it would also be equal to the multiplication of each column product, *i.e.* $(-1) \times (-1) \times (-1) = -1$, which is a contradiction. On the other hand, it is possible with Pauli matrices (recall the definition in eq. (2.1)) to have $\mathbb{I}_4$ for any row product and $-\mathbb{I}_4$ for any column product:

$$\begin{array}{|c|c|c|l} \hline \mathbb{I}_2 \otimes \sigma_z & \sigma_z \otimes \mathbb{I}_2 & \sigma_z \otimes \sigma_z & \rightarrow +\mathbb{I}_4 \\ \hline \sigma_x \otimes \mathbb{I}_2 & \mathbb{I}_2 \otimes \sigma_x & \sigma_x \otimes \sigma_x & \rightarrow +\mathbb{I}_4 \\ \hline -\sigma_x \otimes \sigma_z & -\sigma_z \otimes \sigma_x & \sigma_y \otimes \sigma_y & \rightarrow +\mathbb{I}_4 \\ \hline \downarrow & \downarrow & \downarrow & \\ -\mathbb{I}_4 & -\mathbb{I}_4 & -\mathbb{I}_4 & \end{array}$$

In the game, the Referee samples a row number $R_a$ and a column number $C_b$ for some $a, b \in \{1, 2, 3\}$ and sends $a, b$ as questions to Alice and Bob. In return, Alice and Bob want to mimic the fact that they are able to have a magic square. So they answer with an assignment for each cell of their row/column, and the Referee verifies if the assignment on the intersection $R_a \cap C_b$ is the same for Alice and Bob (which should be the case if they indeed have a magic square). If so, Alice and Bob win the game, otherwise, they lose. From the above discussion, we know that they cannot classically win the game with probability 1 (more precisely, the classical value is $\mathfrak{w}_{\mathcal{L}} = \frac{8}{9} < 1$). In contrast, with quantum entanglement, they can perfectly win the game. Indeed, consider the following entangled state:

$$\rho := \omega \otimes \omega \in \mathcal{D}\big((\mathcal{H}_{\mathsf{A}} \otimes \mathcal{H}_{\mathsf{B}}) \otimes (\mathcal{H}_{\mathsf{A}'} \otimes \mathcal{H}_{\mathsf{B}'})\big)\,,$$

where $\omega := |\Omega\rangle\langle\Omega|$ is the maximally entangled over $\mathbb{C}^2 \otimes \mathbb{C}^2$, where $|\Omega\rangle := \frac{1}{\sqrt{2}}\big(|00\rangle + |11\rangle\big)$, and where both registers $\mathsf{A}, \mathsf{A}'$ (resp. $\mathsf{B}, \mathsf{B}'$) are given to Alice (resp. Bob). When they receive their row/column, Alice and Bob measure the three corresponding observables described in the above table with Pauli matrices—see Example 2.15 for Pauli matrix measurement. On each row/column, note that the three observables commute. Therefore, they are diagonalizable on a common basis, which means that they can be measured simultaneously—see Remark 2.18 about measurement incompatibility. From their measurements, they obtain some values $\pm 1$, which necessarily multiply to $+1$ for Alice because her three measurements are equivalent to measuring the observable $\mathbb{I}_4$, and $-1$ for Bob because his measurements are equivalent to measuring $-\mathbb{I}_4$. Hence, they perfectly win at the Mermin-Peres magic square game with quantum resource and there is a *quantum advantage*:

$$\mathfrak{w}_{\mathcal{L}} = \frac{8}{9} < \mathfrak{w}_{\mathcal{Q}} = 1\,.$$

This is a special example of *quantum pseudo-telepathy* [BBT03; BBT05; BCT99], where a task can perfectly be accomplished with quantum resource but not classically (assuming no communication). Note that the quantum winning probability at this game was later experimentally confirmed [Xu+22]. A known variant of this game is the *magic pentagram game*, also demonstrating quantum pseudo-telepathy [KM17; Mer93], and was generalized in [Ark12].

**XOR Games.** The class of *XOR games* generalizes the CHSH game. They are defined as any nonlocal game with binary outputs ($\mathcal{A} = \mathcal{B} = \{0, 1\}$) such that the predicate $\mathscr{V}$ depends at most on $a \oplus b$, $x$, and $y$, but not on $a$ and $b$ independently (like in the CHSH where we want $a \oplus b = xy$). This class of games also generalizes the odd cycle game introduced in Remark 3.26. For more on XOR games, find details and proofs in [Cle+04; PV16].

As for any binary output game, its quantum value is always achieved (at least) by some local PVMs on a pure state. What is more, computing the quantum value of any XOR game can be cast into a semidefinite program [Cle+04; Tsi87] and unless $P = NP$, is easier to compute than the classical value (integer quadratic program, MAXSNP hard [AN04]). Surprisingly, as is the case for any binary output game, if the quantum value is $1$, then so is the classical value [Cle+04]:

$$\mathfrak{w}_{\mathcal{Q}} = 1 \qquad \Leftrightarrow \qquad \mathfrak{w}_{\mathcal{L}} = 1\,,$$

which implies that no binary output game can display quantum pseudo-telepathy (including the CHSH game). These values can be expressed in terms of tensor norms such as the injective norm $\|\cdot\|_\varepsilon$ introduced at page 37, see [PV16]. Furthermore, the quantum and classical values can be compared using Grothendieck's constant [Gro53] as follows [Tsi87]:

$$\mathfrak{w}_{\mathcal{Q}}(\mathsf{G}) - \tau(\mathsf{G}) \leqslant K_g^{\mathbb{R}} \Big(\mathfrak{w}_{\mathcal{L}}(\mathsf{G}) - \tau(\mathsf{G})\Big)\,, \tag{3.28}$$

where $\tau(\mathsf{G})$ is the winning probability with a trivial random strategy (that does not depend on the inputs $x, y$ and that produces $a, b$ uniformly at random), and where $K_g^{\mathbb{R}} \approx 1.7$ is Grothendieck's constant. For instance, when G is the CHSH game, we have the relation:

$$\mathfrak{w}_{\mathcal{Q}}(\text{CHSH}) - \tau(\text{CHSH}) = \sqrt{2}\Big(\mathfrak{w}_{\mathcal{L}}(\text{CHSH}) - \tau(\text{CHSH})\Big)\,,$$

where $\tau(\text{CHSH}) = \frac{1}{2}$ and where $\sqrt{2} \approx 1.4 \leqslant K_g^{\mathbb{R}}$ indeed. Finally, we mention that there exist known bounds on entanglement for XOR games: There exists an optimal quantum strategy in which the players share a maximally entangled state on $\lceil N/2 \rceil$ qubits, where $N = \min\big\{|\mathcal{X}|, |\mathcal{Y}|\big\}$ [Cle+04].

**Constraint Satisfaction Problem Games.** The class of *constraint satisfaction problem games* (CSPs) generalizes both the graph coloring game (page 98) and the magic square game (page 101). A CSP is a decision problem where one must verify whether or not a given set of constraints admits a solution, such as in graph coloring or magic square problems. We can also think of a set of equations for which we want to know if there exists an assignment for all the variables such that all the equations hold. As in the games before, even though the constraints are not actually satisfiable, the player might mimic that they know a global assignment satisfying all the constraints. Mainly, there exist three variants of the game.

(i) First, the *constraint-constraint* CSP *game*, in which each of the players receives a constraint and answers with an assignment for each variable involved in their constraint. They win the game if they agree on all assignment values for the variables they have in common. This is exactly the setting of the magic square game [Ark12]. Find other examples in [MS24; PS25].

(ii) Second, the *constraint-variable* CSP *game*, in which Alice receives a constraint, Bob a variable, with the promise that this variable is involved in Alice's constraint. Then, they answer with an assignment to the variables, and they win if they choose the same assignment for Bob's variable. Examples of this variant include [CM14; Ji13].

(iii) Last, the $2$-*constraint system game*, in which the Referee samples a constraint, and sends some variables $x, y$ involved in this constraint to Alice and Bob. They answer with an assignment $a, b$ for their respective variable, and they win if there exists a global assignment for the constraint such that both $x \mapsto a$ and $y \mapsto b$ are valid. The graph coloring game is an example of this variant [Cam+07b]. See also [CMS24; Har24].

We refer to [CM25] for more materials on this topic.

**Remark 3.33** (Link with Graph Isomorphism Game) — Interestingly, this class of games was used by Atserias, Mančinska, Roberson, Šámal, Severini, and Varvitsiotis to build an example for the graph isomorphism game of two graphs $(\mathcal{G}, \mathcal{H})$ that are quantum isomorphic ($\mathcal{G} \cong_{\text{qc}} \mathcal{H}$) but not classically isomorphic ($\mathcal{G} \ncong \mathcal{H}$) [Ats+19]. More precisely, the authors prove that a set of constraint $F$ is quantum satisfiable (*i.e.* there is a perfect quantum strategy for the constraint-constraint CSP game) *if, and only if,* the associated graph $\mathcal{G}_F$ is quantum isomorphic to the graph $\mathcal{G}_{F_0}$ associated to the homogenization $F_0$ of $F$. In their example, they rely on the fact that the magic square game admits a perfect quantum strategy but no perfect classical one (*i.e.* that it displays pseudo-telepathy).

**Generalized Nonlocal Games.** Nonlocal games can be generalized in many different ways to tackle more specific problems. We present below three directions to generalize nonlocal games: extended nonlocal games, no-cloning games, and semi-quantum games. A fourth generalization can be found in Section 4.1.1, allowing a classical channel with a finite capacity between the players.

(i) First, there is the class of *extended nonlocal games* introduced by Johnston, Mittal, Russo, and Watrous in [Joh+16]. Here, the common resource (*e.g.* a quantum state or a nonlocal box) is shared not only by Alice (A) and Bob (B) but also with the Referee (R). Say that the resource is $\rho_{\mathsf{RAB}}$ for instance. Then, as in nonlocal games, given classical questions $x \in \mathcal{X}$ and $y \in \mathcal{Y}$ sampled by a distribution $\pi$, Alice and Bob use the shared resource $\rho_{\mathsf{RAB}}$ to produce their classical outputs $a \in \mathcal{A}$ and $b \in \mathcal{B}$. Now, the predicate (or rule) $\mathcal{V}$ is different than in nonlocal games: it is no longer a Boolean function, but rather the result of the Referee's local measurement on $\rho_{\mathsf{RAB}}$ depending on $a, b, x, y$. Find a representation of this game in Figure 3.7. Important examples of games in this class are the *monogamy-of-entanglement games* (MoE games), introduced by Tomamichel, Fehr, Kaniewski, and Wehner in [Tom+13] to leverage the MoE principle (Section 2.2.5). We provide more details on this type of game in Chapter 5. See also [Cul22].

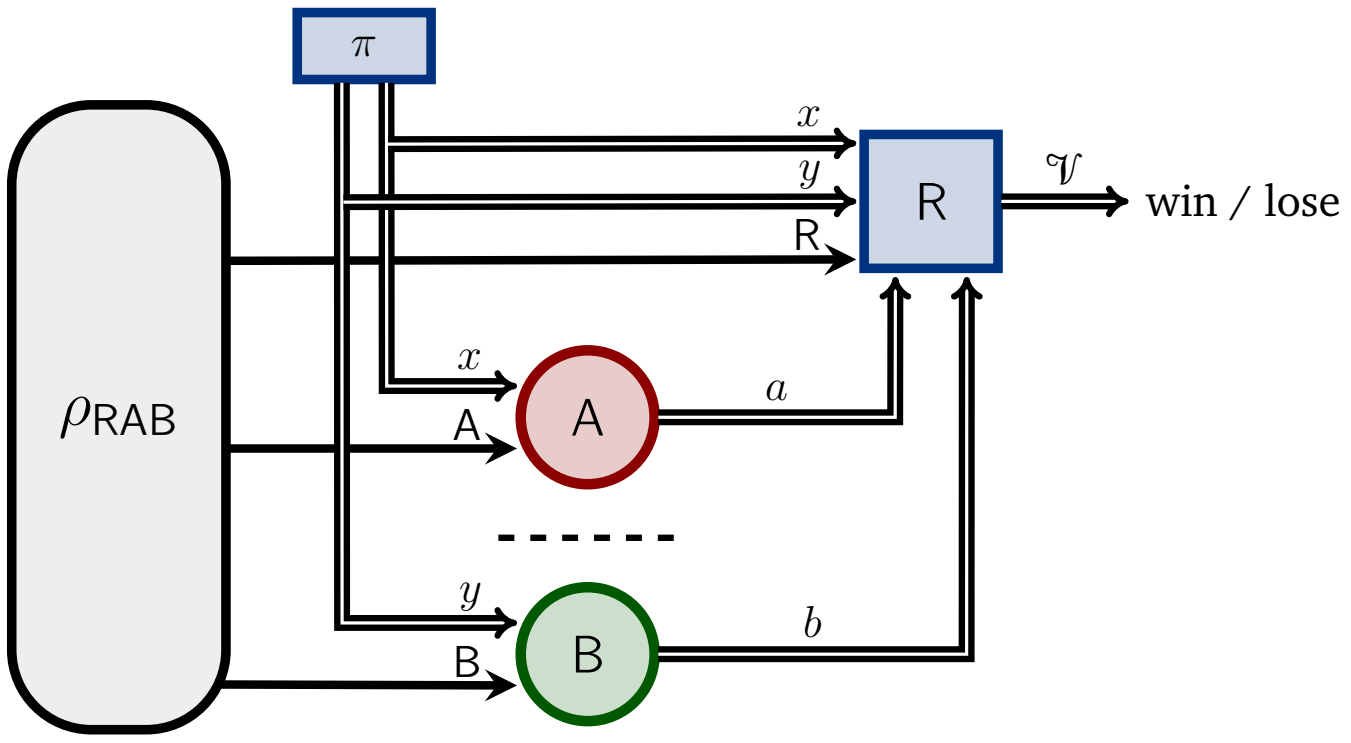

**Figure 3.7 —** *Extended games. This diagram is inspired from [Cul22].*

(ii) Another generalization is given by the family of *no-cloning games* implicitly introduced by Broadbent and Lord in [BL20]. These games have applications in quantum cryptography and leverage the quantum no-cloning theorem (Theorem 2.37), or more precisely the quantum no-broadcasting theorem (Theorem 2.39). In these games, a third player called the *Pirate* (P) cooperates with Alice (A) and Bob (B). This third player is different from the others: as opposed to Alice and Bob who receive classical questions $x \in \mathcal{X}$ and $y \in \mathcal{Y}$ (called *keys*), the Pirate only receives a quantum state $\rho_{m|x,y}$ from the Referee (called the *quantum encryption* of a classical bit $m \in \{0,1\}$), applies some quantum channel $\Phi$, and sends one quantum register to Alice and another to Bob. Then, from the question they received and their share of the state $\Phi(\rho_{m|x,y})$, Alice and Bob give their classical answers $a \in \mathcal{A}$ and $b \in \mathcal{B}$ (called *guesses*) to the Referee. We say that the players $(\mathsf{A}, \mathsf{B}, \mathsf{P})$ win if both of the classical guesses match the original bit $m$, *i.e.* if $a = b = m$. The scenario can be represented as in Figure 3.8. This type of game is useful to study an open problem raised by Broadbent and Lord about the existence of an *uncloneable bit* [BL20]. See Chapter 5 for more details and Chapter 8 [Bot+24b] for some results in this scenario.

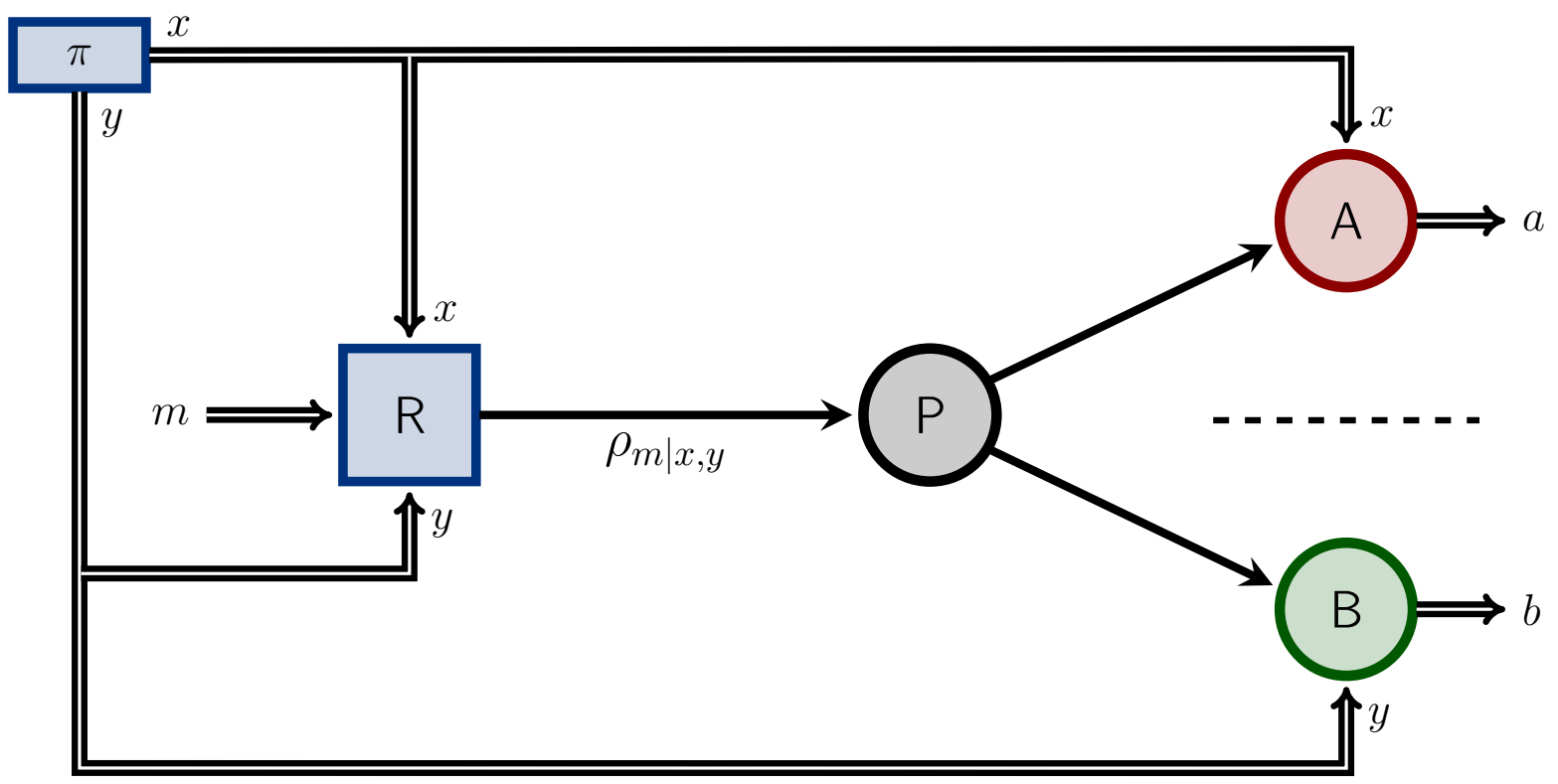

**Figure 3.8 —** *No-cloning games. A similar diagram appears in [Bot+24b].*

(iii) Lastly, there is also a variant of nonlocal games called *semi-quantum games* introduced by Buscemi in [Bus12]. There, instead of having classical inputs and outputs, the game can rather have quantum inputs [BGH22; Bra+23; Lin24; TT20], or quantum outputs [SRB20], or both.

## 3.3 Applications

Nonlocal boxes as well as nonlocal games have a wide range of applications in many fields. In this section, we present five research lines to which they apply, namely self-testing (Section 3.3.1), complexity theory (Section 3.3.2), operator algebra (Section 3.3.3), quantum cryptography (Section 3.3.4), and physical principles (Section 3.3.5).

Let us mention that other directions can also be considered. For instance, they have experimental applications [ADR82; Hen+15; Wei+98; Xu+22], and they connect to the monogamy-of-entanglement principle [CMR25; PB09; RH14; Ton08; TV06].

### 3.3.1 *Self-Testing*

A significant application of nonlocal boxes and nonlocal games is *self-testing*. This is the capacity to identify a specific quantum state from the statistics of a nonlocal box or the winning probability at a game. Find good reviews in [Bru+14; ŠB20].

**Tsirelson's Bound as a Self-Testing.** For instance, in the CHSH-scenario with two parties and binary input-outputs, Tsirelson's Bound $S \leqslant 2\sqrt{2}$ (see eq. (3.12)) is self-testing because achieving the value $2\sqrt{2}$ (or equivalently achieving the winning probability $\cos^2(\pi/8)$ at the CHSH game) with a quantum correlation implies that the underlying state is $\omega \in \mathcal{D}(\mathbb{C}^2 \otimes \mathbb{C}^2)$ up to local isometries [Tsi80] and that the local measurements of Alice (resp. Bob) are anticommuting Pauli operators [BMR92; PR92; SW87]. Note that the first use of the phrasing *self-testing* comes from [MY04], which also sets the terminology and formalism used in later works.

**Self-Testing of Bipartite States.** It is also possible to self-test partially entangled bipartite states [Bac+20; Wag+20; YN13], qudit states [CGS17; YN13], or $k$ copies of the maximally entangled state with sequential methods [RUV13] and parallel methods [CN16; Col17; McK17; Wu+16].

**Self-Testing of Multipartite States.** In the multipartite setting, some methods were introduced to construct permutationally invariant Bell inequalities for the purpose of self-testing [PVN14; Sek+18]. It is also possible to reduce to bipartite methods [Šup+18; Wu+14] or to self-test graph states [HH18; McK14].

**Robust Self-Testing.** A self-testing is said to be *robust* if it is stable against noise, in the sense that being close to the self-testing value implies having a close quantum state. This notion is crucial for practical implementation since we all need to deal with noise. For the maximally entangled state $\omega \in \mathcal{D}(\mathbb{C}^2 \otimes \mathbb{C}^2)$, it was first developed by Magniez, Mayers, Mosca, and Ollivier [Mag+06] and later developed by Bardyn, Liew, Massar, McKague, and Scarani [Bar+09] and McKague, Yang, and Scarani [MYS12].

**Self-Testing Via Nonlocal Games.** Examples of self-testing based on the CHSH game can be found in [Col17; McK17; RUV13], on XOR games in [Cui+24; MS13; Slo11], on an extended version of the CHSH game in [Kan17], on the magic square game in [CN16; Wu+16], on pseudo-telepathy games in [Man14], on linear constraint system game in [CS19], on the GHZ game in [CK18].

**Other Examples of Seft-Testings.** One can also self-test quantum measurements [Kan17; MYS12], quantum computations [Mag+06; RUV13; vD+00], or have device-independent witness of genuine multipartite entanglement [Ban+11; Ban13].

### 3.3.2 *Complexity Theory*

Nonlocal games also have strong connections to complexity theory. For instance, let us describe the celebrated work "MIP*=RE" by Ji, Natarajan, Vidick, Wright, and Yuen [Ji+21]. A consequence of this result is that the set of quantum (tensor) correlations $\mathcal{Q}$ differ from the quantum commuting one $\mathcal{Q}_c$, see below.

**Multiprover Interactive Proofs (MIP).** Suppose we have a problem to solve like showing that a certain equation admits a solution in a certain set. Nevertheless, assume that we have limited computational power and that we need to have an efficient algorithm to verify it (more precisely, a polynomial-time Turing machine). To achieve it, we may interact with all-powerful computers Alice and Bob, called *provers*, who cannot communicate together.

This might be seen as a nonlocal game where we take the viewpoint of the Referee, also called the *verifier* or *challenger* in this context, and we ask questions to the two provers Alice and Bob to decide, after several rounds, whether we accept that there is a solution for our given equation or not. However, we need to be careful because, as for other nonlocal games, Alice and Bob may be malicious and try to mimic a "yes" answer to the problem even when they know that it is "no", and vice-versa. So we need to design a good sequence of questions to be convinced that they are honest. Note that we may accept to make mistakes, but we should be correct often: every correct statement should be accepted with probability at least $\frac{2}{3}$ (*completeness*), while no wrong statement should accepted with probability larger than $\frac{1}{3}$ (*soundness*). All of this framework forms the *MIP complexity class*: it consists of all languages with multipartite, interactive, randomized polynomial-time verification procedures.

Interestingly, Babai, Fortnow, and Lund established that this class is equal to the class NEXP of languages that admit exponentially long "traditional" proofs verifiable in exponential time [BFL91].

**The Class MIP*.** Now, like in nonlocal games, the provers Alice and Bob may share quantum entanglement to better coordinate their answers and thus improve their chance to mislead us, the verifier, limited to polynomial-time power. This gives rise to the complexity class MIP*. At first sight, it is unclear how this class compares to the previous one: MIP* could a priori be smaller, larger, or incomparable to MIP. Nevertheless, Ito and Vidick showed the first non-trivial lower bound NEXP $\subseteq$ MIP* [IV12], and when combined with the above-cited result from [BFL91], it holds that:

$$\text{MIP} \subseteq \text{MIP}^* .$$

Then, this inclusion was strengthened by Natarajan and Wright to NEEXP $\subseteq$ MIP* [NW19], where NEEXP stands for non-deterministic doubly exponential time. Therefore, as NEXP is strictly contained in NEEXP, it yields:

$$\text{MIP} \neq \text{MIP}^* .$$

Finally, building on the top of this sequence of works, it was established by Ji, Natarajan, Vidick, Wright, and Yuen that MIP* is actually equal to RE [Ji+21]:

$$\text{MIP}^* = \text{RE} . \tag{3.29}$$

RE is the class of recursively enumerable languages, *i.e.* the class of all languages for which there is a Turing machine $\mathcal{M}$ such that a statement is correct in the language *if, and only if,* the machine $\mathcal{M}$ halts and accepts the statement. (Actually, the Halting problem is complete for RE.) A surprising consequence of the equality MIP*=RE is that there is a verification procedure describing a physical experiment in Alice's and Bob's laboratories (time-bounded) that could be used to certify that a Turing machine halts (no time-bound).

**Consequence to Quantum Correlations.** In Section 3.1.1, we presented several variants for the definition of quantum correlations, notably the quantum set $\mathcal{Q}$ and the quantum commuting set $\mathcal{Q}_c$. The question of equality between these two sets was raised as *Tsirelson's problem* [Tsi06]. Fritz, Netzer, and Thom achieved a first milestone by proving that the undecidability of MIP* implies a negative answer (more precisely, it implies that the two sets are finitely separated) [FNT14]. Thus, a corollary of eq. (3.29) is the negative answer [Ji+21], that is:

$$\mathcal{Q} \neq \mathcal{Q}_c \,.$$

We will mention another consequence for operator algebra in the next subsection, about Connes' Embedding Problem.

### 3.3.3 Operator Algebra

There is also a natural connection with operator algebra since, by definition, quantum correlations involve operators. Below, we present two connections, one with Connes' embedding problem, and another one with the Grothendieck constant.

**Connes' Embedding Problem.** Formulated by Alain Connes in the seventies [Con76], *Connes' embedding problem* is a major problem in von Neumann's algebra. It asks whether every type $\mathrm{II}_1$ factor on a separable Hilbert space can be embedded into some ultrapower $R^\omega$, where $R$ is the hyperfinite type $\mathrm{II}_1$ factor and $\omega$ the free ultrafilter on the natural numbers.

This problem was shown to be equivalent to many other [Gol22; Had01; Oza13], for instance: Kirchberg's QWEP conjecture in C*-algebra theory, the predual of any (separable) von Neumann algebra is finitely representable in the trace class, or Tsirelson's problem. It is also connected to microstates in free entropy theory [Shl03; Voi93].

Now, as mentioned in the former subsection, the result from Ji, Natarajan, Vidick, Wright, and Yuen implies a negative answer to Tsirelson's problem, thus leading as well to a negative answer to Connes' embedding problem [Ji+21].

As a consequence, this result may lead to the construction of interesting objects in other areas of mathematics.

**Grothendieck Constant.** In eq. (3.28), we mentioned that Grothendieck constant $K_g^{\mathbb{R}}$ [Gro53] appears in a relation between the classical and quantum values of an XOR game G [Tsi87]:

$$\mathfrak{w}_{\mathcal{Q}}(\mathsf{G}) - \tau(\mathsf{G}) \leqslant K_g^{\mathbb{R}} \Big( \mathfrak{w}_{\mathcal{L}}(\mathsf{G}) - \tau(\mathsf{G}) \Big) ,$$

where $\tau(\mathsf{G})$ is the winning probability with a trivial random strategy (that does not depend on the inputs $x, y$ and that produces $a, b$ uniformly at random). Grothendieck constant is defined as the smallest universal constant (independent of $n$) such that for all integer $n \geqslant 2$ and all real matrix $A = (a_{ij}) \in \mathcal{M}_n(\mathbb{R})$, if

$$\left| \sum_{ij} a_{ij}\, x_i\, y_j \right| \leqslant 1 ,$$

for all scalars $x_i, y_j \in [-1, 1]$, then

$$\left| \sum_{ij} a_{ij}\, \vec{x_i} \cdot \vec{y_j} \right| \leqslant K_g^{\mathbb{R}} ,$$

for all unit vectors $\vec{x_i}, \vec{y_j} \in \mathbb{R}^n$. The exact value is currently unknown, we only have lower and upper bounds [Bra+11]:

$$1.67696 \leqslant K_g^{\mathbb{R}} \leqslant 1.78221 .$$

The above-mentioned connection with XOR games can be used to lower bound Grothendieck constant. For example, the CHSH game gives the trivial lower bound $\sqrt{2} \approx 1.41 \leqslant K_g^{\mathbb{R}}$, but better lower bounds can be found using other nonlocal games, going arbitrarily close to the value $1.5$ [FR94]. See also [AGT06] for a connection between the Grothendieck constant and Tsirelson's bound.

### 3.3.4 *Quantum Cryptography*

Both nonlocal boxes and nonlocal games find applications in quantum cryptography. Here, we present only the general idea of these applications. We elaborate more in Chapter 5.

**Device-Independent Cryptography.** Nonlocal boxes are *device-independent* tools, meaning that they provide insight into the statics of outputs given some inputs, but do not tell us what is happening inside of the box. This leads to *device-independent cryptography*, a device-independent approach of quantum cryptography. A first example is the influential *quantum key distribution* protocol (QKD) from Bennett and Brassard [BB84] or its variant from Ekert [Eke91]. This protocol allows two distant parties Alice and Bob to generate a shared secret key in the presence of an eavesdropper, often called Eve, using quantum mechanics. This protocol can be turned into a device-independent variant based on correlations and violation of the CHSH inequality [Ací+07; AGM06; BHK05]. Device-independent protocols can also certify, for instance, *random number generation* [Col11; Pir+10], or *qubit teleportation* [HBS13] (see the“usual” quantum teleportation at page 49). Find reviews on this topic in [Bru+14; BS16; Sca12].

**Security via Nonlocal Games.** Nonlocal games can be used as a framework to represent a cryptographic scenario, in which case the winning probability of the players (the *adversary team*) can be used to define the notion of security. From the Referee's point of view, designing a secure protocol could mean that the players have a low winning probability, no matter what strategy they employ, where low probability means exponentially close to the uniformly random winning probability. An example is given by *no-cloning games* (Figure 3.8), in which the Referee wants to avoid the players broadcasting an encrypted message. Find more details in Chapters 5 and 8.

**Quantum Homomorphic Encryption in Nonlocal Games.** Imagine we want to interact with an untrusted server to perform a difficult computation. We want it to compute what we ask without learning any information about what is computed. This is the purpose of *homomorphic encryption*: the user encrypts some data, sends it to the server, the server performs the computation on the encrypted data, returns the output to the user, and finally the user decrypts it. When additionally quantum cryptography is used in the protocol, we speak of *quantum homomorphic encryption*. Note that such a protocol can be constructed from the hardness of the standard learning with errors (LWE) problem [Bra18; Mah18]. Now, quantum homomorphic encryption can be used in combination with nonlocal games, giving rise to the notion of *compiled game*. There, instead of interacting with two or more players, the Referee communicates with one party only, through quantum homomorphic encryption to ensure "nonlocality" (the player does not know the actual question, they only know the encryption of it, so even if they receive the two encryptions, they do not actually know the two questions). For any nonlocal game $G$, there is a compiled version $G_{comp}$. It is shown that the quantum value of $G_{comp}$ is at least as good as the one of $G$, up to negligible term, for any game $G$ [Kal+23] and that it is at most as good for the CHSH game [NZ23], for XOR games [Bar+24; Cui+24], for the tilted CHSH game [MPW24], and more generally for any game [Kul+24]. This is called *quantum soundness*[11] and has consequences to self-testing and parallel repetitions of games [Cui+24; Kul+24; NZ23].

### 3.3.5 Physical Principles

Current descriptions of quantum correlations rely on Hilbert spaces, quantum states, and quantum measurements (Section 3.1.1). Nevertheless, quantum correlations are device-independent (they are just statistics), so the intuition is that we should be able to describe them in terms of information processing.

To this end, many information-based principles have been introduced but, to this day, none of them completely rules out the set of quantum correlations. Among them, there is the principle of *communication complexity* [Yao79], stating that no physical correlation should enable two distant parties Alice and Bob to perform *any* distributed computation with only one bit of communication. We detail this physical principle, as well as information causality, nonlocal computation, macroscopic locality, local orthogonality, nonlocality swapping, and many-box locality in Chapter 4. This perspective is that the heart of our contributions described in Chapters 6 and 7 [BBP24; Bot+24a; BW24].

[11] Note that classical soundness also holds for any compiled game [Kal+23].

# Chapter 4

# Communication Complexity

In this chapter, we continue introducing the background material. We introduce the principle of communication complexity, then present advances and limitations to characterize the set of quantum correlations, and finally describe other principles with a similar aim.

## Chapter Contents

## 4.1 The Principle of Communication Complexity

In this section, we present the principle of *communication complexity* (CC). It relies on distributed computations with a low number of communication bits, and it is strongly "believed" that a violation of this principle is impossible in nature. Therefore, it can be used as a physical principle to discard unrealistic theories.

Below, we introduce two variants of the communication complexity, first the deterministic CC (Section 4.1.1) and then the randomized CC (Section 4.1.2), which ultimately leads us to the definition of the collapse of communication complexity (Section 4.1.3).

### *4.1.1 Deterministic Communication Complexity*

Communication complexity (CC) quantifies the difficulty of computing a function $f(X, Y)$ where $X$ and $Y$ are given to two parties with limited communication ability. The goal is to correctly estimate the function $f$ (constraint) while minimizing the number of communication bits (cost). This notion was introduced by Yao [Yao79], later expanded by Cleve and Buhrman to a quantum-assisted version [CB97] (our case, see also [BCv01]), and reviewed in [KN96; RY20]. Links between nonlocality and communication complexity are reviewed in [Buh+10].

Below, after introducing CC as a game, we formally define this notion and give several examples.

**Communication Complexity Game.** We can view the scenario of communication complexity as a generalized nonlocal game (Section 3.2), in which an additional classical channel with finite capacity is allowed between the players. This game involves two players, Alice and Bob, and a Referee. Before the game starts, a Boolean function $f : \{0,1\}^n \times \{0,1\}^m \to \{0,1\}$ is publicly broadcasted. Based on this function $f$, the players Alice and Bob can establish a common strategy. Then, the game starts and they are separated in a way that communication between them is impossible unless they use a *classical channel* supervised by the Referee. The Referee provides Alice and Bob with respective strings $X \in \{0,1\}^n$ and $Y \in \{0,1\}^m$. During the game phase, Alice and Bob deploy their strategy. They may use not only the classical channel, but also *nonlocal boxes* (Section 3.1) as in usual nonlocal games. The Referee counts the number of classical bits exchanged in both ways. Eventually, as the game ends, Alice answers a bit $a \in \{0,1\}$ to the Referee, and the Referee declares that the players collaboratively won the game *if, and only if,* we have $a = f(X,Y)$. The situation is illustrated in Figure 4.1.

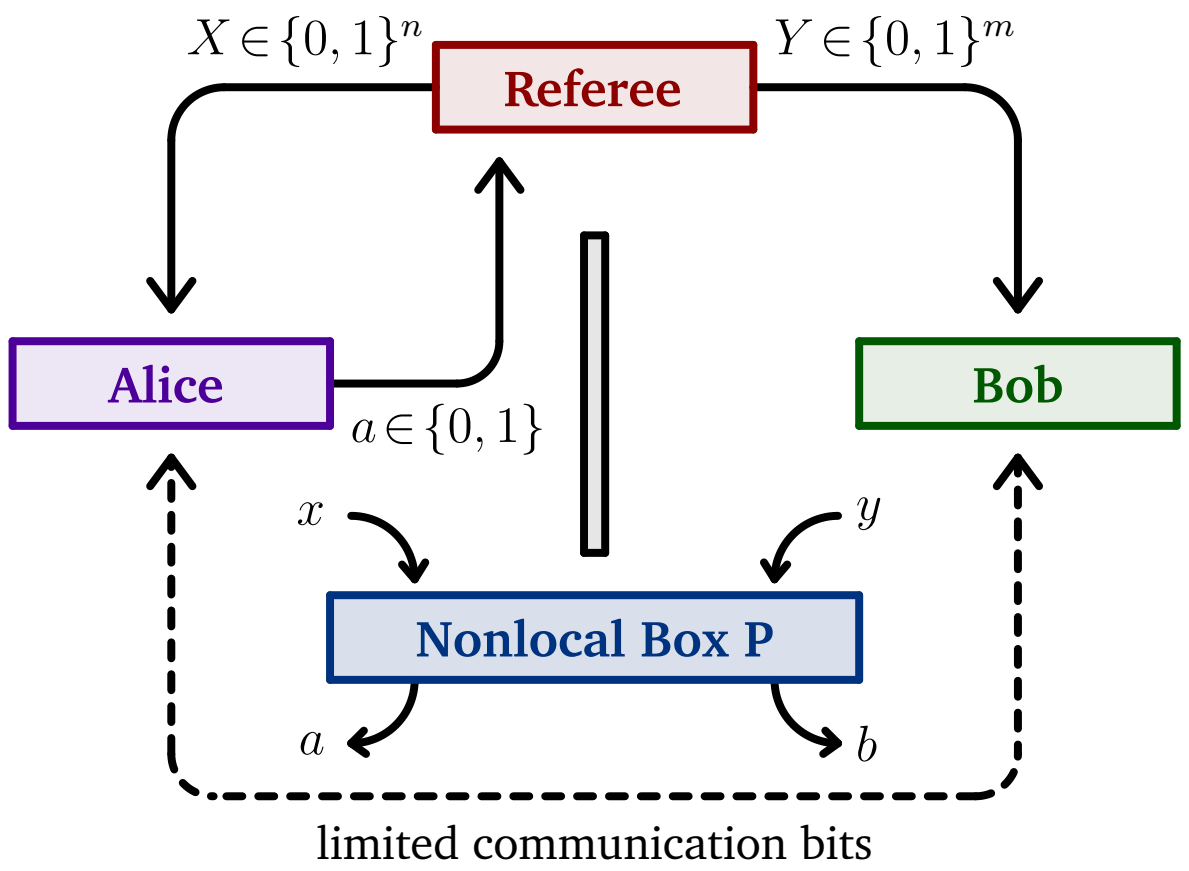

**Figure 4.1** — *Communication complexity game. Alice and Bob win the game if $a = f(X,Y)$. A similar diagram appears in [BBP24].*

**Deterministic CC.** In the above game, Alice and Bob want to win the game with a minimal number of communication bits, which is viewed as a cost. It is like minimizing the number of communication bits under the constraint of winning the game. We denote $R \subseteq \mathcal{NS}$ the set of nonlocal boxes that Alice and Bob are allowed to use in their protocol, called *resource set*. Most of the time, in this thesis, we will consider $R = \mathcal{Q}$, $\mathcal{NS}$, or $\{\mathbf{PR}\}$. This yields the following definition:

**Definition 4.1** (Deterministic Communication Complexity) — *The* deterministic communication complexity *of a Boolean function* $f : \{0,1\}^n \times \{0,1\}^m \to \{0,1\}$ *using nonlocal boxes in a set* $R \subseteq \mathcal{NS}$*, denoted:*

$$\mathrm{CC}^R(f)\,,$$

*is the* minimal *number of classical bits that need to be exchanged between the players so that Alice correctly answers* $a = f(X,Y)$ *in the worst case of* $X$ *and* $Y$*. We simply write* $\mathrm{CC}(f)$ *if no nonlocal box is needed.*

**Remark 4.2** (Symmetric Version) — Here, we require only Alice to know the value $f(X, Y)$, but note that it would cost at most only one additional bit of communication if, instead, we require that both players know the value $f(X, Y)$ (*i.e.* Alice can send the bit $a$ to Bob as the last step of their strategy).

**Remark 4.3** (Trivial Upper Bounds) — There is a trivial (but costly) strategy that Alice and Bob can always perform, no matter the function $f$. Bob can send his full string $Y \in \{0,1\}^m$ to Alice, that is $m$ communication bits, and then Alice can correctly evaluate the function $f$. This gives the following trivial upper bound on the deterministic CC:

$$\mathsf{CC}(f) \leqslant m\,.$$

Nevertheless, in many cases, Alice and Bob can find a clever way to compute $f$ with much fewer communication bits. Furthermore, if $R \subseteq S \subseteq \mathcal{NS}$, another trivial upper bound is the following one:

$$\mathsf{CC}^{\mathcal{NS}}(f) \leqslant \mathsf{CC}^{S}(f) \leqslant \mathsf{CC}^{R}(f) \leqslant \mathsf{CC}(f)\,,$$

because the more resources the players have access to, the better they can win the game with less or as many communication bits.

**Remark 4.4** (Same Input Strings) — Without loss of generality, we can always assume that Alice's and Bob's input strings have the same length $n = m$ because, otherwise, one can enlarge the shortest string with zeros.

**Remark 4.5** (One-Way Variant) — Here, we consider the two-way CC because both Alice and Bob can use the classical channel to send bits to each other. There is a variant where only Bob can send bits to Alice, requiring in general more communication bits for Alice to compute $f(X, Y)$.

**Remark 4.6** (No Computational Assumption) — In contrast with *computation complexity*, here we do not make any assumption on the computational power or memory size of Alice and Bob. They can have unlimited local computational power as well as use as many copies of nonlocal boxes as they wish. We refer to [Kap+11] for a different approach with access to a limited number of nonlocal boxes.

**Distributed Computing.** Given a Boolean function $f : \{0,1\}^n \times \{0,1\}^m \to \{0,1\}$ and two strings $X \in \{0,1\}^n$ and $Y \in \{0,1\}^m$ that are "distributed" to Alice and Bob respectively, a way to compute $f(X,Y)$ is to achieve the following decomposition:

$$a_X \oplus b_Y \; = \; f(X,Y)\,,$$

where $a_X$ and $b_Y$ are bits produced from $X$ and $Y$ respectively and possibly from some communication bits (as few as possible). If such a decomposition is achieved, then with only one additional communication bit $b_Y$, Alice is able to correctly evaluate the function $f(X,Y)$. Below, we consider three examples in the easiest non-trivial case where $n = m = 2$, two for which $\mathrm{CC} = 1$ and one for which $\mathrm{CC} = 2$. (Recall from Remark 4.3 that $\mathrm{CC}(f) \leqslant 2$ in this case.)

**Example 4.7** (Sum) — Consider the function that sums all its entry bits modulo 2:

$$f(x_1, x_2;\, y_1, y_2) \; := \; x_1 \oplus y_1 \oplus x_2 \oplus y_2\,.$$

Upon receiving $Y = (y_1, y_2)$, Bob can compute the bit $b = y_1 \oplus y_2$ and send it to Alice. Then, using $X = (x_1, x_2)$ and $b$, Alice can compute $a = x_1 \oplus x_2 \oplus b$ and she correctly evaluates $f(X,Y)$. Hence, for this function $f$, one bit of communication is enough, so $\mathrm{CC}(f) \leqslant 1$. Finally, as at least one communication bit is necessary for this function, we obtain:

$$\mathrm{CC}(f) \; = \; 1\,.$$

**Example 4.8** (Local Products) — Consider the function that computes the sum of local products:

$$g(x_1, x_2;\, y_1, y_2) \; := \; (x_1 \cdot x_2) \oplus (y_1 \cdot y_2)\,.$$

As in Example 4.7, Bob can send the bit $b := y_1 \cdot y_2$ to Alice, and then she can correctly evaluate the function $g$ via defining $a := (x_1 \cdot x_2) \oplus b$. Again, we obtain that the communication complexity is $1$:

$$\mathrm{CC}(g) \; = \; 1\,.$$

**Example 4.9** (Nonlocal Products) — We present a slight variation of Example 4.8. Consider the function that sums nonlocal products, called *inner product function*:

$$\mathrm{IP}_2(x_1, x_2;\, y_1, y_2) \; := \; (x_1 \cdot y_1) \oplus (x_2 \cdot y_2)\,.$$

One can show that there is no communication protocol with $1$ communication bit or less so that Alice can correctly evaluate $\text{IP}_2$ for any $x_1, x_2, y_1, y_2$. Thus $\text{CC}(\text{IP}_2) \geqslant 2$, and using the trivial upper bound, we infer that:

$$\text{CC}(\text{IP}_2) = 2 .$$

More generally, it can be shown that the inner product function $\text{IP}_n$ with $n$-bit strings has deterministic communication equal to $n$ [CG88] (Proposition 4.13). It means that nonlocal products are difficult to compute in a distributive way. This is where the **PR** box (see page 63) can be very useful: it is designed to perfectly transform any nonlocal product $x \cdot y$ into a distributed sum $a \oplus b$. Hence, if we use one copy of the **PR** box for each of the nonlocal products of $\text{IP}_n$, this function can be turned into a sum function like $f$ in Example 4.7, yielding a communication complexity equal to one:

$$\forall n \geqslant 1 , \qquad \text{CC}^{\text{PR}}(\text{IP}_n) = 1 .$$

This is one of the two key ideas used in the protocol from van Dam [vD99] to show that the **PR** collapses communication complexity (Theorem 4.20).

### 4.1.2 Randomized Communication Complexity

In the *randomized* variant of CC also introduced by Yao [Yao79], Alice and Bob are allowed to use a shared random string $Z \in \{0,1\}^*$ of unbounded length (shared randomness). They can use this common data to better correlate their behavior and enhance their strategy to compute $f(X, Y)$ with fewer communication bits. Here, we are interested in the probability of Alice correctly guessing the value $f(X, Y)$ (we want it to be as high as possible with the lowest number of communication bits). This gives rise to the following definition:

**Definition 4.10** (Randomized Communication Complexity) — *The randomized communication complexity of a Boolean function* $f : \{0,1\}^n \times \{0,1\}^m \to \{0,1\}$ *with parameter* $p \in [0,1]$ *and using nonlocal boxes in a set* $R \subseteq \mathcal{NS}$*, denoted:*

$$\text{CC}_p^R(f) ,$$

*is the* minimal *number of classical bits that need to be exchanged between the players so that Alice correctly answers* $a = f(X, Y)$ *with probability at* *least $p$ in the worst case of $X$ and $Y$:*

$$\min_{X,Y} \mathbb{P}(a = f(X,Y)) \geqslant p\,.$$

*We simply write* $\mathrm{CC}_p(f)$ *if no nonlocal box is needed.*

We emphasize that $p$ does not depend on $X$ nor $Y$ but may depend on $f$.

**Remark 4.11** (Trivial Upper Bounds) — The randomized CC is a relaxation of the deterministic CC, so we have:

$$\mathrm{CC}_p^R(f) \leqslant \mathrm{CC}^R(f)\,,$$

for any $f$, $p$, and $R$. Moreover, notice that it is trivial when $p \leqslant 1/2$, since Alice can simply sample a uniformly random bit $a \in \{0,1\}$ to be correct with probability $1/2$ with no communication. It yields:

$$\forall p \leqslant \frac{1}{2}, \qquad \mathrm{CC}_p^R(f) = 0\,,$$

for any $f$ and $R$.

We give two comparisons between the deterministic and randomized communication complexity. On the other hand, surprisingly, there may be a large difference between the two variants:

**Proposition 4.12** (Comparison Between Deterministic and Randomized CC [KN96]) — *The equality function* $\mathrm{EQ}_n(X,Y) := \mathbb{1}_{X=Y}$ *has maximal deterministic communication complexity, while "trivial" randomized communication complexity:*

$$\mathrm{CC}\big(\mathrm{EQ}_n\big) = m \qquad \textit{and} \qquad \mathrm{CC}_{2/3}\big(\mathrm{EQ}_n\big) = 1\,.$$

On the other hand, interestingly, it may happen that both the deterministic and randomized variants have high complexity, of the order $\Omega(n)$, even when quantum entanglement is allowed:

**Proposition 4.13** (CC with Quantum Resources [CG88; Cle+99]) — *Consider the inner product function* $\mathrm{IP}_n(X,Y) := x_1y_1 \oplus \cdots \oplus x_ny_n$*, also mentioned in Example 4.9. It has the following communication complexity:*

$$\begin{aligned} \mathrm{CC}\big(\mathrm{IP}_n\big) &= n\,, & \mathrm{CC}_p\big(\mathrm{IP}_n\big) &= n - \mathcal{O}\big(\log(1/p)\big)\,, \\ \mathrm{CC}^{\mathcal{Q}}\big(\mathrm{IP}_n\big) &= n\,, & \mathrm{CC}_p^{\mathcal{Q}}\big(\mathrm{IP}_n\big) &\geqslant \max\Big(\tfrac{1}{2}(2p-1)^2, (2p-1)^4\Big) \times n - \tfrac{1}{2}\,. \end{aligned}$$

**Remark 4.14** (Local Randomness Variant) — There exists also a variant with local (private) randomness. Newman showed that any shared (public) randomness protocol with a shared string of size $n$ can be turned into a local randomness protocol that uses at most $\mathcal{O}(\log(n))$ additional communication bits [New91].

**Remark 4.15** (Quantum Variant) — Yao later introduced a variant where classical bits are replaced by quantum bits in the communication protocol [Yao]. Results on this topic include [BCW98; Bd01; Bri+15; HLGM25; Kre95; LMd23]. For instance, in comparison with the bit CC (Proposition 4.13), the qubit deterministic CC of the inner product function is $\lceil n/2 \rceil$ and the randomized one is at least $\frac{1}{2}\,(2p-1)^2\,n - \frac{1}{2}$ [Cle+99]. Further comparisons between the bit and the qubit models can be found in [Lal25].

### *4.1.3 The Collapse of Communication Complexity*

Interestingly, it happens sometimes that there is a *collapse* of the (randomized) communication complexity. This means that only 1 bit of communication is enough to distributively compute *any* function $f$ with arbitrary large input size. This is very strong, and as discussed below, a collapse of CC is believed to be unachievable in nature. First, here is a formal definition:

**Definition 4.16** (Collapse of Communication Complexity) — *We say that a nonlocal box* $\mathbf{P} \in \mathcal{NS}$ collapses communication complexity *if there exists a universal constant* $\mathfrak{p} > 1/2$ *such that:*

$$\forall f : \{0,1\}^n \times \{0,1\}^m \to \{0,1\}, \qquad \mathrm{CC}_{\mathfrak{p}}^{\{\mathbf{P}\}}\big(f\big) \leqslant 1\,.$$

We stress that $\mathfrak{p}$ may depend on $\mathbf{P}$ but not on $f$, $X$, $Y$, $n$, nor $m$. Note that we require $\mathfrak{p} > 1/2$ because of Remark 4.11. Moreover, there is a variant of the principle stating that $\mathrm{CC}_{\mathfrak{p}}^{\{\mathrm{P}\}}(f)$ should be bounded.

**Remark 4.17 —** One can also find the phrasing *non-trivial communication complexity* in the literature [Bra+06; BS09; Nav+15]. Here, we rather use the wording *collapse of communication complexity* as in [Bru+14] because we find it more meaningful.

**Communication Complexity as a Physical Principle.** Such a collapse is strongly believed to be unachievable in nature [BG15; Bra+06; BS09; EWC23a; vD99] since it would imply the absurdity that a single bit of communication would be sufficient to distantly estimate any value of any Boolean function $f$ with arbitrary large input size and with high success probability. From this observation, one may view the collapse of CC as an information-based principle to discard "unphysical" correlations, in working toward a perfect characterization of the set of quantum correlations $\mathcal{Q}$ (or an approximation of it, see the discussion in footnote 1).

> **Open Question 4.18 —** *What are all nonlocal correlations that collapse communication complexity?*

**Advances.** As elaborated on in Section 4.2 and showcased in Figure 4.2, the **PR** box is known to collapse CC [vD99], so this correlation is physically unfeasible according to the principle of communication complexity. More generally, it is known that some noisy versions of the **PR** box also collapse CC for different types of noise [BBP24; Bot+24a; Bra+06; Bri+19; BS09; BW24; EWC23a]. On the other hand, it is known that quantum correlations do *not* collapse communication complexity [Cle+99], and neither does a slightly wider set named *almost quantum correlations* [Nav+15] (see Remark 3.5).[1] By way of compensation, we will also present some limiting results in Section 4.3.

[1] From this result, we can infer that the principle of communication complexity cannot perfectly characterize the quantum set $\mathcal{Q}$, but it may still do so if combined with another information-based principle. Moreover, it may be true that CC perfectly characterizes the set of almost quantum correlations $\widetilde{\mathcal{Q}}$, which is an approximation of $\mathcal{Q}$.

## 4.2 Advances

In this section, we overview the advances related to the collapse of communication complexity in chronological order. The diverse contributions are summarized in Figure 4.2.

Below, we begin by presenting the result that quantum boxes are non-collapsing (Section 4.2.1) and that the **PR** box is collapsing (Section 4.2.2). After this, we elaborate on two results extending the collapse of CC, one in terms of the CHSH value (Section 4.2.3) and one in terms of a convex combination of **PR** and **SR** (Section 4.2.4). Finally, we describe some recent advances in the collapse of CC (Section 4.2.5).

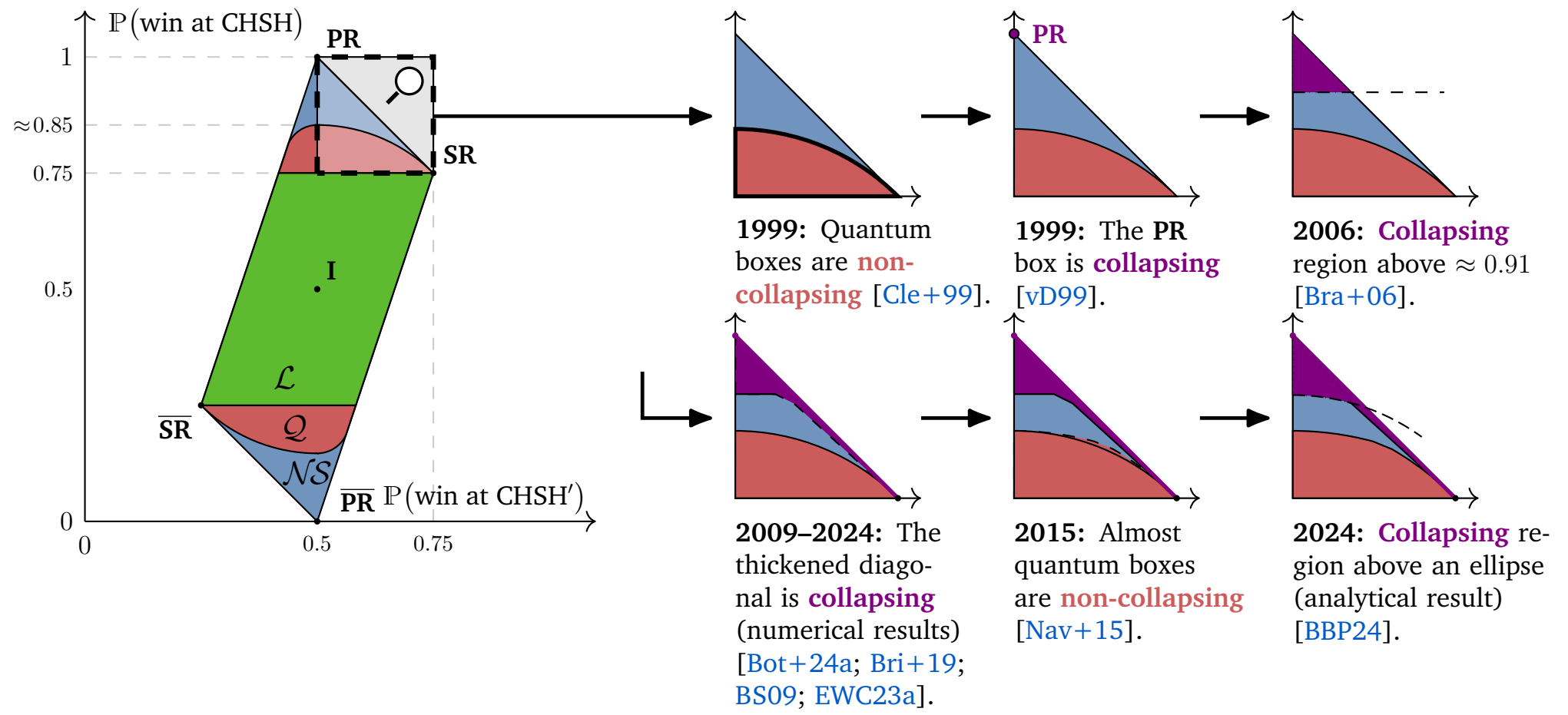

**Figure 4.2** — Historical overview of collapsing boxes, *drawn in the slice of $\mathcal{NS}$ passing through* **PR** *and* **SR** *and* **I**. *In* **red** *and* **purple** *are represented respectively the non-collapsing and the collapsing boxes. In* **blue** *is drawn the region of boxes for which we do not know yet if they collapse communication complexity. See [AIR25; Bri+19; BW24; Mor16; SWH20] for related results. A similar diagram appears in [BBP24].*

### *4.2.1 Quantum Boxes do not Collapse CC*

The first result in this line of research is that quantum boxes cannot collapse CC. This result is due to Cleve, van Dam, Nielsen, and Tapp [Cle+99] and it is a consequence of the above-presented Proposition 4.13. Find an illustration in Figure 4.2.

**Theorem 4.19** (Quantum Boxes are not Collapsing [Cle+99]) — *There exist functions $f$ for which the randomized communication complexity is of order $\Omega(n)$ even with prior-entanglement:*

$$\exists f,\ \forall p > \frac{1}{2}, \qquad \mathrm{CC}_p^{\mathcal{Q}}(f) = \Omega(n)\,.$$

*For instance, consider the inner production function* $\mathrm{IP}_n$ *(Proposition 4.13).*

*Sketch of the Proof.* Denote by $\mathrm{IP}_n(X,Y) := x_1y_1 \oplus \cdots \oplus x_ny_n$ the inner product function. First, the authors argue that any $k$-bit protocol for the function $\mathrm{IP}_n$ can be converted into an $k$-qubit protocol for the function $\mathrm{IP}_{2n}$ using *superdense coding* [BW92]. Then, they prove the lower bound in the quantum variant of CC where qubits are communicated instead of classical bits (see Remark 4.15). To achieve this qubit lower bound, they use a consequence of Holevo's theorem [Hol73], stating that if quantum entanglement and qubit communication are available, then for Alice to obtain $k$ bits of mutual information with respect to Bob's $n$ bits, Bob must send at least $\lceil k/2 \rceil$ qubits. These ideas work both in the deterministic and randomized model of CC in a clever way, hence the result. ∎

### *4.2.2 The* PR *Box Collapses CC*

> « *The solution of all possible distributed functions with a single bit of communication surely does contradict our experiences in computer science.* » — van Dam [vD99]

The first proof of the collapse of CC is due to van Dam in his Ph.D. thesis [vD99], showing that the **PR** box collapses the deterministic communication complexity (with universal constant $\mathfrak{p} = 1$). Find an illustration in Figure 4.2.

**Theorem 4.20** (**PR** Collapses CC [vD99]) — *The* **PR** *box collapses communication complexity:*

$$\forall f : \{0,1\}^n \times \{0,1\}^m \to \{0,1\}, \qquad \mathrm{CC}^{\mathrm{PR}}(f) \leqslant 1\,.$$

*Sketch of the Proof.* The proof relies on two observations. First, as mentioned in Example 4.9, the inner product function $\mathrm{IP}_n(X, Y) := x_1 y_1 \oplus \cdots \oplus x_n y_n$ can be efficiently distributively computed using $n$ copies of the **PR** box. Indeed, by definition, the **PR** box perfectly turns any product $xy$ into the sum $a \oplus b$. Hence, the inner product function can be rewritten as $\mathrm{IP}_n(X, Y) = \big(a_1 \oplus \cdots \oplus a_n\big) \oplus \big(b_1 \oplus \cdots \oplus b_n\big)$, which can be computed by Alice if Bob sends her only one bit, namely $b_1 \oplus \cdots \oplus b_n$. It yields:

$$\forall n \geqslant 1\,, \qquad \mathrm{CC}^{\mathbf{PR}}\big(\mathrm{IP}_n\big) \;=\; 1\,.$$

The second observation is that the distributed computing of any Boolean function $f : \{0,1\}^n \times \{0,1\}^n \to \{0,1\}$ can be reduced to the distributed computing of the function $\mathrm{IP}_{2^n}$ using Lagrange interpolation polynomial [Lag95] over the ring $\mathbb{Z}/2\mathbb{Z}$. Hence the result for any function. ■

**Remark 4.21** (Need of Unbounded Resources) **—** Note that the proof relies on the fact that Alice and Bob can use $2^n$ copies of the **PR** box. In this approach, we are not concerned about resource bounds, see Remark 4.6.

### *4.2.3 Boxes with CHSH Value Above* $\approx 0.91$ *Collapse CC*

> « *Most computer scientists would consider a world in which communication complexity is trivial to be as surprising as a modern physicist would find the violation of causality.* » — Brassard, Buhrman, Linden, Méthot, Tapp, and Unger [Bra+06]

Then, van Dam's result was extended to any noisy **PR** box winning the CHSH game with probability at least $\frac{3+\sqrt{6}}{6} \approx 0.91$ by Brassard, Buhrman, Linden, Méthot, Tapp, and Unger [Bra+06]. This encompasses any nonlocal box $\mathbf{P}(ab|xy) \in \mathcal{NS}$ such that $a \oplus b = xy$ with probability greater than $\frac{3+\sqrt{6}}{6}$. In particular, the result holds for the **PR** box since its winning probability is $1$. Find an illustration in Figure 4.2. We refer to Section 3.2.2 for an introduction to the CHSH game.

**Theorem 4.22** (Collapse of CC Above $\approx 0.91$ [Bra+06]) — *Consider any box* $\mathbf{P} \in \mathcal{NS}$ *such that:*

$$\mathbb{P}\Big(\mathbf{P} \textit{ win at } \mathrm{CHSH}\Big) > \frac{3+\sqrt{6}}{6} \approx 0.91 \,.$$

*Then, the box* $\mathbf{P}$ *collapses communication complexity.*

*Sketch of the Proof.* Let $f : \{0,1\}^n \times \{0,1\}^m \to \{0,1\}$ be any Boolean function, and $X \in \{0,1\}^n$ and $Y \in \{0,1\}^m$ be some of its inputs given to Alice and Bob respectively. The proof consists in finding an initial protocol that computes $f(X,Y)$ distributively, *i.e.* that finds some bits $a$ on Alice's side and $b$ on Bob's side such that:

$$a \oplus b = f(X,Y) \,,$$

with probability greater than $1/2$, using shared randomness but no communication bit. Then, the idea is to amplify the winning probability using several times a majority function:

$$\mathtt{Maj}_3 : \{0,1\}^3 \to \{0,1\} \,, \tag{4.1}$$

that is defined as outputting the most-frequent bit (*e.g.* $\mathtt{Maj}_3(0,1,0) = 0$ and $\mathtt{Maj}_3(1,1,1) = 1$). Then, the end of the proof relies on the following two observations. On the one hand, they show that if Alice and Bob know a protocol to distributively compute the majority function of some distributed bits with probability $q > \frac{5}{6}$, then communication complexity collapses. On the other hand, they prove that if Alice and Bob have access to nonlocal boxes $\mathbf{P} \in \mathcal{NS}$ winning the CHSH with some probability $p$, then Alice and Bob can achieve this distributed computation of the majority function with probability $p^2 + (1-p)^2$. Now, combining the two observations, we see that communication complexity collapses whenever:

$$q = p^2 + (1-p)^2 > \frac{5}{6} \,, \quad \text{that is} \quad p > \frac{3+\sqrt{6}}{6} \quad \text{or} \quad p < 1 - \frac{3+\sqrt{6}}{6} \,.$$

Hence the result. ■

**Remark 4.23** — In [BBP24], we build upon this proof to obtain a better sufficient condition to collapse CC. See Chapter 6.

**Remark 4.24** (Optimalitiy of $\frac{3+\sqrt{6}}{6}$) — As elaborated on in Section 4.3, the threshold of $\frac{3+\sqrt{6}}{6}$ remains the same even if we replace the majority function $\mathtt{Maj}_3$ in this protocol by any functions in a large class of relevant functions [Mor16; SWH20]. This highlights the difficulty of improving this result.

### 4.2.4 Correlated Boxes are Collapsing

> « *This result provides a partial answer to the question of why quantum nonlocality is also bounded below Tsirelson's bound, in regions of the polytope close to the local set of correlations.* » — Brunner and Skrzypczyk [BS09]

*Correlated boxes* are defined as any convex combination between the **PR** box and the shared randomness box **SR**:

$$\mathbf{P}_\alpha := \alpha\,\mathbf{PR} + (1-\alpha)\,\mathbf{SR}\,,$$

where $0 \leqslant \alpha \leqslant 1$, and where $\mathbf{SR}(ab|xy) := \frac{1}{2}\,\mathbb{1}_{a=b}$ is the box that answers either $(0,0)$ or $(1,1)$ with probability $1/2$ independently of the inputs $x$ and $y$. It is shown by Brunner and Skrzypczyk that for any $\alpha > 0$, the correlated boxes collapse CC [BS09]. These boxes are drawn in the diagonal joining **PR** and **SR** in Figure 4.2. This region was also enlarged numerically to the area drawn below the diagonal [Bot+24a; BS09; EWC23a].

> **Theorem 4.25** (Correlated Boxes are Collapsing [BS09]) — *Any box of the form $\alpha\,\mathbf{PR} + (1-\alpha)\,\mathbf{SR}$ with $0 < \alpha \leqslant 1$ collapses communication complexity.*

*Sketch of the Proof.* The authors introduce a specific wiring of boxes (see $\mathsf{W}_{\mathsf{BS}}$ in Figure 3.2) such that, given two copies of a correlated box $\mathbf{P}_{\alpha_0}$, produces another correlated box $\mathbf{P}_{\alpha_1}$ with $\alpha_1 > \alpha_0$. We say that this wiring *distills nonlocality*. Then, repeating this wiring procedure on two copies of $\mathbf{P}_{\alpha_1}$, they build a sequence of boxes $\left(\mathbf{P}_{\alpha_k}\right)_k$ that converges to the **PR** box as $k \to \infty$. But, these wiring procedures do not make use of communication bits, so they do not increase the CC of a function. Moreover, using Theorem 4.22, we know that there is an open neighborhood of **PR** that collapses CC. Hence, any correlated box is collapsing. ■

**Remark 4.26 —** In [Bot+24a], we build upon this proof to obtain wider sets of boxes that collapse CC. See Chapter 6.

**Remark 4.27 —** Of course, when $\alpha = 0$ the correlated box is not collapsing because it is $\mathbf{P}_{\alpha=0} = \mathbf{SR}$ and it belongs to $\mathcal{L} \subseteq \mathcal{Q}$, which is not collapsing (Section 4.2.1). Moreover, this result is interesting since it means that the principle of communication complexity is very precise in a neighborhood of **SR** (in this slice of $\mathcal{NS}$): Close to **SR**, the only non-collapsing boxes are approximatively the quantum boxes. The result of Theorem 4.25 was a lot improved in [Bri+19], where the authors show that many faces of $\mathcal{NS}$ collapse CC with similar distillation techniques. Nevertheless, we present in Section 4.3 some limitations on such a technique [BG15; EWC23a].

**Remark 4.28** (Alternative Proof) **—** An alternative proof of this result is provided in the M.Sc. thesis of Proulx [Pro18]. This proof is not based on distillation, but rather on a variant of the protocol from Pawłowski, Paterek, Kaszlikowski, Scarani, Winter, and Żukowski in [Paw+09] that distributively computes the address function.

### *4.2.5 Recent Results on the Collapse of CC*

As already mentioned in the former sections, we stress that some generalizations of the above protocols were recently found. Theorem 4.22 was improved in [BBP24] with a better necessary condition, and Theorem 4.25 was extended to many faces of $\mathcal{NS}$ [Bot+24a; Bri+19]. Moreover, the principle of CC was proven to perfectly characterize the quantum best-winning value $\mathfrak{w}_{\mathcal{Q}}(\mathcal{G})$ of some nonlocal game $\mathcal{G}$ [SWH20] and was studied in the context of graph games [BW24]. In addition, several algorithms were developed to identify new collapsing boxes [Bot+24a; EWC23a].

In contrast, we list some limitations of the principle of CC to characterize quantum correlations in Section 4.3.

As for now, the question of the collapse of CC is still open for the remaining non-signaling correlations drawn in blue in Figure 4.2.

## 4.3 Limitations

In this section, we present some limitations for the principle of communication complexity to characterize quantum correlations.

We begin by presenting a wider set than the quantum one, called *almost quantum* set, that does not collapse the one-way variant of CC where only one player can send communication bits to the other (Section 4.3.1). Then, we present no-go results for generalizations of the above-presented protocols from [Bra+06] (Section 4.3.2) and from [BS09] (Section 4.3.3). Finally, we present complications in generalizing the principle of CC to the multipartite setting (Section 4.3.4).

### 4.3.1 Almost Quantum Boxes are not 1-Way Collapsing

The almost quantum correlations were introduced by Navascués, Guryanova, Hoban, and Acín in [Nav+15]. They form a set $\widetilde{\mathcal{Q}}$ that is strictly between the quantum set and the non-signaling set:

$$\mathcal{Q} \subsetneq \widetilde{\mathcal{Q}} \subsetneq \mathcal{NS}\,.$$

We already gave two equivalent definitions of these correlations, one in terms of commutative PVMs on a certain state $|\psi\rangle$ (Remark 3.5), and another one in terms of the Navascués-Pironio-Acín hierarchy (Remark 3.10). Find an illustration in Figure 4.2. The authors prove the following striking result in the one-way variant of communication complexity, where only Bob can send a bounded number of bits to Alice:

**Theorem 4.29** (Almost Quantum Boxes are not Collapsing [Nav+15]) — *Almost quantum correlations cannot collapse the one-way communication complexity.*

*Sketch of the Proof.* The proof is an extension of the one for quantum correlations from [Cle+99] (Theorem 4.19). The authors use again the counterexample of the inner product function $\mathrm{IP}_n(X,Y) := x_1y_1 \oplus \cdots \oplus x_ny_n$. By contradiction, assume that there is a collapse of communication complexity. Then, they show that if a universal constant $\mathfrak{p} > 1/2$ exists for the collapse of CC, it must satisfy the following inequality:

$$\forall n \in \mathbb{N}\,, \qquad \frac{1}{2^{n-m}} \geqslant \left(2\,\mathfrak{p} - 1\right)^2,$$

where $n$ is the size of Alice's and Bob's strings, and where $m$ is the number of classical bits that Bob is allowed to send to Alice. But, this inequality cannot be true for $\mathfrak{p} \neq \frac{1}{2}$ and for all $n \in \mathbb{N}$, this is a contradiction. Hence, almost quantum correlations cannot collapse CC. ■

**Remark 4.30** (Contrast) — Let us balance this result with regard to the goal of ruling out the quantum set with the principle of CC. To the best of our knowledge, nothing is known for these almost quantum correlations in the *two-way* communication complexity, where both Alice and Bob can communicate finitely many bits. Moreover, the best-winning probability with quantum correlations $\mathfrak{w}_{\mathcal{Q}}(\mathrm{CHSH})$ at the CHSH game is the same as the almost quantum one:

$$\mathfrak{w}_{\mathcal{Q}}(\mathrm{CHSH}) = \mathfrak{w}_{\widetilde{\mathcal{Q}}}(\mathrm{CHSH}),$$

so CC can still characterize the quantum best-winning value at CHSH, at least. Furthermore, the combination of CC with another information-based principle (find examples in Section 4.4) may have a correct depiction of $\mathcal{Q}$. Finally, the question of whether the principle of CC characterizes the set of almost quantum correlations $\widetilde{\mathcal{Q}}$ is interesting on its own since $\widetilde{\mathcal{Q}}$ is a good approximation of $\mathcal{Q}$ in the sense that for many nonlocal games the best-winning probability is the same for these two correlation sets.

### 4.3.2 *The Threshold* $\approx 0.91$ *Seems Optimal*

> « *The absurdity of a world in which any nonlocal binary function could be evaluated with a constant amount of communication in turn provides a tantalizing way to distinguish quantum mechanics from incorrect theories of physics.* » — Shutty, Wootters, and Hayden [SWH20]

Recall from Theorem 4.22 that Brassard, Buhrman, Linden, Méthot, Tapp, and Unger obtained a sufficient condition for the collapse of CC with the threshold [Bra+06]:

$$\frac{3+\sqrt{6}}{6} \approx 0.91.$$

To obtain such a threshold, the protocol relies on boosting the winning probability with the majority function $\mathtt{Maj}_3$ defined in eq. (4.1). We present two results shedding light on evidence that this threshold might actually be optimal.

**Optimality for XOR Functions.** In [Mor16], Mori considers other functions to boost the winning probability and obtain a better threshold. Nevertheless, they show that it is not possible to improve the result by replacing $\texttt{Maj}_3$ by any *XOR function* $f$ in an *adaptive* **PR**-*correct protocol*. It builds on ideas from [Paw+09]. Let us briefly define these notions before stating their theorem.

**Definition 4.31** (XOR Function, Adaptive **PR**-Correct Protocol) — *A function* $f : \{0,1\}^n \times \{0,1\}^n \to \{0,1\}$ *is said to be* XOR *if there exists a function* $g : \{0,1\}^n \to \{0,1\}$ *such that:*

$$f(X,Y) = g^{\oplus}(X,Y) := g(x_1 \oplus y_1, \dots, x_n \oplus y_n)\,.$$

*Using this notation, the authors of [Bra+06] want to distributively compute* $\texttt{Maj}_3^{\oplus}$. *A protocol is said to be* adaptive *if it employs nonlocal boxes* $\mathbf{P} \in \mathcal{NS}$ *and a wiring (Section 3.1.4) that connects some box outputs to some box inputs. A protocol is said to be* **PR**-correct *if it achieves a distributed computation perfectly using* **PR** *boxes (like for the majority function* $\texttt{Maj}_3$ *with* $2$ *nonlocal boxes).*

Using techniques from discrete Fourier analysis, the author proves the following result:

**Theorem 4.32** (The Threshold $\approx 0.91$ is Optimal for XOR Functions [Mor16]) — *The threshold* $\frac{3+\sqrt{6}}{6} \approx 0.91$ *cannot be improved by using the same methods as in [Bra+06] and replacing the three-input majority function* $\texttt{Maj}_3$ *by any XOR function in an adaptive* **PR**-*correct protocol.*

*Moreover, the function* $\texttt{Maj}_3$ *is essentially the only function achieving this threshold in this class of functions, in the sense that the other optimal functions are exactly the majority of some fixed three-input variables and ignore the other* $n-3$ *input variables.*

**Optimality via Noisy Circuits.** Leaning towards a similar conclusion, Shutty, Wootters, and Hayden [SWH20] provide complementary evidence that the threshold $\frac{3+\sqrt{6}}{6} \approx 0.91$ is optimal. The authors view the protocol from [Bra+06] as a circuit involving two gates: noisy AND gates $\wedge_\varepsilon$, producing an incorrect answer with probability $\varepsilon < \frac{1}{6}$, and perfect XOR gates $\oplus$. This can be generalized to circuits with noisy AND gates $\wedge_\varepsilon$ and noisy XOR gates $\oplus_\tau$, where $\oplus_\tau$ produces an incorrect output with probability $\tau$. They obtain the following result:

**Theorem 4.33** (The Threshold $\approx 0.91$ is Optimal for Some Noisy Circuits [SWH20]) — *Assuming a conjecture (Remark 4.34), the threshold* $\frac{3+\sqrt{6}}{6} \approx 0.91$ *is tight for any circuit using noisy AND gates and perfect XOR gates.*

*Sketch of the Proof.* First, the authors prove that the collapse of CC can be characterized in terms of *reliable computation* in classical circuit models [SWH20, Proposition 2.2], allowing them to rephrase the threshold result from [Bra+06] as:

$$\varepsilon < \tfrac{1}{6} \quad \text{and} \quad \tau = 0 \qquad \Longrightarrow \qquad \text{Collapse of CC}\,.$$

Second, in their main result [SWH20, Theorem 2.4], they prove that the class of noisy *formula* on gates $\{\wedge_\varepsilon, \oplus_\tau\}$ does not support reliable computation for any $\varepsilon > 1/6$ and $\tau > 0$. Third, they use a conjecture (Remark 4.34) to apply this result to the set of *circuits* on gates $\{\wedge_\varepsilon, \oplus_\tau\}$, which is larger. Finally, using some topological result on the closure of sets [SWH20, Theorem 2.9], they sharpen the result to parameters $\varepsilon \geqslant 1/6$ and $\tau \geqslant 0$. This shows that:

$$\varepsilon \geqslant \tfrac{1}{6} \quad \text{and} \quad \tau \geqslant 0 \qquad \Longrightarrow \qquad \text{No Collapse of CC}\,.$$

Hence, whenever XOR gates are noiseless (*i.e.* $\tau = 0$), the collapse of CC is characterized by $\varepsilon < 1/6$, which exactly corresponds to the CHSH winning probabilities $p > \frac{3+\sqrt{6}}{6}$. ∎

**Remark 4.34** (Conjecture) — The proof relies on the conjecture [SWH20, Conjecture 5.4] stating that if certain bounds are valid for formulas on gates $\{\wedge_\varepsilon, \oplus_\tau\}$, then they can be extended to general circuits on gates $\{\wedge_\varepsilon, \oplus_\tau\}$. An analogous conjecture can be found in other works, for instance in [EP98; Pip88; Ung07].

**Remark 4.35** (Open Avenues) — There is still room for improvement. For example, the case $(\varepsilon < 1/6, \tau > 0)$ is not treated. Moreover, as mentioned by the authors, the conjecture may be false, or there might be circuits involving more gates than $\wedge_\varepsilon$ and $\oplus_\tau$ that yield a collapse of CC. So, this result is not a strict no-go theorem.

**Remark 4.36 —** Interestingly, going in a different direction, the authors of this paper also construct a different game G for which the quantum best-winning probability $\mathfrak{w}_{\mathcal{Q}}(\mathrm{G})$ is characterized by the collapse of communication complexity. They put this game G in contrast with the CHSH game for which they believe the same result cannot be achieved.

### *4.3.3 Limits on Nonlocality Distillation*

Recall that in Theorem 4.25, Brunner and Skrzypczyk [BS09] proved the collapse of CC by *distilling* nonlocal boxes. Given several copies of weakly nonlocal boxes, the protocol consists in wiring them in such a way that the resulting box is more nonlocal than the original ones. Sometimes, it is possible to repeat this process until reaching the collapsing area above the threshold $\frac{3+\sqrt{6}}{6} \approx 0.91$ from [Bra+06], thus showing that the initial box is collapsing since these wirings do *not* increasing the number of communication bits. Below, we present some limiting results on nonlocality distillation.

**Isotropic Boxes Cannot be Distilled.** In [BG15], Beigi and Gohari introduce a measure $\mu_{\text{box}} : \mathcal{NS} \to [0,1]$, called *maximal correlation*, with the good property of being *decreasing under wirings* (find more details on this measure $\mu_{\text{box}}$ in Section 3.1.5):

**Theorem 4.37** (Decreasing Measure Under Wirings [BG15]) **—** *For any wiring* W *and any two nonlocal boxes* $\mathbf{P}, \mathbf{Q} \in \mathcal{NS}$*, the measure of the wired box cannot exceed the measure of each box:*

$$\mu_{\text{box}}\big(\mathbf{P} \boxtimes_{\mathrm{W}} \mathbf{Q}\big) \;\leqslant\; \max\big\{\mu_{\text{box}}(\mathbf{P}), \mu_{\text{box}}(\mathbf{Q})\big\}\,. \tag{4.2}$$

This result has interesting consequences to *isotropic boxes*. These boxes correspond to the vertical line in Figure 4.2 and are defined as:

$$\mathbf{P}_\alpha \;:=\; \alpha\,\mathbf{PR} + (1-\alpha)\,\mathbf{I}\,,$$

where $0 \leqslant \alpha \leqslant 1$, and where $\mathbf{PR}(ab|xy) := \frac{1}{2}\,\mathbb{1}_{a\oplus b = xy}$ and $\mathbf{I}(ab|xy) := \frac{1}{4}$. They have the following measure value:

$$\mu_{\text{box}}\big(\mathbf{P}_\alpha\big) \;=\; \alpha\,.$$

Hence, using eq. (4.2), we see that two copies of $\mathbf{P}_\alpha$ cannot produce a box $\mathbf{P}_\beta$ with $\alpha < \beta$ via wirings since:

$$\mu_{\text{box}}\left(\mathbf{P}_\alpha \boxtimes_{\text{W}} \mathbf{P}_\alpha\right) \leqslant \alpha < \beta = \mu_{\text{box}}\left(\mathbf{P}_\beta\right) .$$

Here, we stated it for wirings of two boxes only, but it holds more generally for any number of copies of $\mathbf{P}_\alpha$. In particular, this shows that these isotropic boxes cannot be distilled using wirings only. The result was even extended in [BG15, Theorem 10] to the case where shared randomness is also allowed (in addition to multi-copy wirings), showing that post-quantum isotropic boxes cannot be distilled (*i.e.* those for which $\alpha > \frac{1}{\sqrt{2}}$).

This is unfortunate for the study of the collapse of CC since isotropic boxes are highly important. Indeed, Masanes, Acín, and Gisin showed that any non-signaling box that is not isotropic can be projected to the isotropic line via a protocol called *depolarization* [MAG06], using shared randomness and local operations. As a consequence, if an isotropic box $\mathbf{P}_\alpha$ is shown to be collapsing, then we can infer that the whole half-space "above" this box is also collapsing. Nevertheless, it is difficult to show the collapse of CC for those boxes due to the above-mentioned result [BG15], since we cannot find a distilling protocol as in [BS09] to reach a previously-known collapsing area that is further "above."

**Need of Multi-Copy Wirings.** Wirings involving two boxes (*a.k.a. depth-*2 *wirings*) are convenient to use because simpler to characterize and easier to investigate. Nevertheless, their power in distillation protocols is limited. Indeed, Eftaxias, Weilenmann, and Colbeck showed the following result:

**Theorem 4.38** (Depth-3 Wiring are Better for Distillation [EWC23a]) — *There exist nonlocal boxes* $\mathbf{P} \in \mathcal{NS}$ *that cannot be distilled using any depth-*2 *wirings but that can using a depth-*3 *wiring (drawn in Figure 3.2).*

This suggests that, in order to distill more boxes and show the collapse of CC for more boxes via wirings, one needs to consider multi-copy wirings, which have the drawbacks of being much more complex and much more difficult to manipulate.

Other results on distillation limitations can be found in [DW08; For11; HR10; LVN14; Sho09].

### 4.3.4 Not Well-Suited to Multipartite Settings

The principle of CC presented in this chapter is defined for two parties Alice and Bob. A trivial extension to the multipartite setting could consist in imposing the principle of CC for every pair of parties, *i.e.* for every bipartition.

Nevertheless, in [Gal+11], Gallego, Würflinger, Acín, and Navascués argue that if the principle of communication complexity—or any other bipartite information principle—is extended in such a trivial way to the multipartite setting, then it cannot perfectly single out the set of quantum correlations. In other words, it means that we would need an intrinsically multipartite principle to possibly characterize multipartite quantum correlations. Here is the formal statement:

**Theorem 4.39** (Need of a Genuine Multipartite Principle [Gal+11]) **—** *Any bipartite principle applied to the bipartition of* $n \geqslant 3$ *parties cannot characterize the set of quantum correlations.*

*Sketch of the Proof.* The authors exhibit tripartite correlations that, on the one hand, fulfill any information principle based on bipartite concepts and, on the other hand, are post-quantum. The family of nonlocal boxes they consider is included in the one called *time-ordered bi-local* (TOBL), also studied in [PBS11], defined as any box $\mathbf{P}(a_1a_2a_3|x_1x_2x_3) \in \mathcal{NS}$ satisfying:

$$\begin{aligned} \mathbf{P}\big(a_1, a_2, a_3 \,\big|\, x_1, x_2, x_3\big) &= \sum_\lambda p_\lambda^{i|jk}\, \mathbf{P}(a_i \,|\, x_i, \lambda)\, \mathbf{P}_{j\to k}(a_j, a_k \,|\, x_j, x_k, \lambda)\,, \\ &= \sum_\lambda p_\lambda^{i|jk}\, \mathbf{P}(a_i \,|\, x_i, \lambda)\, \mathbf{P}_{j\leftarrow k}(a_j, a_k \,|\, x_j, x_k, \lambda)\,, \end{aligned}$$

for all $(i,j,k) \in \big\{(1,2,3),(2,3,1),(3,1,2)\big\}$, where the distributions $\mathbf{P}_{j\to k}$ and $\mathbf{P}_{j\leftarrow k}$ satisfy the following partial non-signaling conditions on their marginals:

$$\begin{aligned} \mathbf{P}_{j\to k}\big(a_j \,|\, x_j, \lambda\big) &= \sum_{a_k} \mathbf{P}_{j\to k}\big(a_j, a_k \,\big|\, x_j, x_k, \lambda\big)\,, \\ \mathbf{P}_{j\leftarrow k}\big(a_k \,|\, x_k, \lambda\big) &= \sum_{a_j} \mathbf{P}_{j\leftarrow k}\big(a_j, a_k \,\big|\, x_j, x_k, \lambda\big)\,. \end{aligned}$$

As indicated by the arrow, the distribution $\mathbf{P}_{j\to k}$ does not forbid signaling from the party $\mathsf{A}_j$ to $\mathsf{A}_k$, *i.e.* it is only a one-way non-signaling distribution, and symmetrically for $\mathbf{P}_{j \leftarrow k}$. These nonlocal boxes $\mathbf{P}$ are compatible with any bipartite information principle since they behave classically under any system bipartition. To show that some of these correlations are not quantum, the authors utilize a tripartite Bell inequality called *guess your neighbor's input* [Alm+10] that is upper bounded by $1$ for quantum correlations but for which the value $\frac{7}{6}$ is achievable using TOBL boxes. ■

**Remark 4.40** (Genuine Multipatite Extensions of Communication Complexity) — Fortunately, it is possible to generalize the principle of CC to the multipartite setting in different manners than the trivial bipartition extension. For instance, in [Buh+99], Buhrman, Dam, Høyer, and Tapp study the multipartite generalization where we search for the minimal number of classical bits that needs to be *broadcasted* by the $n$ parties so that *each* of them get to know the value of $f$. Another example is given by Marcovitch and Reznik in [MR08] where we look for the minimal overall number of classical bits that need to be one-to-one communicated between the parties so that at least one party is able to compute $f$. In particular, in this paper, the authors adapt the protocol from [Bra+06] (Theorem 4.22) in order to find multipartite boxes that collapse CC.

Find an example of a principle that is intrinsically multipartite in Section 4.4.3, namely the principle of *local orthogonality* [Fri+13].

## 4.4 Other Principles

In this section, we provide other examples of information-based principles than CC that have been or are being developed to possibly characterize the set of quantum correlations $\mathcal{Q}$. Other principles not presented below include *no-advantage in nonlocal computation* [ABL09; Lin+07], the *uncertainty* principle [OW10], *local quantum mechanics* [Ací+10; Bar+10b], *mutual information* [Per+21], *nonlocality swapping* [SBP09], and *many-box locality* [Zho+17]. Find a review on some of these principles in [Bru+14]. Nevertheless, to the best of our knowledge, no information-based principle is able to fully characterize the set of quantum correlations to this day.

Below, we present three information-based principles: *information causality*, which is to this day the closest principle to characterize the quantum set (Section 4.4.1), *macroscopic locality* (Section 4.4.2), and *local orthogonality* (Section 4.4.3).

### 4.4.1 Information Causality

The principle of *information causality* (IC) was introduced by Pawłowski, Paterek, Kaszlikowski, Scarani, Winter, and Żukowski in [Paw+09]. Informally, they state it as follows: “The information gain that Bob can reach about a previously unknown to him data set of Alice, by using all his local resources and $m$ classical bits communicated by Alice, is at most $m$ bits.” A review on this topic can be found in [PS15].

**Information Causality Game.** One can view the scenario as an asymmetric nonlocal game with limited classical communication. (Recall the definition of a *nonlocal game* in Section 3.2.1.) When the game begins, Alice receives $N$ bits $a_1, .., a_N \in \{0,1\}$ while Bob receives an integer $b \in \{1, .., N\}$. Alice is allowed to send at most $m$ classical bits of communication to Bob, and then Bob answers a bit $\beta$. We say that Alice and Bob win the information causality game if:

$$\beta = a_b \,,$$

that is if Bob correctly guessed Alice’s $b$-th bit. Of course, as in nonlocal games, the players Alice and Bob can use nonlocal boxes in their strategy to improve their winning probability. Note that the setup is similar to the one of *oblivious transfer* (Section 5.2.4).

**Information Causality Principle.** Using the above notations of the IC game, the IC principle states that any physical theory should satisfy the following inequality:

$$\sum_{k=1}^{N} I(a_k \, : \, \beta \,|\, b = k) \leqslant m \,,$$

where $I(a_k \, : \, \beta \,|\, b = k)$ denotes the Shannon mutual information between $a_k$ and $\beta$, computed under the condition that Bob has received $b = k$. The authors prove the following striking result:

**Theorem 4.41** ([Paw+09]) — *The IC principle is satisfied by all quantum correlations $\mathcal{Q}$ but violated in any non-signaling theory that violates Tsirelon’s bound (Equation (3.12)).*

**A Link with Communication Complexity.** For instance, using **PR** boxes, only $m = 1$ of communication is sufficient for Bob to perfectly compute $a_b$. Indeed, if we consider the string $(y_1, .., y_N)$ with all zeros but $y_b = 1$, then computing $a_b$ amounts to computing the following Boolean function:

$$f\Big(a_1, .., a_N, y_1, .., y_N\Big) := \bigoplus_{k=1}^{N} a_k\, y_k\,,$$

which can be done perfectly with **PR** boxes and one bit of communication using van Dam's protocol [vD99] (Theorem 4.20). Find another way to achieve it via a simulation of *oblivious transfer* with **PR** boxes [WW05].

**Remark 4.42 —** For $m = 0$, the IC principle is equivalent to the non-signaling conditions eqs. (3.4) to (3.6).

Later, it was shown that the IC principle allows one to recover part of the boundary of the quantum set $\mathcal{Q}$ [All+09b]. More recently, an infinite family of Tsirelson-type inequalities was also derived from this principle [Gac+22]. Numerical approaches were also developed to test if a given nonlocal box satisfies the IC principle [MP21]. These two results are even improved in [JGM24], with easier methods to derive Tsirelson-type inequalities, including one that is stronger than Uffink inequality [Uff02]. Moreover, note that a multipartite reformulation of this principle is proposed in [PCR23], and that a quantum variant of IC with quantum communication is proposed in [PG13]. Other results on this topic can be found in [ASS11; Bar+10a; BG13; OT24; XR11; Yan+12; YS22].

The question of whether the IC principle completely characterizes the quantum set $\mathcal{Q}$ is still open to this day.

**Remark 4.43 —** Concerning almost quantum correlations (Theorem 4.29), to the best of our knowledge it is still unknown whether they all satisfy the IC principle. Nevertheless, Navascués, Guryanova, Hoban, and Acín precise that their "numerical results strongly suggest that almost quantum correlations satisfy IC" [Nav+15].

## *4.4.2 Macroscopic Locality*

The principle of *macroscopic locality* (ML) was introduced by Navascués and Wunderlich in [NW09].

**Idea Behind the ML Principle.** In short, the idea is to consider nonlocal boxes with continuous output variables. Alice and Bob are not given one pair of entangled particles (microscopic perspective) but rather a beam of $N \gg 1$ entangled particles (macroscopic perspective). When Alice and Bob interact with particles of the beam, they interact with all of them at the same time. Finally, instead of discrete individual measurements, Alice and Bob obtain the distributions of the intensities of the beams. The principle states that, in the asymptotic regime where $N \to \infty$, the set of marginals should admit a classical description in terms of a local hidden variable model (like $\mathcal{L}$ for usual nonlocal boxes $\mathbf{P}$, see page 60). Such a scenario was also studied before by Bancal, Branciard, Brunner, Gisin, Popescu, and Simon [Ban+08].

**Characterizing the ML Principle.** In the same paper, the authors characterize the ML principle in terms of the first level $\mathcal{Q}^{(1)}$ of the Navascués–Pironio–Acín hierarchy [NPA07] (Section 3.1.3):

> **Theorem 4.44** ([NW09]) — *The non-signaling correlations that satisfy the ML principle are precisely the ones in* $\mathcal{Q}^{(1)}$.

Back in that time, the authors already knew that the set $\mathcal{Q}^{(1)}$ is strictly larger than $\mathcal{Q}$, yielding that the ML principle could not perfectly characterize the quantum set. Nevertheless, this principle is still interesting because on the one hand, it provides an example of a principle close to characterize $\mathcal{Q}$, and on the other hand, it is a testable requirement for a theory to be physical. Interestingly, although not tight for every quantum correlation, this principle still allows one to derive tight analytical Tsirelson-type inequalities bounding the quantum set in some directions [Yan+11].

**Remark 4.45** (Closed Under Wirings) — The authors also show that this set $\mathcal{Q}^{(1)}$ is closed under wirings (Definition 3.16).

**ML is Weaker than IC.** As shown by Cavalcanti, Salles, and Scarani in [CSS10], some of the nonlocal boxes $\mathbf{P} \in \mathcal{NS}$ accepted by the ML principle violate the IC principle (Section 4.4.1). As a consequence, the IC principle is more precise in describing the set of quantum correlations.

### 4.4.3 Local Orthogonality

The principle of *local orthogonality* (LO) was introduced by Fritz, Sainz, Augusiak, Brask, Chaves, Leverrier, and Acín in [Fri+13].

**Defining the LO Principle.** Consider the $(n, N, M)$ scenario, with $n$ parties, $N$ inputs, and $M$ outputs. Nonlocal boxes $\mathbf{P} \in \mathcal{NS}$ are of the following form:

$$\mathbf{P}\big(a_1, .., a_n \,\big|\, x_1, .., x_n\big)\,,$$

where $a_i \in \{1, .., M\}$ and $x_i \in \{1, .., N\}$. Two events $e := \big(a_1, .., a_n \,\big|\, x_1, .., x_n\big)$ and $e' := \big(a'_1, .., a'_n \,\big|\, x'_1, .., x'_n\big)$ are said to be *locally orthogonal* if they involve different outputs of the same measurement by at least one party; that is, for some $i$ we have $a_i \neq a'_i$ whereas $x_i = x'_i$. A collection of events $\{e_j\}_j$ is said to be *locally orthogonal* if the events are pairwise orthogonal. Now, the principle of local orthogonality states that a nonlocal box $\mathbf{P} \in \mathcal{NS}$ is physical if, for any set of locally orthogonal events $\{e_j\}_j$, we necessarily have:

$$\sum_j \mathbf{P}(e_j) \leqslant 1\,.$$

**Theorem 4.46** ([Fri+13]) **—** *In the bipartite scenario ($n = 2$), the LO principle is equivalent to non-signaling conditions (*i.e.* no box in $\mathcal{NS}$ violates the LO principle), but some bipartite post-quantum boxes turn out to violate the principle when distributed among several parties.*

*Moreover, in the multipartite setting, this principle reveals the non-quantumness of certain correlations for which any bipartite information-based principle fails.*

**Characterization in Terms of Nonlocal Games.** The authors characterize the LO principle in terms of a *distributed guessing problem* (DGP) game. The Referee samples a vector $(\tilde{a}_1, .., \tilde{a}_n)$ in $\{1, .., M\}^n$ uniformly at random, and applies a publicly-known function $f : \{1, .., M\}^n \to \{1, .., N\}^n$ to it so that he gets a new vector $(x_1, .., x_n)$. He provides each of the parties $\mathsf{A}_i$ with the value $x_i \in \{1, .., N\}$, they do their strategy, and each of them answers a value $a_i \in \{1, .., M\}$. The players win the game if all of them manage to perfectly guess their initial input, *i.e.* if:

$$\forall i \in \{1, .., n\}\,, \qquad a_i = \tilde{a}_i\,.$$

For some choice of function $f$, the best classical winning probability is the random guess, *i.e.* with probability $1/M^n$, in which case the inputs $x_i$ do not bring any useful information to the players. This case is said to be *maximally difficult*. The authors prove that a DGP game is maximally difficult *if, and only if,* its winning probability yields an LO inequality. In other words, in order for the players to win those DGP games with probability greater than $1/M^n$, they need to share a nonlocal box $\mathbf{P}$ that violates LO. (Note that in these games, quantum correlations do not provide any advantage over the trivial random guessing strategy.)

# Chapter 5

# Quantum Cryptography

In this chapter, we continue introducing the necessary background. We begin with fundamental concepts of cryptography, then explore quantum cryptographic constructions, and finally focus on a quantum primitive known as the unclonable bit.

## Chapter Contents

# 5.1 Basics of Cryptography

*Cryptography* is the study of information processing and communication in the presence of adversaries, with *security* as a central concern.

In this section, we first introduce encryption schemes (Section 5.1.1). Then, we present two key security notions, namely perfect security (Section 5.1.2) and computational security (Section 5.1.3). Our presentation follows the formalism of Katz and Lindell [KL20].

## 5.1.1 Encryption Scheme

An *encryption scheme* is defined by three algorithms $\mathsf{Gen}$, $\mathsf{Enc}$, and $\mathsf{Dec}$, together with three finite sets $\mathcal{K}$, $\mathcal{M}$, and $\mathcal{C}$. The *key-generation algorithm* $\mathsf{Gen}$ generates a random key $k \in \mathcal{K}$ according to some distribution, generally the uniform distribution over the *key set* $\mathcal{K}$. For now, we do not need any input in the algorithm $\mathsf{Gen}$, so we can say that the input is always the bit $1$. To emphasize the randomness of the process, instead of writing an equality, we write:

$$k \leftarrow \mathsf{Gen}(1)\,.$$

Once the key is generated, one can encrypt a message $m \in \mathcal{M}$, that can be sampled randomly but independently of the key $k$, where $\mathcal{M}$ is the *message set*. To this end, we use the probabilistic algorithm $\mathsf{Enc}$ called *encryption algorithm* which, on input a key $k \in \mathcal{K}$ and a message, returns a *ciphertext* $c \in \mathcal{C}$, where $\mathcal{C}$ is the *ciphertext set*. To emphasize the randomness of the process, we may write again:

$$c \leftarrow \mathsf{Enc}_k(m)\,.$$

Then, the encrypted message is sent from the *sender* Alice ($\mathsf{A}$) to the *receiver* Bob ($\mathsf{B}$) and can be decrypted using the *decryption algorithm*, which on input a key $k \in \mathcal{K}$ and a ciphertext $c \in \mathcal{C}$ returns a message $m' \in \mathcal{M}$. This algorithm may be assumed deterministic without loss of generality, and we write:

$$m' = \mathsf{Dec}_k(c)\,.$$

All encryption schemes are implicitly assumed to be correct. Here is a first definition of correctness:

**Definition 5.1** (Perfect Correctness) — *We say that an encryption scheme* $(\mathsf{Gen}, \mathsf{Enc}, \mathsf{Dec})$ *has* perfect correctness *if the decryption of an encrypted message is the message itself with probability one:*

$$\forall m \in \mathcal{M}\,, \qquad \mathbb{P}\Big(\mathsf{Dec}_k\big(\mathsf{Enc}_k(m)\big) = m \,\Big|\, k \leftarrow \mathsf{Gen}(1)\Big) \;=\; 1\,.$$

### *5.1.2 Perfect Security*

We assume that both the sender and the receiver are honest, but in between, there may be an *eavesdropper*, often called Eve (E), reading the ciphertext $c$ transferred between the parties. This adversary also knows the three algorithms $(\mathsf{Gen}, \mathsf{Enc}, \mathsf{Dec})$ of the encryption scheme and even the probability that the sender chooses a message $m$ in $\mathcal{M}$, but does not know the secret key $k$ (this is called the *Kerckhoffs' principle* [Ker83]). By reading $c$, we want the eavesdropper to learn no additional information about the message $m$, which is why we have the following first definition of security:

**Definition 5.2** (Perfect Security) — *We say that an encryption scheme* $(\mathsf{Gen}, \mathsf{Enc}, \mathsf{Dec})$ *is* perfectly secure *if for all messages* $m, m' \in \mathcal{M}$ *and all ciphertext* $c \in \mathcal{C}$*, we have:*

$$\mathbb{P}\Big(\mathsf{Enc}_k(m) = c \,\Big|\, k \leftarrow \mathsf{Gen}(1)\Big) \;=\; \mathbb{P}\Big(\mathsf{Enc}_k(m') = c \,\Big|\, k \leftarrow \mathsf{Gen}(1)\Big)\,.$$

One can show that perfect security is equivalent to *perfect indistinguishability*. In perfect indistinguishability, Eve specifies two messages $m_0$ or $m_1$ of her choice in $\mathcal{M}$. Then Alice samples one of them uniformly at random, denoted $m_{\mathsf{A}}$, encrypts it into a ciphertext $c$, and Eve tries to guess $m_{\mathsf{A}}$ from seeing $c$. Eve's guess is denoted $m_{\mathsf{E}}$ and her best winning probability of correctly guessing $m_{\mathsf{A}}$ is at least $\frac{1}{2}$ since this corresponds to the uniform random guess. We say that the encryption scheme $(\mathsf{Gen}, \mathsf{Enc}, \mathsf{Dec})$ is *perfectly indistinguishable* if this is exactly Eve's best winning probability:

$$\mathbb{P}\Big(m_{\mathsf{E}} = m_{\mathsf{A}} \,\Big|\, k \leftarrow \mathsf{Gen}(1),\, c \leftarrow \mathsf{Enc}_k(m_{\mathsf{A}})\Big) \;=\; \frac{1}{2}\,. \tag{5.1}$$

As said above, these two notions are equivalent:

$$\text{Perfect security} \quad \Longleftrightarrow \quad \text{Perfect indistinguishability}\,. \tag{5.2}$$

We stress that no limitations are placed on the computational power of Eve, so this security notion is very strong—we will relax it below to have more concrete applications. But before, here is a celebrated example of an encryption protocol satisfying perfect security (and perfect correctness):

**Example 5.3** (One-Time Pad) — The *one-time pad* protocol is renowned to be one of the strongest since if the keys are used properly, they *cannot* be broken, even in theory. Its invention is often credited to Vernam [Ver26], who filed a patent on it, but recent historical studies [Bel11] show that it was actually used before, in the 19th century, by the banker Miller for telegraphic code. But the proof of perfect security came later with the ground-breaking work of Shannon [Sha49]. For this protocol, consider $\mathcal{M} = \mathcal{K} = \mathcal{C} = \{0,1\}^\ell$ sets of binary strings of length $\ell$. The key-generation algorithm Gen generated $k$ uniformly at random in $\mathcal{K}$. The encryption algorithm Enc, on inputs $k \in \mathcal{K}$ and $m \in \mathcal{M}$, outputs $c := k \oplus m \in \{0,1\}^\ell$ deterministically, where the symbol "$\oplus$" denotes the sum modulo $2$ of each bit in the string (the XOR operation). Finally, the decryption algorithm Dec, on inputs $k \in \mathcal{K}$ and $c \in \mathcal{C}$, produces $m' := k \oplus c$. It is perfectly correct ($m' = k \oplus k \oplus m = m$) and perfectly secure because the ciphertexts are uniformly distributed regardless of what message is encrypted.

Although powerful, this protocol is impractical for several reasons. As for any encryption scheme satisfying perfect security, the key $k$ has to be at least as long as the message and, as the name one-time pad suggests, it should not be reused because the sum of two encrypted messages $m, m'$ (of the same length) with the same key $k$ gives a lot of information:

$$c \oplus c' = (m \oplus k) \oplus (m' \oplus k) = m \oplus m' .$$

This is the reason why we rather use *computation security*, see below. Nevertheless, although inconvenient, this protocol was historically used between nation-intelligence agencies. For instance, the "red phone" linking the White House to the Kremlin during the Cold War used one-time pad encryption, thus requiring exchanging and storing extremely large keys regularly.

### *5.1.3 Computational Security*

The idea behind *computational security* is to only assume that the adversaries have limited computational power (they are "efficient") and have a negligible probability of breaking the scheme–in contrast to perfect security where the adversaries are unbounded and cannot break the scheme. Here, the approach is *asymptotic* and relies on a *security parameter* $\lambda \in \mathbb{N}$ that regulates the efficiency of the adversaries and upper bounds their winning probability. Efficiency is described in terms of PPT algorithms [Gil77]:

**Definition 5.4** (PPT, Efficiency) — *An algorithm is said* probabilistic polynomial-time *(*PPT*) if it can be written as a probabilistic Turing machine in polynomial time in the parameter* $\lambda$*, with an error probability of less than* $\frac{1}{2}$ *for all instances. An adversary is said to be* efficient *if it is restricted to using* PPT *algorithms only.*

For more details on computational complexity, we refer to [AB09]. We also need the notion of negligible function to later bound the winning probabilities:

**Definition 5.5** (Negligible Function) — *A function* $f : \mathbb{N} \to \mathbb{R}_{\geqslant 0}$ *is said to be* negligible *if it decreases asymptotically faster than the inverse of any (positive) polynomial:*

$$\forall p \in \mathbb{N}\,,\ \exists \Lambda > 0\,,\ \forall \lambda \geqslant \Lambda\,, \qquad f(\lambda) \leqslant \frac{1}{\lambda^p}\,.$$

*In such a case, the function* $f(\lambda)$ *is denoted* $\mathrm{negl}(\lambda)$.

Note that summing two negligible functions $\mathrm{negl}_1(\lambda)+\mathrm{negl}_2(\lambda)$ or multiplying it with a (positive) polynomial $P(\lambda)\cdot\mathrm{negl}(\lambda)$ yields again a negligible function.

**Definition 5.6** (Private-Key Encryption Scheme) — *A* private-key encryption scheme *is a tuple of* PPT *algorithms* $(\mathsf{Gen}, \mathsf{Enc}, \mathsf{Dec})$ *such that:*

- $\mathsf{Gen}$ *is the* key-generation algorithm *that, on input the string with* $\lambda$ *ones* $1^\lambda := (1, .., 1)$*, outputs a key* $k \in \mathcal{K}$ *in a randomized way:*

$$k \leftarrow \mathsf{Gen}(1^\lambda)\,;$$

- $\mathsf{Enc}$ *is the* encryption algorithm *that, on inputs* $k \in \mathcal{K}$ *and* $m \in \mathcal{M}$*, outputs a ciphertext* $c \in \mathcal{C}$ *in a randomized way:*

$$c \leftarrow \mathsf{Enc}_k(m)\,;$$

- Dec *is the* decryption algorithm *that, on inputs* $k \in \mathcal{K}$ *and* $c \in \mathcal{C}$*, outputs a message* $m' \in \mathcal{M}$ *or an error* $\perp$ *in a deterministic way:*

$$m' = \mathsf{Dec}_k(c)\,.$$

We present a quantum version of it in Section 5.3.1. Perfect correctness is again implicitly required (Definition 5.1), or a weaker version depending on the context:

$$\forall m \in \mathcal{M}\,, \qquad \mathbb{P}\Big(\mathsf{Dec}_k\big(\mathsf{Enc}_k(m)\big) = m \,\Big|\, k \leftarrow \mathsf{Gen}(1^\lambda)\Big) \geqslant 1 - \mathrm{negl}(\lambda)\,.$$

A variant of this scheme is *public-key encryption scheme*. The idea is broadly the same but there Gen outputs two keys on the receiver's side, a public key $pk$ and a private key $sk$, where $pk$ is sent to the sender and used for the encryption, and $sk$ is kept for the decryption. This is particularly useful for securing large-scale communication like nowadays on the Internet.

Now, the notion of security is a relaxation of the one presented in eqs. (5.1) and (5.2) related to indistinguishability:

**Definition 5.7** (Indistinguishability Security) — *We say that a private-key encryption scheme* (Gen, Enc, Dec) *is* indistinguishable secure *if any efficient adversary can exceed the random guess winning probability only by a negligible term:*

$$\forall \lambda \in \mathbb{N}\,, \quad \mathbb{P}\Big(m_{\mathsf{E}} = m_{\mathsf{A}} \,\Big|\, k \leftarrow \mathsf{Gen}(1^\lambda),\, c \leftarrow \mathsf{Enc}_k(m_{\mathsf{A}})\Big) \leqslant \frac{1}{2} + \mathrm{negl}(\lambda)\,.$$

As a consequence of this definition, it can be shown that no bit of an encrypted string can be learned by an efficient adversary with more than a negligible probability. It means that a ciphertext $c$ leaks no information about individual bits of the plaintext $m$. Moreover, it can be shown that indistinguishability security is equivalent to another type of security called *semantic security*, which roughly protects against an efficient eavesdropper learning any information about the plaintext message $m$ from the ciphertext $c$, which is what we want. This shows that this notion of security is strong enough for what an encryption scheme should guarantee. We refer to [KL20] for more details.

## 5.2 Some Quantum Cryptographic Constructions

In quantum cryptography, the goal is to perform cryptographic tasks similar to those in classical cryptography, but with the added advantages of quantum mechanics, such as the No-Cloning Theorem (Theorem 2.37) and the measurement disturbance property (Postulate 2.17).

In this section, we present four quantum cryptographic constructions, namely conjugate coding (Section 5.2.1), quantum money (Section 5.2.2), quantum key distribution (Section 5.2.3), and quantum oblivious transfer and bit commitment (Section 5.2.4). Our presentation follows the review by Broadbent and Schaffner [BS16], with additional references to [VW23] for a recent book on the subject.

### 5.2.1 *Conjugate Coding*

Introduced by Wiesner in his influential paper [Wie83], *conjugate coding* is a fundamental elementary principle widely used in many quantum cryptographic constructions. It consists of encoding classical information in non-commuting bases (more precisely, mutually unbiased bases) to leverage the incompatibility of measurement and uncertainty principle (Remark 2.18). Typically, for qubits in $\mathbb{C}^2$ we employ the $\sigma_z$-basis (computational basis) and the $\sigma_x$-basis:

$$\Big\{|0\rangle, |1\rangle\Big\} \qquad \text{and} \qquad \Big\{|+\rangle, |-\rangle\Big\}.$$

They are called *conjugate bases*. Given a classical plaintext bit $m \in \{0, 1\}$, Alice selects randomly one of the two bases and encodes the bit into a qubit with the following procedure Enc:

| Classical Bit $m$ | Encoding in the $\sigma_z$-Basis | Encoding in the $\sigma_x$-Basis |
|---|---|---|
| 0 | $\vert 0\rangle$ | $\vert +\rangle$ |
| 1 | $\vert 1\rangle$ | $\vert -\rangle$ |

Although simplistic in appearance, this protocol has the powerful property that measuring in the correct basis gives a deterministic result with the plaintext message $m$, whereas measuring in the wrong one gives a uniformly random bit independent of $m$ (see computations in Example 2.21).

Hence, the choice of the basis can be used as classical key $k$, and measurement in the appropriate basis can be used for the decryption protocol Dec for Bob.

The security of this protocol relies on two observations. On the one hand, the Quantum No-Cloning Theorem (Theorem 2.37) prevents an eavesdropper Eve from perfectly copying an unknown state. On the other hand, suppose Alice encodes a plaintext message $m$ with several bits. If Eve intercepts part of it, she may want to measure it. However, she does not have access to the basis choice (key) so she has to make random choices. Then, if at least one basis choice is wrong, the collapse of the way packet (Postulate 2.17) introduces measurement disturbance and, upon receiving the message and applying a check protocol, Bob can detect the errors. This protocol is crucial for instance for *quantum money* and *quantum key distribution*, see below, as well as for *universal quantum computing* [Bar+12b; BFK09].

### 5.2.2 Quantum Money

Building on conjugate coding, *quantum money* is one of the first applications of quantum cryptography, presented in the same paper by Wiesner [Wie83] who was a lot ahead of his time—the original ideas date to 1968 and took several years to be accepted as a formal publication. The idea is to encode classical banknotes (a bit string) into quantum states via conjugate encoding and leverage the Quantum No-Cloning Theorem (Theorem 2.37) to prevent counterfeiting. The quantum state is encoded by the (trusted) bank employee using random bases, representing the key $k$. When a customer has a quantum banknote and wants to use it, they can go to the bank and measure it in the correct basis with the employee to certify its authenticity.

Now, several variants and extensions of this protocol exist. To name but a few, there is a public-key version from Bennett, Brassard, Breidbart, and Wiesner allowing verification without contacting the issuing bank but with additional computational assumptions [Ben+83] (see also [Aar09; AC12; Far+12]), a proof of security of the private-key multiple qubit by Molina, Vidick, and Watrous based on semi-definite programming [MVW13], a variant with classical interactions with the bank to authenticate the state [Gav12; MVW13], and a noise-tolerant variant [Pas+12].

### 5.2.3 Quantum Key Distribution (QKD)

Arguably the most successful protocol in quantum cryptography, *quantum key distribution* (QKD) aims to distribute keys between two parties in a secure way using quantum mechanics. We present below the two main variants of this protocol, namely the *prepare-and-measure* QKD, also called "BB84," from Bennett and Brassard [BB84], and the *entanglement-based* QKD, called "E91," from Ekert [Eke91]. Find surveys on this topic for instance in [BC96; Ben92; Bru+07; Feh10; Gis+02].

**Prepare-and-Measure QKD.** In the BB84 variant, Alice encodes a string of bits $m$ with a *conjugate coding* (see above) and sends the qubits to Bob. Then, for each qubit, Bob measures it in a random basis (either $\sigma_z$- or $\sigma_x$-basis) and keeps secret the outcomes. After, both Alice and Bob publicly announce their choices of bases and discard the qubits for which they had different bases. The outcomes of the remaining qubits on Bob's side should allow him, in theory, to exactly retrieve the corresponding bits of Alice's plaintext $m$. Nevertheless, they verify is there was an eavesdropper by detecting errors: Alice and Bob publicly compare a random subset of the remaining bits. If all of them match (or most of them in practice), then they proceed, otherwise they detect eavesdropping and abort the process. Finally, after discarding these error-detecting bits, they may correct small errors due to noise and amplify privacy by shortening again the string. After this process, this string can serve as a private key for another protocol (like a one-time pad). This protocol was experimentally confirmed in [Ben+92].

**Entanglement-Based QKD.** In the E91 variant, Alice and Bob need to share several copies of a maximally entangled state beforehand:

$$|\Omega\rangle = \frac{|00\rangle + |11\rangle}{\sqrt{2}},$$

one qubit for Alice and one for Bob for each pair. On each of her qubits, Alice performs a measurement in a random basis among $B_0$, $B_{\pi/8}$, and $B_{\pi/4}$, and similarly for Bob in $B_0$, $B_{\pi/8}$, and $B_{-\pi/8}$, where $B_\theta$ denotes the rotated computational basis with angle $\theta$:

$$B_\theta := \left\{ \cos(\theta)\,|0\rangle + \sin(\theta)\,|1\rangle,\ -\sin(\theta)\,|0\rangle + \cos(\theta)\,|1\rangle \right\}.$$

Then, they reveal publicly their choices of bases while keeping secret the outcomes. They discard the qubits for which they have different bases and try to detect an eavesdropper by analyzing the correlation of a subset of their outcomes: they check for violations of Bell's inequalities (eq. (3.11)). If Eve tries to intercept some qubits, she introduces some detectable decoherence and breaks the expected quantum correlation. Finally, as before, if the qubits pass Bell's inequality test, Alice and Bob keep only the non-public bits and perform error correction and privacy amplification to extract a shared secret key that can be used in another protocol.

**Comparison Between BB84 and E91.** These two variants have strengths and weaknesses. On the one hand, security is based on measurement disturbance for BB84 protocol, while on Bell's inequalities violation for E91, which makes it more resistant to side-channel attacks. Moreover, the latter can be implemented over larger distances (even over 1,200 kilometers! [Yin+17]), while the former suffers from photon loss in optical fibers and therefore has limited communication distance. On the other hand, the former (BB84) is easier to implement since it does not require to pre-share several entangled bits.

**Other Variants.** Other variants exist, including an entanglement-based BB84 protocol [BBM92], a device-independent version [Ací+07; AGM06; BHK05], or a "twin fields" version [Luc+18]. Security of QKD was proved in various ways, for instance using quantum error correction [SP00], exploiting the symmetries of the protocol [Ren08], or based on the complementarity of the measurements [Koa09]—see [TL17] for a comprehensive analysis of QKD security.

### 5.2.4 Quantum Oblivious Transfer & Bit Commitment

*Oblivious transfer* (OT) and *bit commitment* (BC) are basic yet significant primitives in cryptography. While OT implies BC but not the converse in the classical setting, both directions hold in the quantum setting. Below, we present OT, BC, and finally the quantum OT.

**Oblivious Transfer (OT).** Oblivious transfer was introduced by Wiesner and later developed under two equivalent variants by Rabin [Rab81] and Even, Goldreich, and Lempel [EGL85]. Let us describe the aim of this protocol. Alice generates and sends two messages $m_0$ and $m_1$ to Bob, but Bob receives only $m_b$ where $b \in \{0, 1\}$ is a bit of Bob's choice unknown by Alice. Security for Alice against dishonest Bob guarantees that Bob receives only one of the two messages, while security for Bob against malicious Alice ensures that $b$ is unknown by Alice.

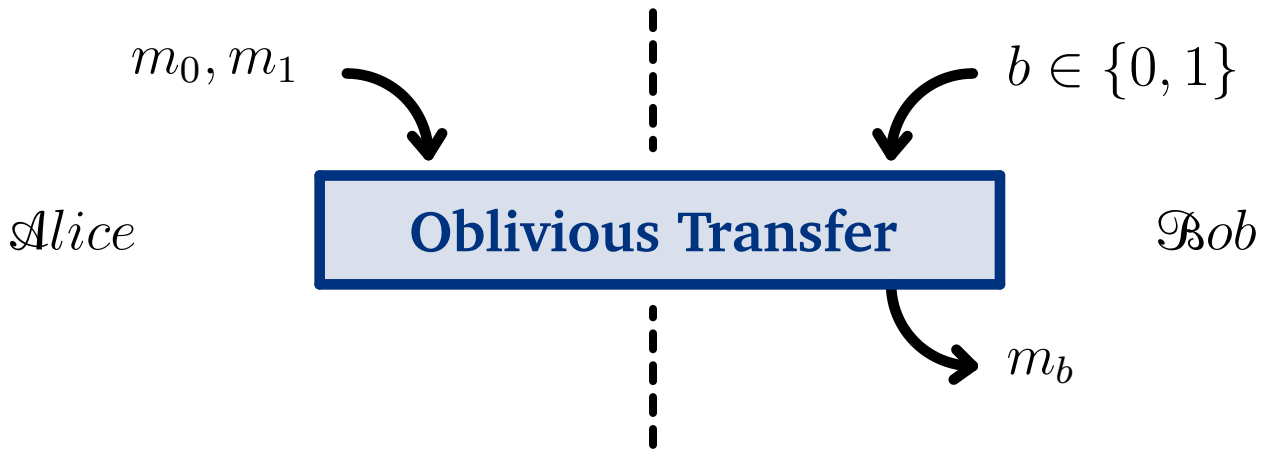

This protocol has significant applications. Notably, it is *universal* for secure two-party computations [Kil88]: any Boolean function $f(x, y)$ can be securely computed using several instances of an OT protocol, where $x$ and $y$ are inputs of Alice and Bob respectively, where each party remains unaware of the other party's input beyond what can be inferred from the result. If such a protocol exists, one can solve *Yao's millionaire's problem* [Yao82]: using the function $f(x, y) = \mathbb{1}_{x \geqslant y}$, two millionaires can compare their fortune without telling the other how much money they own. Note that unconditional OT can be achieved from a single **PR** box, and vice versa [WW05].

**Bit commitment (BC).** Bit commitment was introduced by Blum [Blu83] and is, at first sight, slightly different. Here, Alice wants to commit a bit $a$ to Bob in such a way that Bob can read it only once Alice wants it (*hiding* condition), which guarantees security on Alice's side. Nevertheless, Alice should not be able to modify her bit $a$ once committed (*binding* condition), ensuring Bob's security. Formally, Alice encrypts $a$ in $c = \mathsf{Enc}_k(a)$ with some key $k$, gives the ciphertext $c$ to Bob, and reveals at a later time the bit $a$ and the key $k$ so that Bob can check that indeed $\mathsf{Enc}_k(a) = c$. For a real-world representation, one can think of a safe in which Alice deposits the value of her bit $a$, locks it, and gives it to Bob without the key. When Alice is ready to reveal the bit $a$, she gives Bob $a$ and the details of her encrypting method, so that Bob can create another safe with $a$ inside, and compare the two safes to check if Alice is honest.

**BC implies Quantum OT.** Based on BC, the quantum version of OT was then developed by Bennett, Brassard, Crépeau, and Skubiszewska [Ben+91]. Suppose $m_0, m_1 \in \{0,1\}$ and Bob wants to know $m_b$ with $b \in \{0,1\}$. Preliminarily, Alice prepares a conjugate coding encryption of a string $x \in \{0,1\}^n$ into qubits where the key $k$ is the choice of bases and sends the qubits to Bob. Then, Bob measures each qubit in random conjugate bases $k'$ and obtains a string $x' \in \{0,1\}^n$. In the next step, Alice will reveal her key $k$, so before doing it, she needs to force Bob to indeed perform measurements. This is where BC is used: Bob commits his bases $k'$ and outcomes $x'$ to Alice and she checks a fraction of these commitments before sending the key $k$. After this, Alice reveals the key $k$ and Bob denotes $I_b$ the set of all matching indices, $I_{b\oplus 1}$ the other, so that he obtains a partition of the indices:

$$I_0 \cup I_1 = \{1, .., n\} .$$

Then, Bob informs Alice of $(I_0, I_1)$ in this fixed order (not to reveal $b$), and she uses two hash functions $f_0, f_1 : \{0,1\}^* \to \{0,1\}$ to define:

$$s_i := f_i\big(x|_{I_i}\big) \oplus m_i ,$$

where $x|_I$ denotes the substring of $x$ with bit indices in $I$. Finally, Alice sends both $(s_0, s_1)$ and $(f_0, f_1)$ to Bob, and Bob retrieves $m_b$ by computing:

$$f_b\big(x'|_{I_b}\big) \oplus s_b = m_b ,$$

without knowing $m_{b\oplus 1}$. The security of this protocol was formally proved by Damgård, Fehr, Lunemann, Salvail, and Schaffner [Dam+09], following which Unruh established that this protocol is to BC in the quantum universally composable model [Unr10]. Nevertheless, as explained in the next paragraph, unconditionally secure quantum bit commitment cannot exist.

**Limitation: No Perfect Quantum BC.** Despite increasing attention and excitement in the early 90s [BC90a; Bra+93], quantum bit commitment was shown to be impossible in the unconditionally secure regime by Mayers [May97] and Lo and Chau [LC97]. See also [Bra+97; Chi+13; Win+11].

The core idea is that Alice can cheat and use quantum entanglement and superposition to break the binding condition. For instance, instead of committing only $|0\rangle$ or $|1\rangle$, Alice can prepare a superposition of both:

$$|\psi\rangle = |0\rangle_{\mathsf{A}} \otimes |\phi_0\rangle_{\mathsf{B}} + |1\rangle_{\mathsf{A}} \otimes |\phi_1\rangle_{\mathsf{B}} ,$$

for which Bob reduced state $\mathrm{Tr}_{\mathsf{A}}(|\psi\rangle\langle\psi|)$ is independent of Alice's actual commitment to guarantee hiding. Then, even if Alice had at first the intention to commit $|0\rangle$, she can lie when it is time to reveal the commit and perform a local measurement $|1\rangle\langle 1|$ to match with the commitment. More generally, if there is a quantum bit commitment protocol with *perfect hiding* property, Bob cannot distinguish Alice's commitments to 0 and 1. So, by Uhlmann's theorem about quantum state indistinguishability [Uhl76], the states that Alice commits must be unitarily related. As a consequence, Alice can switch arbitrarily many times her commitment without being detected by Bob, hence breaking the binding condition.

Nevertheless, other lines are explored, showing that quantum bit commitment is possible under assumptions like quantum-secure one-way functions [CLS01; DMS00], relativistic quantum cryptography where Alice and Bob use special relativity to prevent cheating [Ken99], or in the random oracle model [Unr16].

## 5.3 Unclonable Bit

The *unclonable bit* is a quantum primitive that enables a sender, Alice, to encode a single bit $m$ into a quantum state in such a way that it cannot be cloned, making it impossible to retrieve the plaintext $m$ in multiple locations once the key is revealed. Its existence in the plain model, *i.e.* without assumptions on the adversaries, remains an open question within the strong security regime.

In this section, we first formalize quantum encryption of classical messages schemes (Section 5.3.1). We then present this open question through no-cloning games (Section 5.3.2) and describe a connection with monogamy-of-entanglement games (Section 5.3.3). These concepts will be further explored in Chapter 8 in relation to [Bot+24b]. For more details on this topic, we refer to [BL20; Cul22].

### *5.3.1 Quantum Encryption of a Classical Message*

*Quantum encryption of classical messages* (QECM) schemes are encryption protocols with classical plaintext $m \in \{0,1\}^*$, classical key $k \in \{0,1\}^*$ and quantum ciphertext $\rho \in \mathcal{D}(\mathcal{H})$. Preliminarily introduced by Goldreich in [Gol04], this notion is rephrased and much developed by Broadbent and Lord in [BL20]. See also [Lor19]. Below, after defining efficient quantum circuits, we present the definition of the scheme and mention three related notions of security.

**Efficient Circuit.** First, here, the three functions $(\mathsf{Gen}, \mathsf{Enc}, \mathsf{Dec})$ of an encryption scheme are polynomial-time uniform quantum circuits, no longer PPT algorithms like in Section 5.1. We recall the definition and refer to [AB09] for more details on complexity theory:

> **Definition 5.8** (Polynomial-Time Circuit) — *A* polynomial-time (uniform quantum) circuit *is a collection of quantum circuits $\{C_\lambda\}_{\lambda\in\mathbb{N}}$ such that, for any $\lambda \in \mathbb{N}$, there is a deterministic polynomial-time Turing machine $T$ which, on input $1^\lambda$, produces a description of $C_\lambda$.*

**Encryption Scheme.** Quantum encryption of classical messages is formalized as follows:

> **Definition 5.9** (QECM) — *Let $\lambda \in \mathbb{N}$ be the security parameter. A* quantum encryption of classical messages *(*QECM*) scheme is a tuple of polynomial-time circuits* $(\mathsf{Gen}, \mathsf{Enc}, \mathsf{Dec})$ *such that:*
>
> - $\mathsf{Gen}_\lambda : \mathcal{D}(\mathbb{C}) \to \mathcal{D}(\mathcal{H}_{K,\lambda})$ *is the key-generation circuit;*
> - $\mathsf{Enc}_\lambda : \mathcal{D}(\mathcal{H}_{K,\lambda} \otimes \mathcal{H}_{M,\lambda}) \to \mathcal{D}(\mathcal{H}_{C,\lambda})$ *is the encoding circuit;*
> - $\mathsf{Dec}_\lambda : \mathcal{D}(\mathcal{H}_{K,\lambda} \otimes \mathcal{H}_{C,\lambda}) \to \mathcal{D}(\mathcal{H}_{M,\lambda})$ *is the decoding circuit;*
>
> *where the plaintext space $\mathcal{H}_{M,\lambda}$, the key space $\mathcal{H}_{K,\lambda}$, and the ciphertext space $\mathcal{H}_{C,\lambda}$ have polynomial dimension $\ell(\lambda)$, $p(\lambda)$, and $q(\lambda)$ respectively.*

Note that $\mathcal{D}(\mathbb{C})$ in the definition of $\mathsf{Gen}$ is actually set singleton $\{1\}$, meaning that this circuit takes no input. To obtain the classical key $k$ from the circuit $\mathsf{Gen}$, one simply needs to measure the resulting state $\mathsf{Gen}(1)$ in the computational basis. A classical message $m \in \{0,1\}^{\ell(\lambda)}$ is naturally seen as the pure state $|m\rangle\langle m|$. The correctness condition is expressed as follows:

**Definition 5.10** (Correctness) — *For all security parameters* $\lambda \in \mathbb{N}$*, all messages* $m \in \{0,1\}^{\ell(\lambda)}$*, all keys* $k \in \{0,1\}^{p(\lambda)}$ *that are possibly generated by* $\mathsf{Gen}$*, i.e. such that* $\mathrm{Tr}\big(|k\rangle\langle k|\,\mathsf{Gen}_\lambda(1)\big) > 0$*, a QECM scheme* $(\mathsf{Gen}, \mathsf{Enc}, \mathsf{Dec})$ *is said to be* correct *if we have:*

$$\mathrm{Tr}\left[|m\rangle\langle m|\,\mathsf{Dec}_\lambda\Big(|k\rangle\langle k| \otimes \mathsf{Enc}_\lambda\big(|k\rangle\langle k| \otimes |m\rangle\langle m|\big)\Big)\right] = 1\,.$$

**Securities.** For such protocols, Broadbent and Lord [BL20] introduce and study three types of security:

- *indistinguishable security*[1], which generalizes the standard indistinguishability condition (Definition 5.7), and where the sender wants to prevent the adversaries from distinguishing two encrypted messages with high probability;
- *unclonable security*, where the sender wants to prevent the adversaries to clone the encrypted message;
- *unclonable-indistinguishable security*, which is, to some extent, a mix of the former two variants. This is the security notion that we study more in detail in Section 5.3.2.

Moreover, Broadbent and Lord show that the security variants relate a follows [BL20]:

$$\text{0-unclonable security} \implies \text{unclonable-indistinguishable security} \implies \text{indistinguishable security}\,,$$

where the first implication holds for constant-size message schemes. The authors also present an unclonable secure protocol assuming quantum random oracle models.

[1]This first variant is not completely novel, there is a comparable definition in [Ala+16, Def. 7] for quantum keys, quantum messages, and quantum ciphertexts. The other two variants are from [BL20].

### 5.3.2 Security via No-Cloning Games

As mentioned in Section 3.3.4, a useful application of nonlocal games in quantum cryptography is to define security notions. Here, we introduce *unclonable-indistinguishable security* through a family of games called *no-cloning games*, following the original idea from [BL20]. This gives rise to the open question of the existence of the *unclonable bit*, also presented below.

**No-Cloning Games.** Here, in contrast with Section 3.2, the Referee is rather called *challenger* and named Alice (A). The players form an adversary team, composed of a Pirate (P), Bob (B), and Charlie (C), according to the following procedure:

**Definition 5.11** (No-Cloning Games, Figure 5.1) — *Consider a QECM scheme* $(\mathsf{Gen}, \mathsf{Enc}, \mathsf{Dec})$ *and a security parameter* $\lambda$. *The associated* no-cloning game *involves a challenger* A *and three adversaries* $(\mathsf{P}, \mathsf{B}, \mathsf{C})$ *and is defined by the following procedure:*

*(1) A challenger* A *generates a key* $k \leftarrow \mathsf{Gen}_\lambda(1)$ *and a message* $m \in \{0, 1\}$ *uniformly at random, and sends the quantum state* $\rho \leftarrow \mathsf{Enc}_\lambda(m, k)$ *to* P.

*(2) The adversary* P *applies a quantum channel* $\Phi : \mathcal{B}(\mathcal{H}_\mathsf{A}) \to \mathcal{B}(\mathcal{H}_\mathsf{B} \otimes \mathcal{H}_\mathsf{C})$ *on the state* $\rho$ *to obtain the bipartite state* $\Phi(\rho)$, *and sends to* B *and* C *their respective register.*

*(3) The adversaries* B *and* C *receive the secret key* $k$ *and measure their state using two* POVMs $\{B_{m_\mathsf{B}|k}\}_{m_\mathsf{B}}$ *and* $\{C_{m_\mathsf{C}|k}\}_{m_\mathsf{C}}$, *to output* $m_\mathsf{B}, m_\mathsf{C} \in \{0, 1\}$.

*(4) The adversaries* $(\mathsf{P}, \mathsf{B}, \mathsf{C})$ *win if* $m = m_\mathsf{B} = m_\mathsf{C}$.

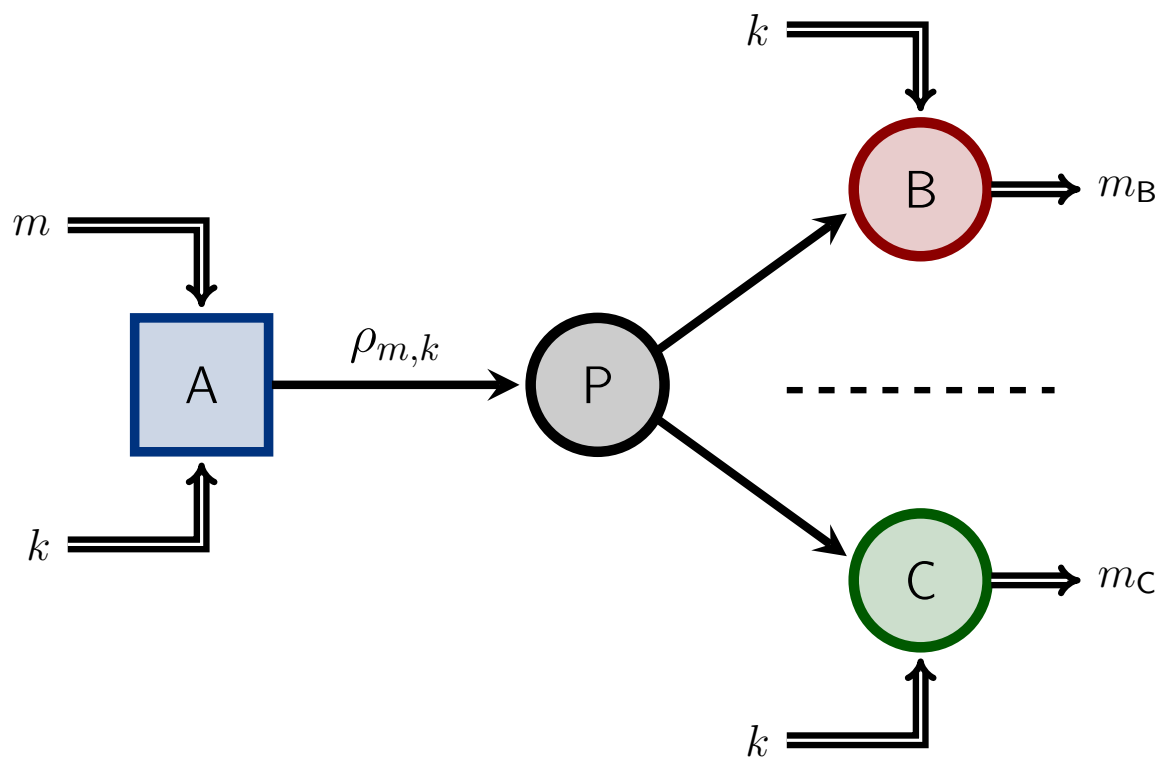

**Figure 5.1** — No-cloning game for a 1-bit message. *Alice encrypts a uniformly random message* $m \in \{0,1\}$ *using key* $k$*, into a quantum state* $\rho_{m,k}$*. She transmits it to a pirate* P *modeled by a quantum channel* $\Phi : \mathcal{B}(\mathcal{H}_\mathsf{A}) \to \mathcal{B}(\mathcal{H}_\mathsf{B} \otimes \mathcal{H}_\mathsf{C})$*. Bob and Charlie are then given the registers for* $\mathcal{H}_\mathsf{B}$ *and* $\mathcal{H}_\mathsf{C}$ *respectively, as well as a copy of* $k$*. They output* $m_\mathsf{B}$*,* $m_\mathsf{C} \in \{0,1\}$ *respectively, and* win *if and only if* $m = m_\mathsf{B} = m_\mathsf{C}$*.* Uncloneable-Indistinguishability *holds if the winning probability is bounded by* $1/2 + \text{negl}(\lambda)$ *for* $\lambda$ *a security parameter. A similar diagram appears in [Bot+24b].*

**Winning probability.** The winning probability at a no-cloning game is expressed as follows:

$$\begin{aligned}
&\mathbb{P}\Big((\mathsf{P},\mathsf{B},\mathsf{C})\text{ win}\Big) \\
&= \sup_{\substack{\Phi \\ B_{m_\mathsf{B}|k}, C_{m_\mathsf{C}|k}}} \mathop{\mathbb{E}}_{\substack{m\in\{0,1\} \\ k\leftarrow \mathsf{Gen}(1)}} \sum_{m_\mathsf{B}, m_\mathsf{C} \in \{0,1\}} \mathbb{1}_{\{m_\mathsf{B}=m_\mathsf{C}=m\}} \operatorname{Tr}\Big[\Phi(\rho_{m,k})\big(B_{m_\mathsf{B}|k} \otimes C_{m_\mathsf{C}|k}\big)\Big] \\
&= \sup_{\Phi,B,C} \mathop{\mathbb{E}}_{m,k} \operatorname{Tr}\Big[\Phi(\rho_{m,k})\big(B_{m|k} \otimes C_{m|k}\big)\Big],
\end{aligned}$$

where $\rho_{m,k} := \mathsf{Enc}(m,k)$, where the expected values are taken with respect to the uniform measures, and where the supremum is taken over all quantum channel $\Phi : \mathcal{B}(\mathcal{H}_\mathsf{A}) \to \mathcal{B}(\mathcal{H}_\mathsf{B} \otimes \mathcal{H}_\mathsf{C})$ (for finite-dimensional Hilbert spaces $\mathcal{H}_\mathsf{B}$ and $\mathcal{H}_\mathsf{C}$) and all families of POVMs $\{B_{m_\mathsf{B}|k}\}_{m_\mathsf{B}}$ and $\{C_{m_\mathsf{C}|k}\}_{m_\mathsf{C}}$.

Note that a trivial strategy for the adversary team could be that the Pirate sends the state $\rho$ to Bob, allowing him to make a perfect guess $m_\mathsf{B}$ with the key $k$, while Charlie can produce a random guess, leading to the trivial winning probability of $\mathbb{P}\big((\mathsf{P},\mathsf{B},\mathsf{C})\text{ win}\big) = 1/2$.

We are interested in upper-bounding the best winning probability of the adversary team $(\mathsf{P},\mathsf{B},\mathsf{C})$ at a no-cloning game. Using the Choi matrix $C_\Phi$ of the quantum channel $\Phi$ (see Definition 2.35), we can rephrase the winning probability as follows:

$$\mathbb{P}\Big((\mathsf{P},\mathsf{B},\mathsf{C})\text{ win}\Big) = \sup_{C_\Phi,B,C} \mathop{\mathbb{E}}_{m,k} \operatorname{Tr}\Big[C_\Phi\big(\rho_{m,k}^\top \otimes B_{m|k} \otimes C_{m|k}\big)\Big],$$

where the supremum is now taken over all $C_\Phi \succcurlyeq \mathbf{0}$ such that $\operatorname{Tr}_{(\mathsf{B},\mathsf{C})}[C_\Phi] = \mathbb{I}_d$, which can be relaxed to $\operatorname{Tr}[C_\Phi] = d$, giving an upper bound on the winning probability:

$$\begin{aligned}\mathbb{P}\Big((\mathsf{P},\mathsf{B},\mathsf{C})\text{ win}\Big) &\leqslant \sup_{C_\Phi,B,C} \mathop{\mathbb{E}}_{m,k} \operatorname{Tr}\Big[\tfrac{1}{d}\cdot C_\Phi\big(d\cdot\rho_{m,k}^\top \otimes B_{m|k} \otimes C_{m|k}\big)\Big] \\ &= \sup_{\sigma,B,C} \mathop{\mathbb{E}}_{m,k} \operatorname{Tr}\Big[\sigma\left(d\cdot\rho_{m,k}^\top \otimes B_{m|k} \otimes C_{m|k}\right)\Big], \qquad (5.3)\end{aligned}$$

where the first supremum is taken over all $C_\Phi \succcurlyeq \mathbf{0}$, and where the last supremum is taken other all $\sigma \succcurlyeq \mathbf{0}$ such that $\operatorname{Tr}[\sigma] = 1$, *i.e.* over all quantum mixed states.

**Security.** The winning probability at this game yields the following definition of unclonable-indistinguishable security [BL20][2]:

> **Definition 5.12** (Unclonable-Indistinguishable Security) — *A QECM scheme* (Gen, Enc, Dec) *satisfies* strong unclonable-indistinguishable security *if for any security parameter $\lambda \in \mathbb{N}$ and for any adversaries* (P, B, C)*, the adversary team cannot exceed the trivial winning probability at the associated no-cloning game by more than a negligible function:*
>
> $$\mathbb{P}\Big((\mathsf{P},\mathsf{B},\mathsf{C})\ \textit{win}\Big) \leqslant \frac{1}{2} + \operatorname{negl}(\lambda)\,.$$
>
> *If instead the adversaries cannot exceed $\frac{1}{2} + f(\lambda)$ for some function $f : \mathbb{R} \to \mathbb{R}$ vanishing at infinity $\lim_\lambda f(\lambda) = 0$, then we simply say that the scheme satisfies* unclonable-indistinguishable security.

[2] Our definition is slightly stronger than the one in [BL20] since we allow for unbounded adversaries.

**Related Work.** In their pioneer paper, Broadbent and Lord [BL20] showed the achievability of this security in the quantum random oracle model. Then, under a less stringent definition called *unclonable security*, unclonable encryption has become an important building block for quantum cryptography, including for private-key quantum money [BL20], preventing storage attacks [BL20], quantum functional encryption [MM24], quantum copy-protection [AK21], quantum position verification [Geo+25], and unclonable decryption [GZ20; KT25; SW22].

Given the importance of unclonable encryption, efforts have focused on its achievability under various models and definitions, including achievability in the quantum random oracle model (QROM) [AKL23; Ana+22; BL20], in an interactive version of the scenario [BC23c], assuming the existence of specific types of obfuscation [AB24; CHV24], in a device-independent variant with variable keys [KT25], and in a variant with quantum keys [AKY24].

Many open questions remain in the study of unclonable cryptography, notably the achievability of *unclonable-indistinguishability* security in the sense originally defined in [BL20]: the security definition considers a game of the form of Figure 5.1, but where a message $m \in \{0,1\}^n$ is selected by the adversary, and the challenge that the adversaries $\mathsf{B}$ and $\mathsf{C}$ face is to identify if the original message $m$, or a fixed message $0^n$ was encrypted, where the two cases happen with equal probability. In this scenario, limitations on possible schemes have been identified [Ana+22; MST21]. Notably, however, achievability in the standard model, even with computational assumptions, is wide open; for further discussion and a candidate scheme, see [CHV24].

**Unclonable Bit.** At the heart of this intriguing open question is the simplest case, called the *unclonable bit*, where $m \in \{0,1\}$ (our case), which, despite its simplicity, has remained unsolved in the plain model.

**Open Question 5.13 —** *Does the unclonable bit exist?*

Its importance is highlighted in [Hir+23], where it is shown that a scheme for an unclonable bit can be transformed into a scheme that encrypts general messages and that satisfies unclonable-indistinguishability[3].

[3] For conventional encryption, encrypting a message bitwise with a single-bit encryption

Our work, presented in Chapter 8, shows advances in this question by demonstrating the weak security in the plain model for small key sizes [Bot+24b]. This result was very recently generalized by Bhattacharyya and Culf to any key size [BC25]. The question remains open in the strong security setting.

### 5.3.3 Link with Monogamy-of-Entanglement Games

There is an interesting connection between no-cloning games and monogamy-of-entanglement (MoE) games. As detailed below, the winning probability of the former game can be upper bounded by the ones of the latter, yielding upper bounds on the unclonable-indistinguishable security. This link arises from the fact that the MoE principle (Section 2.2.5) is closely related to the No-Cloning Theorem (Theorem 2.37), as briefly explained at page 56. Below, after defining MoE games and their winning probability, we provide an explicit link with the no-cloning games and conclude this chapter by giving an example of computation based on conjugate coding and mentioning the more general framework of extended nonlocal games.

**Monogamy-of-Entanglement Games.** This family of games was introduced by Tomamichel, Fehr, Kaniewski, and Wehner in [Tom+13]. As plotted in Figure 5.2, an MoE game involves three parties: a challenger, called Alice (A), and two cooperating players, Bob (B) and Charlie (C). The game proceeds as follows. Initially, Alice has a fixed family of positive operator-valued measures (POVMs) $\{A_{m|k}\}_m$ known by the players. Before the game starts, Bob and Charlie jointly agree upon a strategy, prepare a tripartite quantum state $\sigma_{\mathsf{ABC}}$, and send the corresponding subsytem A to the challenger Alice. Once the game begins, the players are space-like separated, meaning that any form of communication is no longer allowed. The challenger Alice picks a POVM uniformly at random in $\{A_{m|k}\}_m$ and performs a measurement on her quantum state to produce some classical outcome $m_{\mathsf{A}}$. She publicly announces the POVM she used to Bob and Charlie. Their goal is then to independently recover $m_{\mathsf{A}}$. To this end, they perform some measurements $\{B_{m|k}\}_m$ and $\{C_{m|k}\}_m$ on their respective states, resulting in classical outcomes $m_{\mathsf{B}}$ and $m_{\mathsf{C}}$ respectively. Finally, the chal-scheme typically yields secure encryption; however, such a construction is not secure in the context of unclonable encryption.

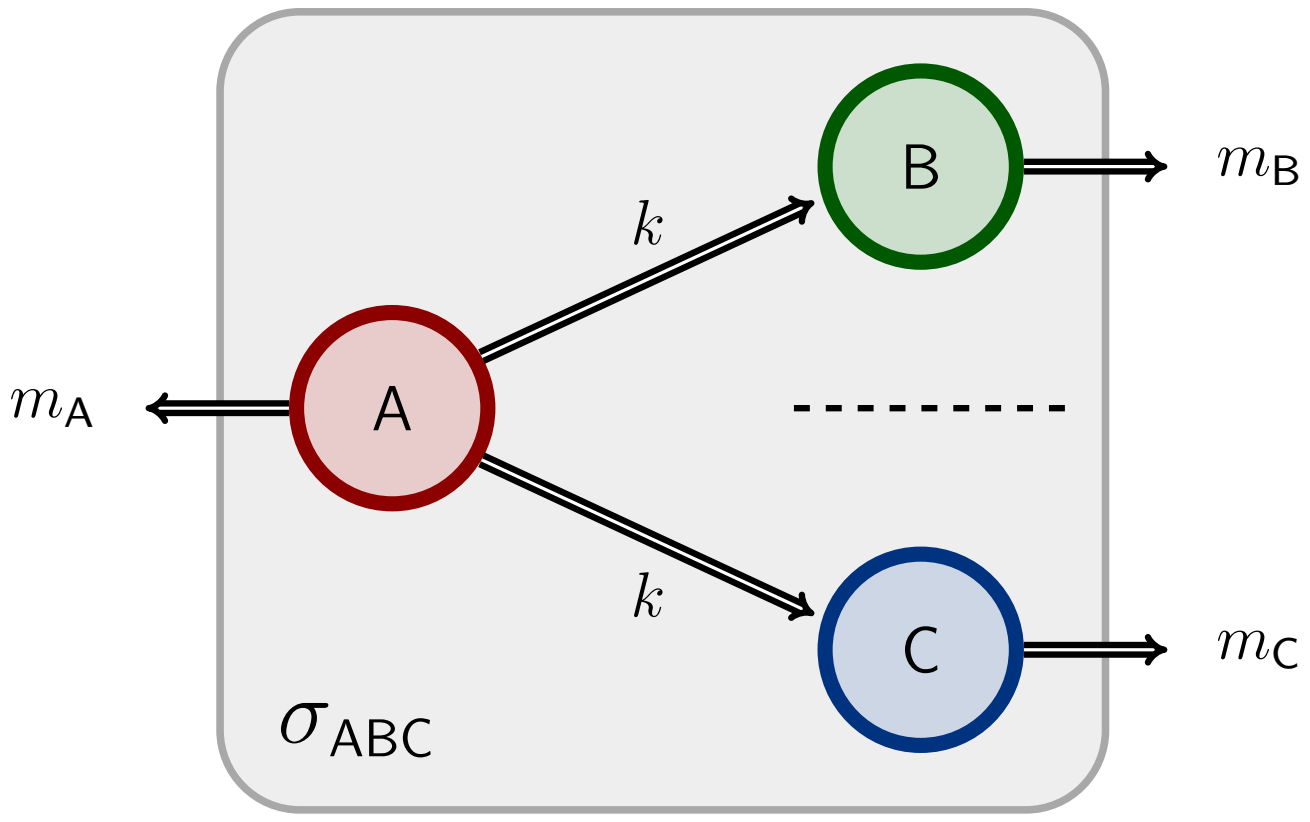

**Figure 5.2 —** Monogamy-of-Entanglement Game. *Alice, Bob, and Charlie share a quantum state* $\sigma_{\mathsf{ABC}}$*. Given a classical random key* $k$*, Alice performs a measurement* $\{A_{m|k}\}_m$ *and obtains* $m_{\mathsf{A}}$*. Using the same key* $k$*, the players Bob and Charlie perform respective measurements* $\{B_{m|k}\}_m$ *and* $\{C_{m|k}\}_m$*. We say that they win the game if both of them recover the exact same message as Alice,* i.e. *if* $m_{\mathsf{A}} = m_{\mathsf{B}} = m_{\mathsf{C}}$*. A similar diagram appears in [Bot+24b].*

lenger A declares that the players Bob and Charlie win the game if both of them recover Alice's outcome, *i.e.* if exactly $m_{\mathsf{A}} = m_{\mathsf{B}} = m_{\mathsf{C}}$.

**Winning Probability.** The winning probability of the adversary team $(\mathsf{B}, \mathsf{C})$ is expressed as follows:

$$\mathbb{P}\Big((\mathsf{B}, \mathsf{C}) \text{ win}\Big) = \sup_{\sigma, B, C} \frac{1}{K} \sum_{m,k} \operatorname{Tr}\big[\sigma\,(A_{m|k} \otimes B_{m|k} \otimes C_{m|k})\big]\,, \tag{5.4}$$

, where $K$ is the number of possible keys, where $\sigma$ is a quantum state in $\mathcal{H}_{\mathsf{A}} \otimes \mathcal{H}_{\mathsf{B}} \otimes \mathcal{H}_{\mathsf{C}}$, and where the sets $\{A_{m|k}\}_k$, $\{B_{m|k}\}_k$, $\{C_{m|k}\}_k$ are POVMs on respectively $\mathcal{H}_{\mathsf{A}}$, $\mathcal{H}_{\mathsf{B}}$, $\mathcal{H}_{\mathsf{C}}$.

**Link Between the Games.** Consider the one-bit message case $m \in \{0, 1\}$, and assume that Alice's measurement is of the form $A_{m|k} = (d/2) \cdot \rho_{m,k}^{\top}$, *i.e.* that the normalization condition $\sum_m \rho_{m,k} = 2\,\mathbb{I}_d/d$ holds for all $k$. Then, the winning probability for the MoE games in eq. (5.4) is precisely the same as the upper bound for the no-cloning (NC) games in eq. (5.3):

$$\mathbb{P}\Big((\mathsf{P}, \mathsf{B}, \mathsf{C}) \text{ win the NC game}\Big) \leqslant \mathbb{P}\Big((\mathsf{B}, \mathsf{C}) \text{ win the MoE game}\Big)\,.$$

Thereby, MoE games provide an upper bound on the notion of unclonable-indistinguishable security (Definition 5.12).

**Example 5.14** (Conjugate Coding Games) — This type of MoE game was studied by Tomamichel, Fehr, Kaniewski, and Wehner in [Tom+13]. Suppose Alice's POVMs arise from the *conjugate coding* protocol [Wie83] (Section 5.2.1), *i.e.* they are obtained from the $\sigma_x$- and $\sigma_z$-basis measurements in the qubit scenario $\mathcal{H}_\mathsf{A} = \mathbb{C}^2$:

$$\begin{aligned} A_{0|0} &:= |0\rangle\!\langle 0| & A_{1|0} &:= |1\rangle\!\langle 1|\,, \\ A_{0|1} &:= |+\rangle\!\langle +| & A_{1|1} &:= |-\rangle\!\langle -|\,. \end{aligned}$$

Then, it is simple to show that the winning probability $\frac{1}{2} + \frac{1}{2\sqrt{2}}$ can be achieved, without even using entanglement. Indeed, the adversaries Bob and Charlie can send to Alice the state $\cos\left(\frac{\pi}{8}\right)|0\rangle + \sin\left(\frac{\pi}{8}\right)|1\rangle$ from the Breidbart basis, and choose to always output $m_\mathsf{B} = m_\mathsf{C} = 0$. This leads them to the following winning probability[4]:

$$\mathbb{P}\Big((\mathsf{B},\mathsf{C}) \text{ win}\Big) = \mathbb{P}\Big(m_\mathsf{A} = 0\Big) = \cos^2\left(\frac{\pi}{8}\right) = \frac{1}{2} + \frac{1}{2\sqrt{2}}\,.$$

Interestingly, this value is actually the optimal one, and if we repeat $n$ times this game in parallel, then the optimal winning probability is precisely $\left(\frac{1}{2} + \frac{1}{2\sqrt{2}}\right)^n$ [Tom+13].

**Remark 5.15** (Extended Nonlocal Games) — More generally, MoE games belong to the class of *extended nonlocal games* [Joh+16]. Extended nonlocal games are generalized nonlocal games in which the predicate $\mathscr{V}$ takes values in positive semi-definite operators instead of $\{0, 1\}$ and where the winning probability is computed as the result of a measurement on quantum state shared between the Referee (R) and the players Alice (A) and Bob (B). See page 105 for a brief description and a diagram.

[4]Alternatively, this can be seen as a consequence of an entropic uncertainty relation due to Deutsch [Deu83].

# Part II

# Contributions

# Chapter 6

# Communication Complexity in the CHSH Game

In this chapter, we present a more detailed version of [BBP24] in Section 6.1 and then rearrange the results of [Bot+24a] in Sections 6.2 to 6.5. Here are the complete references:

## Chapter Contents

# 6.1 New Sufficient Condition to Collapse CC

In this section, we present a new sufficient condition for a nonlocal box to collapse CC, thus extending the known collapsing region. This is the content of the following reference:

Below, after briefly recalling the necessary background definitions and notations (Section 6.1.1), we define a sequence of protocols $(\mathcal{P}_k)_k$ that increasingly better perform distributed computations $a \oplus b = f(X, Y)$ (Section 6.1.2). Then, we present and prove the main result (Section 6.1.3), and finally, we provide two cases of interest deduced from the main result (Section 6.1.4).

### *6.1.1 Background*

Here, we connect with relevant background notions and briefly recall the key definitions.

**CHSH Game.** We consider the CHSH game (Section 3.2.2) and its scenario (page 58): two collaborating but non-communicating players, called Alice and Bob, receive bits $x, y \in \{0,1\}$ and answer bits $a, b \in \{0,1\}$. They win the CHSH game *if, and only if,* they satisfy:

$$a \oplus b = xy\,.$$

There is a variant of this game, called CHSH′, with a different winning condition: $a \oplus b = (x \oplus 1)(y \oplus 1)$.

**Nonlocal Boxes.** We work in the theoretical framework of *nonlocal boxes* (Section 3.1), more specifically in the non-signaling set $\mathcal{NS}$ (page 63). In the CHSH scenario, non-signaling boxes are probability distribution of the form $\mathbf{P}(ab \,|\, xy)$ such that:

$$\begin{aligned} \forall b, x, x', y\,, \qquad & \sum_a \mathbf{P}\big(a, b \,\big|\, x, y\big) \;=\; \sum_a \mathbf{P}\big(a, b \,\big|\, x', y\big) \;=:\; \mathbf{P}(b \,|\, y)\,, \\ \forall a, x, y, y'\,, \qquad & \sum_b \mathbf{P}\big(a, b \,\big|\, x, y\big) \;=\; \sum_b \mathbf{P}\big(a, b \,\big|\, x, y'\big) \;=:\; \mathbf{P}(a \,|\, x)\,. \end{aligned}$$

Consider the following examples of boxes:

$$\begin{array}{ll} \mathbf{PR}\big(a, b \,|\, x, y\big) := \frac{1}{2}\, \mathbb{1}_{a \oplus b = xy}\,, & \mathbf{I}\big(a, b \,|\, x, y\big) := \frac{1}{4}\,, \\ \mathbf{PR}'\big(a, b \,|\, x, y\big) := \frac{1}{2}\, \mathbb{1}_{a \oplus b = (x \oplus 1)(y \oplus 1)}\,, & \mathbf{SR}\big(a, b \,|\, x, y\big) := \frac{1}{2}\, \mathbb{1}_{a = b}\,. \end{array} \tag{6.1}$$

We will also use the notation $\overline{\mathbf{P}} := 1 - \mathbf{P}$ for the inverse box.

**Communication Complexity.** The principle of communication complexity (CC) is introduced in Section 4.1.3. It basically consists in computing a function $f(X, Y)$ with only one bit of communication and with high success probability, where the string $X \in \{0, 1\}^n$ is given to Alice and $Y \in \{0, 1\}^m$ to Bob. In such a case, we say that there is a *collapse* of CC.

### 6.1.2 Protocols

We define by induction a sequence of protocols $(\mathcal{P}_k)_{k \geqslant 0}$ generalizing the BBLMTU protocol, named after Brassard, Buhrman, Linden, Méthot, Tapp, and Unger [Bra+06]. The main difference is that we add local uniformity, see an overview of their protocol in Section 4.2.3.

**Local Uniformization.** We say that a box $\mathbf{P} \in \mathcal{NS}$ is *locally uniform* if on each player's side, the box always outputs uniformly random bits:

$$\mathbf{P}(a \,|\, x) = 1/2 \quad \text{and} \quad \mathbf{P}(b \,|\, y) = 1/2\,,$$

for any $a, b, x, y \in \{0, 1\}$. The local uniformity will be useful many times in later computations. However, some boxes $\mathbf{P}$ are *not* locally uniform, *e.g.* $\mathbf{P} := \frac{\mathbf{PR}+\mathbf{P_{00}}}{2} \in \mathcal{NS}$ where $\mathbf{P_{00}}$ is the box that always outputs $(0, 0)$ independently of the entries $(x, y)$. But one can use shared randomness to "locally uniformize" a nonlocal box. From $\mathbf{P} \in \mathcal{NS}$ and a shared random bit $r$, Alice and Bob simulate another box $\widetilde{\mathbf{P}} \in \mathcal{NS}$ by adding $r$ to the outputs of $\mathbf{P}$ (it is the same idea as in the *one-time pad*, see Example 5.3). This way, the new box $\widetilde{\mathbf{P}}$ is indeed locally uniform, and importantly it has the same *bias* $\eta_{xy}(\widetilde{\mathbf{P}})$ as the initial box $\mathbf{P}$ for all $x, y$:

$$\mathbf{P}\big(a \oplus b = xy \,|\, x, y\big) \;=\; \widetilde{\mathbf{P}}\big(a' \oplus b' = xy \,|\, x, y\big) \;=\; \frac{1 + \eta_{xy}(\mathbf{P})}{2}\,,$$

where $\eta_{xy}(\mathbf{P}) \in [-1, 1]$ is defined as:

$$\eta_{xy}(\mathbf{P}) \;:=\; 2\;\mathbf{P}\big(a \oplus b = xy \,|\, x, y\big) - 1 \;=\; 2\left(\textstyle\sum_{a,b} \mathbf{P}(a, b \,|\, x, y)\, \mathbb{1}_{a \oplus b = xy}\right) - 1\,. \tag{6.2}$$

Below, when the context is clear, we may omit $\mathbf{P}$ and simply write $\eta_{xy}$.

**Protocol $\mathcal{P}_0$.** Fix a Boolean function $f : \{0,1\}^n \times \{0,1\}^m \to \{0,1\}$ and strings $X \in \{0,1\}^n$ and $Y \in \{0,1\}^m$. The goal of the protocol $\mathcal{P}_0$ is to perform a *distributed computation* of $f$ (see Section 4.1.1). In other words, we want to produce bits $a, b \in \{0,1\}$ known by Alice and Bob respectively such that:

$$a \oplus b = f(X,Y)\,. \tag{6.3}$$

Assume Alice and Bob share uniformly random variables $Z \in \{0,1\}^m$ and $r \in \{0,1\}$. Upon receiving her string $X$, Alice produces a bit $a := f(X,Z) \oplus r$. As for Bob, if he receives a string $Y$ that is equal to $Z$, then he sets $b := r$; otherwise, he generates a local random variable $r_{\mathsf{B}}$ and sets $b := r_{\mathsf{B}}$. Now, separating the cases $Y = Z$ and $Y \neq Z$, the distributed computation (6.3) is achieved with the following probability:

$$p_0 := \mathbb{P}\big(\text{“(6.3)”}\big) = \frac{1}{2^m} + \frac{1}{2}\left(1 - \frac{1}{2^m}\right) = \frac{1}{2} + \frac{1}{2^{m+1}} > \frac{1}{2}\,.$$

Due to the shared random bit $r$, note that the bit $a$ is locally uniform:

$$\mathbb{P}(a \,|\, X) = 1/2\,,$$

for all $a, X$, and similarly for $b$. In total, this protocol uses $m+1$ shared random bits and importantly *no communication bit*.

**Protocol $\mathcal{P}_1$.** As in $\mathcal{P}_0$, we fix $f$, $X$, and $Y$, and we try to obtain the distributed computation (6.3) with a better probability $p_1 > p_0$. To that end, we realize four steps:

(a) We use the protocol $\mathcal{P}_0$ independently three times, and obtain three pairs of bits $(a_1,b_1), (a_2,b_2), (a_3,b_3)$ such that:

$$a_i \oplus b_i = \begin{cases} f(X,Y) & \text{with prob. } p_0 \\ f(X,Y) \oplus 1 & \text{with prob. } 1 - p_0\,. \end{cases}$$

for $i = 1,2,3$. Note that this is a repetition code that will be decoded in (b) using a majority vote.

(b) The majority function $\mathtt{Maj} : \{0,1\}^3 \to \{0,1\}$ is the function that outputs the most-frequent bit in its entries, *i.e.* $\mathtt{Maj}(\alpha,\beta,\gamma) = \mathbb{1}_{\alpha+\beta+\gamma \geqslant 2}$, where $\mathbb{1}$ is the indicator function. For instance, we have $\mathtt{Maj}(0,1,0) = 0$ and $\mathtt{Maj}(1,1,1) = 1$. Note that the following equality:

$$f(X,Y) = \mathtt{Maj}\Big(a_1 \oplus b_1,\; a_2 \oplus b_2,\; a_3 \oplus b_3\Big) \tag{6.4}$$

occurs *if, and only if,* at least two of the equations "$f(X,Y) = a_i \oplus b_i$" ($i = 1,2,3$) hold. Denote $e_i := a_i \oplus b_i \oplus f(X,Y)$, and notice that "$e_i = 0$" *if, and only if,* "$a_i \oplus b_i = f(X,Y)$" for any fixed $i$, so that eq. (6.4) is equivalent to having "$\texttt{Maj}(e_1, e_2, e_3) = 0$". But, the $e_i$'s are independent and $\mathbb{P}(e_i = \alpha) = p_0^{1-\alpha}(1-p_0)^{\alpha}$ for any $\alpha = 0,1$, so eq. (6.4) holds with the following probability:

$$\begin{aligned}\mathbb{P}(\text{"(6.4)"}) &= \sum_{\substack{\alpha,\beta,\gamma \in \{0,1\} \\ \text{s.t. } \texttt{Maj}(\alpha,\beta,\gamma)=0}} \mathbb{P}(e_1 = \alpha)\,\mathbb{P}(e_2 = \beta)\,\mathbb{P}(e_3 = \gamma) \\ &= \sum_{\substack{\alpha,\beta,\gamma \in \{0,1\} \\ \text{s.t. } \texttt{Maj}(\alpha,\beta,\gamma)=0}} p_0^{3-\alpha-\beta-\gamma}(1-p_0)^{\alpha+\beta+\gamma}\,.\end{aligned}$$

(c) Now, we try to distributively compute the majority function. Observe that we have:

$$\begin{aligned}&\texttt{Maj}\big(a_1 \oplus b_1,\ a_2 \oplus b_2,\ a_3 \oplus b_3\big) \\ &\qquad\qquad = \texttt{Maj}(a_1,a_2,a_3) \oplus \texttt{Maj}(b_1,b_2,b_3) \oplus r_1\,s_1 \oplus r_2\,s_2\,,\end{aligned}$$

where $r_1 := a_1 \oplus a_2$ and $s_1 := b_2 \oplus b_3$ and $r_2 := a_2 \oplus a_3$ and $s_2 := b_1 \oplus b_2$. To distributively compute the two products $r_j s_j$ ($j = 1,2$), Alice and Bob use two copies of their locally uniform box $\widetilde{\mathbf{P}}$, see Figure 6.1. They obtain pairs of bits $(a_1', b_1')$ and $(a_2', b_2')$ such that $a_j' \oplus b_j' = r_j s_j$ with bias $\eta_{r_j,s_j}$. Consider the following events:

$$\begin{aligned}E_{\alpha,\beta,\gamma} &:= \text{"}e_1 = \alpha, e_2 = \beta, e_3 = \gamma,\text{"} \\ F_{\delta,\varepsilon,\zeta,\theta} &:= \text{"}r_1 = \delta, r_2 = \varepsilon, s_1 = \zeta, s_2 = \theta,\text{"}\end{aligned}$$

where the greek letters designate bits in $\{0,1\}$. On the one hand, under $E_{\alpha,\beta,\gamma}$ and $F_{\delta,\varepsilon,\zeta,\theta}$, we see that the following equality:

$$r_1\,s_1 \oplus r_2\,s_2 = (a_1' \oplus b_1') \oplus (a_2' \oplus b_2') \tag{6.5}$$

holds *if, and only if,* both of the equations "$r_j s_j = a_j' \oplus b_j'$" hold ($j = 1,2$), or that none of them hold (because errors cancel out: $1 \oplus 1 = 0$). Hence this equality holds with a bias $\eta_{\delta,\zeta}\eta_{\varepsilon,\theta}$:

$$\mathbb{P}\Big(\text{"(6.5)"}\,\Big|\,E_{\alpha\beta\gamma}, F_{\delta\varepsilon\zeta\theta}\Big) = \frac{1 + \eta_{\delta,\zeta}\eta_{\varepsilon,\theta}}{2}\,, \tag{6.6}$$

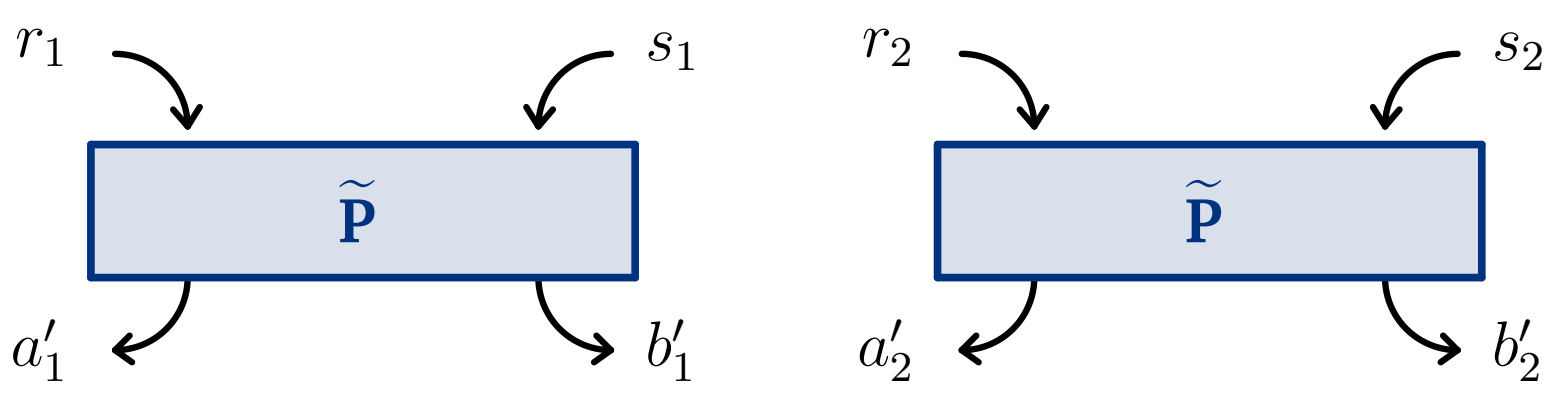

**Figure 6.1 —** *Distributively compute the products $r_1 s_1$ and $r_2 s_2$ with probability bias $\eta_{r_1,s_1}(\mathbf{P})$ and $\eta_{r_2,s_2}(\mathbf{P})$ respectively.*

(conditionally to knowing $X$ and $Y$ as well). On the other hand, seeing that the definitions of $r_j$ and $s_j$ lead to the relations $s_1 = r_2 \oplus e_2 \oplus e_3$ and $s_2 = r_1 \oplus e_1 \oplus e_2$, and using the independence of the $a_i$'s and their local uniform distribution in $\mathcal{P}_0$, direct computations yield that:

$$\mathbb{P}\big(F_{\delta,\varepsilon,\zeta,\theta} \,\big|\, E_{\alpha,\beta,\gamma}\big) = \tfrac{1}{4}\, \mathbb{1}_{\zeta=\beta\oplus\gamma\oplus\varepsilon}\, \mathbb{1}_{\theta=\alpha\oplus\beta\oplus\delta}\,. \tag{6.7}$$

Therefore, summing the products of eqs. (6.6) and (6.7) over all $\delta,\varepsilon,\zeta,\theta \in \{0,1\}$, we obtain:

$$\mathbb{P}\Big(\text{“(6.5)"} \,\Big|\, E_{\alpha\beta\gamma}\Big) = \sum_{\delta,\varepsilon\in\{0,1\}} \frac{1+\eta_{\delta,\beta\oplus\gamma\oplus\varepsilon}\,\eta_{\varepsilon,\alpha\oplus\beta\oplus\delta}}{8}\,. \tag{6.8}$$

Hence, we obtain a distributed computation of the majority function as follows:

$$\begin{aligned} \mathtt{Maj}\Big(a_1 \oplus b_1,\ a_2 \oplus b_2,\ a_3 \oplus b_3\Big) \\ = \underbrace{\Big(\mathtt{Maj}(a_1,a_2,a_3) \oplus a'_1 \oplus a'_2\Big)}_{=:\widetilde{a}} \oplus \underbrace{\Big(\mathtt{Maj}(b_1,b_2,b_3) \oplus b'_1 \oplus b'_2\Big)}_{=:\widetilde{b}}\,, \end{aligned} \tag{6.9}$$

with probability $\sum_{\delta,\varepsilon}\big(1+\eta_{\delta,\beta\oplus\gamma\oplus\varepsilon}\eta_{\varepsilon,\alpha\oplus\beta\oplus\delta}\big)/8$.

(d) Using steps (b) and (c), we obtain that the following equality:

$$f(X,Y) = \widetilde{a} \oplus \widetilde{b} \tag{6.10}$$

holds *if, and only if,* both eqs. (6.4) and (6.9) hold, or that none of them hold (because errors cancel out: $1\oplus 1 = 0$). This happens with the following

probability:

$$p_1 := \mathbb{P}\Big(\text{“(6.10)"}\Big) = \mathbb{P}\Big((6.4) \wedge (6.9)\Big) + \mathbb{P}\Big(\neg(6.4) \wedge \neg(6.9)\Big)$$
$$= \sum_{\alpha,\beta,\gamma,\delta,\varepsilon\in\{0,1\}} p_0^{3-\alpha-\beta-\gamma}\,(1-p_0)^{\alpha+\beta+\gamma}\,\frac{1+(\text{-}1)^{\mathtt{Maj}(\alpha,\beta,\gamma)}\,\eta_{\delta,\beta\oplus\gamma\oplus\varepsilon}\,\eta_{\varepsilon,\alpha\oplus\beta\oplus\delta}}{8}\,.$$

where the sign “+" from eq. (6.8) is now changed into “$(-1)^{\mathtt{Maj}(\alpha,\beta,\gamma)}$" because $\mathbb{P}\big(\neg(6.9)\big) = \sum_{\delta,\varepsilon}\big(1-\eta_{\delta,\beta\oplus\gamma\oplus\varepsilon}\eta_{\varepsilon,\alpha\oplus\beta\oplus\delta}\big)/8$, and this case exactly corresponds to the case where $\mathtt{Maj}(e_1,e_2,e_3)=1$.

Hence, we constructed a protocol $\mathcal{P}_1$ based on $\mathcal{P}_0$, whose probability of achieving distributed computation (eq. (6.3)) is $p_1$. We will find in the Section 6.1.3 a sufficient condition for which $p_1 > p_0$. In total, this protocol $\mathcal{P}_1$ uses $3(m+2)-1$ shared random bits, $2$ copies of $\mathbf{P}$, and importantly *no communication bit*.

**Protocol $\mathcal{P}_{k+1}$ ($k \geqslant 1$).** We proceed as in $\mathcal{P}_1$: We build $\mathcal{P}_{k+1}$ after performing $\mathcal{P}_k$ three times. In total, the protocol $\mathcal{P}_{k+1}$ uses $3^{k+1}(m+2)-1$ shared random bits and $3^{k+1}-1$ copies of $\mathbf{P}$, employs *no communication bit*, and distributively computes $f$ with the following probability:

$$p_{k+1} = \sum_{\alpha,\beta,\gamma,\delta,\varepsilon\in\{0,1\}} p_k^{3-\alpha-\beta-\gamma}\,(1-p_k)^{\alpha+\beta+\gamma}\,\frac{1+(\text{-}1)^{\mathtt{Maj}(\alpha,\beta,\gamma)}\,\eta_{\delta,\beta\oplus\gamma\oplus\varepsilon}\,\eta_{\varepsilon,\alpha\oplus\beta\oplus\delta}}{8}\,.$$

### 6.1.3 *Main Result*

The probability bias associated with $p_{k+1}$ is $\mu_{k+1} := 2\,p_{k+1}-1$ and it can be expressed in terms of $\mu_k$ as $\mu_{k+1} = F_{\mathbf{P}}(\mu_k)$ for any $k\in\mathbb{N}$, where:

$$F_{\mathbf{P}}(\mu) := \frac{\mu}{16}\Big[A(\mathbf{P}) + B(\mathbf{P}) - \mu^2\,\big(A(\mathbf{P}) - B(\mathbf{P})\big)\Big]\,, \tag{6.11}$$

where:

$$A\big(\mathbf{P}\big) := \Big(\eta_{0,0}(\mathbf{P}) + \eta_{0,1}(\mathbf{P}) + \eta_{1,0}(\mathbf{P}) + \eta_{1,1}(\mathbf{P})\Big)^2\,,$$
$$B\big(\mathbf{P}\big) := 2\,\eta_{0,0}(\mathbf{P})^2 + 4\,\eta_{0,1}(\mathbf{P})\,\eta_{1,0}(\mathbf{P}) + 2\,\eta_{1,1}(\mathbf{P})^2\,,$$

and where $\eta_{xy}(\mathbf{P})$ was introduced in eq. (6.2) as the bias of the box $\mathbf{P}$. Note that $0 \leqslant A(\mathbf{P}) \leqslant 16$ and $-8 \leqslant B(\mathbf{P}) \leqslant 8$ because $|\eta_{x,y}(\mathbf{P})| \leqslant 1$ for all $x, y$. Also note that the maximal value of $A(\mathbf{P})$ and $B(\mathbf{P})$ is achieved when $\mathbf{P} = \mathbf{PR}$ because $\eta_{xy}(\mathbf{PR}) = 1$ for all $x, y$, and therefore:

$$A(\mathbf{PR}) + B(\mathbf{PR}) = 24 .$$

In contrast, if we denote by $\mathbf{P}_{\text{quant}}$ the box achieving the best quantum winning probability at the CHSH game (this box is formally defined and computed at page 91), we have $\eta_{xy}(\mathbf{P}_{\text{quant}}) = 2\cos^2(\frac{\pi}{8}) - 1 = \frac{1}{\sqrt{2}}$ for all $x, y$, and therefore:

$$A(\mathbf{P}_{\text{quant}}) + B(\mathbf{P}_{\text{quant}}) = 12 .$$

We obtain the following theorem in terms of $A(\mathbf{P}) + B(\mathbf{P})$ with intermediate values between the former two examples:

**Theorem 6.1** (Sufficient Condition to Collapse CC) — *Let* $\mathbf{P} \in \mathcal{NS}$ *be any nonlocal box such that:*

$$A(\mathbf{P}) + B(\mathbf{P}) > 16 .$$

*Then* $\mathbf{P}$ *collapses communication complexity.*

*Proof.* Fix a nonlocal box $\mathbf{P} \in \mathcal{NS}$ for which $A + B > 16$, where we omit writing $\mathbf{P}$ for simplicity. This inequality yields three interesting consequences:

(1) First, we have $A - B = (A + B) - 2B > 16 - 2B \geqslant 0$. This allows us to compute the fixed points of the polynomial $F_{\mathbf{P}}$ defined in eq. (6.11), *i.e.* the variables $\mu$ such that:

$$F_{\mathbf{P}}(\mu) := \frac{\mu}{16}\left[A + B - \mu^2\,(A - B)\right] = \mu .$$

This equation is equivalent to $\mu\big[A + B - 16 - \mu^2(A - B)\big] = 0$, which is a factorized polynomial because $A + B > 16$ and $A - B \geqslant 0$. The three distinct real roots are:

$$\left\{0,\ \pm\sqrt{\frac{A + B - 16}{A - B}}\right\} =: \{0, \pm\mu_*\} , \tag{6.12}$$

which are thus exactly the three *fixed points* of $F_{\mathbf{P}}$.

(2) Second, as the derivative of $F_{\mathbf{P}}$ satisfies:

$$\frac{dF_{\mathbf{P}}}{d\mu}(\mu) = \frac{1}{16}\Big(A + B - 3\,\mu^2\,(A - B)\Big)\,,$$

the assumption $A + B > 16$ implies that $F_{\mathbf{P}}$ is increasing on $[-\mu_{\max}\,,\mu_{\max}]$, where $\pm\mu_{\max}$ are the two distinct roots of the derivative:

$$\mu_{\max} \;:=\; \sqrt{\frac{A + B}{3(A - B)}} > 0\,. \tag{6.13}$$

Moreover, the assumption gives $\frac{\partial F_{\mathbf{P}}}{\partial \mu}(0) > 1$, so that the fixed point $0$ of $F_{\mathbf{P}}$ is repulsive.

(3) Finally, as $A + B \leqslant 16 + 8 = 24$, we have $\frac{2}{3}\big(A + B\big) \leqslant 16$ and therefore $A + B - 16 = \frac{1}{3}\big(A + B\big) + \frac{2}{3}\big(A + B\big) - 16 \leqslant \frac{1}{3}\big(A + B\big)$. Hence, comparing eqs. (6.12) and (6.13), we obtain:

$$\mu_* \leqslant \mu_{\max} \qquad \text{and} \qquad [0, \mu_*] \subseteq [-\mu_{\max}\,, \mu_{\max}]\,,$$

which means that the function $F_{\mathbf{P}}$ is increasing on the line segment $[0, \mu_*]$.

Now, let us prove the collapse of communication complexity, *i.e.* that there exists a universal constant $\mathfrak{p} > 1/2$ (only depending on the box $\mathbf{P}$) such that any arbitrary Boolean function $f : \{0,1\}^n \times \{0,1\}^m \to \{0,1\}$ can be distributively computed by Alice and Bob with probability $\geqslant \mathfrak{p}$ and with only one bit of communication. We provide Alice and Bob with some strings $X \in \{0,1\}^n$ and $Y \in \{0,1\}^m$ respectively, and as many shared random bits and copies of the box $\mathbf{P} \in \mathcal{NS}$ as they want. Using the protocols from Section 6.1.2, we know that the protocol $\mathcal{P}_0$ enables them to distributively compute $f$ with probability $p_0 = (1 + 1/2^m)/2$, this is with initial bias

$$\mu_0 = 1/2^m > 0\,.$$

Up to adding muted variables to the input strings of $f$, we may assume that $m$ is large enough so that the initial bias $\mu_0$ belongs to the line segment $(0, \mu_*)$. Then, combining the above items (1), (2), and (3), we get that the sequence of biases $(\mu_k)_k$ of protocols $\mathcal{P}_k$ necessarily converges to the fixed point $\mu_* > 0$ defined in eq. (6.12). We set:

$$\mathfrak{p} \;:=\; \frac{1 + \mu_*/2}{2} \;>\; 1/2\,,$$

(or replace $\mu_*/2$ by any choice of constant in $(0, \mu_*)$). Note that this choice of $\mathfrak{p}$ does *not* depend on $f$, $n$, $m$, $X$, or $Y$—it only depends on $\mu_*$, which only depends on the $\eta_{x,y}$'s, which themselves only depend on $\mathbf{P}$. Moreover, as $(\mu_k)_k$ tends to $\mu_*$, we know that there exists a protocol $\mathcal{P}_k$ for some $k$ large enough such that the probability $p_k$ of correctly distributively computing $f$ satisfies $p_k > \mathfrak{p}$, with *no communication*. This means that Alice and Bob are able to construct some bits $a$ and $b$ respectively such that:

$$a \oplus b = f(X, Y)\,,$$

with probability $p_k > \mathfrak{p}$. Finally, Bob sends the bit $b$ to Alice (this is the only bit of communication), so that Alice knows the correct value of $f(X, Y)$ with probability lower-bounded by the universal constant $\mathfrak{p}$, hence the collapse of communication complexity. ■

### 6.1.4 *Cases of Interest*

In this section, we present two corollaries of the main result Theorem 6.1. The first one holds in the two-dimensional slice of $\mathcal{NS}$ passing through the nonlocal boxes **PR**, **PR**′, and **I** (this slice was also studied in [Bra11]), and the other in the slice defined by **PR**, **SR**, and **I** (also studied in [BS09]). Recall that $\mathcal{NS}$ is an eight-dimensional polytope. We stress that these two results were already established in the M.Sc. thesis of our co-author Marc-Olivier Proulx [Pro18].

**Slice Passing Through PR, PR′, and I.** In this case $\eta_{0,0} = \eta_{1,1}$ and $\eta_{0,1} = \eta_{1,0}$, and the condition $A + B > 16$ of Theorem 6.1 reads as:

$$\eta_{0,0}^2 + \eta_{0,0}\,\eta_{0,1} + \eta_{0,1}^2 \;>\; 2\,.$$

We make a change of coordinates using the bias $\sigma = (\eta_{0,0} + \eta_{0,1})/2$ of winning the CHSH game, and $\sigma' = (-\eta_{0,0} + \eta_{0,1})/2$ the one of winning at CHSH′, and we obtain:

$$\sigma^2 + \tfrac{1}{3}\,\sigma'^2 \;>\; \tfrac{2}{3}\,, \qquad \text{or} \qquad \tfrac{1}{3}\,\sigma^2 + \sigma'^2 \;>\; \tfrac{2}{3}\,,$$

where the second equation holds by swapping $\sigma$ and $\sigma'$ in the first one (indeed, we may do it because turning bits $x$ and $y$ into $x \oplus 1$ and $y \oplus 1$ allows one to go from CHSH to CHSH′). These equations give rise to the purple collapsing area drawn in Figure 6.2 (a). Interestingly, on the vertical axis, this allows to retrieve the same result as in [Bra+06]: Taking $\sigma' = 0$, the condition becomes $\sigma > \sqrt{2/3}$, *i.e.* winning at CHSH with probability $\frac{1+\sigma}{2} > \frac{3+\sqrt{6}}{6} \approx 0.91$.

**Slice Passing Through PR, SR, and I.** In this case $\eta_{0,0} = \eta_{0,1} = \eta_{1,0}$, and the condition $A + B > 16$ of Theorem 6.1 reads as:

$$5\,\eta_{0,0}^2 + 2\,\eta_{0,0}\,\eta_{1,1} + \eta_{1,1}^2 \;>\; \frac{16}{3}\,.$$

We make a change of coordinates using $\sigma = (3\eta_{0,0} + \eta_{1,1})/4$ and $\sigma' = (\eta_{0,0} - \eta_{1,1})/4$, and we obtain:

$$\sigma^2 + \sigma'^2 \;>\; \tfrac{2}{3}\,.$$

The induced collapsing area is represented in Figure 6.2 (b). Note that the same results also hold if we replace **SR** by any convex combination of $\mathbf{P_{00}}$ and $\mathbf{P_{11}}$, which are the boxes that always output respectively $(0, 0)$ and $(1, 1)$ independently of the entries $(x, y)$.

**Remark 6.2 —** In Figure 6.2, we compare our result to previously known collapsing regions with analytical methods. But even in comparison to former numerical results, our protocol finds strictly new collapsing boxes. Indeed, for instance, consider boxes in the black region of Figure 6.2 that are close to the vertical axis: They are not distillable by means of the wirings of [BS09; EWC23a], but our result shows that they are still collapsing.

## 6.2 Algebra of Boxes

In this section, we present the framework of the *algebra of boxes*, and in the next sections, its consequences on communication complexity, both numerically and analytically. This is the content of the following reference:

**[Bot+24a]** Pierre Botteron, Anne Broadbent, Reda Chhaibi, Ion Nechita, and Clément Pellegrini. "Algebra of Nonlocal Boxes and the Collapse of Communication Complexity". In: *Quantum* 8 (July 2024), p. 1402. ISSN: 2521-327X. DOI: 10.22331/q-2024-07-10-1402

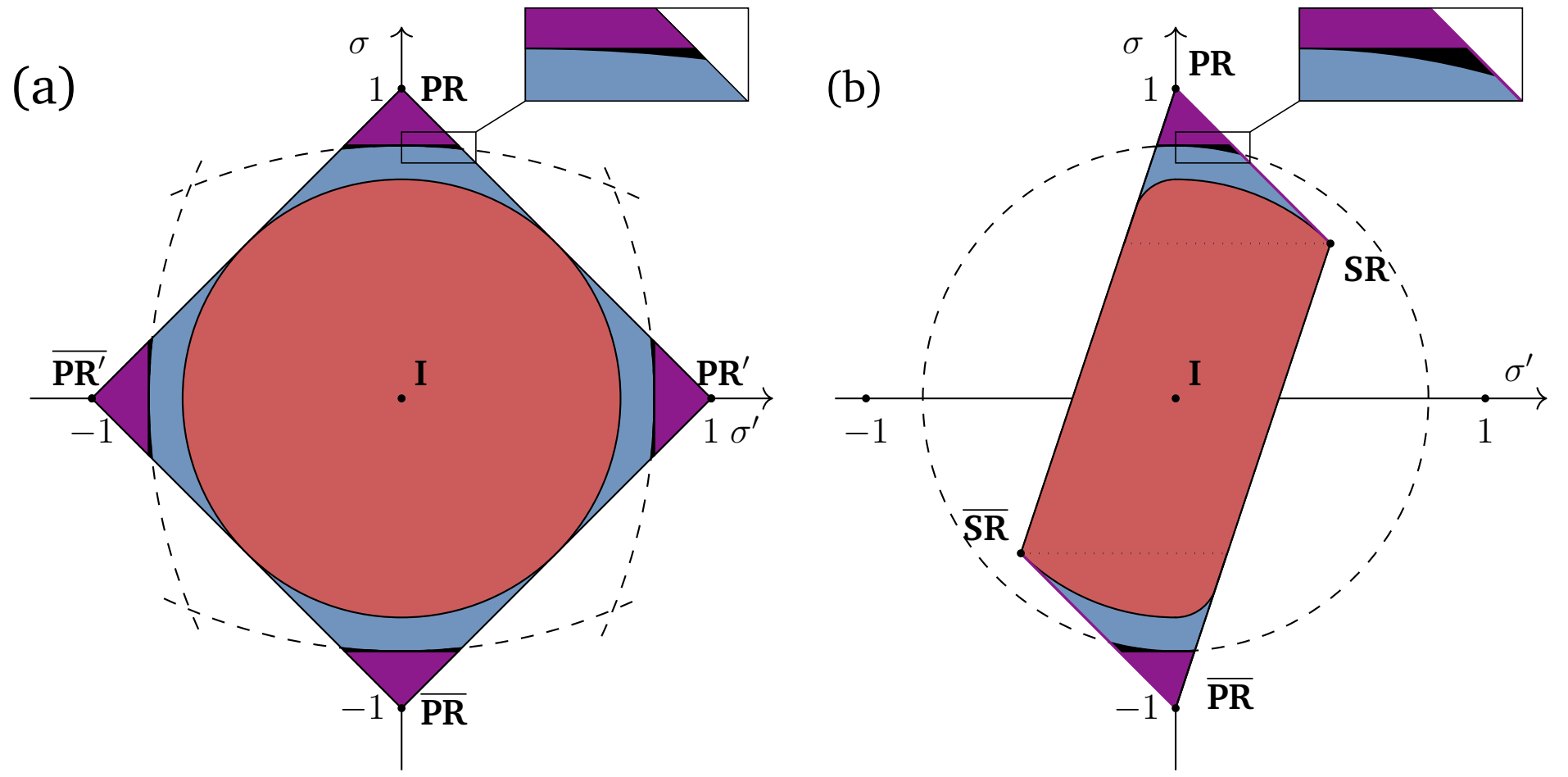

**Figure 6.2 —** *In* **purple** *is drawn the prior (analytically) known collapsing region. We extend it as follows: the* **black** *area is the new analytic collapsing region. The* **red** *area corresponds to the area of non-collapsing boxes. The* **blue** *area is the gap to be filled in red or purple (open problem). Diagrams (a) and (b) represent the slices of $\mathcal{NS}$ passing through respectively $\{\mathbf{PR}, \mathbf{PR}', \mathbf{I}\}$ with $\sigma = \eta_{0,0} + \eta_{0,1}$ and $\sigma' = -\eta_{0,0} + \eta_{0,1}$ (improving [Bra11]) and $\{\mathbf{PR}, \mathbf{SR}, \mathbf{I}\}$ with $\sigma = 3\,\eta_{0,0} + \eta_{1,1}$ and $\sigma' = \eta_{0,0} - \eta_{1,1}$ (improving [BS09]).*

Below, after briefly recalling the necessary background definitions and notations (Section 6.2.1), we introduce the notion of algebra of boxes (Section 6.2.2) and orbit of a box (Section 6.3), then detail our algorithms for finding the "best" wiring given a nonlocal box (Section 6.4), and finally present our numerical and analytical results related to the collapse of communication complexity (Section 6.5).

## *6.2.1 Background*

The background for this work includes the former one (Section 6.1.1). In addition, we consider the following two deterministic boxes:

$$\mathbf{P_{00}}\big(a, b \,|\, x, y\big) := \mathbb{1}_{a=b=0}\,, \qquad \mathbf{P_{11}}\big(a, b \,|\, x, y\big) := \mathbb{1}_{a=b=1}\,, \tag{6.14}$$

which always output the tuples $(0, 0)$ and $(1, 1)$ respectively, independently of the inputs $x$ and $y$. We also need the formalism of wirings of boxes, defined in Section 3.1.4. In short, wirings allow one to create a new box out of two boxes. If the wiring is denoted W and the boxes $\mathbf{P}, \mathbf{Q} \in \mathcal{NS}$, we obtain the following box:

$$\mathbf{P} \boxtimes_{\mathsf{W}} \mathbf{Q} \in \mathcal{NS}\,.$$

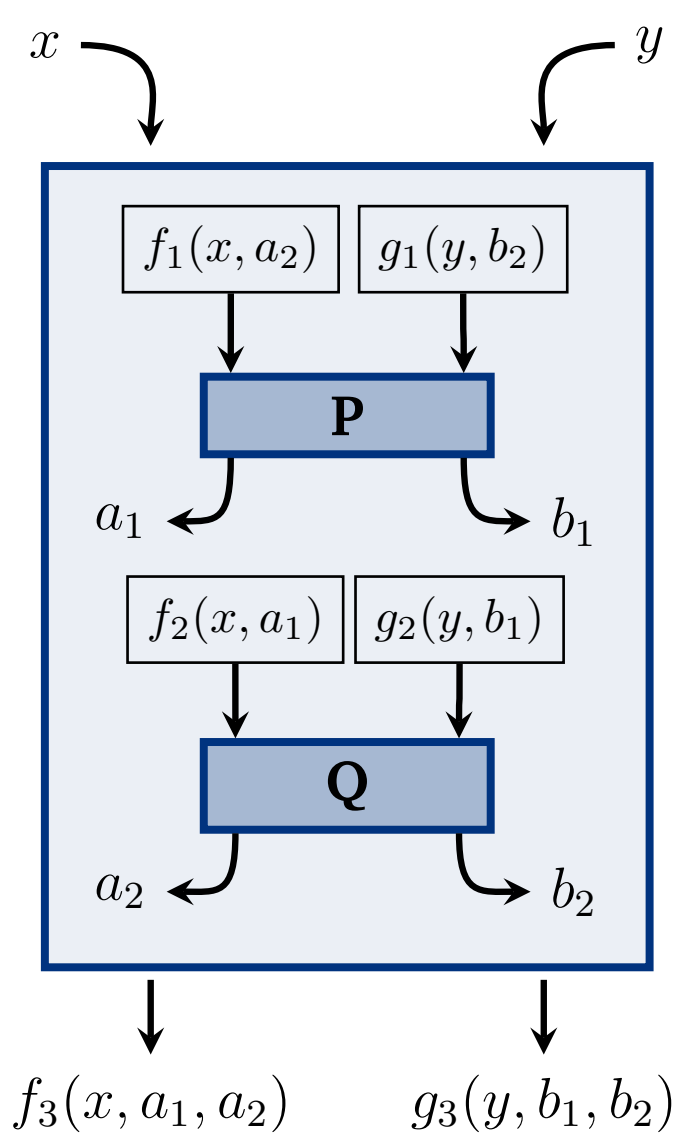

**Figure 6.3** — *General wiring between two boxes* **P** *and* **Q***.*

The explicit expression of $\mathbf{P} \boxtimes_{\mathsf{W}} \mathbf{Q}$ was given in eq. (3.20). In particular, we see that it is linear in $\mathbf{P}$ and $\mathbf{Q}$, hence *bilinear* for any wiring $\mathsf{W} \in \mathcal{W}$. For convenience, we recall the notation for a general wiring $\mathsf{W} = (f_1, g_1, f_2, g_2, f_3, g_3)$ in Figure 6.3, and we recall the expression of $\mathbf{P} \boxtimes_{\mathsf{W}} \mathbf{Q}$ for a deterministic wiring given in eq. (3.17):

$$\mathbf{P} \underset{\mathsf{W}}{\boxtimes} \mathbf{Q}(a, b \,|\, x, y) := \sum_{a_1, a_2, b_1, b_2} \mathbf{P}\Big(a_1,\, b_1 \,|\, f_1(x, a_2),\, g_1(y, b_2)\Big) \\ \times\, \mathbf{Q}\Big(a_2,\, b_2 \,|\, f_2(x, a_1),\, g_2(y, b_1)\Big) \times \mathbb{1}_{a = f_3(x, a_1, a_2)} \times \mathbb{1}_{b = g_3(y, b_1, b_2)}\,. \tag{6.15}$$

As mixed wirings are convex combinations of deterministic wirings (see eq. (3.18)), most of the time it suffices to show results only for deterministic wirings.

### 6.2.2 Algebra of Boxes Induced by a Wiring

Let $\mathcal{B}$ be the vector space of all functions $\{0,1\}^4 \to \mathbb{R}$, and consider a mixed wiring $\mathsf{W} \in \mathcal{W}$. As the operation $\boxtimes_{\mathsf{W}}$ is bilinear, the vector space $\mathcal{B}$ equipped with the product $\boxtimes_{\mathsf{W}}$ is an algebra, which we call the *algebra of boxes* and denote by:

$$\mathcal{B}_{\mathsf{W}} .$$

Its dimension is $\dim(\mathcal{B}_{\mathsf{W}}) = 2^4 = 16$. It contains the non-signaling polytope $\mathcal{NS} \subseteq \mathcal{B}_{\mathsf{W}}$, which has dimension $8$ (eq. (3.14)).

**Multiplication Table.** In order to better understand the behavior of the box product $\boxtimes$, it is interesting to compute the product of some basic boxes: for instance the boxes $\mathbf{PR}$, $\mathbf{P_{00}}$, $\mathbf{P_{11}}$, $\mathbf{I}$ defined in eqs. (6.1) and (6.14). In Figure 6.4, we present the multiplication table derived from the wiring $\mathsf{W}_{\mathrm{BS}}$ from [BS09]. By bilinearity of the box multiplication, this table shows that the convex hull $\mathrm{Conv}\{\mathbf{PR}, \mathbf{P_{00}}, \mathbf{P_{11}}\}$ is stable under $\boxtimes$. On the contrary, observe that the convex hull $\mathrm{Conv}\{\mathbf{PR}, \mathbf{P_{00}}, \mathbf{P_{11}}, \mathbf{I}\}$ is not stable under $\boxtimes$: The product $\mathbf{I} \boxtimes \mathbf{PR}$ gives $\mathbf{Q}_1 := \frac{1}{4}\,\mathbf{PR} - \frac{1}{8}\big(\mathbf{P_{00}} + \mathbf{P_{11}}\big) + \mathbf{I}$ which is out of the convex hull (nevertheless the affine hull $\mathrm{Aff}\{\mathbf{PR}, \mathbf{P_{00}}, \mathbf{P_{11}}, \mathbf{I}\}$ is stable under $\boxtimes$). Notice that we show in Lemma 6.21 that actually $\mathrm{Conv}\{\mathbf{PR}, \mathbf{P_{00}}, \mathbf{P_{11}}\} = \mathcal{NS} \cap \mathrm{Aff}\{\mathbf{PR}, \mathbf{P_{00}}, \mathbf{P_{11}}\}$. From this table, one may postulate that $\mathbf{P_{00}}$ is a right identity in the sense that $\mathbf{P} \boxtimes \mathbf{P_{00}} = \mathbf{P}$ for all $\mathbf{P}$ in $\mathcal{NS}$, and it is indeed true as a simple consequence of formula (6.15). One may similarly verify that $\mathbf{I}$ is a right fixed point, in the sense that $\mathbf{P} \boxtimes \mathbf{I} = \mathbf{I}$ for all $\mathbf{P}$ in $\mathcal{NS}$, as it is possible to guess from the table. See all the multiplication tables of the typical depth-2 wirings in [Bot+24a, Appendix C].

**Non-Commutativity and Non-Associativity.** A direct consequence of the multiplication table in Figure 6.4 is that the algebra $\mathcal{B}_{\mathsf{W}_{\mathrm{BS}}}$ induced by the wiring $\mathsf{W}_{\mathrm{BS}}$ is non-commutative ($\mathbf{P_{00}} \boxtimes \mathbf{PR} \neq \mathbf{PR} \boxtimes \mathbf{P_{00}}$) and non-associative ($(\mathbf{P_{00}} \boxtimes \mathbf{P_{11}}) \boxtimes \mathbf{PR} \neq \mathbf{P_{00}} \boxtimes (\mathbf{P_{11}} \boxtimes \mathbf{PR})$). This non-associativity is at the root of interesting remarks, see drawings of the orbit of a box in the next section, Figure 6.7. Similarly, the algebra induced by the wiring $\mathsf{W}_{\mathrm{dist}}$ is both non-commutative and non-associative, but on the contrary, the algebras induced by $\mathsf{W} \in \{\mathsf{W}_{\mathrm{triv}}, \mathsf{W}_{\oplus}, \mathsf{W}_{\wedge}, \mathsf{W}_{\vee\wedge}\}$ are both commutative and associative. One may wonder if there exist induced algebras that are associative but not commutative, or the converse. To that end, here is a characterization of

| P \ Q | PR | $P_{00}$ | $P_{11}$ | I |
|---|---|---|---|---|
| PR | PR | PR | PR | I |
| $P_{00}$ | $\frac{1}{2}(P_{00} + P_{11})$ | $P_{00}$ | $P_{11}$ | I |
| $P_{11}$ | PR | $P_{11}$ | $P_{00}$ | I |
| I | $Q_1$ | I | I | I |

**Figure 6.4** — *Multiplication table of the operation* $\boxtimes_{W_{BS}}$ *induced by the wiring from [BS09]. Each cell displays the result of* $P \boxtimes Q$*. The box* $Q_1$ *at the bottom left is* $Q_1 := \frac{1}{4}\,PR - \frac{1}{8}(P_{00} + P_{11}) + I$*. Further multiplication tables are available in [Bot+24a, Appendix C].*

commutativity and associativity in a simple case where boxes are set in parallel and with the same input functions:

**Proposition 6.3** (Characterizing Commutativity and Associativity) — *Using notations from Figure 6.3, consider a wiring* W *such that* $f_1 = f_2 = f(x)$ *and* $g_1 = g_2 = g(y)$*. Then:*

(i) $\mathcal{B}_W$ *is commutative* if, and only if, *the functions* $f_3(x, a_1, a_2)$ *and* $g_3(y, b_1, b_2)$ *are "symmetric" in the last two variables, in the sense that* $f_3(x, a_1, a_2) = f_3(x, a_2, a_1)$ *for all* $x, a_1, a_2$*, and similarly for* $g_3$*.*

*If in addition* $f(x) = x$ *and* $g(y) = y$*:*

(ii) $\mathcal{B}_W$ *is associative* if, and only if, *the functions* $f_3(x, a_1, a_2)$ *and* $g_3(y, b_1, b_2)$ *are "associative" in the last two variables, in the sense that* $f_3(x, a_1, f_3(x, a_2, a_3)) = f_3(x, f_3(x, a_1, a_2), a_3)$ *for all* $x, a_1, a_2, a_3$*, and similarly for* $g_3$*.*

*Proof.* (i) First, from the expression in eq. (6.15), see that for all bits

$a, b, x, y$ and any boxes $\mathbf{P}, \mathbf{Q}$ in $\mathcal{B}_{\mathsf{W}}$, we have:

$$\begin{aligned}
&\mathbf{P} \underset{\mathsf{W}}{\boxtimes} \mathbf{Q}(a, b \,|\, x, y) - \mathbf{Q} \underset{\mathsf{W}}{\boxtimes} \mathbf{P}(a, b \,|\, x, y) \\
&\qquad = \sum_{a_1, a_2, b_1, b_2} \mathbf{P}\big(a_1, b_1 \,|\, f(x), g(y)\big) \times \mathbf{Q}\big(a_2, b_2 \,|\, f(x), g(y)\big) \\
&\qquad\qquad \times \Big[ \mathbb{1}_{a = f_3(x, a_1, a_2)} \mathbb{1}_{b = g_3(y, b_1, b_2)} - \mathbb{1}_{a = f_3(x, a_2, a_1)} \mathbb{1}_{b = g_3(y, b_2, b_1)} \Big] .
\end{aligned}$$

Hence, if $f_3$ and $g_3$ are both symmetric in the last two variables, then the difference is zero and the algebra is commutative. Conversely, suppose that $\mathcal{B}_{\mathsf{W}}$ is commutative so that the left-hand side is zero. Taking probability distributions $\mathbf{P}$ and $\mathbf{Q}$ that are always positive (such as the box $\mathbf{I}$), we have that the difference in the right-hand side has to be zero for all $x, y, a, b, a_1, a_2, b_1, b_2$. Fix $x, a_1, a_2$ and consider $a := f_3(x, a_1, a_2)$, and similarly fix $y, b_1, b_2$ and consider $b := g_3(y, b_1, b_2)$. We obtain $1 - \mathbb{1}_{a = f_3(x, a_2, a_1)} \mathbb{1}_{b = g_3(y, b_2, b_1)} = 0$, which means that both indicator functions are equal to 1, and therefore both subscript equalities hold. Hence, this being true for any fixed $x, a_1, a_2$ and $y, b_1, b_2$, we obtain that $f_3$ and $g_3$ are symmetric as wanted.

(ii) From eq. (6.15) again, we have for all bits $a, b, x, y$ and any boxes $\mathbf{P}, \mathbf{Q}, \mathbf{R}$ in $\mathcal{B}_{\mathsf{W}}$:

$$\begin{aligned}
&\mathbf{P} \underset{\mathsf{W}}{\boxtimes} (\mathbf{Q} \underset{\mathsf{W}}{\boxtimes} \mathbf{R})(a, b \,|\, x, y) - (\mathbf{P} \underset{\mathsf{W}}{\boxtimes} \mathbf{Q}) \underset{\mathsf{W}}{\boxtimes} \mathbf{R}(a, b \,|\, x, y) \\
&\qquad = \sum_{a_1, a_2, a_3, b_1, b_2, b_3} \mathbf{P}\big(a_1, b_1 \,|\, x, y\big) \times \mathbf{Q}\big(a_2, b_2 \,|\, x, y\big) \times \mathbf{R}\big(a_3, b_3 \,|\, x, y\big) \\
&\times \Big[ \mathbb{1}_{a = f_3(x, a_1, f_3(x, a_2, a_3))} \mathbb{1}_{b = g_3(y, b_1, g_3(y, b_2, b_3))} - \mathbb{1}_{a = f_3(x, f_3(x, a_1, a_2), a_3)} \mathbb{1}_{b = g_3(y, g_3(y, b_1, b_2), b_3)} \Big] .
\end{aligned}$$

A similar proof with double implication as in (i) applies, hence the associativity criterion follows. ■

Now, it is easier to build an associative non-commutative induced algebra $\mathcal{B}_{\mathsf{W}'}$. Consider the wiring $\mathsf{W}'$ given by $f_1(x, a_2) = f_2(x, a_1) = x$, and $g_1(y, b_2) = g_2(y, b_1) = y$, and $f_3(x, a_1, a_2) := a_1$, and $g_3(y, b_1, b_2) := b_1$. This wiring satisfies the condition (ii) of the proposition and does not satisfy the condition (i), hence it is as wanted. Conversely, with similar arguments, a commutative non-associative algebra $\mathcal{B}_{\mathsf{W}''}$ is induced by the wiring $\mathsf{W}''$ defined by the same $f_1, f_2, g_1, g_2$ and $f_3(x, a_1, a_2) := a_1 a_2 \oplus 1$ and $g_3(y, b_1, b_2) := b_1 b_2 \oplus 1$. Therefore, we obtain the table in Figure 6.5.

| | Associativity | Non-associativity |
|---|---|---|
| Commutativity | $\mathsf{W}_{\text{triv}}, \mathsf{W}_{\oplus}, \mathsf{W}_{\wedge}, \mathsf{W}_{\vee\wedge}$ | $\mathsf{W}''$ |
| Non-commutativity | $\mathsf{W}'$ | $\mathsf{W}_{\text{BS}}, \mathsf{W}_{\text{dist}}$ |

**Figure 6.5 —** *Associativity and commutativity of the induced algebra $\mathcal{B}_{\mathsf{W}}$, depending on the wiring* $\mathsf{W}$ *displayed in the table cell.*

## 6.3 Orbit of a Box

In this section, we stress that most of the results were already reported in the M.Sc. of the author [Bot22]. But for the completeness of the presentation, we still state and prove the results here, since the sequel relies in part on those concepts.

Here, we study the set of all boxes that can be generated given many copies of a starting box $\mathbf{P}$ and a wiring $\mathsf{W}$. After introducing the *orbit* of a box, we provide some consequences to communication complexity. Subsequently, we study a particular example, $\mathsf{W}_{\text{BS}}$, with which we find collapsing boxes in Section 6.5.2, and then we give some general remarks about other orbits. Finally, we conclude this section by giving the technical proof of the theorem stating that the "best" parenthesization is the multiplication on the right.

### 6.3.1 *Definition of an Orbit*

Given multiple copies of a non-signaling box $\mathbf{P} \in \mathcal{NS}$ and of a (mixed) wiring $\mathsf{W}$, Alice and Bob can produce many other boxes, *e.g.* $(\mathbf{P} \boxtimes_{\mathsf{W}} \mathbf{P}) \boxtimes_{\mathsf{W}} \mathbf{P}$ or $\mathbf{P} \boxtimes_{\mathsf{W}} (\mathbf{P} \boxtimes_{\mathsf{W}} \mathbf{P})$. All of these new boxes are again non-signaling because $\mathcal{NS}$ is closed under wirings, see Example 3.17. We call *orbit* of the box $\mathbf{P}$ (induced by the wiring $\mathsf{W}$) the set of all of these possible new boxes:

$$\begin{aligned}\mathrm{Orbit}_{\mathsf{W}}(\mathbf{P}) \;:=& \Big\{\text{boxes } \mathbf{Q} \in \mathcal{NS} \text{ that can be produced by using} \\ &\qquad\qquad \text{finitely many times the box } \mathbf{P} \text{ and the wiring } \mathsf{W}\Big\} \\ =& \bigcup_{k\geqslant 1} \mathrm{Orbit}_{\mathsf{W}}^{(k)}(\mathbf{P}) \subseteq \mathcal{NS}\,,\end{aligned}$$

where $\mathrm{Orbit}^{(k)}(\mathbf{P})$ is called the *orbit of depth* $k$ of $\mathbf{P}$ (or simply $k$-*orbit*), defined as:

$$\mathrm{Orbit}_{\mathsf{W}}^{(k)}(\mathbf{P}) := \Big\{\text{all possible products with } k \text{ times the term } \mathbf{P}, \\ \text{using the multiplication } \boxtimes_{\mathsf{W}}\Big\}.$$

When the context is clear, we do not overload the notation and write $\mathrm{Orbit}$ and $\mathrm{Orbit}^{(k)}$ respectively. In general, these $k$-orbits are not singletons for $k \geqslant 3$ since the algebra $\mathcal{B}_{\mathsf{W}}$ induced by W is not necessarily associative and commutative (see Figure 6.5). Actually, up to multiplicity, the cardinal $\#\,\mathrm{Orbit}^{(k)}$ is exactly the number of parenthesizations with $k$ terms, which is the Catalan number $\frac{1}{k}\binom{2k-2}{k-1}$, which grows exponentially fast. Here are the 3- and 4- orbits:

$$\begin{aligned} \mathrm{Orbit}^{(3)}(\mathbf{P}) &= \Big\{(\mathbf{P}\boxtimes\mathbf{P})\boxtimes\mathbf{P},\ \mathbf{P}\boxtimes(\mathbf{P}\boxtimes\mathbf{P})\Big\}, \\ \mathrm{Orbit}^{(4)}(\mathbf{P}) &= \Big\{\big((\mathbf{P}\boxtimes\mathbf{P})\boxtimes\mathbf{P}\big)\boxtimes\mathbf{P},\ \big(\mathbf{P}\boxtimes(\mathbf{P}\boxtimes\mathbf{P})\big)\boxtimes\mathbf{P},\ (\mathbf{P}\boxtimes\mathbf{P})\boxtimes(\mathbf{P}\boxtimes\mathbf{P}), \\ &\qquad \mathbf{P}\boxtimes\big((\mathbf{P}\boxtimes\mathbf{P})\boxtimes\mathbf{P}\big),\ \mathbf{P}\boxtimes\big(\mathbf{P}\boxtimes(\mathbf{P}\boxtimes\mathbf{P})\big)\Big\}. \end{aligned}$$

Note that a $k$-orbit ($k \geqslant 2$) can be inductively computed using orbits with lower depth:

$$\mathrm{Orbit}^{(k)} = \bigcup_{1\leqslant \ell\leqslant k-1} \mathrm{Orbit}^{(\ell)} \boxtimes \mathrm{Orbit}^{(k-\ell)},$$

which is the same recurrence relation as that of Catalan numbers.

### 6.3.2 *Consequences of Orbits to CC*

Assume Alice and Bob are given infinitely many copies of a nonlocal box $\mathbf{P}$, and assume they want to distantly compute (in finite time) the value of a Boolean function $f(X,Y)$, where $X,Y \in \{0,1\}^n$ are strings that are known by Alice and Bob respectively. Among all the possible protocols they can try to do in order to succeed, they can wire their copies of $\mathbf{P}$ in order to produce a “better" box. For example, starting from a noisy box $\mathbf{P}$, Alice and Bob can try to produce a box that is closer to the “perfect box" $\mathbf{PR}$ which satisfies $a \oplus b = xy$ without noise. Such a protocol is called a *distillation protocol* [BS09]. We call *collapsing box* a nonlocal box that collapses CC.

**Find Collapsing Boxes Using the Orbit.** Imagine Alice and Bob are able to produce a collapsing box $\mathbf{Q}$ after applying wirings to copies of a starting box $\mathbf{P}$. Then they can use that new box $\mathbf{Q}$ to distantly compute the value $f(X,Y)$, which means that they have a protocol to collapse communication complexity and therefore that $\mathbf{P}$ is collapsing. This point of view is particularly interesting since it implies that it is sufficient to find a single collapsing box in the union $\bigcup_{\mathsf{W}} \mathrm{Orbit}_{\mathsf{W}}(\mathbf{P})$ to deduce that $\mathbf{P}$ is collapsing as well. See an illustration in Figure 6.6 (a).

**Find Collapsing Boxes Using a Cone.** Once we find a collapsing box $\mathbf{P}$, we can deduce many other collapsing boxes: there is a convex cone taking origin at $\mathbf{P}$ that is collapsing as well. More precisely, given a box $\mathbf{P}$, denote $\mathcal{C}_{\mathbf{P}}$ the convex cone of boxes $\mathbf{R}$ for which there exists a local correlation $\mathbf{L} \in \mathcal{L}$ such that $\mathbf{P} = \lambda\,\mathbf{R} + (1-\lambda)\,\mathbf{L}$, with $\lambda \in [0,1]$. We claim that if $\mathbf{P}$ is collapsing, then any $\mathbf{R} \in \mathcal{C}_{\mathbf{P}}$ is collapsing as well. Indeed, assume Alice and Bob are given copies of a box $\mathbf{R}$. Then, they can use shared randomness to produce the wanted box $\mathbf{L}$ and the wanted convex coefficient $\lambda$, so that they can generate the box $\mathbf{P}$ with the relation $\mathbf{P} = \lambda\,\mathbf{R} + (1-\lambda)\,\mathbf{L}$. Now, as $\mathbf{P}$ is collapsing, they have a protocol that collapses communication complexity, hence $\mathbf{R}$ is collapsing as well. See an illustration in Figure 6.6 (b). In the study of collapsing boxes, notice that it is standard to assume that shared randomness is a “free" resource—for instance Brassard, Buhrman, Linden, Méthot, Tapp, and Unger made this choice in [Bra+06] in their collapsing protocol.

Combining arguments from these last two paragraphs and the fact that Alice and Bob can make a convex combination of boxes using shared randomness, we deduce a sufficient criterion for a box to collapse communication complexity:

> **Proposition 6.4** (Collapsing orbit) — *Let* $\mathbf{P}$ *be a box in* $\mathcal{NS}$*. If there exists a box* $\mathbf{Q} \in \mathrm{Conv}\left(\mathcal{L} \cup \bigcup_{\mathsf{W}} \mathrm{Orbit}_{\mathsf{W}}(\mathbf{P})\right)$ *that collapses communication complexity, then* $\mathbf{P}$ *is collapsing as well. See Figure 6.6* (c).

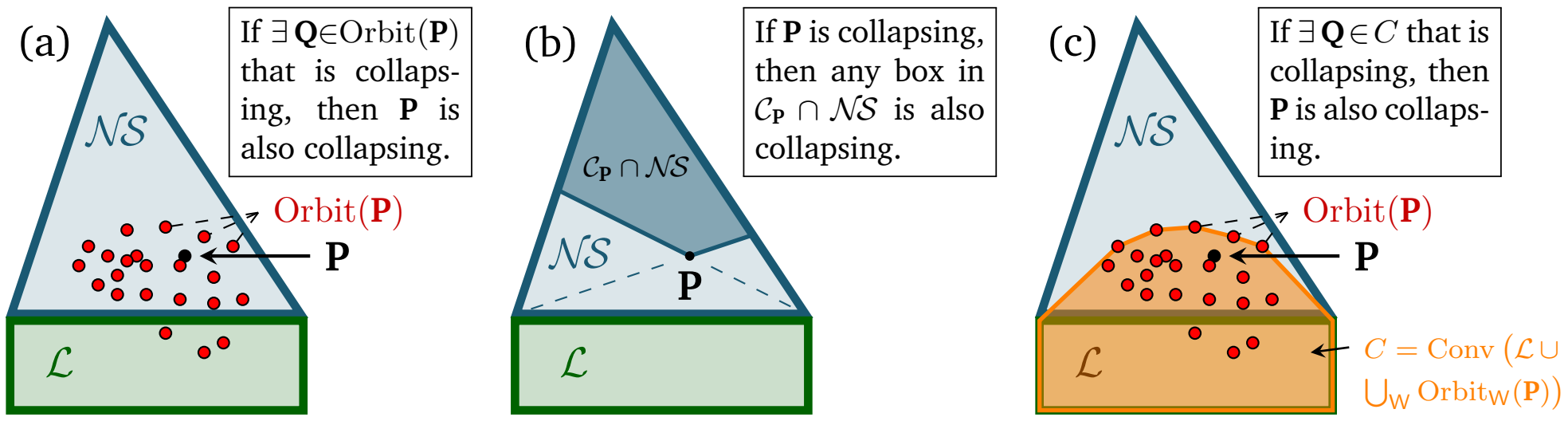

**Figure 6.6 —** *Orbits that collapse communication complexity.*

### 6.3.3 *Case Study: Orbit of* $\mathsf{W}_{\mathrm{BS}}$

In this subsection, we focus our attention on the wiring $\mathsf{W}_{\mathrm{BS}}$ from Brunner and Skrzypczyk [BS09]. Here, we simply write $\boxtimes := \boxtimes_{\mathsf{W}_{\mathrm{BS}}}$ the corresponding box multiplication. Define the *shared randomness* box as $\mathbf{SR} := (\mathbf{P_{00}} + \mathbf{P_{11}})/2$; it is designed to output a couple $(a, b)$ such that $a = b$ uniformly and independently of the inputs. From the multiplication table in Figure 6.4, one can see that the 2-dimensional affine space $\mathcal{A} := \mathrm{Aff}\{\mathbf{PR}, \mathbf{SR}, \mathbf{I}\}$ is stable under $\boxtimes$. As a consequence, the orbit $\mathrm{Orbit}(\mathbf{P})$ of any box $\mathbf{P}$ in $\mathcal{A}$ is itself included in $\mathcal{A}$, and as $\mathcal{A}$ is two-dimensional, it is particularly easy to draw the orbit of a box in that case. We represent an orbit in Figure 6.7.

**Geometry of the Orbits.** By definition of the affine space $\mathcal{A}$, any box $\mathbf{A} \in \mathcal{A}$ can be uniquely written as $\mathbf{A} = c_1(\mathbf{A})\,\mathbf{PR} + c_2(\mathbf{A})\,\mathbf{SR} + c_3(\mathbf{A})\,\mathbf{I}$ for some real coefficients $c_i(\mathbf{A})$ that sum to $1$, called *convex coordinates* of $\mathbf{A}$ in the affine basis $\{\mathbf{PR}, \mathbf{SR}, \mathbf{I}\}$. An interesting aspect of considering convex coordinates is that it gives a simple characterization of the parallelism property of lines:

$$\forall \mathbf{A}, \mathbf{B} \in \mathcal{A}, \qquad \mathrm{Aff}\{\mathbf{A}, \mathbf{B}\} \,||\, \mathrm{Aff}\{\mathbf{PR}, \mathbf{SR}\} \quad \Longleftrightarrow \quad c_3(\mathbf{A}) = c_3(\mathbf{B})\,. \tag{6.16}$$ [1]

Moreover, in our case, we have an additional interesting property of the third convex coordinate:

[1] Indeed, for $\mathbf{A} \neq \mathbf{B} \in \mathcal{A}$ whose convex coefficients are respectively $a_1, a_2, a_3$ and $b_1, b_2, b_3$, saying that the line $\mathrm{Aff}\{\mathbf{A}, \mathbf{B}\}$ is parallel to the line $\mathrm{Aff}\{\mathbf{PR}, \mathbf{SR}\}$ is equivalent to knowing that there exists a scalar $\lambda \in \mathbb{R}^*$ such that $\mathbf{A} - \mathbf{B} = \lambda\,(\mathbf{PR} - \mathbf{SR})$, *i.e.* there exists $\lambda \in \mathbb{R}^*$ such that $a_1 - b_1 = \lambda$ and $a_2 - b_2 = -\lambda$ and $a_3 - b_3 = 0$, *i.e.* we have two equations: $a_1 + a_2 = b_1 + b_2$ and $a_3 = b_3$. Finally, using the normalization condition $\sum_i a_i = \sum_j b_j = 1$, we see that these two equations are equivalent to simply imposing $a_3 = b_3$, as claimed.

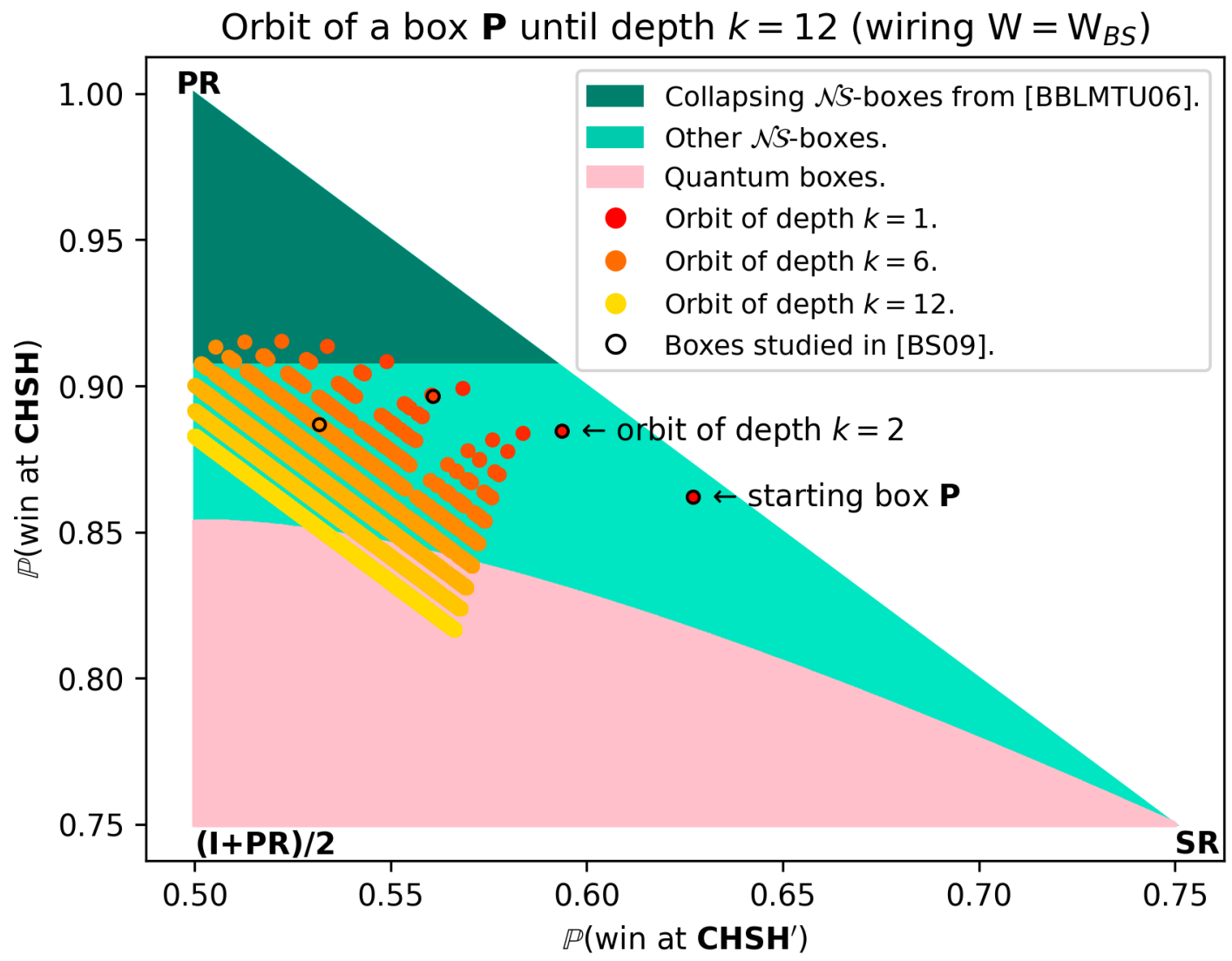

**Figure 6.7 —** *Example of a box orbit, drawn for depth up to $k = 12$, with $\mathrm{W_{BS}}$ from [BS09]. The quantum area $\mathcal{Q}$ (in pink) is drawn using formulas from [Mas03]. Dark green represents the collapsing area that was found by Brassard* et. al. *in [Bra+06], which consists of all the boxes with* CHSH*-value higher than $\frac{3+\sqrt{6}}{6} \approx 0.91$. The orbit is drawn in yellow and orange dots — observe that it intersects the collapsing area in dark green, so Proposition 6.4 tells us that the starting box $\mathbf{P}$ is collapsing. The black circles represent the boxes that were studied in [BS09], doing "pairwise" multiplications: $\mathbf{P}$, $\mathbf{P} \boxtimes \mathbf{P}$, $(\mathbf{P} \boxtimes \mathbf{P}) \boxtimes (\mathbf{P} \boxtimes \mathbf{P})$, etc... Each iteration is the wiring of two copies of the previous iteration, it gives a subset of our orbit. As displayed in the drawing and detailed in the proof of Theorem 6.17, our method allows us to find a larger set of boxes $\mathbf{P}$ that are collapsing.*

**Lemma 6.5 —** *The function* $1 - c_3(\cdot)$ *is multiplicative:*

$$\forall \mathbf{A}, \mathbf{B} \in \mathcal{A}, \qquad 1 - c_3(\mathbf{A} \boxtimes \mathbf{B}) = \big(1 - c_3(\mathbf{A})\big)\,\big(1 - c_3(\mathbf{B})\big)\,.$$

*Proof.* The multiplication table induced by the wiring $W_{BS}$ [BS09] is:

| P \ Q | **PR** | **SR** | **I** |
|---|---|---|---|
| **PR** | **PR** | **PR** | **I** |
| **SR** | $\frac{1}{2}\,\mathbf{PR} + \frac{1}{2}\,\mathbf{SR}$ | **SR** | **I** |
| **I** | $\frac{1}{4}\,\mathbf{PR} - \frac{1}{4}\,\mathbf{SR} + \mathbf{I}$ | **I** | **I** |

, (6.17)

where each cell displays the result of $\mathbf{P} \boxtimes \mathbf{Q}$. For $\mathbf{A}, \mathbf{B} \in \mathcal{A}$ whose coefficients $c_i$ are denoted $a_1, a_2, a_3$ and $b_1, b_2, b_3$ for the sake of readability, we use the bilinearity of the product $\boxtimes$ and we get:

$$\begin{aligned} \mathbf{A} \boxtimes \mathbf{B} \;=\; & \Big[a_1\, b_1 + a_1\, b_2 + \tfrac{1}{2}\, a_2\, b_1 + \tfrac{1}{4}\, a_3\, b_1\Big]\, \mathbf{PR} + \Big[\tfrac{1}{2}\, a_2\, b_1 + a_2\, b_2 - \tfrac{1}{4}\, a_3\, b_1\Big]\, \mathbf{SR} \\ & + \Big[a_1\, b_3 + a_2\, b_3 + a_3\, b_1 + a_3\, b_2 + a_3\, b_3\Big]\, \mathbf{I}\,. \end{aligned}$$

Hence, using the normalization property of coefficients $\sum_i a_i = \sum_j b_j = 1$, the third coefficient simplifies as $b_3 + a_3(1 - b_3)$, which is equal to $1 - (1 - a_3)(1 - b_3)$ as wanted. ∎

Now, interestingly, we observe that the points of a given $k$-orbit are all aligned, and we even know the equation of the line:

**Theorem 6.6** (Alignment) **—** *For any* $k \geqslant 1$ *and* $\mathbf{P} \in \mathcal{A}$*, the points of* $\mathrm{Orbit}^{(k)}(\mathbf{P})$ *are all aligned on a line* $\mathfrak{L}_k$ *whose expression in convex coordinates is given by:*

$$\mathfrak{L}_k \;:=\; \Big\{\mathbf{A} \in \mathcal{A} \,:\, c_3(\mathbf{A}) = 1 - \big(1 - c_3(\mathbf{P})\big)^k\Big\}\,.$$

*Proof.* We prove by induction on $k \geqslant 1$ that $\mathrm{Orbit}^{(k)} \subseteq \mathfrak{L}_k$. For $k = 1$, the 1-orbit contains only one element, namely $\mathbf{P}$, which obviously satisfies $c_3(\mathbf{P}) = 1 - \big(1 - c_3(\mathbf{P})\big)$, so $\mathbf{P}$ indeed belongs to $\mathfrak{L}_1$. Now, assume the result holds *until* some integer $k \geqslant 1$, and let $\mathbf{Q} \in \mathrm{Orbit}^{(k)}$. By definition, the box $\mathbf{Q}$ decomposes as $\mathbf{Q} = \mathbf{Q}_1 \boxtimes \mathbf{Q}_2$, for some $\mathbf{Q}_1 \in \mathrm{Orbit}^{(\ell)}$ and $\mathbf{Q}_2 \in \mathrm{Orbit}^{(k-\ell)}$ for some $1 \leqslant \ell \leqslant k-1$. By the induction hypothesis, we know that $c_3(\mathbf{Q}_1) = 1 - \big(1 - c_3(\mathbf{P})\big)^{\ell}$ and $c_3(\mathbf{Q}_2) = 1 - \big(1 - c_3(\mathbf{P})\big)^{k-\ell}$. Then using Lemma 6.5, we obtain:

$$
\begin{aligned}
c_3(\mathbf{Q}) &= 1 - \big(1 - c_3(\mathbf{Q}_1)\big)\big(1 - c_3(\mathbf{Q}_2)\big) \\
&= 1 - \big(1 - c_3(\mathbf{P})\big)^{\ell}\big(1 - c_3(\mathbf{P})\big)^{k-\ell} \\
&= 1 - \big(1 - c_3(\mathbf{P})\big)^{k},
\end{aligned}
$$

which means that $\mathbf{Q}$ belongs to the line $\mathfrak{L}_k$. ■

As a consequence, we see that all the points of the $k$-orbit have the same third convex coefficient, so using the equivalence given in eq. (6.16), we obtain:

**Corollary 6.7** (Parallelism) — *The supporting line $\mathfrak{L}_k$ of all the orbits $\mathrm{Orbit}^{(k)}$ are parallel to the diagonal line $\mathfrak{L}_D := \mathrm{Aff}\{\mathbf{PR}, \mathbf{SR}\}$:*

$$\forall k \geqslant 1, \qquad \mathrm{Orbit}^{(k)} \,||\, \mathfrak{L}_D\,.$$

*In particular, all the orbits are parallel to each other:*

$$\forall k, \ell \geqslant 1, \quad \mathrm{Orbit}^{(k)} \,||\, \mathrm{Orbit}^{(\ell)}\,.$$ ■

Moreover, looking closely at the sequence of coefficients $1 - \big(1 - c_3(\mathbf{P})\big)^{k}$ and noticing that the diagonal line $\mathfrak{L}_D$ is defined by the equation $c_3(\mathbf{A}) = 0$, we see that:

**Corollary 6.8** (The Orbits Move to the Left) — *Assume $\mathbf{P} \notin \mathfrak{L}_D$. Then the orbits are more and more distant from the diagonal line as $k$ grows. Moreover, the sequence of lines $(\mathfrak{L}_k)_k$ tends to the line $\mathfrak{L}_\infty$ defined by the equation $c_3(\mathbf{A}) = 1$, which is exactly the line passing through $\mathbf{I}$ and parallel to the diagonal $\mathfrak{L}_D$.* ■

**Computing the "Highest" Box of an Orbit.** It takes a lot of computational time to draw $k$-orbits of a box $\mathbf{P}$ as $k$ grows, since it requires to compute $\frac{1}{k}\binom{2k-2}{k-1}$ elements (Catalan number), which grows exponentially. However, our goal is not to compute the whole orbit, but simply to determine whether or not the orbit intersects the known collapsing area (dark green). To that end, one may notice that it is enough to compute the "highest" box of each $k$-orbit in the $y$-coordinate (see Figure 6.7) and to check whether those "highest" boxes intersect the collapsing area (dark green area). This is the purpose of the following proposition, which displays a simple expression of the "highest" box of each $k$-orbit, and which allows much faster tests of a box $\mathbf{P}$ being collapsing or not without computing all the points of the orbit. We prove this result only in a subset of the orbit, that we call *tilted orbit*, which is easier to manipulate in inductions, and which is defined by $\widetilde{\mathrm{Orbit}}^{(1)}(\mathbf{P}) := \{\mathbf{P}\}$ and for $k \geqslant 2$:

$$\begin{aligned}\widetilde{\mathrm{Orbit}}^{(k)} &:= \left(\mathbf{P} \boxtimes \widetilde{\mathrm{Orbit}}^{(k-1)}\right) \cup \left(\widetilde{\mathrm{Orbit}}^{(k-1)} \boxtimes \mathbf{P}\right) \\ &= \bigcup_{\ell \in \{1, k-1\}} \widetilde{\mathrm{Orbit}}^{(\ell)} \boxtimes \widetilde{\mathrm{Orbit}}^{(k-\ell)} \subseteq \mathrm{Orbit}^{(k)} .\end{aligned}$$

Note that the cardinality of that set is $\#\widetilde{\mathrm{Orbit}}^{(k+1)} = 2^k$, up to multiplicity. We call CHSH-value the $y$-coordinate, indicating how "high" is a box:

$$\mathrm{CHSH}(\mathbf{P}) := \mathbb{P}\big(\text{win at CHSH}\big) = \frac{1}{4} \sum_{a \oplus b = xy} \mathbf{P}(a, b \,|\, x, y) .$$

We say that a tilted orbit *distills the* CHSH-*value* if it contains a box $\mathbf{Q}$ such that $\mathrm{CHSH}(\mathbf{Q}) \geqslant \mathrm{CHSH}(\mathbf{P})$. In the following theorem, we present the expression of the best parenthesization in terms of CHSH-value, which explains the numerical observation reported in [EWC23a, Supplementary Material II]:

**Theorem 6.9** (Highest Box) — *Let* $\mathbf{P} \in \widetilde{\mathcal{A}}$ *be a box, and let* $k \geqslant 2$ *an integer such that the tilted* $(k-1)$*-orbit distills the* CHSH*-value. Then the highest* CHSH*-value of* $\widetilde{\mathrm{Orbit}}^{(k)}(\mathbf{P})$ *is achieved at a box whose expression is the product of* $k$ *times* $\mathbf{P}$ on the right*:*

$$\mathbf{P}^{\boxtimes k} := \Big(\big((\mathbf{P} \boxtimes \mathbf{P}) \boxtimes \mathbf{P}\big) \cdots\Big) \boxtimes \mathbf{P} \in \underset{\mathbf{Q} \in \widetilde{\mathrm{Orbit}}^{(k)}(\mathbf{P})}{\mathrm{argmax}} \ \mathrm{CHSH}(\mathbf{Q}) .$$

*Proof.* See Section 6.3.5. ■

**Remark 6.10 —** In the Ph.D. thesis of Giorgos Eftaxias [Eft22, Subsection 5.5.1], the author presents three types of architectures of wirings:

- the exponential architecture, used in [BS09], that we call here pair-wise multiplication;
- the linear architecture, which is the same type as in Theorem 6.9; and
- the Fibonacci architecture.

They present some subsets of $\mathcal{NS}$ for which the linear architecture seems to be the best one among the three [Eft22, Remark 1], and some other subsets of $\mathcal{NS}$ for which it is the Fibonacci architecture [Eft22, Subsection 5.F.2].

**Conjecture 6.11** (Dyck Paths) **—** *We conjecture that the same result actually holds without the tilde,* i.e. *the right multiplication* $\mathbf{P}^{\boxtimes k}$ *gives the highest* CHSH-*value of* $\mathrm{Orbit}^{(k)}(\mathbf{P})$*, as observed numerically. An idea of the proof could be to use Dyck paths. Each time we open/close a parenthesis, the path goes up/down respectively, which produces a certain Dyck path. The statement to be proved is that each time we convert a* $\vee$ *into a* $\wedge$*, the* CHSH-*value is non-decreasing. Then, we would have that the best Dyck path is necessarily the one that always goes up first and then always goes down, which corresponds to the multiplication of boxes on the right.*

**Collapse of Communication Complexity.** In Theorem 6.17, we show that these techniques allow us to find new collapsing boxes.

### 6.3.4 *Other Examples of Orbits*

In Section 6.3.3, we specifically studied the orbit of the wiring $\mathsf{W}_{\mathrm{BS}}$ in the slice of $\mathcal{NS}$ passing through the boxes **PR**, **SR**, **I**. This is slightly restrictive, this is why here, we comment on examples of other orbits in three different ways: (i) it is possible to study the same wiring $\mathsf{W}_{\mathrm{BS}}$ but in different slices of $\mathcal{NS}$; (ii) it is possible to study another wiring than $\mathsf{W}_{\mathrm{BS}}$ but to keep the same slice as in Section 6.3.3; (iii) it is possible to change both the wiring and the slice. Find several illustrations below.

**(i)** We keep the wiring $W_{BS}$ and we consider a different slice of $\mathcal{NS}$, the one passing through $\mathbf{PR}, \mathbf{P_{00}}, \mathbf{P_{11}}$. We draw two examples of such an orbit in Figure 6.8, with two different starting boxes. We observe that both of them seem to recover the alignment and parallelism properties that we showed in Theorem 6.6 and Corollary 6.7, where here, what we called the "diagonal line" $\mathfrak{L}_D$ in Section 6.3.3 is known the line passing through $\mathbf{PR}$ and $\mathbf{SR} = \frac{1}{2}(\mathbf{P_{00}}+\mathbf{P_{11}})$. We show in Corollary 6.20 that all the boxes of this triangle are actually collapsing, except the ones in the segment $\mathrm{Conv}\{\mathbf{P_{00}}, \mathbf{P_{11}}\}$, drawn in pink.

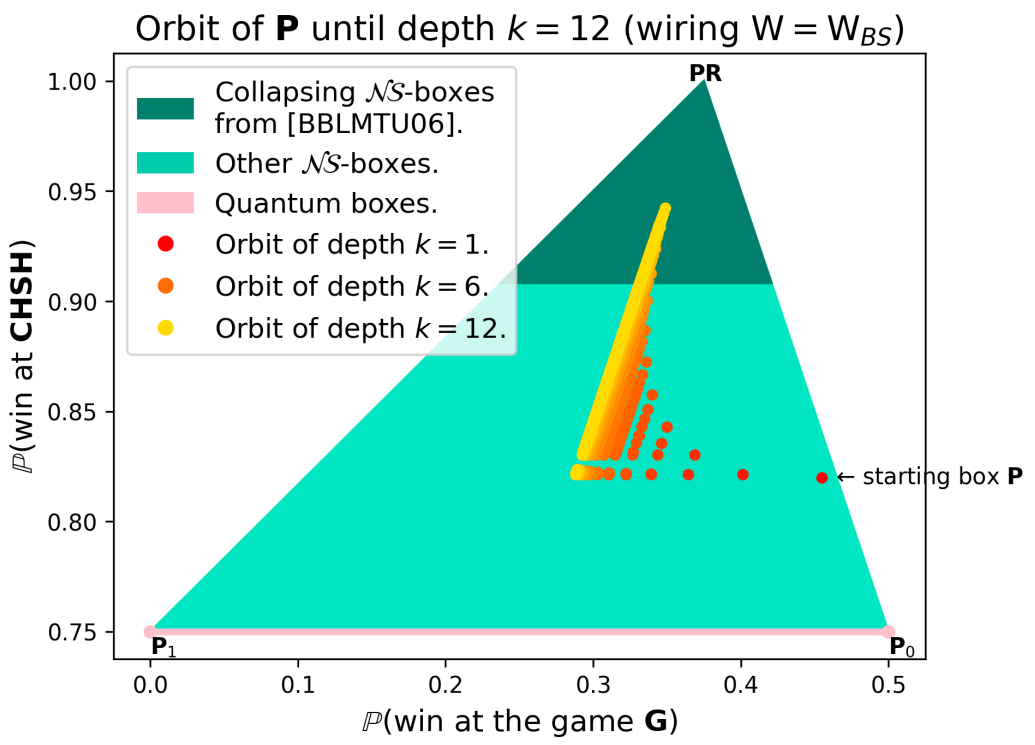

**Figure 6.8** — *Orbit of* $W_{BS}$ *in a different slice than in Section 6.3.3: Here, we consider the slice of* $\mathcal{NS}$ *passing through* $\mathbf{PR}$, $\mathbf{P}_{00}$, *and* $\mathbf{P}_{11}$. *We represent the orbit with two different starting boxes. Each orbit is drawn with depth going until* $k = 12$. *The game* $\mathcal{G}$ *is defined by the winning rule* $a = 0$ *and* $b = y$. *Notice that we give a proof based on CC that this triangle* $\mathrm{Conv}\{\mathbf{PR}, \mathbf{P}_{00}, \mathbf{P}_{11}\}$ *is a quantum void in Corollary 6.22, which is why the only quantum boxes in this triangle are actually local.*

**(ii)** Among the "typical" wirings defined in page 81, the only ones that stabilize the plane $\mathrm{Aff}\{\mathbf{PR}, \mathbf{SR}, \mathbf{I}\}$ are $W_\oplus$ and $W_{BS}$, see [Bot+24a, Appendix C]. This is why, for these two wirings, the orbits are contained in a plane and we can conveniently draw them. The orbit of $W_\oplus$ is drawn below in item (a). We observe that each $k$-orbit contains only one element, which is not surprising since we know from Figure 6.5 that its induced algebra is associative, meaning that the choice of parenthesization does not lead to a different result. In the other items below, we add an illustration of the orbit for three other wirings. Surprisingly, we observe that these three orbits seem to behave in the same way as the orbit of $\mathsf{W}_{\mathrm{BS}}$ (Figures 6.7 and 6.8).

**(iii)** We also illustrate the slice **PR**, $\mathbf{P}_{00}$, $\mathbf{P}_{11}$ in each item below. Interestingly, we observe that the alignment and parallelism properties seem to hold again in those cases. Moreover, we see that item (d) distills the CHSH-value better than the other items in this slice independently of the choice of parenthesization.

**Other Examples of Orbits.** Here are examples of orbits to illustrate the above discussion, using different wirings, each time in two different slices of $\mathcal{NS}$. Each orbit is drawn with depth going until $k = 12$. The game G is defined by the winning rule $a = 0$ and $b = y$.

(a) For W⊕ (see definition in page 81):

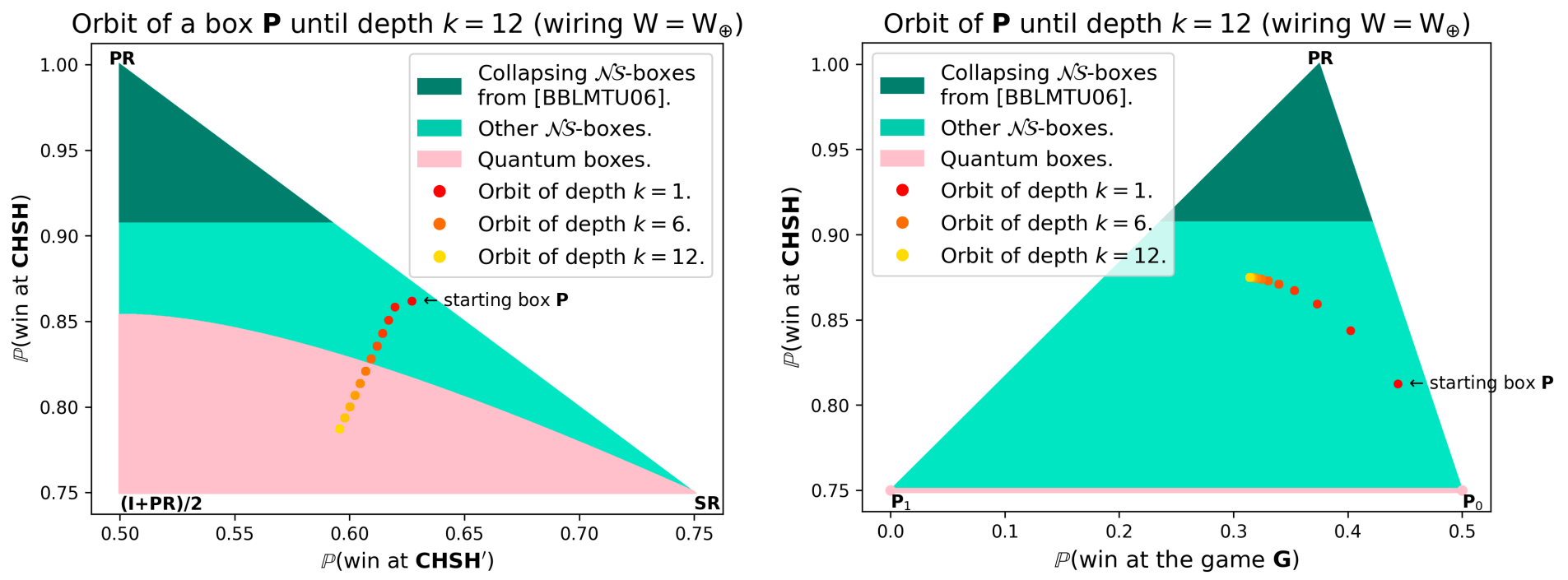

(b) For $\mathsf{W}_{(\mathrm{b})}$:=[0.,0.,1.,1.,0.,0.,1.,1.,0.,0.,0.,1.,0.,0.,0.,1.,1.,0.,0.,1.,1.,0.,0., 1.,1.,0.,0.,1.,1.,0.,0.,1.]:

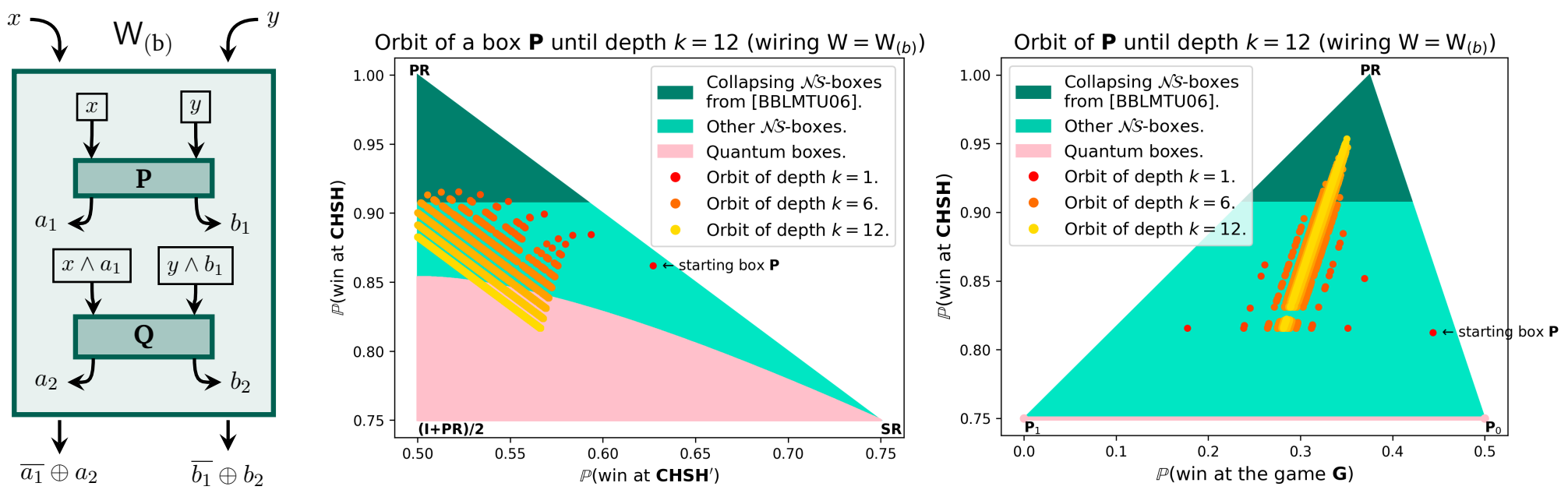

(c) For $W_{(c)}$:=[0.,0.,1.,1.,0.,0.,1.,1.,0.,0.,1.,0.,0.,0.,1.,0.,1.,0.,0.,1.,1.,0.,0., 1.,1.,0.,0.,1.,1.,0.,0.,1.]:

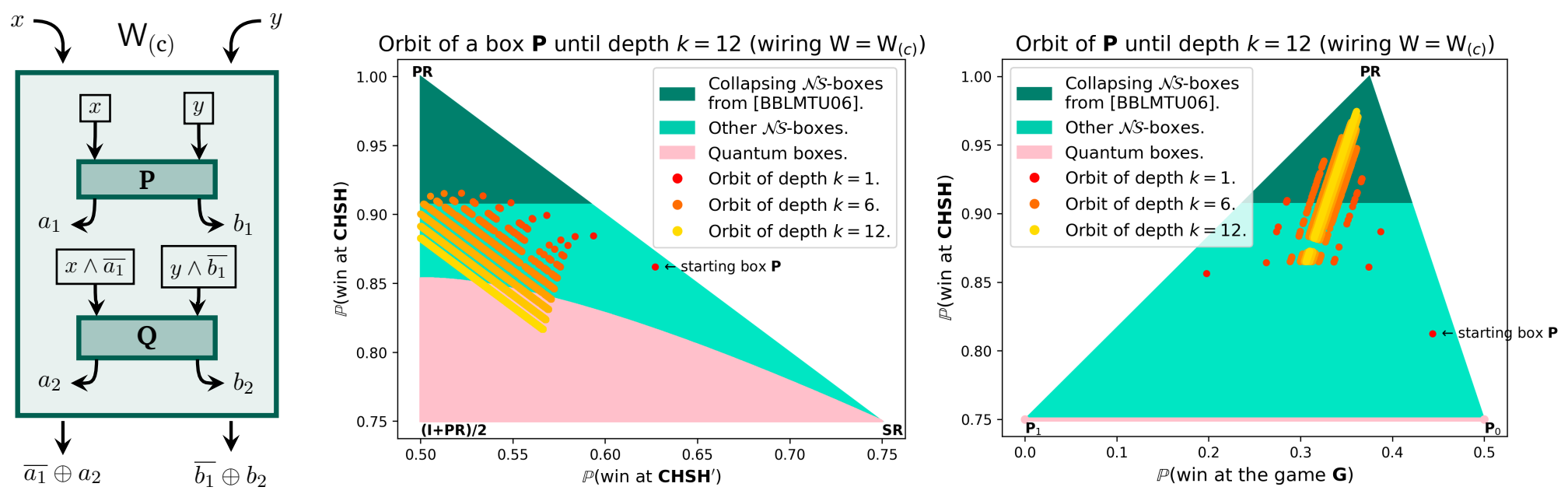

(d) For $W_{(d)}$ :=[0.,0.,1.,1.,0.,0.,1.,1.,0.,0.,1.,0.,0.,0.,1.,0.,0.,1.,1.,0.,0.,1.,1., 0.,0.,1.,1.,0.,0.,1.,1.,0.]:

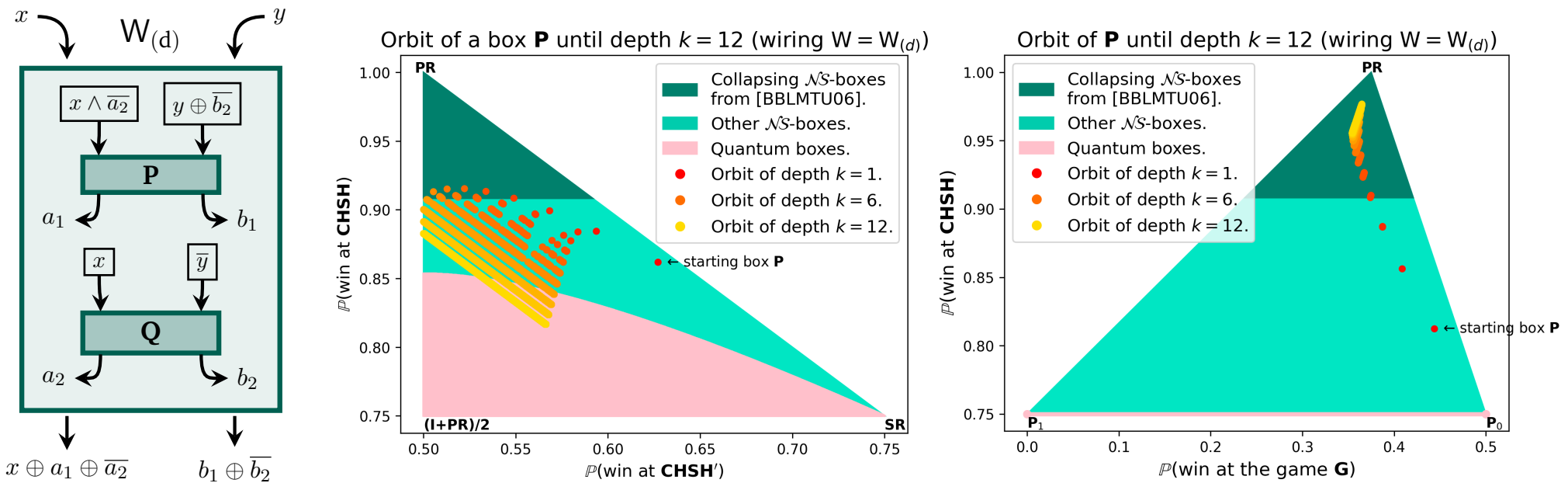

### 6.3.5 *Proof of Theorem 6.9*

Recall that $\mathbf{SR} := (\mathbf{P_{00}} + \mathbf{P_{11}})/2$ is the *shared randomness* box. Given a non-signaling box $\mathbf{P} \in \mathcal{NS}$, its CHSH- and CHSH′-values are defined as follows:

$$\mathrm{CHSH}(\mathbf{P}) := \frac{1}{4} \sum_{a\oplus b = xy} \mathbf{P}(a,b\,|\,x,y)\,, \qquad \mathrm{CHSH}'(\mathbf{P}) := \frac{1}{4} \sum_{\substack{a\oplus b \\ =(x\oplus 1)(y\oplus 1)}} \mathbf{P}(a,b\,|\,x,y)\,.$$

For example, we have $\mathrm{CHSH}(\mathbf{PR}) = 1$ and $\mathrm{CHSH}(\mathbf{SR}) = \frac{3}{4}$ and $\mathrm{CHSH}(\mathbf{I}) = \frac{1}{2}$. Denote $\mathcal{A}$ the affine space $\mathcal{A} := \mathrm{Aff}\{\mathbf{PR}, \mathbf{SR}, \mathbf{I}\}$, and denote $\widetilde{\mathcal{A}} \subseteq \mathcal{A}$ the set of boxes $\mathbf{P}$ in the convex hull $\mathrm{Conv}\{\mathbf{PR}, \mathbf{SR}, \mathbf{I}\}$ whose CHSH-value is $\geqslant 3/4$. We will prove our results in $\widetilde{\mathcal{A}}$; by the symmetry of the problem, similar results also hold in other areas, such as $2\mathbf{I} - \widetilde{\mathcal{A}}$ the symmetric of $\widetilde{\mathcal{A}}$ by $\mathbf{I}$.

**Lemma 6.12** (Multiplying by $\mathbf{P}$ Preserves the CHSH-Value Order) — *Let $\mathbf{P} \in \widetilde{\mathcal{A}}$, and let $\mathbf{Q} \neq \mathbf{R} \in \mathcal{A}$ such that the line $\mathrm{Aff}\{\mathbf{Q}, \mathbf{R}\}$ is parallel to the diagonal line $\mathfrak{L}_D := \mathrm{Aff}\{\mathbf{PR}, \mathbf{SR}\}$. We have:*

$$\mathsf{CHSH}(\mathbf{Q}) \geqslant \mathsf{CHSH}(\mathbf{R}) \quad \Longrightarrow \quad \begin{cases} \mathsf{CHSH}(\mathbf{Q} \boxtimes \mathbf{P}) \geqslant \mathsf{CHSH}(\mathbf{R} \boxtimes \mathbf{P}) \,, \\ \mathsf{CHSH}(\mathbf{P} \boxtimes \mathbf{Q}) \geqslant \mathsf{CHSH}(\mathbf{P} \boxtimes \mathbf{R}) \,. \end{cases}$$

*Proof.* As the box $\mathbf{P}$ lies in $\widetilde{\mathcal{A}}$, it is of the form $\mathbf{P} = p_1\, \mathbf{PR} + p_2\, \mathbf{SR} + (1-p_1-p_2)\, \mathbf{I}$ for some coefficients $p_1, p_2 \geqslant 0$ such that $p_1 + p_2 \leqslant 1$. Rewrite it as $\mathbf{P} = \left(\begin{smallmatrix} p_1 \\ p_2 \end{smallmatrix}\right)$, and similarly denote $\mathbf{Q} = \left(\begin{smallmatrix} q_1 \\ q_2 \end{smallmatrix}\right)$ and $\mathbf{R} = \left(\begin{smallmatrix} r_1 \\ r_2 \end{smallmatrix}\right)$ for some coefficients $q_i, r_j \in \mathbb{R}$. By the parallelism assumption, vectors $\mathbf{Q} - \mathbf{R}$ and $\mathbf{PR} - \mathbf{SR}$ have to be colinear, *i.e.* there must exist some $\lambda \in \mathbb{R}^*$ such that $\mathbf{Q} - \mathbf{R} = \lambda(\mathbf{PR} - \mathbf{SR}) = \lambda\left(\begin{smallmatrix} 1 \\ -1 \end{smallmatrix}\right)$, so we may rewrite the second coefficient of $\mathbf{R}$ as $r_2 = q_1 + q_2 - r_1$. With this notation, we can use the linearity of the function $\mathsf{CHSH}(\cdot)$ to see that condition $\mathsf{CHSH}(\mathbf{Q}) \geqslant \mathsf{CHSH}(\mathbf{R})$ simplifies to $(q_1 - r_1) \geqslant 0$:

$$\begin{aligned} \mathsf{CHSH}(\mathbf{Q}) - \mathsf{CHSH}(\mathbf{R}) &= (q_1 - r_1) \times \mathsf{CHSH}(\mathbf{PR}) + (q_2 - r_2) \times \mathsf{CHSH}(\mathbf{SR}) \\ &\qquad + \Big((1-q_1-q_2) - (1-r_1-r_2)\Big) \times \mathsf{CHSH}(\mathbf{I}) \,, \\ &= (q_1 - r_1) \times 1 - (q_1 - r_1) \times \tfrac{3}{4} + 0 \times \tfrac{1}{2} \,, \\ &= \tfrac{1}{4}\,(q_1 - r_1) \,. \end{aligned}$$

Now, using the multiplication table from Figure 6.4 and bilinearity of $\boxtimes$, we may compute the following expressions:

$$\begin{aligned} \mathsf{CHSH}(\mathbf{Q} \boxtimes \mathbf{P}) - \mathsf{CHSH}(\mathbf{R} \boxtimes \mathbf{P}) &= \tfrac{1}{8}(p_1 + 2p_2)(q_1 - r_1) \geqslant 0 \,, \\ \mathsf{CHSH}(\mathbf{P} \boxtimes \mathbf{Q}) - \mathsf{CHSH}(\mathbf{P} \boxtimes \mathbf{R}) &= \tfrac{1}{16}(1 - p_1 + p_2)(q_1 - r_1) \geqslant 0 \,, \end{aligned}$$

which gives the desired result. ■

**Lemma 6.13** (The Right Multiplication Gives Better CHSH-Value) — *For any $\mathbf{P} \in \widetilde{\mathcal{A}}$ and $\mathbf{Q} \in \widetilde{\mathrm{Orbit}}(\mathbf{P})$, we have:*

$$\mathsf{CHSH}(\mathbf{Q}) \geqslant \mathsf{CHSH}(\mathbf{P}) \quad \Longrightarrow \quad \mathsf{CHSH}\Big(\mathbf{Q} \boxtimes \mathbf{P}\Big) \geqslant \mathsf{CHSH}\Big(\mathbf{P} \boxtimes \mathbf{Q}\Big).$$

*Proof.* Use the coordinate system $(x, y)$ given by the $\mathsf{CHSH}'$- and $\mathsf{CHSH}$-values respectively in order to write $\mathbf{P}$ and $\mathbf{Q}$ as taking coordinates $(x_{\mathbf{P}}, y_{\mathbf{P}})$ and $(x_{\mathbf{Q}}, y_{\mathbf{Q}})$. For instance we have $\mathbf{PR} : (\tfrac{1}{2}, 1)$ and $\mathbf{SR} : (\tfrac{3}{4}, \tfrac{3}{4})$ and $\mathbf{I} : (\tfrac{1}{2}, \tfrac{1}{2})$.

Use the multiplication table from eq. (6.17) and apply the bilinearity of $\boxtimes$ in order to obtain the following expression:

$$\begin{aligned}&\mathrm{CHSH}\big(\mathbf{Q}\boxtimes\mathbf{P}\big)-\mathrm{CHSH}\big(\mathbf{P}\boxtimes\mathbf{Q}\big)\\&\qquad=\tfrac{1}{8}(12x_{\mathbf{P}}y_{\mathbf{Q}}-12y_{\mathbf{P}}x_{\mathbf{Q}}-7x_{\mathcal{P}}+7y_{\mathcal{P}}+7x_{\mathbf{Q}}-7y_{\mathbf{Q}})\;=:\;f_{\mathbf{P}}(x_{\mathbf{Q}},y_{\mathbf{Q}})\,.\end{aligned}$$

For any fixed $\mathbf{P}\in\widetilde{\mathcal{A}}$, we want to show that $f_{\mathbf{P}}(x_{\mathbf{Q}},y_{\mathbf{Q}})\geqslant 0$. By construction, we know that $\mathbf{P}\in\mathfrak{L}_1$ and $\mathbf{Q}\in\mathfrak{L}_k$ for some $k\geqslant 1$, so by Corollary 6.8 we have $x_{\mathbf{Q}}+y_{\mathbf{Q}}\leqslant x_{\mathbf{P}}+y_{\mathbf{P}}$, which we may rewrite as $x_{\mathbf{Q}}\leqslant x_{\mathbf{P}}+y_{\mathbf{P}}-y_{\mathbf{Q}}$. As $\mathbf{P}$ lies in $\widetilde{\mathcal{A}}$, we have $y_{\mathbf{P}}\geqslant\frac{3}{4}$, so the first partial derivative is non-positive: $\frac{\partial}{\partial x_{\mathbf{Q}}}f_{\mathbf{P}}(x_{\mathbf{Q}},y_{\mathbf{Q}})=\frac{1}{8}(7-12y_{\mathbf{P}})\leqslant-1/4\leqslant 0$, which means that the function $f_{\mathbf{P}}(\cdot,y_{\mathbf{Q}})$ is decreasing over $\mathbb{R}$ for any fixed $y_{\mathbf{Q}}$. It yields the following inequalities:

$$f_{\mathbf{P}}(x_{\mathbf{Q}},y_{\mathbf{Q}})\;\geqslant\;f_{\mathbf{P}}(x_{\mathbf{P}}+y_{\mathbf{P}}-y_{\mathbf{Q}},y_{\mathbf{Q}})\;=\;\tfrac{3}{2}\big(y_{\mathbf{Q}}-y_{\mathbf{P}}\big)\big(x_{\mathbf{P}}+y_{\mathbf{P}}-\tfrac{7}{6}\big)\;\geqslant\;0\,,$$

since both factors are non-negative: the first one is non-negative using the hypothesis $\mathrm{CHSH}(\mathbf{Q})\geqslant\mathrm{CHSH}(\mathbf{P})$, and the second one is non-negative using $x_{\mathbf{P}}\geqslant 1/2$ and $y_{\mathbf{P}}\geqslant 3/4$ since $\mathbf{P}\in\widetilde{\mathcal{A}}$. Hence $f_{\mathbf{P}}$ is non-negative and we obtain the wanted result. ■

Recall that the set $\widetilde{\mathrm{Orbit}}^{(k)}(\mathbf{P})$ is called the tilted $k$-orbit of the box $\mathbf{P}$ and contains some boxes $\mathbf{Q}$ that are generated by applying a wiring to copies of $\mathbf{P}$. We say that this tilted $k$-orbit *distills the* CHSH*-value* if it contains a box $\mathbf{Q}$ such that $\mathrm{CHSH}(\mathbf{Q})\geqslant\mathrm{CHSH}(\mathbf{P})$. In that distilling scenario, we can compute the expression of a box achieving the best CHSH-value:

*Proof (Theorem 6.9).* We prove the result by induction on $k\geqslant 2$. It is obviously true for $k=2$ since $\widetilde{\mathrm{Orbit}}^{(2)}(\mathbf{P})$ only contains $\mathbf{P}\boxtimes\mathbf{P}$. Now, fix $k\geqslant 2$ and assume $\mathrm{CHSH}(\mathbf{P}^{\boxtimes k})\geqslant\mathrm{CHSH}(\mathbf{Q})$ for any $\mathbf{Q}$ in the tilted $k$-orbit (induction hypothesis). Assume as well that $\mathrm{CHSH}(\mathbf{P}^{\boxtimes k})\geqslant\mathrm{CHSH}(\mathbf{P})$ (distillation hypothesis). We want to show that:

$$\mathrm{CHSH}(\mathbf{P}^{\boxtimes k+1})\geqslant\mathrm{CHSH}(\mathbf{Q}\boxtimes\mathbf{P})\quad\text{and}\quad\mathrm{CHSH}(\mathbf{P}^{\boxtimes k+1})\geqslant\mathrm{CHSH}(\mathbf{P}\boxtimes\mathbf{Q})\,,$$

for all $\mathbf{Q}$ in the tilted $k$-orbit. The first inequality follows from Lemma 6.12 using the relation $\mathbf{P}^{\boxtimes k+1}=\mathbf{P}^{\boxtimes k}\boxtimes\mathbf{P}$ and the induction assumption. For the other inequality, start from $\mathrm{CHSH}(\mathbf{P}^{\boxtimes k+1})=\mathrm{CHSH}(\mathbf{P}^{\boxtimes k}\boxtimes\mathbf{P})$ and apply Lemma 6.13 in order to get $\geqslant\mathrm{CHSH}(\mathbf{P}\boxtimes\mathbf{P}^{\boxtimes k})$. Then conclude using Lemma 6.12 and the induction hypothesis in order to obtain $\geqslant\mathrm{CHSH}(\mathbf{P}\boxtimes\mathbf{Q})$ for any $\mathbf{Q}$ in the tilted $k$-orbit. ■

## 6.4 Numerical Optimization on the Set of Wirings

We saw in the previous section that, given a non-signaling box **P**, there may exist a wiring W that sufficiently distills the box **P** in order to collapse communication complexity. The question we address in this section is the following: if the box **P** is fixed, how to find a wiring W good enough to collapse communication complexity (when it is possible)? The difficulty is that, for each input $x, y \in \{0, 1\}$, there are $82$ possible deterministic wirings [SPG06], leading to a total number of $82^4 \approx 10^8$ possible deterministic wirings. So a naive discrete optimization over deterministic wirings seems inefficient. To that end, we present two optimization algorithms: (i) an algorithm that tests many different combinations of wirings and that is suitable for numerical simulations, and (ii) another one that finds a "uniform" collapsing wiring W in a whole region of boxes, which is appropriate for deriving an analytical proof (Section 6.5). This section might be skipped at first reading as it is more technical. See our GitHub page for the details of the algorithms [BC23a].

**Remark 6.14** (Comparison with [Bri+19; EWC23a]) — We now compare and contrast our methods with two recent works that also study optimization over wirings:

(i) In [Bri+19], the authors suggest reducing the $82^4$ possible deterministic wirings for Alice and Bob to only $3152$ by simply considering the ones that preserve the **PR** box, *i.e.* wirings W such that $\mathbf{PR} \boxtimes_{\mathsf{W}} \mathbf{PR} = \mathbf{PR}$, and then doing a discrete optimization over that smaller set. This smaller set encompasses for instance the wirings $\mathsf{W}_{\mathrm{BS}}$, $\mathsf{W}_{\mathrm{dist}}$ but discards $\mathsf{W}_{\oplus}$, $\mathsf{W}_{\wedge}$, $\mathsf{W}_{\vee\wedge}$ (see definitions in page 81); see also [EWC23a, Supplementary Material I] which mentions that even some optimal wirings are discarded. This technique allows them to analytically prove that many new areas of boxes are collapsing.

(ii) In [EWC23a], the authors use a mix of exhaustive search and linear programming. For each of Bob's $82^2$ extremal half-wirings, they apply linear programming to optimize Alice's half-wiring, and then they select the best pair of half-wirings. This allows them to numerically find optimal wirings for any pair of boxes, which leads them to discover new collapsing boxes.

(iii) In our work, we use an efficient variant of the Gradient Descent algorithm, based on Line Search methods, frequent resets, and parallel descents. A limitation in the method from [Bri+19] could come from the fact that many wirings are discarded; this is why we choose to take our feasible set to be the entire set of mixed wirings $\mathcal{W} \subseteq [0,1]^{32}$. In [EWC23a, Supplementary Material II], the authors implement a sequence of different optimal wirings, which we do similarly in what we call later Task A, but we also implement a uniform version of it in Task B, which allows us to find a single optimal wiring for a whole region of boxes (instead of a sequence of wirings) and then to prove by hand the collapse of communication complexity for those boxes. In this manner, we recover both the numerical results of [EWC23a] (Section 6.5.1) and the analytical results of [Bri+19] (Section 6.5.3).

### 6.4.1 *Goals of the Algorithms*

We present two possible algorithm tasks:

**Task A: Adaptive Wiring.** To prove that a box **P** is collapsing, a particular case of Proposition 6.4 says that it is enough to find a finite sequence of wirings $(\mathsf{W}_1, \dots, \mathsf{W}_N)$ such that the following box is collapsing:

$$\mathbf{P}_{N+1} := \Big( \big( (\mathbf{P} \underset{\mathsf{W}_1}{\boxtimes} \mathbf{P}) \underset{\mathsf{W}_2}{\boxtimes} \mathbf{P} \big) \underset{\mathsf{W}_3}{\boxtimes} \dots \Big) \underset{\mathsf{W}_N}{\boxtimes} \mathbf{P} \,.$$

Note that we need to specify the parenthetization because the different products $\boxtimes_{\mathsf{W}_i}$ are potentially non-associative. Among the numerous possibilities, we choose the parenthesization on the left because it is easy to implement and because it is the best one when the wiring is $\mathsf{W}_{\mathsf{BS}}$, see Theorem 6.9. This algorithm will consist in an iterative construction of the sequence $(\mathsf{W}_i)_i$: first, find a wiring $\mathsf{W}_1$ such that the CHSH-value of the box $\mathbf{P}_2 := \mathbf{P} \boxtimes_{\mathsf{W}_1} \mathbf{P}$ is the highest possible, then find $\mathsf{W}_2$ such that the CHSH-value of the box $\mathbf{P}_3 := \mathbf{P}_2 \boxtimes_{\mathsf{W}_2} \mathbf{P}$ is the highest possible, so on and so forth until the $N$-th iteration. If the CHSH-value of the box $\mathbf{P}_{N+1}$ is above the threshold $\frac{3+\sqrt{6}}{6} \approx 0.91\%$, we know that communication complexity collapses [Bra+06], so the starting box **P** is collapsing as well. Otherwise, we cannot conclude whether **P** is collapsing or not.

**Task B: Constant Wiring.** The goal of this algorithm is essentially the same as the first one, but we add a strong constraint: we want all the $\mathsf{W}_i$ to be the same wiring $\mathsf{W}$:

$$\Big(\big((\mathbf{P} \underset{\mathsf{W}}{\boxtimes} \mathbf{P}) \underset{\mathsf{W}}{\boxtimes} \mathbf{P}\big) \underset{\mathsf{W}}{\boxtimes} \dots\Big) \underset{\mathsf{W}}{\boxtimes} \mathbf{P} \;=:\; \mathbf{P}^{\boxtimes_{\mathsf{W}} N+1}\,.$$

In that sense, this is a “uniform" version of the first algorithm. The interest of this algorithm is that it helps to give analytical proofs (Section 6.5): if the value $\mathrm{CHSH}(\mathbf{P}^{\boxtimes_{\mathsf{W}} N})$ is above the threshold $\frac{3+\sqrt{6}}{6} \approx 0.91\%$ for some $N$, then by continuity of $\boxtimes_{\mathsf{W}}$, there is an open neighborhood around $\mathbf{P}$ such that for any $\mathbf{Q}$ close enough to $\mathbf{P}$ we also have that $\mathrm{CHSH}(\mathbf{Q}^{\boxtimes_{\mathsf{W}} N})$ is above the threshold, and therefore the whole neighborhood of $\mathbf{P}$ is collapsing. This technique will help to discover wide collapsing areas and to provide analytical proofs by hand.

### 6.4.2 *Toy Example* ($N = 1$)

In this subsection, we treat the case when there is only one product $\boxtimes_{\mathsf{W}}$ between two boxes $\mathbf{Q}, \mathbf{P} \in \mathcal{NS}$. We detail the maximization algorithm we use: Projected Gradient Descent. The optimization problem consists in finding $\mathsf{W}^*$ as follows:

$$\mathsf{W}^* \;=\; \underset{\mathsf{W}\in\mathcal{W}}{\mathrm{argmax}} \;\; \Phi(\mathsf{W})\,. \tag{6.18}$$

where the objective function is $\Phi(\mathsf{W}) := \mathrm{CHSH}(\mathbf{Q} \boxtimes_{\mathsf{W}} \mathbf{P})$ for some fixed non-signaling boxes $\mathbf{Q}, \mathbf{P}$, and where $\mathcal{W}$ is the set of mixed wirings introduced in Definition 3.15, which we recall below.

**The Constraint** $\mathsf{W} \in \mathcal{W}$**.** As mentioned in eqs. (3.15) and (3.16), recall that a *mixed wiring* $\mathsf{W}$ between two boxes $\mathbf{Q}, \mathbf{P} \in \mathcal{NS}$ is the data of six functions $f_1, f_2, g_1, g_2 : \{0,1\}^2 \to [0,1]$ and $f_3, g_3 : \{0,1\}^3 \to [0,1]$ satisfying the following *non-cyclicity conditions*:

$$\forall x, \qquad \big(f_1(x,0) - f_1(x,1)\big)\big(f_2(x,0) - f_2(x,1)\big) \;=\; 0\,, \tag{6.19}$$

$$\forall y, \qquad \big(g_1(y,0) - g_1(y,1)\big)\big(g_2(y,0) - g_2(y,1)\big) \;=\; 0\,. \tag{6.20}$$

Recall that the corresponding diagram can be found in Figure 6.3, and that mixed wirings form a set that we denote $\mathcal{W}$. In our algorithms, we view $\mathsf{W}$ a real vector with $4 \times 2^2 + 2 \times 2^3 = 32$ variables. This vector stores each value of each function:

$$\mathsf{W} = [f_1(0,0)\ \ f_1(0,1)\ \ f_1(1,0)\ \ f_1(1,1)\ \ g_1(0,0)\ \ \cdots]^\top \in \mathbb{R}^{32}. \tag{6.21}$$

To satisfy the normalization constraint that the $f_i, g_j$ take value in $[0,1]$, and the non-cyclicity conditions (6.19) and (6.20) (which are non-linear conditions), we implement a projection function $\texttt{proj} : \mathbb{R}^{32} \to \mathbb{R}^{32}$ in Algorithm 6.1. Notice that our real code is written in a vectorized fashion and is difficult to read as such, so we only present the idea here. Moreover, we use the package PyTorch [Pas+17] for automatic differentiation.

**Algorithm 6.1:** Projection function `proj` on the feasible set $\mathcal{W}$. Vectorized version in our GitHub page [BC23a].

**Data:** $\mathsf{W} = \begin{bmatrix} w_1 & \dots & w_{32} \end{bmatrix} = \begin{bmatrix} f_1(0,0) & f_1(0,1) & f_1(1,0) & f_1(1,1) & g_1(0,0) & \dots & g_3(1,1,1) \end{bmatrix} \in \mathbb{R}^{32}$.

**Result:** $\texttt{proj}(\mathsf{W}) \in \mathcal{W}$.

```
W ← [max{w_1, 0}  ...  max{w_32, 0}] ;
W ← [min{w_1, 1}  ...  min{w_32, 1}] ;
for x ∈ {0,1} do
    if |f_1(x,0) − f_1(x,1)| ⩽ |f_2(x,0) − f_2(x,1)| then
        f_1(x,0), f_1(x,1) ← (f_1(x,0) + f_1(x,1))/2 ;
    else  f_2(x,0), f_2(x,1) ← (f_2(x,0) + f_2(x,1))/2 ;
end
for y ∈ {0,1} do
    if |g_1(y,0) − g_1(y,1)| ⩽ |g_2(y,0) − g_2(y,1)| then
        g_1(y,0), g_1(y,1) ← (g_1(y,0) + g_1(y,1))/2 ;
    else  g_2(y,0), g_2(y,1) ← (g_2(y,0) + g_2(y,1))/2 ;
end
return W
```

**On the Objective Function.** In eq. (6.18), we mentioned that the objective function is:

$$\Phi(\mathsf{W}) := \mathsf{CHSH}(\mathbf{Q} \boxtimes_{\mathsf{W}} \mathbf{P}).$$

In our algorithms, we view a box $\mathbf{P}$ as a $2 \times 2 \times 2 \times 2$-tensor (see page 66): $\mathbf{P}[a,b,x,y] := \mathbf{P}(a,b\,|\,x,y)$ with $a,b,x,y \in \{0,1\}$, whose entries are float numbers between $0$ and $1$. Two things need to be computed separately: $\mathbf{Q} \boxtimes_{\mathsf{W}} \mathbf{P}$ and then $\mathrm{CHSH}(\cdot)$. On the one hand, the product $\mathbf{P} \boxtimes_{\mathsf{W}} \mathbf{P}$ is computed using eq. (3.20), which we vectorized in our algorithm using five types of operations: tensor transposition $\top$, tensor sum $+$, tensor product $\otimes$, contraction of tensors $\cdot$, and entry-wise multiplication $*$; see details in the pdf document of our GitHub page [BC23a]. On the other hand, the function $\mathrm{CHSH}(\cdot)$ is a linear function that computes the CHSH-value of a box, implemented with a dot product as follows:

$$\mathrm{CHSH}(\mathbf{R}) := \frac{1}{4} \sum_{a \oplus b = xy} \mathbf{R}(a, b \,|\, x, y) \;=\; \langle \mathbf{R}, \mathbf{T} \rangle \,,$$

where $\mathbf{T}$ is the $2 \times 2 \times 2 \times 2$-tensor defined as $\mathbf{T}[a, b, x, y] = \frac{1}{4}$ if $a \oplus b = xy$, and $= 0$ otherwise.

#### 6.4.2.1 Naive Gradient Descent

To gain insight into the complexity of the optimization problem, we begin by studying a basic algorithm, the Projected Gradient Descent, with a small learning rate ($\alpha \ll 1$) and a lot of iterations ($K \gg 1$). We will obtain a histogram of the frequency of the different results we obtain, see Figure 6.9 (a).

**Projected Gradient Descent.** We implement a “projected" version of the Gradient Descent algorithm in order to satisfy the constraint $\mathsf{W} \in \mathcal{W}$ at each step. It simply means that each iteration is projected on the feasible set:

$$\mathsf{W}^{k+1} = \mathtt{proj}\big(\mathsf{W}^k + \alpha \, \nabla\Phi(\mathsf{W}^k)\big) \,,$$

where $\alpha \in \mathbb{R}$ is the learning rate. Our implementation can be found in Algorithm 6.2. We compute the gradient of the objective function using the automatic differentiation Python package `torch.autograd` that provides us with the commands `backward` and `grad`. As we do not have a good intuition of what could be a good wiring $\mathsf{W}$ in $\mathcal{W}$ to start with given a fixed box $\mathbf{P}$, we take a random initialization: $\mathsf{W}^0$ is uniformly generated in the hypercube $[0, 1]^{32}$. As such, the vector $\mathsf{W}^0$ is not necessarily a well-defined mixed wiring since it does not necessarily satisfy the non-cyclicity conditions (6.19) and (6.20), but this problem is fixed after one iteration in the Projected Gradient Descent algorithm since the wiring is then projected.

Otherwise, one can also directly apply `proj` to $\mathsf{W}^0$. The notation $\mathsf{W} \sim \mathcal{U}(X)$ means that we uniformly generate $\mathsf{W}$ in the set $X$.

**Algorithm 6.2:** Projected Gradient Descent. More details on our GitHub page [BC23a].

**Data:** $\Phi : \mathbb{R}^{32} \to \mathbb{R}$ objective function, $\alpha \in \mathbb{R}$ learning rate, $K \in \mathbb{N}$ number of iterations, $\varepsilon > 0$ tolerance.

**Result:** $\mathsf{W}_{\text{out}} \approx \operatorname{argmax}_{\mathsf{W}} \Phi(\mathsf{W}) \in \mathcal{W} \subseteq \mathbb{R}^{32}$.

$\mathsf{W} \sim \mathcal{U}([0,1]^{32})$ ;
**for** $k \in \{0, \ldots, K-1\}$ **do**
  $\mathsf{W}_{\text{old}} \leftarrow \mathsf{W}$ ;
  $\mathsf{W} \leftarrow \texttt{proj}\big(\mathsf{W} + \alpha\, \nabla\Phi(\mathsf{W})\big)$ using Algorithm 6.1 ;
  **if** $||\mathcal{W} - \mathsf{W}_{\text{old}}||_\infty < \varepsilon$ **then** break ;
**end**
**return** $\mathsf{W}_{\text{out}} := \mathsf{W} \in \mathbb{R}^{32}$.

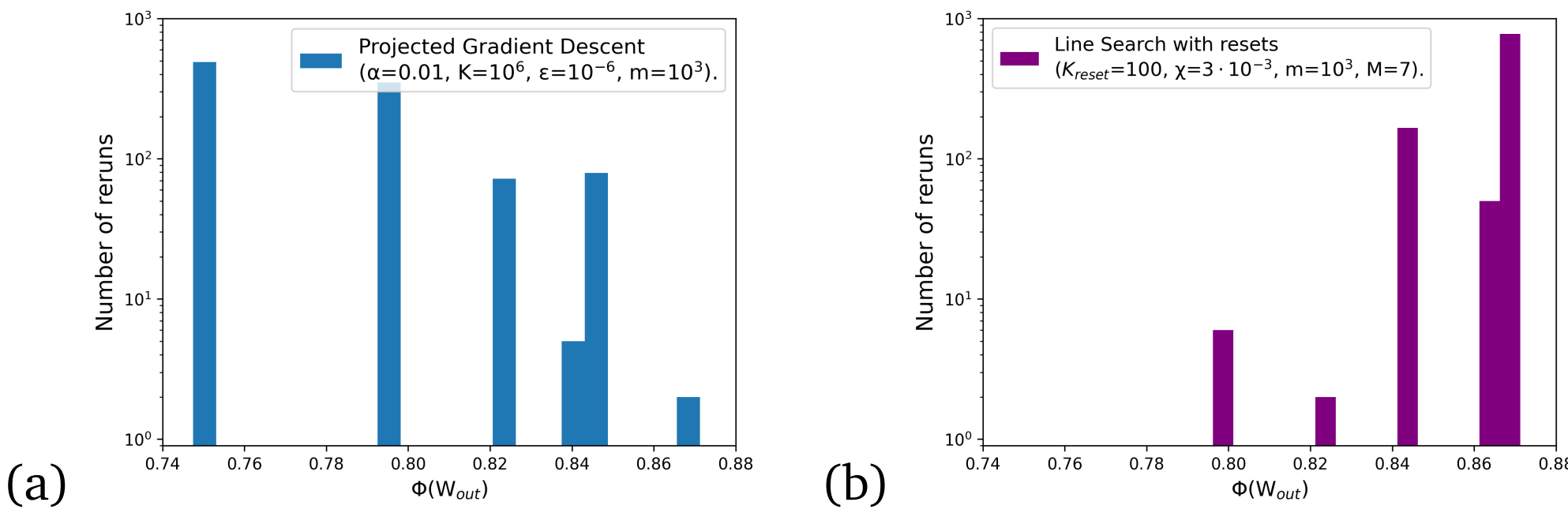

**Figure 6.9 —** *Histograms of the evaluations of the objective function* $\Phi$ *applied at the output* $\mathsf{W}_{\text{out}}$ *of (a) Algorithm 6.2 and (b) Algorithm 6.3. As expected, we observe that the latter is more efficient than the former in maximizing* $\Phi$*, for equivalent computation duration.*

**Estimating the Proportion of "Good" Outputs.** We use Algorithm 6.2 with a learning rate $\alpha = 0.01$, a number of iterations of $K = 10^6$, a tolerance of $\varepsilon = 10^{-6}$, and we obtain the histogram presented in Figure 6.9 (a). Recall that the objective function is $\Phi(\mathsf{W}) := \text{CHSH}(\mathbf{Q} \boxtimes_{\mathsf{W}} \mathbf{P})$; this histogram is drawn with $\mathbf{Q} = \mathbf{P} = p\,\mathbf{PR} + q\,\mathbf{SR} + (1-p-q)\,\mathbf{I}$, where $p = 0.39$ and $q = 0.6$.

The number of reruns is $m = 10^3$, done simultaneously in parallel, which is faster than doing $m$ descents one after another[2]. We observe that the results concentrate on certain discrete values. These values correspond to different attractive points in different basins of attraction (recall that the initial W is taken uniformly at random in $[0,1]^{32}$). As we want to maximize $\Phi$, we are interested in the highest concentrated value $\approx 0.87$. In that example, we observe that the proportion of starting wirings such that $\mathsf{W}_{\text{out}}$ is only $\chi \approx \frac{2}{10^3} = 0.2\%$ using this basic Gradient Descent algorithm. This information tells us that the function $\Phi$ is difficult to maximize, which is why we present a more efficient algorithm in the following subsection.

#### 6.4.2.2 More Efficient Algorithm: Line Search with Resets

In this subsubsection, we present a variant of the Gradient Descent algorithm called Line Search, which we enhance with frequent resets of bad outcomes. See [NW99] for a standard reference book in numerical optimization. The idea of this algorithm is, instead of always keeping the same $\alpha$, to estimate the best coefficient $\alpha_k$ at each step of the descent:

$$\begin{cases} \alpha_k = \operatorname{argmax}_{\alpha \in \mathcal{R}} \Phi\big(\mathsf{W}^k + \alpha \nabla\Phi(\mathsf{W}^k)\big) \,, \\ \mathsf{W}^{k+1} = \texttt{proj}\big(\mathsf{W}^k + \alpha_k \nabla\Phi(\mathsf{W}^k)\big) \,. \end{cases}$$

As we observed in the previous subsection, the proportion $\chi$ of “good" starting wirings is very weak, which is why we apply frequent resets: we do $m = 10^3$ descents in parallel but only $K_{\text{reset}} = 100$ steps, then we keep only the best $m \cdot \chi$ wirings and we reset all the others to a new random initialization. Then we repeat that procedure but we reset fewer wirings (say, at the $j$-th repetition, keep for instance the best $j \cdot m \cdot \chi$ wirings), and we repeat this procedure $\frac{1}{\chi}$ times. In the end, most of the wirings should be in the good basin of attraction, so we can apply one final run of line search, with many more steps so that it converges to the attractor. See Algorithm 6.3, and we obtain results of Figure 6.9 (b).

---

[2] In order to do $m$ gradient descents in parallel efficiently, we “parallelize" Algorithm 6.2: instead of viewing W as a 32-vector, we view it as a $(32 \times m)$-matrix, where each column represents a different wiring. Comparing this method with a naive FOR loop, we observe a factor of 100 in the speed gain. Notice that when the descent is done, one can post-select the best wiring among the $m$ columns of $\mathsf{W}_{\text{out}}$. See our GitHub page [BC23a].

**Algorithm 6.3:** Line Search with Resets. More details on our GitHub page [BC23a].

**Data:** $\Phi : \mathbb{R}^{32} \to \mathbb{R}$ objective function, $K_{\text{reset}} \in \mathbb{N}$ number of iterations before reset, $\chi \in [0, 1]$ proportion of "good" random wirings, $m \in \mathbb{N}$ number of descents in parallel, $M \in \mathbb{N}$ number of iterations to compute the best learning rate $\alpha^*$.

**Result:** $\mathsf{W}_{\text{out}} \in \mathbb{R}^{32\times m}$, where each column is $\approx \operatorname{argmax}_{\mathsf{W}} \Phi(\mathsf{W})$.

$\mathsf{W} = (32 \times m$ zero matrix), whose columns are denoted $\mathsf{W}_i$ ;
**for** $j \in \{0, \dots, \lfloor \frac{1}{\chi} \rfloor - 1\}$ **do**
  $\mathsf{W} \leftarrow$ among the $m$ columns of W, keep the $j \cdot m \cdot \chi$ ones giving the highest values for the objective function $\Phi$, and reset all the other columns randomly with $\mathcal{U}([0, 1]^{32})$ ;
  **if** $j = \lfloor \frac{1}{\chi} \rfloor - 1$ **then** $K_{\text{reset}} \leftarrow 10 \cdot K_{\text{reset}}$;
  **for** $k \in \{0, \dots, K_{\text{reset}} - 1\}$ **do**
    **for** $i \in \{0, \dots, m-1\}$ **do** $\alpha_i^* \leftarrow \operatorname{argmax}_{\alpha>0} \Phi\big(\mathsf{W}_i + \alpha \nabla\Phi(\mathsf{W}_i)\big)$ using $M$ iterations ;
    $\mathsf{W} \leftarrow \Big[\texttt{proj}\big(\mathsf{W}_i + \alpha_i^* \nabla\Phi(\mathsf{W}_i)\big)\Big]_i$ using Algorithm 6.1 ;
  **end**
**end**
**return** $\mathsf{W}_{\text{out}} = \mathsf{W} \in \mathbb{R}^{32\times m}$.

### 6.4.3 *Task A: Adaptive Wiring*

Algorithm A is presented in Algorithm 6.4; it simply consists in applying the toy case $\mathbf{Q} \boxtimes \mathbf{P}$ from the previous subsection recursively $N$ times. We want to find a sequence of wirings $\mathsf{W}_1, \dots, \mathsf{W}_N$ such that the CHSH-value is above the following threshold:

$$\mathrm{CHSH}\Bigg(\underbrace{\Big(\big((\mathbf{P} \underset{\mathsf{W}_1}{\boxtimes} \mathbf{P}) \underset{\mathsf{W}_2}{\boxtimes} \mathbf{P}\big) \underset{\mathsf{W}_3}{\boxtimes} \dots\Big) \underset{\mathsf{W}_N}{\boxtimes} \mathbf{P}}_{=: \mathbf{P}_{N+1}}\Bigg) > \frac{3+\sqrt{6}}{6}.$$

Using [Bra+06], a consequence is that the box $\mathbf{P}$ collapses communication complexity (Section 4.2.3). Notice that for some boxes $\mathbf{P} \in \mathcal{NS}$, it might not be possible to find such a sequence of wirings because it is impossible to distill them by any means. This algorithm is used in Section 6.5.1 in order to plot the new regions of collapsing nonlocal boxes.

**Algorithm 6.4:** Task A. More details on our GitHub page [BC23a].

**Data:** $\mathbf{P} \in [0,1]^{2+2+2+2}$ box, $N \in \mathbb{N}$ number of box products, $(K_{\text{reset}}, \chi, m, M)$ parameters for Algorithm 6.3.
**Result:** $[\mathsf{W}_1^*, \dots, \mathsf{W}_n^*] \in \mathcal{W}^n$ for some $n \leqslant N$.

$\mathbf{P}_{11} \leftarrow \mathbf{P}$ ;
**for** $n \in \{1, \dots, N\}$ **do**
  $\mathsf{W}_n^* \in [0,1]^{32} \leftarrow \operatorname{argmax}_{\mathsf{W}} \mathrm{CHSH}(\mathbf{P}_n \boxtimes_{\mathsf{W}} \mathbf{P})$ by picking the best column among the $m$ columns of the output $\mathsf{W}_{\text{out}} \in \mathbb{R}^{32\times m}$ of Algorithm 6.3 ;
  $\mathbf{P}_{n+1} \leftarrow \mathbf{P}_n \boxtimes_{\mathsf{W}_n^*} \mathbf{P}$ ;
  **if** $\mathrm{CHSH}(\mathbf{P}_{n+1}) > \frac{3+\sqrt{6}}{6}$ **then return** $[\mathsf{W}_1^*, \dots, \mathsf{W}_n^*]$ ;
**end**
**return** *"Nothing found."*

**Remark 6.15 —** Going further, once we find $(\mathsf{W}_1^*, ..., \mathsf{W}_N^*)$, it is possible to do a "backward" process: for all $i \in \{1, ..., N\}$, fix $\mathsf{W}_j^*$ for $j \neq i$, optimize the function $\mathsf{W}_i \mapsto \mathrm{CHSH}(\mathbf{P}_{N+1})$ and update $\mathsf{W}_i^*$. It is also possible to use neural network methods to optimize all the $\mathsf{W}_i$ "at the same time".

### 6.4.4 *Task B: Constant Wiring*

Task B is a "uniform" version of task A, in the sense that we want all the $\mathsf{W}_i$'s to be the same. In order words, we want to find a wiring $\mathsf{W}$ and an integer $N$ such that:

$$\mathrm{CHSH}\Bigg(\underbrace{\Big(\big((\mathbf{P} \underset{\mathsf{W}}{\boxtimes} \mathbf{P}) \underset{\mathsf{W}}{\boxtimes} \mathbf{P}\big) \underset{\mathsf{W}}{\boxtimes} \dots\Big) \underset{\mathsf{W}}{\boxtimes} \mathbf{P}}_{=:\, \mathbf{P}^{\boxtimes_{\mathsf{W}}(N+1)}}\Bigg) > \frac{3+\sqrt{6}}{6}.$$

This algorithm is used in the proof of Corollary 6.20 in order to find appropriate collapsing wirings for the analytical proof.

**Idea of the Algorithm.** First, find a wiring $\mathsf{W}_1^* = \operatorname{argmax}_{\mathsf{W}} \mathrm{CHSH}(\mathbf{P} \boxtimes_{\mathsf{W}} \mathbf{P})$ with a Gradient Descent algorithm, and then evaluate the powers of $\mathbf{P}$ with that wiring $\mathsf{W}_1^*$ until $N+1$, *i.e.* compute $\mathrm{CHSH}\big(\mathbf{P}^{\boxtimes_{\mathsf{W}_1^*} n}\big)$ for $n = 1, \dots, N+1$. If one of those evaluations is greater than the threshold $(3+\sqrt{6})/6$ from [Bra+06], then we can stop the algorithm, it means that the wiring $\mathsf{W}_1^*$ achieves the goal. Otherwise, compute $\mathsf{W}_2^* = \operatorname{argmax}_{\mathsf{W}} \mathsf{CHSH}\big(\mathbf{P}^{\boxtimes_{\mathsf{W}} 3}\big)$ and repeat the same evaluation process of the powers of $\mathbf{P}$ as in the previous step. Proceed like this until computing $\mathsf{W}_M^*$, where $M \in \mathbb{N}$ is some hyper-parameter. Typically, we take $M \leqslant N$ because it is a lot faster to evaluate the $N$-th power of $\mathbf{P}$ than to optimize the $N$-th power of $\mathbf{P}$. See the details in Algorithm 6.5.

**Algorithm 6.5:** Task B. More details on our GitHub page [BC23a].

**Data:** $\mathbf{P} \in [0,1]^{2+2+2+2}$ box, $N \in \mathbb{N}$ maximal tested box power, $L \in \mathbb{N}$ maximal optimized box power, $(K_{\text{reset}}, \chi, m, M)$ parameters for Algorithm 6.3.
**Result:** $\mathsf{W}^* \in \mathcal{W}$.

**for** $\ell \in \{1, \dots, L\}$ **do**
  $\mathsf{W}_\ell^* \leftarrow \operatorname{argmax}_{\mathsf{W}} \mathsf{CHSH}\big(\mathbf{P}^{\boxtimes_{\mathsf{W}} \ell+1}\big)$ using Algorithm 6.3 ;
  **for** $n \in \{1, \dots, N+1\}$ **do**
    **if** $\mathsf{CHSH}\big(\mathbf{P}^{\boxtimes_{\mathsf{W}_\ell^*} n}\big) > \frac{3+\sqrt{6}}{6}$ **then return** $\mathsf{W}_\ell^*$ ;
  **end**
**end**
**return** *"Nothing found."*

## 6.5 Collapse of CC from the Algebra of Boxes

In this section, we present collapsing boxes found in two different ways. (i) First with a numerical approach, using the algorithms (Section 6.4). (ii) Then with an analytical approach, using the algebra of boxes (Section 6.2) and the orbit of a box (Section 6.3).

### *6.5.1 Numerical Regions that Collapse CC*

Using Algorithm 6.4 that addresses Task A, we obtain many collapsing boxes. Some samples are drawn in Figure 6.10 on some slices of the non-signaling set $\mathcal{NS}$, but note that this algorithm also applies more generally to any desired slice. As previously mentioned, this work is concurrent and independent of the work of [EWC23a]. In the drawings, some boxes are denoted $\mathbf{P_L}$ and $\mathbf{P_{NL}}$, let us recall their definition here (more details on page 67). The local set $\mathcal{L}$ and the non-signaling set $\mathcal{NS}$ are polytopes, *i.e.* the convex hull of a finite number of extremal points. The first set $\mathcal{L}$ admits exactly $16$ extremal points, called *local extreme points* and denoted $\mathbf{P}_\mathbf{L}^{\mu,\nu,\sigma,\tau}$, where $\mu,\nu,\sigma,\tau \in \{0,1\}$. These $16$ points are also extremal points of $\mathcal{NS}$, together with $8$ additional extremal points, called *non-local extreme points* and denoted $\mathbf{P}_\mathbf{NL}^{\mu,\nu,\sigma}$. They are defined as follows [All+09b; Bar+05]:

$$\bullet\ \text{Local:} \quad \mathbf{P}_\mathbf{L}^{\mu,\nu,\sigma,\tau}(a,b\,|\,x,y) := \begin{cases} 1 & \text{if } a = \mu\, x \oplus \nu \ \text{ and } \ b = \sigma\, y \oplus \tau, \\ 0 & \text{otherwise}, \end{cases}$$
$$\bullet\ \text{Nonlocal:} \quad \mathbf{P}_\mathbf{NL}^{\mu,\nu,\sigma}(a,b\,|\,x,y) := \begin{cases} 1/2 & \text{if } a \oplus b = xy \oplus \mu\, x \oplus \nu\, y \oplus \sigma, \\ 0 & \text{otherwise}. \end{cases} \tag{6.22}$$

Note that $\mathbf{PR} = \mathbf{P}_\mathbf{NL}^{000}$ and $\mathbf{P_{00}} = \mathbf{P}_\mathbf{L}^{0000}$ and $\mathbf{P_{11}} = \mathbf{P}_\mathbf{L}^{0101}$, where we remove the commas for simplicity of notations.

Observe in Figure 6.10 that, depending on the chosen slice, the collapsing area does not always have the same “shape" nor the same “area." Moreover, notice in the graphs that there seems to exist a collapsing area in the neighborhood below the diagonal segments joining $\mathbf{PR}$ and respectively $\mathbf{SR}$, $\mathbf{P_{00}}$, $\mathbf{P_{11}}$. This is actually true. Indeed, we analytically show below in Corollary 6.20 that those three segments are collapsing, and we also know that the box product $\mathbf{P} \boxtimes_{\mathrm{W}} \mathbf{Q}$ is continuous in $\mathbf{P}$ and $\mathbf{Q}$ for any W (it is even bilinear, recall the expression in eq. (3.20)), so distillation protocols are continuous and in some sense the orbits are also “continuous," hence there exists an open neighborhood below these diagonal segments that collapses communication complexity.

**Remark 6.16** (Continuous Extension of a Finite Collapsing Set) — The algorithm only provides us with finitely many collapsing boxes, but we can still deduce a continuous “extension" of that collapsing set. Indeed, as explained in Section 6.3.2, if we know that a box $\mathbf{P} \in \mathcal{NS}$ collapses communication complexity, then we also know that the cone $\mathcal{C}_\mathbf{P}$ is collapsing, where $\mathcal{C}_\mathbf{P}$ denotes a certain cone taking origin at $\mathbf{P}$ represented in Figure 6.6 (b). As a consequence, if we list the collapsing boxes $\{\mathbf{P}_k\}_{1\leqslant k\leqslant K}$ obtained by numerical means, we can deduce what follows:

$$\text{The union of cones } \bigcup_{k=1}^{K} \mathcal{C}_{\mathbf{P}_k} \text{ is a collapsing set.}$$

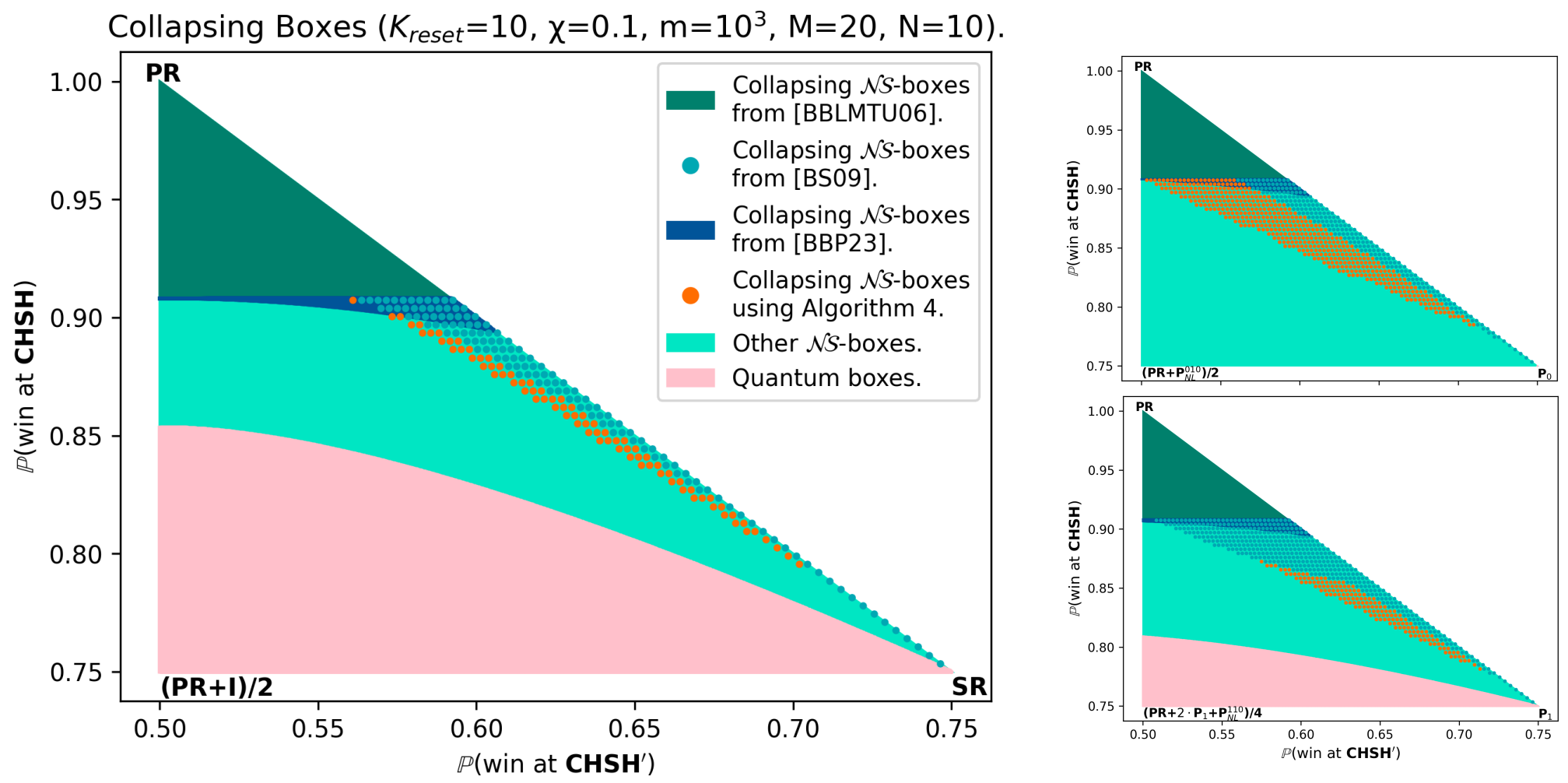

**Figure 6.10 —** *In orange are drawn the collapsing boxes outputted by Algorithm 6.4. Each drawing represents a different slice of $\mathcal{NS}$; the extreme points of the triangles indicate which slice is drawn and the definition of the boxes $\mathbf{P}_{\text{NL}}$ can be found in eq. (6.22). The three graphs have the same color legend, displayed at the center, and they are all configured with the same algorithm parameters $(K_{\text{reset}}, \chi, m, M, N)$, detailed at the top. We adopt the following convention: (i) boxes that are numerically determined are drawn with dots, (ii) boxes that are analytically determined are drawn in plain regions (there exist explicit equations describing those regions). Notice that the left drawing is similar to [EWC23a, Figure 3], which was found using a different algorithm as detailed in Remark 6.14. The quantum set $\mathcal{Q}$ (in pink) is drawn using formulas from [Mas03]. References: [BBLMTU06]=[Bra+06], [BS09]=[BS09], [BBP23]=[BBP24].*

## 6.5.2 *Collapse of CC from the Orbit of a Box*

In this section, we present an example of a new nonlocal box that collapses CC using its orbit. As previously mentioned, the next theorem is concurrent and independent of the work of [EWC23a]. A similar result was also established in the M.Sc. thesis of the author [Bot22].

**Theorem 6.17** (New Collapsing Boxes) — *The techniques of box orbits (Section 6.3.2) allow us to discover new collapsing boxes. See new collapsing areas in Figure 6.10.*

*Proof.* See Figure 6.7 for an intuition of the proof. Take the starting box $\mathbf{P}$ with coordinates $(0.627, 0.862)$ in the affine plane $\mathcal{A} = \mathrm{Aff}\{\mathbf{PR}, \mathbf{SR}, \mathbf{I}\}$, where the coordinate system is given by the CHSH′- and CHSH-values of $\mathbf{P}$. On the one hand, the tilted orbit of $\mathbf{P}$ intersects the collapsing area that was found in [Bra+06] (in dark green), since for instance $\mathrm{CHSH}(\mathbf{P}^{\boxtimes 5}) \approx 0.913 > 0.908 \approx \frac{3+\sqrt{6}}{6}$, so $\mathbf{P}$ is collapsing by Proposition 6.4. On the other hand, this box $\mathbf{P}$ does not lie in any of the previously-known collapsing areas from [BBP24; Bra+06; Bri+19; BS09; vD99] (to the best of our knowledge, these five references are the only previous results showing a collapse of communication complexity, in addition to [EWC23a] which concurrently and independently found a similar result to ours as mentioned before). Indeed, it is not in the collapsing areas from [Bra+06; vD99] since $\mathrm{CHSH}(\mathbf{P}) = 0.862 < 0.908$, nor is it in the collapsing area from [BBP24] since $A(\mathbf{P}) + B(\mathbf{P}) \approx 14.13 < 16$ (using the authors' notation). The box $\mathbf{P}$ neither is in any of the collapsing regions found in [Bri+19] since it does not belong to the boundary $\partial\mathcal{NS}$ of the non-signaling set. The last area to check is the one from [BS09], which was numerically found. From a box $\mathbf{P}$, they define a sequence of boxes using "pairwise" multiplications: $\mathbf{Q}_1 := \mathbf{P}$ and $\mathbf{Q}_{n+1} := \mathbf{Q}_n \boxtimes \mathbf{Q}_n$ for $n \geqslant 1$, and they check whether or not there exists an integer $n$ such that $\mathrm{CHSH}(\mathbf{Q}_n) > \frac{3+\sqrt{6}}{6}$. But, for our starting box $\mathbf{P}$, none of the $\mathbf{Q}_n$ satisfy this inequality: indeed, for $1 \leqslant n \leqslant 5$, it possible to check it by hand, for $n = 6$ we have $\mathbf{Q}_6 \in \mathcal{L}$, and for $n \geqslant 7$ we also have $\mathbf{Q}_n \in \mathcal{L}$ since $\mathcal{L}$ is closed under wirings [All+09a]. Hence our example $\mathbf{P}$ is a new collapsing box. ■

### 6.5.3 *Collapse of CC from Multiplication Tables*

Here, we prove a technique to show the collapse of CC from a multiplication table. Recall that the *convex hull* of a set $\{\mathbf{Q}_1, \dots, \mathbf{Q}_N\}$ is the set of all possible convex combinations of these $\mathbf{Q}_i$:

$$\mathrm{Conv}\{\mathbf{Q}_1, \dots, \mathbf{Q}_N\} := \left\{ \sum_{i=1}^{N} q_i\, \mathbf{Q}_i \text{ such that } q_i \geqslant 0 \text{ and } \textstyle\sum_i q_i = 1 \right\},$$

and the *affine hull* of $\{\mathbf{Q}_1, \ldots, \mathbf{Q}_N\}$ has the same definition but without the non-negativity constraint:

$$\begin{aligned}\mathrm{Aff}\{\mathbf{Q}_1, \ldots, \mathbf{Q}_N\} \;&:=\; \left\{ \sum_{i=1}^{N} q_i\, \mathbf{Q}_i \text{ such that } q_i \in \mathbb{R} \text{ and } \textstyle\sum_i q_i = 1 \right\}\\ &\supseteq\; \mathrm{Conv}\{\mathbf{Q}_1, \ldots, \mathbf{Q}_N\}\,.\end{aligned}$$

**Theorem 6.18 —** *Let* $\mathbf{Q}, \mathbf{R} \in \mathcal{NS}$ *be boxes. Assume there exists a wiring* $\mathsf{W} \in \mathcal{W}$ *that induces the following multiplication table:*

| | **PR** | **Q** | **R** |
|---|---|---|---|
| **PR** | PR | PR | PR |
| **Q** | $\frac{1}{2}$ (Q + R) | Q | R |
| **R** | PR | R | Q |

*Then the triangle* $\mathrm{Conv}\{\mathbf{PR}, \mathbf{Q}, \mathbf{R}\} \setminus \mathrm{Conv}\{\mathbf{Q}, \mathbf{R}\}$ *is collapsing.*

*Proof.* Denote $\mathcal{T} := \mathrm{Conv}\{\mathbf{PR}, \mathbf{Q}, \mathbf{R}\} \setminus \mathrm{Conv}\{\mathbf{Q}, \mathbf{R}\}$, and consider a convex combination of the form $\mathbf{P}_{\alpha,\beta} := \alpha\, \mathbf{PR} + \beta\, \mathbf{Q} + (1-\alpha-\beta)\, \mathbf{R} \in \mathcal{T}$ with $\alpha, \beta \geqslant 0$ and $\alpha \neq 0$ and $\alpha + \beta \leqslant 1$. For any fixed $\mathbf{P}_{\alpha_0,\beta_0}$ in the triangle $\mathcal{T}$, we want to show the collapse of CC. We want to build a sequence $(u_k)_k = \big((\alpha_k, \beta_k)\big)_k$ such that $(\mathbf{P}_{u_k})_k$ tends to the **PR** box. Denote $\boxtimes$ the box product induced by the wiring $\mathcal{W}$ from the assumptions. By bilinearity of $\boxtimes$ and using the multiplication table, computations lead to $\mathbf{P}_{\alpha,\beta} \boxtimes \mathbf{P}_{\alpha_0,\beta_0} = \mathbf{P}_{\tilde{\alpha},\tilde{\beta}}$ where:

$$\begin{bmatrix}\tilde{\alpha}\\ \tilde{\beta}\end{bmatrix} = A \begin{bmatrix}\alpha\\ \beta\end{bmatrix} + b\,,\ \ A := \begin{bmatrix} 1-\alpha_0 & -\alpha_0 \\ -1+\alpha_0+\beta_0 & -1+\frac{3}{2}\,\alpha_0 + 2\,\beta_0 \end{bmatrix},\ \ b := \begin{bmatrix}\alpha_0 \\ 1-\alpha_0-\beta_0\end{bmatrix}.$$

From this remark, we define the following sequence:

$$u_0 := (\alpha_0, \beta_0)\,, \qquad\qquad u_{k+1} = A\, u_k + b\,.$$

We easily identify that $\ell := (1, 0)$ is a fixed point of $x \mapsto A\,x + b$, so it yields:

$$u_{k+1} - \ell \;=\; \big(A\,u_k + b\big) - \big(A\ell + b\big) \;=\; A\,(u_k - \ell) \;=\; A^{k+1}\,(u_0 - \ell)\,,$$

where the last equality follows from an induction on $k$. But the matrix $A$ admits exactly two distinct[3] eigenvalues $\lambda_1 = 1 - a/2$ and $\lambda_2 = -1 + a + 2b$. So $A$ is diagonalizable, and its power $A^k = P \begin{bmatrix} \lambda_1^k & 0 \\ 0 & \lambda_2^k \end{bmatrix} P^{-1}$ tends to the zero matrix because $|\lambda_1|, |\lambda_2| < 1$, where $P$ is an invertible matrix. Hence, from the above equation, the sequence $(u_k)_k$ tends to $\ell$, and by continuity we have that the sequence of boxes $(\mathbf{P}_{u_k})_k \subseteq \mathbb{R}^{16}$ converges to $\mathbf{P}_\ell = \mathbf{PR}$. But Brassard, Buhrman, Linden, Méthot, Tapp, and Unger showed that there is an open neighbor around **PR** that collapses communication complexity [Bra+06]. Therefore, we know that the sequence $(\mathbf{P}_{u_k})_k$ reaches this collapsing neighbor for some $k$ large enough, and using Proposition 6.4 we conclude that any starting box $\mathbf{P}_{u_0} \in \mathcal{T}$ is indeed collapsing. ∎

**Remark 6.19** (Why is $\mathrm{Conv}\{\mathbf{Q}, \mathbf{R}\}$ Not Necessarily Collapsing?) **—** It can be that $\mathbf{Q}, \mathbf{R}$ are local boxes in $\mathcal{L}$. In such a case, we know that the line segment does not collapse CC because it belongs in $\mathcal{Q}$, which cannot collapse CC [Cle+99].

We give some examples of such collapsing triangles. This allows us to recover some results from [Bri+19] with a new proof, based on the algebra of boxes:

**Corollary 6.20 —** *All the triangles drawn in Figure 6.11 are collapsing.*

*Proof.* The proof follows directly from Theorem 6.18 applied to what follows:

| Triangle | P | Q | R | Wiring $\mathsf{W} \in \mathcal{W} \subseteq \mathbb{R}^{32}$ |
|---|---|---|---|---|
| $\mathcal{T}_0$ | **PR** | $\mathbf{P_{00}}$ | $\mathbf{P_{11}}$ | $\mathsf{W}_{\mathrm{BS}} = [0,0,1,1,0,0,1,1,0,0,0,1,0,0,0,1,0,1,1,0,0,1,1,0,0,1,1,0,0,1,1,0]$ |
| $\mathcal{T}_1$ | **PR** | $\mathbf{P_L^{0111}}$ | $\mathbf{P_{00}}$ | $\mathsf{W}_1 = [1,0,0,1,0,0,1,0,1,1,0,0,0,0,1,1,1,0,0,1,1,0,0,1,1,0,0,1,0,1,1,0]$ |
| $\mathcal{T}_2$ | **PR** | $\mathbf{P_{00}}$ | $\mathbf{P_L^{1101}}$ | $\mathsf{W}_2 = [0,0,1,0,0,1,1,0,0,0,1,1,1,1,0,0,0,1,1,0,0,0,1,1,0,1,1,0,0,1,1,0]$ |
| $\mathcal{T}_3$ | **PR** | $\mathbf{P_L^{0010}}$ | $\mathbf{P_L^{1011}}$ | $\mathsf{W}_3 = [0,1,1,1,1,0,1,0,0,0,1,1,1,1,0,0,0,0,1,1,0,1,1,0,0,0,1,1,0,0,1,1]$ |
| $\mathcal{T}_4$ | **PR** | $\mathbf{P_L^{1000}}$ | $\mathbf{P_L^{1110}}$ | $\mathsf{W}_4 = [0,1,0,1,0,1,1,1,1,1,0,0,0,0,1,1,0,0,1,1,0,0,1,1,0,1,1,0,1,0,0,1]$ |

where the notation $\mathsf{W} \in \mathbb{R}^{32}$ comes form eq. (6.21). See a representation of these wirings in Figure 6.11, bottom row. ∎

[3]The eigenvalues $\lambda_1$ and $\lambda_2$ are distinct because otherwise we would have $2 = \frac{3}{2}(a + b) + \frac{b}{2}$, which is achieved only if both $a + b = 1$ and $b = 1$, which is equivalent to $a = 0$ and $b = 1$ and which contradicts the assumption $a \neq 0$.

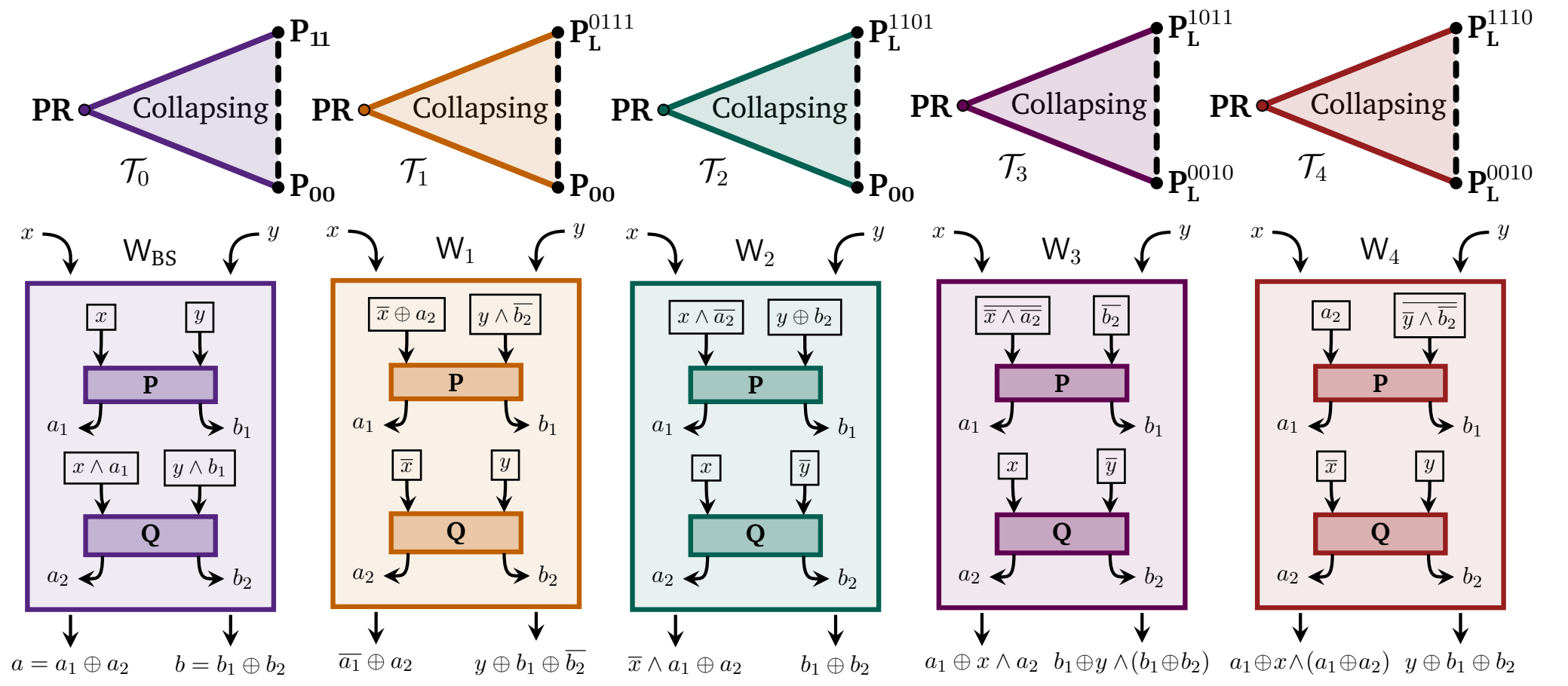

**Figure 6.11** — *Examples of collapsing triangles, together with wirings that satisfy the multiplication table of Theorem 6.18. The definition of the boxes* $\mathbf{P}_{\mathrm{L}}$ *and* $\mathbf{P}_{\mathrm{NL}}$ *can be found in eq. (6.22).*

The wirings of Figure 6.11 are arbitrary examples of collapsing wirings that were obtained using Algorithm 6.5—many more wirings can be found in other triangles of $\mathcal{NS}$ using the same algorithm, which is accessible via our GitHub page [BC23a]. Notice that these wirings are all different from the ones used in the proof of [Bri+19]. Note also that the triangle $\mathcal{T}_0$ of Figure 6.11 extends the result from Brunner and Skrzypczyk [BS09], who showed the collapse of CC in the open line segment joining the boxes **PR** and **SR** $:= (\mathbf{P}_{00} + \mathbf{P}_{11})/2$.

An interesting problem would be to understand better the structure of the set $\mathcal{W}$ so that, given a triangle in $\mathcal{NS}$, we know how to construct a collapsing wiring W without using a search algorithm.

## 6.5.4 *Application to Quantum Voids*

In this section, we recover a result about *quantum voids* from [Rai+19] with a new proof, based on communication complexity. The notion of the quantum void was introduced by Rai, Duarte, Brito, and Chaves in [Rai+19] and is defined as a subset of the boundary of $\mathcal{NS}$ for which all quantum correlations are actually local. (This notion was also studied in [Bri+19].) Let us first prove the following lemma:

**Lemma 6.21** — *The (closed) triangles drawn in Figure 6.11 are faces of* $\mathcal{NS}$.

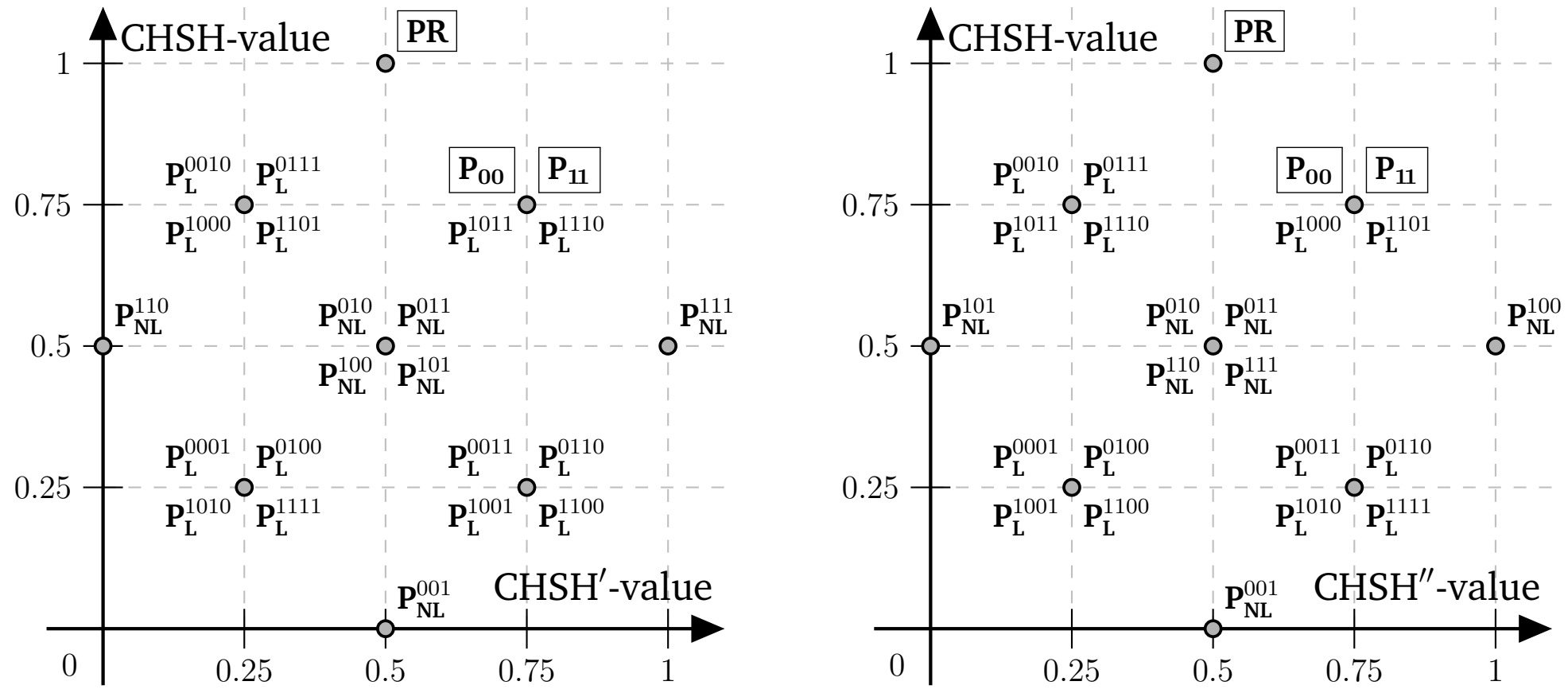

**Figure 6.12 —** *Computation of the* CHSH-, CHSH′- *and* CHSH″- *values of the* 24 *extremal points of* $\mathcal{NS}$ *[Bar+05], where by definition* $\mathrm{CHSH}(\mathbf{P}) := \frac{1}{4}\sum_{a\oplus b=xy}\mathbf{P}(a,b\,|\,x,y)$*, and* CHSH′ *and* CHSH″ *are defined similarly but with respective summand conditions* $a\oplus b=(x\oplus 1)\cdot(y\oplus 1)$ *and* $a\oplus b = x\cdot(y\oplus 1)$.

*Proof.* We prove the result for the triangle $C := \mathrm{Conv}\{\mathbf{PR},\mathbf{P_{00}},\mathbf{P_{11}}\}$—the proof is similar for the other ones. First, we prove the equality between the sets $C$ and $A := \mathcal{NS}\cap\mathrm{Aff}\{\mathbf{PR},\mathbf{P_{00}},\mathbf{P_{11}}\}$, meaning that the convex hull $C$ is actually a slice of $\mathcal{NS}$. The first inclusion $C\subseteq A$ is trivial because the three points $\mathbf{PR},\mathbf{P_{00}},\mathbf{P_{11}}$ are in $\mathcal{NS}$ and because $\mathcal{NS}$ is a polytope so it is stable under taking convex combination. Conversely, recall that by definition $\mathrm{CHSH}(\mathbf{P}) := \frac{1}{4}\sum_{a\oplus b=xy}\mathbf{P}(a,b\,|\,x,y)$, and $\mathrm{CHSH}'(\cdot)$ and $\mathrm{CHSH}''(\cdot)$ are defined similarly but with respective summand conditions $a\oplus b=(x\oplus 1)\cdot(y\oplus 1)$ and $a\oplus b = x\cdot(y\oplus 1)$. As these three functions are linear, they preserve alignment and convexity. It means that if a box $\mathbf{P}$ is of the form $\mathbf{P}=\sum_i q_i\,\mathbf{Q}_i$ for some reals $q_i$ and boxes $\mathbf{Q}_i$, then $\mathrm{CHSH}(\mathbf{P})=\sum_i q_i\,\mathrm{CHSH}(\mathbf{Q}_i)$, and similarly with CHSH′ and CHSH″. We apply the preservation of alignment in Figure 6.12 representing the 24 extremal points of $\mathcal{NS}$ [Bar+05], and we obtain the following inclusions:

(a) $\quad A\subseteq \mathrm{Conv}\{\mathbf{PR},\mathbf{P_{00}},\mathbf{P_{11}},\mathbf{P_L^{1011}},\mathbf{P_L^{1110}},\mathbf{P_{NL}^{111}}\}\,;$

(b) $\quad A\subseteq \mathrm{Conv}\{\mathbf{PR},\mathbf{P_{00}},\mathbf{P_{11}},\mathbf{P_L^{1000}},\mathbf{P_L^{1101}},\mathbf{P_{NL}^{100}}\}\,.$

Now, taking the intersection, we obtain $A\subseteq C$, which yields the wanted equality $A=C$, and $C$ is indeed a slice of $\mathcal{NS}$.

It remains to show that the slice $C$ is included in the boundary $\partial\mathcal{NS}$

so it is indeed a face. Assume that there is a point $\mathbf{P}$ in $C$ of the form $\mathbf{P} = q\,\mathbf{Q}_1 + (1-q)\,\mathbf{Q}_2$ with $q > 0$ and $\mathbf{Q}_1, \mathbf{Q}_2 \in \mathcal{NS}$. Applying the convexity preservation property of CHSH$(\cdot)$, CHSH$'(\cdot)$, CHSH$''(\cdot)$ in Figure 6.12, we obtain the following two necessary conditions:

$$\text{(a)} \qquad \mathbf{Q}_1, \mathbf{Q}_2 \in \mathrm{Conv}\{\mathbf{PR}, \mathbf{P_{00}}, \mathbf{P_{11}}, \mathbf{P_L^{1011}}, \mathbf{P_L^{1110}}, \mathbf{P_{NL}^{111}}\}\,;$$

$$\text{(b)} \qquad \mathbf{Q}_1, \mathbf{Q}_2 \in \mathrm{Conv}\{\mathbf{PR}, \mathbf{P_{00}}, \mathbf{P_{11}}, \mathbf{P_L^{1000}}, \mathbf{P_L^{1101}}, \mathbf{P_{NL}^{100}}\}\,.$$

Then, taking the intersection, we get $\mathbf{Q}_1, \mathbf{Q}_2 \in C$ and therefore $C \subseteq \partial\mathcal{NS}$ as wanted. ∎

A direct consequence of Corollary 6.20 allows us to single out the quantum correlations of the face $C = \mathrm{Conv}\{\mathbf{PR}, \mathbf{P_{00}}, \mathbf{P_{11}}\}$: They are exactly the ones in the segment $\mathrm{Conv}\{\mathbf{P_{00}}, \mathbf{P_{11}}\}$ because quantum correlations *cannot* collapse communication complexity [Cle+99]. This allows us to recover the following statement from [Rai+19] with a new proof, based on communication complexity:

**Corollary 6.22 —** *The (closed) triangles $\overline{\mathcal{T}_i}$ draw in Figure 6.11 are quantum voids:*

$$\mathcal{Q} \cap \overline{\mathcal{T}_i} \subseteq \mathcal{L}\,.$$ ∎

# Chapter 7

# Communication Complexity in Graph Games

In this chapter, we include the following reference and add complementary material (*e.g.* Figure 7.3 or Remark 7.45):

## Chapter Contents

## 7.1 Graph Isomorphism Game

In this section, we show the collapse of communication complexity for some perfect strategies for the graph isomorphism game.

Below, after briefly recalling the definition of the game (Section 7.1.1), we prove a key preliminary result (Section 7.1.2), then we introduce the new family of symmetric strategies (Section 7.1.3), and finally, we state and prove our main theorems for this game (Sections 7.1.4 and 7.1.5).

### *7.1.1 Background*

We refer to Sections 3.1.1 and 3.1.2 for a background on nonlocal boxes, and to Section 3.2.1 for a background on general nonlocal games. We recall the following notations:

$$\begin{array}{ll} \mathbf{PR}(a,b\,|\,x,y) := \frac{1}{2}\,\mathbb{1}_{a\oplus b=xy}\,, & \mathbf{I}(a,b\,|\,x,y) := \frac{1}{4}\,, \\ \mathbf{P_{00}}(a,b\,|\,x,y) := \frac{1}{2}\,\mathbb{1}_{a=b=0}\,, & \mathbf{P_{11}}(a,b\,|\,x,y) := \frac{1}{2}\,\mathbb{1}_{a=b=1}\,. \end{array} \tag{7.1}$$

Moreover, we refer to Section 3.2.3 for a detailed definition of the graph isomorphism game. In short, in this game, the players Alice and Bob want to mimic to the Referee that two graphs $\mathcal{G}$ and $\mathcal{H}$ are isomorphic, although they might not actually be. Assume that they perfectly win the game, *i.e.* with probability $1$. If they have access to local resources only, then $\mathcal{G}$ and $\mathcal{H}$ are isomorphic in the usual sense and we write $\mathcal{G} \cong \mathcal{H}$. Now, if the players can use quantum (commuting) correlations, then the two graphs are said *quantum (commuting) isomorphic* and we write $\mathcal{G} \cong_{\text{qc}} \mathcal{H}$. Similarly, with non-signaling correlations, the graphs are said *non-signaling isomorphic* and we write $\mathcal{G} \cong_{\text{ns}} \mathcal{H}$.

Finally, the principle of communication complexity is introduced in Section 4.1.3. We will repeatedly use the fact that the **PR** box collapses communication complexity (van Dam [vD99]) as well as its noisy versions winning the CHSH game with probability at least $\frac{3+\sqrt{6}}{6} \approx 0.91$ (Brassard, Buhrman, Linden, Méthot, Tapp, and Unger [Bra+06]). Recall also that quantum boxes *cannot* collapse CC (Cleve, van Dam, Nielsen, and Tapp [Cle+99]). More details on these results can be found in Section 4.2.

### *7.1.2 Key Ideas*

In this subsection, we work on a simple case to present the key ideas that will be generalized in the theorem of Section 7.1.4. The assumption here is that $\mathcal{H}$ admits exactly two connected components that are both complete.

**Definition 7.1 —** *We denote by $\mathcal{C}_n$ the* cycle *of size $n \geqslant 1$,* i.e. *the finite undirected graph with vertices $v_1, \ldots, v_n$ and edges $v_i \sim v_{i+1}$ for $1 \leqslant i \leqslant n-1$ and $v_n \sim v_1$. We denote by $\mathcal{K}_n$ the* complete graph *of size $n \geqslant 1$,* i.e. *the finite undirected graph with vertices $v_1, \ldots, v_n$ and edges $v_i \sim v_j$ for any $i \neq j$. We also define the* path graph*, denoted $\mathcal{P}_n$, as the cycle $\mathcal{C}_n$ from which we remove one edge. The distance $d$ between two vertices $v_1, v_2$ in a graph $\mathcal{G}$ is defined as the smallest number of edges of a path connecting $v_1$ to $v_2$ over all possible paths. By convention, it is taken to be $\infty$ when there is no path connecting the vertices. Here, we call* diameter *of a graph $\mathcal{G}$, denoted $\mathrm{diam}(\mathcal{G})$, the largest distance between two connected vertices $g_1, g_2$ of $\mathcal{G}$ (in this definition, the diameter of a finite graph is finite, even if it is not connected).*

**Theorem 7.2** (Collapse of CC) **—** *Let $\mathcal{G}$ and $\mathcal{H}$ be two graphs such that $\mathrm{diam}(\mathcal{G}) \geqslant 2$, and that $\mathcal{H} = \mathcal{K}_n \sqcup \mathcal{K}_m$ where $\mathcal{K}_n, \mathcal{K}_m$ are complete graphs. Then any perfect strategy $\mathcal{S}$ for the graph isomorphism game $\mathcal{G} \cong_{\mathrm{ns}} \mathcal{H}$ collapses communication complexity.*

*Proof.* The proof simply consists in simulating the **PR** box from the strategy $\mathcal{S}$, in the sense that from inputs $x, y \in \{0,1\}$, we want to produce outputs $a, b \in \{0,1\}$ with the same behavior as the **PR** would do, *i.e.* to obtain $a \oplus b = xy$ in a non-signaling way. It is enough to simulate **PR** since this nonlocal box is known to collapse communication complexity [vD99]. As the diameter is $\mathrm{diam}(\mathcal{G}) \geqslant 2$, the graph $\mathcal{G}$ admits the path graph with three vertices $\mathcal{G}_0 = \mathcal{P}_3$ as a subgraph. We will use the protocol described in Figure 7.1. In this protocol, bits $x$ and $y$ are given to Alice and Bob respectively. They apply the respective processes $A_1$ and $B_1$ as described in item (b) and obtain vertices $g_{\mathsf{A}}$ and $g_{\mathsf{B}}$ in $V(\mathcal{G}_0)$. They use these two vertices in the graph isomorphism game of $(\mathcal{G}, \mathcal{H})$ and receive outputs $h_{\mathsf{A}}, h_{\mathsf{B}} \in V(\mathcal{H})$. Finally, they process these vertices with $A_2$ and $B_2$ as described in item (c) to obtain their output $a, b \in \{0,1\}$. Let us prove that this protocol indeed simulates the **PR** box. On the one hand, if $xy = 1$, then $x = 1 = y$, which gives $g_A \not\simeq g_B$ and therefore $h_A \not\simeq h_B$. It yields that the vertices $h_A$ and $h_B$ are in different components $\mathcal{H}_i$ and $\mathcal{H}_{i\oplus 1}$, so $a \oplus b = i \oplus i \oplus 1 = 1 = xy$ as wanted. On the other hand, if $xy = 0$, we have $x = 0$ or $y = 0$, so necessarily $g_A = g_B$ or $g_A \sim g_B$, and therefore $h_A = h_B$ or $h_A \sim h_B$. It follows that the vertices $h_A$ and $h_B$ are both in the same component $\mathcal{H}_i$, and $a \oplus b = i \oplus i = 0 = xy$ as expected as well. In addition, note that this protocol does *not* signal between Alice and Bob. Hence the **PR** is perfectly simulated, and there is a collapse of communication complexity. ∎

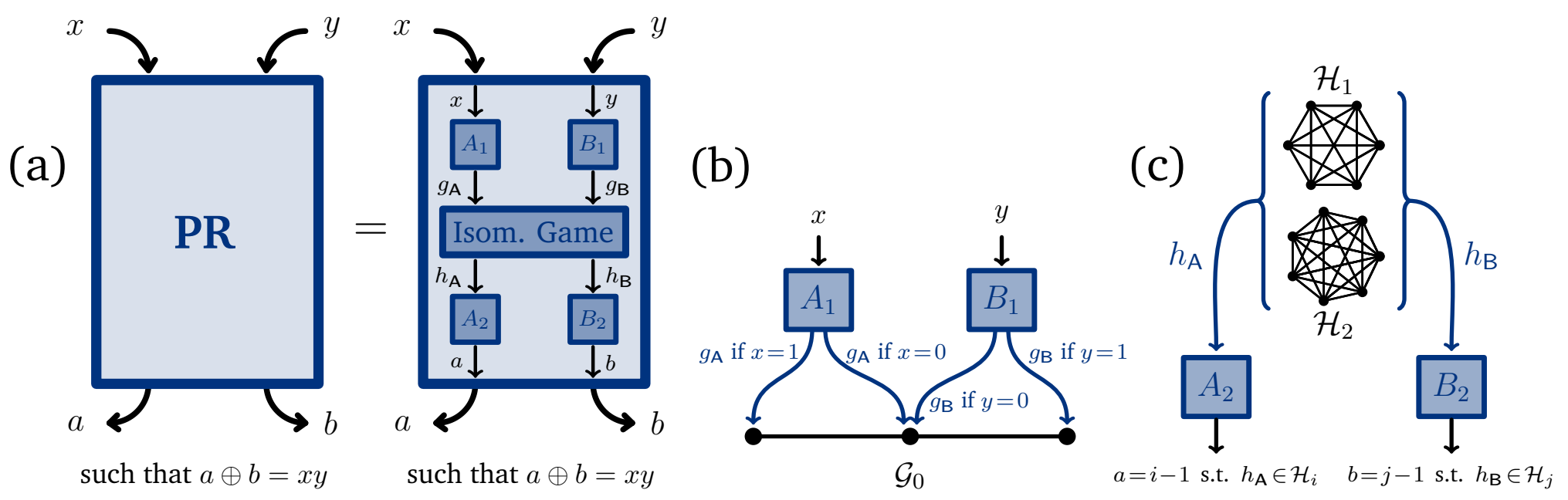

**Figure 7.1 —** *Illustration of the proof of Theorem 7.2. In item (a), we simulate a* **PR** *box from a perfect $\mathcal{NS}$-strategy for the graph isomorphism game, called "Isom. Game" in the figure, together with the local processes $A_1, A_2, B_1, B_2$ that are described in items (b) and (c). In item (b), the graph $\mathcal{G}_0$ is a subgraph of $\mathcal{G}$, in which Alice and Bob choose some input vertices $g_A$ and $g_B$. In item (c), the graphs $\mathcal{H}_1$ and $\mathcal{H}_2$ are the two connected components of $\mathcal{H}$, from which Alice and Bob receive some output vertices $h_A$ and $h_B$.*

**Example 7.3 —** Any perfect strategy for the isomorphism game $\mathcal{C}_6 \cong_{\mathrm{ns}} \mathcal{K}_3 \sqcup \mathcal{K}_3$ allows to perfectly simulate the **PR** box and to collapse CC. [1]

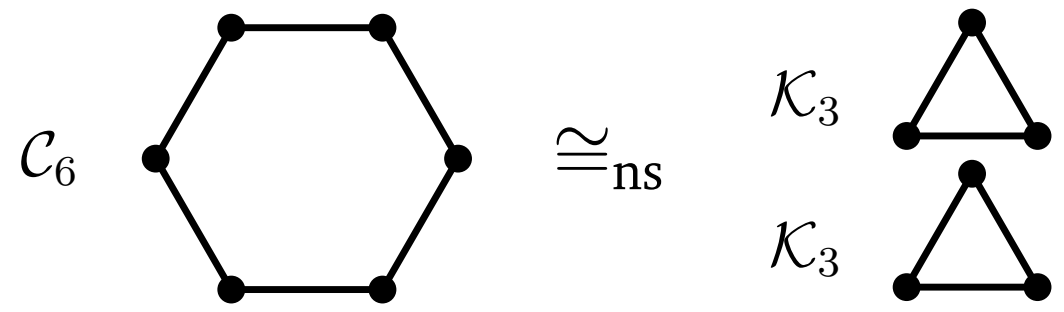

[1] As later detailed in Section 7.1.3, two finite undirected graphs $\mathcal{G}$ and $\mathcal{H}$ with the same number of vertices are $\mathcal{NS}$-isomorphic *if, and only if,* they admit a common equitable partition. A sufficient condition for the latter condition is that the graphs are regular with the same degree, that is each vertex has a fixed constant number of neighbors. In particular, it holds $\mathcal{C}_6 \cong_{\mathrm{ns}} \mathcal{K}_3 \sqcup \mathcal{K}_3$.

This result can be generalized to non-perfect strategies $\mathcal{S}$ as follows:

**Corollary 7.4** (Collapse With Non-Perfect Strategies) **—** *With the same assumptions as in Theorem 7.2, any strategy $\mathcal{S}$ that succeeds with probability $p > \frac{3+\sqrt{6}}{6} \approx 0.91$ at the graph isomorphism game $\mathcal{G} \cong_{\text{ns}} \mathcal{H}$ collapses communication complexity.*

*Proof.* We use the same protocol as in Theorem 7.2. It simulates a **PR** box with probability $p$, which is known to collapse communication complexity from Brassard, Buhrman, Linden, Méthot, Tapp, and Unger [Bra+06]. ∎

These results can be generalized to more than two connected components in $\mathcal{H}$, based on the assumption that Alice and Bob are given access to a perfect $\mathcal{NS}$-strategy for the 2-coloring game of $\mathcal{K}_N$, which is always granted when $N = 2$, see Theorem 7.23.

### 7.1.3 Symmetric Strategies

We define a new type of perfect strategy for the graph isomorphism game, that we call *symmetric strategies*, for which we show a collapse of communication complexity in the next subsection.

**Definition 7.5 —** *An $\mathcal{NS}$-strategy $\mathcal{S}$ for the graph isomorphism game $(\mathcal{G}, \mathcal{H})$ is said* symmetric *from $\mathcal{G}$ to the components of $\mathcal{H}$ if it is perfect (*i.e. *the winning probability is $1$) and if there exist a disjoint decomposition $\mathcal{H} = \mathcal{H}_1 \sqcup \mathcal{H}_2$ and some constants $\eta, \nu_{g_A, g_B} \in [0,1]$ such that, for all $g_A, g_B \in \mathcal{G}$, we have:*

$$\begin{cases} \mathbb{P}_{\mathcal{S}}\big(h_A \in \mathcal{H}_1,\ h_B \in \mathcal{H}_2 \,\big|\, g_A,\ g_B\big) = \mathbb{P}_{\mathcal{S}}\big(h_A \in \mathcal{H}_2,\ h_B \in \mathcal{H}_1 \,\big|\, g_A,\ g_B\big) =: \nu_{g_A, g_B} \\ \mathbb{P}_{\mathcal{S}}\big(h_A \in \mathcal{H}_1,\ h_B \in \mathcal{H}_1 \,\big|\, g_A,\ g_B\big) = \eta - \nu_{g_A, g_B} \,. \end{cases}$$

Examples of graphs admitting symmetric strategies can be found in Example 7.8. For the sake of the next proposition, we recall that two finite undirected graphs $\mathcal{G}$ and $\mathcal{H}$ are $\mathcal{NS}$-isomorphic *if, and only if,* they are fractionally isomorphic [Ats+19], *if, and only if,* they admit a common equitable partition [RSU94], whose definition is recalled below.

**Common Equitable Partition [RSU94].** Given a graph $\mathcal{G}$, define a partition $\mathscr{C} = \{C_1, \ldots, C_k\}$ of its vertices, that is subsets $C_i \subseteq V(\mathcal{G})$ such that every vertex $g$ of $\mathcal{G}$ belongs to exactly one set $C_{i_g}$, which may be viewed as assigning a (unique) color $C_{i_g}$ to each vertex. We say that this partition is *equitable* if, for all $i, j \in [k]$, any vertex of color $C_i$ admits precisely a fixed number, denoted $c_{ij}$, of neighbors colored with $C_j$. Note that $c_{ij}$ and $c_{ji}$ may differ, but the equality $c_{ij}|C_i| = c_{ji}|C_j|$ always holds (see Lemma 7.36 for a proof of a generalized result). A trivial example of an equitable partition is the minimal partition $\mathscr{C} = \{\{g\} : g \in V(\mathcal{G})\}$, where to each vertex a different color is assigned and where the matrix $[c_{ij}]_{i,j}$ is the adjacency matrix. Another example is the maximal partition $\mathscr{C} = \{V(\mathcal{G})\}$, which is equitable *if, and only if,* the graph $\mathcal{G}$ is regular of degree $c_{11}$. Now, we say that two graphs $\mathcal{G}$ and $\mathcal{H}$ admit a *common equitable partition* if they admit equitable partitions of same length $\mathscr{C} = \{C_1, \ldots, C_k\}$ and $\mathscr{D} = \{D_1, \ldots, D_k\}$ such that the cells have same size $|C_i| = |D_i| =: n_i$ and the partition parameters coincide $c_{ij} = d_{ij}$ for all $i, j \in [k]$. When such partitions exist, we may concisely describe them in terms of the parameters $\big(k, (n_i), (c_{ij})\big)$ and, when the context is clear, we may consider $\mathscr{C}$ as an equitable partition of both $\mathcal{G}$ and $\mathcal{H}$. As mentioned above, we will use the fact that $\mathcal{G} \cong_{\text{ns}} \mathcal{H}$ *if, and only if,* the graphs admit a common equitable partition [RSU94]. For instance, the graphs $\mathcal{G} = \mathcal{C}_6$ and $\mathcal{H} = \mathcal{C}_3 \sqcup \mathcal{C}_3$ are both 2-regular, so they admit admit a common equitable partition $\big(k = 1, (n_1 = 6), (c_{11} = 2)\big)$, which is why they are $\mathcal{NS}$-isomorphic.

**Proposition 7.6** (Existence of Symmetric Strategies) — *Let $\mathcal{G} \cong_{\text{ns}} \mathcal{H}$ such that $\mathcal{H}$ is not connected: $\mathcal{H} = \mathcal{H}_1 \sqcup \mathcal{H}_2$. Denote the partitions $\mathscr{C} = \{C_1, \ldots, C_k\}$ and $\mathscr{D} = \{D_1, \ldots, D_k\}$ forming a common equitable partition for $\mathcal{G}$ and $\mathcal{H}$, and assume that the proportion of vertices of $\mathcal{H}_1$ assigned to $D_i$ is independent of $i$:*

$$\forall i, j \in [k], \qquad \frac{|D_i \cap \mathcal{H}_1|}{|D_i|} = \frac{|D_j \cap \mathcal{H}_1|}{|D_j|}\,. \tag{H}$$

*Then the isomorphism game of $(\mathcal{G}, \mathcal{H})$ admits a symmetric strategy.*

*Proof.* Let $\big(k,\, [n_1, \ldots, n_k],\, [c_{ij}]_{1 \leqslant i,j \leqslant k}\big)$ be the parameters of a common equitable partition for $\mathcal{G}$ and $\mathcal{H}$, and consider $\overline{c_{ij}} := n_j - c_{ij} - \delta_{ij}$ the number of non-neighbours a vertex of $C_i$ has in $C_j$ for all $i, j$, where $\delta_{ij}$ is the Kro-necker delta function. We define the following strategy $\mathcal{S}$ as in [Ats+19, Lemma 4.4] for which the authors prove it is perfect for the isomorphism game of $(\mathcal{G},\mathcal{H})$:

$$\mathbb{P}_{\mathcal{S}}(h_{\mathsf{A}},h_{\mathsf{B}}\,|\,g_{\mathsf{A}},g_{\mathsf{B}}) \;:=\; \begin{cases} 1/n_i & \text{if } g_{\mathsf{A}}=g_{\mathsf{B}} \text{ and } h_{\mathsf{A}}=h_{\mathsf{B}} \text{ and } (\star),\\ 1/n_i c_{ij} & \text{if } g_{\mathsf{A}}\sim g_{\mathsf{B}} \text{ and } h_{\mathsf{A}}\sim h_{\mathsf{B}} \text{ and } (\star),\\ 1/n_i \overline{c_{ij}} & \text{if } g_{\mathsf{A}}\not\simeq g_{\mathsf{B}} \text{ and } h_{\mathsf{A}}\not\simeq h_{\mathsf{B}} \text{ and } (\star),\\ 0 & \text{otherwise,}\end{cases} \tag{7.2}$$

where $(\star)$ denotes the condition "$g_{\mathsf{A}}\in C_i$, $g_{\mathsf{B}}\in C_j$, $h_{\mathsf{A}}\in D_i$, $h_{\mathsf{B}}\in D_j$". Note that $\mathbb{P}_{\mathcal{S}}$ is well defined because in each case the division by zero is prevented using the condition of occurrence. Let us show that $\mathcal{S}$ is symmetric in two steps.

• First, we compute the constants $\nu_{g_{\mathsf{A}},g_{\mathsf{B}}} := \mathbb{P}_{\mathcal{S}}\big(h_{\mathsf{A}}\in\mathcal{H}_1,\, h_{\mathsf{B}}\in\mathcal{H}_2\,\big|\,g_{\mathsf{A}},\, g_{\mathsf{B}}\big)$. Notice that $\nu_{g_{\mathsf{A}},g_{\mathsf{B}}}=0$ when $g_{\mathsf{A}}=g_{\mathsf{B}}$ or $g_{\mathsf{A}}\sim g_{\mathsf{B}}$ by disconnectedness of $\mathcal{H}_1$ and $\mathcal{H}_2$. Now if $g_{\mathsf{A}}\not\simeq g_{\mathsf{B}}$ for some $g_{\mathsf{A}}\in C_i$ and $g_{\mathsf{B}}\in C_j$, then:

$$\begin{aligned}\mathbb{P}_{\mathcal{S}}\big(h_{\mathsf{A}}\in\mathcal{H}_1,h_{\mathsf{B}}\in\mathcal{H}_2\,\big|\,g_{\mathsf{A}}\not\simeq g_{\mathsf{B}}\big) &= \sum_{(h_{\mathsf{A}},h_{\mathsf{B}})\in\mathcal{H}_1\times\mathcal{H}_2} \mathbb{P}_{\mathcal{S}}(h_{\mathsf{A}},h_{\mathsf{B}}\,|\,g_{\mathsf{A}}\not\simeq g_{\mathsf{B}})\\ &= \sum_{\substack{(h_{\mathsf{A}},h_{\mathsf{B}})\in\mathcal{H}_1\times\mathcal{H}_2\\ \text{s.t. } h_{\mathsf{A}}\in D_i,h_{\mathsf{B}}\in D_j}} \frac{1}{n_i\,\overline{c_{ij}}}\\ &= \frac{|D_i\cap\mathcal{H}_1|\times|D_j\cap\mathcal{H}_2|}{n_i\,\overline{c_{ij}}}\,.\end{aligned}$$

But as $|D_i|=|D_i\cap\mathcal{H}_1|+|D_i\cap\mathcal{H}_2|$ for all $i$, we see that the assumption (H) is equivalent to saying $|D_i\cap\mathcal{H}_1|\times|D_j\cap\mathcal{H}_2| = |D_i\cap\mathcal{H}_2|\times|D_j\cap\mathcal{H}_1|$. Therefore, the above quantity also equals $\mathbb{P}_{\mathcal{S}}\big(h_{\mathsf{A}}\in\mathcal{H}_2,h_{\mathsf{B}}\in\mathcal{H}_1\,\big|\,g_{\mathsf{A}},g_{\mathsf{B}}\big)$, and we obtain:

$$\nu_{g_{\mathsf{A}},g_{\mathsf{B}}} = \begin{cases}\dfrac{|D_i\cap\mathcal{H}_1|\times|D_j\cap\mathcal{H}_2|}{n_i\,\overline{c_{ij}}} & \text{if } g_{\mathsf{A}}\not\simeq g_{\mathsf{B}} \text{ and } g_{\mathsf{A}}\in C_i \text{ and } g_{\mathsf{B}}\in C_j,\\ 0 & \text{otherwise.}\end{cases}$$

• Second, we compute $\eta := \mathbb{P}_{\mathcal{S}}\big(h_{\mathsf{A}}\in\mathcal{H}_1,\, h_{\mathsf{B}}\in\mathcal{H}_1\,\big|\,g_{\mathsf{A}},\, g_{\mathsf{B}}\big)+\nu_{g_{\mathsf{A}},g_{\mathsf{B}}}$, which should be independent of $g_{\mathsf{A}},g_{\mathsf{B}}$. Let $g_{\mathsf{A}}\in C_i$ and $g_{\mathsf{B}}\in C_j$. We split the study into three cases.

$(i)$ If $g_\mathsf{A} = g_\mathsf{B}$, then:

$$\begin{aligned}\mathbb{P}_\mathcal{S}\big((h_\mathsf{A}, h_\mathsf{B}) \in {\mathcal{H}_1}^2 \,\big|\, g_\mathsf{A} = g_\mathsf{B}\big) &= \sum_{(h_\mathsf{A},h_\mathsf{B})\in{\mathcal{H}_1}^2} \mathbb{P}_\mathcal{S}(h_\mathsf{A}, h_\mathsf{B} \,|\, g_\mathsf{A} = g_\mathsf{B}) \\ &= \sum_{\substack{h\in\mathcal{H}_1 \\ \text{s.t. } h\in D_i}} \underbrace{\mathbb{P}_\mathcal{S}(h, h \,|\, g_\mathsf{B} = g_\mathsf{B})}_{=1/n_i} \\ &= \frac{|D_i \cap \mathcal{H}_1|}{n_i} \quad = \quad \eta - \nu_{g_\mathsf{A},g_\mathsf{B}}\,,\end{aligned}$$

where we fixed $\eta := \frac{|D_i\cap\mathcal{H}_1|}{n_i}$, and where we have $n_i \neq 0$ because $g_\mathsf{A} \in C_i$. Let us verify that this $\eta$ is appropriate in the other cases as well.
$(ii)$ If $g_\mathsf{A} \sim g_\mathsf{B}$, then:

$$\begin{aligned}\mathbb{P}_\mathcal{S}\big((h_\mathsf{A}, h_\mathsf{B}) \in {\mathcal{H}_1}^2 \,\big|\, g_\mathsf{A} \sim g_\mathsf{B}\big) &= \sum_{(h_\mathsf{A},h_\mathsf{B})\in{\mathcal{H}_1}^2} \mathbb{P}_\mathcal{S}(h_\mathsf{A}, h_\mathsf{B} \,|\, g_\mathsf{A} \sim g_\mathsf{B}) \\ &= \sum_{\substack{(h_\mathsf{A},h_\mathsf{B})\in{\mathcal{H}_1}^2 \\ \text{s.t. } h_\mathsf{A}\in D_i, \\ h_\mathsf{B}\in D_j\cap N(h_\mathsf{A})}} \underbrace{\mathbb{P}_\mathcal{S}(h_\mathsf{A}, h_\mathsf{B} \,|\, g_\mathsf{A} \sim g_\mathsf{B})}_{=1/n_i c_{ij}}\,,\end{aligned}$$

where $N(h_\mathsf{A})$ is the set of adjacent vertices to $h_\mathsf{A}$ and where $c_{ij} \neq 0$ because $g_\mathsf{A} \sim g_\mathsf{B}$, so:

$$= \sum_{h_\mathsf{A}\in D_i\cap\mathcal{H}_1} \frac{\big|D_j \cap N(h_\mathsf{A}) \cap \mathcal{H}_1\big|}{n_i\, c_{ij}}\,.$$

But by disconnectedness of $\mathcal{H}_1$ and $\mathcal{H}_2$ we have $|D_j \cap N(h_\mathsf{A}) \cap \mathcal{H}_1| = |D_j \cap N(h_\mathsf{A})|$ , which is equal to $c_{ij}$ by definition. Hence, it simplifies with the coefficient in the denominator, so we obtain:

$$= \sum_{h_\mathsf{A}\in D_i\cap\mathcal{H}_1} \frac{1}{n_i} \quad = \quad \frac{|D_i \cap \mathcal{H}_1|}{n_i} \quad = \quad \eta - \nu_{g_\mathsf{A},g_\mathsf{B}}\,.$$

$(iii)$ If $g_\mathsf{A} \not\simeq g_\mathsf{B}$:

$$\begin{aligned}\mathbb{P}_\mathcal{S}\big((h_\mathsf{A}, h_\mathsf{B}) \in {\mathcal{H}_1}^2 \,\big|\, g_\mathsf{A} \not\simeq g_\mathsf{B}\big) &= \sum_{(h_\mathsf{A},h_\mathsf{B})\in{\mathcal{H}_1}^2} \mathbb{P}_\mathcal{S}(h_\mathsf{A}, h_\mathsf{B} \,|\, g_\mathsf{A} \not\simeq g_\mathsf{B}) \\ &= \sum_{\substack{(h_\mathsf{A},h_\mathsf{B})\in{\mathcal{H}_1}^2 \\ \text{s.t. } h_\mathsf{A}\in D_i, \\ h_\mathsf{B}\in D_j\setminus(N(h_\mathsf{A})\cup\{h_\mathsf{A}\})}} \underbrace{\mathbb{P}_\mathcal{S}(h_\mathsf{A}, h_\mathsf{B} \,|\, g_\mathsf{A} \not\simeq g_\mathsf{B})}_{=1/n_i\overline{c_{ij}}}\,,\end{aligned}$$

where $\overline{c_{ij}} \neq 0$ because $g_{\mathsf{A}} \not\simeq g_{\mathsf{B}}$, so:

$$= \sum_{h_{\mathsf{A}} \in D_i \cap \mathcal{H}_1} \frac{\big|D_j \cap \mathcal{H}_1 \backslash (N(h_{\mathsf{A}}) \cup \{h_{\mathsf{A}}\})\big|}{n_i \, \overline{c_{ij}}} \, .$$

But by definition $n_j := |D_j| = |D_j \cap \mathcal{H}_2| + |D_j \cap \mathcal{H}_1 \backslash (N(h_{\mathsf{A}}) \cup \{h_{\mathsf{A}}\})| + |D_j \cap \mathcal{H}_1 \cap (N(h_{\mathsf{A}}) \cup \{h_{\mathsf{A}}\})|$, where the last term is $c_{ij} + \delta_{ij}$. After reordering the terms, it yields that the second term is $|D_j \cap \mathcal{H}_1 \backslash (N(h_{\mathsf{A}}) \cup \{h_{\mathsf{A}}\})| = \overline{c_{ij}} - |D_j \cap \mathcal{H}_2|$, therefore:

$$\begin{aligned} &= \sum_{h_{\mathsf{A}} \in D_i \cap \mathcal{H}_1} \frac{\overline{c_{ij}} - |D_j \cap \mathcal{H}_2|}{n_i \, \overline{c_{ij}}} \\ &= \frac{|D_i \cap \mathcal{H}_1|}{n_i} - \frac{|D_i \cap \mathcal{H}_1| \times |D_j \cap \mathcal{H}_2|}{n_i \, \overline{c_{ij}}} \\ &= \eta - \nu_{g_{\mathsf{A}}, g_{\mathsf{B}}} \, . \end{aligned}$$

Hence, the coefficient $\eta$ is the same in the three cases and independent of $g_{\mathsf{A}}, g_{\mathsf{B}}$, so we proved that $\mathcal{S}$ is indeed symmetric. ■

**Corollary 7.7 —** *Let $\mathcal{G}, \mathcal{H}$ be two graphs of degree $d$ such that $\mathcal{H}$ is not connected. Then $\mathcal{G} \cong_{\text{ns}} \mathcal{H}$ and this isomorphism game admits a symmetric strategy.*

*Proof.* Consider the common equitable partition given by the parameters $\big(k = 1,\, n_1 = |V(\mathcal{G})|,\, c_{11} = d\big)$. Deduce that $\mathcal{G} \cong_{\text{ns}} \mathcal{H}$ by [Ats+19], and then conclude using Proposition 7.6, because the above hypothesis (H) is obviously satisfied when $k = 1$. ■

**Example 7.8 —** For any integer decomposition $M = m_1 + \cdots + m_K$ with $m_i \geqslant 3$ and $K \geqslant 2$, the previous corollary tells us that the following isomorphism of cycles holds:

$$\mathcal{C}_M \quad \cong_{\text{ns}} \quad \mathcal{C}_{m_1} \sqcup \cdots \sqcup \mathcal{C}_{m_K} \, ,$$

and that the associated isomorphism game admits a symmetric strategy.

### *7.1.4 Existence of Perfect $\mathcal{NS}$-Strategies that Collapse CC*

In this subsection, under some conditions, we prove that the graph isomorphism game $(\mathcal{G}, \mathcal{H})$ admits a perfect non-signaling strategy that collapses communication complexity. In the next subsection, we will add conditions on $\mathcal{H}$ to obtain that all perfect non-signaling strategies are collapsing.

**Theorem 7.9** (Collapse of CC) — *Let $\mathcal{G} \cong_{\text{ns}} \mathcal{H}$ such that $\operatorname{diam}(\mathcal{G}) \geqslant 2$ and such that $\mathcal{H}$ is not connected: $\mathcal{H} = \mathcal{H}_1 \sqcup \mathcal{H}_2$, where each of $\mathcal{H}_1$ and $\mathcal{H}_2$ may possibly be decomposed in several connected components. Denote the partitions $\mathscr{C} = \{C_1, \dots, C_k\}$ and $\mathscr{D} = \{D_1, \dots, D_k\}$ forming a common equitable partition for $\mathcal{G}$ and $\mathcal{H}$, and assume the hypothesis (H) as in Proposition 7.6. Then the isomorphism game of $(\mathcal{G}, \mathcal{H})$ admits a perfect strategy that collapses communication complexity.*

To prove the theorem, we define a noisy version of the **PR** box, that we denote $\mathbf{PR}_{\alpha,\beta} := \alpha\,\mathbf{PR} + \beta\,\mathbf{P_{00}} + (1-\alpha-\beta)\,\mathbf{P_{11}}$, which is the convex combination with coefficients $\alpha, \beta \geqslant 0$ of the boxes **PR**, $\mathbf{P_{00}}$ and $\mathbf{P_{11}}$ defined in eq. (7.1). This noisy box is known to collapse communication complexity as long as $\alpha > 0$ [Bot+24b; Bri+19], so the idea is to first prove that $\mathbf{PR}_{\alpha,\beta}$ can be generated from a perfect strategy $\mathcal{S}$ for the isomorphism game:

**Lemma 7.10** — *Let $\mathcal{G}, \mathcal{H}$ two graphs such that $\operatorname{diam}(\mathcal{G}) \geqslant 2$ and such that $\mathcal{H}$ is not connected: $\mathcal{H} = \mathcal{H}_1 \sqcup \mathcal{H}_2$. Assume $\mathcal{G} \cong_{\text{ns}} \mathcal{H}$ for some strategy $\mathcal{S}$ that is symmetric from $\mathcal{G}_0$ to the components of $\mathcal{H}$, and suppose that:*

$$\nu_{g_1,g_3} > 0\,. \tag{7.3}$$

*Then the box $\mathbf{PR}_{\alpha,\beta}$ is perfectly simulated with $\alpha = 2\,\nu_{g_1,g_3} > 0$ and some $\beta \geqslant 0$.*

*Proof.* The protocol in this proof is inspired by the one of Theorem 7.2. As the diameter is $\operatorname{diam}(\mathcal{G}) \geqslant 2$, the graph $\mathcal{G}$ admits the path graph with three vertices $\mathcal{G}_0 = \mathcal{P}_3$ as a subgraph, whose vertices are called $g_1, g_2, g_3$ each one being connected to the next one. We proceed with the same protocol as described in Figure 7.1, the only difference being that $\mathcal{H}_1$ and $\mathcal{H}_2$ are not necessarily complete and not even connected. Denote $\mathbf{P}(a,b\,|\,x,y)$ the nonlocal box induced by this protocol. Let us prove that $\mathbf{P} = \mathbf{PR}_{\alpha,\beta}$ for some $\alpha, \beta \in [0,1]$. To this end, we compare the correlation tables of **P** and $\mathbf{PR}_{\alpha,\beta}$ (see definition in page 66), defined as follows:

$$M_{\mathbf{P}} := \begin{bmatrix} \mathbf{P}(0,0\,|\,0,0) & \mathbf{P}(0,1\,|\,0,0) & \mathbf{P}(1,0\,|\,0,0) & \mathbf{P}(1,1\,|\,0,0) \\ \mathbf{P}(0,0\,|\,0,1) & \mathbf{P}(0,1\,|\,0,1) & \mathbf{P}(1,0\,|\,0,1) & \mathbf{P}(1,1\,|\,0,1) \\ \mathbf{P}(0,0\,|\,1,0) & \mathbf{P}(0,1\,|\,1,0) & \mathbf{P}(1,0\,|\,1,0) & \mathbf{P}(1,1\,|\,1,0) \\ \mathbf{P}(0,0\,|\,1,1) & \mathbf{P}(0,1\,|\,1,1) & \mathbf{P}(1,0\,|\,1,1) & \mathbf{P}(1,1\,|\,1,1) \end{bmatrix},$$

and similarly defined for $\mathbf{PR}_{\alpha,\beta}$. On the one hand, by symmetry of the strategy $\mathcal{S}$, the correlation table of **P** can be computed explicitly:

$$M_{\mathbf{P}} = \begin{bmatrix} \eta - \nu_{g_2,g_2} & \nu_{g_2,g_2} & \nu_{g_2,g_2} & 1-\eta-\nu_{g_2,g_2} \\ \eta - \nu_{g_2,g_3} & \nu_{g_2,g_3} & \nu_{g_2,g_3} & 1-\eta-\nu_{g_2,g_3} \\ \eta - \nu_{g_1,g_2} & \nu_{g_1,g_2} & \nu_{g_1,g_2} & 1-\eta-\nu_{g_1,g_2} \\ \eta - \nu_{g_1,g_3} & \nu_{g_1,g_3} & \nu_{g_1,g_3} & 1-\eta-\nu_{g_1,g_3} \end{bmatrix}.$$

But, we see that if $x = 0$ or $y = 0$, then the inputs $g_{\mathsf{A}}, g_{\mathsf{B}} \in V(\mathcal{G})$ of Alice and Bob in $\mathcal{S}$ are either equal or adjacent. It turns out that the outputs $h_{\mathsf{A}}, h_{\mathsf{B}} \in V(\mathcal{H})$ of $\mathcal{S}$ are in the same connected component, therefore $a = b$ almost surely and $\mathbf{P}(a \neq b\,|\,x = 0 \text{ or } y = 0) = 0$. Hence, in the first three lines of the matrix, we have $\nu_{g_2,g_2} = \nu_{g_2,g_3} = \nu_{g_1,g_2} = 0$, which gives:

$$M_{\mathbf{P}} = \begin{bmatrix} \eta & 0 & 0 & 1-\eta \\ \eta & 0 & 0 & 1-\eta \\ \eta & 0 & 0 & 1-\eta \\ \eta - \nu_{g_1,g_3} & \nu_{g_1,g_3} & \nu_{g_1,g_3} & 1-\eta-\nu_{g_1,g_3} \end{bmatrix}.$$

On the other hand, by linearity, the correlation table of the box $\mathbf{PR}_{\alpha,\beta}$ is the convex combination of the correlation tables of **PR**, $\mathbf{P}_{00}$, $\mathbf{P}_{11}$ with coefficients $\alpha, \beta \in [0,1]$ fixed above, so:

$$M_{\mathbf{PR}_{\alpha,\beta}} = \begin{bmatrix} \frac{\alpha}{2}+\beta & 0 & 0 & 1-\frac{\alpha}{2}-\beta \\ \frac{\alpha}{2}+\beta & 0 & 0 & 1-\frac{\alpha}{2}-\beta \\ \frac{\alpha}{2}+\beta & 0 & 0 & 1-\frac{\alpha}{2}-\beta \\ \beta & \frac{\alpha}{2} & \frac{\alpha}{2} & 1-\alpha-\beta \end{bmatrix}.$$

Now, taking $\alpha := 2\,\nu_{g_1,g_3}$ and $\beta := \eta - \nu_{g_1,g_3}$, we obtain $\mathbf{P} = \mathbf{PR}_{\alpha,\beta}$ as wanted. ■

*Proof (Theorem 7.9).* To invoke the former lemma, we need to prove the existence of a symmetric strategy, which was the purpose of Proposition 7.6.

We can apply this proposition because all its assumptions are found in the theorem as well. It yields a symmetric strategy from $\mathcal{G}$ to the components of $\mathcal{H}$, and in particular, its restriction to $\mathcal{G}_0$ is also symmetric. Moreover, assumption (7.3) in Lemma 7.10 is also satisfied because, using the computations in the proof of Proposition 7.6, we have:

$$\nu_{g_1,g_3} = \frac{|D_i \cap \mathcal{H}_1| \times |D_j \cap \mathcal{H}_2|}{n_i\, \overline{c_{ij}}} > 0\,,$$

where $i$ and $j$ are such that $g_1 \in C_i$ and $g_3 \in C_j$. Thus we can apply Lemma 7.10 and we can simulate $\mathbf{PR}_{\alpha,\beta}$ for some $\alpha > 0$. But the box $\mathbf{PR}_{\alpha,\beta}$ with $\alpha > 0$ is known to be collapsing [Bot+24b; Bri+19]. Hence we deduce the existence of a protocol that collapses CC. ■

**Corollary 7.11** (Collapse of CC) — *Each of the $\mathcal{NS}$-isomorphisms given in Example 7.8 admits a perfect strategy $\mathcal{S}$ that allows one to perfectly produce a box $\mathbf{PR}_{\alpha,\beta}$ with $\alpha > 0$ and therefore to collapse CC.* ■

## *7.1.5 All Perfect Non-Signaling Strategies Collapse CC*

In Theorem 7.9 above, the statement was that *some* perfect strategies collapse communication complexity. Now, if we add regularity and transitivity conditions on $\mathcal{H}$, we obtain that *every* perfect strategy collapses communication complexity. First, we recall the definition of the automorphism group of a graph.

### 7.1.5.1 Automorphism Group

The *automorphism group* of $\mathcal{H}$, denoted $\mathrm{Aut}(\mathcal{H})$, is the set of all bijective maps $\varphi : \mathcal{H} \to \mathcal{H}$ that preserve the adjacency relation, meaning that $h_1 \sim h_2$ *if, and only if,* $\varphi(h_1) \sim \varphi(h_2)$. As a consequence, any automorphism $\varphi \in \mathrm{Aut}(\mathcal{H})$ also preserves the relation "$\not\simeq$", and therefore a graph $\mathcal{H}$ and its complement $\mathcal{H}^c$ have the same automorphism group: $\mathrm{Aut}(\mathcal{H}) = \mathrm{Aut}(\mathcal{H}^c)$. Moreover, the composition of two automorphisms is again an automorphism, and $\varphi^{-1} \in \mathrm{Aut}(\mathcal{H})$, which endows the set $\mathrm{Aut}(\mathcal{H})$ with a group structure. For instance, the automorphism group of the complete graph $\mathcal{K}_N$ is the symmetric group $\mathrm{Aut}(\mathcal{K}_N) = \mathfrak{S}_N$ of order $N!$, and the one of the cycle $\mathcal{C}_N$ is the dihedral group $\mathrm{Aut}(\mathcal{C}_N) = D_N$ of order $2N$. We refer to [GR01] for more details on automorphism groups.

#### 7.1.5.2 Graph Transitivity

The graph $\mathcal{H}$ is said to be *vertex-transitive* if for all vertices $h, h' \in V(\mathcal{H})$, there exists a graph automorphism $\varphi \in \mathrm{Aut}(\mathcal{H})$ such that $\varphi(h) = h'$. Similarly, we say that $\mathcal{H}$ is *edge-transitive* if for all edges $h_1 \sim h_2$, $h_1' \sim h_2' \in E(\mathcal{H})$, there exists a graph automorphism $\varphi \in \mathrm{Aut}(\mathcal{H})$ such that $\varphi(h_1) = h_1'$ and $\varphi(h_2) = h_2'$. We refer to [GR01] for more details on graph transitivity, and to [Gau97; SVW19] for related notions. In the definition below, we strengthen the vertex- and edge-transitivity of a graph $\mathcal{H}$ and its complement $\mathcal{H}^c$ in what we call the *strong transitivity*:

**Definition 7.12** (Strongly Transitive) — *We say that a graph $\mathcal{H}$ is* strongly transitive *if there exists a subset of the automorphism group $S \subseteq \mathrm{Aut}(\mathcal{H})$ such that the three following conditions hold:*

*(1) There is a constant $d_1 \geqslant 1$ such that for all vertices $h, h' \in V(\mathcal{H})$, there exist exactly $d_1$ automorphisms $\varphi \in S$ such that $\varphi(h) = h'$.*

*(2) There is a constant $d_2 \geqslant 1$ such that for all edges $h_1 \sim h_2$, $h_1' \sim h_2' \in E(\mathcal{H})$, there exist exactly $d_2$ automorphisms $\varphi \in S$ such that $\varphi(h_1) = h_1'$ and $\varphi(h_2) = h_2'$.*

*(3) There is a constant $d_3 \geqslant 1$ such that for all edges in the complement graph $h_1 \sim h_2$, $h_1' \sim h_2' \in E(\mathcal{H}^c)$, there exist exactly $d_3$ automorphisms $\varphi \in S$ such that $\varphi(h_1) = h_1'$ and $\varphi(h_2) = h_2'$.*

Notice that strong transitivity implies vertex- and edge-transitivity of both $\mathcal{H}$ and its complement $\mathcal{H}^c$, since it is possible to pick one among the respective $d_1$, $d_2$, $d_3$ automorphisms $\varphi \in S$ satisfying the wanted condition for each vertices and edges. Note that if $\mathcal{H}$ is strongly transitive, then its complement $\mathcal{H}^c$ is also strongly transitive. Note also that $S$ cannot be the empty set $\emptyset$ because of item (1), unless the graph $\mathcal{H}$ is itself empty.

Let us prove the following characterization of strong transitivity, and then provide some examples of strongly transitive graphs. First recall that a graph $\mathcal{H}$ is called *distance-transitive* if for any $d \in \mathbb{N}$ and any two pairs $(h_1, h_2)$ and $(h_1', h_2')$ of vertices $h_1, h_2, h_1', h_2' \in V(\mathcal{H})$ with distance $d(h_1, h_2) = d(h_1', h_2') = d$, there is an automorphism $\varphi$ of $\mathcal{H}$ such that $\varphi(h_1) = h_1'$ and $\varphi(h_2) = h_2'$.

**Lemma 7.13** (Characterization of Strong Transitivity) — *A graph $\mathcal{H}$ is strongly transitive* if, and only if, *it is distance-transitive and its diameter* *satisfies* $\mathrm{diam}(\mathcal{H}) \leqslant 2$. *Moreover, we may always choose* $S = \mathrm{Aut}(\mathcal{H})$ *in Definition 7.12.*

*Proof.* Assume that $\mathcal{H}$ is strongly transitive in the sense of Definition 7.12. First, we prove distance-transitivity for three instances of $d \in \mathbb{N}$: vertex-transitivity ($d = 0$) is a particular case of item (1) of the definition; edge-transitivity ($d = 1$) is a particular case of item (2); and in any other case ($d \geqslant 2$), vertices at distance $d$ in $\mathcal{H}$ are adjacent in the complement graph $\mathcal{H}^c$, so the existence of automorphism $\varphi$ follows from item (3); hence the distance-transitivity. We then prove that all vertices in $\mathcal{H}$ have distance at most 2. Assume by contradiction that there are two vertices $h_1, h_2 \in V(\mathcal{H})$ with $d(h_1, h_2) > 2$. Hence, there is a path from $h_1$ to $h_2$ of length greater than two, which needs to pass through a vertex $h_3 \in V(\mathcal{H})$ with $d(h_1, h_3) = 2$. Now, as $h_1 \sim h_2$ and $h_1 \sim h_3$ are edges of the complement graph $\mathcal{H}^c$, item (3) of Definition 7.12 tells us that we can find an automorphism $\varphi \in S$ with $\varphi(h_1) = h_1$ and $\varphi(h_2) = h_3$, and as automorphisms preserve distances, we have $d(h_1, h_2) = d(\varphi(h_1), \varphi(h_2)) = d(h_1, h_3)$. But this contradicts $d(h_1, h_2) > d(h_1, h_3)$, so we have $\mathrm{diam}(\mathcal{H}) \leqslant 2$ as claimed.

Conversely, assume that $\mathcal{H}$ is distance-transitive with $\mathrm{diam}(\mathcal{H}) \leqslant 2$, and choose $S = \mathrm{Aut}(\mathcal{H})$. We prove the three items of Definition 7.12 in the canonical order. Given vertices $h, h' \in V(\mathcal{H})$ denote by $a_{h,h'}$ the number of automorphisms $\varphi \in \mathrm{Aut}(\mathcal{H})$ with $\varphi(h) = h'$. It is nonzero since $\mathcal{H}$ is distance-transitive so in particular vertex-transitive. Now, given two further vertices $k, k' \in V(\mathcal{H})$, there are automorphisms $\varphi_1, \varphi_2 \in \mathrm{Aut}(\mathcal{H})$ with $\varphi_1(h) = k$ and $\varphi_2(h') = k'$. For any automorphism $\psi$ with $\psi(k) = k'$, the map $\varphi_2^{-1} \circ \psi \circ \varphi_1$ yields an automorphism mapping $h$ to $h'$. This shows $a_{k,k'} \leqslant a_{h,h'}$. Now, by the symmetry of the argument, this is actually equality, and we may set $d_1 := a_{h,h'} \geqslant 1$, thus proving item (1). Then, for item (2), the proof is very similar, using the edge-transitivity of $\mathcal{H}$. Finally, for item (3), note that edges in the complement graph $\mathcal{H}^c$ correspond to pairs of vertices in $\mathcal{H}$ at distance 2, so once again distance-transitivity allows us to conclude with the same argument as for item (1). ■

**Example 7.14 —** From this characterization, we deduce that the following graphs are examples of strongly transitive graphs, among others: the complete graphs $\mathcal{K}_N$ and their complement $\mathcal{K}_N^c$ for any $N \geqslant 0$ (note that the empty graph $\mathcal{K}_N^c$ has diameter 0 with our convention), the path graphs $\mathcal{P}_N$ for $N \leqslant 3$, the cycle graphs $\mathcal{C}_N$ for $N \leqslant 5$, the complete bipartite graph $\mathcal{K}_{3,3}$, and the famous Petersen graph. Moreover, several finite groups may be written as the automorphism group of a distance-transitive graph with diameter $2$, see details in [GR01].

We prove that when $\mathcal{H}$ is strongly transitive, there is a strong connection between cardinalities, which will be useful in the proof of the collapse of CC in Theorem 7.16.

**Lemma 7.15 —** *Let $\mathcal{H}$ be a strongly transitive graph different from the complete graph $\mathcal{K}_N$ and its complement $\mathcal{K}_N^c$, together with its associated subset $S \subseteq \mathrm{Aut}(\mathcal{H})$ and parameters $(d_1, d_2, d_3)$. Then the size of the set $S$ is necessarily*

$$|S| \;=\; d_1 \, |V(\mathcal{H})| \;=\; 2\, d_2 \, |E(\mathcal{H})| \;=\; 2\, d_3 \, |E(\mathcal{H}^c)| \, .$$

*Proof.* We prove the first equality by showing the two inequalities. Let us index $h_1, \ldots, h_n$ the vertices of $\mathcal{H}$, where $n = |V(\mathcal{H})|$. On the one hand, using condition (1), we know that there are exactly $d_1$ automorphisms $\varphi \in S$ sending $h_1$ to itself, and again $d_1$ other automorphisms sending $h_1$ to $h_2$, so on and so forth until $h_n$. Note also that a given automorphism $\varphi \in \mathrm{Aut}(\mathcal{H})$ cannot send $h_1$ to two different vertices. It yields that the set $S$ contains at least $d_1 n$ elements. On the other hand, each element $\varphi$ of $S$ necessarily sends $h_1$ to a vertex $h_i$ of $\mathcal{H}$, so $S$ contains at most $d_1 n$ elements, which gives the first equality. For the other two equalities, proceed similarly using items (2) and (3), where the factor "2" comes from the fact that graphs are undirected, so the two relations $h_1 \sim h_2$ and $h_2 \sim h_1$ are counted as only one edge. Note also that the condition $\mathcal{H} \neq \mathcal{K}_N$ and $\mathcal{H} \neq \mathcal{K}_N^c$ prevents the sets $E(\mathcal{H})$ and $E(\mathcal{H}^c)$ to be empty. Hence the wanted chain of equalities. ■

#### 7.1.5.3 Collapse of Communication Complexity

In the theorem below, we combine this notion of strong transitivity with $d$-regularity of $\mathcal{H}$ to obtain a collapse of CC. The key idea will be to compute an expectation $\mathbb{E}$ over $\varphi$ uniformly sampled in a subset of the automorphism group $S \subseteq \mathrm{Aut}(\mathcal{H})$ and to use the strong transitivity to obtain a symmetric strategy that collapses CC. Recall that $\mathcal{H}$ is said to be $d$-regular if every vertex is connected to exactly $d$ other vertices.

**Theorem 7.16** (All Perfect Strategies Collapse CC) — *Let $\mathcal{G}$ and $\mathcal{H}$ be two graphs such that the conditions of Theorem 7.9 hold. Assume moreover that $\mathcal{H}$ is strongly transitive and $d$-regular, and that the players share randomness. Then* every *perfect non-signaling strategy for the isomorphism game associated with $\mathcal{G} \cong_{\text{ns}} \mathcal{H}$ collapses communication complexity.*

*Proof.* Denote $\mathbf{P}$ an arbitrary perfect strategy for $\mathcal{G} \cong_{\text{ns}} \mathcal{H}$, and denote $S \subseteq \text{Aut}(\mathcal{H})$ and $(d_1, d_2, d_3)$ as given in the definition of strong transitivity. We will post-process $\mathbf{P}$ in order to generate a symmetric strategy, which will allow us to generate a collapsing nonlocal box. The two players Alice and Bob can use their shared randomness to pick uniformly at random an automorphism $\varphi \in S$ that is known by the two of them. They apply the following protocol: once they receive the outputs $h_{\mathsf{A}}, h_{\mathsf{B}}$ from $\mathbf{P}$, they compute the images $h'_{\mathsf{A}} = \varphi(h_{\mathsf{A}})$ and $h'_{\mathsf{B}} = \varphi(h_{\mathsf{B}})$ and they call $\tilde{\mathbf{P}}(h'_{\mathsf{A}}, h'_{\mathsf{B}} \,|\, g_{\mathsf{A}}, g_{\mathsf{B}})$ the new strategy, in other words this is the following expectation:

$$\tilde{\mathbf{P}}(h'_{\mathsf{A}}, h'_{\mathsf{B}} \,|\, g_{\mathsf{A}}, g_{\mathsf{B}}) \;=\; \mathop{\mathbb{E}}_{\varphi \in S} \mathbf{P}\big(\varphi^{-1}(h'_{\mathsf{A}}), \varphi^{-1}(h'_{\mathsf{B}}) \,\big|\, g_{\mathsf{A}}, g_{\mathsf{B}}\big)\,.$$

Observe that $\tilde{\mathbf{P}}$ is non-signaling by the composition of a non-signaling strategy with a non-signaling post-process. Let us compute the expression of $\tilde{\mathbf{P}}$. First, if $g_{\mathsf{A}} = g_{\mathsf{B}}$, then:

$$\begin{aligned}
\tilde{\mathbf{P}}(h'_{\mathsf{A}}, h'_{\mathsf{B}} \,|\, g_{\mathsf{A}}, g_{\mathsf{A}}) &= \frac{1}{d_1\,|V(\mathcal{H})|} \sum_{\varphi \in S} \mathbf{P}\big(\varphi^{-1}(h'_{\mathsf{A}}), \varphi^{-1}(h'_{\mathsf{B}}) \,\big|\, g_{\mathsf{A}}, g_{\mathsf{A}}\big) \\
&= \frac{1}{d_1\,|V(\mathcal{H})|} \sum_{\varphi \in S} \mathbf{P}\big(\varphi^{-1}(h'_{\mathsf{A}}), \varphi^{-1}(h'_{\mathsf{A}}) \,\big|\, g_{\mathsf{A}}, g_{\mathsf{A}}\big)\, \delta_{\varphi^{-1}(h'_{\mathsf{A}}) = \varphi^{-1}(h'_{\mathsf{B}})} \\
&= \frac{1}{d_1\,|V(\mathcal{H})|}\, d_1 \underbrace{\sum_{h \in V(\mathcal{H})} \mathbf{P}\big(h, h \,\big|\, g_{\mathsf{A}}, g_{\mathsf{A}}\big)}_{=1} \delta_{h'_{\mathsf{A}} = h'_{\mathsf{B}}} \;=\; \frac{\delta_{h'_{\mathsf{A}} = h'_{\mathsf{B}}}}{|V(\mathcal{H})|}\,,
\end{aligned}$$

where the first equality follows from the definition of $\tilde{\mathbf{P}}$ combined with Lemma 7.15, the second one from the rules of the isomorphism game, and the third one from item (1) in the definition of strong transitivity of $\mathcal{H}$; moreover, the Kronecker delta condition is changed using the bijectivity property of an automorphism $\varphi$, and the underbrace equality "$= 1$" comes from the rules of the isomorphism game. Second, if $g_\mathsf{A} \sim g_\mathsf{B}$, then similarly:

$$\begin{aligned}\tilde{\mathbf{P}}(h'_\mathsf{A}, h'_\mathsf{B} \,|\, g_\mathsf{A}, g_\mathsf{B}) &= \frac{1}{2\,d_2\,|E(\mathcal{H})|} \sum_{\varphi\in S} \mathbf{P}\big(\varphi^{-1}(h'_\mathsf{A}), \varphi^{-1}(h'_\mathsf{B}) \,\big|\, g_\mathsf{A}, g_\mathsf{B}\big) \\ &= \frac{1}{2\,d_2\,|E(\mathcal{H})|}\, d_2 \underbrace{\sum_{h_1\sim h_2\in V(\mathcal{H})} \mathbf{P}\big(h_1, h_2 \,\big|\, g_\mathsf{A}, g_\mathsf{B}\big)}_{=1} \delta_{h'_\mathsf{A}\sim h'_\mathsf{B}} \\ &= \frac{\delta_{h'_\mathsf{A}\sim h'_\mathsf{B}}}{2\,|E(\mathcal{H})|}\,.\end{aligned}$$

Third, if $g_\mathsf{A} \not\simeq g_\mathsf{B}$, then we proceed similarly, simply replacing "$d_2$" by "$d_3$", "$\mathcal{H}$" by "$\mathcal{H}^c$", and "$\sim$" by "$\not\simeq$", and we obtain $\tilde{\mathbf{P}} = \delta_{h'_\mathsf{A}\not\simeq h'_\mathsf{B}}/2|E(\mathcal{H}^c)|$. To sum up, we have:

$$\tilde{\mathbf{P}}(h'_\mathsf{A}, h'_\mathsf{B} \,|\, g_\mathsf{A}, g_\mathsf{B}) = \begin{cases} 1/|V(\mathcal{H})| & \text{if } g_\mathsf{A} = g_\mathsf{B} \text{ and } h'_\mathsf{A} = h'_\mathsf{B}\,, \\ 1/2|E(\mathcal{H})| & \text{if } g_\mathsf{A} \sim g_\mathsf{B} \text{ and } h'_\mathsf{A} \sim h'_\mathsf{B}\,, \\ 1/2|E(\mathcal{H}^c)| & \text{if } g_\mathsf{A} \not\simeq g_\mathsf{B} \text{ and } h'_\mathsf{A} \not\simeq h'_\mathsf{B}\,, \\ 0 & \text{otherwise}\,. \end{cases}$$

Now, when comparing this expression of $\tilde{\mathbf{P}}$ with the expression of $\mathbb{P}_\mathcal{S}$ in eq. (7.2) of the proof of Proposition 7.6, we see that they coincide in the case of the maximal partition $\mathcal{D}' = \{V(\mathcal{H})\}$ with parameters $(k = 1, n_1 = |V(\mathcal{H})|, c_{11} = d)$, where $d$ is the parameter of regularity of $\mathcal{H}$ by assumption. Then, following the same proof, it turns out that the strategy $\tilde{\mathbf{P}}$ is perfect for the isomorphism game associated with $\mathcal{G} \cong_{\text{ns}} \mathcal{H}$, and that it is symmetric with parameter:

$$\nu_{g_1,g_3} = \frac{|\mathcal{H}_1| \times |\mathcal{H}_2|}{|V(\mathcal{H})| \times |E(\mathcal{H}^c)|} > 0\,,$$

where the denominator is not zero because $\mathcal{H}$ contains several connected components, so it is not complete and $|E(\mathcal{H}^c)| > 0$. Therefore, we can apply Lemma 7.10 and we can simulate $\mathbf{PR}_{\alpha,\beta}$ for some $\alpha > 0$. Hence, as in the proof of Theorem 7.9 we conclude the existence of a non-signaling protocol that collapses communication complexity. ■

Finally, as quantum correlations cannot collapse CC [Cle+99], it yields:

**Corollary 7.17** (These Strategies are not Quantum) — *A perfect $\mathcal{NS}$-strategy $\mathcal{S}$ for the graph isomorphism game satisfying the conditions of Theorem 7.16 cannot be quantum.* ∎

## 7.2 Graph Coloring Game

In this section, we prove similar results of the collapse of communication complexity but for a different game, the *graph coloring game*.

Below, after connecting to background notions (Section 7.2.1), we provide examples of results in the collapse of communication complexity for this game (Section 7.2.2) and finally give results combining this game with the previous one, the graph isomorphism game (Section 7.2.3).

### 7.2.1 *Background*

We refer to Section 3.2.3 for a detailed definition of the graph homomorphism game and graph coloring game. As in the graph isomorphism game, in this game, the players Alice and Bob pretend to have a homomorphism from a graph $\mathcal{G}$ to a graph $\mathcal{H}$ or a coloring of $\mathcal{G}$. (Recall that the graph coloring game is a particular case of the graph homomorphism game, corresponding to the cases where $\mathcal{H} = \mathcal{K}_M$ is complete.) If there exists a perfect strategy for those games (*i.e.* winning with probability 1), then we write:

$$\mathcal{G} \to \mathcal{H}\,, \qquad \mathcal{G} \to_{\mathrm{qc}} \mathcal{H}\,, \qquad \mathcal{G} \to_{\mathrm{ns}} \mathcal{H}\,,$$

corresponding to perfect classical, quantum (commuting), and non-signaling strategies respectively.

### 7.2.2 *Link with Communication Complexity*

We begin with the following simple result about a protocol generating a perfect **PR** box. Note that this protocol is slightly different than the one we used for the isomorphism game, but it has the same taste.

**Lemma 7.18** (Simulation of a **PR** Box) — *Any perfect non-signaling strategy for the 2-coloring game of $\mathcal{K}_3$ allows for perfectly simulating the* **PR** *box.*

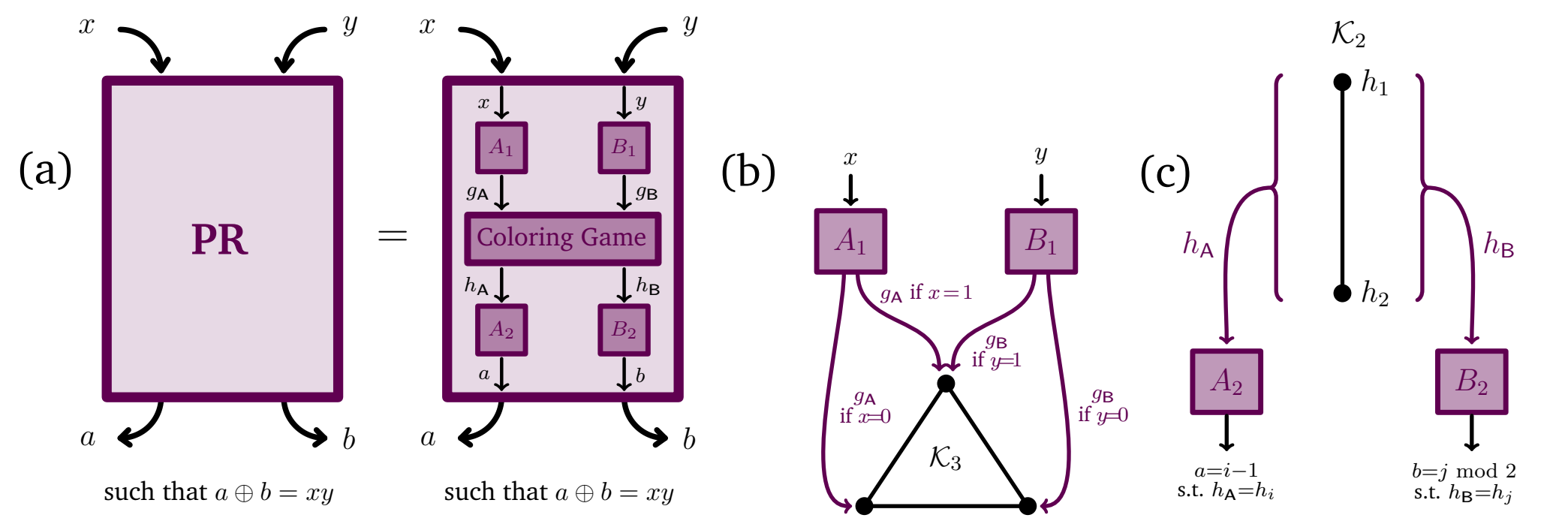

**Figure 7.2 —** *Illustration of the proof of Lemma 7.18. In item (a), we simulate a* **PR** *box from a perfect $\mathcal{NS}$-strategy for the graph coloring game, called "Coloring Game" in the figure, together with the local processes $A_1, A_2, B_1, B_2$ that are described in items (b) and (c). In item (b), given $x$ and $y$, Alice and Bob choose some input vertices $g_A$ and $g_B$ in $\mathcal{K}_3$. In item (c), Alice and Bob receive some output vertices $h_A$ and $h_B$ from $\mathcal{K}_2$, and they choose $a$ and $b$ accordingly.*

*Proof.* Consider the protocol described in Figure 7.2. We check that it indeed produces a **PR** box. On the one hand, if $x = 1 = y$, then Alice and Bob input the same vertex. Therefore they obtain the same vertex in $\mathcal{K}_2$ and it yields $a \oplus b = 1 = xy$ as expected. On the other hand, if $x = 0$ or $y = 0$, then Alice and Bob input adjacent vertices. So they receive two different vertices in $\mathcal{K}_2$ and $a \oplus b = 0 = xy$ as wanted. ■

More generally, note that if we factorize the homomorphism $\mathcal{K}_3 \to_{\text{ns}} \mathcal{K}_2$ by a graph $\mathcal{G}$, *i.e.* if we have $\mathcal{K}_3 \to_{\text{ns}} \mathcal{G} \to_{\text{ns}} \mathcal{K}_2$, we obtain from the lemma:

**Proposition 7.19** (Simulation of a **PR** Box) **—** *Given a perfect strategy for each of the homomorphism games $\mathcal{K}_3 \to_{\text{ns}} \mathcal{G}$ and $\mathcal{G} \to_{\text{ns}} \mathcal{K}_2$, the* **PR** *box can be perfectly simulated.* ■

Recall from [Bra+06] that whenever the **PR** box is simulated with probability $> \frac{3+\sqrt{6}}{6}$, there is a collapse of communication complexity. Hence, combining this fact with the former proposition, we obtain:

**Theorem 7.20** (Collapse of CC) — *Let $\mathcal{G}$ be a finite undirected graph, and let $0 \leqslant p, q \leqslant 1$ such that $pq > \frac{3+\sqrt{6}}{6} \approx 0.91$. Then, any strategy winning the homomorphism game $\mathcal{K}_3 \to_{\text{ns}} \mathcal{G}$ with probability $p$, combined with a non-signaling strategy winning the 2-coloring game of $\mathcal{G}$ with probability $q$, induces a collapse of communication complexity.* ■

**Example 7.21** — Let $p > \frac{3+\sqrt{6}}{6}$. The following examples are consequences of the theorem:

- As cycles of even length $\mathcal{C}_{2N}$ and path graphs $\mathcal{P}_N$ are 2-colorable for any $N \geqslant 1$, we have that any strategy winning the homomorphism game $\mathcal{K}_3 \to_{\text{ns}} \mathcal{C}_{2N}$ or $\mathcal{K}_3 \to_{\text{ns}} \mathcal{P}_N$ with probability at least $p$ allows to collapse CC.
- As $\mathcal{K}_3$ is $N$-colorable for any $N \geqslant 3$, we have that any non-signaling strategy winning the 2-coloring game of $\mathcal{K}_N$ with probability $p$ allows to collapse CC.

**Proposition 7.22** (Graph Sequence) — *If we have the following decomposition:*

$$\mathcal{G} =: \mathcal{G}_1 \twoheadrightarrow \cdots \twoheadrightarrow \mathcal{G}_n \twoheadrightarrow \mathcal{K}_3 \to_{\text{ns}} \mathcal{H}_1 \to_{\text{ns}} \ldots \to_{\text{ns}} \mathcal{H}_m \to_{\text{ns}} \mathcal{K}_2 \,,$$

*where "$\mathcal{G} \twoheadrightarrow \mathcal{H}$" denotes surjectivity, then the* **PR** *box can be perfectly simulated and therefore there is a collapse of CC.*

*Proof.* Denote $g_1, g_2, g_3$ the three vertices of $\mathcal{K}_3$. By surjectivity of the first $n$ maps, there exist some vertices $a_1, a_2, a_3$ in $\mathcal{G}_1$ that are (deterministically) mapped to $g_1, g_2, g_3$ respectively in $\mathcal{K}_3$. We do a similar protocol as in Lemma 7.18: Upon receiving $x$, Alice chooses $a_1$ if $x = 0$ or $a_2$ otherwise, and upon receiving $y$ Bob chooses $a_3$ if $y = 0$ or $a_2$ otherwise. It produces in $\mathcal{K}_3$ the same scenario as in the protocol of Lemma 7.18. Then, the composition of the last $(m+1)$ morphisms simulates a morphism $\mathcal{K}_3 \to_{\text{ns}} \mathcal{K}_2$, so **PR** is perfectly simulated. ■

### 7.2.3 *Combining with Graph Isomorphism Strategies*

We present a result that generalizes Corollary 7.4 to more than two connected components in $\mathcal{H}$, based on the assumption that Alice and Bob are given access to a perfect $\mathcal{NS}$-strategy for the 2-coloring game of $\mathcal{K}_N$, which is possible thanks to Example 3.32.

**Theorem 7.23** (Collapse of CC) — *Let $\mathcal{G}$ and $\mathcal{H}$ be such that $\mathrm{diam}(\mathcal{G}) \geqslant 2$, and that $\mathcal{H}$ admits exactly $N$ connected components $\mathcal{H}_1, \dots, \mathcal{H}_N$, all being complete. Then, given any strategy $\mathcal{S}$ winning the graph isomorphism game $\mathcal{G} \cong_{\mathrm{ns}} \mathcal{H}$ with probability $p$, combined with an $\mathcal{NS}$-strategy winning the 2-coloring game of $\mathcal{K}_N$ with probability $q$ such that $pq > \frac{3+\sqrt{6}}{6} \approx 0.91$, there is a collapse of communication complexity.*

*Proof.* We proceed similarly as in the proof of Theorem 7.2, but here the choice of $a$ and $b$ is given by the coloring of the components $\mathcal{H}_i$, $\mathcal{H}_j$ containing respectively $h_\mathsf{A}$, $h_\mathsf{B}$. By assumption, Alice and Bob are given access to an $\mathcal{NS}$-strategy at the 2-coloring game of $\mathcal{K}_N$, so they can use it to simulate a coloring of the components of $\mathcal{H}$: They can assign different colors to two different components and to assign simultaneously the same color if they are given the same component. Based on this ability, if the component $\mathcal{H}_i$ containing $h_\mathsf{A}$ is of the first color, Alice assigns $a = 0$, otherwise, she assigns $a = 1$, and similarly for Bob. It yields $a \neq b$ *if, and only if,* Alice and Bob have different colors, *if, and only if,* $h_\mathsf{A}$ and $h_\mathsf{B}$ are in different connected components of $\mathcal{H}$ with probability $q$, *if, and only if,* $g_\mathsf{A} \not\simeq g_\mathsf{B}$ with probability $p$ because of the completeness of the components of $\mathcal{H}$, *if, and only if,* $x = y = 1$ in the protocol of Figure 7.1. Hence, the relation $a \oplus b = xy$ is satisfied with probability $pq$, the **PR** box is simulated with the same probability and thanks to [Bra+06], we conclude that there is a collapse of communication complexity. ∎

## 7.3 Vertex Distance Game

In this section, we introduce and study a generalization of the graph isomorphism game (Section 7.1). We name it the *vertex distance game*.

Below, after thoroughly introducing this new game (Section 7.3.1), we characterize its perfect classical and quantum strategies (Section 7.3.2) as well as its perfect non-signaling strategies (Section 7.3.3), provide an example of two graphs that are $D$-isomorphic but not $(D+1)$-isomorphic (Section 7.3.4), and finally give applications to the collapse of communication complexity (Section 7.3.5).

### 7.3.1 *Definition of the Game*

We introduce a new nonlocal game that we call *vertex distance game* with parameter $D \in \mathbb{N}$, or simply *D-distance game*. Given two graphs $\mathcal{G}$ and $\mathcal{H}$ with disjoint vertex sets, two question vertices are chosen by the Referee $x_\mathsf{A}, x_\mathsf{B} \in V = V(\mathcal{G}) \cup V(\mathcal{H})$ and distributed to space-like separated players Alice and Bob who are not allowed to communicate. See Remark 3.29 to understand why we choose the inputs $x_\mathsf{A}, x_\mathsf{B}$ in $V$ and not simply in $V(\mathcal{G})$. In order to win the game, Alice and Bob try to output vertices $y_\mathsf{A}, y_\mathsf{B} \in V$ satisfying two conditions. The first one is the same first rule as eq. (3.24) for the graph isomorphism game:

$$x_\mathsf{A} \in V(\mathcal{G}) \Leftrightarrow y_\mathsf{A} \in V(\mathcal{H}) \quad \text{and} \quad x_\mathsf{B} \in V(\mathcal{G}) \Leftrightarrow y_\mathsf{B} \in V(\mathcal{H})\,. \tag{7.4}$$

Assuming that this relation holds, we relabel the vertices into $g_\mathsf{A}, g_\mathsf{B}, h_\mathsf{A}, h_\mathsf{B}$ as in the isomorphism game: only one vertex among $x_\mathsf{A}$ and $y_\mathsf{A}$ is in $V(\mathcal{G})$, let us call it $g_\mathsf{A} \in V(\mathcal{G})$ and the other $h_\mathsf{A} \in V(\mathcal{H})$; and similarly for $g_\mathsf{B} \in V(\mathcal{G})$ and $h_\mathsf{B} \in V(\mathcal{H})$. Then, the second condition is that distances are preserved until $D$:

$$d(h_\mathsf{A}, h_\mathsf{B}) = \begin{cases} d(g_\mathsf{A}, g_\mathsf{B}) & \text{if } d(g_\mathsf{A}, g_\mathsf{B}) \leqslant D\,, \\ > D & \text{otherwise}\,. \end{cases} \tag{7.5}$$

Find an example in Figure 7.3. We write $\mathcal{G} \cong^D \mathcal{H}$, and we say that the graphs $\mathcal{G}$ and $\mathcal{H}$ are $D$-isomorphic, if there exists a perfect classical strategy for the $D$-distance game, and similarly $\cong^D_{\text{qc}}$ and $\cong^D_{\text{ns}}$ with perfect quantum and non-signaling strategies. These notations will make sense with regard to Example 7.24 because this game generalizes the isomorphism game. Note that:

$$\cdots \implies \mathcal{G} \cong^{D=2}_s \mathcal{H} \implies \mathcal{G} \cong^{D=1}_s \mathcal{H} \implies \mathcal{G} \cong^{D=0}_s \mathcal{H}\,, \tag{7.6}$$

for any strategy type $s$ such as classical, quantum, non-signaling, or any other type. Note that we do not need to assume that $|V(\mathcal{G})| = |V(\mathcal{H})|$ since it is a consequence of the setting, see eq. (7.7) below. Three cases are noticeable:

**Example 7.24** (Remarkable Cases) — (1) The case $D = 0$ corresponds to the *graph bisynchronous game* [PR21], where we require that same vertices $g_\mathsf{A} = g_\mathsf{B}$ are mapped to same vertices $h_\mathsf{A} = h_\mathsf{B}$, and that different vertices are mapped to different vertices. In particular, if we consider the graphs $\mathcal{G} = \mathcal{K}_M$ and $\mathcal{H} = \mathcal{K}_N$, the case $D = 0$ exactly corresponds to the $N$-coloring game of $\mathcal{K}_M$.

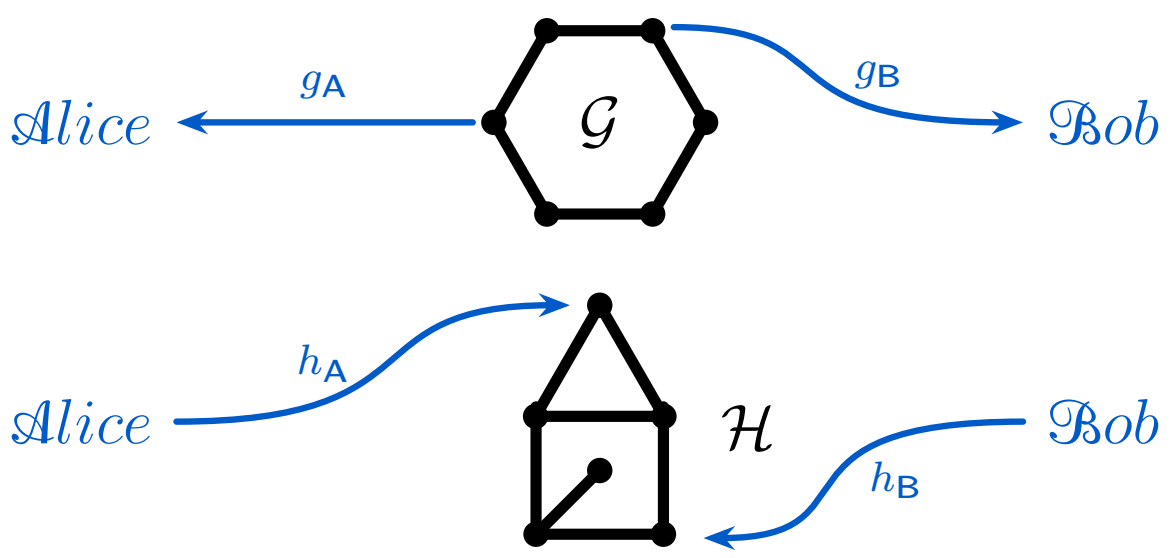

**Figure 7.3 —** *Example of what can happen in the vertex distance game associated with the* 6*-cycle* $\mathcal{G} = \mathcal{C}_6$ *and a graph* $\mathcal{H}$*. There, both inputs* $g_A$ *and* $g_B$ *are in* $V(\mathcal{G})$*, and they are at distance* 2*. The players correctly answer since* $h_A$ *and* $h_B$ *are both in* $V(\mathcal{H})$ *and at distance* 2 *again.*

(2) The case $D = 1$ corresponds precisely to the *graph isomorphism game* introduced in Section 7.1, where the relations $\{=, \sim, \not\simeq\}$ are preserved. Hence the vertex distance game is a generalization of the graph isomorphism game.

(3) The case $D = \mathrm{diam}(\mathcal{H})$ corresponds to:

$$d(h_A, h_B) = \begin{cases} d(g_A, g_B) & \text{if } d(g_A, g_B) \leqslant \mathrm{diam}(\mathcal{H}), \\ \infty & \text{otherwise.} \end{cases}$$

Note that it requires that $\mathcal{H}$ admits at least two connected components if the diameter of $\mathcal{G}$ is larger than the one of $\mathcal{H}$. We will particularly be interested in this third case in what follows.

**Remark 7.25 —** There is another way to define the vertex distance game. From a given graph $\mathcal{G}$, construct $\mathcal{G}_t$ the graph with the same vertex set as $\mathcal{G}$, by putting an edge between two vertices in $\mathcal{G}_t$ if the distance in $\mathcal{G}$ is exactly $t$. Then, observe that winning the $D$-distance game is equivalent to winning the graph isomorphism games of $(\mathcal{G}_t, \mathcal{H}_t)$ for all $t \leqslant D$.

We will need the following lemma that generalizes [Ats+19, Lemma 4.1]:

**Lemma 7.26 —** *If* $\mathbf{P} \in \mathcal{NS}$ *is a perfect non-signaling strategy for the* $D$*-distance game of* $(\mathcal{G}, \mathcal{H})$*, then for all* $g \in V(\mathcal{G})$ *and* $h \in V(\mathcal{H})$*:*

*(1)* $\sum_{h \in V(\mathcal{H})} \mathbf{P}(h, h \,|\, g, g) = 1 = \sum_{g \in V(\mathcal{G})} \mathbf{P}(h, h \,|\, g, g)$*,*

*(2)* $\mathbf{P}(h, h \,|\, g, g) = \mathbf{P}(g, h \,|\, h, g) = \mathbf{P}(h, g \,|\, g, h) = \mathbf{P}(g, g \,|\, h, h)$*.*

*Proof.* The first equality of item (1) follows from the first condition of the game in eq. (7.4), which states that both outputs have to be in $V(\mathcal{H})$ if both inputs are in $V(\mathcal{G})$, combined with the condition in eq. (7.5) with distance $0$ stating that equality is preserved. The second equality is similarly shown. As for item (2), if we denote $V := V(\mathcal{G}) \cup V(\mathcal{H})$, the winning conditions of the game together with the non-signaling condition give rise to:

$$\mathbf{P}(h,h\,|\,g,g) = \sum_{y\in V} \mathbf{P}(y,h\,|\,g,g) = \sum_{y\in V} \mathbf{P}(y,h\,|\,h,g) = \mathbf{P}(g,h\,|\,h,g)\,.$$

We obtain the other equalities with a similar method. ■

In particular, by combining items (1) and (2), we obtain that the vertex sets have the same cardinality:

$$\begin{aligned} |V(\mathcal{G})| = \sum_{g\in V(\mathcal{G})} 1 &= \sum_{g\in V(\mathcal{G})} \sum_{h\in V(\mathcal{H})} \mathbf{P}(h,h\,|\,g,g) \\ &= \sum_{h\in V(\mathcal{H})} \sum_{g\in V(\mathcal{G})} \mathbf{P}(g,g\,|\,h,h) = \sum_{h\in V(\mathcal{H})} 1 = |V(\mathcal{H})|\,, \end{aligned} \tag{7.7}$$

which was not obvious at first glance. (Another way to obtain this equality is to use the fact that $\cong_s^D$ implies $\cong_s$ for any $s \in \{\emptyset, \mathrm{q}, \mathrm{ns}\}$ and any $D \geqslant 1$ (see eq. (7.6)), and that $\cong_s$ satisfies the above equality [Ats+19].)

### 7.3.2 *Characterizing Perfect Classical & Quantum Strategies*

Surprisingly, as shown in the following proposition, the case $D = 1$ may be extended to any $D \geqslant 1$ when strategies are deterministic, classical, or quantum. On the contrary, we will see in Section 7.3.4 that it is not the case for non-signaling strategies.

**Proposition 7.27** (Characterization) — *For any $D \geqslant 1$, perfect deterministic/classical/quantum strategies coincide for the graph isomorphism game and the $D$-distance game:*

$$\begin{aligned} \mathcal{G} \cong \mathcal{H} \quad &\Longleftrightarrow \quad \forall D \geqslant 1,\ \mathcal{G} \cong^D \mathcal{H} \quad \Longleftrightarrow \quad \exists D \geqslant 1,\ \mathcal{G} \cong^D \mathcal{H}\,, \\ \mathcal{G} \cong_{\mathrm{qc}} \mathcal{H} \quad &\Longleftrightarrow \quad \forall D \geqslant 1,\ \mathcal{G} \cong^D_{\mathrm{qc}} \mathcal{H} \quad \Longleftrightarrow \quad \exists D \geqslant 1,\ \mathcal{G} \cong^D_{\mathrm{qc}} \mathcal{H}\,. \end{aligned}$$

*Proof.* Let $D \geqslant 1$. On the one hand, as the isomorphism game consists in preserving distances $\{0, 1, > 1\}$, we see that any perfect strategy for the $D$-distance game is also perfect for the isomorphism game, whether it is deterministic, classical, quantum or non-signaling. On the other hand:

(1) Perfect deterministic strategies for the isomorphism game preserve the distances. Indeed, if $D'$ is the distance separating some vertices $g_{\mathsf{A}}$ and $g_{\mathsf{B}}$ in $\mathcal{G}$, then there is a path $(g_0, g_1, \dots, g_{D'})$ in $\mathcal{G}$, where $g_0 = g_{\mathsf{A}}$, $g_d = g_{\mathsf{B}}$ and $g_i \sim g_{i+1}$ for all $i$. It follows that:

$$\begin{aligned} d(g_{\mathsf{A}}, g_{\mathsf{B}}) &= d(g_0, g_1) + \dots + d(g_{D-1}, g_{D'}) \\ &= d(h_0, h_1) + \dots + d(h_{D'-1}, h_{D'}) \geqslant d(h_0, h_{D'}), \end{aligned}$$

where $h_i$ is the image of $g_i$ under the bijection $\varphi : \mathcal{G} \to \mathcal{H}$ induced by the deterministic strategy, and where the last inequality holds by the triangular inequality. We note that $h_0$, $h_D$ are the respective images of $g_{\mathsf{A}}$, $g_{\mathsf{B}}$. Assume by contradiction that the last inequality is strict. Then there is a path in $\mathcal{H}$ connecting $h_0$ to $h_{D'}$ of length $D'' < D'$. Applying $\varphi^{-1}$, we obtain a path in $\mathcal{G}$ connecting $g_{\mathsf{A}}$ to $g_{\mathsf{B}}$ of length $D'' < D'$, which contradicts the fact that $D'$ is the distance between $g_{\mathsf{A}}$ and $g_{\mathsf{B}}$. Thus $d(g_{\mathsf{A}}, g_{\mathsf{B}}) = d(h_0, h_{D'})$ and we have the desired result.

(2) A classical strategy is a convex combination of deterministic strategies:

$$\mathbf{P}(h_{\mathsf{A}}, h_{\mathsf{B}} \,|\, g_{\mathsf{A}}, g_{\mathsf{B}}) = \sum_i p_i \, \mathbf{P}_i^{\text{det}}(h_{\mathsf{A}}, h_{\mathsf{B}} \,|\, g_{\mathsf{A}}, g_{\mathsf{B}}),$$

with $\sum_i p_i = 1$ and the sum index is finite. It means that we apply the strategy $\mathbf{P}_i^{\text{det}}$ with probability $p_i$. If $\mathbf{P}$ is a perfect classical strategy, then it wins with probability one, and therefore each $\mathbf{P}_i^{\text{det}}$ also has to be perfect. So by item (1), each $\mathbf{P}_i^{\text{det}}$ preserves the distance, thus $\mathbf{P}$ also preserves the distance.

(3) Let $\mathbf{P}$ be a perfect quantum strategy for the isomorphism game. This proof is an easy adaptation of [Sch20, Theorem 1.1]. We want to show that distance is preserved, *i.e.* that the event "$d(h_{\mathsf{A}}, h_{\mathsf{B}}) \neq d(g_{\mathsf{A}}, g_{\mathsf{B}})$" has zero probability. Using [Ats+19, Theorem 5.14], we know that there exists a $C^*$-algebra $\mathcal{A}$ with a tracial state $\tau$ and projections $E_{gh} \in \mathcal{A}$ for $g \in V(\mathcal{G})$ and $h \in V(\mathcal{H})$ such that $u$ is a quantum

permutation matrix and:

$$u_{g_\mathsf{A}h_\mathsf{A}}u_{g_\mathsf{B}h_\mathsf{B}} = 0 \quad \text{if} \quad \mathrm{rel}(g_\mathsf{A},g_\mathsf{B}) \neq \mathrm{rel}(h_\mathsf{A},h_\mathsf{B})\,, \tag{7.8}$$

where the function $\mathrm{rel}(g_\mathsf{A},g_\mathsf{B})$ takes value $0$ if $g_\mathsf{A} = g_\mathsf{B}$, value $1$ if $g_\mathsf{A} \sim g_\mathsf{B}$, and value $2$ if $g_\mathsf{A} \not\simeq g_\mathsf{B}$. Using these notations, the correlation $\mathbf{P}$ is then of the form

$$\mathbf{P}(h_\mathsf{A},h_\mathsf{B}\,|\,g_\mathsf{A},g_\mathsf{B}) = \tau(u_{g_\mathsf{A}h_\mathsf{A}}u_{g_\mathsf{B}h_\mathsf{B}})\,. \tag{7.9}$$

Now, as the isomorphism game is equivalent to the $1$-distance game, we already know that distances $0$ and $1$ are preserved. It remains to show the results for distances $\geqslant 2$. Consider vertices $g_\mathsf{A}, g_\mathsf{B}, h_\mathsf{A}, h_\mathsf{B}$ such that $2 \leqslant d(g_\mathsf{A},g_\mathsf{B}) = t < d(h_\mathsf{A},h_\mathsf{B})$. We will show that then $\mathbf{P}(h_\mathsf{A},h_\mathsf{B}\,|\,g_\mathsf{A},g_\mathsf{B}) = 0$. By definition, there exist a path in $\mathcal{G}$ with adjacent vertices $g_1,\dots,g_t \in V(\mathcal{G})$ going from $g_1 = g_\mathsf{A}$ to $g_t = g_\mathsf{B}$, but no such path exists in $\mathcal{H}$ from $h_\mathsf{A}$ to $h_\mathsf{B}$. So for all $h_1,\dots,h_t \in V(\mathcal{H})$ such that $h_1 = h_\mathsf{A}$ and $h_t = h_\mathsf{B}$, we infer there exists at least one index $s \in \{1,\dots,t-1\}$ such that $h_s \not\sim h_{s+1}$ but $g_s \sim g_{s+1}$. By eq. (7.8), we deduce that $u_{g_sh_s}u_{g_{s+1}h_{s+1}} = 0$, and therefore:

$$\begin{aligned} u_{g_\mathsf{A}h_\mathsf{A}}u_{g_\mathsf{B}h_\mathsf{B}} &= u_{g_\mathsf{A}h_\mathsf{A}} \cdot \mathbb{I} \cdot \ldots \cdot \mathbb{I} \cdot u_{g_\mathsf{B}h_\mathsf{B}} \\ &= \sum_{h_2,\dots,h_{t-1}\in V(\mathcal{H})} u_{g_\mathsf{A}h_\mathsf{A}}u_{g_2h_2}\dots u_{g_{t-1}h_{t-1}}u_{g_\mathsf{B}h_\mathsf{B}} = 0\,. \end{aligned}$$

Thus using eq. (7.9), we obtain that the event "$d(h_\mathsf{A},h_\mathsf{B}) \neq d(g_\mathsf{A},g_\mathsf{B})$" has probability probability zero, and similarly for the event with the opposite inequality. ■

Now, combining Proposition 7.27 with references [Ats+19; CV93; LMR20; Lov67; MR20] (see Figure 3.5), we obtain the following lists of characterizations:

**Corollary 7.28** (Classical Strategies) — *The followings are equivalent:*

*(i)* $\exists D \geqslant 1,\ \mathcal{G} \cong^D \mathcal{H}$.

*(ii)* $\forall D \geqslant 1,\ \mathcal{G} \cong^D \mathcal{H}$.

*(iii)* $\mathcal{G} \cong \mathcal{H}$.

*(iv) There exists a permutation matrix $P$ such that $A_\mathcal{G}P = PA_\mathcal{H}$.*

*(v) For any graph $\mathcal{F}$, we have* $\#\,\mathrm{Hom}(\mathcal{F},\mathcal{G}) = \#\,\mathrm{Hom}(\mathcal{F},\mathcal{H})$.

*(vi) For any graph $\mathcal{F}$, we have* $\#\,\mathrm{Hom}(\mathcal{G},\mathcal{F}) = \#\,\mathrm{Hom}(\mathcal{H},\mathcal{F})$.

**Corollary 7.29** (Quantum Strategies) — *The followings are equivalent:*

*(i)* $\exists D \geqslant 1,\, \mathcal{G} \cong_{\text{qc}}^{D} \mathcal{H}$.
*(ii)* $\forall D \geqslant 1,\, \mathcal{G} \cong_{\text{qc}}^{D} \mathcal{H}$.
*(iii)* $\mathcal{G} \cong_{\text{qc}} \mathcal{H}$.
*(iv)* *There exists a quantum permutation matrix* $P$ *such that* $A_{\mathcal{G}} P = P A_{\mathcal{H}}$.
*(v)* *For all planar graph* $\mathcal{P}$*, we have* $\#\operatorname{Hom}(\mathcal{P},\mathcal{G}) = \#\operatorname{Hom}(\mathcal{P},\mathcal{H})$.

### 7.3.3 Characterizing Perfect Non-Signaling Strategies

In this subsection, we generalize the results that $\mathcal{G} \cong_{\text{ns}} \mathcal{H}$ *if, and only if,* $\mathcal{G}$ and $\mathcal{H}$ are fractionally isomorphic [Ats+19], *if, and only if,* they admit a common equitable partition [RSU94]. Along this subsection, we relax the definitions of fractional isomorphism and common equitable partition with a parameter $D$ (see Definition 7.33 and Definition 7.35), and the combination of all lemmata leads to the following theorem:

**Theorem 7.30** (Characterization of Perfect $\mathcal{NS}$-Strategies) — *Let* $\mathcal{G}$ *and* $\mathcal{H}$ *be two graphs and* $D \geqslant 0$ *be an integer. The followings are equivalent:*

*(1)* $\mathcal{G} \cong_{\text{ns}}^{D} \mathcal{H}$,
*(2)* $\mathcal{G} \cong_{\text{frac}}^{D} \mathcal{H}$,
*(3) There exists a* $D$*-common equitable partition of* $(\mathcal{G},\mathcal{H})$.

**Remark 7.31** (Characterization for the Game with Inputs in $\mathcal{G}$) — If we consider the similar—yet different—$D$-distance game where the inputs $x_{\mathsf{A}}, x_{\mathsf{B}}$ are always in $V(\mathcal{G})$ instead of $V$, then a similar proof shows the following statement. Let $\mathcal{G},\mathcal{H}$ two graphs with the same number of vertices, and let $D \geqslant 0$ be an integer. Then, the followings are equivalent:

(1') This version of the game admits a perfect strategy $\mathbf{P} \in \mathcal{NS}$ such that the flipped correlation $\mathbf{P}'(g_{\mathsf{A}}, g_{\mathsf{B}} \,|\, h_{\mathsf{A}}, h_{\mathsf{B}}) := \mathbf{P}(h_{\mathsf{A}}, h_{\mathsf{B}} \,|\, g_{\mathsf{A}}, g_{\mathsf{B}})$ is also in $\mathcal{NS}$.
(2) $\mathcal{G} \cong_{\text{frac}}^{D} \mathcal{H}$.
(3) There exists a $D$-common equitable partition of $(\mathcal{G},\mathcal{H})$.

#### 7.3.3.1 Generalized Fractional Isomorphism

A *bistochastic* matrix is a matrix $u \in \mathcal{M}_n(\mathbb{R})$ whose entries are non-negative, and whose rows and columns sum to one. We generalize the notion of fractional isomorphism as follows:

**Definition 7.32** ($D$-Fractional Isomorphism) — *Let $D \geqslant 0$. Two graphs $\mathcal{G}$ and $\mathcal{H}$ are said to be $D$-*fractionally isomorphic*, denoted $\mathcal{G} \cong^D_{\text{frac}} \mathcal{H}$, if there exists a bistochastic matrix $u \in \mathcal{M}_n(\mathbb{R})$ such that for all distances $t \leqslant D$ we have:*

$$\forall\, g \in V(\mathcal{G}),\ \forall\, h \in V(\mathcal{H}), \qquad \sum_{h' \in C(h,t)} u_{gh'} = \sum_{g' \in C(g,t)} u_{g'h}\,, \tag{7.10}$$

*where $C(g,t)$ is the circle of radius $t$ in $\mathcal{G}$ centered at $g$,* i.e. *the set of neighbors of $g$ in $\mathcal{G}$ at distance exactly $t$.*

Note that in the case $D = 1$, we retrieve the usual notion of fractional isomorphism, because the condition in the equation amounts to $\sum_{h' \sim h} u_{gh'} = \sum_{g' \sim g} u_{g'h}$, which is equivalent to saying that the adjacency matrices satisfy $u\, A_{\mathcal{G}} = A_{\mathcal{H}}\, u$. We can rephrase eq. (7.10) in terms of a generalization of the adjacency matrix. We call the matrix $A^{(D)}_{\mathcal{G}}$ the $D$-*adjacency matrix* of a graph $\mathcal{G}$, whose coefficients $a_{ij}$ are $1$ if the distance between $g_i$ and $g_j$ satisfies $d(g_i, g_j) = D$, and $0$ otherwise. We have the equivalence:

$$\text{Equation (7.10)} \qquad \Longleftrightarrow \qquad u\, A^{(t)}_{\mathcal{G}} = A^{(t)}_{\mathcal{H}}\, u\,. \tag{7.11}$$

This leads to the following equivalent definition of $D$-fractional isomorphism:

**Definition 7.33** ($D$-Fractional Isomorphism, bis) — *Let $D \geqslant 0$. Two graphs $\mathcal{G}$ and $\mathcal{H}$ are said to be $D$-*fractionally isomorphic*, denoted $\mathcal{G} \cong^D_{\text{frac}} \mathcal{H}$, if there exists a bistochastic matrix $u \in \mathcal{M}_n(\mathbb{R})$ such that for all distances $t \leqslant D$ we have:*

$$u\, A^{(t)}_{\mathcal{G}} = A^{(t)}_{\mathcal{H}}\, u\,,$$

*where $A^{(t)}_{\mathcal{G}}$ and $A^{(t)}_{\mathcal{H}}$ are the $t$-adjacency matrices of $\mathcal{G}$ and $\mathcal{H}$ respectively.*

Note that the $D$-adjacency matrix may be expressed in terms of the (usual) powers $A_{\mathcal{G}}^t$ of the adjacency matrix. The coefficients of the latter matrix may be interpreted as taking value $1$ *if, and only if,* there exists a path of length $t$ in $\mathcal{G}$ joining the corresponding vertices. We see that a coefficient is $1$ in the $D$-adjacency matrix $A_{\mathcal{G}}^{(D)}$ *if, and only if,* there exists a path of length $D$ in $\mathcal{G}$ joining the corresponding vertices, but no path of length $t < D$. In other words $A_{\mathcal{G}}^{(t)}$ is the adjacency matrix of the graph $\mathcal{G}_t$ as described in Remark 7.25. We have the following relation:

$$A_{\mathcal{G}}^{(D)} = \left(A_{\mathcal{G}}^{D} \div A_{\mathcal{G}}^{D}\right) \circledast \left(\mathbf{1} - A_{\mathcal{G}}^{D-1} \div A_{\mathcal{G}}^{D-1}\right) \circledast \dots \circledast \left(\mathbf{1} - A_{\mathcal{G}}^{0} \div A_{\mathcal{G}}^{0}\right),$$

where $\div$ and $\circledast$ are the element-wise division and multiplication of matrices (*a.k.a.* the Hadamard division and product, or Schur product), and where $\mathbf{1}$ is the matrix with all entries $1$ of the same size as $A_{\mathcal{G}}$. Observe that $A_{\mathcal{G}}^{(0)} = \mathbb{I}$ the identity matrix, and $A_{\mathcal{G}}^{(1)} = A_{\mathcal{G}}$ the adjacency matrix, and $A_{\mathcal{G}}^{(D)} = \mathbf{0}$ the zero matrix for all $D > \operatorname{diam}(\mathcal{G})$ because the graph $\mathcal{G}$ admits no vertices with such a distance $D$. Note that the $D$-adjacency matrix may be equivalently recursively defined:

$$A_{\mathcal{G}}^{(D)} = \left(A_{\mathcal{G}}^{D} \div A_{\mathcal{G}}^{D}\right) \circledast \left(\mathbf{1} - \sum_{t=0}^{D-1} A_{\mathcal{G}}^{(t)}\right),$$

because two vertices of $\mathcal{G}$ have distance $D$ *if, and only if,* there is a path of length $D$ joining them and they do not have distance $t \leqslant D-1$. Note that we will provide in Section 7.3.4 an example of sequence of graphs $(\mathcal{G}_D, \mathcal{H}_D)$ such that $\mathcal{G}_D \cong_{\text{frac}}^{D} \mathcal{H}_D$ but $\mathcal{G}_D \ncong_{\text{frac}}^{(D+1)} \mathcal{H}_D$. Here is a sufficient condition in order to have a $D$-fractional isomorphism:

**Lemma 7.34 —** *If $\mathcal{G} \cong_{\text{ns}}^{D} \mathcal{H}$ for some integer $D \geqslant 0$, then $\mathcal{G}$ and $\mathcal{H}$ are $D$-fractionally isomorphic:*

$$\mathcal{G} \cong_{\text{ns}}^{D} \mathcal{H} \quad \Longrightarrow \quad \mathcal{G} \cong_{\text{frac}}^{D} \mathcal{H}.$$

*Proof.* This proof generalizes [Ats+19, Lemma 4.2]. We want to construct a bistochastic matrix $u$ such that eq. (7.10) holds for all $t \leqslant D$. We will index the elements of the matrix $u$ by the vertices of $\mathcal{G}$ and $\mathcal{H}$, for instance $u_{gh}$. As the strategy $\mathbf{P}$ is a valid probability distribution, we can define $u_{gh} = \mathbf{P}(h, h \,|\, g, g) \geqslant 0$, and using Lemma 7.26 (1), we have that rows and columns sum to one, $\sum_h u_{gh} = 1$ and $\sum_g u_{gh} = 1$, so the matrix $u$ is bistochastic. Let us verify the equality of eq. (7.10) for an arbitrary integer $t \leqslant D$ and vertices $g \in V(\mathcal{G})$ and $h \in V(\mathcal{H})$. We have:

$$\sum_{h' \in C(h,t)} u_{gh'} = \sum_{h' \in C(h,t)} \mathbf{P}(h', h' \,|\, g, g) = \sum_{h' \in C(h,t)} \sum_{g' \in V(\mathcal{G})} \mathbf{P}(h', h' \,|\, g, g')\,,$$

which holds because $\mathbf{P}$ is perfect so it satisfies the rule that the outputs need to be equal *if, and only if,* the inputs $g$ and $g'$ are equal. Now, using the non-signaling condition of $\mathbf{P}' \in \mathcal{NS}$ on Bob's marginal, we obtain:

$$= \sum_{h' \in C(h,t)} \sum_{g' \in V(\mathcal{G})} \mathbf{P}(h', h \,|\, g, g') = \sum_{h' \in C(h,t)} \sum_{g' \in C(g,t)} \mathbf{P}(h', h \,|\, g, g')\,,$$

where the last equality holds because $\mathbf{P}$ is perfect so it satisfies the rule that the distance $t$ is the same for the outputs $h', h$ and the inputs $g, g'$. Then, we swap the two sums and we use that $\mathbf{P} \in \mathcal{NS}$ and similar arguments as before to derive what follows:

$$\begin{aligned}
&= \sum_{g' \in C(g,t)} \sum_{h' \in C(h,t)} \mathbf{P}(h', h \,|\, g, g') = \sum_{g' \in C(g,t)} \sum_{h' \in V(\mathcal{H})} \mathbf{P}(h', h \,|\, g, g')\,, \\
&= \sum_{g' \in C(g,t)} \sum_{h' \in V(\mathcal{H})} \mathbf{P}(h', h \,|\, g', g') = \sum_{g' \in C(g,t)} \mathbf{P}(h, h \,|\, g', g')\,, \\
&= \sum_{g' \in C(g,t)} u_{g'h}\,.
\end{aligned}$$

Hence, eq. (7.10) holds, and the graphs $\mathcal{G}$ and $\mathcal{H}$ are $D$-fractionally isomorphic. ■

#### 7.3.3.2 Generalized Common Equitable Partition

For Theorem 7.30, we also generalize the notion of common equitable partition. Recall that the usual notion of common equitable partition was defined on page 222.

> **Definition 7.35** ($D$-Common Equitable Partition) — *Let $D \geqslant 0$. We say that two graphs $\mathcal{G}$ and $\mathcal{H}$ admit a $D$-*common equitable partition *if they* *admit respective partitions* $\mathscr{C} = (C_1, \ldots, C_k)$ *and* $\mathscr{D} = (D_1, \ldots, D_\ell)$ *with the following common parameters:*

$$\begin{aligned} & k = \ell\,, \\ \forall i \in \{1,..,k\}, \quad & |C_i| = |D_i| =: n_i\,, \\ \begin{matrix}\forall t \leqslant D,\ \forall i,j \in \{1,\ldots,k\}, \\ \forall g \in C_i,\ \forall h \in D_i,\end{matrix} \quad & |C_j \cap C(g,t)| = |D_j \cap C(h,t)| =: c_{ij}^{(t)}\,. \end{aligned}$$

Note that the case $D = 1$ corresponds exactly to the usual notion of common equitable partition. Note that $c_{ij}^{(0)} = \delta_{ij}$ is the Kronecker delta, and that $c_{ij}^{(t)} = 0$ when $t > \min\big\{\operatorname{diam}(\mathcal{G}), \operatorname{diam}(\mathcal{H})\big\}$. We do not necessarily have $c_{ij}^{(t)} = c_{ji}^{(t)}$, but we always have the following relation:

**Lemma 7.36 —** *If the graph* $\mathcal{G}$ *admits a* $D$*-equitable partition with parameters as above, then:*

$$n_i\, c_{ij}^{(t)} \;=\; n_j\, c_{ji}^{(t)}\,.$$

*Proof.* This proof is a generalization of [RSU94, Section 2.1]. Up to reordering the rows and columns of the $t$-adjacency matrix $A_{\mathcal{G}}^{(t)}$, this matrix can be decomposed in blocks $A_{ij}^{(t)}$ of size $n_i \times n_j$ such that the rows sum to $c_{ij}^{(t)}$. By symmetry of the $t$-adjacency matrix, the blocks satisfy $A_{ij}^{(t)\top} = A_{ji}^{(t)}$, so the columns of $A_{ij}^{(t)}$ sum to $c_{ji}^{(t)}$. Now, as the sum of all the elements of $A_{ij}^{(t)}$ equals both the sum of its rows and the sum of its columns, we obtain $n_i\, c_{ij}^{(t)} = n_j\, c_{ji}^{(t)}$, hence the wanted result. ∎

We prove that $D$-fractional isomorphic is a sufficient condition for the graphs to admit a $D$-common equitable partition:

**Lemma 7.37 —** *If* $\mathcal{G} \cong_{\text{frac}}^{D} \mathcal{H}$ *for some integer* $D \geqslant 0$*, then there exists a* $D$*-common equitable partition of* $(\mathcal{G}, \mathcal{H})$.

*Proof.* This proof generalizes [RSU94, Theorem 2.2]. We use an equivalent characterization of $\cong_{\text{frac}}^{D}$ as the one given in eq. (7.11), *i.e.* there exists a bistochastic matrix $u$ such that for all $t \leqslant D$:

$$A_{\mathcal{G}}^{(t)}\, u = u\, A_{\mathcal{H}}^{(t)}\,. \tag{7.12}$$

From this matrix $u$, we define a partition $\mathscr{C}$ on $V(\mathcal{G})$ based on the following equivalence relation:

$$g \leftrightarrow g'$$

*if, and only if,* there exists a "link" from $g$ to $g'$, *i.e.* $\exists n, g_1, ..., g_n, h_1, ..., h_n$ such that:

$$g_1 = g, \qquad g_n = g', \qquad u_{g_1h_1} \cdot u_{g_2h_1} \cdot u_{g_2h_2} \cdot ... \cdot u_{g_nh_n} > 0\,,$$

and we similarly define a partition $\mathscr{D}$ on $V(\mathcal{H})$. By construction, up to reordering the rows and columns of $u$, these partitions $\mathscr{C}$ and $\mathscr{D}$ are in correspondence with a block decomposition $u = U_1 \oplus \cdots \oplus U_k$, where each $U_i$ is an indecomposable $n_i \times m_i$ bistochastic matrix for some $n_i, m_i$. In particular, we have that both partitions $\mathscr{C}$ and $\mathscr{D}$ have $k$ cells and that each cell $C_i$ and $D_i$ has respective size $n_i$ and $m_i$. Using the fact that $u$ is bistochastic, we have:

$$m_i \;=\; \sum_{h\in D_i} 1 \;=\; \sum_{h\in D_i}\sum_{g\in C_i} u_{gh} \;=\; \sum_{g\in C_i}\sum_{h\in D_i} u_{gh} \;=\; \sum_{g\in C_i} 1 \;=\; n_i\,,$$

hence $|C_i| = |D_i| = n_i$ as wanted. It remains to prove that $\mathscr{C}$ and $\mathscr{D}$ admit common parameters $c_{ij}^{(t)}$. Let $t \leqslant D$ and denote $A^{(t)} := A_{\mathcal{G}}^{(t)}$ and $B := A_{\mathcal{H}}^{(t)}$ the $t$-adjacency matrices of the graphs $\mathcal{G}$ and $\mathcal{H}$. Write $A^{(t)}$ with blocks $A_{ij}^{(t)}$ of size $n_i \times n_j$ induced by the partition $\mathscr{C}$, and similarly for $B^{(t)}$ with blocks $B_{ij}^{(t)}$ of the same size. From eq. (7.12), we deduce the following relations:

$$\forall i,j \in [k], \qquad A_{ij}^{(t)}\,U_j \;=\; U_i\,B_{ij}^{(t)}\,. \tag{7.13}$$

As well, after swapping $i$ and $j$, we have $A_{ji}^{(t)}\,U_i = U_j\,B_{ji}^{(t)}$, and taking the transpose we obtain:

$$\forall i,j \in [k], \qquad U_i^\top A_{ij}^{(t)} \;=\; B_{ij}^{(t)}\,U_j^\top\,. \tag{7.14}$$

Let $v_{ij}^{(t)} := A_{ij}^{(t)}\,\mathbf{1}$ be the vector such that each coordinate corresponds to a $g \in C_i$ and represents the number of $g' \in C_j$ at distance exactly $t$ of $g$, where $\mathbf{1}$ denotes the vector of appropriate size with all entries $1$. In order to have a "$D$-equitable" partition, we want all the coordinates of the vector $v_{ij}^{(t)}$ to have the same value $c_{ij}^{(t)} \in \mathbb{R}$, *i.e.* that $v_{ij}^{(t)} = c_{ij}^{(t)}\,\mathbf{1}$. Similarly, we define the vector $w_{ij}^{(t)} := B_{ij}^{(t)}\mathbf{1}$, and in order to have a $D$-“common” equitable partition, we want to prove that:

$$v_{ij}^{(t)} = w_{ij}^{(t)} = c_{ij}^{(t)}\,\mathbf{1}\,. \tag{7.15}$$

Using the fact that $U_j$ is bistochastic and then eq. (7.13), we have:

$$v_{ij}^{(t)} \;:=\; A_{ij}^{(t)}\,\mathbf{1} \;=\; A_{ij}^{(t)}\,U_j\,\mathbf{1} \;=\; U_i\,B_{ij}^{(t)}\,\mathbf{1} \;=\; U_i\,w_{ij}^{(t)}\,,$$

and similarly, using the fact that $U_j^\top$ is bistochastic and then eq. (7.14), we get:

$$w_{ij}^{(t)} \;:=\; B_{ij}^{(t)}\,\mathbf{1} \;=\; B_{ij}^{(t)}\,U_j^\top\,\mathbf{1} \;=\; U_i^\top\,A_{ij}^{(t)}\,\mathbf{1} \;=\; U_i^\top\,v_{ij}^{(t)}\,.$$

Now, from those two relations, we can apply [RSU94, Lem 2.3] and conclude that there exists a constant $c_{ij}^{(t)} \in \mathbb{R}$ such that eq. (7.15) is satisfied. Moreover, by construction of $v_{ij}^{(t)}$, we know that $c_{ij}^{(t)} = |C_j \cap C(g,t)| \in \mathbb{N}$ for any $g \in C_i$. This yields the desired $D$-common equitable partition. ■

#### 7.3.3.3 $D$-Common Equitable Partition Implies $D$-$\mathcal{NS}$-Isomorphism

Lastly, we prove that the generalized notion of common equitable partition is sufficient in order to have a perfect non-signaling strategy for the $D$-distance game:

**Lemma 7.38 —** *Let $D \geqslant 0$. If $(\mathcal{G},\mathcal{H})$ admits a $D$-common equitable partition, then $\mathcal{G} \cong_{\mathrm{ns}}^{D} \mathcal{H}$.*

*Proof.* This proof generalizes [Ats+19, Lemma 4.4]. Denote:

$$\big(k, [n_1,\dots,n_k], [c_{ij}^{(t)}]\big)\,,$$

the parameters of the given $D$-common equitable partition of $(\mathcal{G},\mathcal{H})$, and consider $\overline{c_{ij}} := n_j - \sum_{t=0}^{D} c_{ij}^{(t)}$ the number of elements in $C_j$ that are at distance $> D$ of a fixed element in $C_i$, wher $t \in \{0,\dots,D\}$. We consider the following correlation:

$$\mathbf{P}(h_{\mathsf{A}},h_{\mathsf{B}}\,|\,g_{\mathsf{A}},g_{\mathsf{B}}) \;=\; \begin{cases} 1/n_i c_{ij}^{(t)} & \text{if } d(h_{\mathsf{A}},h_{\mathsf{B}}) = d(g_{\mathsf{A}},g_{\mathsf{B}}) = t \leqslant D \text{ and } (\star)\,,\\ 1/n_i\overline{c_{ij}} & \text{if } d(h_{\mathsf{A}},h_{\mathsf{B}}) > D,\ d(g_{\mathsf{A}},g_{\mathsf{B}}) > D \text{ and } (\star)\,,\\ 0 & \text{otherwise}\,,\end{cases}$$

where the condition $(\star)$ stands for "$g_{\mathsf{A}} \in C_i, g_{\mathsf{B}} \in C_j, h_{\mathsf{A}} \in D_i, h_{\mathsf{B}} \in D_j$". Moreover, define:

$$\mathbf{P}(h, h' \,|\, g, g') = \mathbf{P}(g, h' \,|\, h, g') = \mathbf{P}(h, g' \,|\, g, h') = \mathbf{P}(g, g' \,|\, h, h')\,,$$

for all $g, g' \in V(\mathcal{G})$ and all $h, h' \in V(\mathcal{H})$, and set $\mathbf{P} = 0$ in all the cases not yet accounted for. By construction, this is a perfect strategy for the $D$-distance game because the probability of losing is zero, so it remains to show that $\mathbf{P} \in \mathcal{NS}$. First of all, the non-negativity condition is satisfied by the construction of $\mathbf{P}$. Let $V = V(\mathcal{G}) \cup V(\mathcal{H})$. Let us check the marginal condition in the case where both inputs $x_{\mathsf{A}}, x_{\mathsf{B}}$ are in $V(\mathcal{G})$: fix $y_{\mathsf{A}} = h_{\mathsf{A}} \in D_i$ and $x_{\mathsf{A}} = g_{\mathsf{A}} \in C_i$ and $x_{\mathsf{B}} = g_{\mathsf{B}} \in C_j$ . On the one hand, if $d(g_{\mathsf{A}}, g_{\mathsf{B}}) = t \leqslant D$, then:

$$\sum_{y \in V} \mathbf{P}(h_{\mathsf{A}}, y \,|\, g_{\mathsf{A}}, g_{\mathsf{B}}) \;=\; \sum_{h_{\mathsf{B}} \in D_j \cap C(h_{\mathsf{A}}, t)} \frac{1}{n_i\, c_{ij}^{(t)}} \;=\; \frac{|D_j \cap C(h_{\mathsf{A}}, t)|}{n_i\, c_{ij}^{(t)}} \;=\; \frac{1}{n_i}\,,$$

because $c_{ij}^{(t)} = |D_j \cap C(h_{\mathsf{A}}, t)|$. On the other hand, if $d(g_{\mathsf{A}}, g_{\mathsf{B}}) > D$, then:

$$\sum_{y \in V} \mathbf{P}(h_{\mathsf{A}}, y \,|\, g_{\mathsf{A}}, g_{\mathsf{B}}) \;=\; \sum_{h_{\mathsf{B}} \in D_j \cap B(h_{\mathsf{A}}, D)^c} \frac{1}{n_i\, \overline{c_{ij}}} \;=\; \frac{|D_j \cap B(h_{\mathsf{A}}, D)^c|}{n_i\, \overline{c_{ij}}} \;=\; \frac{1}{n_i}\,,$$

because $\overline{c_{ij}} = |D_j \cap B(h_{\mathsf{A}}, D)^c|$, where $B(h_{\mathsf{A}}, D)^c$ is the complement of the ball centered at $h_{\mathsf{A}}$ of radius $D$, *i.e.* the element of $V(\mathcal{H})$ that are at distance $> D$ of $h_{\mathsf{A}}$. In both equations, the result does not depend on $g_{\mathsf{B}}$, hence Alice's marginal $\mathbf{P}(h_{\mathsf{A}} \,|\, g_{\mathsf{A}})$ is well-defined. Similarly, Bob's marginal $\mathbf{P}(h_{\mathsf{B}} \,|\, g_{\mathsf{B}}) = 1/n_j$ is also well-defined using the relation $n_i\, c_{ij}^{(t)} = n_j\, c_{ji}^{(t)}$ from Lemma 7.36. A similar proof works in all the other choices of $x_{\mathsf{A}}, x_{\mathsf{B}} \in V$, using the fact that the parameters $n_i$, and $c_{ij}^{(t)}$, and $\overline{c_{ij}}$ are "common" for $\mathcal{G}$ and $\mathcal{H}$, and we have $\mathbf{P}(g_{\mathsf{A}} \,|\, h_{\mathsf{A}}) = 1/n_i$ and $\mathbf{P}(g_{\mathsf{B}} \,|\, h_{\mathsf{B}}) = 1/n_j$. Finally, for any $x_{\mathsf{A}} \in C_i \subseteq V(\mathcal{G})$ and $x_{\mathsf{B}} \in V$, the normalization condition is verified by summing the marginals:

$$\sum_{y_{\mathsf{A}}, y_{\mathsf{B}} \in V} \mathbf{P}(y_{\mathsf{A}}, y_{\mathsf{B}} \,|\, x_{\mathsf{A}}, x_{\mathsf{B}}) \;=\; \sum_{y_{\mathsf{A}} \in D_i} \mathbf{P}(y_{\mathsf{A}} \,|\, x_{\mathsf{A}}) \;=\; \sum_{y_{\mathsf{A}} \in D_i} \frac{1}{n_i} \;=\; \frac{|D_i|}{n_i} \;=\; 1\,,$$

and similarly in the case $x_{\mathsf{A}} \in D_i \subseteq V(\mathcal{H})$. We therefore obtain the wanted result. ■

The above Lemmas 7.34, 7.37 and 7.38 prove the respective implications (1)⇒(2), (2)⇒(3), and (3)⇒(1) of Theorem 7.30, hence we obtain the wanted characterization of perfect non-signaling strategies for the $D$-distance game in terms of $D$-fractional isomorphism and of $D$-common equitable partition.

### 7.3.4 Example of $D$- but not $(D+1)$-Isomorphic Graphs

In this subsection, we construct a sequence of graphs $(\mathcal{G}_D, \mathcal{H}_D)$ that are $D$-isomorphic but not $(D+1)$-isomorphic in the sense of the generalized fractional isomorphism defined on Section 7.3.3.1.

We label the vertices of the cycle $\mathcal{C}_n$ from $0$ to $n-1$ clockwise. The adjacency matrix of this graph is the matrix $A_n := (a_{ij})$ such that $a_{ij} = 1$ if $j = i \pm 1\,[n]$, and $a_{ij} = 0$ otherwise, where $[n]$ denotes the congruence modulo $n$. More generally, for any $t < \frac{n}{2}$, its $t$-adjacency matrix is

$$A_n^{(t)} := (a_{ij}^{(t)})_{i,j=0,\dots,n-1}\,, \quad \text{where} \quad a_{ij}^{(t)} := \begin{cases} 1 & j = i + t\,[n]\,, \\ 1 & j = i - t\,[n]\,, \\ 0 & \text{otherwise}\,. \end{cases}$$

Note that for $t = 1$ we recover $A_n = A_n^{(1)}$. Denote with $\mathcal{C}'_{2n} := \mathcal{C}_n \sqcup \mathcal{C}_n$ the disjoint union of two cycles $\mathcal{C}_n$ on $n$ points. Using block matrix notation, its adjacency and $t$-adjacency matrices are:

$$B_{2n} := \begin{pmatrix} A_n & \mathbf{0} \\ \mathbf{0} & A_n \end{pmatrix}, \quad B_{2n}^{(t)} := \begin{pmatrix} A_n^{(t)} & \mathbf{0} \\ \mathbf{0} & A_n^{(t)} \end{pmatrix},$$

for any $t < \frac{n}{2}$. Finally, let $u_{2n}$ be the following bistochastic matrix:

$$u_{2n} := \frac{1}{2} \begin{pmatrix} \mathbb{I}_n & \mathbb{I}_n \\ \mathbb{I}_n & \mathbb{I}_n \end{pmatrix},$$

where $\mathbb{I}_n$ is the $n \times n$ identity matrix.

**Lemma 7.39 —** *Denoting by $e_0, \dots, e_{2n-1} \in \mathbb{C}^{2n}$ the canonical basis of $\mathbb{C}^{2n}$, the matrices $A_{2n}^{(t)}$, $B_{2n}^{(t)}$, and $u_{2n}$, act as follows, for any $t < \frac{n}{2}$ and $k = 0, \dots, 2n-1$.*

*(1)* $A_{2n}^{(t)} e_k = e_{k+t\,[2n]} + e_{k-t\,[2n]}$.

*(2)* $B_{2n}^{(t)} e_k = \begin{cases} e_{k+t\,[n]} + e_{k-t\,[n]} & k < n\,, \\ e_{(k+t\,[n])+n} + e_{(k-t\,[n])+n} & k \geqslant n\,. \end{cases}$

*(3)* $u_{2n} e_k = \frac{1}{2}(e_k + e_{k+n\,[2n]})$*.*

*(4) In particular* $u_{2n} e_k = u_{2n} e_{k+n} = \frac{1}{2}(e_k + e_{k+n})$*, if* $k < n$*.*

*Proof.* Items (1) to (3) follow directly from the definitions of the matrices. For item (4), the result follows from item (3) and using the fact that $k + n + n\,[2n] \equiv k\,[2n]$. ■

We need the following simple facts from number theory.

**Lemma 7.40 —** *For* $n, t \in \mathbb{N}$ *with* $t < \frac{n}{2}$ *and* $k \in \{0, \ldots, 2n-1\}$*, we have*

*(1)* $\{k + t\,[n], (k + t\,[n]) + n\} = \{k + t\,[2n], k + t + n\,[2n]\}$*,*

*(2)* $\{k - t\,[n], (k - t\,[n]) + n\} = \{k - t\,[2n], k - t + n\,[2n]\}$*.*

*Proof.* For item (1), we check by case distinction:

| | $1 \leqslant k+t < n$ | $n \leqslant k+t < 2n$ | $2n \leqslant k+t < 3n$ |
|---|---|---|---|
| $k + t\,[n]$ | $k+t$ | $k+t-n$ | $k+t-2n$ |
| $(k + t\,[n]) + n$ | $k+t+n$ | $k+t$ | $k+t-n$ |
| $k + t\,[2n]$ | $k+t$ | $k+t$ | $k+t-2n$ |
| $k + t + n\,[2n]$ | $k+t+n$ | $k+t-n$ | $k+t-n$ |

Similarly for item (2). ■

The next proposition is a consequence of the preceding two lemmata.

**Proposition 7.41 —** *For all* $n, t \in \mathbb{N}$ *with* $t < \frac{n}{2}$*, we have* $u_{2n} A_{2n}^{(t)} = B_{2n}^{(t)} u_{2n}$*, which means:*

$$\mathcal{C}_{2n} \cong^{D}_{\text{frac}} \mathcal{C}_n \sqcup \mathcal{C}_n\,,$$

*for all* $D < \frac{n}{2}$*.*

*Proof.* Let $k \in \{0, \ldots, 2n-1\}$ with $k < n$. Then, by Lemma 7.39:

$$\begin{aligned} u_{2n} A_{2n}^{(t)} e_k &= u_{2n}\big(e_{k+t\,[2n]} + e_{k-t\,[2n]}\big) \\ &= \tfrac{1}{2}\big(e_{k+t\,[2n]} + e_{k+t+n\,[2n]} + e_{k-t\,[2n]} + e_{k-t+n\,[2n]}\big)\,, \end{aligned}$$

and:

$$\begin{aligned} B_{2n}^{(t)} u_{2n} e_k &= \tfrac{1}{2} B_{2n}^{(t)} (e_k + e_{k+n}) \\ &= \tfrac{1}{2} \big( e_{k+t\,[n]} + e_{k-t\,[n]} + e_{(k+n+t\,[n])+n} + e_{(k+n-t\,[n])+n} \big) \\ &= \tfrac{1}{2} \big( e_{k+t\,[n]} + e_{(k+t\,[n])+n} + e_{k-t\,[n]} + e_{(k-t\,[n])+n} \big) \,, \end{aligned}$$

which by Lemma 7.40 shows the relation $u_{2n} A_{2n}^{(t)} e_k = B_{2n}^{(t)} u_{2n} e_k$. In the case $k \geqslant n$, we apply Lemma 7.39 (4) and obtain the same result. ■

We now give a criterion, when two graphs fail to be fractionally $D$-isomorphic.

**Lemma 7.42 —** *Let $\mathcal{G}$ and $\mathcal{H}$ be two finite graphs with the same number of vertices. If $\operatorname{diam}(\mathcal{G}) \geqslant D > \operatorname{diam}(\mathcal{H})$, then $\mathcal{G}$ and $\mathcal{H}$ are not $D$-fractionally isomorphic.*

*Proof.* Let $(u_{gh})$ a bistochastic matrix indexed by the vertices $g \in V(\mathcal{G})$ and $h \in V(\mathcal{H})$. On the one hand, as $D > \operatorname{diam}(\mathcal{H})$, the $D$-adjacency matrix of $\mathcal{H}$ is zero $A_{\mathcal{H}}^{(D)} = \mathbf{0}$, therefore:

$$u\, A_{\mathcal{H}}^{(D)} = \mathbf{0}\,.$$

On the other hand, we can find in $\mathcal{G}$ two vertices $g_1$ and $g_2$ at a distance exactly $D$, and as $\sum_h u_{g_2 h} = 1$ by bistochasticity, we know there exists at least one vertex $h_2 \in V(\mathcal{H})$ such that $u_{g_2 h_2} > 0$. It yields that the matrix $A_{\mathcal{G}}^{(D)} u$ admits a non-zero element:

$$\big[ A_{\mathcal{G}}^{(D)}\, u \big]_{g_1, h_2} = \sum_{g' \in C(g_1, D)} u_{g' h_2} \geqslant u_{g_2 h_2} > 0\,.$$

Hence $u\, A_{\mathcal{H}}^{(D)} \neq A_{\mathcal{G}}^{(D)} u$, and the graphs fail to be $D$-fractionally isomorphic. ■

Now, combining Proposition 7.41 and Lemma 7.42, we obtain:

**Proposition 7.43 —** *For any $D \in \mathbb{N}$, the graphs $\mathcal{C}_{2(2D+1)}$ and $\mathcal{C}'_{2(2D+1)} = \mathcal{C}_{2D+1} \sqcup \mathcal{C}_{2D+1}$ are $D$-fractionally isomorphic but not $(D+1)$-fractionally isomorphic:*

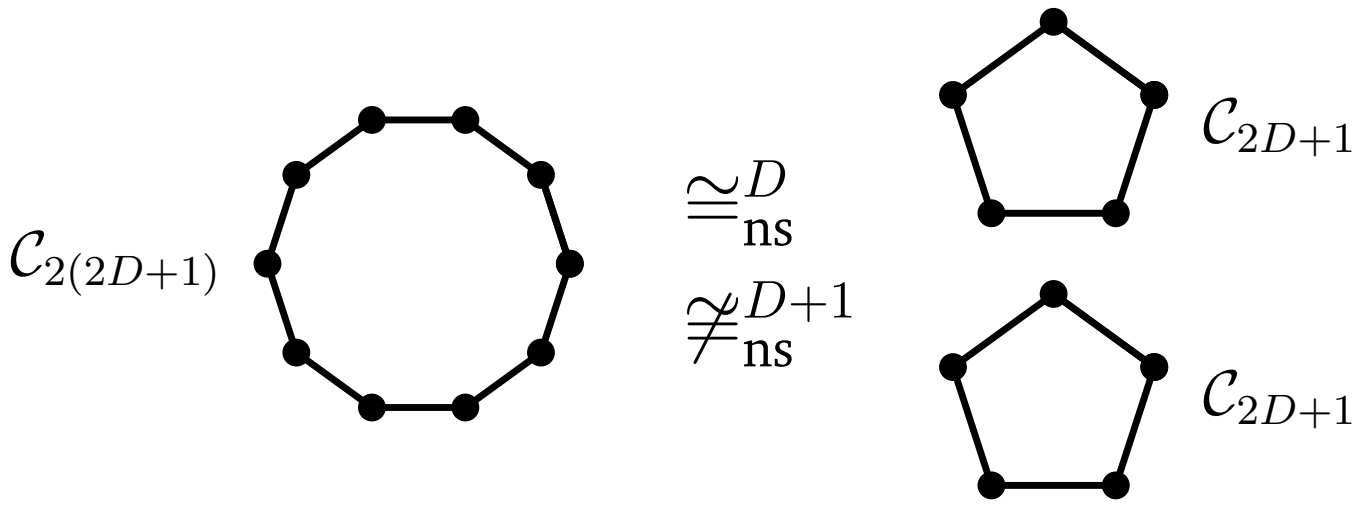

**Remark 7.44 —** More precisely, the graphs $\mathcal{G} = \mathcal{C}_{2(2D+1)}$ and $\mathcal{H} = \mathcal{C}_{2D+1} \sqcup \mathcal{C}_{2D+1}$ are $t$-fractionally isomorphic for $t \leqslant D = \mathrm{diam}(\mathcal{H})$ and $t > 2D + 1 = \mathrm{diam}(\mathcal{G})$, and are not $t$-fractionally isomorphic for $D < t \leqslant 2D + 1$.

**Remark 7.45 —** There exist graphs that are $D$-fractionally isomorphic for all $D \in \mathbb{N}$ but not quantum isomorphic [Sch24]. Indeed, for instance, consider $\mathcal{G}$ to be the Shrikhande graph and $\mathcal{H}$ the $4 \times 4$ Rook's graph. (Both are strongly regular graphs of parameters $(16, 6, 2, 2)$.) On the one hand, both of them have diameter $2$, so their $2$-adjacency matrices are exactly the adjacency matrix of the complement graph: $A^{(2)}_{\mathcal{G}} = A_{\mathcal{G}^c}$ and $A^{(2)}_{\mathcal{H}} = A_{\mathcal{H}^c}$. But, we know that $\mathcal{G} \cong_{\mathrm{frac}} \mathcal{H}$ and $\mathcal{G}^c \cong_{\mathrm{frac}} \mathcal{H}^c$. Therefore $\mathcal{G} \cong^{D}_{\mathrm{frac}} \mathcal{H}$ for all $D \in \mathbb{N}$. On the other hand, we know that they do not admit the same number of homomorphisms from the planar complete graph $\mathcal{K}_4$: there is no such homomorphism to $\mathcal{G}$, while one to $\mathcal{H}$ exists. Hence, using the homomorphism counts characterization of the quantum isomorphism [MR20], we deduce that $\mathcal{G} \ncong_{\mathrm{qc}} \mathcal{H}$, as wanted.

We obtain the chain of *strict* implications drawn in Figure 7.4. Note that if $\mathcal{G} \cong^{D_0}_{\mathrm{frac}} \mathcal{H}$ for $D_0 := \max\big(\mathrm{diam}(\mathcal{G}), \mathrm{diam}(\mathcal{H})\big)$, then $\mathcal{G} \cong^{D}_{\mathrm{frac}} \mathcal{H}$ for all $D \in \mathbb{N}$.

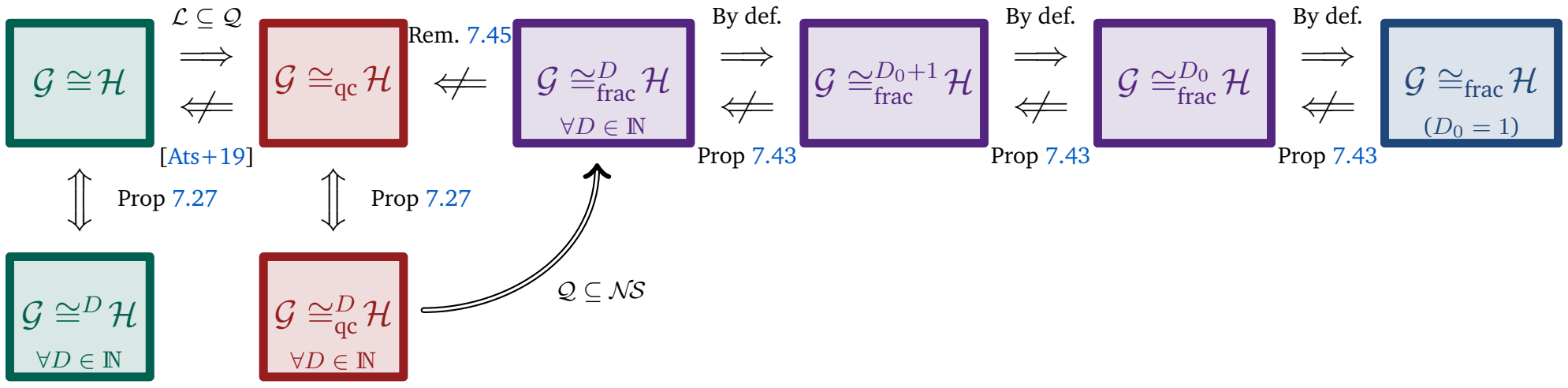

**Figure 7.4 —** *Chain of strict implications, with fixed* $D_0 \geqslant 2$.

We conclude the subsection with the following theorem, which is a consequence of Theorem 7.30 and Figure 7.4, which tells us that we can distinguish perfect non-signaling strategies between the $D$-distance game and the graph isomorphism game, as opposed to the classical and quantum cases:

**Theorem 7.46** ($\mathcal{NS}$ is Finer Than $\mathcal{L}$ and $\mathcal{Q}$) — *Let $D \geqslant 2$. As opposed to the classical and quantum cases (Section 7.3.2), the set of perfect non-signaling strategies for the $D$-distance game is strictly included in the set of perfect non-signaling strategies for the isomorphism game.* ■

## 7.3.5 Links with Communication Complexity

In this subsection, we give statements showing the collapse of communication complexity in various cases. These statements are mainly generalizations of results from the first two sections about the isomorphism and coloring games.

### 7.3.5.1 Existence of Collapsing Non-Signaling Strategies

To show the existence of a perfect collapsing strategy, we want to adapt Theorem 7.9 to the $D$-distance game. To this end, we generalize two results from the isomorphism game to the $D$-distance game. First, see that Proposition 7.6 can be easily generalized using the characterization of perfect strategies in terms of $D$-common equitable partition (Theorem 7.30), and it gives:

**Lemma 7.47** — *Let $\mathcal{G} \cong_{\text{ns}}^{D} \mathcal{H}$ such that $\mathcal{H}$ is not connected: $\mathcal{H} = \mathcal{H}_1 \sqcup \mathcal{H}_2$. Denote the partitions $\mathscr{C} = \{C_1, \ldots, C_k\}$ and $\mathscr{D} = \{D_1, \ldots, D_k\}$ forming a $D$-common equitable partition for $\mathcal{G}$ and $\mathcal{H}$, and assume that the proportion of vertices of $\mathcal{H}_1$ assigned to $D_i$ is independent of $i$:*

$$\forall i, j \in [k], \qquad \frac{|D_i \cap \mathcal{H}_1|}{|D_i|} = \frac{|D_j \cap \mathcal{H}_1|}{|D_j|}\,. \tag{H'}$$

*Then the $D$-distance game of $(\mathcal{G}, \mathcal{H})$ admits a symmetric perfect strategy of the*

*following form:*

$$\mathbb{P}_{\mathcal{S}}(h_{\mathsf{A}}, h_{\mathsf{B}} \,|\, g_{\mathsf{A}}, g_{\mathsf{B}}) := \begin{cases} 1/n_i c_{ij}^{(t)} & \text{if } d(g_{\mathsf{A}}, g_{\mathsf{B}}) = d(h_{\mathsf{A}}, h_{\mathsf{B}}) = t \leqslant D \text{ and } (\star), \\ 1/n_i \overline{c_{ij}} & \text{if } d(g_{\mathsf{A}}, g_{\mathsf{B}}) > D,\ d(h_{\mathsf{A}}, h_{\mathsf{B}}) > D, \text{ and } (\star), \\ 0 & \text{otherwise}, \end{cases}$$

*where* $(\star)$ *denotes the condition "*$g_{\mathsf{A}} \in C_i,\ g_{\mathsf{B}} \in C_j,\ h_{\mathsf{A}} \in D_i,\ h_{\mathsf{B}} \in D_j$*".* ■

Then, using similar arguments as in the proof of Proposition 7.50, we observe that Lemma 7.10 can be generalized straightforwardly as follows:

**Lemma 7.48 —** *Let* $\mathcal{G}, \mathcal{H}$ *two graphs such that* $1 \leqslant D < \operatorname{diam}(\mathcal{G})$ *and such that* $\mathcal{H}$ *is not connected:* $\mathcal{H} = \mathcal{H}_1 \sqcup \mathcal{H}_2$*. There exists a path graph* $\mathcal{P} \subseteq \mathcal{G}$ *of length* $D + 1$*, for which we call* $g_1$ *and* $g_3$ *the extremal vertices. Assume* $\mathcal{G} \cong^D_{\text{ns}} \mathcal{H}$ *for some strategy* $\mathcal{S}$ *that is symmetric from* $\mathcal{P}$ *to the components of* $\mathcal{H}$*, and suppose that:*

$$\nu_{g_1,g_3} > 0\,.$$

*Then the box* $\mathbf{PR}_{\alpha,\beta}$ *is perfectly simulated with* $\alpha = 2\,\nu_{g_1,g_3} > 0$ *and some* $\beta \geqslant 0$*.* ■

Now, using these two generalized lemmata, the exact same proof as the one of Theorem 7.9 also gives the result for the $D$-distance game:

**Theorem 7.49** (Existence of Collapsing Strategies) **—** *Let* $\mathcal{G} \cong^D_{\text{ns}} \mathcal{H}$ *for some* $1 \leqslant D < \operatorname{diam}(\mathcal{G})$ *and such that* $\mathcal{H}$ *is not connected:* $\mathcal{H} = \mathcal{H}_1 \sqcup \mathcal{H}_2$*, where each of* $\mathcal{H}_1$ *and* $\mathcal{H}_2$ *may possibly be decomposed in several connected components. Denote the partitions* $\mathscr{C} = \{C_1, \dots, C_k\}$ *and* $\mathscr{D} = \{D_1, \dots, D_k\}$ *forming a* $D$*-common equitable partition for* $\mathcal{G}$ *and* $\mathcal{H}$*, and assume that condition (H′) holds. Then the* $D$*-distance game of* $(\mathcal{G}, \mathcal{H})$ *admits a perfect strategy that collapses communication complexity.* ■

#### 7.3.5.2 All Perfect Non-Signaling Strategies Collapse CC

Now, we want to prove sufficient conditions so that all perfect strategies for the $D$-distance game collapse communication complexity. We begin the study with the simple case where the graph $\mathcal{H}$ has a smaller diameter than $\mathcal{G}$. First, we assume that $\mathcal{H}$ admits exactly $2$ connected components, and then more generally $N$ connected components.

**Proposition 7.50** (Collapse of CC) — *If* $\mathrm{diam}(\mathcal{G}) > \mathrm{diam}(\mathcal{H}) \geqslant D \geqslant 1$ *and if* $\mathcal{H}$ *admits exactly two connected components, then any perfect* $\mathcal{NS}$*-strategy for the* $D$*-distance game collapses communication complexity.*

*Proof.* By assumption, there exist vertices $g_1, g_3$ in $\mathcal{G}$ whose distance is exactly $\mathrm{diam}(\mathcal{H}) + 1$. In a minimal path joining $g_1$ to $g_3$ in $\mathcal{G}$, consider $g_2$ at distance $D$ of $g_1$ and distance $\mathrm{diam}(\mathcal{H}) + 1 - D$ of $g_3$. Assume that there exists a perfect strategy $\mathcal{S}$ for the $D$-distance game. Similarly to the proof of Theorem 7.2, Alice and Bob will use this perfect strategy $\mathcal{S}$ as a black box to generate a **PR** box, which is known to collapse communication complexity [vD99]. Suppose Alice and Bob are given respective bits $x, y \in \{0,1\}$. They want to produce $a, b \in \{0,1\}$ without signaling such that $a \oplus b = xy$. If $x = 0$, Alice chooses $g_{\mathsf{A}} = g_2$, and if $x = 1$, she chooses $g_{\mathsf{A}} = g_1$. As for Bob, given respectively $y = 0, 1$, he chooses $g_{\mathsf{B}} = g_2, g_3$. Alice and Bob input their choice $(g_{\mathsf{A}}, g_{\mathsf{B}})$ in the strategy $\mathcal{S}$, which outputs some vertices $(h_{\mathsf{A}}, h_{\mathsf{B}})$ of $\mathcal{H}$ satisfying the conditions of the $D$-distance game. Notice that $h_{\mathsf{A}}$ and $h_{\mathsf{B}}$ are in different connected components of $\mathcal{H} = \mathcal{H}_1 \sqcup \mathcal{H}_2$ if and only if $x = y = 1$. Upon receiving $h_{\mathsf{A}} \in \mathcal{H}_i$, Alice produces the bit $a = i$, and similarly for Bob with $h_{\mathsf{B}} \in \mathcal{H}_j$ and $b = j$. It follows that the relation $a \oplus b = xy$ is always satisfied, thus the **PR** box is perfectly simulated, and there is a collapse of communication complexity. ∎

**Remark 7.51 —** Actually, it is enough to have a noisy $\mathcal{NS}$-strategy winning the $D$-distance game with probability $p > \frac{3+\sqrt{6}}{6}$, since the same proof would generate a **PR** box with probability $p$ and therefor collapse CC by [Bra+06].

**Proposition 7.52** (Collapse of CC) — *If* $\mathrm{diam}(\mathcal{G}) > \mathrm{diam}(\mathcal{H}) \geqslant D \geqslant 1$ *and if* $\mathcal{H}$ *admits exactly* $N$ *connected components, then any perfect* $\mathcal{NS}$*-strategy for the* $D$*-distance game, combined with a perfect* $\mathcal{NS}$*-strategy for the* 2*-coloring game of* $\mathcal{K}_N$*, collapses communication complexity.*

*Proof.* Proceed as in the proof of Theorem 7.23 combined with Proposition 7.50. ∎

**Remark 7.53 —** Again, we can generalize this result to a noisy version: it is enough that we have $p$ and $q$ such that the product satisfies $pq > \frac{3+\sqrt{6}}{6}$, that the $\mathcal{NS}$-strategy for the $D$-distance game wins with probability $p$, and that the $\mathcal{NS}$-strategy for the 2-coloring game of $\mathcal{K}_N$ wins with probability $q$.

Finally, the following statement is a particular case of Theorem 7.16. Indeed, any perfect strategy for the $D$-distance game is perfect for the isomorphism game. In the theorem, we gave sufficient conditions on graphs so that all perfect strategies for the isomorphism game collapse CC. Hence, with the same conditions, we have that the result also holds for the $D$-distance game for any $D \geqslant 1$. Recall that the proof of this theorem was based on tools from graph automorphism theory and graph transitivity notions.

**Theorem 7.54** (Collapse of CC) — *Let $D \geqslant 1$. Let $\mathcal{G} \cong^D_{\text{ns}} \mathcal{H}$ such that $2 \leqslant \text{diam}(\mathcal{G})$ and such that $\mathcal{H}$ is not connected: $\mathcal{H} = \mathcal{H}_1 \sqcup \mathcal{H}_2$, where each of $\mathcal{H}_1$ and $\mathcal{H}_2$ may possibly be decomposed in several connected components. Let $\mathscr{C} = \{C_1, \dots, C_k\}$ and $\mathscr{D} = \{D_1, \dots, D_k\}$ form a 1-common equitable partition for $\mathcal{G}$ and $\mathcal{H}$ such that condition (H′) holds. Assume moreover that $\mathcal{H}$ is strongly transitive and $d$-regular, and that the players share randomness. Then* every *perfect non-signaling strategy for the $D$-distance game of $(\mathcal{G}, \mathcal{H})$ collapses communication complexity.* ■

# Chapter 8

# Unclonable Bit in No-Cloning Games

In this chapter, we include and rearrange the following reference:

**[Bot+24b]** Pierre Botteron, Anne Broadbent, Eric Culf, Ion Nechita, Clément Pellegrini, and Denis Rochette. *Towards Unconditional Uncloneable Encryption*. 2024. arXiv: 2410.23064 [quant-ph]

## Chapter Contents

## 8.1 Preliminaries

In this chapter, we propose a candidate for the unconditional unclonable bit problem and provide strong evidence that the adversary's success probability in the related security game converges quadratically as:

$$\frac{1}{2} + \frac{1}{2\sqrt{K}},$$

where $K$ is polynomial in the encoding size, representing the number of keys, and where $\frac{1}{2}$ is trivially achievable. We prove this bound's validity for $K$ ranging from $2$ to $7$ and demonstrate the validity up to $K = 17$ using computations based on the NPA hierarchy. We furthermore provide compelling heuristic evidence towards the general case. In addition, we prove an asymptotic upper bound of $\frac{5}{8}$ and give a numerical upper bound of $\sim 0.5980$, which to our knowledge is the best-known value in the unconditional model.[1]

Below, after pointing to the relevant background materials (Section 8.1.1) and presenting a preliminary upper bound (Section 8.1.2), we expose our candidate scheme based on Clifford algebra (Section 8.2) and finally present our analytical and numerical results on security bounds (Section 8.3).

### 8.1.1 *Background*

This reference mainly relies on Chapter 5 about quantum cryptography. More specifically, we refer to Section 5.3 for a definition of quantum encryption of classical messages schemes, the no-cloning games, and their connection with monogamy-of-entanglement (MoE) games. For convenience, we illustrate the no-cloning game again in Figure 8.1.

We also refer to Theorem 2.37 for a statement of the No-Cloning Theorem, and to Section 2.2.5 for background materials on the principle of MoE.

We will also use the NPA hierarchy and SoS decompositions, both introduced in Section 3.1.3.

[1]Since then, a recent result from Bhattacharyya and Culf shows that the adversary's winning probability is indeed upper bounded by $\frac{1}{2} + \widetilde{\mathcal{O}}(\frac{1}{\lambda})$ for any security parameter $\lambda$ [BC25]. The authors use different techniques based on Haar-measure games.

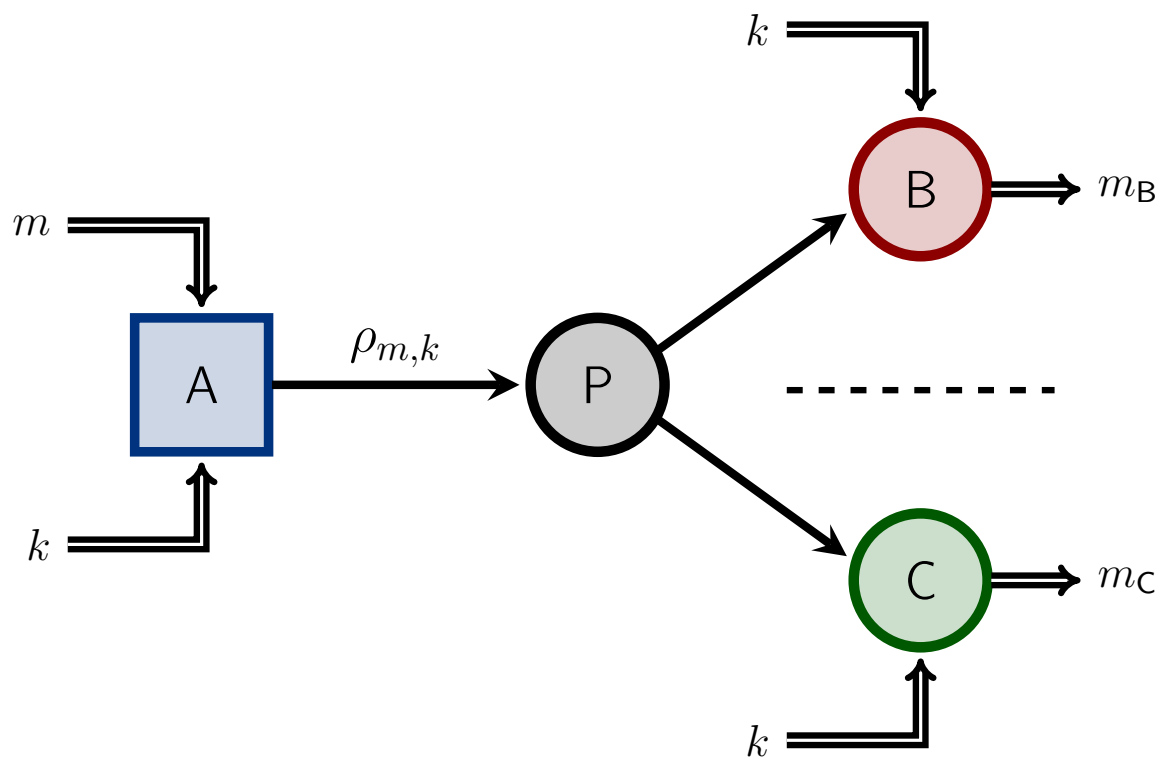

**Figure 8.1 —** No-cloning game for a 1-bit message. *Alice encrypts a uniformly random message $m \in \{0,1\}$, using key $k$, into a quantum state $\rho_{m,k}$. She transmits it to a pirate* P *modeled by a quantum channel $\Phi : \mathcal{B}(\mathcal{H}_{\mathsf{A}}) \to \mathcal{B}(\mathcal{H}_{\mathsf{B}} \otimes \mathcal{H}_{\mathsf{C}})$. Bob and Charlie are then given the registers for $\mathcal{H}_{\mathsf{B}}$ and $\mathcal{H}_{\mathsf{C}}$ respectively, as well as a copy of $k$. They output $m_{\mathsf{B}}, m_{\mathsf{C}} \in \{0,1\}$ respectively, and* win *if and only if $m = m_{\mathsf{B}} = m_{\mathsf{C}}$.* Unclonable-indistinguishability *holds if the winning probability is bounded by $1/2 + \mathrm{negl}(\lambda)$ for $\lambda$ a security parameter.*

### *8.1.2 Preliminary Upper Bound on the Winning Probability*

The security notion that we want to achieve is defined in terms of an upper bound on the winning probability of the adversary team $(\mathsf{P},\mathsf{B},\mathsf{C})$:

$$\mathbb{P}\Big((\mathsf{P},\mathsf{B},\mathsf{C})\text{ win}\Big) \leqslant \frac{1}{2} + f(\lambda),$$

for some function $f : \mathbb{R} \to \mathbb{R}$ vanishing at infinity $\lim_{\lambda} f(\lambda) = 0$ (see Definition 5.12). Furthermore, recall from eq. (5.3) that we already have the following upper bound:

$$\mathbb{P}\Big((\mathsf{P},\mathsf{B},\mathsf{C})\text{ win}\Big) \leqslant \sup_{\sigma,B,C} \mathop{\mathbb{E}}_{m,k} \mathrm{Tr}\Big[\sigma\left(d \cdot \rho_{m,k}^{\top} \otimes B_{m|k} \otimes C_{m|k}\right)\Big], \tag{8.1}$$

where the supremum is taken other all operators $\sigma \succcurlyeq \mathbf{0}$ such that $\mathrm{Tr}[\sigma] = 1$, *i.e.* over all quantum mixed states, and POVMs $\{B_{m|k}\}_m$ and $\{C_{m|k}\}_m$.

Now, we can refine these conditions in order to have an upper bound that depends on the operator norm $\|\cdot\|_{\mathrm{op}}$ (the largest absolute value of the eigenvalues). As pure quantum states form the extreme points of the convex set of mixed states (see page 27) and as the optimization in eq. (8.1)

is linear in $\sigma$, the upper bound is saturated on pure quantum states $|\psi\rangle\langle\psi|$ and we have:

$$\mathbb{P}\Big((\mathsf{P},\mathsf{B},\mathsf{C})\text{ win}\Big) \leqslant \sup_{\psi,B,C} \langle\psi| \mathop{\mathbb{E}}_{m,k}\left[d\cdot\rho_{m,k}^{\top}\otimes B_{m|k}\otimes C_{m|k}\right]|\psi\rangle \tag{8.2}$$

$$\leqslant \sup_{\psi,B,C}\left|\langle\psi| \mathop{\mathbb{E}}_{m,k}\left[d\cdot\rho_{m,k}^{\top}\otimes B_{m|k}\otimes C_{m|k}\right]|\psi\rangle\right| \tag{8.3}$$

$$= \sup_{B,C}\left\|\mathop{\mathbb{E}}_{m,k}\left[d\cdot\rho_{m,k}^{\top}\otimes B_{m|k}\otimes C_{m|k}\right]\right\|_{\text{op}}, \tag{8.4}$$

were the first two suprema are taken over all $\|\psi\| = 1$, and where the last equality holds because $d\cdot\rho_{m,k}^{\top}\otimes B_{m|k}\otimes C_{m|k}$ is a Hermitian operator.

Finally, we can further rephrase the upper bound as follows. By Naimark's Dilation Theorem (Theorem 2.25), we may always assume that the POVMs are in fact PVMs up to increasing the dimensions. Moreover, any adversaries $(\mathsf{P},\mathsf{B},\mathsf{C})$ can be assumed to be *symmetric*, *i.e.* $\mathcal{H}_{\mathsf{B}} = \mathcal{H}_{\mathsf{C}}$ and $\{B_{i|k}\} = \{C_{j|k}\}$, by taking Bob's and Charlie's spaces to be the direct sum $\mathcal{H}_{\mathsf{B}}\oplus\mathcal{H}_{\mathsf{C}}$, the PVM operators to be $\{B_{i|k}\oplus C_{i|k}\}$,[2] and the CPTP map $\Phi$ that sends Bob's part of the output to the first component of the direct sum space and Charlie's to the second component. Hence, the upper bound in eq. (8.4) becomes:

$$\mathbb{P}\Big((\mathsf{P},\mathsf{B},\mathsf{C})\text{ win}\Big) \leqslant \sup_{M}\left\|\mathop{\mathbb{E}}_{m,k}\left[d\cdot\rho_{m,k}^{\top}\otimes M_{m|k}\otimes M_{m|k}\right]\right\|_{\text{op}}, \tag{8.5}$$

where the supremum is taken over all PVMs $\{M_{i|k}\}$.

## 8.2 Candidate Scheme

In this section, we propose a new encryption scheme based on the Clifford algebra. In the next section, we show that it allows us to recover previously known results for $K = 2$, and we provide strong numerical evidence that this scheme achieves the unclonable security for large $K$.

### 8.2.1 *Clifford Algebra*

For our unclonable encryption scheme, we use the structure of the *Clifford algebra* [DL03; Lou01]. It is widely used in many areas of quantum information theory, including for a generalization of the Bloch sphere representation [Die06; WW08], in operator algebras theory [BN18; Pis03], and in non-local games [Ost16; Slo11].

[2]We do not need to consider the cases where $i \neq j$ because then the winning probability of the adversaries is zeros.

**Clifford Algebra.** For any integer $n \in \mathbb{N}$, the free real associative algebra generated by $\Gamma_1, \dots, \Gamma_n$, subject to the following anti-commutation relations:

$$\{\Gamma_i, \Gamma_j\} \coloneqq \Gamma_i\Gamma_j + \Gamma_j\Gamma_i = 2\,\delta_{ij}\,\mathbb{I}\,,$$

is called the *Clifford algebra*, and denoted $\mathbf{CL}_n$. Any irreducible representation of the Clifford algebras $\mathbf{CL}_{2n}$ and $\mathbf{CL}_{2n+1}$ can be constructed explicitly by the following $2n+1$ Pauli string operators acting on $\mathbb{C}^{2^n}$, called the Jordan-Wigner transformation [JW93]:

$$\begin{aligned}
\sigma_{n,2i-1} &= \sigma_x^{\otimes(i-1)} \otimes \sigma_y \otimes \mathbb{I}^{\otimes(n-i)}, \quad i \in \{1, \dots, n\}\,,\\
\sigma_{n,2i} &= \sigma_x^{\otimes(i-1)} \otimes \sigma_z \otimes \mathbb{I}^{\otimes(n-i)}, \quad i \in \{1, \dots, n\}\,,\\
\sigma_{n,2n+1} &= \sigma_x^{\otimes n}\,,
\end{aligned}$$

These operators are traceless, and as any Hermitian unitary, they have spectrum included in $\{\pm 1\}$. Mapping the generators $\Gamma_1, \dots, \Gamma_{2n}$ to $\sigma_{n,1}, \dots, \sigma_{n,2n}$ gives the irreducible representation of the even Clifford algebra $\mathbf{CL}_{2n}$. Mapping the generators $\Gamma_1, \dots, \Gamma_{2n+1}$ to $\sigma_{n,1}, \dots, \sigma_{n,2n+1}$ or $\sigma_{n,1}, \dots, \sigma_{n,2n}, -\sigma_{n,2n+1}$ gives the two inequivalent irreducible representations of the odd Clifford algebra $\mathbf{CL}_{2n+1}$. Below, we will consider the following security parameter $\lambda$ and number of keys $K$:

$$\lambda = n \qquad \text{and} \qquad K = 2\lambda \text{ or } 2\lambda + 1\,.$$

**Operator Norm of the Sum.** An important property of the Clifford algebra is that $2n$ Hermitian anti-commuting operators $\Gamma_1, \dots, \Gamma_{2n}$ can be viewed as orthogonal vectors, for the normalized Frobenius inner product, forming a basis for the $2n$-dimensional real vector space they span. Indeed, given $u, v \in \mathbb{R}^{2n}$, we have:

$$\begin{aligned}
\left\langle \sum_{i=1}^{2n} u_i\Gamma_i, \sum_{j=1}^{2n} v_j\Gamma_j \right\rangle &= \frac{1}{2n}\left(\sum_{i=1}^{2n} u_i v_i \underbrace{\mathrm{Tr}\big(\Gamma_i^2\big)}_{=2n} + \sum_{\substack{i,j=1\\ i<j}}^{2n} (u_i v_j - u_j v_i) \underbrace{\mathrm{Tr}\big(\Gamma_i\Gamma_j\big)}_{=0}\right)\\
&= \langle u, v\rangle\,.
\end{aligned}$$

But, given any unit vector $v \in \mathbb{R}^{2n}$, the operator $\sum_{i=1}^{2n} v_i \Gamma_i$ is a Hermitian unitary. Therefore, it yields:

$$\forall v \in \mathbb{R}^{2n}\,, \qquad \left\| \sum_{i=1}^{2n} v_i \Gamma_i \right\|_{\text{op}} = \|v\|_2\,. \tag{8.6}$$

In particular, by taking the all-ones vector $v = (1, .., 1)$, we get:

$$\left\| \sum_{i=1}^{2n} \Gamma_i \right\|_{\text{op}} = \sqrt{2n}\,.$$

### 8.2.2 *Definition of the Scheme*

Now, based on the Clifford algebra (Section 8.2.1), we present our candidate scheme for the (weak) unclonable bit problem:

**Definition 8.1** (Candidate Unclonable Bit Encryption) — *Let* $\lambda \in \mathbb{N}$. *Consider the following scheme:*

- $\mathsf{Gen}(1^\lambda)$ *samples a key* $k \in \{1, .., K\}$ *uniformly at random, with* $K = 2\lambda$ *(even) or* $K = 2\lambda + 1$ *(odd);*
- $\mathsf{Enc}_k(m)$ *is described in Algorithm 8.1;*
- $\mathsf{Dec}_k(\rho)$ *is described in Algorithm 8.2.*

---

**Algorithm 8.1:** The encryption $\mathsf{Enc}(m, k)$.

**Input** : A message $m \in \{0, 1\}$ and a key $k \in \{1, .., K\}$.
**Output:** A quantum state $\rho_{m,k}$ acting on $\mathbb{C}^d$.
Compute $\rho_{m,k} = \frac{2}{d} \frac{\mathbb{I}_d + (-1)^m \Gamma_k}{2}$, the normalized projector onto the $(-1)^m$-eigenspace of $\Gamma_k$ ;
**return** $\rho_{m,k}$

---

**Algorithm 8.2:** The decryption $\mathsf{Dec}(\rho, k)$.

**Input** : A quantum state $\rho$ acting on $\mathbb{C}^d$ and a key $k \in \{1, .., K\}$.
**Output:** A message $m \in \{0, 1\}$.
Measure $\rho$ in the eigenbasis of $\Gamma_k$, with the PVM $\{\frac{1}{2}(\mathbb{I}_d + (-1)^i \Gamma_k)\}_i$. Call the outcome $m$ ;
**return** $m$

---

**Remark 8.2 —** As mentioned in Section 8.2.1, when $\lambda$ is odd, we have the choice between two irreducible representations. We choose $\sigma_{\lambda,1}, \ldots, \sigma_{\lambda,2\lambda+1}$ although the other also yield a valid scheme.

**Remark 8.3 —** The correctness is immediate since the operators $\frac{\mathbb{I}_d+\Gamma_k}{2}$ and $\frac{\mathbb{I}_d-\Gamma_k}{2}$ are orthogonal to each other. Indeed, the measurement of $\rho_{m,k}$ is:

$$\mathrm{Tr}\left[\frac{\mathbb{I}_d + (\text{-}1)^i\Gamma_k}{2}\,\rho_{m,k}\,\frac{\mathbb{I}_d + (\text{-}1)^i\Gamma_k}{2}\right] = \begin{cases} 1 & \text{if } i = m\,, \\ 0 & \text{otherwise}\,, \end{cases}$$

hence $\mathbb{P}\Big[\mathsf{Dec}\big(\mathsf{Enc}(m,k),k\big) = m\Big] = 1$.

**Remark 8.4 —** When $K = 2$, we have $\Gamma_1 \coloneqq \sigma_x$ and $\Gamma_2 \coloneqq \sigma_z$, and then:

$$\frac{\mathbb{I}_d + (\text{-}1)^m\Gamma_1}{2} = H|m\rangle\langle m|H^* \qquad \text{and} \qquad \frac{\mathbb{I}_d + (\text{-}1)^m\Gamma_2}{2} = |m\rangle\langle m|\,,$$

which is exactly the encryption used by the scheme defined in [BL20]. It should also be noted that in this case, the state $\rho_{m,k}$ is pure. However, it is no longer pure for larger $K$, matching the requirement from the impossibility results of [Ana+22; MST21].

### 8.2.3 *Conjecture*

Recall the upper bound on the winning probability for the no-cloning game from eq. (8.5):

$$\mathbb{P}[\text{win}] \leqslant \sup_M \left\| \mathop{\mathbb{E}}_{m,k}\left[d \cdot \rho_{m,k}^\top \otimes M_{m|k} \otimes M_{m|k}\right]\right\|_{\text{op}}\,,$$

where the supremum is taken over all PVMs $\{M_{i|k}\}_i$. For all the keys $k \in \{1,..,K\}$, the PVMs $\{M_{i|k}\}_i$ are binary measurement operators of dimension $D$. We can write:

$$M_{i|k} = \frac{\mathbb{I}_D + (\text{-}1)^i\,U_k}{2}\,,$$

for some Hermitian unitaries observables $U_k := M_{0|k} - M_{1|k}$. Then the upper bound becomes:

$$\begin{aligned}
\mathbb{P}[\text{win}] &\leqslant \sup_{\{U_k\}} \left\| \mathop{\mathbb{E}}_{m,k} \left[ d \cdot \frac{2}{d} \frac{\mathbb{I}_d + (-1)^m \Gamma_k}{2} \otimes \frac{\mathbb{I}_D + (-1)^m U_k}{2} \otimes \frac{\mathbb{I}_D + (-1)^m U_k}{2} \right] \right\|_{\text{op}} \\
&= \sup_{\{U_k\}} \frac{1}{2K} \left\| \sum_{m,k} d \cdot \frac{2}{d} \frac{\mathbb{I}_d + (-1)^m \Gamma_k}{2} \otimes \frac{\mathbb{I}_D + (-1)^m U_k}{2} \otimes \frac{\mathbb{I}_D + (-1)^m U_k}{2} \right\|_{\text{op}} \\
&= \sup_{\{U_k\}} \frac{1}{2K} \left\| \frac{1}{4} \sum_{m,k} \Big( \mathbb{I}_d \otimes \mathbb{I}_D \otimes \mathbb{I}_D \right. \\
&\qquad + (-1)^m \left( \Gamma_k \otimes \mathbb{I}_D \otimes \mathbb{I}_D + \mathbb{I}_d \otimes U_k \otimes \mathbb{I}_D + \mathbb{I}_d \otimes \mathbb{I}_D \otimes U_k \right) \\
&\qquad + (-1)^{2m} \left( \Gamma_k \otimes U_k \otimes \mathbb{I}_D + \Gamma_k \otimes \mathbb{I}_D \otimes U_k + \mathbb{I}_d \otimes U_k \otimes U_k \right) \\
&\qquad \left. + (-1)^{3m}\, \Gamma_k \otimes U_k \otimes U_k \Big) \right\|_{\text{op}} \\
&= \frac{1}{4} + \frac{1}{4K} \sup_{\{U_k\}} \left\| \sum_k \Big( \Gamma_k \otimes (U_k \otimes \mathbb{I}_D + \mathbb{I}_D \otimes U_k) + \mathbb{I}_d \otimes U_k \otimes U_k \Big) \right\|_{\text{op}},
\end{aligned} \tag{8.7}$$

where the supremum is taken over all Hermitian unitaries $U_k \in \mathcal{U}(\mathbb{C}^D)$. Note that a naive triangular inequality of the operator norm yields the upper bound $\mathbb{P}[\text{win}] \leqslant \frac{1}{4} + \frac{3K}{4K} = 1$, which is trivial. We formulate the following conjecture in terms of the operator norm:

> **Conjecture 8.5 —** *Let $\Gamma_1, \ldots, \Gamma_K$ be any pairwise anti-commuting Hermitian unitaries (Hermitian representation of some Clifford algebra) of dimension $d$. Then, for all Hermitian unitaries $\{U_k\}$ of dimension $D$, the following upper bound holds:*
>
> $$\left\| \sum_{k \in \{1,..,K\}} \Big( \Gamma_k \otimes (U_k \otimes \mathbb{I}_D + \mathbb{I}_D \otimes U_k) + \mathbb{I}_d \otimes U_k \otimes U_k \Big) \right\|_{\text{op}} \leqslant K + 2\sqrt{K}\,. \tag{8.8}$$

Under this conjecture, the upper bound in eq. (8.7) gives:

$$\mathbb{P}[\text{win}] \leqslant \frac{1}{2} + \frac{1}{2\sqrt{K}}, \tag{8.9}$$

with the asymptotic limit $\lim_{K\to\infty}\mathbb{P}[\text{win}]=\frac{1}{2}$ at a rate $\mathcal{O}(\frac{1}{\sqrt{K}})$. This would prove that the Clifford unclonable encryption is secure for the (weak) unclonable encryption security.[3]

### 8.2.4 Basic Properties

In this section, we prove basic properties related to our conjecture. First, we prove that the identity matrix saturates the bound of the conjecture, which means that if the wanted upper bound holds, then it is tight (Proposition 8.6). Then we prove that the conjecture is true if the adversary's strategy is reduced to commuting operators $U_k$ (Proposition 8.8) or if they only use product states (Proposition 8.10).

In what follows, we will refer to the family of operators involved in Conjecture 8.5 as:

$$W_K(U_1,\dots,U_K) := \sum_{k\in\{1,..,K\}} \Big(\Gamma_k\otimes(U_k\otimes\mathbb{I}_D+\mathbb{I}_D\otimes U_k)+\mathbb{I}_d\otimes U_k\otimes U_k\Big)\,. \quad (8.10)$$

We simply write $W_K := W_K(U_1,\dots,U_K)$ if there are no ambiguities the Hermitian unitaries $U_1,\dots,U_K$.

**Proposition 8.6** (Tight Upper Bound) — *If we take $U_k=\mathbb{I}_D$ for all $k\in\{1,..,K\}$, then $\|W_K\|_{\text{op}}=K+2\sqrt{K}$. As a consequence:*

$$\sup_{\{U_k\}}\|W_K\|_{\text{op}} \geqslant K+2\sqrt{K}\,,$$

*and the upper bound in Conjecture 8.5 can only be tight.*

*Proof.* We have $\|\sum_i\Gamma_i+I\|_{\text{op}} = \|\sum_i\Gamma_i\|_{\text{op}}+1$ because the spectrum of $\sum_i\Gamma_i$ is symmetric [Hel+19]. Hence, using eq. (8.6) we obtain the value $K+2\sqrt{K}$. ■

**Remark 8.7** (Low Rank Operators Also Saturate the Bound) — Actually, the value $K+2\sqrt{K}$ is also achieved by taking any $U_k\in\mathbb{C}^D$ that is a

[3] Note that the convergence rate is quadratic and not negligible, so the conjecture cannot be used to prove strong unclonable encryption security.

projector onto the 1-eigenspace of rank at most $r \leqslant \frac{D-1}{K}$. Indeed, write:

$$U_k = 2\left( \sum_{i=1}^{r} |u_k^{(i)}\rangle\langle u_k^{(i)}| \right) - \mathbb{I}_D \,,$$

for all $k \in \{1, .., K\}$ and for some pure states $|u_k^{(i)}\rangle \in \mathbb{C}^D$ (or the zero vector). Then, using the condition $D \geqslant K\,r+1$, there exists a quantum state $|u^\perp\rangle \in \mathbb{C}^D$ that is orthogonal to all the $|u_k^{(i)}\rangle$ for $i \in [r]$ and $k \in \{1, .., K\}$. Note that this vector satisfies $\langle u^\perp | U_k | u^\perp\rangle = 2(\sum_i 0) - 1 = -1$ for all $k$. It follows that we have:

$$\begin{aligned}
&\left\| W_K \right\|_{\mathrm{op}} \\
&= \sup_{\substack{|\psi\rangle \in \mathbb{C}^d \otimes \mathbb{C}^D \otimes \mathbb{C}^D, \\ \||\psi\rangle\| = 1}} \left| \langle \psi \,|\, \textstyle\sum_k \left( \Gamma_k \otimes U_k \otimes \mathbb{I}_D + \Gamma_k \otimes \mathbb{I}_D \otimes U_k + \mathbb{I}_d \otimes U_k \otimes U_k \right) |\, \psi \rangle \right| \\
&\geqslant \sup_{\substack{|a\rangle \in \mathbb{C}^d, \\ \||a\rangle\| = 1}} \left| \langle a \otimes u^\perp \otimes u^\perp \,|\, \textstyle\sum_k \left( \Gamma_k \otimes U_k \otimes \mathbb{I}_D + \Gamma_k \otimes \mathbb{I}_D \otimes U_k + \mathbb{I}_d \otimes U_k \otimes U_k \right) \right. \\
&\qquad\qquad \left. |\, a \otimes u^\perp \otimes u^\perp \rangle \right| \\
&= \sup_{\substack{|a\rangle \in \mathbb{C}^d, \\ \||a\rangle\| = 1}} \left| 2 \langle a \,|\, \textstyle\sum_k -\Gamma_k \,|\, a \rangle + K \right| \;=\; 2 \left\| \textstyle\sum_k -\Gamma_k \right\|_{\mathrm{op}} + K \\
&= \; 2 \left\| \begin{bmatrix} -1 \\ \vdots \\ -1 \end{bmatrix} \right\|_2 + K \;=\; 2\sqrt{K} + K \,,
\end{aligned}$$

using the symmetry of the spectrum of $\sum_k \Gamma_k$ [Hel+19] in the third last equality and then using eq. (8.6) in the second last one. Hence the claimed result.

**Proposition 8.8** (True for Commuting Operators) — *If the operators $U_k$ commute, then Conjecture 8.5 holds.*

*Proof.* If the operators $U_k$ commute, then they are diagonalizable in a common basis. But their eigenvalues are $\pm 1$ because they are Hermitian and unitaries, so we may assume that they are of the form

$$U_k \simeq \begin{pmatrix} \pm 1 & & \\ & \ddots & \\ & & \pm 1 \end{pmatrix} .$$

Then, using the triangular inequality, we obtain:

$$\begin{aligned}\|W_K\|_{\text{op}} &\leqslant \left\|\sum_{k=1}^{K}\Gamma_k\otimes(\pm1)\otimes 1\right\|_{\text{op}} + \left\|\sum_{k=1}^{K}\Gamma_k\otimes 1\otimes(\pm1)\right\|_{\text{op}} + \sum_{k=1}^{K}\left\|\begin{pmatrix}\pm1 & & \\ & \ddots & \\ & & \pm1\end{pmatrix}\right\|_{\text{op}}\\ &= \sqrt{K}+\sqrt{K}+K\,,\end{aligned}$$

because $\left\|\sum_{k=1}^{K}\Gamma_k\otimes 1\otimes(\pm1)\right\|_{\text{op}} = \left\|\sum_{k=1}^{K}\Gamma_k\right\|_{\text{op}} = \sqrt{K}$ using eq. (8.6). ■

We give a connection between the operator norm of $W_K$ and the supremum of $\langle\psi|W_K|\psi\rangle$:

**Lemma 8.9** (Expression of the Operator Norm) — *For all $K\in\mathbb{N}$, we have the two following equalities:*

$$\sup_U \sup_{\|\psi\|=1}\langle\psi|W_K|\psi\rangle \overset{(1)}{=} \sup_U \sup_{\|\psi\|=1}|\langle\psi|W_K|\psi\rangle| \overset{(2)}{=} \sup_U \|W_K\|_{\text{op}}\,,$$

*where the suprema are taken over all Hermitian unitaries $U_k\in\mathcal{U}(\mathbb{C}^D)$.*

*Proof.* Let $W'_K := W_K + K\cdot\mathbb{I}_d\otimes\mathbb{I}_D\otimes\mathbb{I}_D$. Then, we can write:

$$W'_K = \sum_k\Big(\Gamma_k\otimes\mathbb{I}_D\otimes\mathbb{I}_D + \mathbb{I}_d\otimes U_k\otimes\mathbb{I}_D\Big)\Big(\Gamma_k\otimes\mathbb{I}_D\otimes\mathbb{I}_D+\mathbb{I}_d\otimes\mathbb{I}_D\otimes U_k\Big)\,.$$

Each $\Gamma_k$ is a Hermitian unitary with a symmetric spectrum $\pm1$. There exist an invertible matrix $H_k$ that can diagonalize $\Gamma_k$, *i.e.*

$$\Gamma_k = H_k\begin{pmatrix}-\mathbb{I}_{d/2} & 0\\ 0 & \mathbb{I}_{d/2}\end{pmatrix}H_k^{-1}\,.$$

Thus, under those changes of basis, the operator $W'_K$ is expressed as follows:

$$W'_K = \sum_k \tilde{H}_k\begin{pmatrix}(\mathbb{I}-U_k)^{\otimes2} & 0\\ 0 & (\mathbb{I}+U_k)^{\otimes 2}\end{pmatrix}\tilde{H}_k^{-1}\,,$$

with $\tilde{H}_k := H_k\otimes\mathbb{I}_D\otimes\mathbb{I}_D$. Since for all $U_k$, the inequalities $-\mathbb{I}_D\preccurlyeq U_k\preccurlyeq\mathbb{I}_D$ hold, the two operators $\mathbb{I}+U_k$ and $\mathbb{I}-U_k$ are both positive semi-definite, so is each term of the sum, and thus $W'_K\succcurlyeq 0$.

The first equality holds since,

$$\sup_{\|\psi\|=1}|\langle\psi|W_K|\psi\rangle| = \max\left\{\sup_{\|\psi\|=1}\langle\psi|W_K|\psi\rangle, -\inf_{\|\psi\|=1}\langle\psi|W_K|\psi\rangle\right\},$$

but because $W'_K \succcurlyeq 0$, we have that $-\sup_U \inf_{\|\psi\|=1}\langle\psi|W_K|\psi\rangle = K$, and when all $U_k$ are equal to $\mathbb{I}_D$, we already know that $\sup_{\|\psi\|=1}\langle\psi|W_K|\psi\rangle = K + 2\sqrt{K}$ from Proposition 8.6.

Finally, the second equality is always true for Hermitian operators [Zim90, Lemma 3.2.4], and $W_K$ is a Hermitian operator, as a sum of tensor products of Hermitian operators $\Gamma_k$ and $U_k$. ∎

Hence, the three upper bounds eqs. (8.2) to (8.4) of the winning probability of the no-cloning game for three adversaries $(\mathsf{P}, \mathsf{B}, \mathsf{C})$, are all equal, and Conjecture 8.5 can also be stated as the largest eigenvalue of the operator $W_K$.

From Lemma 8.9, see that an equivalent formulation of Conjecture 8.5 is:

$$\sup_U \sup_{\|\psi\|=1} \langle\psi|W_K|\psi\rangle \leqslant K + 2\sqrt{K}\,,$$

where the quantum state $|\psi\rangle$ is taken in $\mathbb{C}^d \otimes \mathbb{C}^D \otimes \mathbb{C}^D$. We show that the conjecture holds in the restricted case where $|\psi\rangle$ is a product state between Alice and {Bob, Charlie}:

**Proposition 8.10** (True for Product State) — *If the state is of the form* $|\psi\rangle = |\alpha_{\mathsf{A}}\rangle \otimes |\varphi_{\mathsf{BC}}\rangle$*, then:*

$$\sup_U \langle\psi|W_K|\psi\rangle \leqslant K + 2\sqrt{K}\,.$$

*Proof.* We have:

$$\begin{aligned}
&\sum_{k=1}^{K} \big\langle \alpha_{\mathsf{A}} \otimes \varphi_{\mathsf{BC}} \,\big|\, \Big(\Gamma_k \otimes (U_k \otimes \mathbb{I}_D + \mathbb{I}_D \otimes U_k) + \mathbb{I}_d \otimes U_k \otimes U_k\Big) \,\big|\, \alpha_{\mathsf{A}} \otimes \varphi_{\mathsf{BC}} \big\rangle \\
&= \sum_{k=1}^{K} \big\langle \alpha_{\mathsf{A}} \,\big|\, \Gamma_k \,\big|\, \alpha_{\mathsf{A}} \big\rangle \underbrace{\big\langle \varphi_{BC} \,\big|\, (U_k \otimes \mathbb{I}_D + \mathbb{I}_D \otimes U_k) \,\big|\, \varphi_{BC} \big\rangle}_{=:c_k} \\
&\qquad\qquad + \sum_{k=1}^{K} \big\langle \alpha_{\mathsf{A}} \otimes \varphi_{\mathsf{BC}} \,\big|\, \mathbb{I}_d \otimes U_k \otimes U_k \,\big|\, \alpha_{\mathsf{A}} \otimes \varphi_{\mathsf{BC}} \big\rangle \\
&= \sum_{k=1}^{K} \big\langle \alpha_{\mathsf{A}} \,\big|\, c_k \Gamma_k \,\big|\, \alpha_{\mathsf{A}} \big\rangle + \sum_{k=1}^{K} \underbrace{\big\langle \alpha_{\mathsf{A}} \otimes \varphi_{\mathsf{BC}} \,\big|\, \mathbb{I}_d \otimes U_k \otimes U_k \,\big|\, \alpha_{\mathsf{A}} \otimes \varphi_{\mathsf{BC}} \big\rangle}_{\leqslant 1} \\
&\leqslant \big\|(c_1, \ldots, c_K)\big\|_2 + K \\
&= 2\sqrt{K} + K \,,
\end{aligned}$$

using eq. (8.6) in the second last line. ■

**Remark 8.11 —** When $|\psi\rangle$ is a product state in all its tensors, *i.e.* when $|\psi\rangle = |\alpha_{\mathsf{A}} \otimes \beta_{\mathsf{B}} \otimes \gamma_{\mathsf{C}}\rangle$, then this result is equivalent to the one in Proposition 8.8.

### 8.2.5 *Indistinguishability*

Analogously to the two variants of unclonable-indistinguishability security presented in Definition 5.12, namely those with a strong convergence rate and those with an arbitrary convergence rate, we can similarly define two corresponding notions of indistinguishability security.

The *strong indistinguishability security* requires that any adversary, upon receiving the encryption of a message $m$ randomly chosen from a pair of messages, cannot predict the value of $m$ with a probability greater than negligibly close to $\frac{1}{2}$. If the adversary's probability of correctly predicting the encrypted message converges to $\frac{1}{2}$ at any arbitrary rate, we refer to this security notion as simply *indistinguishability security*.

In this section, we prove that our candidate scheme Definition 8.1 satisfies the latter indistinguishability security. Indeed, the success probability of such an adversary is bounded by:

$$\mathbb{P}[\text{win}] \leqslant \sup_{\substack{\Phi \\ \{M_0,M_1\}}} \mathop{\mathbb{E}}_{\substack{m\in\{0,1\} \\ k\leftarrow\mathsf{Gen}(1^\lambda)}} \mathrm{Tr}\left[\Phi(\rho_{m,k})M_m\right],$$

with $\rho_{m,k} := \mathsf{Enc}(m,k)$, and where the expected values are taken with respect to the uniform measures, and the supremum is taken over all CPTP maps $\Phi : \mathcal{B}(\mathcal{H}_d) \to \mathcal{B}(\mathcal{H}_D)$ (for all finite-dimensional Hilbert spaces $\mathcal{H}_D$), and all binary POVMs $\{M_0, M_1\}$. We now follow the same sequence of equations as presented in eq. (5.3) to derive the result, which is outlined below:

$$\mathbb{P}[\text{win}] \leqslant \sup_M \left\| \mathop{\mathbb{E}}_{m,k} \left[d \cdot \rho_{m,k}^\top \otimes M_m\right] \right\|_{\text{op}},$$

where the supremum is taken over all binary POVMs $\{M_0, M_1\}$. We proceed by adapting eq. (8.7), starting from the observable form $M_i = \frac{\mathbb{I}_D + (-1)^i U}{2}$ for some Hermitian unitary $U$, to derive the following inequalities,

$$\begin{aligned}
\mathbb{P}[\text{win}] &\leqslant \sup_U \left\| \mathop{\mathbb{E}}_{m,k} \left[ d \cdot \frac{2}{d} \frac{\mathbb{I}_d + (-1)^m \Gamma_k}{2} \otimes \frac{\mathbb{I}_D + (-1)^m U}{2} \right] \right\|_{\text{op}} \\
&\leqslant \sup_U \frac{1}{2K} \left\| \frac{1}{2} \sum_{m,k} \Big( \mathbb{I}_d \otimes \mathbb{I}_D + (-1)^m \big( \Gamma_k \otimes \mathbb{I}_D + \mathbb{I}_d \otimes U \big) + (-1)^{2m}\, \Gamma_k \otimes U \Big) \right\|_{\text{op}} \\
&\leqslant \frac{1}{2} + \frac{1}{2K} \sup_U \left\| \sum_k \Gamma_k \otimes U \right\|_{\text{op}},
\end{aligned}$$

where the supremum is taken over all Hermitian unitaries $U_k \in \mathcal{U}(\mathbb{C}^D)$. By applying the sub-multiplicative property of the operator norm (*i.e.* , $\|A \otimes B\|_{\text{op}} \leqslant \|A\|_{\text{op}} \cdot \|B\|_{\text{op}}$ for all $A$ and $B$), along with the fact that the operator norm of a unitary is one, and the inequality $\|\sum_k \Gamma_k\|_{\text{op}} = \sqrt{K}$, we have the following upper bound:

$$\mathbb{P}[\text{win}] \leqslant \frac{1}{2} + \frac{1}{2\sqrt{K}}.$$

This ensures a quadratic convergence rate for the indistinguishability security of our Clifford encryption scheme.

## 8.3 Analytical and Numerical Results

Although we have been unable to fully prove Conjecture 8.5, we present in this section several analytical and numerical results on the first values of $K \in \mathbb{N}$.

Specifically, we prove the conjecture for $K \leqslant 7$ (Sections 8.3.1 and 8.3.2), provide numerical confirmation for $K \leqslant 17$ (Section 8.3.3), and present numerical evidence for $K = 18$ (Section 8.3.4). Additionally, we establish a weaker bound than the conjecture (Theorem 8.17), which holds this time for all $K \in \mathbb{N}$.

**Inapplicability of the Triangular Inequality.** As we saw in Section 8.2.1, a naive triangular inequality of the operator norm in eq. (8.7) yields a trivial upper bound $\mathbb{P}[\text{win}] \leqslant \frac{1}{4} + \frac{3K}{4K} = 1$. In light of the property of eq. (8.6), one might consider that a slightly more refined triangular inequality could be sufficient to address Conjecture 8.5, and that

$$\Big\| \sum_{k\in\{1,..,K\}} \Gamma_k \otimes \big(U_k \otimes \mathbb{I}_D + \mathbb{I}_D \otimes U_k\big) \Big\|_{\text{op}} + \Big\| \sum_{k\in\{1,..,K\}} \mathbb{I}_d \otimes U_k \otimes U_k \Big\|_{\text{op}},$$

would be smaller than $K + 2\sqrt{K}$, but this is not true for $K > 2$. One should notice that $\|\sum_{k\in\{1,..,K\}} \Gamma_k \otimes U_k\|_{\text{op}} \leqslant K$, with equality by taking all $U_k = \Gamma_k$. With those $U_k$, the tensor products $U_k \otimes U_k$ are pairwise commuting and thus $\|\sum_{k\in\{1,..,K\}} \mathbb{I}_d \otimes U_k \otimes U_k\|_{\text{op}} = K$, but the left-hand side of the triangular inequality $\|\sum_{k\in\{1,..,K\}} \Gamma_k \otimes (U_k \otimes \mathbb{I}_D + \mathbb{I}_D \otimes U_k)\|_{\text{op}}$ is in general larger than $2\sqrt{K}$, with first values:

$$2\sqrt{2},\ 2\sqrt{4},\ 2\sqrt{6},\ 2\sqrt{9},\ 2\sqrt{12},\ 2\sqrt{16},\ 2\sqrt{20},\ 2\sqrt{25},\ \ldots$$

In comparison, when all $U_k = \Gamma_k$, the complete operator norm $\|\sum_{k\in\{1,..,K\}} \Gamma_k \otimes (U_k \otimes \mathbb{I}_D + \mathbb{I}_D \otimes U_k) + \mathbb{I}_d \otimes U_k \otimes U_k\|_{\text{op}}$ is smaller than $K + 2\sqrt{K}$, with first values:

$$4,\ 3,\ 6,\ 7,\ 8,\ 9,\ 10,\ 11,\ \ldots$$

### 8.3.1 Elementary Proofs for K=2

We present two distinct proofs showing that Conjecture 8.5 is true for $K = 2$, a first one using the equivalence with the [BL20] scheme, and another one based on a Sum-of-Squares method. We will demonstrate later that a Sum-of-Squares certificate applies to larger values of $K$. A third proof can also be obtained by exploiting the anti-commutation of the two pairs of matrices in $W_2$ and the recent results in [GHG23; HO21; MH24; XSW24] regarding uncertainty relations.

#### 8.3.1.1 First Proof, with the BB84 MoE Game

As we saw in Section 8.2.1, for $K = 2$ our candidate scheme is the same as the scheme defined in [BL20], thus the upper bound of the winning probability of the no-cloning game for three adversaries $(\mathsf{P}, \mathsf{B}, \mathsf{C})$ is the same as in [BL20; Tom+13], *i.e.* $\frac{1}{2} + \frac{1}{2\sqrt{2}}$. Thus, by Lemma 8.9, we have:

$$\mathbb{P}[\text{win}] \leqslant \frac{1}{4}+\frac{1}{4K} \sup_U \Big\| \sum_k \Big(\Gamma_k \otimes (U_k \otimes \mathbb{I}_D + \mathbb{I}_D \otimes U_k) + \mathbb{I}_d \otimes U_k \otimes U_k\Big) \Big\|_{\text{op}} = \frac{1}{2}+\frac{1}{2\sqrt{2}} .$$

This is equivalent to $\|W_2\|_{\text{op}} \leqslant 2 + 2\sqrt{2}$.

#### 8.3.1.2 Second Proof, with Sum-of-Squares

From eq. (8.2), the upper bound of the winning probability of the no-cloning game for three adversaries $(\mathsf{P}, \mathsf{B}, \mathsf{C})$ is

$$\mathbb{P}[\text{win}] \leqslant \sup_{\psi, B, C} \langle \psi | \mathop{\mathbb{E}}_{m,k} \left[ d \cdot \rho_{m,k}^{\top} \otimes B_{m|k} \otimes C_{m|k} \right] |\psi\rangle ,$$

where the supremum is taken over all $\|\psi\| = 1$, all families of PVM $\{B_{i|k}\}$ and $\{C_{j|k}\}$, as well as their respective dimensions. Following the same steps as Section 8.2, we found

$$\mathbb{P}[\text{win}] \leqslant \frac{1}{4}+\frac{1}{4K} \sup_{\psi, B, C} \langle \psi | \sum_k \Big(\Gamma_k \otimes (B_k \otimes \mathbb{I}_D + \mathbb{I}_D \otimes C_k) + \mathbb{I}_d \otimes B_k \otimes C_k\Big) |\psi\rangle , \tag{8.11}$$

where the supremum is now taken over all families of observables $\{B_k\}$ and $\{C_k\}$. Note that this time, we do not assume the adversaries $(\mathsf{P}, \mathsf{B}, \mathsf{C})$ to be symmetric, *i.e.* $\mathsf{B}$ and $\mathsf{C}$ may have different observables. The last part of the upper bound eq. (8.11) can be stated as the optimization problem:

$$\begin{aligned} \sup_{\psi,B,C} \quad & \langle\psi| \sum_k \Big(\Gamma_k \otimes (B_k \otimes \mathbb{I}_D + \mathbb{I}_D \otimes C_k) + \mathbb{I}_d \otimes B_k \otimes C_k\Big)|\psi\rangle\,, \\ \text{subject to} \quad & \cdot\ \|\psi\| = 1\,, \\ & \cdot\ B_i^* = B_i \quad \cdot\ C_i^* = C_i \quad \cdot\ B^2 = C^2 = \mathbb{I}_D \quad \forall i\,. \end{aligned} \tag{8.12}$$

This problem can be relaxed through the use of what is called commuting operator strategies, in which the tensor product structure between Alice's and Bob's operators is replaced by the assumption that these operators commute:

$$\begin{aligned} \sup_{\psi,b,c} \quad & \langle\psi| \sum_k \Big(\Gamma_k \otimes (b_k + c_k) + \mathbb{I}_d \otimes b_k \cdot c_k\Big)|\psi\rangle\,, \\ \text{subject to} \quad & \cdot\ \|\psi\| = 1\,, \\ & \cdot\ b_i^* = b_i \quad \cdot\ c_i^* = c_i \quad \cdot\ b_i^2 = c_i^2 = \mathbb{I}_D \quad \forall i\,, \\ & \cdot\ [b_i, c_j] = 0 \quad \forall i,j\,. \end{aligned} \tag{8.13}$$

An optimal value for the problem eq. (8.12) is immediately a lower bound for the problem eq. (8.13) by taking $b_k := B_k \otimes \mathbb{I}_D$ and $c_k := \mathbb{I}_D \otimes C_k$. The question of the equality of the two optimization problems eqs. (8.12) and (8.13) (and in general of the tensor-based versus the commuting-based models) was a long-standing problem that was refuted only recently by [Ji+21] (Remark 3.4). However, in the case of finite-dimensional Hilbert spaces, the equality holds as an inductive consequence of Tsirelson's theorem [RXL24; Tsi93].

If the optimal value of the problem eq. (8.13) is $K + 2\sqrt{K}$, then the Hermitian operator

$$P_K := (K + 2\sqrt{K}) \cdot \mathbb{I}_d \otimes \mathbb{I}_{D^2} - \sum_k \Big(\Gamma_k \otimes (b_k + c_k) + \mathbb{I}_d \otimes b_k \cdot c_k\Big)\,, \tag{8.14}$$

must be positive semi-definite under the constraints of eq. (8.13). If we can find a family $\{H_i\}_i$ such that $P_k = \sum_i H_i^* H_i$, then the operator $P_K$ would be guaranteed to be positive semi-definite, as a sum of positive semi-definite operators $H_i^* H_i$. If all $H_i$ are also Hermitian, then we can write $P_K = \sum_i H_i^2$. This constitutes the fundamental principle of the Sum-of-Squares (SoS) decomposition technique, which is used to establish bounds on a variety of quantum correlations, including the CHSH Bell inequality [Bel64; CHSH69] and its associated Tsirelson bound [Tsi80]. Find a discussion on SoS decompositions in Section 3.1.3.

Let $P(\mathbf{X})$ be an Hermitian polynomial in non-commutative variables $\mathbf{X} = (X_i)_i$, then $P(\mathbf{X})$ is a positive semi-definite polynomial (*i.e.* $P(\mathbf{X}) \succcurlyeq 0$ for all evaluations of $P$ on matrices $\mathbf{X}$) if and only if $P(\mathbf{X})$ is a Hermitian Sum-of-Squares [Hel02; McC01]:

$$P(\mathbf{X}) = \sum_i H_i(\mathbf{X})^* \, H_i(\mathbf{X}) \, .$$

This result can be used to maximize the operator norm of a non-commutative Hermitian polynomial. If the polynomial is constrained to an archimedean semi-algebraic set (a condition that is satisfied in our case), a hierarchy of converging upper bounds can be obtained based on semi-definite optimization programs (SDP) [HM04]. The dual problem of those SDPs form the Navascués-Pironio-Acín (NPA) hierarchy [NPA08] and correspond to the non-commutative variant of the Lasserre's hierarchy [Las01].

For $K = 2$ such SoS decomposition was already known [BC23b]:

$$\begin{aligned} P_2 \; = \; & \frac{1}{2\sqrt{2}}\big(\sigma_x \otimes b_1 + \sigma_z \otimes c_2 - \sqrt{2}\cdot \mathbb{I}\big)^2 + \frac{1}{2\sqrt{2}}\big(\sigma_x \otimes c_1 + \sigma_z \otimes b_2 - \sqrt{2}\cdot \mathbb{I}\big)^2 \\ & + \frac{1}{2}\big(b_1 - c_1\big)^2 + \frac{1}{2}\big(b_2 - c_2\big)^2 , \end{aligned}$$

where we took $\Gamma_1 \coloneqq \sigma_x$ and $\Gamma_2 \coloneqq \sigma_z$. This implies, using Lemma 8.9, that $\|W_2\|_{\text{op}} \leqslant 2 + 2\sqrt{2}$.

## *8.3.2 Proofs for $K \leqslant 7$ and Asymptotic Upper Bound*

We prove that Conjecture 8.5 is true for all $K \leqslant 7$ in two manners: first using a family of SoS decompositions, then based on its dual SDP problem, the NPA hierarchy.

### 8.3.2.1 First Proof, with Sum-of-Squares

When $K \in \{2, \ldots, 7\}$, the non-negativity certificates for $P_K$ (eq. (8.14)) are given by the parameterized family of SoS in terms of the coefficient $\alpha_K$:

$$P_K \; = \; \frac{K - \sqrt{K}}{2K(K-1)} \sum_{i=1}^{K} \Big( Q_K + (\sqrt{K}+1)\, \Gamma_i \otimes (c_i - b_i) \Big)^2 + \alpha_K \, Q_K^2 \, , \quad (8.15)$$

where $Q_K := \sqrt{K}\,\mathbb{I}\otimes\mathbb{I} - \sum_{j=1}^{K}(\Gamma_j \otimes c_j)$. The values of the real coefficient $\alpha_K$ are determined by solving the equality and are of the following form:

$$\alpha_K = \frac{(3K-2)\sqrt{K} - K^2}{2K(K-1)} \qquad \text{for } K \geqslant 2\,.$$

Notice that the coefficient $\alpha_K$ is positive for $K \leqslant 7$, so eq. (8.15) is a SoS decomposition for $P_2$ through $P_7$, thus providing certificates for the validity of Conjecture 8.5 for $K \leqslant 7$. However, for $K \geqslant 8$, the coefficient $\alpha_K$ is negative, so the decomposition presented in eq. (8.15) no longer provides a valid non-negativity certificate. Notice that it does not exclude the possibility of finding another SoS decomposition valid until larger values of $K$. It is also worth noting that the SoS decomposition given in eq. (8.15) can be readily symmetrized with respect to the variables $b_i$ and $c_i$—the current formulation has been chosen for its simplicity.

#### 8.3.2.2 Second Proof, with NPA Level-1

We introduce two scenario algebras and prove that the optimal value of our conjecture is equivalent to the supremum of the operator norm of a new problem.

**Definition 8.12** (Scenario Algebra) — *The* scenario algebra $\mathcal{A}(K)$ *is the* $*$*-algebra generated by* $b_1, c_1, \ldots, b_K, c_K$ *such that* $b_i^2 = c_i^2 = 1$*,* $b_i^* = b_i$*,* $c_i^* = c_i$*, and* $b_i c_j = c_j b_i$ *for all* $i, j \in \{1, .., K\}$*.*
*The* anticommuting scenario algebra $\mathcal{A}_{ac}(K)$ *is the* $*$*-algebra generated by* $\hat{b}_1, \hat{c}_1, \ldots, \hat{b}_K, \hat{c}_K$ *such that* $\hat{b}_i^2 = \hat{c}_i^2 = 1$*,* $\hat{b}_i^* = \hat{b}_i$*,* $\hat{c}_i^* = \hat{c}_i$*,* $\hat{b}_i\hat{c}_i = \hat{c}_i\hat{b}_i$ *for all* $i$*, and* $\hat{b}_i\hat{c}_j = -\hat{c}_j\hat{b}_i$ *for all* $i \neq j \in \{1, .., K\}$*.*

In our case, we can take, as a representation, $\Gamma_i \otimes B_i \otimes \mathbb{I}$ for $\hat{b}_i$ and $\Gamma_i \otimes \mathbb{I} \otimes C_i$ for $\hat{c}_i$. The game polynomial can be seen as an element $p_K \in \mathcal{M}_d \otimes \mathcal{A}(K)$, with $p_K = \sum_k \Gamma_k \otimes (b_k + c_k) + \mathbb{I} \otimes 1$ so that the winning probability is $\frac{1}{4} + \frac{1}{4K}\sup_\pi \|(\mathrm{id}\otimes\pi)(p_K)\|$, where the supremum is over all finite-dimensional representations of $\mathcal{A}(K)$.

**Proposition 8.13** — *Let* $\hat{p}_K = \sum_k(\hat{b}_k + \hat{c}_k + \hat{b}_k\hat{c}_k) \in \mathcal{A}_{ac}(K)$*. Then,*

$$\sup_{\pi}\|(\mathrm{id}\otimes\pi)(p_K)\|_{\mathrm{op}} = \sup_{\hat{\pi}}\|\hat{\pi}(\hat{p}_K)\|_{\mathrm{op}}\,,$$

*where the suprema are over finite-dimensional representations $\pi, \hat{\pi}$ of $\mathcal{A}(K)$, $\mathcal{A}_{ac}(K)$ respectively.*

*Proof.* Let $\pi$ be a finite-dimensional representation of $\mathcal{A}(K)$. Then, let $\hat{\pi}$ be the representation of $\mathcal{A}_{ac}(K)$ defined by $\hat{\pi}(\hat{b}_k) = \Gamma_k \otimes \pi(b_k)$ and $\hat{\pi}(\hat{c}_k) = \Gamma_k \otimes \pi(c_k)$. It is direct to see that this satisfies the relations of $\mathcal{A}_{ac}(K)$ and $(\mathrm{id} \otimes \pi)(p_K) = \hat{\pi}(\hat{p}_K)$, and hence that $\|(\mathrm{id} \otimes \pi)(p_K)\|_{\mathrm{op}} = \|\hat{\pi}(\hat{p}_K)\|_{\mathrm{op}}$. Taking suprema, we get $\sup_\pi \|(\mathrm{id} \otimes \pi)(p_K)\|_{\mathrm{op}} \leqslant \sup_{\hat{\pi}} \|\hat{\pi}(\hat{p}_K)\|_{\mathrm{op}}$.

For the other direction, let $\hat{\pi}$ be a finite-dimensional representation of $\mathcal{A}_{ac}(K)$. Then, define a representation $\pi$ of $\mathcal{A}(K)$ by $\pi(b_k) = \Gamma_k^\top \otimes \hat{\pi}(\hat{b}_k)$ and $\pi(c_k) = \Gamma_k^\top \otimes \hat{\pi}(\hat{c}_k)$. As above, it is direct to see that this satisfies the relations of $\mathcal{A}(K)$. Let $|\psi\rangle$ be a unit vector such that $\|\hat{\pi}(\hat{p}_K)\|_{\mathrm{op}} = |\langle\psi|\hat{\pi}(\hat{p}_K)\rangle\psi|$, and write $|\Psi_d\rangle \in \mathcal{C}^d \otimes \mathcal{C}^d$ for the maximally entangled state. Then, $\langle\Psi_d|\Gamma_k \otimes \Gamma_k^\top\rangle\Psi_d = 1$, so

$$\begin{aligned} &\|(\mathrm{id} \otimes \pi)(p_K)\|_{\mathrm{op}} \\ \geqslant\ &\left|\left\langle \Psi_d \otimes \psi \,\middle|\, \sum_k \left(\Gamma_k \otimes \Gamma_k^\top \otimes (\hat{\pi}(\hat{b}_k) + \hat{\pi}(\hat{c}_k)) + \mathbb{I}_{d^2} \otimes \hat{\pi}(\hat{b}_k \hat{c}_k)\right) \,\middle|\, \Psi_d \otimes \psi\right\rangle\right| \\ =\ &\left|\langle\psi|\hat{\pi}(\hat{p}_K)\rangle\psi\right| \ = \ \left\|\hat{\pi}(\hat{p}_K)\right\|_{\mathrm{op}}, \end{aligned}$$

giving the inequality $\sup_\pi \|(\mathrm{id} \otimes \pi)(p_K)\|_{\mathrm{op}} \geqslant \sup_{\hat{\pi}} \|\hat{\pi}(\hat{p}_K)\|_{\mathrm{op}}$. ∎

Thus, the winning probability can be found via optimization over representations of $\mathcal{A}_{ac}(K)$. Let the bias of a strategy be $4K\mathbb{P}[\text{win}] - K$. Then the optimal bias is $\beta_K = \sup_{\hat{\pi}} \|\hat{\pi}(\hat{p}_K)\|$. So, the optimal bias is the solution to the optimization

$$\sup_{\psi, b, c} \quad \langle\psi| \sum_k \Big(b_k + c_k + b_k \cdot c_k\Big) |\psi\rangle\,,$$

subject to
- $\|\psi\| = 1\,,$
- $b_i^* = b_i$ • $c_i^* = c_i$ • $b_i^2 = c_i^2 = \mathbb{I}_D \quad \forall i\,,$
- $b_i c_i = c_i b_i \quad \forall i\,,$
- $b_i c_j = -c_j b_i \quad \forall i \neq j\,.$

By our conjecture, the value of the optimal bias should be $K + 2\sqrt{K}$. We consider the first level of the NPA hierarchy on this algebra [NPA08]. To do so, we write $|u_i\rangle = b_i|\psi\rangle$, and $|v_i\rangle = c_i|\psi\rangle$, and dilate the parameter spaces so that the only relations on these vectors are those that follow directly from the relations on the operators. For example, we have $\langle u_i | v_j \rangle = \langle \psi | b_i c_j \rangle \psi = -\langle \psi | c_j b_i \rangle \psi = -\langle v_j | u_i \rangle$ for all $i \neq j$. We have the level-1 of NPA optimization:

$$\begin{aligned} \sup_{\psi,u,v} \quad & \sum_i \Big( \langle \psi | u_i \rangle + \langle \psi | v_i \rangle + \langle u_i | v_i \rangle \Big)\,, \\ \text{subject to} \quad & \cdot\ \langle \psi | \psi \rangle = \langle u_i | u_i \rangle = \langle v_i | v_i \rangle = 1 \quad \forall\, i\,, \\ & \cdot\ \langle \psi | u_i \rangle = \langle u_i | \psi \rangle \qquad \cdot\ \langle \psi | v_i \rangle = \langle v_i | \psi \rangle \quad \forall\, i\,, \\ & \cdot\ \langle u_i | v_i \rangle = \langle v_i | u_i \rangle \quad \forall\, i\,, \\ & \cdot\ \langle u_i | v_j \rangle = -\langle v_j | u_i \rangle \quad \forall\, i \neq j\,. \end{aligned}$$

It is useful to express the value of the first level using the Gram matrix $G$ of the vectors,

$$\Big\{ |\psi\rangle, |u_1\rangle, \ldots, |u_K\rangle, |v_1\rangle, \ldots, |v_K\rangle \Big\}\,.$$

The first constraint on the optimization gives that diagonal elements of $G$ are all $1$. Also, since $G$ is positive semidefinite, the second and third constraints specify elements of $G$ that are real; and the fourth constraint specifies elements of $G$ that are imaginary. Taking

$$H = \frac{1}{2} \begin{pmatrix} \mathbf{0} & \langle 1_K | & \langle 1_K | \\ |1_K\rangle & \mathbf{0} & \mathbb{I} \\ |1_K\rangle & \mathbb{I} & \mathbf{0} \end{pmatrix}\,,$$

where $|1_K\rangle \in \mathcal{C}^K$ is the column vector of ones, the optimization becomes

$$\begin{aligned} \sup_{G} \quad & \mathrm{Tr}(HG)\,, \\ \text{subject to} \quad & \cdot\ G(\psi,\psi) = G(u_i,u_i) = G(v_i,v_i) = 1 \quad \forall\, i\,, \\ & \cdot\ G(\psi,u_i), G(\psi,v_i), G(u_i,v_i) \in \mathbb{R} \quad \forall\, i\,, \\ & \cdot\ G(u_i,v_j) \in i\mathbb{R} \quad \forall\, i \neq j\,, \\ & \cdot\ G \succcurlyeq \mathbf{0}\,. \end{aligned} \tag{8.16}$$

**Lemma 8.14 —** *The optimal value of the SDP eq. (8.16) is equal to the*

*value of the optimization*

$$\begin{aligned} \sup_{g} \quad & 2Kg_1 + Kg_3 \,, \\ \textit{subject to} \quad & \cdot \begin{pmatrix} 1 & g_1\langle 1_K| & g_1\langle 1_K| \\ g_1|1_K\rangle & \mathbb{I} + g_2(|1_K\rangle\langle 1_K| - \mathbb{I}) & g_3\mathbb{I} \\ g_1|1_K\rangle & g_3 I & \mathbb{I} + g_2(|1_K\rangle\langle 1_K| - \mathbb{I}) \end{pmatrix} \succcurlyeq \mathbf{0} \,, \\ & \cdot \; g_1, g_2, g_3 \in \mathbb{R} \,. \end{aligned}$$

*Proof.* The simplification exploits the symmetries of $H$, which can be used to induce symmetries on $G$. First, note that $H$ has real components, and therefore $G$ can be assumed to be real, yielding the optimization

$$\begin{aligned} \sup_{G} \quad & \mathrm{Tr}(HG) \,, \\ \text{subject to} \quad & \cdot \; G(\psi,\psi) = G(u_i,u_i) = G(v_i,v_i) = 1 \,, \\ & \cdot \; G(u_i,v_j) = 0 \quad \forall\, i \neq j \,, \\ & \cdot \; G \succcurlyeq \mathbf{0} \,. \end{aligned}$$

Next, for any permutation $\sigma \in S_K$, $H$ is invariant under the permutation of the indices $u_i \mapsto u_{\sigma(i)}$, $v_j \mapsto v_{\sigma(j)}$. $H$ is also invariant under the permutation $u_i \mapsto v_i$, $v_i \mapsto u_i$. Thus, $G$ can be supposed to be invariant under those permutations, so we can add the following constraints to the optimization: $G(\psi,u_i) = G(\psi,u_j) = G(\psi,v_i) = G(\psi,v_j)$ and $G(u_i,v_i) = G(u_j,v_j)$ for all $i,j$, and $G(u_i,u_j) = G(u_k,u_l) = G(v_i,v_j) = G(v_k,v_l)$ for all $i \neq j$, $k \neq l$. These additional constraints take the SDP to the form given in the statement. ■

By the previous lemma, we can find the optimal value by first finding the set of feasible points $(g_1, g_2, g_3)$ explicitly, and then finding the optimum of the objective function $2\,K\,g_1 + K\,g_3$ over that set.

**Lemma 8.15 —** *Let $A = \sum_i a_i|i\rangle\langle i|$ and $B = \sum_i b_i|i\rangle\langle i|$ be commuting Hermitian matrices. Then we have the eigendecomposition*

$$\begin{pmatrix} A & B \\ B & A \end{pmatrix} = \sum_i \Big( (a_i + b_i)|i+\rangle\langle i+| + (a_i - b_i)|i-\rangle\langle i-| \Big) \,,$$

*where $|i+\rangle = \frac{1}{\sqrt{2}}\begin{pmatrix} |i\rangle \\ |i\rangle \end{pmatrix}$ and $|i-\rangle = \frac{1}{\sqrt{2}}\begin{pmatrix} |i\rangle \\ -|i\rangle \end{pmatrix}$.*

*Proof.* We can express $\left(\begin{smallmatrix} A & B \\ B & A \end{smallmatrix}\right) = \sum_i \left(\begin{smallmatrix} a_i & b_i \\ b_i & a_i \end{smallmatrix}\right) \otimes |i\rangle\langle i|$. Now, $\left(\begin{smallmatrix} a_i & b_i \\ b_i & a_i \end{smallmatrix}\right) = (a_i + b_i)\,|+\rangle\langle+| + (a_i - b_i)\,|-\rangle\langle-|$, giving the result. ■

**Lemma 8.16 —** *Let $H = \sum_i \lambda_i |v_i\rangle\langle v_i|$ be a Hermitian matrix. Then the Hermitian matrix $\begin{pmatrix} 1 & \omega\langle v_1| \\ \omega|v_1\rangle & H \end{pmatrix}$ has eigenvalues $\lambda_i$ for $i > 1$, and $\frac{1+\lambda_1}{2} \pm \sqrt{\left(\frac{1-\lambda_1}{2}\right)^2 + \omega^2}$.*

*Proof.* Noting that $\left(\begin{smallmatrix} 0 \\ |v_i\rangle \end{smallmatrix}\right)$ is an eigenvector with eigenvalue $\lambda_i$ for all $i > 1$, the remaining two eigenvalues are the eigenvalues of $\left(\begin{smallmatrix} 1 & \omega \\ \omega & \lambda_1 \end{smallmatrix}\right)$, which are of the above form. ■

**Theorem 8.17 —** *The value of the first level of the NPA hierarchy for the Clifford unclonable encryption MoE game is $\frac{1}{2} + \frac{1}{2\sqrt{K}}$ for $K \leqslant 7$, and $\frac{5}{8} + \frac{1}{2(K-2)} - \frac{1}{4K}$ for $K > 7$. In particular, we obtain the upper bound of $5/8 = 0.625$ on the winning probability of the no-cloning game in the limit $K \to \infty$.*

*Proof.* First, we want to find the feasible points of the optimization in Lemma 8.14 by calculating the eigenvalues of the matrices $G$ of that form. Using Lemma 8.15, we see that the eigenvalues of

$$\begin{pmatrix} \mathbb{I} + g_2\left(|1_K\rangle\langle 1_K| - \mathbb{I}\right) & g_3\,\mathbb{I} \\ g_3\,I & \mathbb{I} + g_2(|1_K\rangle\langle 1_K| - \mathbb{I}) \end{pmatrix}$$

are $Kg_2 + (1 - g_2) + g_3$ with eigenvector $\frac{1}{\sqrt{2K}}|1_{2K}\rangle$, and $Kg_2 + (1 - g_2) - g_3$, $1 - g_2 + g_3$, $1 - g_2 - g_3$ with eigenvectors orthogonal to $\frac{1}{\sqrt{2K}}|1_{2K}\rangle$. Using Lemma 8.16, the eigenvalues of $G$ are $Kg_2+(1-g_2)-g_3$, $1-g_2+g_3$, $1-g_2-g_3$ and $\frac{1}{2}(1+Kg_2+(1\text{-}g_2)+g_3)\pm\sqrt{\frac{1}{4}(1\text{-}(Kg_2 + (1\text{-}g_2) + g_3))^2 + 2Kg_1^2}$. For this to be a feasible point of the SDP, all of these eigenvalues must be positive. This means that $1+(K-1)g_2 \geqslant g_3$, $1-g_2 \geqslant \pm g_3$, and $1+(K-1)g_2+g_3 \geqslant 2Kg_1^2$.

To find the optimal point, consider a new parametrisation $x = 1-g_2+g_3$, $y = 1-g_2-g_3$, $\lambda = 2g_1+g_3$. Then, $g_1 = \lambda/2-(x-y)/4$, $g_2 = 1-(x+y)/2$, $g_3 = (x-y)/2$ so the objective function becomes $K\lambda$ and the constraints become $x, y \geqslant 0$, $2 \geqslant x + \frac{K\text{-}2}{K}\,y$, and $\lambda^2-(x-y)\lambda+\frac{(x\text{-}y)^2}{4}+\frac{K\text{-}2}{K}\,x+y-2 \leqslant 0$. As $\lambda$ is to be maximised, we get

$$\lambda = \frac{x\text{-}y}{2} + \sqrt{2 - \frac{K\text{-}2}{K}\,x - y}\,.$$

Since $\lambda$ decreases with $y$, it is maximised at $y = 0$, giving $x \in [0,2]$ and $\lambda = \frac{x}{2} + \sqrt{2 - \frac{K\text{-}2}{K}\,x}$. We have

$$\frac{d\lambda}{dx} = \frac{1}{2} - \frac{K\text{-}2}{2K\sqrt{2 - \frac{K\text{-}2}{K}\,x}}\ ,$$

which is $0$ if $x = \frac{2K}{K\text{-}2} - \frac{K\text{-}2}{K}$ and positive for smaller $x$. If $K \leqslant 7$, then this value of $x$ is greater than $2$, so $\lambda$ is maximised at $x = 2$, giving $\lambda = 1 + \frac{2}{\sqrt{K}}$, and hence winning probability

$$w_1 = \frac{1}{4} + \frac{\lambda}{4} = \frac{1}{2} + \frac{1}{2\sqrt{K}}\ .$$

If $K \geqslant 8$, the maximum value is attained in the interval $[0,2]$ so the optimum $\lambda = \frac{K}{K\text{-}2} + \frac{K\text{-}2}{2K}$, and therefore the winning probability

$$w_1 = \frac{5}{8} + \frac{1}{2(K-2)} - \frac{1}{4K}\ . \qquad \blacksquare$$

### *8.3.3 Numerical Results with NPA Level-2 for* $K \leqslant 17$

To get a better upper bound on the game, we make use of a higher level of the NPA hierarchy. Here, we use level $2$ of the NPA hierarchy, where we optimize over Gram matrices indexed by quadratic terms in the generators of $\mathcal{A}_{ac}(K)$. Even using the symmetries of the problem, as in Lemma 8.14, this becomes a vastly more difficult optimization and requires us to do it numerically. However, we observe the same behavior as for NPA level 1: the optimal value of the SDP matches the conjectured value exactly, until a certain $K$, where it starts to diverge towards a higher limit than $1/2$. Here, the point of divergence is $K = 18$. We give some of the optimal values for the NPA level 1 and 2 optimizations in the table below, highlighting the points of divergence: Note also that the NPA level 2 upper bound of $w_2 = 0.5980$ for $K = 35$ is, to the best of our knowledge, the best known unconditional upper bound on the security of an unclonable encryption scheme.

To finish this section, we outline the construction of the SDP that allows us to find the NPA level 2 optimal values more efficiently. In particular, it allows us to reduce the number of free parameters from $1 + 3K^2$ to $18$.

| K | NPA level 1 | NPA level 2 | Conjecture |
|---|---|---|---|
| 2 | 08536 | 0.8536 | 0.8536 |
| 4 | 0.7500 | 0.7500 | 0.7500 |
| 7 | 0.6890 | 0.6890 | 0.6890 |
| 8 | 0.6771 | 0.6768 | 0.6768 |
| 12 | 0.6542 | 0.6443 | 0.6443 |
| 16 | 0.6451 | 0.6250 | 0.6250 |
| 17 | 0.6436 | 0.6213 | 0.6213 |
| 18 | 0.6424 | 0.6182 | 0.6179 |
| 25 | 0.6367 | 0.6062 | 0.6000 |
| 35 | 0.6330 | 0.5980 | 0.5845 |

**Figure 8.2 —** *Comparing the levels* $1$ *and* $2$ *of the NPA hierarchy and our conjecture for different values of* $K$.

First, the second level of the NPA hierarchy can be derived similarly to the first by considering the optimization of the value over vectors $|u_i\rangle = b_i|\psi\rangle$, $|v_i\rangle = c_i|\psi\rangle$, $|u_iv_j\rangle = b_ic_j|\psi\rangle$ for all $i, j$, and $|u_iu_j\rangle = b_ib_j|\psi\rangle$, $|v_iv_j\rangle = c_ic_j|\psi\rangle$ for $i \neq j$ (noting that $|u_iu_i\rangle = |v_iv_i\rangle = |\psi\rangle$), and taking the relations on the vectors to be those that follow immediately from the relations on the operators. As for the first level, we may assume that the Gram matrix is real and that it is invariant under the permutations of the indices $i \mapsto \sigma(i)$ and $u_i \leftrightarrow v_i$ as in Lemma 8.14. Then, this gives us that that the optimization is over $18$ independent parameters $g_1, \dots, g_{18} \in \mathbb{R}$, where the inner products of the vectors satisfy the conditions of Figure 8.3 for all distinct $1 \leqslant i, j, k, l \leqslant K$. Then, the SDP simplifies to the form:

$$\begin{aligned} \sup_{g} \quad & 2\,K\,g_1 + K\,g_2\,, \\ \text{subject to} \quad & \cdot\; G_0 + g_1\,G_1 + \ldots + g_{18}\,G_{18} \succcurlyeq 0\,, \\ & \cdot\; g_1, g_2, \ldots, g_{18} \in \mathbb{R}\,, \end{aligned}$$

where $G_0$ is the $1+3K^2$-dimensional matrix that has components $1$ where the Gram matrix $G$ is $1$ and zeros elsewhere, $G_1$ is the matrix that has components $\pm 1$ where $G$ is $\pm g_1$, $G_2$ is the matrix that has components $\pm 1$ where $G$ is $\pm g_2$, and so on. We solve this SDP numerically to find the winning probabilities displayed in the second column of Figure 8.2.

$$
\begin{aligned}
1 &= \langle\psi|\psi\rangle = \langle u_i|u_i\rangle = \langle v_i|v_i\rangle = \langle u_iv_i|u_iv_i\rangle = \langle u_iv_j|u_iv_j\rangle \\
&= \langle u_iu_j|u_iu_j\rangle = \langle v_iv_j|v_iv_j\rangle \\
0 &= \langle u_i|v_j\rangle = \langle\psi|u_iv_j\rangle = \langle u_iv_i|u_iu_j\rangle = \langle u_iv_i|u_ju_i\rangle = \langle u_iv_j|u_iu_k\rangle \\
&= \langle u_iv_j|u_ku_i\rangle = \langle u_iv_j|u_ku_l\rangle = \langle u_iv_i|v_iv_j\rangle = \langle u_iv_i|v_jv_i\rangle = \langle u_iv_j|v_kv_j\rangle \\
&= \langle u_iv_j|v_jv_k\rangle = \langle u_iv_j|v_kv_l\rangle = \langle u_iu_j|v_kv_i\rangle = \langle u_iu_j|v_jv_k\rangle \\
g_1 &= \langle\psi|u_i\rangle = \langle\psi|v_i\rangle = \langle u_i|u_iv_i\rangle = \langle v_i|u_iv_i\rangle = \langle u_i|u_iv_j\rangle = -\langle v_i|u_jv_i\rangle \\
&= \langle u_i|u_iu_j\rangle = \langle v_i|v_iv_j\rangle \\
g_2 &= \langle u_i|v_i\rangle = \langle\psi|u_iv_i\rangle = \langle u_iv_j|u_iu_j\rangle = -\langle u_iv_j|v_jv_i\rangle \\
g_3 &= \langle u_i|u_j\rangle = \langle v_iv_j| =\rangle\langle u_iv_i|u_iv_j\rangle = -\langle u_iv_i|u_jv_i\rangle = \langle\psi|u_iu_j\rangle \\
&= \langle\psi|v_iv_j\rangle = \langle u_iv_j|u_iv_k\rangle = \langle u_iv_j|u_kv_j\rangle = \langle u_iu_j|u_iu_k\rangle = \langle v_iv_j|v_iv_k\rangle \\
g_4 &= \langle u_i|u_jv_j\rangle = -\langle u_i|u_jv_i\rangle = \langle v_i|u_jv_j\rangle = \langle v_i|u_iv_j\rangle = \langle v_i|u_iu_j\rangle \\
&= -\langle v_i|u_ju_i\rangle = \langle u_i|v_iv_j\rangle = -\langle u_i|v_jv_i\rangle \\
g_5 &= \langle u_i|u_jv_k\rangle = -\langle v_i|u_jv_k\rangle = \langle v_i|u_ju_k\rangle = \langle u_i|v_jv_k\rangle \\
g_6 &= \langle u_iv_i|u_jv_j\rangle = -\langle u_iu_j|v_jv_i\rangle \\
g_7 &= \langle u_iv_j|u_jv_i\rangle = -\langle u_iu_j|v_iv_j\rangle \\
g_8 &= \langle u_iv_j|u_kv_i\rangle = -\langle u_iu_j|v_iv_k\rangle = \langle u_iu_j|u_kv_k\rangle \\
g_9 &= \langle u_iv_j|u_kv_l\rangle = \langle u_iu_j|v_kv_l\rangle \\
g_{10} &= \langle u_i|u_ju_i\rangle = \langle v_i|v_jv_i\rangle \\
g_{11} &= \langle u_i|u_ju_k\rangle = \langle v_i|v_jv_k\rangle \\
g_{12} &= \langle u_iv_j|u_ju_i\rangle = -\langle u_iv_j|v_iv_j\rangle \\
g_{13} &= \langle u_iv_i|u_ju_k\rangle = \langle u_iv_j|u_ku_j\rangle = \langle u_iv_i|v_jv_k\rangle = -\langle u_iv_j|v_kv_i\rangle \\
g_{14} &= \langle u_iv_j|u_ju_k\rangle = -\langle u_iv_j|v_iv_k\rangle \\
g_{15} &= \langle u_iu_j|u_ju_i\rangle = \langle v_iv_j|v_jv_i\rangle \\
g_{16} &= \langle u_iu_j|u_ku_i\rangle = \langle v_iv_j|v_kv_i\rangle \\
g_{17} &= \langle u_iu_j|u_ku_j\rangle = \langle v_iv_j|v_kv_j\rangle \\
g_{18} &= \langle u_iu_j|u_ku_l\rangle = \langle v_iv_j|v_kv_l\rangle
\end{aligned}
$$

**Figure 8.3 —** *Constraints in the level* 2 *of NPA hierarchy.*

### 8.3.4 *Heuristic Numerical Results for* $K \leqslant 18$

We present in this subsection a heuristic algorithm for optimizing the largest eigenvalue of the operator:

$$W_K = \sum_{k=1}^{K} \Big( \Gamma_k \otimes B_k \otimes \mathbb{I} + \Gamma_k \otimes \mathbb{I} \otimes C_k + \mathbb{I} \otimes B_k \otimes C_k \Big) ,$$

over self-adjoint contractions $B_1, C_1, \ldots, B_K, C_K$. We shall use the *alternating optimization* (or seesaw) method [BH02]. Note that our goal is to maximize a convex function over a convex set, that is:

$$\max_{\substack{\|z\|=1 \\ \|B_k\|_{\text{op}} \leqslant 1 \\ \|C_k\|_{\text{op}} \leqslant 1}} \langle z | W_K | z \rangle ,$$

hence our problem does not fall in the classical convex optimization framework.

Our algorithm optimizes iteratively over each of the variables $z$, $\{B_k\}_{k=1}^K$, and $\{C_k\}_{k=1}^K$ for a pre-determined number of iterations $M$. The variables $z, B_k, C_k$ are initialized with random values ($z$ uniform on the unit sphere of $\mathbb{C}^d \otimes \mathbb{C}^D \otimes \mathbb{C}^D$, and $B_k$, $C_k$ i.i.d. with Haar-distributed eigenvectors and half of eigenvalues $\pm 1$). Here are the details for of optimization step:

(1) Optimizing $z$: compute the largest eigenvalue of the current operator $P$ and assign to $z$ the corresponding eigenvector.

(2) Optimizing the contractions $B$: permute the first two tensors to rewrite the problem as:

$$\langle z | \sum_{k=1}^{K} \Gamma_k \otimes \mathbb{I} \otimes C_k | z \rangle + \max_{\|B_k\|_{\text{op}} \leqslant 1} \langle z_{1\leftrightarrow 2} | \sum_{k=1}^{K} B_k \otimes (\Gamma_k \otimes \mathbb{I} + \mathbb{I} \otimes C_k) | z_{1\leftrightarrow 2} \rangle ,$$

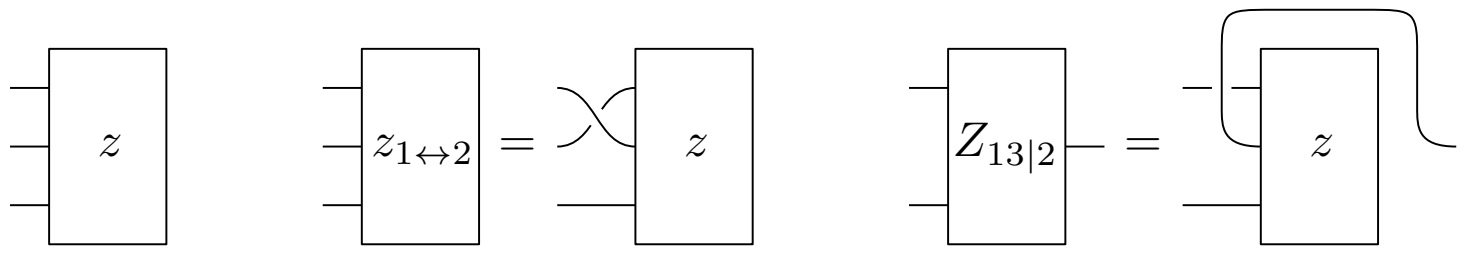

**Figure 8.4 —** *Graphical representations of the tensors* $z$*,* $z_{1\leftrightarrow 2}$*, and* $Z_{13|2}$*.*

where $z_{1\leftrightarrow 2}$ is the 3-tensor $z$ with the first two factors permuted. Clearly, the maximum above is equal to the sum of maxima over individual $B_k$'s that we can further decompose as:

$$\begin{aligned} &\max_{\|B_k\|_{\text{op}}\leqslant 1} \langle z_{1\leftrightarrow 2}|B_k \otimes (\Gamma_k \otimes \mathbb{I} + \mathbb{I} \otimes C_k)|z_{1\leftrightarrow 2}\rangle \\ =&\max_{\|B_k\|_{\text{op}}\leqslant 1} \langle B_k, Z_{13|2}^*(\Gamma_k \otimes \mathbb{I} + \mathbb{I} \otimes C_k) Z_{13|2}\rangle \\ =&\ \|Z_{13|2}^*(\Gamma_k \otimes \mathbb{I} + \mathbb{I} \otimes C_k) Z_{13|2}\|_1 \,, \end{aligned}$$

where $Z_{13|2} \in \mathcal{M}_{Dd\times D}$ is the reshaping of the 3-tensor $z_{1\leftrightarrow 2}$, see Figure 8.4. In the last equality above we have used the following fact:

$$\max_{\|B\|_{\text{op}}\leqslant 1} \langle B, \sigma_x\rangle = \|\sigma_x\|_1 \,,$$

with the maximum being attained for:

$$B_{\text{opt}} = \sum_i \text{sign}(\lambda_i) |x_i\rangle\langle x_i| \qquad \text{for} \qquad \sigma_x = \sum_i \lambda_i |x_i\rangle\langle x_i| \,.$$

We apply this procedure for all the maximization problems corresponding to the $B_k$'s and update the matrices $B_k$ accordingly.

(3) Optimizing the contractions $C$: similar procedure as above, up to tensor permutation.

We present in Figure 8.5 the results of numerical experiments in the cases $K = 3$ and $K = 18$ for different matrix sizes $D$. The results we find agree with Conjecture 8.5.

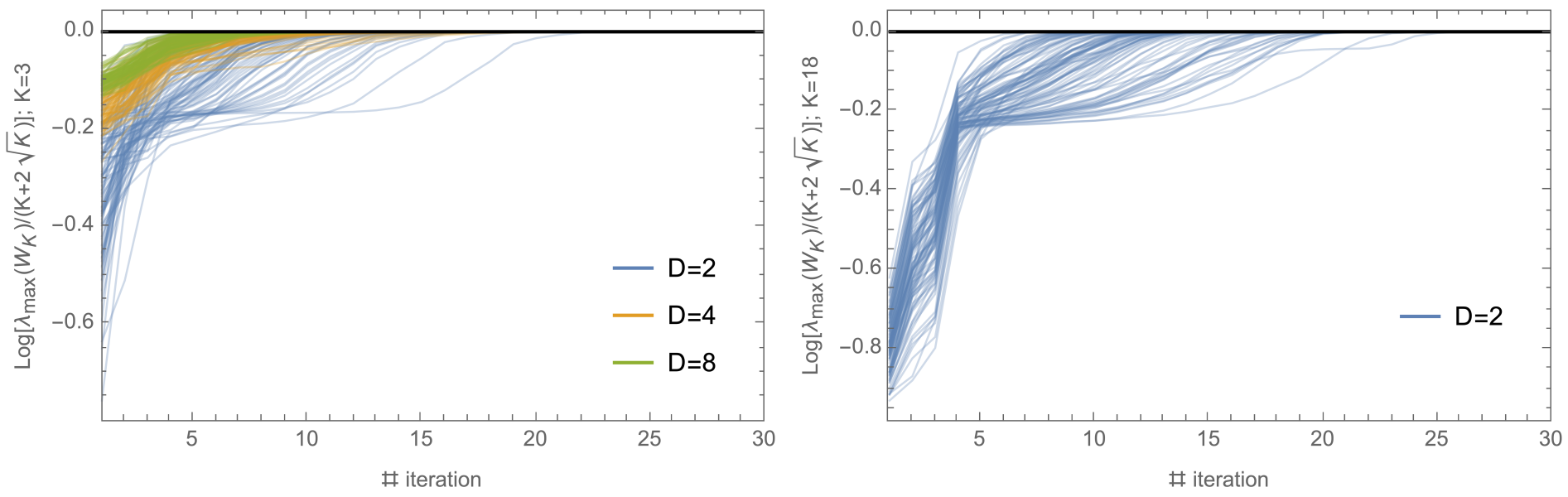

**Figure 8.5 —** *Optimizing the maximum eigenvalue of the operator $W_K$ using a see-saw algorithm. Left panel: $K = 3$; right panel: $K = 18$. We consider different matrix sizes $D$ and we run $100$ instances of the algorithm (with random initializations) for each dimension, for $M = 10$ iterations of the three steps. The $x$ axis tracks the intermediate step of the algorithm, ranging from $1$ to $3M = 30$. The $y$ axis tracks the log-relative error $\log[\lambda_{\max}(W_K)/(K + 2\sqrt{K})]$. Note that all the curves are below $0$, providing evidence towards Conjecture 8.5.*

# Chapter 9

# Conclusion

We conclude this thesis by discussing our contributions and related open questions.

## Chapter Contents

## 9.1 Discussion and Perspectives

In this section, we present some perspectives regarding our four contributions [BBP24; Bot+24a; Bot+24b; BW24].

We begin with the collapse of communication complexity in the CHSH game (Section 9.1.1), then the collapse in graph games (Section 9.1.2), and finally the unclonable bit problem (Section 9.1.3).

### 9.1.1 Collapse of CC in the CHSH Game

In [BBP24; Bot+24a] (Chapter 6), we presented ways to obtain new nonlocal boxes that collapse communication complexity. The strength of these results is emphasized by considering the many known impossibility results (Section 4.3). Furthermore, our new algebraic perspective on nonlocal boxes allowed us to discover a surprising structure of what we called the *orbit of a box*, with some strong alignment and parallelism properties (Figure 6.7).

However, there is still a gap to be filled, notably in terms of CHSH winning probability. On the one hand, Tsirelson's bound [Tsi80] (eq. (3.12)) implies that quantum strategies cannot outperform the following value:

$$\cos^2\left(\frac{\pi}{8}\right) = \frac{1}{2} + \frac{1}{\sqrt{8}} \approx 85\% \,.$$

On the other hand, from Brassard, Buhrman, Linden, Méthot, Tapp, and Unger [Bra+06], it is known that there is a collapse of CC for winning probabilities greater than:

$$\frac{3+\sqrt{6}}{6} = \frac{1}{2} + \frac{1}{\sqrt{6}} \approx 91\% \,.$$

Nevertheless, between the values $85\%$ and $91\%$, there is an open gap for which we do not know yet whether there is a collapse of CC (Open Question 9.1).

### 9.1.2 Collapse of CC in Graph Games

We introduced in [BW24] (Chapter 7) methods to collapse communication complexity in various graph games, leading to the first connection of this kind between these two notions. More precisely, we found several sufficient criteria for a graph game to collapse CC. Moreover, we characterized the perfect classical/quantum/non-signaling strategies for the $D$-distance game. As a consequence of these three characterizations, we showed the surprising fact that only non-signaling strategies display a difference with perfect strategies for the celebrated graph isomorphism game. Therefore, the non-signaling set yields a finer distinction of these nonlocal games than quantum and classical sets.

It would be interesting to extend this study and to prove that any two graphs $(\mathcal{G}, \mathcal{H})$ that admit a perfect non-signaling strategy, but no perfect quantum strategy, necessarily collapse communication complexity (Open Question 9.7), like in the CHSH game with the **PR** box.

### 9.1.3 *Unclonable Bit Problem*

In [Bot+24b] (Chapter 8), we proposed a candidate scheme in working towards an unconditional solution for the (weak) unclonable bit problem, and we proved the unclonable-indistinguishable security for small key sizes. Moreover, as suggested by the numerical evidence, we believe that this result should hold for any key size. Nevertheless, even if the proof were extended to any key size, we stress that our protocol only achieves weak security (Equation (8.9)) and weak indistinguishability (Section 8.2.5). Thus, the avenue is still wide open for finding an unconditional encryption protocol that is both strongly indistinguishable and strongly secure (Open Question 9.9).

Moreover, note that we reduced the cryptographic problem to solving an operator algebra inequality (Open Question 9.10). But since we used several upper bounds to derive this sufficient inequality, this reduction might not be tight. Therefore, it may be that, with the same encryption scheme based on the Clifford algebra, one could find more clever upper bounds in computations, thus leading to an improvement on the security strength.

## 9.2 Related Open Questions

We leave here a list of open questions related to this thesis, that we find both interesting and challenging.

The topics are organized as follows: physical principles (Section 9.2.1), nonlocal boxes (Section 9.2.2), nonlocal games (Section 9.2.3), graph theory (Section 9.2.4), quantum cryptography (Section 9.2.5), and operator algebra (Section 9.2.6).

### 9.2.1 On Physical Principles

One of the main questions of this thesis is the following one, remaining open to this day:

> **Open Question 9.1 —** *What are all nonlocal boxes* $\mathbf{P} \in \mathcal{NS}$ *that collapse communication complexity?*

As illustrated in Figure 4.2, there is some progress in this question, but the gap still needs to be filled. Notably, considering all the known limiting results (Section 4.3), an improvement of the collapsing threshold on the vertical line of the above-mentioned figure (*i.e.* on isotropic boxes $\alpha\,\mathbf{PR} + (1-\alpha)\,\mathbf{I}$) would be extremely significant.

We presented several examples of other principles in Section 4.4. It raises the following question:

> **Open Question 9.2 —** *Is there an information-based principle that perfectly characterizes the set of quantum correlations* $\mathcal{Q}$*?*

To the best of our knowledge, none of them perfectly characterizes the quantum set $\mathcal{Q}$ to this day. The nearest principle to achieve this breakthrough might surely be *information causality* as it already characterizes quantum correlations in some slices of $\mathcal{NS}$ (Section 4.4.1).

### 9.2.2 On Nonlocal Boxes

In the usual definition of communication complexity, we are not concerned about the number of nonlocal boxes used in Alice's and Bob's protocol (Remark 4.6). For instance, van Dam's protocol (Theorem 4.20) requires $2^n$ copies of the **PR** to distributively compute a function with entries size $n$. What happens now if, like in the real world, the shared resource is limited? This question was asked to us by Andreas Bluhm in personal communication:

**Open Question 9.3 —** *Given a finite number of* **PR** *boxes (such that each of them can be used only once), what are all Boolean functions* $f : \{0,1\}^n \times \{0,1\}^m \to \{0,1\}$ *that can be distributively computed with only one bit of communication?*

We refer to [Kap+11] for some ideas in this direction.

Many collapsing results rely on box distillation through wirings [Bot+24a; Bri+19; BS09; EWC23a] (wirings were defined in Section 3.1.4). Moreover, in [BG15], Beigi and Gohari introduced a measure of boxes $\mu_{\text{box}} : \mathcal{NS} \to [0,1]$ (page 86) that is remarkably decreasing under wirings, implying that its sublevel sets $\{\mathbf{P} \in \mathcal{NS} : \mu_{\text{box}}(\mathbf{P}) \leqslant x\}$ are closed under wirings (Definition 3.16) for any $x \in [0,1]$. As a consequence, no box from such a sublevel set can be distilled out of it, thus imposing strong limitations on distillation methods. Nevertheless, one drawback of this measure is that it does not permit the distinction of some local boxes from the **PR** box, for instance $\mu_{\text{box}}(\mathbf{SR}) = \mu_{\text{box}}(\mathbf{PR}) = 1$. So, we ask:

**Open Question 9.4 —** *Is there another measure* $\mu'_{\text{box}} : \mathcal{NS} \to [0,1]$ *that is, as* $\mu_{\text{box}}$*, semi-continuous, efficiently computable, and decreasing under wirings, but also that vanishes on classical boxes* $\mathcal{L}$ *and takes value* $1$ *precisely on the* **PR** *box (unique up to symmetry)?*

Such a measure $\mu'_{\text{box}}$ could give rise to interesting sets closed under wirings via the sublevel sets. (Note that a similar question was raised by Beigi in [Bei13] in order to study the problem of entanglement distillation under LOCC maps. This question was addressed by themselves in [Bei14] with the quasi-convexification $\nu_{\text{quant}}$ of $\mu_{\text{quant}}$ (Section 3.1.5), vanishing exactly on separable states, albeit not monotone under classical communication.) A natural guess for the above question could be to define the quasi-convexification $\nu_{\mathcal{NS}}$ of $\mu_{\text{box}}$ as it was done for $\nu_{\text{quant}}$ from $\mu_{\text{quant}}$. This measure $\nu_{\mathcal{NS}}$ has the benefit of being upper semi-continuous, decreasing under wirings, and vanishing exactly on $\mathcal{L}$, but has the drawback of being difficult to compute.

In Theorem 6.18, we proved a sufficient criterion in order to have a 2-dimensional convex subset of $\mathcal{NS}$ that is distillable to the **PR** box. Moreover, in [Bri+19], Brito, Moreno, Rai, and Chaves find 3-dimensional sets that are distillable, but not necessarily until the **PR** box. They raise the following question:

**Open Question 9.5** ([Bri+19]) — *Are there* $3$*-dimensional convex subsets of* $\mathcal{NS}$ *that are distillable to the* **PR** *box?*

If so, then such a set collapses communication complexity. Note that, in [Bri+19], the authors also prove that for dimensions $d \geqslant 4$, no quantum void can be distilled due to the presence of isotropic boxes, which are not distillable [BG15].

### 9.2.3 On Nonlocal Games

The collapse of communication complexity is generally studied in the CHSH game scenario. As presented in Chapter 7 [BW24], it is possible to extend it to some graph games. Furthermore, Shutty, Wootters, and Hayden also defined a different game G for which communication complexity perfectly characterizes the quantum best-winning probability [SWH20]. It yields:

**Open Question 9.6** — *Are there other interesting nonlocal games for which one can show the collapse of communication complexity?*

Other examples of nonlocal games are provides in Section 3.2.4.

Moreover, we know that any perfect non-signaling strategy for the CHSH game (*i.e.* exactly the **PR** box) collapses CC. Can we extend this observation to graph games?

**Open Question 9.7** — *Assume that two graphs* $(\mathcal{G}, \mathcal{H})$ *admit a perfect non-signaling strategy for the* $D$*-distance game (Section 7.3.1), but no perfect quantum strategy, that is:*

$$\mathcal{G} \cong^{D}_{\text{ns}} \mathcal{H} \qquad \textit{but} \qquad \mathcal{G} \ncong^{D}_{\text{qc}} \mathcal{H} \,.$$

*Then, do we necessarily collapse communication complexity? Or, relaxing the question, do such graphs* $(\mathcal{G}, \mathcal{H})$ *admit at least one non-signaling strategy that collapses communication complexity?*

### 9.2.4 On Graph Theory

In Figure 3.5, we presented several Lovasz-type characterizations of certain variants of graph isomorphisms in terms of homomorphism counts:

the classical isomorphism is characterized by homomorphism counts from all graphs, the quantum isomorphism by homomorphism counts on planar graphs, and the non-signaling/fractional isomorphism by homomorphism counts on trees. This gives rise to the following question:

> **Open Question 9.8 —** *Is our generalized notion of fractional isomorphism isomorphism $\cong^D_{\text{frac}}$ (Definition 7.33) characterized in terms of Lovász-type homomorphism counts?*

Intuitively, if so, this should be in terms of a class of graphs that lies between planar graphs and trees, with a dependency on the parameter $D \in \mathbb{N}$.

### 9.2.5 On Quantum Cryptography

Another main question of this thesis is the following one, remaining open to this day:

> **Open Question 9.9 —** *Does the unconditional unclonable bit exist? That is, is there a quantum encryption scheme of a bit $m \in \{0, 1\}$ in the plain model that is unclonable-indistinguishable secure in the strong sense (Definition 5.12) and that achieves strong indistinguishability (Section 8.2.5)?*

Related works on these questions are showcased in page 161. We stress that there is a recent positive answer in the weak security regime by Bhattacharyya and Culf [BC25], but the avenue is still wide open for strong security.

### 9.2.6 On Operator Algebra

A question in operator algebra arose from our study of the *unclonable bit* problem in [Bot+24b] (Chapter 8). Though simple in appearance, we only proved it for small $K$ (Section 8.3) and left the general question open:

> **Open Question 9.10 —** *Consider $\Gamma_1, .., \Gamma_K$ to be pairwise anti-commuting Hermitian unitaries of dimension $d$, and $U_1, .., U_K$ Hermitian unitaries of dimension $D$. Is it true that the following upper bound holds:*
>
> $$\left\| \sum_{k=1}^{K} \Big( \Gamma_k \otimes U_k \otimes \mathbb{I}_D + \Gamma_k \otimes \mathbb{I}_D \otimes U_k + \mathbb{I}_d \otimes U_k \otimes U_k \Big) \right\|_{\text{op}} \leqslant K + 2\sqrt{K} \ ?$$

A positive answer to this question implies an unconditional weak security in Open Question 9.9.

# Index

RÉPUBLIQUE FRANÇAISE
*Liberté*
*Égalité*
*Fraternité*

Université de Toulouse

**Titre :** Jeux non-locaux au travers de la complexité de la communication et de la cryptographie quantique

**Résumé :** Cette thèse explore des aspects fondamentaux de la théorie de l'information quantique et de la cryptographie quantique.
D'une part, nous étudions les corrélations quantiques dans des contextes interactifs, notamment les jeux de CHSH et d'isomorphisme de graphes. Notre objectif est de distinguer les corrélations quantiques des corrélations non-signalantes en nous appuyant sur le principe de complexité de la communication.
Pour cela, nous utilisons des techniques telles que le calcul distribué, l'amplification de biais grâce à la fonction majorité, les propriétés algébriques et géométriques des câblages de boîtes non-locales, ainsi que des variantes de certaines propriétés de graphes comme l'isomorphisme, la transitivité et les partitions équitables. Cette étude fait progresser notre compréhension des corrélations non-physiques.
D'autre part, nous abordons un problème ouvert majeur en cryptographie : la faisabilité du chiffrement non-clonable. Notre objectif est de construire un schéma de chiffrement qui empêche deux réceptionneurs distants l'un de l'autre d'obtenir simultanément de l'information sur un message chiffré partagé.
Nous introduisons un candidat au chiffrement non-clonable dans le modèle standard, c'est-à-dire sans hypothèse, en vue d'obtenir une preuve inconditionnelle de la sécurité.
Notre protocole repose sur l'algèbre de Clifford et utilise des matrices unitaires hermitiennes à coefficients complexes qui anti-commutent. Pour des tailles de clés réduites, nous prouvons rigoureusement la sécurité à l'aide de méthodes de sommes de carrés, tandis que pour des tailles de clés plus grandes, nous fournissons des validations numériques solides via la hiérarchie NPA.

**Title:** Nonlocal Games Through Communication Complexity and Quantum Cryptography

**Abstract:** This thesis explores foundational aspects of quantum information theory and quantum cryptography.
First, we investigate quantum correlations in interactive settings, including the CHSH and graph isomorphism games. We aim to distinguish quantum correlations from non-signaling correlations by leveraging the principle of communication complexity. To this end, we employ techniques such as distributed computation, majority-function-based distillation protocols, the algebraic and geometric properties of nonlocal box wirings, and variations of some graph properties such as isomorphism, transitivity, and equitable partitions. This inquiry advances our understanding of non-physical correlations.
Second, we address a key open problem in cryptography: the feasibility of unclonable encryption. We aim to construct an encryption scheme that prevents two distant parties from simultaneously obtaining information about a shared encrypted message.
We introduce a candidate for unclonable encryption in the plain model, i.e., without assumptions, in working towards an unconditional proof.
Our protocol is based on Clifford algebra, utilizing complex Hermitian unitary matrices that anti-commute. For small key sizes, we rigorously prove security using sum-of-squares methods, while for larger key sizes, we provide strong numerical evidence via the NPA hierarchy.